\documentclass[11pt,a4paper,oneside]{book}

\usepackage{abstract}
\usepackage[left=1.5cm,right=1.5cm,top=2.5cm,bottom=2.5cm]{geometry}

\usepackage[numbers,square]{natbib}
\usepackage[T1]{fontenc}
\usepackage{tikz-cd}
\usepackage[english]{babel}
\usepackage{layout}
\usepackage{amsmath}
\usepackage{amssymb,amstext, amsthm, simplewick, amsfonts}
\usepackage{framed}
\usepackage{amsfonts}
\usepackage{color} 
\usepackage{braket}
\usepackage{multicol} 
\usepackage{epsfig}
\usepackage{cancel}
\usepackage{caption}
\usepackage{epsf}
\usepackage{feynmf} 
\usepackage{frontespizio}
\usepackage{rotating}
\usepackage{pstricks}
\usepackage{type1ec}
\usepackage{lettrine}
\usepackage{bbold}
\usepackage{calligra}
\usepackage{tikz}
\usepackage{float}
\usepackage{slashed}
\usepackage{mathrsfs}
\usepackage{curve2e}
\usepackage{setspace}
\usepackage{indentfirst}
\usepackage{emptypage}
\usepackage{relsize}
\usepackage{mathrsfs}
\usepackage{stackengine}
\usepackage{calc}
\usepackage{hyperref}
\hypersetup{
	colorlinks,
	citecolor=black,
	filecolor=black,
	linkcolor=black,
	urlcolor=black
}

\usepackage{multicol}  
\usepackage{mathtools}
\usepackage{float}
\usepackage{booktabs}
\usepackage{array}
\usepackage{tabularx}
\newcolumntype{Y}{>{\centering\arraybackslash}X}
\usepackage{booktabs}
\usepackage{array}
\usepackage{booktabs}
\usepackage{amssymb,amstext, amsthm, simplewick, amsfonts}
\usepackage{longtable}
\usepackage{pdflscape}
\usepackage{lipsum}
\newcommand{\pa}{\partial}

\def\nbox#1#2{\vcenter{\hrule \hbox{\vrule height#2in
			\kern#1in \vrule} \hrule}}

\let\ul=\underline
\newcommand{\3}[1]{C_{
		\ifthenelse{\equal{\ThreePt}{\empty}}{#1}{
			\ifthenelse{\equal{#1}{\empty}}{\ThreePt}{\ThreePt,#1}}}}
\newcommand{\redef}[1]{{C'}_{
		\ifthenelse{\equal{\ThreePt}{\empty}}{#1}{
			\ifthenelse{\equal{#1}{\empty}}{\ThreePt}{\ThreePt,#1}}}}
\newcommand{\ren}[1]{C_{
		\ifthenelse{\equal{\ThreePt}{\empty}}{#1}{
			\ifthenelse{\equal{#1}{\empty}}{\ThreePt}{\ThreePt,#1}}}}
\newcommand{\sd}[1]{D_{
		\ifthenelse{\equal{\ThreePt}{\empty}}{#1}{
			\ifthenelse{\equal{#1}{\empty}}{\ThreePt}{\ThreePt,#1}}}}

\numberwithin{equation}{section} 
\makeatletter
\@addtoreset{equation}{section}
\newcommand{\bea}{\begin{eqnarray}}
	\newcommand{\eea}{\end{eqnarray}}
\makeatother
\newcommand{\beqa}{\begin{eqnarray}}
	\newcommand{\eeqa}{\end{eqnarray}}

\newcommand{\sm}{\mathcal{S}}
\newcommand{\nn}{\nonumber}
\let\a=\alpha   \let\b=\beta   \let\g=\gamma   \let\d=\delta
\let\e=\epsilon \let\z=\zeta   \let\h=\eta     
\let\i=\iota    \let\k=\kappa  \let\l=\lambda  \let\m=\mu
\let\n=\nu           \let\p=\pi      \let\r=\rho
\let\s=\sigma  \let\t=\tau      
\let\c=\chi         \let\w=\omega
\def\pd{\partial}
\def\si{\sigma}
\newcommand{\bann}{\begin{eqnarray*}}
	\newcommand{\eann}{\end{eqnarray*}}
\def\eps{\epsilon}
\newcommand{\del}{\delta}

\newcommand{\bmi}{\begin{minipage}}
	\newcommand{\emi}{\end{minipage}}

\newcommand{\psl}{p \! \! \!  /}
\newcommand{\qsl}{q \! \! \!  /}
\newcommand{\lsl}{l \! \! \!  /}

\newcommand{\mD}{\mathcal{D}}
\let\G=\Gamma  \let\D=\Delta   
         \let\S=\Sigma  
    \let\Y=\Psi

\newcommand{\la}{\langle}
\newcommand{\ra}{\rangle}
\newcommand{\bs}[1]{\boldsymbol{#1}}
\newcommand{\sq}{\square}

\newcommand{\vf}{\varphi}
\newcommand{\be}{\begin{equation}}
	\newcommand{\ee}{\end{equation}}
\newcommand{\sdfrac}[2]{\mbox{\small$\displaystyle\frac{#1}{#2}$}}

\renewcommand{\Re}{\textrm{Re}}
\renewcommand{\Im}{\textrm{Im}}
\newcommand{\beq}{\begin{equation}}
	\newcommand{\eeq}{\end{equation}}
\newcommand{\figref}[1]{Fig.~\ref{#1}}			% for figures
\newcommand{\secref}[1]{Section~\ref{#1}}		% for sections
\newcommand{\Tr}{\text{Tr}}

\newcommand{\Z}{\mathds{Z}}
\newcommand{\ThreePt}{\empty}
\usepackage{accents}
\makeatletter
\newcommand{\xLine}[2][]{\ext@arrow 0359\Rightarrowfill@{#1}{#2}}
\makeatother
\xdefinecolor{darkgreen}{RGB}{102, 204, 70}
\xdefinecolor{darkblue}{RGB}{0, 0, 153}
\usepackage{amssymb}
\usepackage{pifont}
\newcommand{\bes}{\begin{subequations}}
	\newcommand{\ees}{\end{subequations}}
\newcommand{\brak}[1]{\left({#1}\right)}

\newcommand{\na}{\nabla}

\def\sqb{\,\raise.5pt\hbox{$\overline{\nbox{.09}{.09}}$}\,}

\newcommand{\cA}{\mathcal{A}}

\newcommand{\cL}{\mathcal{L}}

\newcommand{\cS}{\mathcal{S}}

\usepackage{tikz-feynman}
\usetikzlibrary{arrows}
\usetikzlibrary{snakes}
\usetikzlibrary{decorations,decorations.pathmorphing,decorations.markings}

\tikzset{graviton/.style={decorate, decoration={snake}, double}}
\tikzset{gluon/.style={decorate, decoration={coil, segment length=8, aspect=1.2, amplitude=3 }}}

\usepackage{comment}
\usepackage{graphicx}
\usepackage{appendix}
\usepackage{braket}
\usepackage{subcaption} 

\usepackage{fancyhdr}

\usepackage{textcase}

\newcommand{\expval}[1]{\langle #1 \rangle}

\usepackage{bibunits}
\defaultbibliography{part1,TJJdilatonHprime} 
\defaultbibliographystyle{jhep}

\newcommand{\KBessel}{%
	K_0\!\left(bQ\sqrt{(\alpha_1-1)(\alpha_2-1)}\right)
	-
	K_0\!\left(bQ\sqrt{\alpha_1\alpha_2 }\right)
}

\newcommand{\Den}{\alpha_1+\alpha_2-1}

\newcommand{\Cz}{%
	C_0\!\left(
	( 1-\alpha_1-\alpha_2+\alpha_1\alpha_2)Q^2,\,
	\alpha_1\alpha_2 Q^2,\,
	Q^2,\,
	0,0,0
	\right)
}

\newcommand{\Lq}{\log\!\left(\frac{\mu^2}{Q^2}\right)}
\newcommand{\La}{\log\!\left(\frac{\mu^2}{\alpha_1\alpha_2 Q^2}\right)}
\newcommand{\Lb}{%
	\log\!\left(
	\frac{\mu^2}{
		(1-\alpha_1-\alpha_2+\alpha_1\alpha_2)Q^2
	}
	\right)
}
\newcommand{\mel}[3]{\bra{#1}{#2}\ket{#3}}

\begin{document}
	
	\frontmatter
	\begin{titlepage}
		%\hspace{-\parindent}
		\begin{figure}[t]
			\centering
			\begin{center}
				\hspace*{-1cm} 
				\includegraphics[scale=0.45]{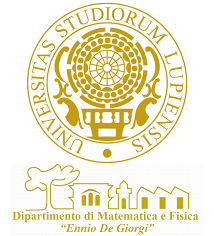} 
			\end{center}
		\end{figure}

		\begin{center}
			
			\begin{minipage}[c]{10.4cm}
				\begin{center}%\small
					\textsc{\LARGE Università del Salento}\\
					\textsc{\large{Dipartimento di Matematica e Fisica \\"Ennio De Giorgi"}}
				\end{center}
			\end{minipage}
		\end{center}
		
		\vspace{2mm}

		\begin{center}
			
			\rule[0.1cm]{16.5cm}{0.1mm}
			\rule[0.5cm]{16.5cm}{0.6mm}
			
			\vspace{0.5cm}
			
			{\textsc{\Large\bfseries Dottorato di Ricerca in Fisica e Nanoscienze}}
			
			\vspace{2.5cm}
			
			{\scshape\fontsize{28}{32}\selectfont
				\textbf{Aspects of Anomalies}
				\par}
			
			\vspace{1.cm}
			
			{\huge
				Fermion Family Constraints,\\[0.25cm]
				Gravitational Form Factors\\[0.25cm]
				and Scattering Amplitudes
				\par}
			
			\vspace{1cm}
			
		\end{center}

		\vspace{25mm}
		%\par
		%\noindent
		
		\textsc{\textbf{Supervisor:}} \hfill \textsc{\textbf{Candidato:\ \ \ \ \ \ }}
		
		\textsc{Prof. Claudio Corianò}  \hfill \textsc{Dario Melle\ \ \ \ \ \ }
		
		%\hfill
		
		%\textbf{Co-Supervisor:}  \hfill 
		
		%Prof. Paul Howard Frampton

		%\vspace{6mm}

		\vspace{15mm}
		
		\begin{center}
			
			\rule[0.1cm]{16.5cm}{0.1mm}
			\rule[0.5cm]{16.5cm}{0.6mm}
			
			{{\large{\bf {Anno Accademico 2025  -- 2026} }}}
		\end{center}
	\end{titlepage}

	%\cleardoublepage
	\thispagestyle{empty}
	
 	\newpage
 	\thispagestyle{empty}
 	\newpage
	\vspace*{\stretch{1}}
	\begin{flushright}
		\itshape ``Nulla ars imitari sollertiam naturae potest"\\
		Cicerone, De Natura Deorum, 1.92
	\end{flushright}
	\vspace{\stretch{2}}
	%\cleardoublepage
	\newpage
	\thispagestyle{empty}
	\newpage
	%\begin{abstract}
		
		%------
		
	%\end{abstract}
	\tableofcontents
	\frontmatter
	
	  \chapter*{List of Publications}
   \markboth{LIST OF PUBLICATIONS}{}
  \addcontentsline{toc}{chapter}{\protect\large List of Publications}
  The Chapters of this thesis are based on the following research papers:
  \begin{itemize}
  	
 \item C.~Corian\`o, H.-N.~Li, D.~Melle and L.~Torcellini,
 \textit{The Gravitational Form Factor of the Pion in Perturbative QCD with a Dilaton Interaction},
 \href{https://arxiv.org/abs/2606.10993}{arXiv:2606.10993 [hep-ph]} (2026).

\item C.~Corian\`o, H.-N.~Li and D.~Melle,
\textit{The Pion Gravitational Form Factors and the Trace Anomaly in QCD Factorization},
\href{https://arxiv.org/abs/2606.03055}{arXiv:2606.03055 [hep-ph]} (2026).

\item C.~Corian\`o, S.~Lionetti, D.~Melle and L.~Torcellini,
\textit{Anomaly-mediated Scalar Gravitational Interactions and the
	Coupling of Conformal Sectors},
\href{https://arxiv.org/abs/2603.28966}{arXiv:2603.28966 [hep-th]} (2026).

\item G.~Corcella, C.~Corian\`o, P.~H.~Frampton and D.~Melle,
\textit{Heavy quark decays in the bilepton model},
\textit{Phys.\ Lett.\ B} \textbf{873} (2026) 140188,
\href{https://doi.org/10.1016/j.physletb.2026.140188}{doi:10.1016/j.physletb.2026.140188},
\href{https://arxiv.org/abs/2507.12556}{arXiv:2507.12556 [hep-ph]}.

\item C.~Corian\`o, S.~Lionetti, D.~Melle and L.~Torcellini,
\textit{A dilaton sum rule for the conformal anomaly form factor
	in QCD at order $\alpha_s$},
\textit{Eur.\ Phys.\ J.\ C} \textbf{85} (2025) no.~9, 983,
\href{https://doi.org/10.1140/epjc/s10052-025-14686-w}{doi:10.1140/epjc/s10052-025-14686-w},
\href{https://arxiv.org/abs/2504.01904}{arXiv:2504.01904 [hep-ph]}.

\item C.~Corian\`o, S.~Lionetti and D.~Melle,
\textit{Topological sum rules and spectral flows of chiral and
	gravitational axionlike interactions},
\textit{Phys.\ Rev.\ D} \textbf{112} (2025) no.~1, 015031,
\href{https://doi.org/10.1103/qd7t-4dhl}{doi:10.1103/qd7t-4dhl},
\href{https://arxiv.org/abs/2502.03182}{arXiv:2502.03182 [hep-ph]}.

\item C.~Corian\`o, P.~H.~Frampton and D.~Melle,
\textit{The 33311 left-right bilepton model},
\textit{Phys.\ Lett.\ B} \textbf{866} (2025) 139498,
\href{https://doi.org/10.1016/j.physletb.2025.139498}{doi:10.1016/j.physletb.2025.139498},
\href{https://arxiv.org/abs/2411.08078}{arXiv:2411.08078 [hep-ph]}.

\item C.~Corian\`o, S.~Lionetti, D.~Melle and R.~Tommasi,
\textit{The gravitational form factors of hadrons from CFT in
	momentum space and the dilaton in perturbative QCD},
\textit{Eur.\ Phys.\ J.\ C} \textbf{85} (2025) no.~5, 498,
\href{https://doi.org/10.1140/epjc/s10052-025-14104-1}{doi:10.1140/epjc/s10052-025-14104-1},
\href{https://arxiv.org/abs/2409.05609}{arXiv:2409.05609 [hep-ph]}.

\item C.~Corian\`o and D.~Melle,
\textit{The $\mathrm{SU}(3)_C \times \mathrm{SU}(3)_L \times
	\mathrm{U}(1)_X$ (331) Model: Addressing the Fermion Families
	Problem within Horizontal Anomalies Cancellation},
\textit{Entropy} \textbf{26} (2024) no.~5, 420,
\href{https://doi.org/10.3390/e26050420}{doi:10.3390/e26050420},
\href{https://arxiv.org/abs/2404.05821}{arXiv:2404.05821 [hep-ph]}.

\end{itemize}
\vspace{0.5cm}

The following papers are related to the topics presented in this thesis but are not discussed in detail:
 \begin{itemize}
 	\item C.~Corian\`o, M.~Cret\`\i, S.~Lionetti, D.~Melle and R.~Tommasi,
 	\textit{Axion-Like Interactions and CFT in Topological Matter,
 		Anomaly Sum Rules and the Faraday Effect},
 	\textit{Adv.\ Phys.\ Res.} \textbf{4} (2025) no.~7, 2400043,
 	\href{https://doi.org/10.1002/apxr.202400043}{doi:10.1002/apxr.202400043},
 	\href{https://arxiv.org/abs/2403.15641}{arXiv:2403.15641 [hep-ph]}.
 	
\item C.~Corian\`o, P.~H.~Frampton, D.~Melle, T.~W.~Kephart and
T.-C.~Yuan,
\textit{An $\mathrm{SU}(15)$ approach to bifermion classification},
\textit{Mod.\ Phys.\ Lett.\ A} \textbf{38} (2023) no.~26n27, 2350124,
\href{https://doi.org/10.1142/S0217732323501249}{doi:10.1142/S0217732323501249},
\href{https://arxiv.org/abs/2301.02425}{arXiv:2301.02425 [hep-ph]}.

\end{itemize}

\section*{Peer-reviewed conference proceedings}
 \begin{itemize}
 	
 	\item C.~Corian\`o, S.~Lionetti, D.~Melle and L.~Torcellini,
 	\textit{Dilaton Sum Rules of Gravitational Form Factors in QCD at
 		Order $\alpha_s$},
 	\href{https://arxiv.org/abs/2603.24303}{arXiv:2603.24303 [hep-ph]} (2026).
 	
 	\item C.~Corian\`o, S.~Lionetti, D.~Melle, L.~Torcellini and R.~Tommasi,
 	\textit{Gravitational Form Factors and the QCD Dilaton at Large
 		Momentum Transfer},
 	\textit{PoS} \textbf{CORFU2024} (2025) 148,
 	\href{https://doi.org/10.22323/1.490.0148}{doi:10.22323/1.490.0148},
 	\href{https://arxiv.org/abs/2504.20884}{arXiv:2504.20884 [hep-ph]}.
 	
\item C.~Corian\`o, S.~Lionetti, D.~Melle and R.~Tommasi,
\textit{The gravitational form factor of the pion and proton and
	the conformal anomaly},
\textit{EPJ Web Conf.} \textbf{314} (2024) 00030,
\href{https://doi.org/10.1051/epjconf/202431400030}{doi:10.1051/epjconf/202431400030},
\href{https://arxiv.org/abs/2409.19586}{arXiv:2409.19586 [hep-ph]}.

\item C.~Corian\`o, S.~Lionetti, D.~Melle, R.~Tommasi and L.~Torcellini,
\textit{Conformal Backreaction, Chiral and Conformal Anomalies in
	the Early Universe},
in \textit{17th Marcel Grossmann Meeting},
\href{https://arxiv.org/abs/2409.18004}{arXiv:2409.18004 [astro-ph.CO]}  (2024).
\end{itemize}

	%\addcontentsline{toc}{chapter}{\protect\large Introduction}
	\chapter*{Introduction}\label{ch:intro}
\markboth{INTRODUCTION}{}
\addcontentsline{toc}{chapter}{\protect\large Introduction}
%=========================================================================

\section*{The Power of Anomalies}\label{sec:power}
\addcontentsline{toc}{section}{The Power of Anomalies}
%=========================================================================

Symmetries occupy a central place in physics, and nowhere more so than in
quantum field theory (QFT). A symmetry of a classical theory is a transformation
of the fields that leaves the action invariant; by Noether's theorem, each
continuous symmetry is associated with a conserved current. A natural and
consequential question is whether such an invariance survives quantisation. It
does not always: a {quantum anomaly} occurs when a symmetry of the classical
action fails to be a symmetry of the quantum theory. In the path-integral
formulation the mechanism is transparent  the action may be invariant under
the transformation while the functional integration measure is not, so that the
symmetry cannot be preserved by any regularisation. 

Anomalies have several properties that give them exceptional theoretical power. The anomaly coefficient is determined solely by the representation content of the massless chiral fields, their gauge quantum numbers and chirality. For a non-Abelian gauge symmetry, the anomaly coefficient is given by the trace over the generator matrices in the fermion representation, $d^{abc} = \operatorname{Tr}\!\left[T^{a}\{T^{b},T^{c}\}\right]$,
whereas for an Abelian \(U(1)\) symmetry the anomaly coefficient reduces simply to a sum over the cubic (or linear in the case of mixed anomalies) charges of the chiral fermions, \(\sum_i q_i^3\). In both cases, the topological origin of the anomaly is stated by the Atiyah--Singer index theorem: the integrated anomaly equals the index of the Dirac operator in a non-trivial gauge background, a topological invariant that depends only on the global structure of the gauge field configuration and not on its local fluctuations.

Perturbatively, the anomaly coefficient is one-loop exact: the Adler--Bardeen theorem establishes that any higher-loop correction to the one-loop result would require a tree-level violation of the symmetry under which the action is, by assumption, invariant, so all such contributions vanish identically to all orders. The form of the anomaly is further constrained by the Wess--Zumino consistency conditions, which restrict the possible counterterms and ensure that the structure of the anomalous Ward identity is fixed independently of the regularisation scheme.

Of all anomalies, the chiral and the conformal ones are among the most studied.
The chiral, or Adler--Bell--Jackiw (ABJ), anomaly manifests as the failure of
the classically conserved axial current $J^{\mu}_{A}=\bar{\psi}\gamma^{\mu}\gamma^{5}\psi$
to remain conserved,
\begin{equation*}
	\nabla_{\mu}\langle J^{\mu}_{A}\rangle
	= a_{1}\,\varepsilon^{\mu\nu\rho\sigma}F_{\mu\nu}F_{\rho\sigma}
	+ a_{2}\,\varepsilon^{\mu\nu\rho\sigma}R^{\alpha}{}_{\beta\mu\nu}R^{\beta}{}_{\alpha\rho\sigma}\,,
\end{equation*}
where $a_{1}$ and $a_{2}$ are numerical coefficients and the two terms encode,
respectively, a gauge and a gravitational contribution. Although discovered in
high-energy physics its
significance extends far beyond. Its implications reach from cosmology to
condensed-matter physics, underlying phenomena such as the quantum Hall
effect, the chiral magnetic effect, and applications to topological insulators.

The conformal, or trace, anomaly is the analogous statement for scale
invariance: the classically traceless energy--momentum tensor of a
scale-invariant theory develops a non-vanishing trace at the quantum level,
\begin{equation*}
	g_{\mu\nu}\langle T^{\mu\nu}\rangle
	= b_{1}E_{4}
	+ b_{2}\,C_{\mu\nu\rho\sigma}C^{\mu\nu\rho\sigma}
	+ b_{3}\,\nabla^{2}R
	+ b_{4}\,F_{\mu\nu}F^{\mu\nu}\,,
\end{equation*}
where $C_{\mu\nu\rho\sigma}$ is the Weyl tensor, $E_{4}$ the Gauss--Bonnet
(Euler) density, and the $b_{i}$ are numerical coefficients directly related to the particle content of the theory. The
$b_{4}F^{2}$ term is proportional to the beta function of the gauge coupling, so
that the trace anomaly directly measures the breaking of scale invariance by
quantum corrections. 

Anomalies enter the formulation of a quantum field theory in two distinct ways,
and this thesis is organised around this distinction.

In the first role, anomalies act as a consistency condition. The gauge
symmetries that are dynamically realized must be free of anomalies, since an
anomalous gauge symmetry spoils the decoupling of unphysical polarisations and
thereby unitarity and renormalisability. The requirement of gauge-anomaly
cancellation is therefore, a constraint on the admissible field content of a
theory. This constraint is highly predictive. In the Standard Model (SM) the
cancellation of gauge anomalies relates the hypercharges of quarks and leptons
and implies the quantisation of electric charge. In ten-dimensional superstring
theory, the Green--Schwarz mechanism restricts the gauge group to $SO(32)$ or
$E_8\times E_8$. A related principle is 't~Hooft anomaly matching: because the
anomaly coefficients of global symmetries are invariant under the
renormalisation-group flow, the values computed from the ultraviolet degrees of
freedom must coincide with those computed from the infrared degrees of freedom.
This provides a non-perturbative constraint on the spectrum of strongly coupled
theories and on the formation of composite states. The same reasoning, through
the mechanism of anomaly inflow, constrains the boundary modes of systems with a
gapped bulk and underlies the robustness of phenomena such as the quantum Hall
effect and the transport properties of Weyl semimetals.

In the second role, anomalies are dynamical effects with observable
consequences. When a symmetry is anomalous but not gauged, the anomaly is
physical: examples include the $\pi^0$ decay rate, the resolution of the
$U(1)_A$ problem and the mass of the $\eta'$ meson, and anomaly-induced
transport such as the chiral magnetic effect. At the level of correlation
functions, anomalies produce a characteristic kinematic feature, namely a
massless $1/q^2$ pole in the longitudinal or trace sector of the relevant
vertex. This anomaly pole can, in appropriate kinematic limits, be interpreted
as the exchange of a massless composite state: an effective pseudoscalar in the
chiral case and an effective scalar, or dilaton, in the conformal case.
Determining whether such poles correspond to propagating degrees of freedom, and
how their spectral weight is distributed and constrained, is directly relevant to
the structure of hadrons and to gravitational interactions.

These two roles, anomalies as constraints on the field content of a theory and
anomalies as dynamical effects encoded in the analytic structure of correlators,
provide the common framework of this thesis. Each part of this Thesis addresses one of these two roles, making manifest the power of anomalies and their centrality in many domains of physics.

%=========================================================================
\section*{Anomaly Cancellation and the Family Problem}\label{sec:part1}
\addcontentsline{toc}{section}{Anomaly Cancellation and the Family Problem}
%=========================================================================

The first part of this thesis develops the role of anomalies as a
{constructive principle} for physics beyond the Standard Model. The SM is,
by any measure, an extraordinarily successful theory: its gauge structure
$SU(3)_C\times SU(2)_L\times U(1)_Y$ accounts for all confirmed laboratory
phenomena, and the discovery of the Higgs boson completed its particle content.
Yet it is manifestly incomplete. It offers no dynamical explanation for the
stabilization of the electroweak scale against the much higher scales of new
physics (the gauge hierarchy problem); it does not accommodate non-zero neutrino
masses without ad-hoc additions; it contains no candidate for dark matter and no
mechanism for the observed baryon asymmetry; and, perhaps most strikingly, it
provides no theoretical reason for the most conspicuous regularity of the matter
spectrum the existence of exactly {three} replicated families of quarks
and leptons. In the SM the number of generations is simply an experimental input.
The replication of families, and the bizarre hierarchy of their masses and
mixings, constitute the {flavour}, or {family}, problem.

Within the SM, gauge anomalies cancel {vertically}, generation by
generation: each family is separately anomaly-free, so any number of identical
copies would yield a consistent theory. The number three is therefore left
entirely unexplained. The central idea pursued in Part~I is that this state of
affairs can be turned into a prediction by enlarging the gauge symmetry so that
anomaly cancellation becomes {horizontal} intertwining the different
families rather than respecting each one individually.

This is realized in the minimal  $331$ Model, based on the gauge group
\begin{equation*}
	SU(3)_C \times SU(3)_L \times U(1)_X\,,
\end{equation*}
introduced by Frampton, Pisano and Pleitez.
Its defining feature is an asymmetric embedding of the quark generations: the
first two families transform as triplets of $SU(3)_L$ while the third transforms
as an antitriplet, the leptons being treated democratically as antitriplets. As
a consequence, several anomaly coefficients, notably those of the
$SU(3)_L^2\,U(1)_X$ and $U(1)_X^3$ combinations, no longer vanish family by
family, but only upon summing over all three generations at once. Anomaly
freedom then {requires} the number of families to be a multiple of three.
When this constraint is combined with the requirement of QCD asymptotic
freedom, which caps the number of quark triplets, the family number is fixed to
be exactly three. A fact that the SM can only record, the $331$ construction
explains from first principles. This is anomaly cancellation operating as a generative principle, and the thesis argues that it can serve as a
guide for model building aimed at the flavour puzzle more generally.

The price and the phenomenological opportunity of this elegant accounting is
a rich spectrum of new physics tied to the TeV scale. The enlarged symmetry
demands three exotic quarks and an extended gauge sector whose hallmark is the
doubly-charged {bileptons} $Y^{\pm\pm}$, vector bosons carrying lepton
number $L=\pm2$, alongside an additional neutral $Z'$ and singly-charged states.
The embedding of $SU(2)_L$ into $SU(3)_L$ ties the $U(1)_X$ coupling to the
Weinberg angle and imposes the bound $\sin^2\theta_W\le 1/4$, whose saturation
drives a Landau pole into the running of the coupling at a scale of a few TeV.
The model is thus intrinsically a TeV-scale theory, with its decisive tests
within reach of present and future hadron colliders. An extended scalar sector,
including a sextet of $SU(3)_L$ required to generate realistic charged-lepton
masses, gives rise to doubly-charged Higgses, and the non-universal family
embedding produces tree-level flavour-changing neutral currents mediated by the
$Z'$ and by neutral scalars, together with new sources of CP violation.

Part~I develops these ideas along three lines. First, the structure of the
minimal $331$ model is laid out in detail, from its cross-family anomaly
cancellation and symmetry-breaking pattern to its gauge, scalar and flavour
sectors. Second, the collider phenomenology of the exotic quark sector is
studied through the pair production and decay of heavy quarks into bileptons,
yielding spectacular same-sign dilepton signatures; a viable benchmark point is
identified with a full spectrum-generation and constraint pipeline, and the
discovery prospects are assessed at the LHC and at a prospective $100$~TeV
collider, where the signal is found to be cleanly observable. Third, the
anomaly-cancellation logic is pushed further in the left--right symmetric
\mbox{$33311$} extension, which embeds the $331$ structure into
$SU(3)_C\times SU(3)_L\times SU(3)_R\times U(1)_{X_L}\times U(1)_{X_R}$,
inherits the family-number constraint, and addresses the origin of parity
violation while avoiding the proton-decay problem of conventional grand unified
theories. Throughout, the guiding theme is that the consistency requirement of
anomaly freedom, far from being a mere bookkeeping device, can dictate the
particle content, fix the number of generations, and predict new physics at
accessible energies.

%=========================================================================
\section*{Anomalies in Conformal Field Theory: Sum Rules, Form Factors and Gravitational Scattering Amplitudes}\label{sec:part2}
\addcontentsline{toc}{section}{Anomalies in Conformal Field Theory: Sum Rules, Form Factors and Gravitational Scattering Amplitudes}
%=========================================================================

The second part of the thesis turns to the complementary face of anomalies: not
their cancellation, but their physical {manifestation} in the structure of
correlation functions, with applications to the internal structure of hadrons
and to gravitational interactions. The focus shifts from chiral gauge anomalies
to the {conformal} (trace) anomaly of QCD and to its chiral and
gravitational relatives, and the natural language becomes that of conformal
field theory in momentum space ($\mathrm{CFT}_p$).

The central objects of study are the anomaly three-point functions: the
$\langle TJJ\rangle$ correlator of one stress-energy tensor with two vector
currents, which carries the conformal anomaly; the $\langle AVV\rangle$
correlator, which carries the chiral anomaly; the $\langle ATT\rangle$
correlator of an axial current with two stress tensors, which carries the mixed
gravitational chiral anomaly; and the $\langle TTT\rangle$ vertex of three
stress tensors. In all of these, the conformal Ward identities of the
$\mathrm{CFT}_p$ formalism are remarkably constraining: the inclusion of a
single anomaly pole in the longitudinal or trace sector, together with the
transversality and trace Ward identities, suffices to reconstruct the entire
correlator including its transverse-traceless part up to the value of the
anomaly coefficient. The anomaly coefficient thus acts not as the prefactor of
one isolated tensor structure but as a structural factor permeating the whole
vertex, and the correlator vanishes identically when it is set to zero.

A unifying physical question runs through these Chapters: {in what sense is
the anomaly pole a particle?} A $1/q^2$ singularity in the longitudinal or trace
sector is reminiscent of the exchange of a massless Goldstone-like state an
effective axion for the chiral anomaly, an effective dilaton for the conformal
anomaly. The thesis sharpens this intuition through a dispersive analysis that
moves systematically away from the conformal point by giving the internal
fermions a mass. The key finding is that the correct, kinematics-independent
characterisation of an anomaly vertex is {not} the pole itself but a
{mass-independent sum rule} obeyed by the spectral density of the anomaly
form factor. For the chiral $AVV$ and gravitational $ATT$ correlators, and then
for the non-Abelian conformal $TJJ$ correlator at order $\alpha_s$, the integral
of the anomaly spectral density over the entire cut is shown to equal the
anomaly coefficient $1/(2\pi^2)$ for $AVV$, the gravitational coefficient
$a_g$ for $ATT$, and $-\tfrac13\,\beta(g)/g$ for $TJJ$ independently of the
fermion mass and of the external virtualities. As the mass varies, spectral
weight flows between a $\delta(s)$ spike at the origin and a continuum above the
two-particle threshold $4m^2$, but the total area is conserved: an
{area-law} or spectral-flow phenomenon in which the massless anomaly pole
of the conformal limit dissolves into a branch cut without ever losing its
integrated weight. This establishes that anomaly-induced interactions are
intrinsically {nonlocal}: a local effective vertex reproduces the pole but violates the sum rule, and can be
justified only when additional nonperturbative input (confinement, a Higgs
phase, anomaly matching) is supplied. The light-cone analysis further reveals a
qualitative difference between the chiral and conformal cases: whereas the
chiral pole exhibits ``anomaly dominance'' on the light cone, in the conformal
case a second, non-anomalous pole structure survives, so that the dilaton-like
behaviour is not controlled by the anomaly alone.

These formal developments find a direct phenomenological home in the
{gravitational form factors} (GFFs) of hadrons. The matrix elements of the
QCD energy--momentum tensor between hadronic states are parameterized, for a
spin-$1/2$ target, by the form factors $A(t)$, $B(t)$ and $D(t)$, encoding
respectively the distribution of momentum (with $A(0)=1$), of angular momentum
(with $B(0)=0$), and of internal pressure and shear the elusive ``$D$-term'',
unconstrained at zero momentum transfer and often called the last unknown global
property of the nucleon. These quantities are accessible experimentally through
generalised parton distributions in deeply virtual Compton scattering, and they
are a flagship goal of the JLab~12~GeV programme and of the future
Electron--Ion Collider (EIC). The thesis shows that at large momentum transfer
the GFFs are governed, through QCD factorisation, by the same $TJJ$ vertex whose
trace anomaly was analysed above: the anomaly enters the hard-scattering kernel
as a dilaton-like contribution proportional to the $\beta$-function coefficient
$\beta_0=(11C_A-2n_f)/3$. This connection is made fully concrete in the case of
the pion, the simplest strongly-bound system, whose GFFs $A_\pi(Q^2)$,
$D_\pi(Q^2)$ and trace form factor $\Theta_\pi(Q^2)$ are computed at large $Q^2$
with the $TJJ$ insertion in the hard kernel and with Sudakov resummation of the
endpoint logarithms in the modified perturbative approach of Li and Sterman. The
anomaly is found to feed predominantly  into the
$D$-term and the trace form factor, and its inclusion improves the agreement with
recent lattice QCD data a tangible imprint of the conformal anomaly on the
mechanical structure of a hadron.

Finally, the thesis carries the conformal anomaly into the gravitational scattering amplitudes domain.
By sewing two $TJJ$ vertices through single-graviton exchange in the
perturbative Einstein--Hilbert framework, one asks whether the anomaly can
activate a genuine {scalar} channel in four-point scattering amplitudes a
question with resonances in induced-gravity scenarios \`a~la~Sakharov, in
modified-gravity models predicting scalar gravitational waves, and in the
nonlocal Riegert anomaly-induced action. The analysis identifies a unique
gauge-invariant scalar mode $\Phi_{\rm cov}$ which, although non-propagating in
pure Einstein gravity, is sourced by the trace anomaly through an interaction of
the schematic form $R\,\Box^{-1}F^2$, so that the anomaly activates a distinct
scalar gravitational response without introducing a new fundamental particle. A
double-copy-like structure emerges in graviton--graviton scattering, where the
anomaly vertex factorises into a square of the photon amplitude, and the
discussion extends naturally to dark conformal sectors interacting only through
gravity. As in the hadronic case, the recurring lesson is that the observable
content of the anomaly resides in its global, nonlocal and spectral
properties rather than in any single local term.

%=========================================================================
\section*{Overview of the Thesis}\label{sec:overview}
\addcontentsline{toc}{section}{Overview of the Thesis}
%=========================================================================

The thesis is divided into two parts, bound together by the central role of
quantum anomalies. Part~I treats anomalies as a consistency constraint that
shapes extensions of the Standard Model; Part~II treats anomalies as physical
interactions encoded in the analytic structure of correlators, with
applications to hadron structure and gravity. The Chapters are organised as
follows.

\subsubsection*{Part I --- Constraining the Number of Generations in Physics Beyond the Standard Model through the Cross-Family Anomaly Cancellation Mechanism and its Phenomenology}

Chapter~\ref{ch331} introduces the minimal $331$ model based on
$SU(3)_C\times SU(3)_L\times U(1)_X$, presents its particle content, and
develops in detail the cross-family anomaly-cancellation mechanism that fixes
the number of fermion generations to three. The pattern of spontaneous symmetry
breaking, the gauge-boson and Higgs spectra, the bound $\sin^2\theta_W\le1/4$
and the associated TeV-scale Landau pole, and the flavour structure of the
quark and lepton sectors, including tree-level FCNCs and new CP-violating
phases, are analysed.

{Chapter \ref{ch331ph}} studies the collider phenomenology of the heavy exotic
quarks of the bilepton model. A phenomenologically viable benchmark point is
established through a full spectrum-generation and constraint pipeline, and the
pair-production process leading to a same-sign-dimuon-plus-$b$-jets final state
is simulated. The discovery reach is assessed at the LHC and at a $100$~TeV
collider, where a large statistical significance is obtained.

{Chapter \ref{ch33311}} develops the left--right symmetric $33311$ model, extending
the gauge group to
$SU(3)_C\times SU(3)_L\times SU(3)_R\times U(1)_{X_L}\times U(1)_{X_R}$. Its
fermion content, the systematic cancellation of all triangle anomalies, the
hierarchical pattern of symmetry breaking that reduces it to the $331$ model in
the appropriate limit, and its distinctive features including the absence of
proton decay and a possible dark-matter candidate are discussed.

\subsubsection*{Part II --- Anomaly-Induced Dynamics in Gauge and Gravitational Theories: Hadronic Gravitational Form Factors, Graviton Scattering Amplitudes, and Anomaly Sum Rules}

Chapter~\ref{chtjj} opens the second part by constructing a complete
off-shell parameterisation of the non-Abelian $TJJ$ correlator in perturbative
QCD within the $\mathrm{CFT}_p$ formalism, establishing the emergence of the
dilaton-like trace-anomaly pole, its relation to renormalisation, and its
entrance into the gravitational form factors of hadrons through factorisation.

{Chapter \ref{chsr1}} carries out a dispersive analysis of the chiral and
gravitational anomaly form factors of the $AVV$ and $ATT$ correlators with
massive fermions, deriving the mass-independent topological sum rules and the
spectral-flow picture, and showing that the anomaly pole is not, in general, an
asymptotic particle state.

{Chapter \ref{chsr2}} extends this analysis to the conformal anomaly, establishing
the dilaton sum rule $-\tfrac13\,\beta(g)/g$ for the $TJJ$ form factor at order
$\alpha_s$, analysing its spectral structure including the gluon sector, and
clarifying the difference between conformal and chiral anomaly interactions on
the light cone.

{Chapter \ref{chpionff}} applies the framework to a concrete observable: the pion
gravitational form factors at large momentum transfer, computed with the $TJJ$
insertion and Sudakov resummation, and compared with lattice QCD. The trace
anomaly is shown to leave a visible imprint on the $D$-term and the trace form
factor.

{Chapter \ref{chgsampl}} investigates anomaly-mediated scalar gravitational
interactions, showing how the conformal anomaly activates a gauge-invariant
scalar channel in graviton exchange, computing photon--photon and
graviton--graviton scattering amplitudes, and extending the discussion to
gravitationally coupled dark conformal sectors.

To ensure readability, each Chapter begins with a synopsis summarizing key results and future directions. Furthermore, every chapter is structured with its own introduction, conclusion, and appendix, allowing it to stand as a self-sustained unit.

	\mainmatter
	\part{Constraining the Number of Generations in Physics Beyond the Standard Model through the Cross-Family Anomaly Cancellation Mechanism and its Phenomenology}
	\mainmatter

	\begin{bibunit}[jhep]
		\chapter{The  331 Model:  Addressing the Fermion Families Problem within Horizontal Anomalies Cancellation}\label{ch331}
\addcontentsline{toc}{section}{Synopsis}
\section*{Synopsis}

This Chapter introduces the minimal 331 model, based on the gauge group $SU(3)_C \times SU(3)_L \times U(1)_X$, as a theoretically motivated extension of the Standard Model (SM). The central motivation is twofold: to address known shortcomings of the SM, including the gauge hierarchy problem, the origin of neutrino masses, and the absence of a theoretical principle fixing the number of fermion generations and to explore potential new physics accessible at the TeV scale probed by the LHC.

The Chapter begins by presenting the particle content of the minimal 331 model, originally proposed by Frampton, Pisano, and Pleitez. A defining structural feature is the asymmetric embedding of the three quark generations: the first two families transform as triplets under $SU(3)_L$, while the third transforms as an antitriplet. All three lepton families, by contrast, are treated democratically as antitriplets. This unconventional assignment necessitates the introduction of three exotic quarks and gives rise to a richer gauge boson spectrum, most notably the bileptons $Y^{\pm\pm}$ doubly charged gauge bosons carrying lepton number $L = \pm 2$ which are a hallmark of the model in its minimal formulation ($\beta = \sqrt{3}$).

A central focus of the Chapter is the mechanism of cross-family anomaly cancellation, which constitutes one of the most distinctive and attractive features of the model. Unlike the SM, where anomalies cancel generation by generation (vertically), certain anomalies in the 331 model, in particular the $SU(3)_L^2\, U(1)_X$ and $U(1)_X^3$ contributions, only vanish upon summing over all three generations simultaneously. This horizontal cancellation mechanism inherently constrains the number of fermion families to exactly three, offering a first-principles rationale for a fact that in the SM is fixed only by experiment. The Chapter argues that this approach can serve as a guiding principle for constructing other beyond-the-SM theories aimed at addressing the flavor puzzle.

The spontaneous symmetry breaking (SSB) pattern is then analyzed in detail. The symmetry is broken in two stages: first from $SU(3)_C \times SU(3)_L \times U(1)_X$ down to the SM gauge group, then to $SU(3)_C \times U(1)_{em}$. This requires three scalar triplets ($\rho$, $\eta$, $\chi$) and a scalar sextet ($\sigma$) of $SU(3)_L$. The gauge boson mass spectrum is derived explicitly, including the masses of the $W^\pm$, $Z$, $Z'$, the exotic charged bosons $V^\pm$, and the bileptons $Y^{\pm\pm}$. A notable structural consequence of the embedding of the SM into the 331 gauge group is a relation between the $U(1)_X$ coupling $g_1$ and the Weinberg angle, which imposes the bound
\begin{equation}
	\sin^2\theta_W \leq \frac{1}{4}\ .
\end{equation}
The saturation of this bound signals a Landau pole in the renormalization group evolution of $g_1$ at an energy scale of a few TeV, placing a hard upper limit on the validity of the perturbative description and effectively bounding the model to the TeV regime.

The Higgs sector is then examined, including the rich spectrum of physical states that emerges after electroweak symmetry breaking: five CP-even scalars, three pseudoscalars, four charged Higgses, and three doubly-charged Higgses. The necessity of the scalar sextet is established from group-theoretical considerations in the lepton Yukawa sector: without it, the antisymmetry of the leptonic Yukawa coupling in flavor space prevents the generation of three independent charged lepton masses. The sextet resolves this by providing a symmetric Yukawa coupling, and its neutral component can also generate small Majorana neutrino masses. An interesting phenomenological consequence is that the doubly-charged scalar decay $H^{\pm\pm} \to l^\pm l^\pm$ is enhanced for heavier leptons, particularly the tau, a feature largely unique to this model.

The flavor structure of the model is analyzed in both the quark and lepton sectors. Because the three quark generations transform under different representations of $SU(3)_L$, flavor rotation matrices do not cancel uniformly across all interaction vertices. This leads to flavor-changing neutral currents (FCNCs) mediated by neutral scalars and by the $Z'$ boson in the left-handed quark sector, in contrast to the SM where the GIM mechanism suppresses such effects at tree level. The independent appearance of the rotation matrix $V_L$ beyond the standard CKM combination introduces additional mixing angles and CP-violating phases with observable consequences. A similar structure emerges in the lepton sector, where the presence of both triplet and sextet Yukawa contributions generates a charged lepton mass matrix with both symmetric and antisymmetric flavor components, resulting in a non-trivial PMNS structure and additional leptonic mixing matrices entering the interactions with the exotic gauge bosons $V^\pm$ and $Y^{\pm\pm}$.

The Chapter concludes by emphasizing that the 331 model, through its cross-family anomaly cancellation mechanism, provides a concrete and theoretically grounded framework for understanding the three-generation structure of the SM fermion spectrum, while simultaneously predicting a variety of new phenomena exotic gauge bosons, doubly-charged scalars, FCNCs, and additional CP violation all potentially accessible at TeV-scale colliders.

The work presented in this Chapter is based primarily on, and adapted from, the publication \cite{corianoSU3CSU3LU1X2024}, co-authored with C.~Corianò.

\section{Introduction}
In the quest to unveil new physics governing fundamental interactions at the Large Hadron Collider (LHC), resolving several crucial questions remains a challenge within the Standard Model (SM). These include the gauge hierarchy problem in the Higgs sector and the origin of light neutrino masses.\\
Addressing these issues often requires theories involving larger gauge groups and a broader spectrum of particles. Grand Unified Theories (GUTs) offer promising avenues, but their high energy scales (around $10^{12}$ to $10^{15}$ GeV) far exceed the electroweak scale probed by the LHC.\\
Bridging the gap between the GUT scale and the TeV scale, where the LHC operates, to identify signatures of symmetry breaking presents a significant challenge due to the increased complexity of these extended theories.\\
However, specific scenarios exist where evidence for enlarged gauge symmetries might be discovered or excluded at the LHC scale, suggesting alternative exploration paths.\\
One such example is the 331 model $(SU(3)_c \times SU(3)_L \times U(1)_X)$, where the constraint of real gauge couplings significantly restricts the parameter space for potential signal searches.
This model was proposed as a potential extension to the SM in order to address certain theoretical and experimental shortcomings, as well as to provide explanations for phenomena not accounted for within the SM.\\
The 331 model introduces a new gauge group, $SU(3)_L$, which is isomorphic to the color gauge group $SU(3)_C$. This implies that the strong force acting between quarks within hadrons is now governed by the $SU(3)_L$ symmetry, in addition to the color symmetry. \\
The fermion content of the 331 model differs from the SM due to the extended gauge symmetry. Typically, in the 331 model, the quarks and leptons are organized into multiplets that transform under the representations of the $SU(3)_L$ gauge group. For example, quarks and leptons may be arranged in triplets or antitriplets of $SU(3)_L$, depending on their electric charge and other quantum numbers.
One notable feature of the 331 model is the presence of new gauge bosons called "bileptons." These are bosons carrying both lepton number and electric charge, with charges $Q=\pm 2$ and $L=\pm 2$. Bileptons arise due to the extended gauge symmetry and can have significant implications for various phenomena, including neutrino masses and decays of heavy particles.\\
Similar to the SM, the 331 model also involves spontaneous symmetry breaking, where the gauge symmetries are broken at some energy scale. This results in the generation of particle masses and the emergence of the familiar gauge bosons such as the $W^\pm$ and $Z^0$ bosons.\\
The 331 model offers potential explanations for various phenomena beyond the scope of the SM, including neutrino masses and mixing, and the unification of fundamental forces at high energies.\\
Overall, the 331 model represents an intriguing extension of the SM, offering new avenues for exploring fundamental physics beyond the established framework. However, it remains subject to experimental scrutiny and theoretical refinement to fully ascertain its validity and implications for our understanding of the fundamental forces and particles in nature.\\
In the model, the constraint of real gauge couplings significantly restricts the parameter space for potential signal searches. 
This property establishes the vacuum expectation values (vevs) of the Higgs bosons, responsible for symmetry breaking from the 331 scale to the electroweak scale, around the TeV region.
The model under consideration incorporates the presence of bileptons, denoted as gauge bosons $(Y^{--}, Y^{++})$, possessing a charge $Q=\pm 2$ and lepton number $L=\pm 2$. Consequently, we refer to this framework as the "bilepton model." Within the array of 331 models, the existence of bileptons within the spectrum arises only through specific embeddings of the $U(1)_X$ symmetry, as well as the charge $(Q)$ and hypercharge $(Y)$ generators within the local gauge structure.\\
An additional noteworthy aspect of this model is its departure from the conventions of the Standard Model or typical chiral models seen thus far. In contrast to merely extending the SM spectrum and symmetries, the determination of the (chiral) fermion generations hinges on the interfamily cancellation of gauge anomalies. Remarkably, gauge anomalies cancel across distinct fermion families, thereby pinpointing the number of generations to be 3. From this vantage point, the model emerges as distinctly singular. Furthermore, in the framework outlined by Frampton, which we adopt henceforth, the treatment of the third fermion family is asymmetrical in comparison to the initial two families.\\
Our work is organized as follows: We will first discuss the general structure of the model, starting with the anomaly constraints that crucially characterize the charge assignments of the spectrum. We then move on to characterize the gauge boson spectrum, turning afterwards to the Higgs sector. The structure of the potential is thoroughly examined, both in its triplet and sextet contributions. We discuss the energy bound present in the model, induced by the structure of the gauge coupling relation coming from the embedding of the Standard Model into the 331. This bound on the energy scale at which the model is characterized by real (as opposed to complex) values of the couplings is one of the salient features of this theory. A second important feature is the presence of a Landau pole in the renormalization group equations (RGEs) of the gauge couplings, which we briefly illustrate numerically. The results are based on an analysis of the RGEs, which we derive and present in Appendix B. A list of results for the mass eigenstates of the Higgs sector is contained in Appendix A. Notice that an important feature of the model is the identification of the electric charge operator in terms of the diagonal generators of the fundamental gauge symmetry. In general this is given by 
\begin{equation}
	Q=T^3+ \beta\,T^8+X \mathbb{1}\ 
\end{equation}
where $\beta$ is a parameter of the model, with $T^3$ and $T^8$ generators of $SU(3)_L$. Our discussion will be focused on the choice $\beta=\sqrt{3}$, as in Pisano, Pleitez and Frampton original formulation. This choice induces the presence of bileptons, which are doubly charged gauge bosons carrying lepton number $L=\pm 2$. We will refer to this version of the model as to the minimal one.

\section{The particle content of the minimal 331 Model}\label{331}
One of the most important questions that arise in particle physics is why there are only three families of quarks and leptons. There are many observables in particle physics that depend on the family number and they all agree to constrain this number to three. In the Standard Model there is no mechanism that prohibits the existence of more families than those observed. To answer this question P. H. Frampton, F. Pisano and V. Pleitez \cite {frampton_chiral_1992} \cite {pisano_su3ensuremathbigotimesu1_1992} have proposed a new model that extends the Standard Model and could provide an elegant answer to this fundamental question. Specifically it is called the 331 Model, and it is built on the gauge group
\begin{equation}
	SU(3)_C\times SU(3)_L\times U(1)_X\ ,
\end{equation}
which enlarges the $SU(2)_L$ symmetry of the Standard Model. Three exotic quarks must be added to the particle content of the Standard Model to allow for $ SU(3)_L $ symmetry in the quark sector. The 331 Model democratically treats leptons in each of the three families, in fact, color singlets are $ SU (3)_L $ anti-triplets
\begin{equation}
	\begin{pmatrix}
		e\\-\n_e\\ e^c
	\end{pmatrix}_L,
	\begin{pmatrix}
		\m\\-\n_\m\\ \m^c
	\end{pmatrix}_L,
	\begin{pmatrix}
		\t\\-\n_\t\\ \t^c
	\end{pmatrix}_L\rightarrow \brak{\textbf{1},\bar{\textbf{3}},0}
\end{equation}
each with $X=0$. Where with $e^c\ ,\m^c$ and $\t^c$ we are denoting the left-handed Weyl spinors of the relative charge conjugate field. On the contrary, in the quark sector the three families are treated differently, the first two generations are in triplets of $ SU (3)_L $ with the corresponding left-handed field of exotic quark
\begin{equation}
	\begin{pmatrix}
		u\\d\\ D
	\end{pmatrix}_L,
	\begin{pmatrix}
		c\\s\\ S
	\end{pmatrix}_L\rightarrow \brak{\textbf{3},\textbf{3},-\frac{1}{3}}\ ,
\end{equation}
both with $X=-\frac{1}{3}$. On the other hand, the third generation in the quark sector is embedded in anti-triplets of $SU(3)_L$
\begin{equation}
	\begin{pmatrix}
		b\\-t\\ T
	\end{pmatrix}_L
	\rightarrow \brak{\textbf{3},\bar{\textbf{3}},\frac{2}{3}}\ ,
\end{equation}
with $X=\frac{2}{3}$. To each left handed field there is a corresponding right handed field singlet under $SU(3)_L$ 
\begin{equation}
	\begin{matrix}
		(u^c)_L\\
		(c^c)_L\\ 
		(t^c)_L
	\end{matrix} \rightarrow \brak{\textbf{1},\textbf{1},-\frac{2}{3}}\ ,
\end{equation}

\begin{equation}
	\begin{matrix}
		(d^c)_L\\
		(s^c)_L\\
		(b^c)_L
	\end{matrix} \rightarrow \brak{\textbf{1},\textbf{1},\frac{1}{3}}\ ,
\end{equation}

\begin{equation}
	\begin{matrix}
		(D^c)_L\\(S^c)_L
	\end{matrix} \rightarrow \brak{\textbf{1},\textbf{1},\frac{4}{3}}\ ,
\end{equation}

\begin{equation}
	\begin{matrix}
		(T^c)_L
	\end{matrix} \rightarrow \brak{\textbf{1},\textbf{1},-\frac{5}{3}}\ ,
\end{equation}
The $U(1)_X$ charges are respectively $-\frac{2}{3}$, $\frac{1}{3}$ and $\frac{4}{3}$ for $u^c$, $d^c$, $D^c$ and the corresponding field of the second generation. The $X$ quantum numbers of the third family instead are respectively  $\frac{1}{3}$, $-\frac{2}{3}$ and $-\frac{5}{3}$ for $b^c$, $t^c$, $T^c$. 
Three scalar triplets under $SU(3)_L$ are necessary to ensure the spontaneous symmetry breaking
\begin{equation}
	\r=\begin{pmatrix}
		\r^{++}\\ \r^+\\ \r^0
	\end{pmatrix}
	\eta=\begin{pmatrix}
		\eta^{+}_1\\ \eta^0\\ \eta^-_2
	\end{pmatrix}
	\chi=\begin{pmatrix}
		\chi^{0}\\ \chi^-\\ \chi^{--}
	\end{pmatrix}
\end{equation}
respectively with $X$ charge $X=1,0,-1$, and a scalar sextet 
\begin{equation}
	\s=\begin{pmatrix}
		\s^{++}_1&\frac{\s_1^+}{\sqrt{2}}&\frac{\s_1^0}{\sqrt{2}}\\
		\frac{\s_1^+}{\sqrt{2}}&\s_2^0&\frac{\s^-_2}{\sqrt{2}}\\ 
		\frac{\s_1^0}{\sqrt{2}}&\frac{\s^-_2}{\sqrt{2}}&\s_2^{--}\\
	\end{pmatrix}\ ,
\end{equation}
in order to generate physical masses for leptons as we will see in Section \ref{331lep}.
All quantum numbers under $SU(3)_C\times SU(3)_L\times U(1)_X$ in the 331 Model can be found in Table \ref{Tab:minimal331_qn}. This unconventional assignment of quantum numbers in the model ensures that gauge anomalies are not canceled vertically for each family, as in the Standard Model. It is necessary to add the contribution of each quark in the triangle anomaly to obtain total cancellation. This is one of the most important and attractive feature of the model because it provides a possible explanation for the number of generations. This could provide a first step towards understanding the flavor puzzle and perhaps serve as a guide for new models inspired by it.

\begin{table}
	\centering
	\begin{tabular}{cccc}
		\hline
		\hline
		\rule[-4mm]{0mm}{1cm}
		& $SU(3)_C$ & $SU(3)_L$ & $U(1)_X$\\
		\hline\\
		$u^c\ \ \ \ c^c$  & $\bar{\textbf{3}}$ & $\textbf{1}$ & $-\frac{2}{3}$\\\\
		
		$t^c$  & $\bar{\textbf{3}}$ & $\textbf{1}$ & $\frac{1}{3}$\\\\
		
		$d^c\ \ \ \ s^c$ & $\bar{\textbf{3}}$ & $\textbf{1}$ & $\frac{1}{3}$\\\\
		
		$b^c$ & $\bar{\textbf{3}}$ & $\textbf{1}$ & $-\frac{2}{3}$\\\\
		
		$D^c\ \ \ \ S^c$ & $\bar{\textbf{3}}$ & $\textbf{1}$ & $\frac{4}{3}$\\\\
		
		$T^c$ & $\bar{\textbf{3}}$ & $\textbf{1}$ & $-\frac{5}{3}$\\\\
		
		$ \begin{pmatrix}
			u\\d\\ D
		\end{pmatrix}_L\ \ \ \ \begin{pmatrix}
			c\\s\\ S
		\end{pmatrix}_L\ \ \ \ $ & $\textbf{3}$ & $\textbf{3}$ & $-\frac{1}{3}$\\\\
		
		$\begin{pmatrix}
			b\\t\\ T
		\end{pmatrix}_L$& $\textbf{3}$ & $\bar{\textbf{3}}$ & $\frac{2}{3}$\\\\\\
		
		%$e^c\ \ \ \ \mu^c\ \ \ \ \tau^c$ & $\textbf{1}$ & $\textbf{1}$ & $-1$\\\\
		
		$\begin{pmatrix}
			e^-\\\n_e\\ e^+
		\end{pmatrix}_L\ \ \ \ \begin{pmatrix}
			\m^-\\\n_\m\\ \m^+
		\end{pmatrix}_L\ \ \ \ \begin{pmatrix}
			\t^-\\\n_\t\\ \t^+
		\end{pmatrix}_L$ & $\textbf{1}$ & $\bar{\textbf{3}}$ & $0$\\\\
		\\
		$X_\mu$ & $\textbf{1}$ & $\textbf{1}$ & $0$\\\\
		
		$W^a_\mu$ & $\textbf{1}$ & $\textbf{8}$ & $0$\\\\
		
		$G^b_\mu$ & $\textbf{8}$ & $\textbf{1}$ & $0$\\\\\\
		
		$\r$ & $\textbf{1}$ & $\textbf{3}$ & $1$\\\\
		
		$\eta$ & $\textbf{1}$ & $\textbf{3}$ & $0$\\\\
		
		$\chi$ & $\textbf{1}$ & $\textbf{3}$ & $-1$\\\\
		
		$\s$ & $\textbf{1}$ & $\textbf{6}$ & $0$\\\\
		\hline
		\hline
	\end{tabular}
	\caption{Quantum numbers of the particle spectrum of the minimal 331 Model}
	\label{Tab:minimal331_qn}
\end{table}
\section{Cross-Family Anomaly cancellation and the Flavor question}
In this section, we will analyze the non-trivial cancellation of anomalies in the 331 Model and discuss why this method of eliminating them could represent an initial step towards addressing the flavor question.

There are six types of anomalies that occur in the 331 Model
\begin{align}
	SU(3)_C^3\ ,&\ \ \ \ \ \ \ \ \ \ \ \ SU(3)_C^2U(1)_X\ ,\ \ \ \ \ \ \ \ \ \ \ \ SU(3)_L^3\ ,\nonumber\\ SU(3)_L^2U&(1)_X\ ,\ \ \ \ \ \ \ \ \ \ \ \ U(1)_X^3\ ,\ \ \ \ \ \ \ \ \ \ \ \  \textrm{grav}^2\,U(1)_X,\ 
\end{align}
where $\textrm{grav}^2\,U(1)_X$ is the mixed chiral anomaly with two gravitons and a chiral $U(1)_X$ gauge current, i.e. the gravitational chiral anomaly. \\
Each vertex collects a factor 
\begin{equation}
	2tr(T^a_R\{T_R^b,T_R^c\})=A(R)\, d^{abc}\ ,\ \ \ \ \ \ \ \textnormal{with  }A(\textrm{fund})=1
\end{equation}
from the gauge current group generators, where $A(R)$ is the representation dependent anomaly coefficient, with $A(\textrm{fund})$ in the fundamental representation of $SU(3)$, and $d^{abc}$ is the totally symmetric invariant $SU(3)$ tensor.

The anomaly involving only gluons is obviously zero as in the Standard Model.  Another anomaly arises from the mixed $SU(3)_C^2\, U(1)_X$ vertex, which imposes the following constraint on the $X$ charges  
\begin{equation}
	SU(3)_C^2U(1)_X\rightarrow  3X_{Q_i}+X_{u^c_i}+X_{d^c_i}+X_{J^c_i}=0.\ 
\end{equation}
Here, the index $i$ ranges over families, but there is no summation over it, $Q_i$ stands for quark triplets, while $J_i^c$ denotes the complex conjugate of the exotic quark fields. Indeed, for this reason, the anomaly cancellation in this case occurs vertically between families, as in the Standard Model.\\
The same cannot be said for another anomaly, which involves only the gauge group $SU(3)_L$. In the Standard Model, the $SU(2)_L$ group possesses a vanishing anomaly coefficient, rendering it unnecessary. However, in the case of $SU(3)$ or, in general, $SU(N)$ with $N>2$, it exhibits a non-zero anomaly coefficient. In the 331 Model, such $SU(3)_L^3$ anomaly cancels due to the equal number of fermions in the $\textbf{3}$ and $\bar{\textbf{3}}$ representations of $SU(3)_L$.\\
In the minimal 331 Model, the $SU(3)_L^2U(1)_X$ anomaly can only be canceled by accounting for contributions from all three families. This type of anomaly cancellation is referred to as a horizontal cancellation. If we consider anomaly cancellation on a generation-by-generation basis, it does not vanish, and requires the summation over different families. The relative constraint is
\begin{equation}
	SU(3)_L^2\, U(1)_X\rightarrow \sum_{i=1}^3 X_{Q_i}=0\ .
\end{equation}
The same motivations lead to the cancellation of the cubic anomaly
\begin{equation}
	U(1)_X^3\rightarrow \sum_{i=1}^3 \brak{3X_{Q_i}^3+X_{u^c_i}^3+X_{d^c_i}^3+X_{J^c_i}^3}=0\, ,
\end{equation}
with the relative constraint.

\begin{table}
	\centering
	\begin{tabular}{cccc}
		\hline
		\hline
		\rule[-4mm]{0mm}{1cm}
		Anomaly &&\\
		\hline\\
		$SU(3)_C^2U(1)_X $&&$3X_Q+X_{u^c_i}+X_{d^c_i}+X_{J^c_i}=0$\\\\
		
		$SU(3)_L^3 $&&$\textnormal{Equal number of $\textbf{3}_L$ and $\bar{\textbf{3}}_L$ representations} $\\\\
		
		$ SU(3)_L^2U(1)_X $&& $\sum_{i=1}^3 X_{Q_i}=0$\\\\
		
		$ U(1)_X^3$&&$\sum_{i=1}^3 \brak{3X_{Q_i}^3+X_{u^c_i}^3+X_{d^c_i}^3+X_{J^c_i}^3}=0 $\\\\
		
		$  grav^2U(1)_X$&&$ 3X_{Q_i}+X_{u^c_i}+X_{d^c_i}+X_{J^c_i}=0$\\\\
		\hline
		\hline
	\end{tabular}
	\caption{Anomaly cancellation constraints on the fermion charges in the minimal 331 Model}
	\label{Tab:331_anomaly}
\end{table}
The last anomaly that needs to be checked is the gravitational anomaly, but it is not difficult to show that it leads to the same constraint as for $SU(3)^2_C\,U(1)_X$
\begin{equation}
	\textrm{grav}^2\, U(1)_X\rightarrow 3X_{Q_i}+X_{u^c_i}+X_{d^c_i}+X_{J^c_i}=0\ .
\end{equation}
A summary of anomalies and relative constraints can be found in Table \ref{Tab:331_anomaly}.\\
The non-trivial cancellation of anomalies in the model is arguably one of its most intriguing and distinctive feature, first discussed in \cite{singer_canonical_1980}. The horizontal approach, involving all three generations of quarks and leptons, inherently constrains the number of families to three. Unlike the Standard Model, which lacks a mechanism to limit the number of fermion generations, the 331 model provides such a constraint. \\
Notice that in the Standard Model the number of generations is fixed experimentally by the annihilation process $e^+e^-\rightarrow \text{hadrons}$, which is sensitive to the number of families. Experimental constraints based on these observables restrict the number to 3, nevertheless, the model lacks a theoretical guiding principle for predicting such a number. The horizontal approach to anomaly cancellation of the 331, on the other hand, as already mentioned, can serve as a guiding principle for investigating other beyond the Standard Model (BSM)  UV-completions, that aim to address the flavor question.

\section{Spontaneous symmetry breaking}
Below the electroweak scale, in the Standard Model, the gauge symmetry is $SU(3)_C \times U(1)_{em}$. Therefore, also in the 331 model, spontaneous symmetry breaking (SSB) must occur, in order to reduce  the $SU(3)_C\times SU(3)_L\times U(1)_X$ gauge symmetry. The breaking can be divided into two stages. Initially, at energy scales greater than $246 \,\textnormal{GeV}$, the gauge symmetry of the 331 model can be broken down to that of the Standard Model, and subsequently to $SU(3)_C \times U(1)_{em}$. This can be represented as
\begin{equation}
	SU(3)_C\times SU(3)_L\times U(1)_X\rightarrow SU(3)_C\times SU(2)_L\times U(1)_{Y}\rightarrow SU(3)_C\times U(1)_{em}\ .
\end{equation}
Achieving this requires a more intricate Higgs sector comprising three scalar triplets and a scalar sextet of $SU(3)_L$. In the following two sections, we will analyze the pattern of SSB and explore how it predicts the existence of bileptons, specifically massive double-charged gauge bosons which carry lepton number of $L=\pm 2$ which can be classified as elementary bifermions (including also leptoquarks and biquarks) in the framework based on $SU(15)$ noted recently in \cite{coriano_su15_2023}.

\subsection{The breaking $SU(3)_C\times SU(3)_L\times U(1)_X\rightarrow SU(3)_C\times SU(2)_L\times U(1)_{Y}$}
The spontaneous symmetry breaking to the Standard Model gauge group can be accomplished by means of a vacuum expectation value of a scalar triplet belonging to $SU(3)_L$, denoted as $\r$,
\begin{equation}
	\r=\begin{pmatrix}
		\r^{++}\\ \r^+\\ \r^0
	\end{pmatrix}
\end{equation}
which carries charge under $U(1)_X$, namely $X=1$
\begin{equation}
	\expval{\r}=\begin{pmatrix}
		0\\0\\v_\r
	\end{pmatrix}\ .
\end{equation}
The $SU(3)_L\times U(1)_X$ covariant derivative can be written as follow
\begin{align}\label{covdev}
	D_\m=&\ \partial_\m -ig_1XX_\m-ig_2\frac{\l^a}{2}W^a_\m=\nonumber\\
	=&\ \brak{\partial_\m-i\sqrt{\frac{2}{3}}g_1X X_\m}
	\begin{pmatrix}
		1&0&0\\
		0&1&0\\
		0&0&1
	\end{pmatrix}-ig_2
	\begin{pmatrix}
		\frac{   {W^3_\m}   }{2}+\frac{   {W^8_\m}   }{2 \sqrt{3}} & \frac{   {W^1_\m}   }{2}-\frac{i    {W^2_\m}   }{2} & \frac{   {W^4_\m}   }{2}-\frac{i    {W^5_\m}   }{2} \\
		\frac{   {W^1_\m}   }{2}+\frac{i    {W^2_\m}   }{2} &    -\frac{   {W^3_\m}   }{2}+\frac{   {W^8_\m}   }{2 \sqrt{3}} & \frac{   {W^6_\m}   }{2}-\frac{i    {W^7_\m}   }{2} \\
		\frac{   {W^4_\m}   }{2}+\frac{i    {W^5_\m}   }{2} & \frac{   {W^6_\m}   }{2}+\frac{i    {W^7_\m}   }{2} &    -\frac{   {W^8_\m}   }{\sqrt{3}} 
	\end{pmatrix}\ ,
\end{align}
where $X$ is the charge under $U(1)_X$ of the fermion, $X_\mu$ is the corresponding gauge boson and 
$W_\mu^a$ are the generators of $SU(3)_L$. 
$\l^a$ are the Gell-Mann matrices normalized as $\Tr(\l^a \l^b)=2\d^{ab}$. Once the Higgs triplet $\rho$ acquires the vacuum expectation value, its kinetic term gives
\begin{align}\label{d}
	\brak{D_\m\begin{pmatrix}
			0\\0\\v_\r
	\end{pmatrix}}^\dagger\brak{D_\m\begin{pmatrix}
			0\\0\\v_\r
	\end{pmatrix}}=-\frac{1}{6} v_\r^2 \biggl(&-4 \sqrt{2} g_1g_2 X_\m W^{\m 8} +4g_1^2 X^\m X_\m+3 g_2^2Y^{++\m} Y^{--}_\m\nonumber \\&+3 g_2^2 V^{\m+}V_\m^- + 2g_2^2 W^8_\m W^{\m 8}\biggl)\ .
\end{align}
From equation \eqref{d} it is immediate to observe that the mass terms obtained are expressed in a basis that is not completely diagonal. This implies that in order to obtain the mass eigenstates of the bosons it is necessary to perform an orthogonal rotation on the corresponding states. Before doing this, we mention that the $ W^\pm $ bosons, given by
\begin{equation}
	W^\pm_\m=\frac{1}{\sqrt{2}}\brak{W^1_\m\mp W^2_\m}\ ,
\end{equation}
remain massless. This is due to the fact the residual symmetry $SU(2)_L$ remains unbroken at this stage.  We recognise also two correctly diagonalized kinetic energy contributions, given in terms of the charge operator eigenstates. As we will see, the charge operator in the 331 Model is embedded as $Q=\frac{1}{2}\l^3+\frac{\sqrt{3}}{2}\l^8+X\mathbb{1}$. We have  
\begin{equation}
	Y^{\pm\pm}_\m=\frac{1}{\sqrt{2}}\brak{W^4_\m\mp i W^5_\m}\ ,
\end{equation}
for the bileptons and 
\begin{equation}
	V^{\pm}_\m=\frac{1}{\sqrt{2}}\brak{W^6_\m\mp i W^7_\m}\ ,
\end{equation}
for the exotic charged gauge bosons. The mass matrix that needs to be diagonalized in the $\{X,W^8\}$ bases is the following
\begin{equation}
	\begin{pmatrix}
		\frac{g_2^2v_\r^2}{3}&-\frac{\sqrt{2}}{3}g_1g_2v_\r^2\\
		-\frac{\sqrt{2}}{3}g_1g_2v_\r^2&\frac{2}{3}g_1^2v_\r^2
	\end{pmatrix}\ .
\end{equation}
The diagonalization can be easily achieved through the orthogonal transformations
\begin{equation}\label{cb1}
	Z'_\m=\frac{1}{\sqrt{g_2^2+2g_1^2}}\brak{g_2W^8_\m+\sqrt{2}g_1X_\m}\ ,
\end{equation}
\begin{equation}\label{cb2}
	B_\m=\frac{1}{\sqrt{g_2^2+2g_1^2}}\brak{\sqrt{2}g_1W^8_\m-g_2X_\m}\ ,
\end{equation}
which can be thought as a rotation from the basis $\{X,W^8\}$ to $\{B,Z'\}$, with an angle 
\begin{equation}
	\sin{\th_{331}}=\frac{g_2}{\sqrt{g_2^2+\frac{g_1^2}{2}}}\ ,
\end{equation}
that gives
\begin{equation}\label{m}
	\begin{pmatrix}
		0&0\\
		0&\frac{1}{3} v_\r^2 \left(g_2^2+2 g_1^2\right)
	\end{pmatrix}\ .
\end{equation}
From the matrix in equation \eqref{m}, we can read off the squared masses of the mass eigenstates $\{B,Z'\}$
\begin{equation}
	M_B^2=0\ \ \ \ \ \ \ \ \ \ M_{Z'}^2=\frac{1}{3} v_\r^2 \left(g_2^2+2 g_1^2\right)\ .
\end{equation}
It is clear that we obtain a massless boson related to the $U(1)_Y$ symmetry of the Standard Model. It is not difficult to identify the embedding of the $Y$ charge operator in the 331 Model, namely
\begin{equation}
	\frac{Y}{2}=\sqrt{3}T^8+X\mathbb{1}\ , 
\end{equation}
where $T^8$ is the eighth generator of $SU(3)$. The matching condition between the $U(1)_X$ coupling and the Standard Model hypercharge can be easily computed, leading to the relation
\begin{equation}
	\frac{1}{g_Y^2}=\frac{6}{g_1^2}+\frac{3}{g_2^2}\ .
\end{equation}

\subsection{The breaking $SU(3)_C\times SU(2)_L\times U(1)_{Y}\rightarrow SU(3)_C\times U(1)_{em}$}
Once the gauge symmetry has been decomposed into that of the Standard Model, another symmetry breaking is necessary to end up with the residual $SU(3)_C\times U(1)_{em}$ gauge symmetry. To realize the correct breaking scheme, we require two Higgs triplets, $\eta$ and $\chi$, which acquire the vacuum expectation values

\begin{equation}
	\expval{\eta}=\begin{pmatrix}
		0\\ \frac{v_\eta}{\sqrt{2}}\\0
	\end{pmatrix} \ \ \ \ \ \ \textnormal{with\ } X=0\ ,
\end{equation}
and
\begin{equation}
	\expval{\chi}=\begin{pmatrix}
		\frac{v_\chi}{\sqrt{2}}\\0\\0
	\end{pmatrix} \ \ \ \ \ \ \textnormal{with\ } X=-1\ ,
\end{equation}
and a sextet of $SU(3)_L$ 
\begin{equation}
	\expval{\s}=\begin{pmatrix}
		0&0&\frac{v_\s}{2}\\
		0&0&0\\
		\frac{v_\s}{2}&0&0
	\end{pmatrix} \ \ \ \ \ \ \textnormal{with\ } X=0\ .
\end{equation}
After the first symmetry breaking, the covariant derivative can be written in terms of the mass eigenstate fields, namely in the basis $\{B,Z'\}$. Inserting the inverse of the equations \eqref{cb1} and\eqref{cb2} into \eqref{covdev} gives
\begin{equation}
	D_\m=
	\begin{pmatrix}
		\partial_\m-ig_2\frac{ {W^3_\m}   }{2}+K_1 & -ig_2\brak{\frac{   {W^1_\m}   }{2}-\frac{i    {W^2_\m}   }{2}} & -ig_2\brak{\frac{   {W^4_\m}   }{2}-\frac{i    {W^5_\m}   }{2}} \\
		-ig_2\brak{\frac{   {W^1_\m}   }{2}+\frac{i    {W^2_\m}   }{2}} &\partial_\m    -ig_2\frac{   {W^3_\m}   }{2}+K_1 & -ig_2\brak{\frac{   {W^6_\m}   }{2}-\frac{i    {W^7_\m}   }{2}} \\
		-ig_2\brak{\frac{   {W^4_\m}   }{2}+\frac{i    {W^5_\m}   }{2}} &-ig_2\brak{ \frac{   {W^6_\m}   }{2}+\frac{i    {W^7_\m}   }{2}} & \partial_\m   +K_2 \\
	\end{pmatrix}
\end{equation}
where $K_1$ and $K_2$ are given by 
\begin{equation}
	K_1=\frac{i \left(\sqrt{2} B_\m g_2 {g_1} (2 X-1)-Z'_\m \left(g_2^2+4 {g_1}^2 X\right)\right)}{2 \sqrt{3}}\ ,
\end{equation}
and
\begin{equation}
	K_2=\frac{i \left(\sqrt{2} B_\m g_2 {g_1} (X+1)+Z'_\m \left(g^2_2-2 {g_1}^2 X\right)\right)}{\sqrt{3}}\ .
\end{equation}
Once all the scalar field acquire a vacuum expectation value, the gauge fields  $ W^\pm $ and $ Z $ become massive too, while $ Y^{\pm \pm} $, $ V^\pm $ and $ Z'$ get more involved mass terms.
The squared masses of $ W^\pm $, $ V^\pm $ and $ Y^{\pm \pm} $ are given by 
\begin{equation}
	M^2_W= \frac{g_2^2 v_\eta^2}{4}+\frac{g_2^2 v_\chi^2}{4}+\frac{g_2^2 v_\sigma^2}{4} \ ,
\end{equation}

\begin{equation}
	M^2_V= \frac{g_2^2 v_\rho^2}{4}+\frac{g_2^2 v_\eta^2}{4}+\frac{g_2^2 v_\sigma^2}{4}\ ,
\end{equation}

\begin{equation}
	M^2_Y= \frac{g_2^2 v_\rho^2}{4}+\frac{g_2^2 v_\chi^2}{4}+g_2^2 v_\sigma^2 \ ,
\end{equation}
where we recall that  
\begin{equation}
	W^{\pm}_\m=\frac{1}{\sqrt{2}}\brak{W^1_\m\mp i W^2_\m}\ ,\ \ \ \ \ \ V^{\pm}_\m=\frac{1}{\sqrt{2}}\brak{W^6_\m\mp i W^7_\m}\ ,\ \ \ \ \ \ Y^{\pm\pm}_\m=\frac{1}{\sqrt{2}}\brak{W^4_\m\mp i W^5_\m}\ .
\end{equation}
On the other side, the neutral gauge bosons gain also non-diagonal mass terms, that in the $\{W^3,W^8,X\}$ basis are given by the following matrix
\begin{equation}
	\left(
	\begin{array}{ccc}
		g_1^2 v_\r^2+g_1^2 v_\chi^2 & -\frac{g_1 g_2 v_\r^2}{\sqrt{3}}-\frac{g_1 g_2 v_\chi^2}{2 \sqrt{3}} & -\frac{1}{2} g_1 g_2 v_\chi^2 \\
		-\frac{g_1 g_2 v_\r^2}{\sqrt{3}}-\frac{g_1 g_2 v_\chi^2}{2 \sqrt{3}} & \frac{g_2^2 v_\r^2}{3}+\frac{g_2^2 v_\eta^2}{12}+\frac{g_2^2 v_\chi^2}{12}+\frac{g_2^2 v_\sigma^2}{12} & -\frac{g_2^2 v_\eta^2}{4 \sqrt{3}}+\frac{g_2^2 v_\chi^2}{4 \sqrt{3}}-\frac{g_2^2 v_\sigma^2}{4 \sqrt{3}} \\
		-\frac{1}{2} g_1 g_2 v_\chi^2 & -\frac{g_2^2 v_\eta^2}{4 \sqrt{3}}+\frac{g_2^2 v_\chi^2}{4 \sqrt{3}}-\frac{g_2^2 v_\sigma^2}{4 \sqrt{3}} & \frac{g_2^2 v_\eta^2}{4}+\frac{g_2^2 v_\chi^2}{4}+\frac{g_2^2 v_\sigma^2}{4} \\
	\end{array}
	\right)
\end{equation}
In a first step, the two neutral gauge bosons $W^8$ and $X$ mix, giving rise to the two bosons $B$ and $Z$. The mixing angle is denoted by $\theta_{331}$ and is given by 
\begin{equation}\label{th331}
	\sin{\th_{331}}=\frac{g_2}{\sqrt{g_2^2+\frac{g_1^2}{2}}}\ ,
\end{equation}
which was the rotation discussed in the previous Section.
Then we can proceed in complete analogy with the Standard Model, where $B$ mixes with $W^3$ through the Weinberg angle $\theta_W$, which in the minimal 331 model takes the form
\begin{equation}\label{thw}
	\sin{\th_{W}}=\frac{g_Y}{\sqrt{g^2+g_Y^2}}\ .
\end{equation}
Using \eqref{thw} is straightforward to show that we can express the $\th_{331}$ angle in terms of the Weinberg angle, namely
\begin{equation}
	\cos{\th_{331}}=\sqrt{3}\tan{\th_W}\ ,
\end{equation}
therefore, the photon field $A$ and the two massive neutral gauge bosons $Z$ and $Z'$ are identified 
as
\begin{equation}\label{aa}
	A=\sin{\th_W}W_3+\cos{\th_W}\brak{\sqrt{3}\tan{\th_W} W_8+\sqrt{1-3\tan^2{\th_W}}X} \ ,
\end{equation}
\begin{equation}\label{bbhh}
	Z=\cos{\th_W}W_3-\sin{\th_W}\brak{\sqrt{3}\tan{\th_W} W_8+\sqrt{1-3\tan^2{\th_W}}X}\ ,
\end{equation}
\begin{equation}\label{cc}
	Z'=-\sqrt{1-3\tan^2{\th_W}} W_8+\sqrt{3}\tan{\th_W}X\ .
\end{equation}
The Weinberg angle also relates the coupling of $SU(3)_L$ which by matching is equal to the $SU(2)_L$ of the Standard Model, to the $U(1)_X$ coupling, through
\begin{equation}
	\label{boundd}
	\frac{g^2_1}{g^2_2}=\frac{6\sin^2{\th_W}}{1-4\sin^2{\th_W}}\ ,
\end{equation}
which can be obtained from equations \eqref{th331} and \eqref{thw}.
From equations \eqref{aa} \eqref{bbhh} and \eqref{cc} it is evident that there is a residual mixing between the massive gauge boson, which is given by the matrix
\begin{equation}
	\left(
	\begin{array}{cc}
		C_{ZZ} & C_{ZZ'} \\
		C_{ZZ'}& C_{Z'Z'} \\
	\end{array}
	\right)
\end{equation}
with entries 
\begin{equation}
	C_{ZZ}= g_2^2(2 \theta_W) \left(v_\r^2+v_\eta^2+v_\s^2\right)\ ,
\end{equation}
\begin{equation}
	C_{Z'Z'}=\frac{g_2^2\sin^2\th_W \left(\csc ^2(\theta_W) \left(v_\r^2+v_\eta^2+4 v_\chi^2+v_\s^2\right)+9 \sec ^2(\theta_W) \left(v_\r^2+v_\eta^2+v_\s^2\right)-4 \left(v_\r^2+4 v_\eta^2+v_\chi^2+4 v_\s^2\right)\right)}{24 \cos (2 \theta_W)-12} \ ,
\end{equation}
\begin{equation}
	C_{ZZ'}= \frac{g_2^2 \sec ^3(\theta_W) \left(-\cos (2 \theta_W) \left(v_\r^2+2 \left(v_\eta^2+v_\s^2\right)\right)+2 v_\r^2+v_\eta^2+v_\s^2\right)}{4 \sqrt{12-9 \sec ^2(\theta_W)}}\ .
\end{equation}
Therefore, a further rotation is needed 
\begin{equation}
	\begin{pmatrix}
		Z_1\\Z_2
	\end{pmatrix}=\begin{pmatrix}
		\cos{\th_Z}&-\sin{\th_Z}\\
		\sin{\th_Z}&\cos{\th_Z}
	\end{pmatrix} \begin{pmatrix}
		Z\\Z'
	\end{pmatrix}
\end{equation}
in order to obtain the masses of propagating gauge bosons, namely
\begin{align}
	M_Z^2=&\frac{1}{6} (3 g_1^2 (v_\rho^2+v_\chi^2)-(9 g_1^4 (v_\rho^2+v_\chi^2)^2+6 g_1^2 g_2^2 (v_\rho^4-v_\rho^2 (v_\eta^2+v_\sigma^2)\nonumber\\&+v_\chi^2 (-v_\eta^2+v_\chi^2-v_\sigma^2))+ g_2^4 (v_\rho^4-v_\rho^2 (v_\eta^2+v_\chi^2+v_\sigma^2)+v_\eta^4+v_\sigma^2 (2 v_\eta^2-v_\chi^2)\nonumber\\ &-v_\eta^2 v_\chi^2+v_\chi^4+v_\sigma^4)^\frac{1}{2})+g_2^2 (v_\rho^2+v_\eta^2+v_\chi^2+v_\sigma^2))\ ,
\end{align}
and
\begin{align}
	M_{Z'}^2=&\frac{1}{6} (3 g_1^2 (v_\rho^2+v_\chi^2)+(9 g_1^4 (v_\rho^2+v_\chi^2)^2+6 g_1^2 g_2^2 (v_\rho^4-v_\rho^2 (v_\eta^2+v_\sigma^2)\nonumber\\&+v_\chi^2 (-v_\eta^2+v_\chi^2-v_\sigma^2))+g_2^4 (v_\rho^4-v_\rho^2 (v_\eta^2+v_\chi^2+v_\sigma^2)+v_\eta^4+v_\sigma^2 (2 v_\eta^2-v_\chi^2)\nonumber\\&-v_\eta^2 v_\chi^2+v_\chi^4+v_\sigma^4))^{\frac{1}{2}}+g_2^2 (v_\rho^2+v_\eta^2+v_\chi^2+v_\sigma^2))\ .
\end{align}

\begin{figure}[]
	\centering
	\begin{subfigure}[t]{0.45\textwidth}
		\centering
		\includegraphics[width=\linewidth,height=50mm, keepaspectratio]{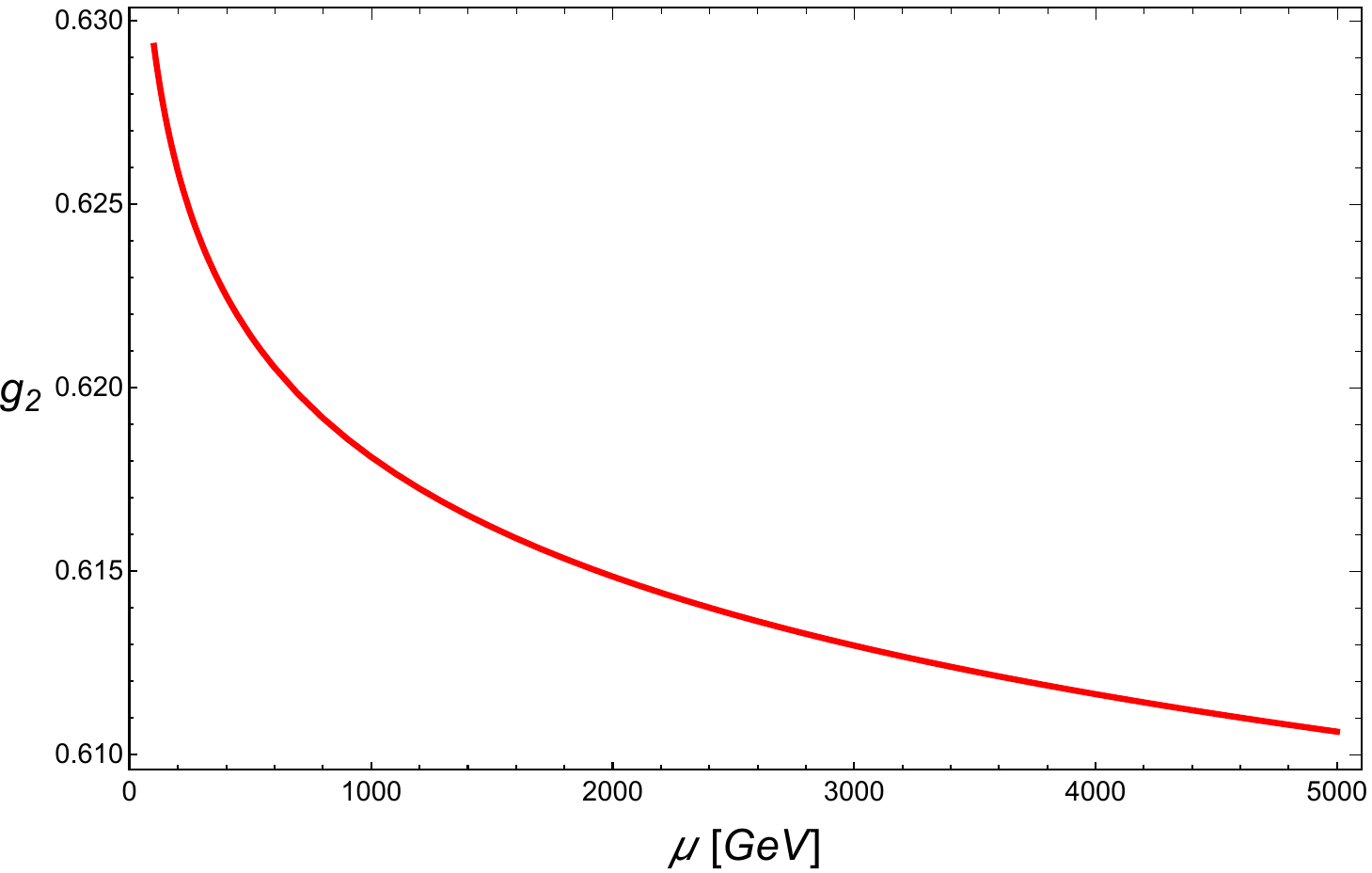}
		\caption{Running of the coupling $g_2$ in the Standard Model.}
		\label{fig:fig-a}
	\end{subfigure}
	\quad
	\begin{subfigure}[t]{0.45\textwidth}
		\centering
		\includegraphics[width=\linewidth,height=50mm, keepaspectratio]{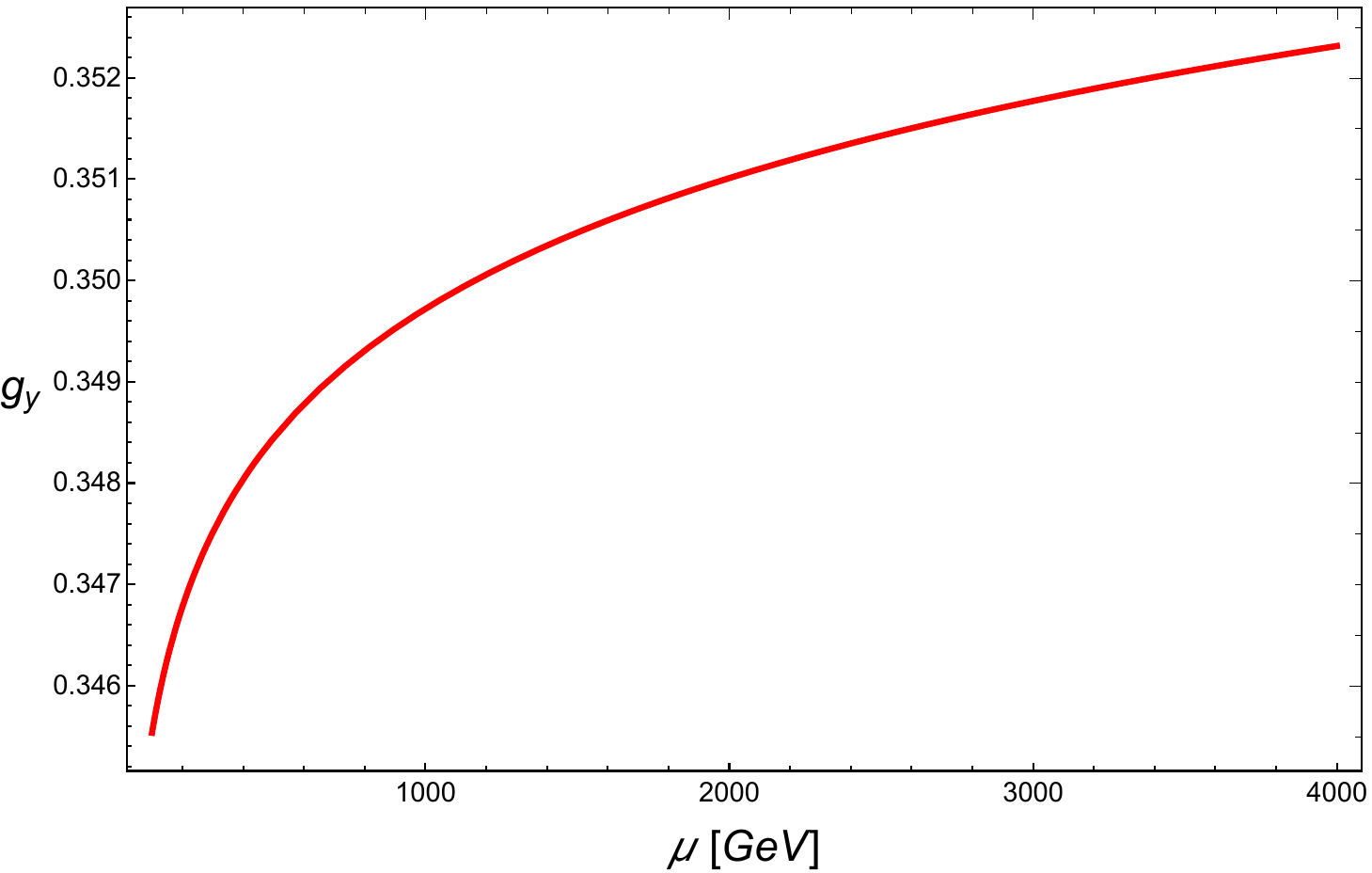}
		\caption{Running of the coupling $g_y$ in the Standard Model.}
		\label{fig:fig-b}
	\end{subfigure}
	\par\bigskip
	\begin{subfigure}[t]{0.45\textwidth}
		\centering
		\includegraphics[width=\linewidth,height=50mm, keepaspectratio]{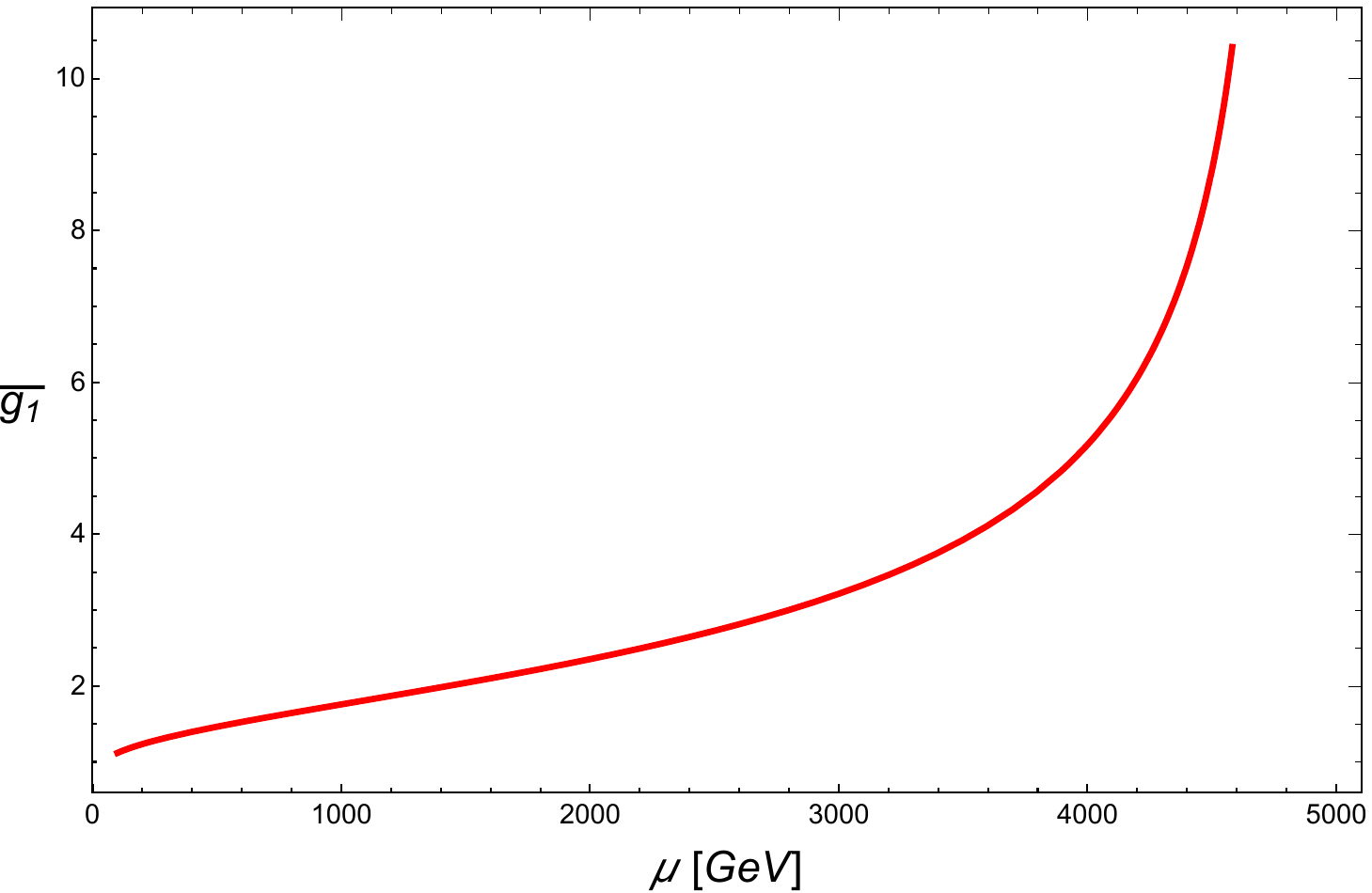}
		\caption{Evolution of the matching condition of the coupling $g_1$ of the 331 Model with the Standard Model coupling constants.}
		\label{fig:fig-c}
	\end{subfigure}
	\quad
	\begin{subfigure}[t]{0.45\textwidth}
		\centering
		\includegraphics[width=\linewidth,height=50mm, keepaspectratio]{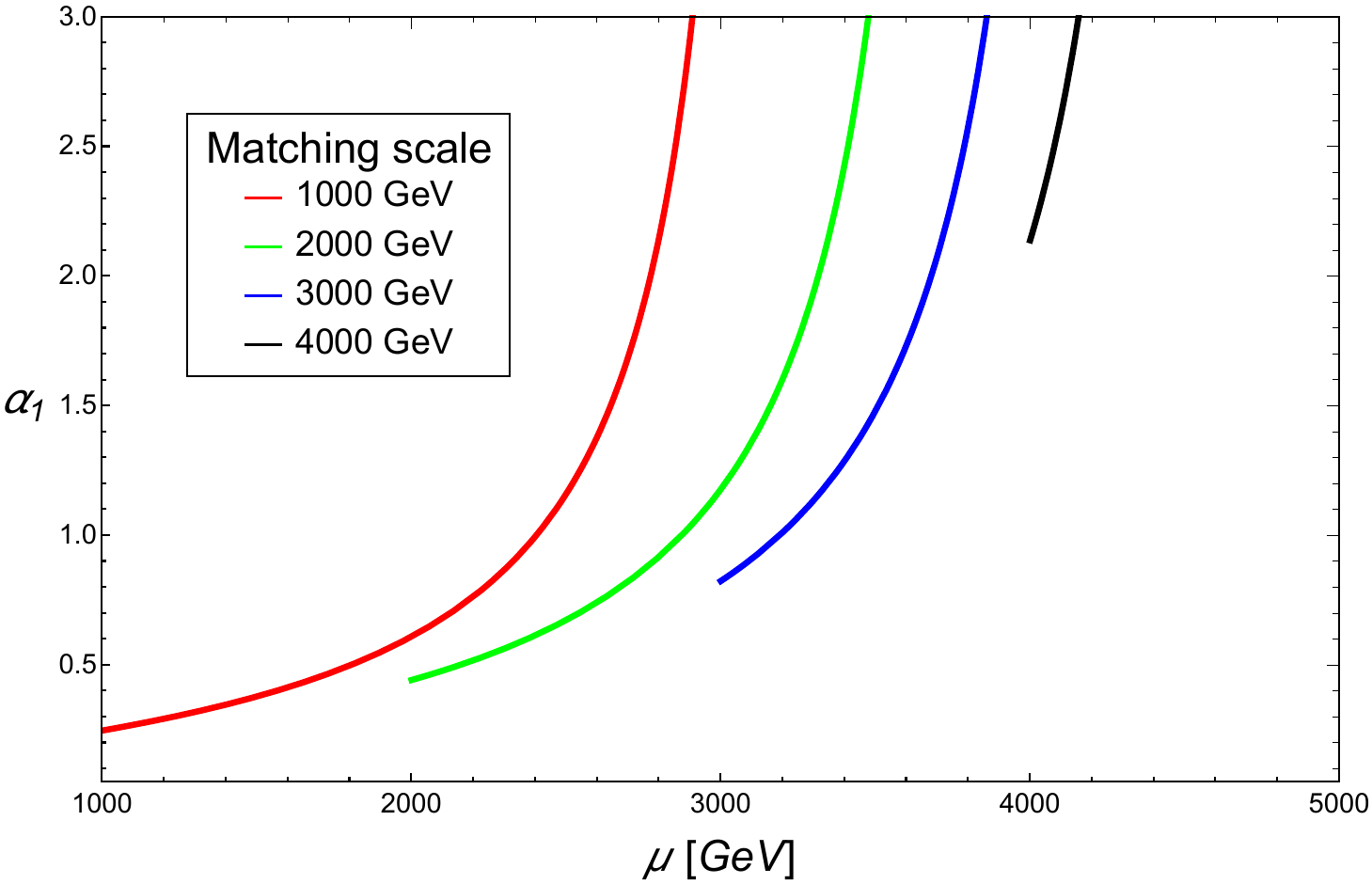}
		\caption{Running of the coupling $\a_1$ in the minimal 331 Model.}
		\label{fig:fig-d}
	\end{subfigure}
	\caption{Behaviour of the matching condition between the Standard Model and 331 Model, which make it clear how the Landau pole at the TeV scale emerge from the matching with the Standard Model.}
	\label{fig:my-fig}
\end{figure}

Once the orthogonal rotations have been performed and the Lagrangian has been written in terms of mass eigenstates of bosonic fields, it is possible to extract the values of the couplings. In particular, it possible to derive the expression of the electric charge in terms of the couplings of the minimal 331  
\begin{equation}
	e=\frac{g_1 g_2}{\sqrt{ g^2_2+4 g_1^2}}\ ,
\end{equation}
with the embedding of the charge operator given by
\begin{equation}
	Q=T^3+\sqrt{3}T^8+X \mathbb{1}\ .
\end{equation}
One of the most interesting features of the model is that the embedding of the Standard Model gauge group into the 331 Model gauge group induces a bound on the UV completion of the model \cite{martinez_landau_2007}. From \eqref{boundd}
it is clear that we need to satisfy the following condition
\begin{equation}\label{dis}
	\sin^2\theta_W\leq \frac{1}{4}\ ,
\end{equation}
in order to guarantee that the $g_1$ coupling of the minimal 331 Model is finite.
When $\sin^2\theta_W(\mu)=1/4$, the coupling constant $g_1(\mu)$ diverges, indicating a Landau pole in the renormalization group evolution of the model, which causes the theory to lose its perturbative character even at energy scales lower than $\mu$. With the particle content of the Standard Model the condition $\sin^2\theta_W(\mu)=1/4$ is reached at an energy scale of around $4$ TeV, the presence of additional particle at the TeV scale can make this behaviour even faster, loosing perturbative character before the $4$ TeV. The alternative scenario where $g_2$ tends towards zero is disregarded, as $g_2$ coincides with the Standard Model's $SU(2)_L$ coupling, $g_2$, due to the full embedding of $SU(2)_L$ into $SU(3)_L$.  We show the running of the Standard Model couplings $g_2$ in Figure \ref{fig:fig-a} and $g_y$ in Figure \ref{fig:fig-b} and in Figure \ref{fig:fig-c} we show how the matching condition of the coupling $g_1$ evolve in terms of the matching scale. Here we have defined 
\begin{equation}
	\overline{g_1}=\frac{g_2 g_y}{\sqrt{g_2^2-3 g_y^2}}  \ ,
\end{equation}
which express the value of the $U(1)_X$ coupling in terms of the Standard Model. Finally the evolution of $g_1$ in the context of the 331 Model is showed in Figure \ref{fig:fig-d} where we have plotted the quantity
\begin{equation}
	\a_1=\frac{g_1^2}{4\p}
\end{equation}
which make it clear that when $\a_1>1$ the theory looses perturbativity. Some more recent study was performed in \cite{w_barela_3-3-1_2024}.

The occurrence of a Landau-like pole in the minimal 331 model is not surprising, as many non 
asymptotically free theories exhibit similar behavior. What distinguishes some of these models is the possibility of encountering this behavior at energies as low as a few TeVs. Consequently, the cutoff scale, $\Lambda_{_{cutoff}}$, cannot be removed by taking $\Lambda_{_{cutoff}} \to\infty$, as in other renormalizable theories. 

From a phenomenological perspective, this result is not overly concerning. The necessity of embedding QED within the electroweak theory at energies of a few hundred GeVs, along with the requirement to account for weak and strong corrections in calculations of physical observables, is already acknowledged. Nevertheless, as a mathematical exercise, studying pure QED at infinitesimal distances proves intriguing. Lattice calculations suggest that chiral symmetry breaking within QED mitigates the Landau pole issue by shifting it above the cutoff scale. Interestingly, the potential existence of the Landau pole or the triviality of the theory arises even at low orders in perturbation theory, suggesting that this phenomenon is not merely a perturbative artifact.\\
The renormalization group offers qualitative insights into the asymptotic behavior of theories at very high energies, even when coupling constants at the relevant scale prohibit the use of perturbation theory. However, it's essential to remember that both QED and the Standard Model are effective, not fundamental, theories. Consequently, effective operators with dimensions higher than $d=4$ must be considered for a realistic continuum limit in lattice calculations. Thus, employing the pure versions of these models remains inconclusive, and the renormalization group may provide valuable insights into this issue within minimal 331 Model.

\section{Higgs Sector}
The inclusion of the sextet representation in the potential enriches the phenomenology of the model and enlarges the number of physical states in the spectrum. In fact we now have, after electroweak symmetry breaking (EWSB) $SU(3)_L\times U(1)_X \to SU(2)_L\times U(1)_Y\to   U(1)_{\rm{em}}$,
five scalar Higgses, three pseudoscalar Higgses, four charged Higgses and three doubly-charged Higgses. The (lepton-number conserving) potential of the model is given by  \cite{tully_scalar_2003}
\begin{align}\label{pott}
	V=&m_1\r^\dagger\r+m_2\eta^\dagger\eta+m_3\chi^\dagger\chi+\l_1(\rho^\dagger\r)^2+\l_2(\eta^\dagger\eta)^2+\l_3(\chi^\dagger\chi)^2\nonumber\\
	&+\l_{12}\r^\dagger\r\eta^\dagger\eta+\l_{13}\r^\dagger\r\chi^\dagger\chi+\l_{23}\chi^\dagger\chi\eta^\dagger\eta+\z_{12}\r^\dagger\eta\eta^\dagger\r+\z_{13}\r^\dagger\chi\chi^\dagger\r+\z_{23}\eta^\dagger\chi\chi^\dagger\eta\nonumber\\
	&+m_4\Tr(\s^\dagger\s)+\l_4(\Tr(\s^\dagger\s))^2+\l_{14}\r^\dagger\r\Tr(\s^\dagger\s)+\l_{24}\eta^\dagger\eta\Tr(\s^\dagger\s)+\l_{34}\chi^\dagger\chi\Tr(\s^\dagger\s)\nonumber\\
	&+\l_{44}\Tr(\s^\dagger\s\s^\dagger\s)+\z_{14}\r^\dagger\s\s^\dagger\rho+\z_{24}\eta^\dagger\s\s^\dagger\eta+\z_{34}\chi^\dagger\s\s^\dagger\chi\nonumber\\
	&+(\sqrt{2}f_{\r\eta\chi}\epsilon^{ijk}\r_i\eta_j\chi_k+\sqrt{2}f_{\r\s\chi}\r^T\s^\dagger\chi\nonumber\\
	&+\xi_{14}\epsilon^{ijk}\r^{*l}\s_{li}\r_j\eta_k+\xi_{24}\epsilon^{ijk}\epsilon^{lmn}\eta_i\eta_l\s_{jm}\s_{kn}+\xi_{34}\epsilon^{ijk}\chi^{*l}\s_{li}\chi_j\eta_k )+\textnormal{h.c.}\ .
\end{align}
In principle it is possible to extend the potential with additional lepton number violating terms which are singlet under 331 gauge group \cite{tully_scalar_2003}, this possibility has been largely discussed in \cite{fonseca_lepton_2016}.
The EWSB mechanism will cause a mixing among the Higgs fields \cite{costantini_theoretical_2020}. From Eq. (\ref{pott}) it is possible to obtain the explicit expressions of the mass matrices of the scalar, pseudoscalar, charged and doubly-charged Higgses, by using standard procedures. In the broken Higgs phase, the minimization conditions
\be\label{mincond}
\frac{\partial V}{\partial v_\phi}=0, \quad \langle \phi^0\rangle=v_\phi, \quad \phi=\rho, \eta, \chi, \sigma
\ee
will define the tree-level vacuum, one loop contributions to the vacuum stability have been recently analyzed in \cite{dorsch_vacuum_2024} for a simpler model version namely the economical 331 Model \cite{foot_su3_mathitlensuremathbigotimesu1_mathitn_1994,ponce_analysis_2002}. We remind that we are considering massless neutrinos by choosing the vev of the neutral field $\sigma_2^0$ to be zero. This was the choice in Frampton's original formulation. It can be 
generalized in order to give a small Majorana neutrino mass to the neutrinos \cite{tully_generating_2001}.

The explicit expressions of the minimization conditions are then given by
\begin{align}\label{minpot1}
	m_1 v_\rho + \lambda_1 v_\rho^3 + \frac{1}{2}\lambda_{12}v_\rho v_\eta^2-f_{\rho\eta\chi} v_\eta v_\chi+\frac{1}{2}\lambda_{13}v_\rho v_\chi^2 - \frac{1}{\sqrt2}\xi_{14}v_\rho v_\eta v_\sigma + f_{\rho\sigma\chi}v_\chi v_\sigma&\\
	+ \frac{1}{2}\lambda_{14}v_\rho v_\sigma^2 + \frac{1}{4}\zeta_{14}v_\rho v_\sigma^2&=0\nonumber\\
	m_2 v_\eta + \frac{1}{2}\lambda_{12}v_\rho^2 v_\eta +\lambda_2 v_\eta^3 - f_{\rho\eta\chi} v_\rho v_\chi +\frac{1}{2}\lambda_{23} v_\eta v_\chi^2 - \frac{1}{2\sqrt2}\xi_{14}v_\rho^2 v_\sigma+ \frac{1}{2\sqrt2}v_\chi^2 v_\sigma&\\
	+\frac{1}{2}\lambda_{24} v_\eta v_\sigma^2-\xi_{24} v_\eta v_\sigma^2&=0\nonumber\\
	m_3 v_\chi + \lambda_3 v_\chi^3 + \frac{1}{2} \lambda_{13} v_\rho^2 v_\chi - f_{\rho\eta\chi} v_\rho v_\eta +\frac{1}{2}\lambda_{23}v_\eta^2 v_\chi +\frac{1}{\sqrt2}\xi_{34}v_\eta v_\chi v_\sigma + f_{\rho\sigma\chi} v_\rho v_\sigma&\\
	+\frac{1}{2}\lambda_{34} v_\chi v_\sigma^2 + \frac{1}{4}\zeta_{34} v_\chi v_\sigma^2&=0\nonumber\\
	\label{minpot2}
	m_4 v_\sigma + \frac{1}{2}\lambda_{14}v_\rho^2 v_\sigma + \lambda_{44} v_\sigma^3 + \frac{1}{2}\lambda_4 v_\sigma^3 + f_{\rho\sigma\chi} v_\rho v_\chi - \frac{1}{2\sqrt2} \xi_{14} v_\rho^2 v_\eta + \frac{1}{2\sqrt2} \xi_{34} v_\eta v_\chi^2&\\
	+\frac{1}{2}\lambda_{14}v_\rho^2 v_\sigma + \frac{1}{4}\zeta_{14} v_\rho^2 v_\sigma + \frac{1}{2}\lambda_{24} v_\eta^2 v_\sigma - \xi_{24} v_\eta^2 v_\sigma + \frac{1}{2} \lambda_{34} v_\chi^2 v_\sigma + \frac{1}{4} \zeta_{34} v_\chi^2 v_\sigma&=0
	\nonumber
\end{align}
These conditions are inserted into the tree-level mass matrices of the CP-even and CP-odd Higgs sectors, derived from $M_{ij}=\left.{\partial^2 V}/{\partial \phi_i\partial \phi_j}\right|_{vev}$, where $V$ is the potential in
Eq.~(\ref{pott}). The mass eigenstates are defined as follow
\begin{equation}
	h=R^S \begin{pmatrix}
		{\rm Re}\, \rho^0\\ {\rm Re}\, \eta^0\\ {\rm Re}\, \chi^0 \\ {\rm Re}\, \sigma^0_1\\ \s^{0}_2
	\end{pmatrix}
	Ah=R^P   \begin{pmatrix}
		{\rm Im}\, \rho^0\\ {\rm Im}\, \eta^0\\ {\rm Im}\, \chi^0 \\ {\rm Im}\, \sigma^0_1
	\end{pmatrix}
	H^+=R^C \begin{pmatrix}
		\r^{+}\\ \chi^{+}\\ \eta^{+}_1\\ \eta^{+}_2\\ \s^{+}_1 \\ \s^{+}_2
	\end{pmatrix}
	H^{++}=R^{2C} \begin{pmatrix}
		\r^{++}\\ \chi^{++}\\ \s^{++}_1 \\ \s^{++}_2
	\end{pmatrix}
\end{equation}
where the explicit expressions of the mass matrices
are too cumbersome to be presented here, and are given in Appendix \ref{massmat}. \\
In this case we have 5 scalar Higgs bosons, one of them will be the SM
Higgs of mass about 125 GeV, along with 4 neutral pseudoscalar Higgs bosons, out of which, 2  are the Goldstones of the $Z$ and the $Z^\prime$ massive vector bosons. In addition, there are 6 charged Higgses, 2 of which are the charged Goldstones, and 3 are doubly-charged Higgses, one of which is a Goldstone boson.\\
Hereafter we shall give the schematic expression of the physical Higgs states,
after EWSB, in terms of the gauge eigenstates, whose expressions contain only the vev of the various fields.
In the following equations, ${\rm R}_{ij}^K\equiv{\rm R}_{ij}^K(m_1,m_2,m_3,\lambda_1,\lambda_2,\ldots)$ refers to the rotation matrix of each Higgs sector that depends on all the parameters of the potential in Eq.~(\ref{pott}). Starting from the scalar (CP-even) Higgs bosons we have
\bea
H_i = {\rm R}_{i1}^S {\rm Re}\, \rho^0 + {\rm R}_{i2}^S {\rm Re}\, \eta^0 +
{\rm R}_{i3}^S {\rm Re}\, \chi^0 +{\rm R}_{i4}^S {\rm Re}\, \sigma^0_1 +
{\rm R}_{i5}^S {\rm Re}\, \sigma_2^0,
\eea 
expressed in terms of the rotation matrix of the scalar components
${\rm R}^S$. There are similar expressions for the pseudoscalars 
\bea
Ah_i = {\rm R}_{i1}^P {\rm Im}\, \rho^0 + {\rm R}_{i2}^P {\rm Im}\,
\eta^0 +{\rm R}_{i3}^P{\rm Im}\, \chi^0 +{\rm R}_{i4}^P {\rm Im}\, \sigma^0_1 +{\rm R}_{i5}^P {\rm Im}\, \sigma_2^0
\eea 
in terms of the rotation matrix of the pseudoscalar components ${\rm R}^P$.
Here, however, we have two Goldstone bosons responsible for the generation of the masses of the neutral gauge bosons $Z$ and $Z^\prime$ given by
\begin{align}
	A_0^1 &= \frac{1}{N_1}\left(v_\rho {\rm Im}\,\rho^0 -v_\eta {\rm Im}\,\eta^0 + v_\sigma {\rm Im}\,\sigma^0_1\right),\,\qquad N_1=\sqrt{v_\rho^2+v_\eta^2+v_\sigma^2}\ ;\\
	A_0^2 &= \frac{1}{N_2}\left(-v_\rho {\rm Im}\,\rho^0+ v_\chi {\rm Im}\,\chi^0\right),\,\qquad N_2=\sqrt{v_\rho^2+v_\chi^2}.
\end{align}

For the charged Higgs bosons the interaction eigenstates are
\bea
H_i^+ = {\rm R}^{C}_{i1}\rho^{+} + {\rm R}^{C}_{i2}(\eta^{-})^* + {\rm R}^{C}_{i3}\eta^{+} + {\rm R}^{C}_{i4}(\chi^{-})^* + {\rm R}_{i5}^C \sigma_1^+ + {\rm R}_{i6}^C (\sigma_2^-)^*,
\eea 
with ${\rm R}^C$ being a rotation matrix of the charged sector. 
We recall that even in this case two $H^+_i$ are massless  Goldstones bosons, because in the minimal 331 model there are the $W^\pm$ and the $V^\pm$ gauge bosons that both become massive after EWSB. The explicit expressions of the Goldstones are
\begin{align}
	H_{W}^+ &= \frac{1}{N_W} \left(-v_\eta\eta^+ + v_\chi (\chi^-)^* + v_\sigma (\sigma_2^-)^*\right),\,\qquad N_W=\sqrt{v_\eta^2 + v_\chi^2 + v_\sigma^2}; \\
	H_{V}^+ &= \frac{1}{N_V} \left(v_\rho \rho^+ - v_\eta (\eta^-)^* + v_\sigma \sigma_1^+\right),\,\qquad N_V=\sqrt{v_\rho^2 + v_\eta^2 + v_\sigma^2}.
\end{align}
In particular, we are interested in the doubly-charged Higgses, where the number of physical states, after EWSB, is three, whereas 
we would have had only
one physical doubly-charged Higgs, if we had not included the
sextet. The physical doubly-charged Higgs states are expressed in
terms of the gauge eigenstates and the elements of the
rotation matrix ${\rm R}^C$ as
\bea
H_i^{++} ={\rm R}^{2C}_{i1}\rho^{++} + {\rm R}^{2C}_{i2}(\chi^{--})^* + {\rm R}^{2C}_{i3}\sigma_1^{++} + {\rm R}^{2C}_{i4}(\sigma_2^{--})^*.
\eea
In particular, the structure of the corresponding Goldstone boson is
\bea
H_0^{++} =\frac{1}{N}\left(-v_\rho\rho^{++} + v_\chi(\chi^{--})^* - \sqrt2v_\sigma\sigma_1^{++} + \sqrt2v_\sigma(\sigma_2^{--})^*\right)
\eea
where $N=\sqrt{v_\rho^2+v_\chi^2+4v_\sigma^2}$ is a normalization factor.

\section{The Yukawa Sector}\label{331lep}
The model presented in the previous section exhibits
the interesting feature of having both scalar and vector doubly-charged
bosons, which is a peculiarity of the minimal version of the 331 model.
In fact it is possible to consider various versions of the $SU(3)_c\times SU(3)_L\times U(1)_X$ gauge symmetry, usually parametrized by $\beta$ \cite{cao_diphoton_2016,buras_anatomy_2013}. We discuss the case of $\beta=\sqrt3$, corresponding to the minimal version presented here \cite{frampton_chiral_1992,pisano_su3ensuremathbigotimesu1_1992}, leading to vector bosons with electric charge equal to $\pm 2$.\\
Doubly-charged states hold particular interest due to their potential for unique characteristics in terms of permissible decay channels, such as the production of same-sign lepton pairs \cite{corcella_exploring_2018,corcella_bilepton_2017,corcella_non-leptonic_2022,corcella_vector-like_2021,Calabrese:2023ryr}. Within the framework of the minimal 331 model, an even more intriguing prospect arises. It becomes possible to discern whether a same-sign lepton pair originates from either a scalar or a vector boson. As we will elaborate, this distinction also offers insights into the existence of a higher representation within the $SU(3)_c\times SU(3)_L\times U(1)_X$ gauge group, notably the sextet.

\subsection{The triplet sector}
In the previous section we have seen that the EWSB mechanism is realized in the 331 model by giving a vev to the neutral component of the triplets $\rho$, $\eta$ and $\chi$. The Yukawa interactions for SM and exotic quarks
are obtained by means of these scalar fields and are given by 
\begin{align}
	\mathcal{L}^{Y}_{q,\ triplet}=\ &- \overline{Q}_m\brak{Y^d_{m\a} \eta^*d_{\a R}+Y_{m\a}^u\chi^*u_{m\a}}
	-\overline{Q}_3( Y^d_{3\a}\chi d_{\a R}+Y^u_{3\a}\eta u_{m\a})+\nonumber\\
	&-\overline{Q}_{m}(Y^J_{mn}\chi J_{nR})-\overline{Q}_{3}Y^J_{3}\chi J_{3R} +\textnormal{h.c.}\ 
\end{align}
where $y^i_{d}$, $y^i_u$ and $y^i_E$ are the Yukawa couplings for down-,
up-type and exotic quarks, respectively. The masses of the exotic quarks are related to the vev of the neutral component of $\rho=(0,0,v_\rho)$ via the invariants
\begin{eqnarray}
	Q_1\,  \rho^*  D_R^*, Q_1\,  \rho^*  S_R^*&\sim & (3,3,-1/3)\times (1,\bar{3},-1)\times (\bar{3},1,4/3) \nonumber \\
	Q_3\,  \rho  T_R^* &\sim& (3,\bar{3},2/3)\times (1,{3},1)\times (\bar{3},1,-5/3), 
\end{eqnarray}
responsible of the breaking $SU(3)_c\times SU(3)_L\times U(1)_X \to SU(3)_c\times SU(2)_L\times U(1)_Y$. It is clear that, being
$v_\rho\gg v_{\eta,\chi}$, the masses of the exotic quarks are
${\cal O}(\rm{TeV})$ whenever the relation $Y^J\sim \mathbb{1}$ is satisfied. As we are going to see, the model also needs a sextet.

\subsection{The sextet Yukawa coupling}
The need for introducing a sextet sector can be summarised as follows. 
A typical Dirac mass term for the leptons in the SM
is associated with the operator $\bar{l}_LH e_R$, with $l_L=(v_{eL},e_L)$ being the $SU(2)_L $ doublet, with the representation content $(\bar{2},1/2)\times (2,1/2)\times(1,-1)$ (for $l, H$ and $e_R$, respectively) in $SU(2)_L\times U(1)_Y$. In the 331 model, the $L$ and $R$ components of the lepton $(e)$ belong to the same multiplet. Consequently, identifying an $SO(1,3)\times SU(3)_L$ singlet requires two leptons in the same representation. This can be achieved (at least partially) with the operator
\begin{eqnarray}
	\mathcal{L}_{l,\, triplet}^{Yuk}&=& G^\eta_{a b}( l^i_{a \alpha}\epsilon^{\alpha \beta} l^j_{b \beta})\eta^{* k}\epsilon^{i j k} + \rm{h. c.}\nonumber\\
	&=& G^\eta_{a b}\, l^i_{a}\cdot l^j_{b}\,\eta^{* k}\epsilon^{i j k} + \rm{h. c.} 
\end{eqnarray}
where the indices $a$ and $b$ run over the three generations of flavour, $\alpha$ and $\beta$ are Weyl indices contracted in order to generate  an $SO(1,3)$ invariant ($l^i_{a}\cdot l^j_{b}\equiv l^i_{a \alpha}\epsilon^{\alpha \beta} l^j_{b \beta}$) from two Weyl fermions, and $i,j,k=1,2,3$, are $SU(3)_L$ indices. \\
The use of 
$\eta$ as a Higgs field is mandatory, since the components of the multiplet $l^j$ are $U(1)_X$ singlets. 
The representation content of the operator $l^i_a l^j_b$ according to $SU(3)_L$ is given by $3\times 3= 6 + \bar{3}$, with the 
$\bar{3}$ extracted by an anti-symmetrization over $i$ and $j$ via $\epsilon^{i j k}$. This allows one to identify 
$l^i_a l^j_b \eta^{*k}\epsilon^{i j k}$ as an $SU(3)_L$ singlet. Considering that the two leptons are anticommuting Weyl spinors, and that the $\epsilon^{\alpha\beta}$ (Lorentz) and $\epsilon^{i j k}$ ($SU(3)_L$) contractions introduce two sign flips under the $a\leftrightarrow b$ exchange, the combination 
\begin{equation}
	M_{a b}=( l^i_{a}\cdot l^j_{b })\eta^{* k}\epsilon^{i j k} 
\end{equation}
is therefore, antisymmetric under the exchange of the two flavours, implying
that even $G_{a b}$ has to be antisymmetric.  However, an antisymmetric $G^\eta_{a b}$ is not sufficient to provide mass to all the leptons.\\
In fact, the diagonalization of $G^\eta$ by means a unitary matrix 
$U$, namely $G^\eta=U \Lambda U^\dagger$, with $G^\eta$ antisymmetric in
flavour space, implies that its 3 eigenvalues are given by
$\Lambda=(0,\lambda_{22}, \lambda_{33})$, with $\lambda_{22}=-\lambda_{33}$,
i.e. one eigenvalue is null and the other two are equal in magnitude.
At the minimim of $\eta$, i.e. $\eta=(0,v_\eta,0)$, one has
\begin{equation} 
	G^\eta_{a b}M^{a b}=-Tr (\Lambda\, U M U^\dagger)=
	2 v_{\eta}\lambda_{22}\, U_{2 a}\, l^{1}_{a}\cdot l^{3}_{b}\,U_{2 b}^* + 2 v_{\eta}\lambda_{33}\, U_{3 a}\, l^{1}_{a}\cdot l^{3}_{b}\,U_{3 b}^*, 
\end{equation}
with $ l^1_a=e_{a L}$ and  $l^3_b=e_{b R}^{c}$. Introducing the linear combinations 
\begin{equation} 
	E_{2 L}\equiv U_{2 a}\, l^1_a=U_{2 a} \, '\, e_{a L} \qquad U_{2 b}^*\,
	l^3_b=U_{2 b}^*\, e_{b R}^{c} = i\sigma_2(U_{2 b} \,e_{b R})^*\equiv
	E_{2 R}^{c},
\end{equation}
then the antisymmetric contribution in flavour space becomes 
\begin{equation}
	\mathcal{L}_{l, \, triplet}^{Yuk}= 2 v_{\eta}\lambda_{22} \left( E_{2 L}E_{2 R}^{c}  - E_{3 L}E_{3 R}^{c}\right),
\end{equation}
which is clearly insufficient to generate the lepton masses of three
non-degenerate
lepton families. We shall solve this problem by introducing a second 
invariant operator, with the inclusion of a sextet $\sigma$
\begin{equation}
	\sigma=\left(
	\renewcommand*{\arraystretch}{1.5}
	\begin{array}{ccc}
		\sigma_1^{++}&\sigma_1^+/\sqrt2&\sigma^0/\sqrt2\\
		\sigma_1^+/\sqrt2&\sigma_1^0&\sigma_2^-/\sqrt2\\
		\sigma^0/\sqrt2&\sigma_2^-/\sqrt2&\sigma_2^{--}
	\end{array}
	\right)\in(1,6,0),
\end{equation}
leading to the Yukawa term
\begin{equation}\label{lag}
	\mathcal{L}_{l, sextet}^{{Yuk.}}= G^\sigma_{a b} l^i_a\cdot l^j_b \sigma^*_{i,j},
\end{equation}
which allows one to build a singlet out of the representation
$6$ of $SU(3)_L$, contained in $l^i_a\cdot l^j_b$, by combining it with
the flavour-symmetric
$\sigma^*$, i.e. $\bar{6}$.
Notice that $G^\sigma_{a b}$ is symmetric in flavour space. \\
It's interesting to note that without considering the sextet, a doubly-charged scalar wouldn't be able to decay into same-sign leptons. This is because, without the sextet, the interaction responsible for the leptons only involves the scalar triplet, denoted as $\eta$, which doesn't contain a doubly-charged state. 
%So, if we detect a decay like $H^{\pm\pm}\to l^\pm l^\pm$, it would directly indicate the presence of the sextet in the 331 model.

\subsection{Lepton mass matrices}
Let's now come to discuss the lepton mass matrices in the model. They are related to the Yukawa interactions by the Lagrangian 
\begin{equation}
	\mathcal{L}_{l}^{{Yuk.}}=\mathcal{L}_{l, sextet}^{{Yuk.}} + \mathcal{L}_{l, triplet}^{{Yuk.}} + \rm{h. c.}
\end{equation}
and are combinations of triplet and sextet contributions.
The structure of the mass matrix that emerges
from the vevs of the neutral components of  $\eta$ and $\sigma$ is thus
given by
\begin{equation} 
	\mathcal{L}_{l}^{{Yuk.}}=\left(\sqrt{2} \sigma_0 G_{a,b}^\sigma  +2 v_\eta G^\eta_{a b}\right) (e_{a L}\cdot e_{b R}^{c}) 
	+\sigma_1^0 G^\sigma_{a b} \left(\nu_L^T i \sigma_2 \nu_L\right)  +\rm{h. c.},
\end{equation}
which generates a Dirac mass matrix for the charged leptons $M_{ab}^l$
and a Majorana mass matrix for neutrinos $M_{a b}^{\nu_l}$
\begin{equation}\label{mlgen}
	M^l_{a b} =\sqrt{2} \langle\sigma_0\rangle\, G_{a,b}^\sigma  +2 v_\eta\, G^\eta_{a b} \qquad , \qquad M_{a b}^{\nu_l}=\langle \sigma^0_1\rangle \, G^\sigma_{a b}. 
\end{equation}
In the expression above $\langle\sigma^0\rangle$ and $\langle\sigma_1^0\rangle$ are the vacuum expectation values of the neutral components of $\sigma$. For a vanishing $G^\sigma$, as we have already discussed, we will not be able to generate the lepton masses consistently, nor any mass for the neutrinos, i.e.   
\begin{equation}\label{mltip}
	M^l_{a b}=2 v_\eta\, G^\eta_{a b} \qquad , \qquad M^{\nu_l}=0.
\end{equation} 
On the contrary, in the limit $G^\eta\to0$, Eq.~(\ref{mlgen}) becomes 
\begin{equation}\label{mlsext}
	M^l_{a b}=\sqrt{2} \langle\sigma_0\rangle\, G_{ab}^\sigma \qquad,
	\qquad M^{\nu_l}_{a,b}=\frac{\langle\sigma_1^0\rangle}{\sqrt2}G^\sigma_{a b},
\end{equation}
which has some interesting consequences.
Since the Yukawa couplings are the same for both leptons and neutrinos,
we have to require $\langle\sigma_1^0\rangle\ll\langle\sigma^0\rangle$, in order to obtain small neutrino masses. For the goal of our analysis, we will
assume that the vev of $\sigma_1^0$ vanishes, i.e.
$\langle\sigma_1^0\rangle\equiv0$. Clearly,
if the matrix $G^\sigma$ is diagonal in flavour space, from Eq.
(\ref{mlsext}) we will
immediately conclude that the Yukawa coupling $G^\sigma$ has
to be chosen to be proportional to the masses of the SM leptons.
An interesting consequence of this is that the decay $H^{\pm\pm}\to l^\pm l^\pm$, which is also proportional to $G^\sigma$, and therefore, to the lepton masses,
will be enhanced for the heavier leptons, in particular for the $\tau$,
as thoroughly discussed in  \cite{tonasse_decay_2012}.
This is an almost unique situation which is not encountered in other models with doubly-charged scalars decaying into same-sign leptons \cite{muhlleitner_note_2003}.

\section{Flavor Physics in the minimal 331 Model}
One of the features of the minimal 331 Model lies in its arrangement of fermions within triplets of $SU(3)_L$. However, to maintain anomaly cancellation, it becomes necessary to assign one of the quark families to a different representation than the other two, ensuring an equal number of triplets and anti-triplets in the fermion sector.\\
This introduces several complexities, particularly concerning flavor physics within the model. First, to achieve the spontaneous symmetry breaking of the 331 gauge symmetry to the $SU(3)_C\times U(1)_{em}$ symmetry, at least three scalar triplets of $SU(3)_L$ must be introduced. While these are sufficient to impart masses to quarks in the quark sector, the flavor structure becomes intricate due to the differing group representations of the three quark families.\\
Conversely, in the lepton sector, a fundamentally different situation arises. Realistic masses cannot be obtained using only three triplets. As previously argued, the introduction of a scalar sextet belonging to $(\textbf{1},\textbf{6},0)$ becomes necessary to generate appropriate masses for charged leptons.\\
A general feature of models of this kind, where mass terms arise from different scalar fields, is the introduction of flavor-changing neutral currents mediated by neutral scalars \cite{pleitez_challenges_2022}.

\subsection{Quark sector}

Let's revisit the presence of three scalar triplets within the model. The first triplet,

\begin{equation}
	\rho=\begin{pmatrix}
		\rho^{++}\\ \rho^+\\ \rho^0
	\end{pmatrix}\in (\textbf{1},\textbf{3},1)\ ,
\end{equation}

acquires a vacuum expectation value (vev) on the order of the spontaneous symmetry breaking of the 331 symmetry. The other two triplets,

\begin{equation}
	\eta=\begin{pmatrix}
		\eta^{+}_1\\ \eta^0\\ \eta^-_2
	\end{pmatrix}\in (\textbf{1},\textbf{3},0)\ ,
\end{equation}

\begin{equation}
	\chi=\begin{pmatrix}
		\chi^{0}\\ \chi^-\\ \chi^{--}
	\end{pmatrix}\in (\textbf{1},\textbf{3},-1)\ ,
\end{equation}

acquire non-zero vevs at the electroweak scale. In the quark sector, the Yukawa interactions are described by

\begin{align}\label{yuklagq}
	\mathcal{L}^{Y}_{q,\ triplet}=\ &- \overline{Q}_m\brak{Y^d_{m\a} \eta^*d_{\a R}+Y_{m\a}^u\chi^*u_{m\a}}
	-\overline{Q}_3( Y^d_{3\a}\chi d_{\a R}+Y^u_{3\a}\eta u_{m\a})+\nonumber\\
	&-\overline{Q}_{m}(Y^J_{mn}\chi J_{nR})-\overline{Q}_{3}Y^J_{3}\chi J_{3R} +\textnormal{h.c.}\ .
\end{align}

The mass matrices for the up-type and down-type quarks arise when all scalar triplets acquire real vevs: $\rho\rightarrow v_\rho/\sqrt{2}$, $\eta\rightarrow v_\eta/\sqrt{2}$, and $\chi\rightarrow v_\chi/\sqrt{2}$. Both matrices involve contributions from the triplets $\eta$ and $\chi$. Exotic quark mass terms emerge following the initial spontaneous symmetry breaking.

Since there are two quarks, namely $D$ and $S$ with electric charge $Q=-4/3\ e$, they undergo mixing via a Cabibbo-like $2\times2$ matrix
\begin{equation}
	\begin{pmatrix}
		D'_R\\S'_R
	\end{pmatrix}
	=\Tilde{V}_R^{-1}\begin{pmatrix}
		D_R\\S_R
	\end{pmatrix}\ ,\ \ \ \ \ \ \ \begin{pmatrix}
		D'_L\\S'_L
	\end{pmatrix}
	=\Tilde{V}_L^{-1}\begin{pmatrix}
		D_L\\S_L
	\end{pmatrix}\ .
\end{equation}

Notably, only the quark $T$ remains unmixed, being the sole quark with $Q=5/3\ e$. The resulting mass matrices for ordinary quarks from the Lagrangian in Equation \eqref{yuklagq} are as follows

\begin{equation}
	M_u=\begin{pmatrix}
		v_\chi Y^u_{11} & v_\chi Y^u_{12} &v_\chi Y^u_{13} \\
		v_\chi Y^u_{21} & v_\chi Y^u_{22} &v_\chi Y^u_{23}\\
		v_\eta Y^u_{31} & v_\eta Y^u_{32} &v_\eta Y^u_{33}\\
	\end{pmatrix}\ ,
\end{equation}

\begin{equation}
	M_d=\begin{pmatrix}
		v_\eta Y^d_{11} & v_\eta Y^d_{12} &v_\eta Y^d_{13} \\
		v_\eta Y^d_{21} & v_\eta Y^d_{22} &v_\eta Y^d_{23} \\
		v_\chi Y^d_{31} & v_\chi Y^d_{32} &v_\chi Y^d_{33} \\
	\end{pmatrix}\ .
\end{equation}

In the minimal version, there is typically no reason to initially place one type of quark in the diagonal basis, unless some specific additional symmetry is introduced. Therefore, similar to the Standard Model, we proceed by independently rotating down-type and up-type quarks into their mass eigenstates
\begin{equation}\label{rotu}
	\begin{pmatrix}
		d'_L\\s'_L\\b'_L
	\end{pmatrix}
	=V_L^{-1}\begin{pmatrix}
		d_L\\s_L\\b_L
	\end{pmatrix}\ \ \ \ \ \ \begin{pmatrix}
		d'_R\\s'_R\\b'_R
	\end{pmatrix}
	=V_R^{-1} \begin{pmatrix}
		d_R\\s_R\\b_R
	\end{pmatrix}
\end{equation}
\begin{equation}\label{rotd}
	\begin{pmatrix}
		u'_L\\c'_L\\t'_L
	\end{pmatrix}
	=U_L^{-1}\begin{pmatrix}
		u_L\\c_L\\t_L
	\end{pmatrix} \ \ \ \ \ \  \begin{pmatrix}
		u'_R\\c'_R\\t'_R
	\end{pmatrix}
	=U_R^{-1}\begin{pmatrix}
		u_R\\c_R\\t_R
	\end{pmatrix}
\end{equation}
Here, the primed fields represent flavor eigenstates, while unprimed fields denote mass eigenstates. Matrices $U_L$, $U_R$, $V_L$, and $V_R$ are unitary matrices satisfying
\begin{equation}
	V_L^\dagger V_L =V_R^\dagger V_R=U_L^\dagger U_L=U_R^\dagger U_R= \mathbb{1}\ ,
\end{equation}
which diagonalize $M_u$ and $M_d$ respectively via bi-unitary transformations
\begin{equation}
	U_L^\dagger M_u U_R=\hat{M}_u\ \ \ \ \  V_L^\dagger M_d V_R=\hat{M}_d\ ,
\end{equation}
where $\hat{M}_{u}$ and $\hat{M}_{d}$ are diagonal $3\times 3 $ matrices containing quark masses.\\
In the Standard Model all the three families are placed within the same group representation of the gauge symmetry, this results in no family distinction in the Lagrangian apart from the Yukawa sector. Here, non-diagonal terms between quark masses arise if the couplings are non-diagonal. The same field rotation procedure is implemented in the Standard Model to diagonalize Yukawa interactions while leaving the Lagrangian mostly unchanged, the only part sensitive to flavour rotation is the $W$ boson interactions, where the CKM matrix arises as
\begin{equation}
	V_{CKM}=U_L^\dagger V_L\ .
\end{equation}
Every other sector in the Lagrangian remains unaffected. Consequently, the GIM mechanism is naturally implemented, since there are no flavor-changing neutral currents (FCNCs) at tree-level, resulting in a natural suppression of these processes, which can only occur at loop-level. Up-type quark masses can be assumed diagonal from the outset, while down-type quark flavor and mass eigenstates can be related through the CKM matrices, as any other rotation effects cancel out in the Lagrangian.\\
However, in the context of the minimal 331 Model, the situation is fundamentally different. 
As previously emphasized, the appealing feature of constraining the number of families is that it causes every sector of the Lagrangian to be sensitive to flavor, owing to the different group representations to which the quarks are assigned. Consequently, flavor rotation matrices persist in various combinations in fermion interactions, either with scalars or gauge bosons.
From the Yukawa Lagrangian in Equation \eqref{yuklagq}, we can derive the fermion-scalar interactions after the field rotations discussed previously
\begin{equation}
	\mathcal{L}^{cc}_{scalar}=-\overline{\textbf{d}}_LV_L S_{du}^{cc}U_R^\dagger\textbf{u}_R-\overline{\textbf{u}}_LU_L S_{ud}^{cc}V_R^\dagger\textbf{d}_R+\textnormal{h.c.}\ ,
\end{equation}
where $\textbf{d}$ and $\textbf{u}$ represent the down-type and up-type flavor vectors in the mass basis, and
\begin{equation}\label{sdu}
	S_{du}^{cc}=\begin{pmatrix}
		\chi^- Y^u_{11} & \chi^- Y^u_{12} & \chi^- Y^u_{13} \\
		\chi^- Y^u_{21} & \chi^- Y^u_{22} & \chi^- Y^u_{23}\\
		\eta^-_1 Y^u_{31} & \eta^-_1 Y^u_{32} & \eta^-_1 Y^u_{33}\\
	\end{pmatrix}\ ,
\end{equation}

\begin{equation}\label{sud}
	S_{ud}^{cc}=\begin{pmatrix}
		\eta^+_1 Y^d_{11} &  \eta^+_1 Y^d_{12} & \eta^+_1 Y^d_{13} \\
		\eta^+_1 Y^d_{21} &  \eta^+_1 Y^d_{22} & \eta^+_1 Y^d_{23} \\
		\chi^+ Y^d_{31} &  \chi^+ Y^d_{32} & \chi^+ Y^d_{33} \\
	\end{pmatrix}\ .
\end{equation}
In the expression above, all the charged scalars are interaction eigenstates, and we need to perform a rotation in \eqref{sud} and \eqref{sdu} in order to extract the corresponding mass eigenstates.
It's evident that in general, the combination of flavor rotation matrices cannot be further reduced, and they persist in a combination distinct from the CKM matrix. This behavior also manifests in the neutral currents mediated by scalar fields, where the Lagrangian becomes

\begin{equation}
	\mathcal{L}^{nc}_{scalar}=-\overline{\textbf{u}}_L U_L S_{u}^{nc}U_R^\dagger\textbf{u}_R-\overline{\textbf{d}}_LV_L S_{d}^{nc}V_R^\dagger\textbf{d}_R+\textnormal{h.c.}\ ,
\end{equation}
with
\begin{equation}\label{sdu}
	S_{u}^{nc}=\begin{pmatrix}
		\chi^0 Y^u_{11} & \chi^0 Y^u_{12} & \chi^0 Y^u_{13} \\
		\chi^0 Y^u_{21} & \chi^0 Y^u_{22} & \chi^0 Y^u_{23}\\
		\eta^0 Y^u_{31} & \eta^0 Y^u_{32} & \eta^0 Y^u_{33}\\
	\end{pmatrix}
\end{equation}
and
\begin{equation}\label{sud}
	S_{d}^{nc}=\begin{pmatrix}
		\eta^0 Y^d_{11} &  \eta^0 Y^d_{12} & \eta^0 Y^d_{13} \\
		\eta^0 Y^d_{21} &  \eta^0 Y^d_{22} & \eta^0 Y^d_{23} \\
		\chi^0 Y^d_{31} &  \chi^0 Y^d_{32} & \chi^0 Y^d_{33} \\
	\end{pmatrix}
\end{equation}

where, as in the case of charged scalars, also here we need a further rotation in order to extract the mass eigenstates of this sector.

Although the matrix combinations $U_L S_{u}^{nc}U_R^\dagger$ and $V_L S_{d}^{nc}V_R^\dagger$ resemble those in Equations \eqref{rotu} and \eqref{rotd}, they are insufficient to diagonalize the interaction. Hence, flavor-changing neutral currents in the scalar sector are a prediction of the minimal 331 Model. They can be controlled but not entirely avoided.\\
In the minimal 331 Model, eight bosons are present, including the neutral ones: the photon $A^\mu$, with straightforward interactions, $Z^\mu$ and $Z'^\mu$, and the charged ones, $W^\pm$, $V^\pm$, and the doubly charged bileptons $Y^{\pm\pm}$. However, rotations in flavor space do not leave the interactions unaffected. The d-type and u-type quarks interact with the $W^\pm$ boson as in the Standard Model
\begin{equation}
	\mathcal{L}_{duW}=\frac{g_2}{\sqrt{2}}\overline{\textbf{d}_L}\gamma^\mu V_{CKM}\textbf{u}_LW^-_\mu+\text{h.c.}
\end{equation}
Here, the CKM matrix appears as expected. Further interactions are possible between exotic quarks and ordinary quarks mediated by exotic gauge bosons. The interactions of u-type and d-type quarks with the exotic $T$ quark are given by

\begin{equation}
	\mathcal{L}_{uTV}=\frac{g_2}{\sqrt{2}}\overline{T_L}\gamma^\mu (U_{L})_{3j}\textbf{u}_{jL}V^+_\mu +\text{h.c.}
\end{equation}

\begin{equation}
	\mathcal{L}_{dTY}=\frac{g_2}{\sqrt{2}}\overline{T_L}\gamma^\mu ( V_{L})_{3j}\textbf{d}_{jL}Y^{++}_\mu +\text{h.c.}
\end{equation}

Here, only the third column of the rotation matrix appears due to the difference between the third quark generation and the other two. Similar interactions occur between up-type and down-type quarks with the other two exotic quarks $D$ and $S$
\begin{equation}\label{tt1}
	\mathcal{L}_{djY}=\frac{g_2}{\sqrt{2}}\overline{\textbf{j}_L}\gamma^\mu (O^\dagger V_{L})_{mj}\textbf{d}_{jL}V^{-}_\mu +\text{h.c.}\ \ \ \ \ \text{with } j_L=(D_L,\ S_L)
\end{equation}

\begin{equation}\label{tt2}
	\mathcal{L}_{ujY}=\frac{g_2}{\sqrt{2}}\overline{\textbf{j}_L}\gamma^\mu (O^\dagger U_{L})_{mj}\textbf{d}_{jL}Y^{++}_\mu +\text{h.c.}\ \ \ \ \ \text{with } j_L=(D_L,\ S_L)
\end{equation}
where the $O$ matrix in Equations \eqref{tt1} and \eqref{tt2} is a $2\times 2$ Cabibbo-like matrix which mixes the exotic quarks $D$ and $S$.\\
Finally, the interaction of ordinary quarks with the $Z'$ boson can be schematized as follows - omitting an overall coefficient - 
\begin{align}
	\mathcal{L}_{qqZ'}=&\biggl(\overline{\textbf{u}_L}U_L^\dagger\gamma^\mu \Y^{u}_L U_L\textbf{u}_L+\overline{\textbf{d}_L}V_L^\dagger \gamma^\mu\Y^{d}_L V_L\textbf{d}_L\nonumber\\
	&+\overline{\textbf{u}_R}U_R^\dagger \gamma^\mu\Y^{u}_R U_R\textbf{u}_R +\overline{\textbf{d}_R}V_R^\dagger \gamma^\mu\Y^{d}_R V_R\textbf{d}_R \biggl)Z'_\mu
\end{align}
where we have defined the couplings with $\Y^u_L$ and $\Y^d_L$ which are proportional to the following matrix
\begin{equation}
	\Y_L^u=\Y_L^d\propto\begin{pmatrix}
		1-2\sin^2(\theta_W)&0&0\\
		0&1-2\sin^2(\theta_W)&0\\
		0&0&-1\\
	\end{pmatrix}
\end{equation}
In this interaction, there are flavor-changing neutral currents (FCNCs) in the left-handed interactions, but the right-handed neutral currents mediated by the gauge boson $Z'$ are diagonal in flavor space, as $\Y_R^u\propto\mathbb{1}$ and $\Y_R^d\propto\mathbb{1}$ \cite{machado_flavor-changing_2013}. \\
It's noteworthy that in the quark sector, the rotation matrices of the right-handed quarks cancel out from the Lagrangian, similar to the case of the Standard Model. Conversely, the left-handed $U_L, V_L$ matrices not only survive in a combination analogous to the CKM matrix of the Standard Model, but also independently. From a practical standpoint, it is possible to redefine the fields in the interactions - using the unitary condition $U_L^\dagger U_L=\mathbb{1}$ - to construct a Lagrangian for the quark sector in which only two matrices appear: the CKM matrix and $V_L$.\\
To obtain an appropriate parameterization for the matrix $V_L$, it is necessary to initially enumerate the additional parameters present within this matrix. Upon examining all conceivable interaction terms, it becomes apparent that, subsequent to employing phase transformations of the up and down-type quarks to simplify the CKM matrix, three more potential phases emerge from transformations in the $D$, $S$, and $T$ quarks. This results in a total of six supplementary parameters, encompassing three mixing angles and three phases. However, it is evident that only the $\tilde{V}_{3j}$ elements are essential when computing FCNCs, thus allowing for a parameterization that effectively diminishes the number of parameters involved. We obtain \cite{promberger_flavor-changing_2007}

\begin{equation}
	V_L=\begin{pmatrix}
		c_{12}c_{13}            & s_{12}c_{23}e^{i \d_3}-c_{12}s_{13}s_{23}e^{i (\d_1-\d_2)}& c_{12}c_{23}s_{13}e^{i \d_1}+s_{12}s_{23}e^{i (\d_2+\d_3)}\\
		-c_{13}s_{12}e^{-i \d_3}& c_{12}c_{23}+s_{12}s_{13}s_{23}e^{i (\d_1-\d_2-\d_3)}& -s_{12}c_{23}s_{13}e^{i (\d_1-\d_3)}-c_{12}s_{23}e^{i \d_2}\\
		-s_{13}e^{-i \d_1}      & -c_{13}s_{23}e^{-i \d_2}& c_{13}c_{23}\\
	\end{pmatrix}\ .
\end{equation}
where only two additional CP violating quantities $\d_1$ and $\d_2$ appear, that are responsible for the additional CP violating effects.

\subsection{Lepton Sector}
In the minimal 331 Model, similar to quarks, the leptonic sector exhibits a multitude of parameters. Despite the $Z$ and $Z'$ interactions being diagonal in flavor space (owing to the consistent transformation behavior of the three lepton generations under the electroweak symmetries $SU(3)_L\times U(1)_X$), Flavor-Changing Neutral Currents (FCNCs) occur in the scalar sector \cite{machado_lepton_2019}. These parameters result from the diverse contributions to the mass matrices of charged leptons, mirroring the scenario observed in quarks.

The Yukawa interactions in the lepton sector must incorporate the triplet $\eta$ and the sextet, whose components are detailed as follows
\begin{equation}
	\s=\begin{pmatrix}
		\s^{++}_1&\frac{\s_1^+}{\sqrt{2}}&\frac{\s^0_1}{\sqrt{2}}\\
		\frac{\s_1^+}{\sqrt{2}}&\s_2^0&\frac{\s^-_2}{\sqrt{2}}\\ 
		\frac{\s^0_1}{\sqrt{2}}&\frac{\s^-_2}{\sqrt{2}}&\s_2^{--}\\
	\end{pmatrix}\ .
\end{equation}
In the minimal 331 Model, including the sextet in the scalar sector is necessary to assign physical masses to charged leptons. This necessity arises from how the Yukawa interaction is constructed from group theory. When combining three triplets according to $\textbf{3} \otimes \textbf{3} \otimes \textbf{3}$, the resulting invariant structure demands antisymmetry among the triplets. As a consequence, the Yukawa matrix must exhibit antisymmetry in flavor indices to allow for vanishing interactions. However, this interaction pattern leads to eigenvalues of $(0, m, -m)$ of this matrix, which is evidently an unphysical solution.\\
Once spontaneous symmetry breaking occurs, the scalars acquire vacuum expectation values as follows

\begin{equation}
	\eta=\frac{1}{\sqrt{2}}\begin{pmatrix}
		\eta^{+}_1\\\frac{1}{\sqrt{2}} v_\eta +\frac{1}{\sqrt{2}}\Sigma_\eta+\frac{i}{\sqrt{2}}\z_\eta\\ \eta^-_2
	\end{pmatrix}
\end{equation}

\begin{equation}
	\s=\frac{1}{\sqrt{2}}\begin{pmatrix}
		\s^{++}_1&\frac{\s_1^+}{\sqrt{2}}&\frac{v_\s+\Sigma_\s+i\z_\s}{2}\\
		\frac{\s_1^+}{\sqrt{2}}&\s_2^0&\frac{\s^-_2}{\sqrt{2}}\\ 
		\frac{v_\s+\Sigma_\s+i\z_\s}{2}&\frac{\s^-_2}{\sqrt{2}}&\s_2^{--}\\
	\end{pmatrix}\ .
\end{equation}
In the context of the minimal 331 Model, as originally proposed, neutrinos are massless at the tree level. However, a mechanism to generate massive neutrinos can be obtained through the scalar sextet. Indeed, the component $\s_2^0$ can also acquire a vev, which can be used in the Yukawa interactions to construct a Majorana mass term for the neutrinos
\begin{equation}
	\mathcal{L}^Y_\nu=-\overline{\nu_L^c}G^\s \nu_L\s+\text{h.c.}\ .
\end{equation}
This interaction is not invariant under flavor rotation, therefore, an additional rotation matrix appears in this sector in order to diagonalize the neutrino mass terms
\begin{equation}
	V_\nu^T M_\nu V_\nu=\hat{M}_\nu\ ,
\end{equation}
where $\hat{M}_\nu$ is a diagonal matrix containing physical neutrino masses.\\
The inclusion of this additional vev carries further implications such as the mixing between the singly charged gauge bosons $W^\pm$ and $V^\pm$, which can be predicted to be a small value of $v_{\s_2}$ chosen to yield small neutrino masses. The interactions between charged leptons, neutrinos, and $W^\pm$ remain the same as in the Standard Model

\begin{equation}
	\mathcal{L}_{l\nu W}=\frac{ig_2}{2\sqrt{2}}\left(\overline{\nu_L}V_{PMNS}^\dagger\gamma^\mu l_L+\overline{(l^c)_R}V_{PMNS}\gamma^\mu(\nu^c)_R\right)W^+_\mu+\text{h.c.}
\end{equation}
where $V_{PMNS}$ is the Pontecorvo-Maki-Nakagawa-Sakata matrix given by

\begin{equation}
	V_{PMNS}=V_l^\dagger V^\nu\ .
\end{equation}
The model also predicts that the interaction between charged leptons and neutrinos must include additional contributions from $V^\pm$

\begin{equation}\label{aa1}
	\mathcal{L}_{l\nu V}=\frac{ig_2}{2\sqrt{2}}\left(\overline{l^c_L}V_{l\nu}^*\gamma^\mu \nu_L+\overline{\nu^c_R}V_{l\nu}\gamma^\mu(l)_R\right)V^+_\mu+\text{h.c.}
\end{equation}
where the definition
\begin{equation}\label{ab1}
	V_{l\nu}=(V^\nu_L)^\dagger (V_R^l)^*\ .
\end{equation}
In this model, there are also interactions between charged leptons and doubly charged vector bosons given by the Lagrangian
\begin{equation}
	\mathcal{L}_{llY}=\frac{ig_2}{2\sqrt{2}}\left(\overline{l^c}\gamma^\mu\left(\tilde{V}_{l\nu}-\tilde{V}_{l\nu}^T\right)-\gamma^5\gamma^\mu\left(\tilde{V}_{l\nu}+\tilde{V}_{l\nu}^T\right)\right)Y^{++}
\end{equation}
with
\begin{equation}\label{ab2}
	\tilde{V}_{l\nu}=(V^l_R)^T V_L^l\ .
\end{equation}
Finally, leptons couple universally to neutral vector bosons since no distinction has been made between generations of leptons, where the $Z'$ boson also has the property to be Leptophobic \cite{gomez_dumm_leptophobic_1997}.\\
Therefore, in the minimal 331 Model, $V_\nu$ is always not equal to $V_{PMNS}$, and $V^l_L$ and $V^l_R$ both appear separately in the combination shown in Equations \eqref{ab1} and \eqref{ab2}. Because the charged lepton mass matrix includes an anti-symmetric contribution, we cannot assume it is diagonal from the beginning. This means we cannot simply set $V^l_R$ equal to $\mathbb{1}$ in the interactions in Equation \eqref{aa1}. However, if interactions between leptons and the scalar triplet $\eta$ are forbidden by some discrete symmetry, then this simplification becomes possible.\\
In this scenario, the charged lepton mass matrix is diagonalized using the same unitary matrix as the neutrino mass matrix, making the PMNS matrix straightforward. Similarly, just like in the quark sector, determining the values of the entries in the matrices $V_\nu$, $V^l_L$, and $V^l_R$ is crucial for understanding the realistic behavior of this model.

\section{Conclusions}
The 331 model stands as a remarkable embodiment of the anomaly cancellation mechanism, encompassing all fermion generations. It represents a compelling avenue for extending the fermion sector of the Standard Model in an alternative direction. Within the framework of the 331 model, the significance of anomaly constraints cannot be overstated. They serve as pivotal guiding principles, steering us away from the conventional sequential structure of the Standard Model's fermion families. Unlike the traditional model, which lacks a clear rationale regarding the number of fermion families, the 331 model offers a fresh perspective, inviting exploration and deeper understanding of fundamental particle interactions.

		\chapter{Heavy Quark Decays in the Bilepton Model}\label{ch331ph}
\addcontentsline{toc}{section}{Synopsis}
\section*{Synopsis}

This Chapter investigates the collider phenomenology of heavy exotic quarks predicted by the minimal 331 bilepton model, with a focus on their production and decay into vector bileptons at the LHC and the future FCC-$hh$ collider. Building on the theoretical framework established in the previous Chapter, the analysis concentrates on the distinctive signatures that the exotic quark sector of the 331 model can generate in high-energy experiments, completing a series of phenomenological studies on the bilepton model.

The Chapter opens by recalling the key motivational features of the 331 model from a collider perspective. While the LHC has so far found no evidence for physics beyond the SM, the bilepton model remains particularly attractive because its new particle spectrum exotic gauge bosons, heavy quarks, and an extended Higgs sector is predicted to lie near the TeV scale, making it potentially accessible to current and future colliders. The perturbativity bound associated with the embedding of $SU(2)_L$ into $SU(3)_L$, which causes the $U(1)_X$ coupling to develop a singularity near $\mu \simeq 4$~TeV, is acknowledged, and the analysis is explicitly restricted to energy scales safely below this threshold.

A central element of the Chapter is the identification of a phenomenologically viable benchmark point for the model. This is accomplished through a systematic computational pipeline involving the \texttt{SARAH} and \texttt{SPheno} packages for spectrum generation, combined with \texttt{HiggsBounds} and \texttt{HiggsSignals} for confronting the predictions against experimental Higgs exclusion limits and precision measurements. The scalar sector of the model contains 26 free parameters, and given the high dimensionality of the parameter space, a stochastic scan strategy is employed within theoretically motivated ranges. The vacuum expectation values are constrained by the hierarchical condition $v_\rho \gg v_{\eta}, v_\chi, v_\sigma$ to ensure the decoupling of heavy states from the electroweak scale, and by the electroweak consistency condition $v_\eta^2 + v_\chi^2 + v_\sigma^2 = v_W^2 = (246~\text{GeV})^2$. The selected benchmark point features a lightest neutral scalar consistent with the observed Higgs boson at 125 GeV, vector bileptons $Y^{\pm\pm}$ and $X^\pm$ with masses around 1.3~TeV, a $Z'$ boson at approximately 4~TeV, and exotic quarks with masses $m_D \simeq 1.5$~TeV, $m_S \simeq 2$~TeV, and $m_T \simeq 3$~TeV. Two distinct Yukawa benchmark configurations are also identified, differing in the structure of the down-type quark rotation matrix $U_L^d$ and leading to different dominant decay channels for the exotic $T$ quark: one favouring the $T \to Y^{++} b$ mode and the other the $T \to Y^{++} d$ channel.

The phenomenological analysis centers on the pair-production process
\begin{equation*}
	pp \to T\bar{T} \to (Y^{++}b)(Y^{--}\bar{b}) \to (\mu^+\mu^+ b)(\mu^-\mu^- \bar{b}),
\end{equation*}
which yields a distinctive final state consisting of two same-sign muon pairs and two $b$-jets. Hard scattering is simulated at leading order using \texttt{MadGraph}, matched to \texttt{HERWIG~6} for parton showering and hadronization. The leading SM backgrounds considered are $ZZb\bar{b}$ production, where both $Z$ bosons decay into muon pairs, and $t\bar{t}Z$ production in the dilepton channel, where the $Z$ decays into muons. Acceptance cuts are applied on the transverse momenta, pseudorapidities, invariant opening angles $\Delta R$, and missing transverse energy of the final-state objects, together with a flat $b$-tagging efficiency of $\epsilon_b = 0.82$.

Part of the computational workflow developed for this study was optimized for HPC environments, with parallelized implementations adopted for both the benchmark-point search and the Monte Carlo simulations. In particular, accelerated computing techniques, including GPU-enabled numerical workloads and distributed event-generation tasks, were exploited to improve the efficiency of the stochastic scans and \texttt{MadGraph}-based collider simulations.

At $\sqrt{s} = 14$~TeV, the signal cross section is of order $10^{-7}$~pb, rendering detection impossible even at the HL-LHC. At $\sqrt{s} = 100$~TeV with an integrated luminosity of $3000~\text{fb}^{-1}$, the signal cross section rises to approximately $0.02$~fb, yielding around 32\,880 expected signal events after cuts, compared to approximately 1\,288 and 8\,338 background events from $ZZb\bar{b}$ and $t\bar{t}Z$, respectively.

A detailed comparison of kinematic distributions is presented, including the transverse momenta of the hardest and next-to-hardest muons, the invariant mass and opening angle $\Delta R$ of same-sign muon pairs, the pseudorapidity and polar angle of the leading lepton, and the transverse momentum of the hardest $b$-jet together with the $\Delta R$ separation between the leading jet and leading muon. In all cases, the signal distributions are found to be strikingly different from the SM backgrounds, owing to the large mass scales involved: the production of 3~TeV $T\bar{T}$ pairs and subsequent decays via 1.3~TeV bileptons generate leptons and jets at transverse momenta of $\mathcal{O}(1~\text{TeV})$, far above the typical scales of $ZZ$ and $t\bar{t}Z$ processes. The invariant mass distribution of same-sign muon pairs is particularly discriminating, being sharply peaked at $m_{\mu\mu} \simeq m_Y \simeq 1.3$~TeV for the signal while broadly distributed around 80--90~GeV for the backgrounds. A statistical significance of $s \simeq 34$ is obtained at FCC-$hh$, demonstrating that the signal would be observable with very high confidence.

The Chapter concludes by placing these results in the broader context of the 331 model phenomenology program. Together with previous studies on leptonic bilepton-pair production and non-leptonic bilepton decays into heavy quarks, the present analysis of heavy-quark production and decay into bileptons completes a comprehensive picture of the observable signatures of the bilepton model at hadron colliders, and reinforces the case for pursuing these searches at future 100~TeV machines.

The work presented in this Chapter is based primarily on, and adapted from, the publication \cite{Corcella:2025ypl}, co-authored with G.~Corcella, C.~Corianò and P. H.~Frampton.

\section{Introduction}
%One of the most important questions that arise in particle physics is why there are only three families of quarks and leptons. There are many observables in particle physics that depend on the family number and they all agree to constrain this number to three. In the Standard Model there is no mechanism that prohibits the existence of more families than those observed. To answer this question P. H. Frampton, F. Pisano and V. Pleitez \cite {frampton_chiral_1992} \cite {pisano_su3ensuremathbigotimesu1_1992} have proposed a new model that extends the Standard Model and could provide an elegant answer to this fundamental question. Specifically it is called the 331 Model, and it is built on the gauge group
%\begin{equation}
%	SU(3)_C\times   SU(3)_L\times   U(1)_X\ ,
%\end{equation}
%which enlarge the $SU(2)_L$ symmetry of the Standard Model. 
The search for new physics at the Large Hadron Collider (LHC) has revealed persistent gaps in our understanding of fundamental forces, particularly regarding the limitations of the Standard Model (SM). Two of the most pressing issues are the hierarchy problem in the Higgs sector and smallness of the
neutrino masses. To address these challenges, theorists often propose extensions of the SM that include larger gauge groups and additional particles. Grand Unified Theories (GUTs) are a prominent example, offering a unified framework for the forces. However, they operate at energy scales, approximately from
$10^{14}$ to $10^{16}$ GeV, which are far beyond the reach of the LHC and the future colliders
currently under debate, such as  the hadron machine FCC-$hh$ foreseen at 100 TeV.
Bridging this vast energy gap down to the TeV scale, which is accessible to the LHC and ultimately
FCC-$hh$, remains a major challenge due to the complexity and richness of these extended theories.

Nevertheless, certain models predict that signs of new physics could still emerge at the LHC
or FCC-$hh$ scales, particularly in scenarios involving enlarged gauge symmetries. One such model is the bilepton model \cite{frampton_chiral_1992,pisano_su3ensuremathbigotimesu1_1992}, based on the gauge group $SU(3)_C \times   SU(3)_L \times   U(1)_X$, also known as 331 model. This scenario is particularly attractive because it provides solutions to several shortcomings of the SM, including explanations for neutrino masses and charge quantization, while remaining testable at TeV energies. 
This reshuffling necessitates that, already at TeV scale, the three families transform differently under the gauge group; this can explain the failure of $SU(5)$ Grand Unification Theory, where it was assumed that all three families remain sequential up to the GUT scale $\sim 10^{11}-10^{13}$ TeV.
The bilepton model introduces the new gauge group $SU(3)_L$, an extension analogous to the familiar colour gauge group $SU(3)_C$ of QCD, and reorganizes quarks and leptons into multiplets of $SU(3)_L$,
with fermions transforming as triplets or antitriplets, depending on their charge and quantum
properties. This reshuffling of particle content leads to new interactions and potentially observable
phenomena even at current collider energies.

A hallmark of the 331 model is the prediction of novel gauge bosons, known
as bileptons and labelled as $Y^{++/--}$ hereafter, which carry both lepton number and electric charge. These particles, with charges $Q = \pm2$ and
lepton number $L=\pm2$ , open up a new avenue for exploring physics beyond the SM. Bileptons can lead to distinctive signatures, such as lepton-flavour violating processes or new decay channels for heavy particles, which could be detected at the LHC or future colliders.
Spontaneous symmetry breaking features additional complexities: it occurs at an energy scale close to the TeV region, transitioning the gauge group down to the familiar electroweak symmetry. This process
generates the masses of the gauge bosons, including the SM-like $W^\pm$ and $Z^0$, as well as the
new bileptons.

The 331 model provides a framework for addressing key open questions, such as the origin of neutrino masses and the unification of forces at high energies. It also constrains its parameter space through the requirement of real gauge couplings, which limits possible signals and enhances its predictive power. Despite these promising features, the model still faces experimental scrutiny and requires further theoretical refinements to assess its full viability. In the specific realization of the 331 model that includes bileptons, symmetry breaking occurs through Higgs fields whose vacuum expectation values (vevs) are tied to the TeV scale. This process not only gives mass to the exotic bileptons, but also introduces a rich phenomenology of new particles, such as the doubly-charged gauge bosons $Y^{++}$ and $Y^{--}$.
Furthermore, at energies beyond 14 TeV, but accessible to the 100 TeV  FCC-$hh$, QCD will possibly bind 
the new quarks predicted by the bilepton model and produce a plethora of additional baryons and mesons \cite{framptonAdditionalBaryonsMesons2021}.
Overall, the bilepton version of the 331 model stands out as a compelling extension of the SM, with potential observable consequences in high-energy experiments. It offers a unique window into new physics, linking low-energy phenomena to high-energy symmetries. Whether the LHC, including the
high-luminosity HL-LHC, can uncover evidence for bileptons or other exotic particles will play a crucial role in determining the future direction of particle physics beyond the Standard Model.

This work completes a series of papers on the theory and phenomenology of the bilepton model.
In detail,
in Refs.~\cite{corcella_bilepton_2017,corcella_exploring_2018}
we investigated bilepton-pair production and decay into same-sign lepton pairs, such as
$Y^{++}\to \mu^+\mu^+$, with
Ref.~\cite{corcella_exploring_2018} also exploring possible
differences between signals
of vector and scalar bileptons. The conclusion of
Refs.~\cite{corcella_bilepton_2017,corcella_exploring_2018} was that, for a bilepton mass about 900 GeV, consistent with the
exclusion limits at that time, it would have been possible to discover bileptons at LHC,
with the vector signal greatly overwhelming the scalar one.
More recently, Ref.~\cite{corcella_non-leptonic_2022} explored possible nonleptonic decays
of bileptons, namely decays into a heavy TeV-scale quark plus a Standard Model heavy quark, and found
that such a signal will potentially be visible at the future FCC-$hh$, as the LHC statistics are
too low, even in the high luminosity phase.
Ref.~\cite{corcella_non-leptonic_2022} assumed a bilepton mass
around 1.3 TeV, consistently with the findings, based on renormalization group
arguments, of Ref.~\cite{Coriano:2020iiz}.
We underline that, regardless of the decay modes, a general
result of Refs.~\cite{corcella_bilepton_2017,corcella_exploring_2018,corcella_non-leptonic_2022} is that, given the high mass of the bilepton and consequently
of the lepton or heavy-quark pairs produced in the decay, it is relatively straightforward
to discriminate any bilepton signal from the Standard Model backgrounds.
To our knowledge, there is currently no specific experimental analysis on
vector bileptons, while both ATLAS and CMS have searched for doubly-charged
scalar Higgs bosons in some scenarios beyond the Standard Model.
In detail, ATLAS \cite{ATLAS:2022pbd} has set the limit
$m_{H^{\pm\pm}}>1080$~GeV in left-right symmetric type-II see-saw model
$m_{H^{\pm\pm}}>900$~GeV in the Zee--Babu neutrino-mass model at
$\sqrt{s}=13$~TeV and integrated luminosity ${\cal L}=139$~fb$^{-1}$.
CMS set instead limits between 535 and 820 GeV according to the
lepton flavours for leptonic doubly-charged Higgs decays
at 13 TeV and 12.9 fb$^{-1}$ \cite{CMS:2017pet}.

In the present paper we study
primary production of heavy quarks, such as $T^\pm$, charged $\pm 5/3$, and their subsequent decays into
bileptons and Standard Model quarks, with the bileptons again decaying into same-sign leptons. To our knowledge, an analysis on such heavy quarks focused
on the 331 model has never been carried out so far.
Nevertheless, a companion investigation was undertaken in \cite{corcella_vector-like_2021}, where the authors, in the framework of
simplified models,  studied pair production of vector-like heavy quarks
with charge 5/3 and decay into singly- or doubly-charged bosons plus a
top or a bottom quark. A CMS analysis searching for top partners decaying to a top quark and a $W$
boson \cite{CMS:2020ttz} was also recast in Ref.~\cite{corcella_vector-like_2021}
and the results were eventually provided as recast
efficiencies, which allow reinterpretation in any model.
More recently, Ref.~\cite{Liu:2024hvp} explored single production
of heavy quarks with charge 5/3 at the LHC and presented exclusion
limits and discovery reach, typically
in the range of few TeV, for a few values of the integrated luminosity.
From the experimental viewpoint, such heavy quarks, labelled as
$X_{5/3}$, were searched for
in Ref.~\cite{CMS:2018ubm} by the CMS collaboration at $\sqrt{s}=13$~TeV
and integrated luminosity ${\cal L}=35.9$~fb$^{-1}$ and lower limits at 95\%
confidence level on their mass were set at  1.33 and 1.30~TeV
for right-handed and left-handed couplings to $W$ bosons, respectively.
Furthermore, vector-like quarks as well as
singly and  doubly-charged gauge bosons
in the minimal 331 model
were also explored in \cite{Calabrese:2023ryr}, where the authors
considered both single and pair production of bileptons, possibly
accompanied by one extra jet. The results were then compared with the
ATLAS searches for a doubly-charged Higgs boson \cite{ATLAS:2022pbd}.

The present paper focuses on heavy quarks in the bilepton model and is organized
as follows.
In Section 2 we review the particle content of the minimal 331 model, while in Section 3 we scrutinize
Higgs and Yukawa sectors. Section 4 will discuss the choice of the benchmark point for the analysis
and Section 5 present some phenomenological results, mostly at FCC-$hh$.
Section 6 will contain some concluding remarks.

\section{The particle content of the minimal 331 Model}\label{331}

As discussed in the introduction, the 331 model is 
based on a $SU(3)_C \times   SU(3)_L \times   U(1)_X$ gauge symmetry.
In its minimal version,
the symmetry group $SU(3)_L$ requires the addition of three exotic quarks to the Standard Model quark content. 
The first two generations of quarks are placed in triplets of $SU(3)_L$ along with exotic quarks $D$ or $S$: 
\begin{equation}
	\begin{pmatrix}
		u\\d\\ D
	\end{pmatrix}_L,
	\begin{pmatrix}
		c\\s\\ S
	\end{pmatrix}_L\rightarrow \brak{\textbf{3},\textbf{3},-\frac{1}{3}}\ ,
\end{equation}
with a quantum number $X=-\frac{1}{3}$ under $U(1)_X$.
In contrast, the third generation is placed in an anti-triplet: \begin{equation}
	\begin{pmatrix}
		b\\-t\\ T
	\end{pmatrix}_L
	\rightarrow \brak{\textbf{3},\bar{\textbf{3}},\frac{2}{3}}\ ,
\end{equation}
with $X=\frac{2}{3}$.
Conversely, leptons in each of the three families are treated uniformly as $SU(3)_L$ anti-triplets, colour singlets, represented as: 
\begin{equation}
	\begin{pmatrix}
		e\\-\n_e\\ e^c
	\end{pmatrix}_L,
	\begin{pmatrix}
		\m\\-\n_\m\\ \m^c
	\end{pmatrix}_L,
	\begin{pmatrix}
		\t\\-\n_\t\\ \t^c
	\end{pmatrix}_L\rightarrow \brak{\textbf{1},\bar{\textbf{3}},0}\ ,
\end{equation}
with $X=0$. Here, $e^c$, $\mu^c$, and $\tau^c$ are the left-handed Weyl spinors of the charge conjugate fields.

For each left-handed quark field, there is a corresponding right-handed singlet under $SU(3)_L$ with the following quantum numbers:\begin{equation}
	\begin{matrix}
		(u^c)_L\\
		(c^c)_L\\ 
		(t^c)_L
	\end{matrix} \rightarrow \brak{\bar{\textbf{3}},\textbf{1},-\frac{2}{3}}\ ,
	\ \ \ \ \ \ \ \ \ 	\begin{matrix}
		(d^c)_L\\
		(s^c)_L\\
		(b^c)_L
	\end{matrix} \rightarrow \brak{\bar{\textbf{3}},\textbf{1},\frac{1}{3}}\ ,
\end{equation}

\begin{equation}
	\begin{matrix}
		(D^c)_L\\(S^c)_L
	\end{matrix} \rightarrow \brak{\bar{\textbf{3}},\textbf{1},\frac{4}{3}}\ ,
	\ \ \ \ \ \ \ \ \ 
	\begin{matrix}
		(T^c)_L
	\end{matrix} \rightarrow \brak{\bar{\textbf{3}},\textbf{1},-\frac{5}{3}}\ ,
\end{equation}

The $U(1)_X$ charges for the first two families are $-\frac{2}{3}$, $\frac{1}{3}$, and $\frac{4}{3}$ for $u^c$, $d^c$, and $D^c$, respectively, with identical charges for the second generation. For the third generation, the $X$ quantum numbers are $\frac{1}{3}$, $-\frac{2}{3}$, and $-\frac{5}{3}$ for $b^c$, $t^c$, and $T^c$, respectively.

Three scalar triplets under $SU(3)_L$ are introduced to trigger spontaneous symmetry breaking:
\begin{equation}
	\r=\begin{pmatrix}
		\r^{++}\\ \r^+\\ \r^0
	\end{pmatrix}\ ,\ 
	\eta=\begin{pmatrix}
		\eta^{+}_1\\ \eta^0\\ \eta^-_2
	\end{pmatrix}\ ,\ 
	\chi=\begin{pmatrix}
		\chi^{0}\\ \chi^-\\ \chi^{--} 
	\end{pmatrix} \ ,
\end{equation}
with charges $X=1,0,-1$, respectively. Additionally, a scalar sextet is necessary for generating physical lepton masses: 
\begin{equation}
	\s=\begin{pmatrix}
		\s^{++}_1&\frac{\s_1^+}{\sqrt{2}}&\frac{\s_1^0}{\sqrt{2}}\\
		\frac{\s_1^+}{\sqrt{2}}&\s_2^0&\frac{\s^-_2}{\sqrt{2}}\\ 
		\frac{\s_1^0}{\sqrt{2}}&\frac{\s^-_2}{\sqrt{2}}&\s_2^{--}\\
	\end{pmatrix}\ ,
\end{equation}
This unconventional assignment of quantum numbers ensures that gauge anomalies are not cancelled individually for each family, as in the Standard Model. Instead, the contributions from all quarks must be combined to achieve anomaly cancellation, which is one of the key features of the model. This mechanism offers a potential explanation for the existence of three generations of fermions, providing insight into the so-called flavour puzzle and guiding the development of new models in particle physics. The interfamily anomaly cancellation requires the number of families $n_f$  to be a multiple of the number of the QCD colours $(N)$, $n_f= k \, N$, with $k$ an integer, 
but asymptotic freedom of QCD requires $k=1$ and hence $n_f=3$
exactly.

\section{Higgs Sector}
The particle content of the spectrum of the 331 model is very rich due to the presence of three scalar triplets and a scalar sextet. In fact, spontaneous
symmetry breaking happens in two steps. In the first one, namely $SU(3)_L\times   U(1)_X \to SU(2)_L\times   U(1)_Y$, only the scalar triplet $\r$ acquires
a vacuum expectation value (vev), i.e.  
\begin{equation}
	\expval{\r}=\begin{pmatrix}
		0\\0\\\frac{v_\r}{\sqrt{2}}
	\end{pmatrix}\ ,
\end{equation}
while the other scalars get a vev only in the subsequent breaking $SU(2)_L\times   U(1)_Y\to   U(1)_{\rm{em}}$, i.e.
\begin{equation}
	\expval{\eta}=\begin{pmatrix}
		0\\ \frac{v_\eta}{\sqrt{2}}\\0
	\end{pmatrix} \ ,\ \ \ \
	\expval{\chi}=\begin{pmatrix}
		\frac{v_\chi}{\sqrt{2}}\\0\\0
	\end{pmatrix} \ ,\ \ \ \
	\expval{\s}=\begin{pmatrix}
		0&0&\frac{v_\s}{2}\\
		0&0&0\\
		\frac{v_\s}{2}&0&0
	\end{pmatrix}\ .
\end{equation}
In principle, also the component $\s_2^0$ can acquire a vev, a choice which
is particularly relevant for neutrino masses \cite{tully_generating_2001}, and hence contributes to enlarge the phenomenology of the model; in
the following, we shall instead set the vev of $\s_2^0$ to zero.
After electroweak symmetry breaking (EWSB), in the scalar sector one has five scalar neutral  Higgses, three pseudoscalar Higgses, four charged Higgses and three doubly-charged Higgses. The (lepton-number conserving) potential of the model is given by  \cite{tully_scalar_2003}:
\allowdisplaybreaks
\begin{align}\label{pot}
	V=&m_1\r^\dagger\r+m_2\eta^\dagger\eta+m_3\chi^\dagger\chi+\l_1(\rho^\dagger\r)^2+\l_2(\eta^\dagger\eta)^2+\l_3(\chi^\dagger\chi)^2\nonumber\\
	&+\l_{12}\r^\dagger\r\eta^\dagger\eta+\l_{13}\r^\dagger\r\chi^\dagger\chi+\l_{23}\chi^\dagger\chi\eta^\dagger\eta+\z_{12}\r^\dagger\eta\eta^\dagger\r+\z_{13}\r^\dagger\chi\chi^\dagger\r+\z_{23}\eta^\dagger\chi\chi^\dagger\eta\nonumber\\
	&+m_4\Tr(\s^\dagger\s)+\l_4(\Tr(\s^\dagger\s))^2+\l_{14}\r^\dagger\r\Tr(\s^\dagger\s)+\l_{24}\eta^\dagger\eta\Tr(\s^\dagger\s)+\l_{34}\chi^\dagger\chi\Tr(\s^\dagger\s)\nonumber\\
	&+\l_{44}\Tr(\s^\dagger\s\s^\dagger\s)+\z_{14}\r^\dagger\s\s^\dagger\rho+\z_{24}\eta^\dagger\s\s^\dagger\eta+\z_{34}\chi^\dagger\s\s^\dagger\chi\nonumber\\
	&+(\sqrt{2}f_{\r\eta\chi}\epsilon^{ijk}\r_i\eta_j\chi_k+\sqrt{2}f_{\r\s\chi}\r^T\s^\dagger\chi\nonumber\\
	&+\xi_{14}\epsilon^{ijk}\r^{*l}\s_{li}\r_j\eta_k+\xi_{24}\epsilon^{ijk}\epsilon^{lmn}\eta_i\eta_l\s_{jm}\s_{kn}+\xi_{34}\epsilon^{ijk}\chi^{*l}\s_{li}\chi_j\eta_k )+\textnormal{h.c.}\ .
\end{align}
This potential can be extended by incorporating additional lepton-number violating terms that are singlets under the 331 gauge group, as discussed in detail in \cite{tully_scalar_2003} and further elaborated in \cite{fonseca_lepton_2016}. The EWSB mechanism induces a mixing between the Higgs fields, as shown in \cite{costantini_theoretical_2020}. From Eq. (\ref{pot}), one can derive explicit expressions for the mass matrices of scalar, pseudoscalar, charged, and doubly-charged Higgs bosons using standard techniques.
In the broken Higgs phase, the minimization conditions \be\label{mincond} \frac{\partial V}{\partial v_\phi}=0, \quad \langle \phi^0\rangle=v_\phi, \quad \phi=\rho, \eta, \chi, \sigma \ee define the tree-level vacuum structure. One-loop corrections to vacuum stability have been recently analyzed in
\cite{dorsch_vacuum_2024}, for a simpler version of the model, known as the `economical' 331 model \cite{foot_su3_mathitlensuremathbigotimesu1_mathitn_1994,ponce_analysis_2002}.

As pointed out above, in this work we consider massless neutrinos by setting the vacuum expectation value (vev) of the neutral field $\sigma_2^0$ to zero. This choice can be generalized to generate small Majorana neutrino masses, as suggested in \cite{tully_generating_2001}.
The explicit expressions of the minimization conditions are then given by:
\begin{align}\label{minpot1}
	m_1 v_\rho + \lambda_1 v_\rho^3 + \frac{1}{2}\lambda_{12}v_\rho v_\eta^2-f_{\rho\eta\chi} v_\eta v_\chi+\frac{1}{2}\lambda_{13}v_\rho v_\chi^2 - \frac{1}{\sqrt2}\xi_{14}v_\rho v_\eta v_\sigma + f_{\rho\sigma\chi}v_\chi v_\sigma&\\
	+ \frac{1}{2}\lambda_{14}v_\rho v_\sigma^2 + \frac{1}{4}\zeta_{14}v_\rho v_\sigma^2&=0\ ,\nonumber\\
	m_2 v_\eta + \frac{1}{2}\lambda_{12}v_\rho^2 v_\eta +\lambda_2 v_\eta^3 - f_{\rho\eta\chi} v_\rho v_\chi +\frac{1}{2}\lambda_{23} v_\eta v_\chi^2 - \frac{1}{2\sqrt2}\xi_{14}v_\rho^2 v_\sigma+ \frac{1}{2\sqrt2}v_\chi^2 v_\sigma&\\
	+\frac{1}{2}\lambda_{24} v_\eta v_\sigma^2-\xi_{24} v_\eta v_\sigma^2&=0\ ,\nonumber\\
	m_3 v_\chi + \lambda_3 v_\chi^3 + \frac{1}{2} \lambda_{13} v_\rho^2 v_\chi - f_{\rho\eta\chi} v_\rho v_\eta +\frac{1}{2}\lambda_{23}v_\eta^2 v_\chi +\frac{1}{\sqrt2}\xi_{34}v_\eta v_\chi v_\sigma + f_{\rho\sigma\chi} v_\rho v_\sigma&\\
	+\frac{1}{2}\lambda_{34} v_\chi v_\sigma^2 + \frac{1}{4}\zeta_{34} v_\chi v_\sigma^2&=0\ ,\nonumber\\
	\label{minpot2}
	m_4 v_\sigma + \frac{1}{2}\lambda_{14}v_\rho^2 v_\sigma + \lambda_{44} v_\sigma^3 + \frac{1}{2}\lambda_4 v_\sigma^3 + f_{\rho\sigma\chi} v_\rho v_\chi - \frac{1}{2\sqrt2} \xi_{14} v_\rho^2 v_\eta + \frac{1}{2\sqrt2} \xi_{34} v_\eta v_\chi^2&\\
	+ \frac{1}{4}\zeta_{14} v_\rho^2 v_\sigma + \frac{1}{2}\lambda_{24} v_\eta^2 v_\sigma - \xi_{24} v_\eta^2 v_\sigma + \frac{1}{2} \lambda_{34} v_\chi^2 v_\sigma + \frac{1}{4} \zeta_{34} v_\chi^2 v_\sigma&=0.
	\nonumber
\end{align}
These conditions are inserted into the tree-level mass matrices of the CP-even and CP-odd Higgs sectors, derived from $M_{ij}=\left.{\partial^2 V}/{(\partial \phi_i\partial \phi_j)}\right|_{vev}$, where $V$ is the potential in
Eq.~(\ref{pot}). The mass eigenstates are defined as follows:
\begin{equation}
	h=R^S \begin{pmatrix}
		{\rm Re}\, \rho^0\\ {\rm Re}\, \eta^0\\ {\rm Re}\, \chi^0 \\ {\rm Re}\, \sigma^0_1\\ \s^{0}_2
	\end{pmatrix}\,,\ \ \ 
	Ah=R^P   \begin{pmatrix}
		{\rm Im}\, \rho^0\\ {\rm Im}\, \eta^0\\ {\rm Im}\, \chi^0 \\ {\rm Im}\, \sigma^0_1
	\end{pmatrix}\,,\ \ \ 
	H^+=R^C \begin{pmatrix}
		\r^{+}\\ \chi^{+}\\ \eta^{+}_1\\ \eta^{+}_2\\ \s^{+}_1 \\ \s^{+}_2
	\end{pmatrix}\,,\ \ \ 
	H^{++}=R^{2C} \begin{pmatrix}
		\r^{++}\\ \chi^{++}\\ \s^{++}_1 \\ \s^{++}_2
	\end{pmatrix}\,,
\end{equation}
where the explicit expressions of the mass matrices
are too cumbersome to be presented here, and are given in \cite{corianoSU3CSU3LU1X2024}.

As anticipated before, in this scenario we have five scalar Higgs bosons, one of which should correspond to the Standard Model Higgs with a mass of approximately 125 GeV. Additionally, there are four neutral pseudoscalar Higgs bosons:
two of them are the Nambu-Goldstone (NG) bosons associated with
the massive vector bosons $Z$ and $Z'$, while the other two will be labelled as $Ah_1$ and $Ah_2$.
Moreover, the spectrum includes six charged Higgs bosons ($H_1^\pm$, $\dots$,
$H_6^\pm$), two of which are charged NG bosons, along with three doubly-charged Higgses ($H_1^{\pm\pm}$, $H_2^{\pm\pm}$, $H_3^{\pm\pm}$),
one of which is also a NG boson.

\subsection{The Yukawa sector}
The Yukawa interactions for Standard Model and exotic quarks are given by 
\begin{align}
	\mathcal{L}^{Y}_{q,\ triplet}=\ &- \overline{Q}_m\brak{Y^d_{m\a} \eta^*d_{\a R}+Y_{m\a}^u\chi^*u_{m\a}}
	-\overline{Q}_3( Y^d_{3\a}\chi d_{\a R}+Y^u_{3\a}\eta u_{m\a})+\nonumber\\
	&-\overline{Q}_{m}(Y^J_{mn}\chi J_{nR})-\overline{Q}_{3}Y^J_{3}\chi J_{3R} +\textnormal{h.c.}\ , 
\end{align}
where $y^i_{d}$, $y^i_u$ and $y^i_E$ are the Yukawa couplings for down-,
up-type and exotic quarks, respectively. The masses of the exotic quarks are related to the vev of the neutral component of $\rho=(0,0,v_\rho)$ through the invariants
\begin{eqnarray}
	Q_1\,  \rho^*  D_R^*, Q_1\,  \rho^*  S_R^*&\sim & (\textbf{3},\textbf{3},-1/3)\times   (\textbf{1},\bar{\textbf{3}},-1)\times   (\bar{\textbf{3}},\textbf{1},4/3), \nonumber \\
	Q_3\,  \rho  T_R^* &\sim& (\textbf{3},\bar{\textbf{3}},2/3)\times   (\textbf{1},{\textbf{3}},1)\times   (\bar{\textbf{3}},\textbf{1},-5/3), 
\end{eqnarray}
responsible of the breaking $SU(3)_C\times   SU(3)_L\times   U(1)_X \to SU(3)_C\times   SU(2)_L\times   U(1)_Y$. It is clear that, since
$v_\rho\gg v_{\eta,\chi}$, the masses of the exotic quarks are
${\cal O}(\rm{TeV})$ whenever the relation $Y^J\sim \mathbb{1}$ is satisfied. 
In the lepton Yukawa sector, the interactions are more intricate due to the need to introduce an additional interaction with the scalar sextet to reproduce the physical lepton masses.
A typical Dirac mass term for leptons in the Standard Model is associated with the operator $\bar{l}_L H e_R$, where $l_L = (\nu_{eL}, e_L)$ represents the $SU(2)_L$ doublet. This has the representation content $(\bar{\textbf{2}}, 1/2) \times   (\textbf{2}, 1/2) \times   (\textbf{1}, -1)$ for $l_L$, $H$, and $e_R$, respectively, under the $SU(2)_L \times  U(1)_Y$ symmetry.
In the 331 model, however, both left- and right-handed lepton
components, e.g. $e_L$ and $e_R$, belong to the same multiplet. Therefore, to construct a $SO(1,3) \times  SU(3)_L$ singlet, two leptons from the same representation must be combined. This can be achieved, at least partially, through the operator
\begin{eqnarray}
	\mathcal{L}_{l,\, triplet}^{Yuk}&=& G^\eta_{a b}( l^i_{a \alpha}\epsilon^{\alpha \beta} l^j_{b \beta})\eta^{* k}\epsilon^{i j k} + \rm{h. c.}\nonumber\\
	&=& G^\eta_{a b}\, l^i_{a}\cdot l^j_{b}\,\eta^{* k}\epsilon^{i j k} + \rm{h. c.} \ , 
\end{eqnarray}
where the indices $a$ and $b$ run over the three generations of flavour, $\alpha$ and $\beta$ are Weyl indices, contracted to generate  an $SO(1,3)$ invariant ($l^i_{a}\cdot l^j_{b}\equiv l^i_{a \alpha}\epsilon^{\alpha \beta} l^j_{b \beta}$) from two Weyl fermions, and $i,j,k=1,2, 3$ are $SU(3)_L$ indices. 
The use of 
$\eta$ as a Higgs field is mandatory, since the components of the multiplet $l^j$ are $U(1)_X$ singlets. 
The representation content of the operator $l^i_a l^j_b$, according to $SU(3)_L$, is given by $\textbf{3}\times  \textbf{3}= \textbf{6} + \bar{\textbf{3}}$, with the 
$\bar{\textbf{3}}$ extracted by an anti-symmetrization over $i$ and $j$ via $\epsilon^{i j k}$. This allows one to identify 
$l^i_a l^j_b \eta^{*k}\epsilon^{i j k}$ as an $SU(3)_L$ singlet. Considering that the two leptons are anticommuting Weyl spinors, and that the $\epsilon^{\alpha\beta}$ (Lorentz) and $\epsilon^{i j k}$ ($SU(3)_L$) contractions introduce two sign flips under the $a\leftrightarrow b$ exchange, the combination 
\begin{equation}
	M_{a b}=( l^i_{a}\cdot l^j_{b })\eta^{* k}\epsilon^{i j k} 
\end{equation}
is therefore antisymmetric under the exchange of the two flavours, implying
that even $G_{a b}$ has to be antisymmetric.  However, an antisymmetric $G^\eta_{a b}$ Yukawa matrix is not sufficient to provide mass to all the leptons. 
In fact, the diagonalization of $G^\eta$ by means of a unitary matrix 
$U$, namely $G^\eta=U \Lambda U^\dagger$, with $G^\eta$ antisymmetric in
flavour space, implies that its three eigenvalues are given by
$\Lambda=(0,\lambda_{22}, \lambda_{33})$, with $\lambda_{22}=-\lambda_{33}$,
i.e., one eigenvalue is null and the other two are equal in magnitude.
We shall solve this problem by introducing a second 
invariant operator, with the inclusion of a sextet $\sigma$,
\begin{equation}
	\sigma=\left(
	\renewcommand*{\arraystretch}{1.5}
	\begin{array}{ccc}
		\sigma_1^{++}&\sigma_1^+/\sqrt2&\sigma^0_1/\sqrt2\\
		\sigma_1^+/\sqrt2&\sigma_2^0&\sigma_2^-/\sqrt2\\
		\sigma^0_1/\sqrt2&\sigma_2^-/\sqrt2&\sigma_2^{--}
	\end{array}
	\right)\in(\textbf{1},\textbf{6},0),
\end{equation}
leading to the Yukawa term
\begin{equation}\label{lag}
	\mathcal{L}_{l, sextet}^{{Yuk.}}= G^\sigma_{a b} l^i_a\cdot l^j_b \sigma^*_{i,j},
\end{equation}
which allows one to build a singlet out of the representation
$6$ of $SU(3)_L$, contained in $l^i_a\cdot l^j_b$, by combining it with
the flavour-symmetric
$\sigma^*$, i.e. $\bar{6}$.
Notice that $G^\sigma_{a b}$ is symmetric in flavour space. The
Yukawa Lagrangian in the lepton sector
can then be written as a combination of triplet and sextet contributions:
\begin{equation}
	\mathcal{L}_{l}^{{Yuk.}}=\mathcal{L}_{l, sextet}^{{Yuk.}} + \mathcal{L}_{l, triplet}^{{Yuk.}} + \rm{h. c.}\ .
\end{equation}
It is interesting to note that, without considering the sextet, a doubly-charged scalar would not be able to decay into same-sign leptons. This is because,
if one had no sextet, the interaction responsible for the leptons
would only involve the scalar triplet $\eta$, which does not contain any
doubly-charged state. 
%So, if we detect a decay like $H^{\pm\pm}\to l^\pm l^\pm$, it would directly indicate the presence of the sextet in the 331 model.

\section{Benchmark points}

In order to identify some viable benchmark points of the 331 model with $\beta = \sqrt{3}$, we follow a systematic procedure involving model implementation and phenomenological constraints. First, we construct the Lagrangian of the model using the \texttt{SARAH} package \cite{staub_automatic_2011,staub_sarah_2014}, which enables the automatic generation of the \texttt{SPheno} code \cite{porodSPhenoProgramCalculating2003,porodSPheno31Extensions2012}. This code is subsequently used to compute mass spectrum and relevant observables of the model.
To ensure consistency with current experimental data, we employ \texttt{HiggsBounds} \cite{bechtleHiggsBoundsConfrontingArbitrary2010,bechtleHiggsBounds200Confronting2011} and \texttt{HiggsSignals} \cite{bechtleHiggsSignalsConfrontingArbitrary2014} to test a possible benchmark point against the full set of Higgs sector constraints. These include exclusion limits from direct searches as well as precision measurements of the Higgs boson properties. Only the points that pass
all these constraints are considered as viable benchmark candidates. 

To streamline and automate the entire workflow, we employ the \texttt{SSP}
(\texttt{SARAH} Scan and Plot) package \cite{staubToolBoxImplementing2012}, which facilitates parameter space exploration in BSM scenarios. \texttt{SSP} allows for the automatic generation of SLHA input files for \texttt{SPheno}, execution of the spectrum calculation, and direct interfacing with \texttt{HiggsBounds} and \texttt{HiggsSignals} using the \texttt{SPheno} output. 
This automated pipeline enables a systematic and efficient scan over a predefined set of model parameters, verifying at each step whether the generated points are consistent with current Higgs data. Through this setup, one can perform comprehensive phenomenological analyses and identify viable regions of parameter space that satisfy both theoretical and experimental constraints. 

To ensure the consistency of the 331 model with the Standard Model (SM), the three vacuum expectation values (VEVs) must satisfy the hierarchical condition
\begin{equation}
	v_\rho \gg v_\eta\,, \quad v_\rho \gg v_\chi \quad \text{and}\quad v_\rho \gg v_\sigma\ .
\end{equation}
These relations guarantee that the additional heavy particles predicted by the 331 model do not lie near the electroweak scale. Otherwise, such states would have already manifested in collider searches. 

In order to reproduce the correct mass for the $W$ boson, the VEVs must also satisfy the electroweak constraint
\begin{equation}
	v_\eta^2 + v_\chi^2 + v_\sigma^2 = v_W^2 = (246~\text{GeV})^2\ .
\end{equation}
This condition effectively reduces by one the number of independent parameters in the scalar potential. It is convenient to parametrize the three VEVs in terms of the Weinberg scale through two mixing angles $\beta_1$ and $\beta_2$ as follows:
\begin{align}
	v_\eta &= v_W \sin\beta_2 \cos\beta_1\ , \\
	v_\chi &= v_W \sin\beta_2 \sin\beta_1\ , \\
	v_\sigma &= v_W \cos\beta_2\ ,
\end{align}
where we have imposed $\tan\beta_1 = 0.15$ and $\tan\beta_2 = 0.5$ for the scan of the parameter space.  

The stability of the scalar potential restricts the possible values of the couplings. In particular, the self-couplings $\lambda_{\rho,\eta,\chi,\sigma}$ must be positive to guarantee the boundedness of the potential along the $\rho$, $\eta$, $\chi$, and $\sigma$ directions. Conversely, the mixed couplings $\lambda_{ij}$, $\zeta_{ij}$, $\xi_{ij}$, as well as the trilinear terms $f_{\rho\eta\chi}$ and $f_{\rho\sigma\chi}$, can in principle take negative values without violating stability, provided that the overall potential remains bounded from below.

For the numerical scan, we fixed the following parameters:
\begin{align}
	v_\rho &= 4~\text{TeV}\ , \\
	f_{\rho\eta\chi} &= f_{\rho\sigma\chi} = -4~\text{TeV}\ .
\end{align}

Given that the scalar sector of the 331 model contains a large number of free parameters (26 in total), we have adopted a stochastic exploration strategy. The allowed parameter space consistent with vacuum stability and with at least one scalar state identified as the observed Higgs boson is found to be extremely sparse due to the high dimensionality of the scalar potential and the interplay among couplings. Therefore the strategy that we adopted was to generate random points using the \texttt{SSP} framework within the following ranges:
\begin{equation}
	0 < \lambda_{\rho,\eta,\chi,\sigma} < 3\ , \qquad \lambda_{ij},\, \zeta_{ij},\, \xi_{ij} \in [-1,\,1]\ ,
\end{equation}
which we have found to be the region with the major density of allowed points.
We used \texttt{HiggsBounds} to test the compatibility of the selected benchmark points with the current experimental exclusion limits from Higgs searches. The \texttt{HiggsBounds} framework allows one to confront the predictions of extended Higgs sectors with a comprehensive collection of model-independent bounds derived from collider data.

For each benchmark point, we provided as input masses, total widths and branching ratios of all neutral and charged Higgs bosons predicted by the 331 model, together with the relevant Higgs production cross sections normalized to their Standard Model (SM) equivalents. This allows \texttt{HiggsBounds} to internally compute the hadronic and partonic cross section ratios required for the comparison with the experimental analyses. The observables tested include both production and decay channels for neutral and charged Higgs bosons. Representative examples of these observables are \cite{bechtleHiggsBounds5TestingHiggs2020}:
\begin{itemize}
	\item Neutral Higgs production via gluon fusion ($gg \to h_j$), bottom-quark associated production ($pp \to b\bar{b}h_j$), vector boson fusion ($pp \to qqh_j$), associated production with vector bosons ($pp \to Vh_j$, with $V=W,Z$), and top-associated production ($pp \to t\bar{t}h_j$);
	\item Higgs boson decays into Standard Model final states such as $h_j \to b\bar{b}$, $\tau^+\tau^-$, $\gamma\gamma$, $Z\gamma$, $WW$, and $ZZ$, as well as flavor-violating decays like $h_j \to \mu^\pm \tau^\mp$ or invisible final states;
	\item Non-standard decay topologies, including $h_k \to h_i h_j$, $h_j \to h_i Z$, and $h_j \to H_i^\pm W^\mp$, relevant for multi-Higgs models;
	\item Charged Higgs production processes such as $pp \to tbH^\pm$ and $pp \to H^\pm W^\mp$, with subsequent decays $H^\pm \to \tau^\pm \nu_\tau$, $H^\pm \to tb$, or $H^\pm \to W^\pm Z$;
	\item LHC searches in the $\tau^+\tau^-$, $b\bar{b}$, $WW$, $ZZ$, and $t\bar{t}$ final states, as well as resonant di-Higgs production ($pp \to \phi \to hh$) and cascade decays like $pp \to \phi_2 \to Z\phi_1$;
\end{itemize}
Only those benchmark points that passed all \texttt{HiggsBounds} exclusion tests were retained for further phenomenological investigation of exotic decay modes within the 331 mdel.
In particular, we have selected the reference point in Table \ref{table:mass}. 
\begin{table}[h]
	\begin{center}
		\renewcommand{\arraystretch}{1.4}
		\begin{tabular}{|c|c|c|}
			\hline\hline
			\multicolumn{3}{|c|}{Benchmark Point}\\
			\hline
			\hline
			$m_{h_1}=125.3$ GeV & $m_{h_2}=6799.5$ GeV& $m_{h_3}=10489.8$ GeV\\
			\hline
			$m_{h_4}=17673.2$ GeV & $m_{h_5}=17737.2$ GeV& \\
			\hline
			$m_{Ah_1}=10489.5$ GeV& $m_{Ah_2}=17734.9$ GeV& $m_{H^\pm_1}=2566.4$ GeV\\
			\hline
			$m_{H^\pm_2}=10490.1$ GeV & $m_{H^\pm_3}=17673.5$ GeV &$m_{H^\pm_4}=17735.7$ GeV\\
			\hline
			$m_{H^{\pm\pm}_1}=10572.0$ GeV& $m_{H^{\pm\pm}_2}=17673.9$ GeV & $m_{H^{\pm\pm}_3}=17796.0$ GeV \\
			\hline
			$m_{Y^{\pm\pm}}=1325.7$ GeV& $m_{X^\pm}=1328.1$ GeV & $m_{Z'}=4150.4$ GeV\\
			\hline
			$m_D=1500.0$ GeV & $m_S=2000.0$ GeV& $m_T=3000.0$ GeV\\
			\hline
			\hline
		\end{tabular}
		\caption{Benchmark point for our collider study, consistent with the Standard Model Higgs mass and present exclusion limits on BSM physics.}\label{table:mass}
	\end{center}
\end{table}
For the purpose
of our investigation, a few comments on the
benchmark point in Table~\ref{table:mass} are in order.
The lightest neutral scalar $h_1$ has mass consistent with a SM Higgs boson,
while the mass of the lightest charged $H_1^\pm$ is about 2.5 TeV,
hence potentially in the reach of LHC. The other Higgs bosons are instead
too heavy to be searched for at LHC.
Consistently with the findings of \cite{Coriano:2020iiz},
doubly- ($Y^{\pm\pm}$) and singly-charged ($X^{\pm}$) vector bileptons have mass
about 1.3 TeV. Furthermore, the $Z'$ boson is slightly above 4 TeV and
the new heavy quarks feature masses $m_T\simeq 3$~TeV, $m_S\simeq 2$~TeV and
$m_D\simeq 1.5$~TeV, hence, at least from the kinematic viewpoint,
they are within the reach of LHC.

Additionally, within this parameter point of the scalar sector, we identify two benchmark points for the Yukawa couplings that are phenomenologically interesting and compatible with current collider bounds. These benchmarks can serve as reference points for further studies, including collider simulations and
investigation of flavour observables.
At this point, it is important to observe the 331 model is sensible to the Yukawa couplings, since different choices can lead
to the same CKM matrix, but with different phenomenology. This happens because in the Lagrangian of the 331 model the CKM matrix is not the only combination of quark rotation matrices, but they appear also singularly, as in the following interaction term:
\begin{equation}
	\mathcal{L}_{dTY}=\bar{T}_L\g^\m(U_L)_{j3}\textbf{d}_{j}Y_\m^{++}+\text{h.c.}
\end{equation}
where 
\begin{equation}
	\textbf{d}=\begin{pmatrix}
		d\\s\\b
	\end{pmatrix}
\end{equation}
The first benchmark that we have found is
\begin{equation}
	Y_d=\begin{pmatrix}
		2.03\times 10^{-4}	&	2.14\times 10^{-3}	&	2.26\times 10^{-2}	\\
		2.26\times 10^{-3}	&	1.77\times 10^{-2}	&	3.63\times 10^{-1}	\\
		2.26\times 10^{-1}	&	3.06	&	80.1
	\end{pmatrix},
\end{equation} 
which leads to the rotation matrix:
\begin{equation}
	U_{L}^d=\begin{pmatrix}
		-9.74\times 10^{-1}	&	2.27\times 10^{-1}	&	-4.21\times 10^{-3}	\\
		2.27\times 10^{-1}	&	9.74\times 10^{-1}	&	-3.01\times 10^{-2}	\\
		2.72\times 10^{-3}	&	3.03\times 10^{-2}	&	9.99\times 10^{-1}
	\end{pmatrix}.
\end{equation}
Likewise, the second one  
\begin{equation}
	Y_d=\begin{pmatrix}
		1.66\times 10^{-2}	&	8.37\times 10^{-5}	&	3.21\times 10^{-5}	\\
		4.27\times 10^{-4}	&	2.97\times 10^{-4}	&	9.76\times 10^{-5}	\\
		4.60\times 10^{-4}	&	4.08\times 10^{-4}	&	2.74\times 10^{-4}
	\end{pmatrix}
\end{equation}
leads to
\begin{equation}
	U_{L}^d=\begin{pmatrix}
		-1.68\times 10^{-3}	&	2.23\times 10^{-1}	&	-9.75\times 10^{-1}	\\
		-2.61\times 10^{-2}	&	9.75\times 10^{-1}	&	-2.23\times 10^{-1}	\\
		9.99\times 10^{-1}	&	2.59\times 10^{-2}	&	4.18\times 10^{-3}
	\end{pmatrix}
\end{equation}
In fact, these two benchmarks predict two different decay branching ratios for the exotic $T$ quark: the first one favours the decay into bottom quarks,
while in the second one the dominant mode is the $d$-quark channel.
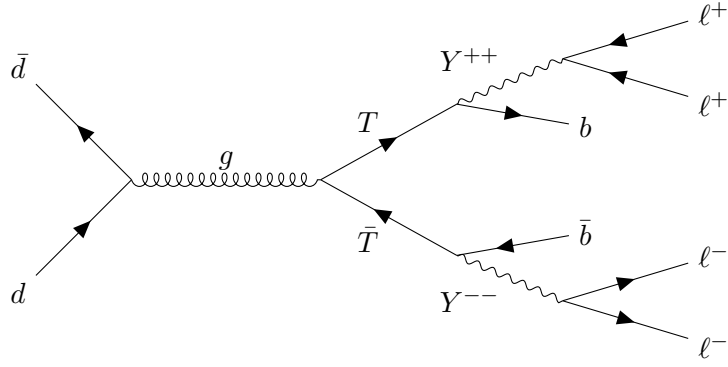
\begin{figure}
	\centering
	\begin{tikzpicture}
		\begin{feynman}
			
			% Vertices
			\vertex (dbar) at (0,1.5) {\(\bar{d}\)};
			\vertex (d) at (0,-1.5) {\(d\)};
			\vertex (v1) at (1.5,0);
			\vertex (v2) at (4,0);
			\vertex (T) at (5.8,1);
			\vertex (Tbar) at (5.8,-1);
			
			% Y++, bbar
			\vertex (Ypp) at (7.2,1.6);
			\vertex (bbar) at (7.5,0.7) {\(b\)};
			\vertex (lp1) at (9.2,2.2) {\(\ell^+\)};
			\vertex (lp2) at (9.2,1.0) {\(\ell^+\)};
			
			% Y--, b
			\vertex (Ymm) at (7.2,-1.6);
			\vertex (b) at (7.5,-0.7) {\(\bar{b}\)};
			\vertex (lm1) at (9.2,-1.0) {\(\ell^-\)};
			\vertex (lm2) at (9.2,-2.2) {\(\ell^-\)};
			
			% Diagram
			\diagram* {
				(v1) -- [fermion] (dbar) ,
				(d)-- [fermion] (v1),
				(v1) -- [gluon, edge label=\(g\)] (v2),
				(v2) -- [fermion,edge label = \(T\)] (T),
				(v2) -- [anti fermion, edge label' = \(\bar{T}\)] (Tbar),
				
				% Upper branch
				(T) -- [fermion] (bbar),
				(T) -- [boson, edge label=\(Y^{++}\)] (Ypp),
				(Ypp) -- [anti fermion] (lp1),
				(Ypp) -- [anti fermion] (lp2),
				
				% Lower branch
				(Tbar) -- [anti fermion] (b),
				(Tbar) -- [boson, edge label'=\(Y^{--}\)] (Ymm),
				(Ymm) -- [fermion] (lm1),
				(Ymm) -- [fermion] (lm2),
			};
			
		\end{feynman}
	\end{tikzpicture}
	\caption{Example of $T\bar{T}$ pair production at LHC or FCC-$hh$,
		with $T$ decaying into a bilepton and a $b$ quark.}
\end{figure}\par
Before investigating the phenomenology of the bilepton model in the
representative point in Table~\ref{table:mass}, we wish to comment on the issue of
the perturbativity of the model. As discussed in \cite{diasConcerningLandauPole2005,martinez_landau_2007} and pointed out in the
phenomenological analysis of \cite{corcella_exploring_2018},
the bilepton model possesses
a singular behaviour associated
with the theoretical embedding of $SU(2)$ in $SU(3)$ and matching conditions
between the couplings. In particular, the theory may lose its
perturbativity whenever the $U(1)$ coupling constant $g_X(\mu)$ is evaluated
at a scale such that $\sin^2\theta_W(\mu)\simeq 1/4$, which corresponds
to $\mu\simeq 4$~TeV. This singularity has often been labelled as
Landau pole, although, strictly speaking,
the Landau pole has a different meaning and
refers to the divergence in the QED coupling
at very high energy due to a positive renormalization-group
$\beta$-function \cite{Landau:1954nau}. More recently,
Ref.~\cite{Morisi:2025svd} found out that it is possible to further shift
the singularity of the bilepton model to higher energies
by adding a fourth family of both quarks and leptons.
As a matter of fact, the
present paper will deal with energy scales for bilepton production
well below 4 TeV, therefore we shall assume that we are still within
the perturbative regime and defer to future work any possible discussion
and comparison with \cite{Morisi:2025svd} 
on how to deal with such a singularity whenever the energy should approach
$\mu\simeq 4$~TeV.

\section{Phenomenology}
In this section, we investigate the phenomenology of heavy-quark production and their
decays into vector bileptons at LHC (14 TeV) and FCC-$hh$ (100 TeV).
In our benchmark point, heavy quarks like $T$ have mass 3 TeV: therefore, one can already
envisage that, as found in \cite{corcella_non-leptonic_2022} for the purpose of
non-leptonic bilepton decays, the production cross section at LHC would be rather low, since
one may need at least a partonic centre-of-mass energy of $2m_T\simeq 6$~TeV.
Therefore, 
the expected number of events is quite suppressed even in the high-luminosity
phase. On the other hand, one may foresee a substantial number
of events at 100 TeV and integrated luminosity ${\cal L} = 3000$~fb$^{-1}$, as expected at
FCC-$hh$.

We focus on the primary production of $T\bar{T}$ pairs, followed by the decays $T \to Y^{++} b$ and $\bar{T} \to Y^{--} \bar{b}$, where the bileptons subsequently decay into same-sign muon pairs:
\begin{equation}
	\label{ttbar}
	pp \to T\bar{T} \to (Y^{++}b)(Y^{--}\bar{b}) \to (\mu^+ \mu^+ b)(\mu^- \mu^- \bar{b}).
\end{equation}
The final state at parton level thus consists of two same-sign muon pairs and two $b$ quarks, which hadronize through $b$-flavoured jets. 
For the chosen benchmark point (see Table~\ref{table:mass}), the heavy quark $T$ has a width of $\Gamma(T) \simeq 32.1$~GeV and the following branching ratios:
\begin{equation}
	\text{BR}(T \to Y^{++} b) \simeq 47.1\%,\quad \text{BR}(T \to X^+ t) \simeq 46.3\%,\quad \text{BR}(T \to H_1^+ t) \simeq 4.6\%.
\end{equation}
Bileptons exhibit nearly democratic decay rates into all three lepton families:
\begin{equation}
	\text{BR}(Y^{++} \to e^+ e^+) \simeq \text{BR}(Y^{++} \to \mu^+ \mu^+) \simeq \text{BR}(Y^{++} \to \tau^+ \tau^+) \simeq \frac{1}{3}.
\end{equation}
The hard scattering process is simulated at leading order (LO) using \texttt{MadGraph}~\cite{Alwall:2011uj}, matched to \texttt{HERWIG 6}~\cite{Corcella:2000bw} for parton showering, hadronization and underlying event modelling. The possible inclusion of higher-order corrections to the
hard scattering, e.g. NLO,
as well as detector effects, is deferred to future work. 

The LO cross section for the signal process~\eqref{ttbar} at $\sqrt{s} = 14$~TeV is approximately $\sigma({\rm LHC}) \simeq 6 \times 10^{-7}$~pb, rendering observation unfeasible even with the full HL-LHC dataset. At $\sqrt{s} = 100$~TeV, however, the cross section increases to $\sigma({\rm FCC}) \simeq 0.02$~fb, making the signal potentially observable at FCC-$hh$.

The dominant Standard Model backgrounds to the signal in Eq.~\eqref{ttbar} include:
\begin{itemize}
	\item $ZZ$ production in association with a $b\bar{b}$ pair, where both $Z$ bosons decay into $\mu^+\mu^-$ (background $b_1$), i.e.
	\begin{equation}
		\label{back1}
		pp \to ZZb\bar{b} \to (\mu^+\mu^-)(\mu^+\mu^-) b\bar{b}; 
	\end{equation}
	\item associated production of $t\bar{t}Z$, with the $Z$ decaying into $\mu^+\mu^-$, and both top quarks decaying leptonically (background $b_2$), namely
	\begin{equation}
		\label{back2}
		pp \to t\bar{t}Z \to (b\mu^+\nu_\mu)(\bar{b} \mu^- \bar{\nu}_\mu)(\mu^+\mu^-).
	\end{equation}
\end{itemize}
The $ZZb\bar b$ background yields the same final state as our signal, however,
the kinematics of the particles in the final state are expected to be
rather different.
In fact, $ZZb\bar b$ processes yield lepton pairs with different sign and invariant mass about
$m_Z\simeq 91.19$~GeV, while the signal produces
same-sign lepton pairs with mass around
$m_Y\simeq 1.3$~TeV.
As for the $t\bar t Z$ background, its final states are different from the signal because of the
presence of neutrinos, however, as already observed in \cite{corcella_bilepton_2017,corcella_non-leptonic_2022}, processes
with top quarks in the dilepton channel can mimic the signal whenever the
missing energy due to neutrinos is small.

In the following, we simulate all backgrounds by means of \texttt{MadGraph},
matched to HERWIG 6, as was done for the signal, and concentrate ourselves on
FCC-$hh$.
The following LO cross sections are obtained at $\sqrt{s} = 100$~TeV:
\begin{equation}
	\sigma(ZZb\bar{b}) \simeq 4.7 \times 10^{-3}~{\rm pb}, \qquad \sigma(t\bar{t}Z) \simeq 1.7 \times 10^{-2}~{\rm pb}.
\end{equation}

In line with Ref.~\cite{corcella_non-leptonic_2022}, we apply the following acceptance cuts to transverse momentum ($p_T$),
pseudorapidity ($\eta$) and invariant opening angle
$\Delta R=\sqrt{(\Delta\eta)^2+(\Delta\phi)^2}$, where $\Delta\eta$ 
and $\Delta\phi$ are azimuthal and pseudorapidity gaps of 
final-state muons and $b$-jets, as well as to the missing transvere energy
(MET)\footnote{As in Ref.~\cite{corcella_non-leptonic_2022}
	the missing transverse energy is defined as ${\rm MET}=\sqrt{\left(\sum_{i=\nu}p_{x,i}\right)^2+  \left(\sum_{i=\nu}p_{y,i}\right)^2}$.}: 
\begin{eqnarray}
	\label{cuts}
	&& p_{T,b} > 30~{\rm GeV},\quad p_{T,\mu} > 20~{\rm GeV},\quad |\eta_b| < 4.5,\quad |\eta_\mu| < 2.5, \nonumber \\
	&& \Delta R_{bb} > 0.4,\quad \Delta R_{\mu\mu} > 0.1,\quad \Delta R_{b\mu} > 0.4,\quad {\rm MET} < 200~{\rm GeV}.
\end{eqnarray}
As in our previous work, these cuts reflect a conservative implementation of the lepton–jet overlap-removal algorithm used in LHC analyses~\cite{ATLAS:2018zzq}, and the MET requirement targets background~\eqref{back2}, which includes final-state neutrinos. However, the MET cut is applied universally to account for neutrinos from hadron decays in all processes.
To ensure realistic projections, we incorporate $b$-tagging efficiency. Although the efficiency depends on the $b$-jet kinematic properties, we adopt a flat approximation, as done in~\cite{corcella_non-leptonic_2022}, and defer a detailed modeling to future work. Following Ref.~\cite{selvaggi}, we assume a tagging efficiency of $\epsilon_b = 0.82$\footnote{This efficiency applies to $b$-jets with pseudorapidity $|\eta_j| < 2.5$ and transverse momentum in the range $10~{\rm GeV} < p_{T,j} < 500~{\rm GeV}$, which covers the dominant phase space of the events considered.}.

After applying the cuts in Eq.~\eqref{cuts} and accounting for $b$-tagging, the expected number of signal events at FCC-$hh$ with ${\cal L} = 3000$~fb$^{-1}$ is approximately $N(2b4\mu)_s \simeq 32880$
\footnote{We also varied the $T$ mass between 2 and 5 TeV, and found an expected
	number events at FCC-$hh$ between 9750 and 60450. The HL-LHC cross sections are instead
	between $2\times 10^{-6}$~pb and $2\times 10^{-9}$~pb, hence too low to see any signal.}.
The corresponding background event yields are $N(2b4\mu)_{b_1} \simeq 1288$ for $ZZb\bar{b}$ and $N(2b4\mu)_{b_2} \simeq 8338$ for $t\bar{t}Z$.

Strictly speaking, one should also consider processes where $Z$ bosons are
accompanied
by light jets which are misidentified as $b$-jets, but one can anticipate that,
after the cuts in Eq.~(\ref{cuts}) and mistag efficiencies are
applied \footnote{In Ref.~\cite{corcella_non-leptonic_2022},
	an average mistag rate $\epsilon_j=0.15$ was used.},
only few dozens event survive, which
makes such a background negligible when searching for heavy quarks and
bileptons at both LHC and FCC-$hh$.
Furthermore, unlike Ref.~\cite{corcella_non-leptonic_2022}, we
do not include the 4-top
Standard Model background, namely $pp\to (t\bar t)(t\bar t)$ followed by
$t\to bW^+$ and $W^+\to \mu^+\nu_\mu$. In fact, this process would yield a final state with 
4 $b$-jets, while our signal exhibits only  2 $b$-jets.
\footnote{In principle, one may even have 2 $b$-jets
	starting from 4 $b$-quarks, e.g., whenever two pairs of
	$B$-hadron are very close in the
	phase space and get clustered in the same jet, but nevertheless,
	as long as one uses a reasonable jet algorithm, such events are very
	much suppressed.}.

\begin{figure}[b]
	\centerline{\resizebox{0.59\textwidth}{!}
		{\includegraphics{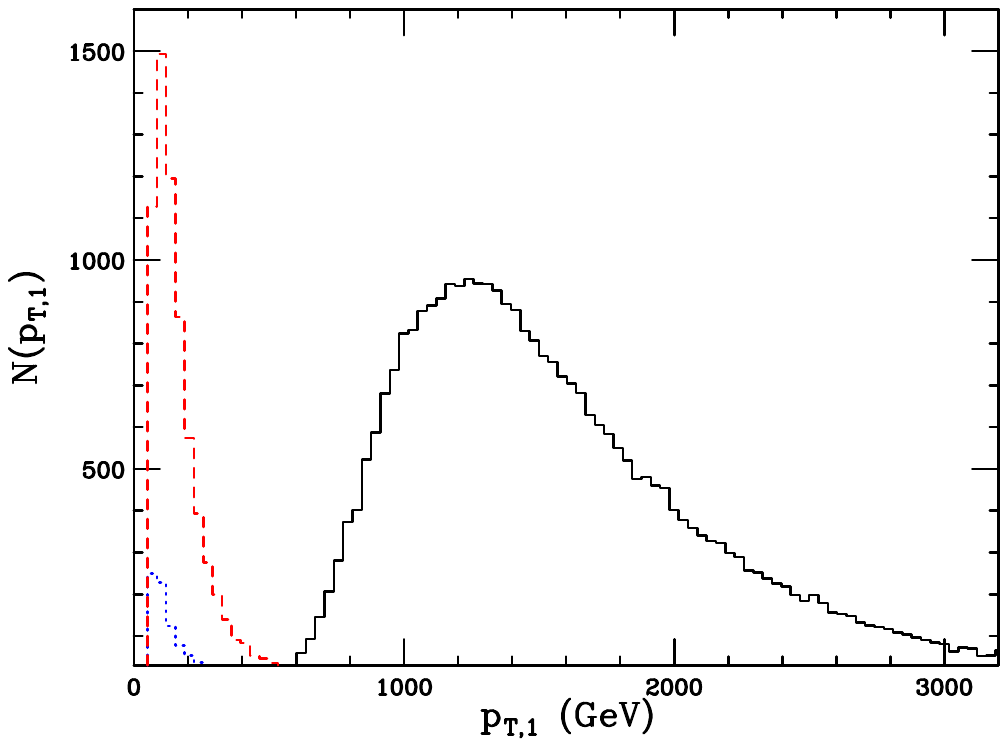}}%
		\hfill%
		\resizebox{0.59\textwidth}{!}{\includegraphics{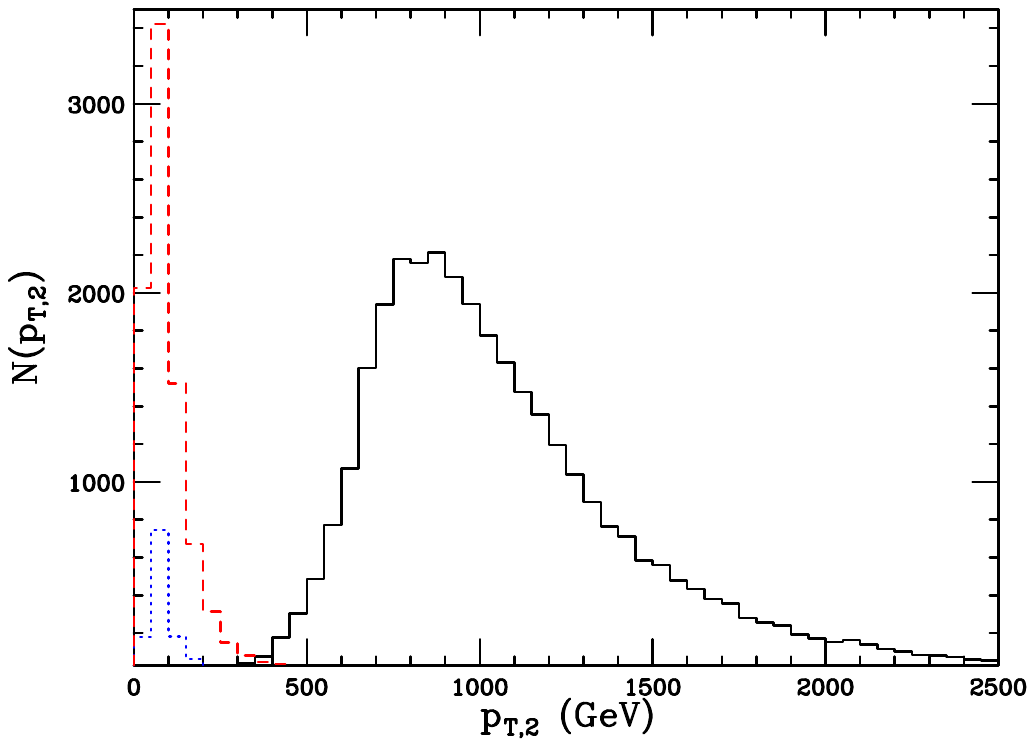}}}
	\caption{Transverse momentum of the hardest (left) and next-to-hardest
		muon (right) for the signal (solid) and the backgrounds $b_1$ (dots) and
		$b_2$ (dashes).}
	\label{ptl}
\end{figure}

As in previous work, we present some distributions according to
our bilepton signal in Eq.~(\ref{ttbar}) and backgrounds
Eqs.~(\ref{back1}) and (\ref{back2}).
All spectra are normalized in such a way that $N(x)$ yields the number
of expected events in the histogram bin around $x$ at FCC-$hh$. 
In Fig.~\ref{ptl}, we present the transverse momentum of the hardest
($p_{T,1}$, left) and next-to-hardest ($p_{T,2}$, right) muon
for the signal (solid) and backgrounds $b_1$ (dots) and $b_2$ (dashes):
overall, signal and background distributions look very different, which makes
their separation potentially straightforward.
The signal starts to be substantial at $p_{T,1}\simeq 600$~GeV and
$p_{T,2}\simeq 400$~GeV, peaks around 1.2 TeV ($p_{T,1}$) and
800 GeV ($p_{T,1}$) and vanishes for $p_{T,1}>3.5$~TeV and 
$p_{T,2}>2.5$~TeV. Background distributions are instead quite sharp:
$b_1$ is substantial only for $p_{T,1}< 200$~GeV and $p_{T,2}< 200$~GeV
and is maximum at low $p_{T,1}$ and $p_{T,2}$; $b_2$ yields a substantial number
of events for $p_{T,1}< 500$~GeV and $p_{T,2}< 500$~GeV
and is very sharply peaked
about 200 GeV ($p_{T,1}$) and 100 GeV ($p_{T,2}$).
Our signal features the production of heavy $T\bar T$ pairs and therefore the partonic centre-of mass energy is typically ${\cal O}(10~{\rm TeV})$
, which leads to bileptons and hence hardest leptons with transverse
momentum about
$p_T\sim {\cal O}(1~{\rm TeV})$. On the other hand, the $ZZb\bar b$ and
$t\bar tZ$ backgrounds are characterized by partonic centre-of-mass
energies which
are typically one order of magnitude lower and therefore the final-state
leptons are much softer than those originating from bileptons.

\begin{figure}
	\centerline{\resizebox{0.59\textwidth}{!}
		{\includegraphics{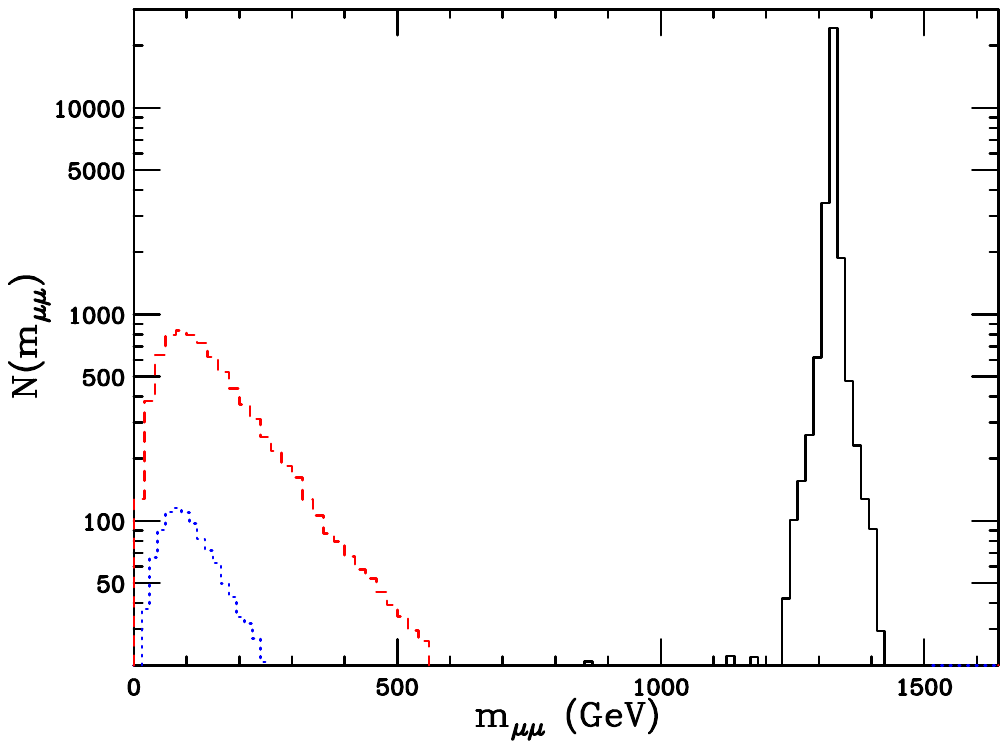}}%
		\hfill%
		\resizebox{0.59\textwidth}{!}{\includegraphics{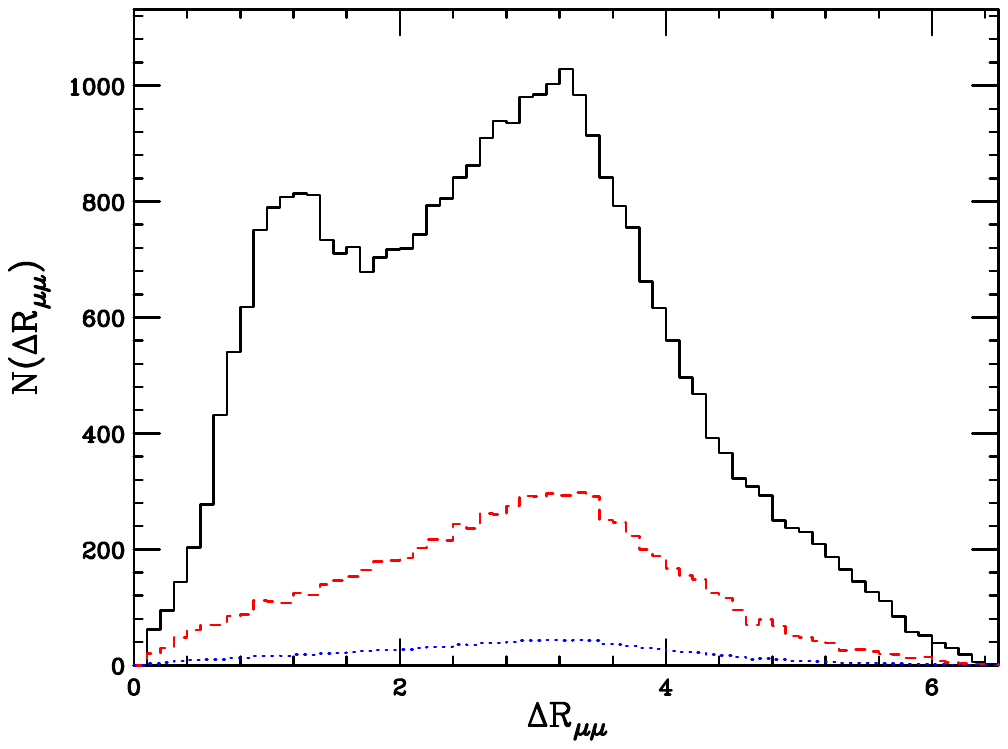}}}
	\caption{Invariant-mass distribution (left) and invariant opening angle
		(right) between the two hardest
		muons, according to the bilepton signal and the SM backgrounds . The histograms are labelled as in Fig.~\ref{ptl}.}
	\label{mlldr}
\end{figure}

Fig.~\ref{mlldr} shows the invariant mass distributions $m_{\mu\mu}$ of same-sign muon pairs (left) and the invariant opening angle $\Delta R_{\mu\mu}$ 
%=\sqrt{(\Delta\phi)^2+(\Delta\eta)^2}$, where $\Delta\phi$ and $\Delta\eta$ are the azimuthal and  pseudorapidity gaps
between the two hardest muons (right).
As observed in previous work as well, the invariant mass distribution yielded by the
bilepton model is strikingly different from the others, since it is sharply peaked
at the bilepton mass, i.e. $m_{\mu\mu}\simeq m_Y\simeq 1.3$~TeV; the background spectra 
are instead broader and peaked at a lower $m_{\mu\mu}\simeq$~80-90 GeV.
In fact, as discussed above, the backgrounds typically feature much softer leptons: furthermore, in order to get two leptons with the same sign, one
needs to pair the decay products of either two $Z$ bosons or
one $Z$ and one $W$ from top decays, which may
be quite distant in the phase space. Such effects are therefore
responsible of the
fact that the background $m_{\mu\mu}$  spectra are much softer and broader than
the signal.

As for $\Delta R_{\mu\mu}$, the bilepton signal is remarkably above the
background over the whole range and exhibits two maxima about
$\Delta R_{\mu\mu}\simeq 1$ and $\Delta R_{\mu\mu}\simeq 3$, with a dip in the
middle. Both backgrounds $b_1$ and $b_2$ increase until they reach a peak
value around $\Delta R_{\mu\mu}\simeq 3$ and rapidly vanish
for $\Delta R_{\mu\mu}>6$. All distributions are rather broad, 
which means that the two hardest leptons come either from the same
bilepton or vector boson (small $\Delta R$) or from two different ones (large $\Delta R$).

\begin{figure}
	\centerline{\resizebox{0.59\textwidth}{!}
		{\includegraphics{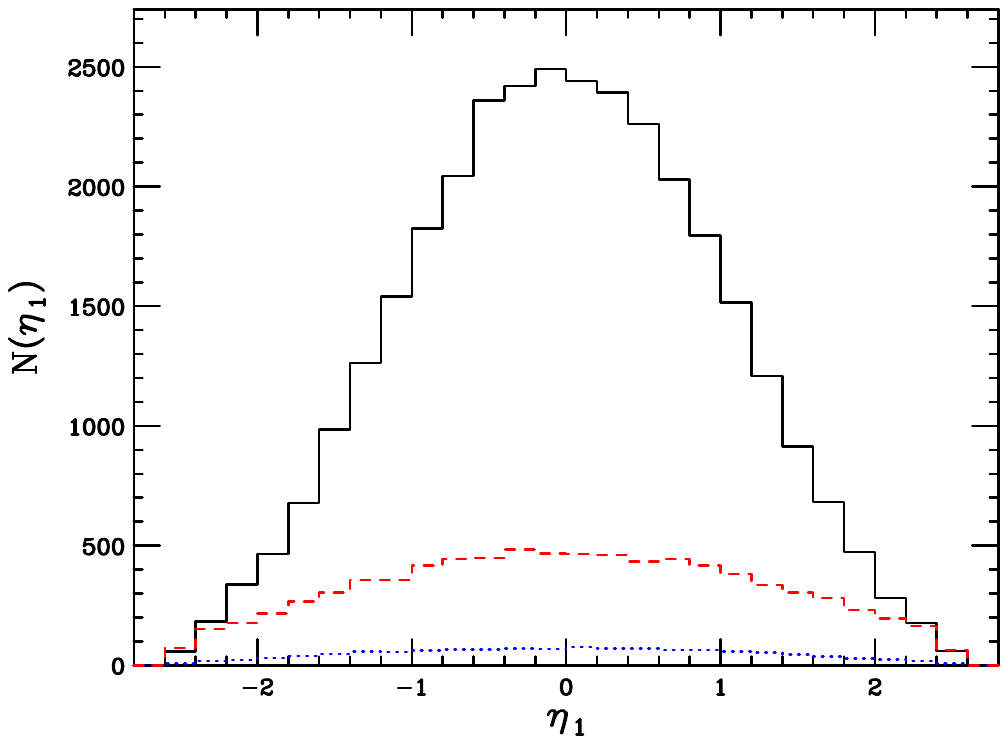}}%
		\hfill%
		\resizebox{0.59\textwidth}{!}{\includegraphics{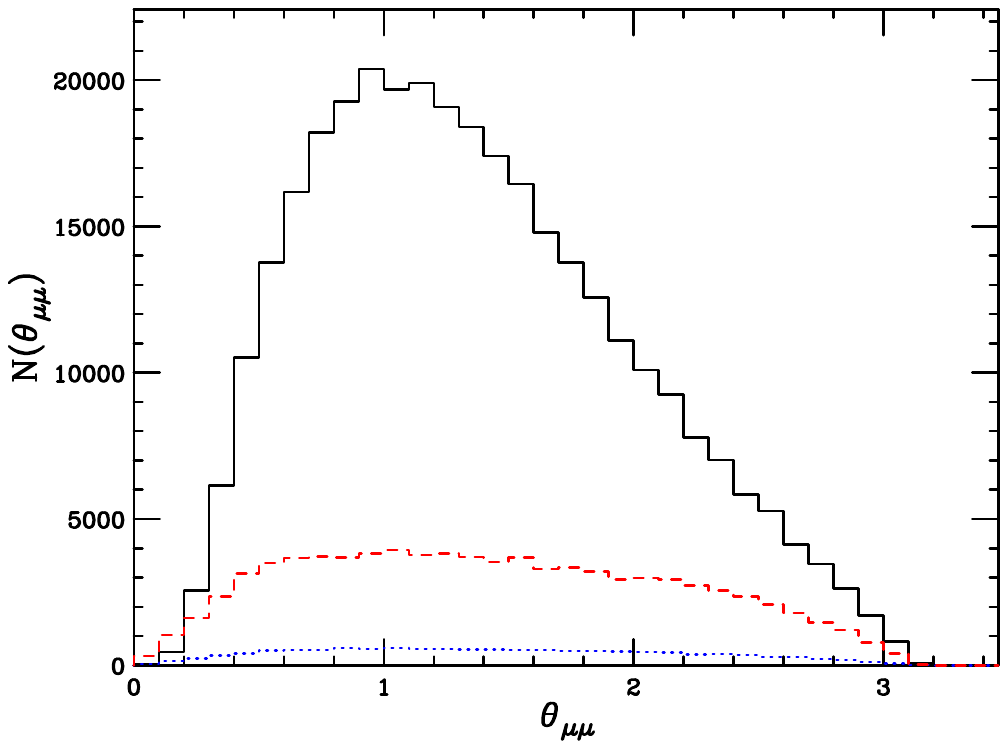}}}
	\caption{As in Fig.~\ref{ptl}, but displaying the pseudorapidity of the
		hardest lepton (left) of the angle (right) between the two hardest ones.}
	\label{etath}
\end{figure}

Fig.~\ref{etath} presents the pseudorapidity of the hardest muon $\eta_1$
and the angle $\theta_{\mu\mu}$ between the two hardest ones.
The pseudorapidity spectra exhibit a rise-and-fall behaviour, peaked at
$\eta_1\simeq 0$, which means that in both signals and backgrounds the
production of the hardest lepton in the central-rapidity region dominates.
The difference between the distributions is mostly due to normalization
and to the largest number of events yielded at FCC-$hh$ by the signal.
One can also notice that the $ZZb\bar b$ pseudorapidity is rather
flat over the full $-3<\eta_1<3$ range.

\begin{figure}
	\centerline{\resizebox{0.59\textwidth}{!}
		{\includegraphics{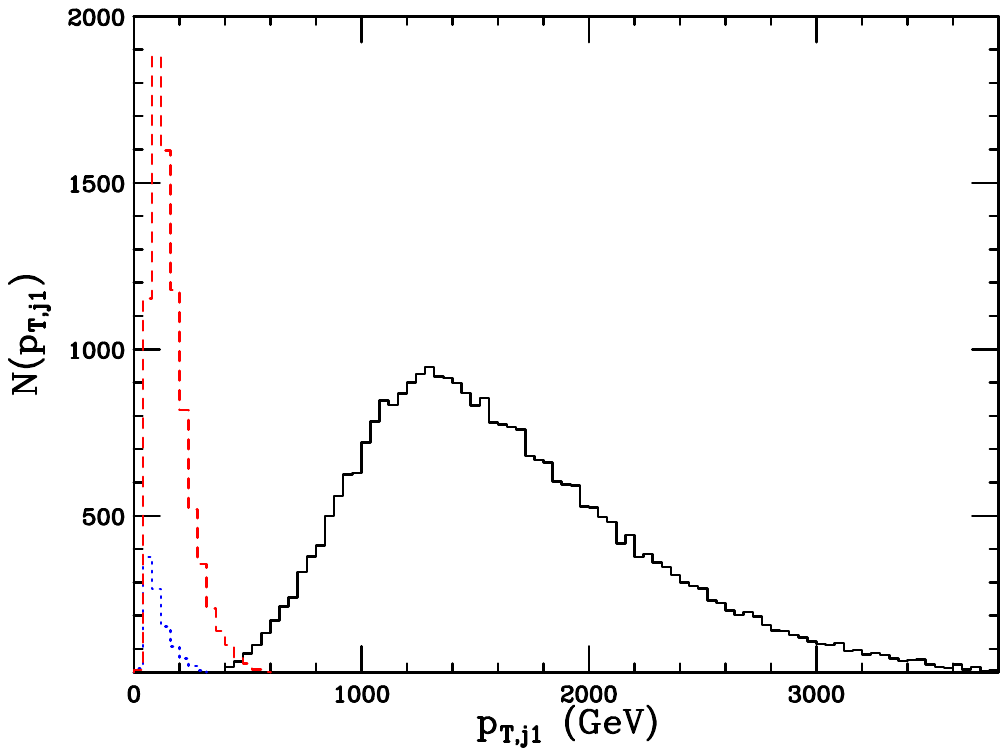}}%
		\hfill%
		\resizebox{0.59\textwidth}{!}{\includegraphics{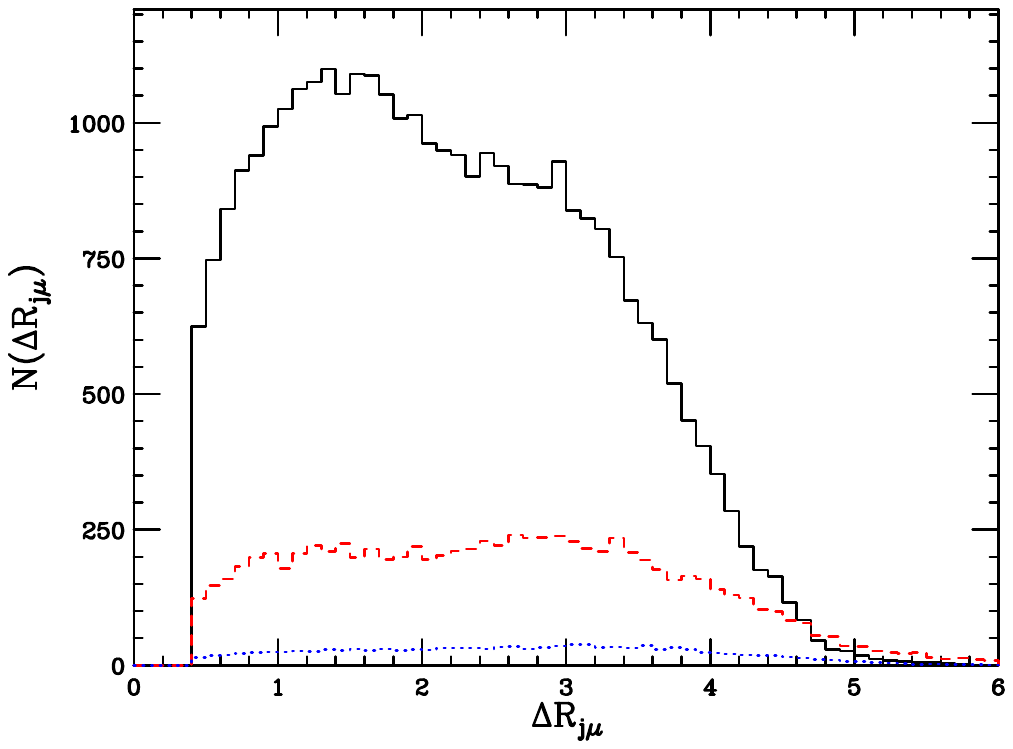}}}
	\caption{Transverse momentum of the hardest jet (left) and invariant opening angle
		among hardest jet and hardest muon. Histograms are labelled like in the
		previous figures.}
	\label{ptjdr}
\end{figure}

Similar observations hold also for $\theta_{\mu\mu}$: the signal is pretty broad and
peaked at $\theta_{\mu\mu}\simeq 1$, in agreement with the finding of
a large number of events in the central pseudorapidity region, 
while the backgrounds are quite flat.
Finally, Fig.~\ref{ptjdr} shows distributions which involve the $b$-jets 
accompanying the four muons as well: the transverse momentum of the hardest jet $p_{T,j1}$
(left) and $\Delta R_{j\mu}$ between hardest jet and hardest muon
(right). 
The $p_{T,j1}$ spectra are similar to the leptonic ones in Fig.~\ref{ptl}, with the
signal predicting jets at large transverse momenta and background jets concentrated
in the low-$p_T$ range. In fact, $b$-quarks and $b$-jets produced in
$T\to Y^{++}b$, where the $T$ mass is about 3 TeV,
are typically much harder that those in $ZZb\bar b$ or top decays in $t\bar tZ$.
As for the $\Delta R_{j\mu}$ predictions, they mostly differ because of the
normalization: the signal spectrum is broad and peaked about
$\Delta R_{j\mu}\simeq 1.5$,
while the backgrounds are flatter, with $b_2$ yielding even more events
at large $\Delta R_{j\mu}>4.6$. As observed in the lepton case, the
broadness of such spectra are due to the fact that the hardest jet and the
hardest lepton can come from either the same (small $\Delta R$) or
different decay chains (large $\Delta R$).

Before concluding this section, as in
\cite{corcella_exploring_2018,corcella_non-leptonic_2022}
we wish to calculate the significance $s$ to discriminate the bilepton signals $S$ through heavy quark
decays from the overall background $B$
\begin{equation}\label{signif}
	s=\frac{S}{\sqrt{B+\sigma_B^2}}.
\end{equation}
In (\ref{signif}), $\sigma_B$ is the systematic error on $B$, which
we roughly estimate as $\sigma_B\simeq 0.1 B$.
We then obtain $s\simeq 34$, which underlines, once again, the sensitivity of
FCC to heavy quarks and bileptons.

\section{Conclusions}
As the LHC experiments have so far revealed no evidence of physics beyond the Standard Model, we revisited the 331 model, which predicts a variety of new particles, including bileptons, i.e. gauge bosons with charge and lepton number
$\pm 2$, heavy quarks with electric charges $5/3$ and $4/3$, a $Z'$ boson, and an extended Higgs sector. Our analysis focused on the production of heavy quarks $T$ with charge $5/3$, which decay into bileptons and bottom quarks:
the production of vector bileptons is in fact
a distinctive feature of the 331 model.
This process results in final states characterized by two $b$-jets and two same-sign lepton pairs, with the leptons arising from the bilepton decays.

In our investigation, we focused on heavy quarks $T$ with
mass around 3 TeV and chose a benchmark point consistent with current
exclusion limits and theoretical expectations for the bilepton mass, i.e. 
about 1.3~TeV, in agreement with the finding of \cite{Coriano:2020iiz}.
In the considered scenario, bileptons are assumed to
decay into same-sign muon pairs.
As for the backgrounds, we explored $ZZb\bar b$ and $t\bar tZ$ production,
with the $Z$ bosons decaying into muon pairs and dilepton top decays, i.e.
$t\to bW^+$ and $W^+\to \mu^+\nu_\mu$, with
small values of the missing energy due to neutrinos.

Under these assumptions, we investigated a number of observables related to
both leptons and jets in the final states, i.e. muon and $b$-jet
transverse momenta, pseudorapidities, production polar angles,
invariant opening angles $\Delta R$, as well as the invariant mass
of same-sign muons.
We generally found that the bilepton signal can be cleanly separated from the Standard Model backgrounds and should be detectable with very high significance at a future 100~TeV hadron collider, such as the FCC-$hh$.
In fact, the kinematic features of leptons and $b$-jets originating from decays
of a 3 TeV heavy quark and a bilepton with mass 1.3 TeV are pretty much
different from those in Standard Model vector boson and top decays.
Unlike FCC-$hh$, the expected cross sections and event yields at the LHC,
even with the projected high-luminosity upgrade, are too low to allow any meaningful detection and therefore our analysis was just concentrated on
FCC-$hh$.

In summary, we argue that the bilepton signal explored in this work — together with the scenarios proposed in our previous studies~\cite{corcella_exploring_2018,corcella_non-leptonic_2022} — represents a compelling case for investigation at future collider experiments, which may shed light
on physics Beyond the Standard Model.


		\chapter{The 33311 Left-Right Bilepton Model}\label{ch33311}
\addcontentsline{toc}{section}{Synopsis}
\section*{Synopsis}

This Chapter introduces the 33311 model, a further extension of the 331 framework that enlarges the electroweak gauge group to $SU(3)_C \times SU(3)_L \times SU(3)_R \times U(1)_{X_L} \times U(1)_{X_R}$, incorporating a left-right symmetric structure alongside the bilepton mechanism. The Chapter is motivated by the observation that, while the 331 model provides an elegant resolution to the fermion family problem through interfamily anomaly cancellation, it still leaves open questions regarding the deeper origin of parity violation and the structure of the electroweak sector. The 33311 model addresses these by embedding the 331 gauge group within a left-right symmetric framework, inheriting its most attractive features while extending its theoretical scope.

The fermion content of the 33311 model is presented in detail. The quark sector follows an assignment analogous to the 331 model: the first two families are placed in triplets of $SU(3)_L$ paired with antitriplets of $SU(3)_R$, while the third family is assigned to antitriplets of $SU(3)_L$ paired with triplets of $SU(3)_R$, preserving the asymmetric inter-family structure that underpins anomaly cancellation. Three exotic quarks are introduced, one per family, with unconventional electric charges $Q = (-4/3, -4/3, +5/3)$. In the lepton sector, the three families are treated democratically, each comprising an $SU(3)_L$ antitriplet and an $SU(3)_R$ antitriplet. Crucially, each family now requires an additional charged lepton and a new neutral lepton beyond those present in the SM and the original 331 model, making the extended lepton content an unavoidable consequence of the left-right symmetric structure. The electroweak gauge boson sector correspondingly expands from the nine gauge bosons of the 331 model to eighteen in the 33311 model, with a mirrored set of right-handed gauge bosons including right-handed analogues of the $W^\pm$, $Z$, $Z'$, and the doubly-charged bileptons $Y^{\pm\pm}$ added to the spectrum.

The anomaly cancellation structure of the 33311 model is analyzed systematically. The thirteen potentially problematic triangle anomalies decompose into two groups. The first five, involving only $SU(3)_C$, $SU(3)_L$, and $U(1)_{X_L}$, cancel by exactly the same mechanism as in the 331 model and again require the number of families to be a multiple of three; QCD asymptotic freedom then excludes any multiple greater than three, fixing the family number to exactly three. The remaining eight anomalies, which involve $SU(3)_R$ and $U(1)_{X_R}$, are shown to cancel individually through explicit counting of the representations and their $U(1)$ charges. A notable simplification arises for the mixed $U(1)_{X_L}^2 U(1)_{X_R}$ and $U(1)_{X_L} U(1)_{X_R}^2$ anomalies, which vanish trivially because no state in the spectrum carries simultaneously non-zero $X_L$ and $X_R$ charges. The overall anomaly cancellation of the 33311 model is thus established, confirming the theoretical consistency of the construction.

The scalar sector and the pattern of spontaneous symmetry breaking are then described. The 33311 model involves three hierarchically separated energy scales. At the highest scale $M_{33311}$, which is unbounded from above, the symmetry $SU(3)_R \times U(1)_{X_L} \times U(1)_{X_R}$ is broken by the vacuum expectation values of a singlet scalar $\Phi(1,1,1)_{+1,-1}$, which reduces the two $U(1)$ factors to a single $U(1)_X$, and two misaligned scalar triplets $\Phi(1,1,3)_{0,0}$ under $SU(3)_R$, which break $SU(3)_R$ completely. At the intermediate scale $M_{331} \sim 4$~TeV, the scalar sector is identical to that of the 331 model: a triplet $\Phi(1,3,1)_{+1,0}$ breaks $SU(3)_L \times U(1)_X$ down to the SM electroweak group $SU(2)_L \times U(1)_Y$. At the electroweak scale $M_W$, three further scalar representations two $SU(3)_L$ triplets with $U(1)_X$ charges $0$ and $-1$, and a scalar sextet $\Phi(1,6,1)_{0,0}$ complete the breaking down to $U(1)_{\rm em}$, reproducing exactly the Higgs sector of the 331 model at the electroweak scale. In the asymptotic limit $M_{33311} \to \infty$, the 33311 model reduces to the 331 model, establishing a clear hierarchical relationship between the two constructions.

The Chapter concludes with a broader discussion of the theoretical implications of the 33311 model. Unlike traditional Grand Unified Theories such as $SU(5)$, which treat all three fermion families as sequential and identical under a single unified gauge group, the 33311 model requires a non-sequential family arrangement to achieve anomaly cancellation, fundamentally distinguishing it from conventional GUT approaches. This structure also avoids the prediction of proton decay, since quarks and leptons are not unified within a single representation, thereby remaining consistent with the experimental stability of the proton. The Chapter further notes that the 33311 model admits a natural extension to include a scalar dark matter candidate by introducing a discrete $Z_2$ symmetry, under which all original 33311 particles are even and the lightest odd scalar becomes a stable dark matter candidate, though the detailed analysis of this possibility is deferred to future work. The 33311 model thus stands as a theoretically consistent and phenomenologically rich framework that shares the family-number constraint of the 331 model while extending its gauge structure in the direction of left-right symmetry.

The work presented in this Chapter is based primarily on, and adapted from, the publication \cite{Coriano:2024qxe}, co-authored with C.~Corianò and P. H. ~Frampton.

\section{Introduction}
A fundamental requirement in any extension of the Standard Model (SM) is the cancellation of gauge anomalies. This principle is crucial because the internal consistency of gauge theories and the renormalisability of their extensions rely on preserving the Ward identities of the quantum action. In the SM, anomaly cancellation is achieved within the electroweak sector on a family-by-family basis, with fermion charge assignments naturally satisfying Diophantine equations that enforce anomaly-free conditions.\\
Despite this success, the SM leaves a central mystery unresolved: the existence of exactly three generations (or families) of quarks and leptons. These families interact directly only through Yukawa couplings, which give rise to the observed mass hierarchies, but the Model does not provide any fundamental explanation for why there are three families. The anomaly cancellation condition, when applied individually to each family, ensures internal consistency but does not inherently account for this replication across three generations.\\
This ambiguity extends to the question of whether, and how, to expand the electroweak sector. While the strong interaction sector, represented by the unbroken \( SU(3)_C \) gauge group of Quantum Chromodynamics (QCD), appears robust and well-understood, the original electroweak gauge group \( SU(2)_L \times U(1)_Y \) offers no explanation for the existence of three sequential fermion families. Consequently, the SM remains incomplete in this regard, motivating the search for extensions that could account for the observed family structure and suggest deeper principles governing the replication of fermion generations.
\noindent
\subsection{Interfamily and the 331 Model} 
The simplest solution to the family replication problem in the Standard Model (SM) involves extending \cite{frampton_chiral_1992,pisano_su3ensuremathbigotimesu1_1992} the electroweak gauge group 
\( SU(2)_L \times U(1)_Y \) to \( SU(3)_L \times U(1)_X \). This extension incorporates the electric charge operator \( Q \) into a more complex structure, defined as
\begin{equation}
	Q \equiv \frac{1}{2} \lambda_L^3 + \frac{\sqrt{3}}{2} \lambda_L^8 + X \sqrt{\frac{3}{2}} \lambda_L^9,
\end{equation}
where \( \lambda_L^a \) are the Gell-Mann matrices, normalized such that \( \text{Tr}(\lambda_L^a \lambda_L^b) \equiv \delta^{ab} \). This extension, when combined with the color gauge group \( SU(3)_C \), defines the 331 Model with the gauge symmetry \( SU(3)_C \times SU(3)_L \times U(1)_X \).\\
In this model, each of the three fermion families contains a new quark with charges 
\( Q =( -\frac{4}{3}, -\frac{4}{3}, +\frac{5}{3} )\) respectively for each family. By requiring asymptotic freedom in Quantum Chromodynamics (QCD), which ensures that the strong interaction coupling weakens at high energies, the 331 Model naturally restricts the number of families to three.\\
This is a stronger statement than the observation by Kobayashi and Maskawa \cite{kobayashi_cp-violation_1973} that CP violation emerges naturally with three families, as their framework only implies that there should be \textit{at least} three families, without explaining why there are exactly three. In contrast, the 331 Model directly links the existence of three generations to its gauge structure and anomaly cancellation requirements. \\
The 331 Model predicts a rich spectrum of new particles, including exotic quarks and leptons with unconventional charges and flavor interactions \cite{corianoSU3CSU3LU1X2024}. Among these are the doubly-charged bileptons, which are gauge bosons carrying lepton number and electric charge. One of the most distinctive predictions of the model is the existence of bileptons with charge \( \pm 2 \), which can manifest as same-sign lepton pairs in high-energy collisions. The production of such bileptons provides a clear experimental signature and represents one of the most accessible tests of the 331 Model at TeV-scale energies \cite{corcella_bilepton_2017,corcella_exploring_2018,corcella_non-leptonic_2022,corcella_vector-like_2021,corcella_hunting_2024}.\\
High-energy colliders, such as the Large Hadron Collider (LHC) and its potential upgrades, or future colliders, are well-suited to search for these bileptons. Their discovery would offer compelling evidence for the 331 Model and provide a window into physics beyond the Standard Model.\\
Unlike in the Standard Model, where anomaly cancellation occurs within each family independently, the 331 Model achieves anomaly cancellation across all three families collectively. This \textit{interfamily anomaly cancellation} is a fundamental feature of the model, distinguishing it from other extensions.

\section{33311-Model : chiral fermions and gauge bosons}
\noindent
In the 331-model, the minimal number of beyond the SM fermions
is three, namely one exotically-charged quark in each family and no
additional lepton. For the 33311-model, we can be equally economical
for the quarks but additional leptons are inevitable.

\noindent
Let us define the electric charge $Q$ embedding by
\begin{equation}
	Q \equiv \frac{1}{2} \lambda_L^3  + \frac{\sqrt{3}}{2} \lambda_L^8 
	+ \frac{1}{2} \lambda_R^3  + \frac{\sqrt{3}}{2} \lambda_R^8 
	+ X_L \sqrt{\frac{3}{2}} \lambda_L^9
	+ X_R \sqrt{\frac{3}{2}} \lambda_R^9
	\label{Q33311}
\end{equation}

\noindent
For the first family of quarks, in a notation
$(3_C, 3_L, 3_R)_{X_L,X_R}$ we include
\begin{equation}
	(3, 3, 1)_{-1/3,0} + (3, 1, 3^*)_{0,+1/3}
	\label{quarks1}
\end{equation}
which correspond to multiplets
\begin{equation}\label{f1}
	\left( \begin{array}{c} u \\d\\D  
	\end{array} \right) +
	\left( \begin{array}{c} \bar{D} \\ \bar{d} \\ \bar{u}  
	\end{array} \right)
\end{equation}
The second family quarks are in the representations of Eq.(\ref{quarks1}),
for the multiplets
\begin{equation}\label{f2}
	\left( \begin{array}{c} c \\s\\S  
	\end{array} \right) +
	\left( \begin{array}{c} \bar{S} \\ \bar{s} \\ \bar{c}  
	\end{array} \right)
\end{equation}

\bigskip

\noindent
For the third family we replace Eq.(\ref{quarks1})  by
\begin{equation}
	(3, 3^*, 1)_{+2/3,0} + (3, 1, 3)_{0,-2/3}
	\label{quarks3}
\end{equation}
and assign quarks to the multiplets
\begin{equation}\label{f3}
	\left( \begin{array}{c} T \\t\\b  
	\end{array} \right) +
	\left( \begin{array}{c} \bar{b} \\ \bar{t} \\ \bar{T}  
	\end{array} \right)
\end{equation}
The three additional quarks in Eq. \eqref{f1}, \eqref{f2} and \eqref{f3} can affect flavor-changing neutral weak currents (FCNCs) but early
work on the 331 model \cite{PhysRevD.50.548} proved that so long as the third family is the one which is treated asymmetrically,
the FCNCs are consistent with experiment, and this remains true in the present 33311 extension.
For every of this field the corresponding right handed singlets are added with the following quantum numbers
\begin{align}
	&\begin{matrix}
		(u^c)_L\\
		(c^c)_L\\ 
		(t^c)_L
	\end{matrix} \rightarrow \brak{{{3}}^*, {1}, {1}}_{-\frac{2}{3},0}\ ,
	&&\begin{matrix}
		(\bar u^c)_L\\
		(\bar c^c)_L\\ 
		(\bar t^c)_L
	\end{matrix} \rightarrow \brak{{ {3}}^*, {1}, {1}}_{0,\frac{2}{3}}\ ,\\\nn\\
	&\begin{matrix}
		(d^c)_L\\
		(s^c)_L\\
		(b^c)_L
	\end{matrix} \rightarrow \brak{{ {3}}^*, {1}, {1}}_{\frac{1}{3},0}\ ,
	&&\begin{matrix}
		({\bar{ d}}^c)_L\\
		(\bar s^c)_L\\
		({\bar b } ^c)_L
	\end{matrix} \rightarrow \brak{{ {3}}^*, {1}, {1}}_{0,-\frac{1}{3}}\ ,\\\nn\\
	&\begin{matrix}
		(D^c)_L\\(S^c)_L
	\end{matrix} \rightarrow \brak{{ {3}}^*, {1}, {1}}_{\frac{4}{3},0}\ ,
	&&\begin{matrix}
		(\bar D^c)_L\\(\bar S^c)_L
	\end{matrix} \rightarrow \brak{{ {3}}^*, {1}, {1}}_{0,-\frac{4}{3}}\ ,\\\nn\\
	&\begin{matrix}
		(T^c)_L
	\end{matrix} \rightarrow \brak{{ {3}}^*, {1}, {1}}_{-\frac{5}{3},0}\ ,
	&&\begin{matrix}
		(\bar T^c)_L
	\end{matrix} \rightarrow \brak{{ {3}}^*, {1}, {1}}_{0,\frac{5}{3}}\ ,
\end{align}

\bigskip

\noindent
Similarly to the  331-model, the leptons are neutral with respect
to $U(1)_{LX} \times U(1)_{RX}$ and assigned to
\begin{equation}
	3[ (1, 3^*, 1)_{0,0} + (1, 1, 3^*)_{0.0}]
	\label{leptons33311}
\end{equation}
\bigskip

\noindent
The lepton multiplets are
\begin{equation}
	\left( \begin{array}{c} e^+ \\\nu_e\\e^- 
	\end{array} \right) +
	\left( \begin{array}{c} E^+\\ N_e\\ E^-
	\end{array} \right) ~~~
	\left( \begin{array}{c} \mu^+ \\ \nu_{\mu}\\ \mu^- 
	\end{array} \right) +
	\left( \begin{array}{c} M^+\\  N_{\mu}\\ M^-
	\end{array} \right) ~~~\left( \begin{array}{c} \tau^+ \\ \nu_{\tau} \\ \tau^-
	\end{array} \right) +
	\left( \begin{array}{c} {\cal T}^+ \\ N_{\tau} \\ {\cal T}^-
	\end{array} \right)
\end{equation}
and for each family we need to include an additional
charged lepton and neutrino.  The three families
are treated democratically.

\noindent
While the strong interaction QCD  remains with its eight gluons,
the electroweak sector has expanded its number of gauge bosons
from four is the standard model to nine in the 331-model to
eighteen in the 33311-model.  We denote by an $R$ subscript
the new gauge bosons as follows
\begin{equation}
	(W^{\pm}, Z, \gamma, Y^{++},Y^+, Y^-,Y^{--}, Z^{'};
	W_R^{\pm}, Z_R, \gamma_R, Y^{++}_R,Y^+_R, Y^-_R,Y^{--}_R, Z^{'}_R)
	\label{gauge}
\end{equation}

\section{Anomaly cancellation in 33311}

There are thirteen potentially troublesome triangle anomalies
which are, in an obvious notation,
\begin{equation}
	SU(3)_C^3, \qquad SU(3)_C^2 U(1)_{XL}, \qquad SU(3)_L^3, \qquad  SU(3)_L^2 U(1)_{XL}, \qquad U(1)_{XL}^3
	\label{anomalies1}
\end{equation}
as well as
\begin{eqnarray}
	SU(3)_C^2U(1)_{XR} \qquad SU(3)_L^2U(1)_{XR}\qquad  SU(3)_R^3 \qquad  SU(3)_R^2U(1)_{XR},\nonumber  \\
	SU(3)_R^2U(1)_{XR}\qquad U(1)_{XL}^2 U(1)_{XR}\qquad U(1)_{XL}U(1)_{XR}^2 \qquad U(1)_{XR}^3
	\label{anomalies2}
\end{eqnarray}
\noindent
The first five in Eq.(\ref{anomalies1}) cancel exactly as in the
331-model, provided that there exist exactly three families. Actually, to be more precise, anomaly cancellation requires a multiple of three families but all higher multiples
of three are excluded by the requirement of not more than sixteen quark flavours due to QCD asymptotic
freedom. In the 331 Model each family has three flavors so that six families, or any higher multiple of three,
is excluded.
\noindent
We must therefore examine the anomalies in Eq.(\ref{anomalies2}). That these all cancel
follows from\\\\
\noindent
{\underline{$SU(3)_C^2U(1)_{XR}$.}\\
	Six colour triplets with $X_R=+1/3$\\
	Three colour triplets with $X_R=-2/3$.\\\\}
\noindent 
\underline{$SU(3)_L^2U(1)_{XR}$.}\\
Two triplets with $X_R=-1/3.$\\
One triplet with $X_R =- 2/3.$\\
One triplet with $X_R=0$\\\\
\noindent
\underline{$SU(3)_R^3$.}\\
Six triplets.\\
Six antitriplets.\\\\
\noindent\\
\underline{$SU(3)_R^2U(1)_{XL}$. }\\
Two triplets with $X_L=-1/3$.\\
One antriplet wuth $X=+2/3$.
One triplet with $X_R=0$.\\\\
\noindent
\underline{$SU(3)_R^2U(1)_{XR}$.}\\
Two antitriplets with $X_R=+1/3$.\\
One triplet with $X_R=-2/3$.\\
One.triplet with $X_R =0$.\\\\
\noindent
\underline{$U(1)_{XL}^2 U(1)_{XR}$.}\\
No states with $X_L \neq 0$ and $X_R \neq 0$.\\\\
\noindent
\underline{$U(1)_{XL} U(1)_{XR}^2$}.\\
No states with $X_L \neq 0$ and $X_R \neq 0$.\\\\
\noindent
\underline{$U(1)_{XR}^3$.}\\
Eighteen states have $X_R =1/3$.\\
Nine states have $X_R = -2/3$.

\section{Scalar representations and stages of symmetry breaking}
In the context of the 331 Model, the scalar sector is responsible for the spontaneous breaking of the gauge symmetry at different energy scales. The scalar fields are organized into representations under the extended gauge symmetry, and their vacuum expectation values (VEVs) trigger symmetry breaking, ultimately reducing the gauge group down to the Standard Model at low energies. Here is an explanation of how the scalar sector works in terms of representations.\\
\noindent
The 33311-model has a minimum of three energy scale
associated with its spontaneous symmetry breaking.\
The lowest one is the weak scale $M_W \simeq 248$ GeV.\\
The middle one is the 331 scale, $M_{331}\sim 4$ TeV \cite{martinez_landau_2007}.
Unlike the middle scale, the highest one has no upper bound $M_{331} < M_{33311}<\infty$.\\
In the asymptotic limit $M_{33311} \rightarrow \infty$,  the 33311-model reduces to the
331-model.
\noindent
Thus, although the 33311-Model shares with the 331-Model the explanation
for the existence of three families, it does not share the attractive property
that all the new physics is constrained to be in the TeV region.
\noindent
In a minimal scenario, the spontaneous symmetry breaking can occur 
at three distinct energy scales.\\
Using again the notation \( (3_C, 3_L, 3_R)_{X_L, X_R} \), we here list a set of scalar fields whose VEVs can be sufficient to spontaneously break the symmetry, as indicated in the previous section.

\noindent
\subsection{Breaking at the Highest Scale \( M_{33311} \)}
At the highest scale \( M_{33311} \), the gauge group \( SU(3)_R \times U(1)_{X_L} \times U(1)_{X_R} \) is broken down. The following scalar fields are introduced with VEVs that trigger this breaking:

\begin{itemize}
	\item \(\Phi(1, 1, 1)_{+1, -1}\): This scalar field is a singlet under \( SU(3)_L \) and \( SU(3)_R \) but carries charges under \( U(1)_{X_L} \) and \( U(1)_{X_R} \). Its VEV breaks the \( U(1)_{X_L} \times U(1)_{X_R} \) symmetry down to a single \( U(1)_X \).
	
	\item \(\Phi(1, 1, 3)_{0,0}\): This is a triplet under \( SU(3)_R \) and is neutral under \( SU(3)_L \) and \( U(1)_X \). Its VEV breaks \( SU(3)_R \) completely, leaving only the unbroken gauge group \( SU(3)_L \times U(1)_X \). The model requires two of these scalar triplets $3_R$  misaligned. 
\end{itemize}

\noindent
At this scale 

\begin{equation}
	U(1)_{X_L} \times U(1)_{X_R} \rightarrow U(1)_X
\end{equation}
\begin{equation}
SU(3)_R \rightarrow \text{nothing}
\end{equation}
Of course, as $M_{33311}$ is increased, there develops a hierarchy issue between $M_{33311}$ and $M_{331}$ although not
generally as severe as the hierarchy problem associated with grand unified theories.

\subsection{At scale \( M_{331} \)}

In the 33311 Model, the scalar sector at the intermediate scale \( M_{331} \) is identical to the scalar sector of the 331 Model \cite{tully_scalar_2003,costantini_theoretical_2020}. The scalar field listed here plays a specific role in this symmetry-breaking process, ensuring that the correct gauge bosons acquire masses while preserving the Standard Model structure at lower energies.

\begin{itemize}
	\item \(\Phi((1, 3, 1)_{+1,0}\): 
	This field is a triplet under \( SU(3)_L \) and a singlet under \( SU(3)_R \), carrying a charge under \( U(1)_X \).
	Its primary function is to break the \( SU(3)_L \) symmetry down to \( SU(2)_L \), the gauge group of the Standard Model electroweak interactions.
\end{itemize}

\noindent
This field contributes to the mass generation of the extended gauge bosons which are also present in the 331 Model, maintains anomaly cancellation, and aligns the vacuum expectation values to ensure that the electroweak \( SU(2)_L \times U(1)_Y \) structure remains intact:
\begin{equation}
SU(3)_L \times U(1)_X \rightarrow SU(2)_L \times U(1)_Y
\end{equation}
Note that provided $M_{331}$ is below the scale of 4 TeV, so that the electroweak mixing angle satisfies $\sin^2 \theta_W <
0.25$, there is no problem with the matching condition. Also no further such problem occurs between $M_{331}$
and $M_{33311}$.

\subsection{At scale \( M_W \)}
At the electroweak scale \( M_W \), the scalar fields are 
\begin{itemize}
	\item \(\Phi(1, 3, 1)_{0,0}\): 
	This is another triplet under \( SU(3)_L \) and a singlet under \( SU(3)_R \), but it is neutral under \( U(1)_X \).
	
	\item \(\Phi(1, 3, 1)_{-1,0}\): 
	This scalar is similar to \(\Phi(1, 3, 1)_{0,0}\) in terms of its \( SU(3)_L \) triplet structure, but it carries a charge of \(-1\) under \( U(1)_X \).
	
	\item \(\Phi(1, 6, 1)_{0,0}\): 
	This field is a symmetric representation under \( SU(3)_L \), often written as a sextet, and is neutral under both \( SU(3)_R \) and \( U(1)_X \).
\end{itemize}
which are necessary to break the symmetry to the electomagnetic $U(1)$
\begin{equation}
SU(2)_L \times U(1)_Y  \to  U(1)_Q .
\end{equation}

\section{Discussion}
\noindent
In traditional GUTs, like $SU(5)$, all three families of fermions are treated as identical, sequential generations under a single unified gauge group \cite{georgi_unity_1974,frampton_cosmological_2002}. However, the 33311 Model breaks away from this assumption by requiring a non-sequential arrangement of families to cancel anomalies. This structure fundamentally differentiates the 33311 Model from typical GUTs in several ways.
This distinction becomes crucial above the TeV scale, where the 33311 Model's non-sequential family organization ensures anomaly cancellation and theoretical consistency.\\
One can readily extend the 33311 Model to include a scalar dark matter candidate by introducing a dicrete
$Z_2$ symmetry, arranging that the original 33311 particles all have $Z_2 = +1$ and then the lightest $Z_2 =-1$
scalar will become a stable dark matter candidate. We defer a detailed discussion to further work.\\
Traditional GUTs predict proton decay as an outcome of unifying quarks and leptons within a single representation. However, the 33311 Model sidesteps this issue by not embedding quarks and leptons in the same way. Thus, it provides an explanation for the observed stability of the proton without conflicting with experimental bounds on proton decay.

\begin{comment}
	\providecommand{\href}[2]{#2}\begingroup\raggedright\endgroup
\end{comment}

		\begin{appendices}

\chapter{Appendix to Chapter 1}
\section{Rotations for the determination of the mass eigenstates}\label{massmat} 
In this appendix we summarize some results concerning the scalar/psesudoscalar sectors of the model \cite{staub_automatic_2011,staub_sarah_2014}. 

\begin{itemize} 
	\item {\bf Mass matrix of the scalar sector}, Basis: \( \left(\rho_1, \rho_2, \rho_3, {\rho}_{4}, {{\sigma}_{2}}^0\right), \left(\rho_1, \rho_2, \rho_3, {\rho}_{4}, {{\sigma}_{2}}^{0,*}\right) \) 
	
	\begin{equation} 
		m^2_{h} = \left( 
		\begin{array}{ccccc}
			m_{\rho_1\rho_1} &m_{\rho_2\rho_1} &m_{\rho_3\rho_1} &m_{{\rho}_{4}\rho_1} &0\\ 
			m_{\rho_1\rho_2} &m_{\rho_2\rho_2} &m_{\rho_3\rho_2} &m_{{\rho}_{4}\rho_2} &0\\ 
			m_{\rho_1\rho_3} &m_{\rho_2\rho_3} &m_{\rho_3\rho_3} &m_{{\rho}_{4}\rho_3} &0\\ 
			m_{\rho_1{\rho}_{4}} &m_{\rho_2{\rho}_{4}} &m_{\rho_3{\rho}_{4}} &m_{{\rho}_{4}{\rho}_{4}} &0\\ 
			0 &0 &0 &0 &m_{{{\sigma}_{2}}^0{{\sigma}_{2}}^{0,*}}\end{array} 
		\right) 
	\end{equation} 
	\begin{align} 
		m_{\rho_1\rho_1} &= \frac{1}{4} \Big(12 \lambda_{1} v_{\rho}^{2}  + 2 \Big(\lambda_{12} v_{\eta}^{2}  + \lambda_{13} v_{\chi}^{2} \Big) -2 \sqrt{2} v_{\eta} v_{\sigma} {\xi}_{14}  + v_{\sigma}^{2} \Big(2 {\lambda}_{14}  + \zeta_{14}\Big)\Big) + {\mu}_{1}\\ 
		m_{\rho_1\rho_2} &= f v_{\chi}  + v_{\rho} \Big(- \frac{1}{\sqrt{2}} v_{\sigma} {\xi}_{14}  + \lambda_{12} v_{\eta} \Big)\\ 
		m_{\rho_2\rho_2} &= \frac{1}{2} \Big(6 \lambda_{2} v_{\eta}^{2}  + \lambda_{12} v_{\rho}^{2}  + \lambda_{23} v_{\chi}^{2}  + v_{\sigma}^{2} \Big(-2 {\xi}_{24}  + {\lambda}_{24}\Big)\Big) + {\mu}_{2}\\ 
		m_{\rho_1\rho_3} &= \frac{1}{\sqrt{2}} {f}_{\sigma} v_{\sigma}  + f v_{\eta}  + \lambda_{13} v_{\rho} v_{\chi} \\ 
		m_{\rho_2\rho_3} &= f v_{\rho}  + v_{\chi} \Big(\frac{1}{\sqrt{2}} v_{\sigma} {\xi}_{34}  + \lambda_{23} v_{\eta} \Big)\\ 
		m_{\rho_3\rho_3} &= \frac{1}{4} \Big(12 \lambda_{3} v_{\chi}^{2}  + 2 \Big(\lambda_{13} v_{\rho}^{2}  + \lambda_{23} v_{\eta}^{2}  + \sqrt{2} v_{\eta} v_{\sigma} {\xi}_{34} \Big) + v_{\sigma}^{2} \Big(2 {\lambda}_{34}  + \zeta_{34}\Big)\Big) + {\mu}_{3}\\ 
		m_{\rho_1{\rho}_{4}} &= \frac{1}{2} \Big(\sqrt{2} {f}_{\sigma} v_{\chi}  + v_{\rho} \Big(- \sqrt{2} v_{\eta} {\xi}_{14}  + v_{\sigma} \Big(2 {\lambda}_{14}  + \zeta_{14}\Big)\Big)\Big)\\ 
		m_{\rho_2{\rho}_{4}} &= \frac{1}{2} \frac{1}{\sqrt{2}} \Big(v_{\chi}^{2} {\xi}_{34}  - v_{\rho}^{2} {\xi}_{14} \Big) + v_{\eta} v_{\sigma} \Big(-2 {\xi}_{24}  + {\lambda}_{24}\Big)\\ 
		m_{\rho_3{\rho}_{4}} &= \frac{1}{2} \Big(\sqrt{2} {f}_{\sigma} v_{\rho}  + v_{\chi} \Big(\sqrt{2} v_{\eta} {\xi}_{34}  + v_{\sigma} \Big(2 {\lambda}_{34}  + \zeta_{34}\Big)\Big)\Big)\\ 
		m_{{\rho}_{4}{\rho}_{4}} &= \frac{1}{4} \Big(2 v_{\eta}^{2} \Big(-2 {\xi}_{24}  + {\lambda}_{24}\Big) + 6 \Big(2 {\lambda}_4  + {\lambda}_{44}\Big)v_{\sigma}^{2}  + v_{\chi}^{2} \Big(2 {\lambda}_{34}  + \zeta_{34}\Big)\nonumber\\ &\ \ \ \ + v_{\rho}^{2} \Big(2 {\lambda}_{14}  + \zeta_{14}\Big)\Big) + {\mu}_{4}\\ 
		m_{{{\sigma}_{2}}^0{{\sigma}_{2}}^{0,*}} &= \frac{1}{2} \Big(2 {\lambda}_4 v_{\sigma}^{2}  + {\lambda}_{14} v_{\rho}^{2}  + {\lambda}_{34} v_{\chi}^{2}  + v_{\eta}^{2} \Big({\lambda}_{24} + \zeta_{24}\Big)\Big) + {\mu}_{4}
	\end{align} 
	This matrix is diagonalized by \(R^{S}\)
	\begin{equation} 
		R^{S} m^2_{h} R^{S,\dagger} = m^{dia}_{2,h} 
	\end{equation} 
	with 
	\begin{align} 
		\rho_1 = \sum_{j}R^{{S},*}_{j 1}h_{{j}}\,, \hspace{1cm} 
		\rho_2 = \sum_{j}R^{{S},*}_{j 2}h_{{j}}\,, \hspace{1cm} 
		\rho_3 = \sum_{j}R^{{S},*}_{j 3}h_{{j}}\\ 
		{\rho}_{4} = \sum_{j}R^{{S},*}_{j 4}h_{{j}}\,, \hspace{1cm} 
		{{\sigma}_{2}}^0 = \sum_{j}R^{{S},*}_{j 5}h_{{j}}
	\end{align} 
	\item {\bf Mass matrix of the pseudoscalar sector}, Basis: \( \left(\sigma_1, \sigma_2, \sigma_3, {\sigma}_{4}\right), \left(\sigma_1, \sigma_2, \sigma_3, {\sigma}_{4}\right) \) 
	
	\begin{equation} 
		\label{ma0}
		m^2_{A^0} = \left( 
		\begin{array}{cccc}
			m_{\sigma_1\sigma_1} &- f v_{\chi}  &- \frac{1}{\sqrt{2}} {f}_{\sigma} v_{\sigma}  - f v_{\eta}  &- \frac{1}{\sqrt{2}} {f}_{\sigma} v_{\chi} \\ 
			- f v_{\chi}  &m_{\sigma_2\sigma_2} &- f v_{\rho}  &m_{{\sigma}_{4}\sigma_2}\\ 
			- \frac{1}{\sqrt{2}} {f}_{\sigma} v_{\sigma}  - f v_{\eta}  &- f v_{\rho}  &m_{\sigma_3\sigma_3} &- \frac{1}{\sqrt{2}} {f}_{\sigma} v_{\rho} \\ 
			- \frac{1}{\sqrt{2}} {f}_{\sigma} v_{\chi}  &m_{\sigma_2{\sigma}_{4}} &- \frac{1}{\sqrt{2}} {f}_{\sigma} v_{\rho}  &m_{{\sigma}_{4}{\sigma}_{4}}\end{array} 
		\right) +  \xi_{Z}m^2(Z) +  \xi_{Z^\prime}m^2(Z^\prime) 
	\end{equation} 
	\begin{align} 
		m_{\sigma_1\sigma_1} &= \frac{1}{4} \Big(2 \Big(\lambda_{12} v_{\eta}^{2}  + \lambda_{13} v_{\chi}^{2} \Big) -2 \sqrt{2} v_{\eta} v_{\sigma} {\xi}_{14}  + 4 \lambda_{1} v_{\rho}^{2}  + v_{\sigma}^{2} \Big(2 {\lambda}_{14}  + \zeta_{14}\Big)\Big) + {\mu}_{1}\\ 
		m_{\sigma_2\sigma_2} &= \frac{1}{2} \Big(2 \lambda_{2} v_{\eta}^{2}  + \lambda_{12} v_{\rho}^{2}  + \lambda_{23} v_{\chi}^{2}  + v_{\sigma}^{2} \Big(2 {\xi}_{24}  + {\lambda}_{24}\Big)\Big) + {\mu}_{2}\\ 
		m_{\sigma_3\sigma_3} &= \frac{1}{4} \Big(2 \Big(\lambda_{13} v_{\rho}^{2}  + \lambda_{23} v_{\eta}^{2}  + \sqrt{2} v_{\eta} v_{\sigma} {\xi}_{34} \Big) + 4 \lambda_{3} v_{\chi}^{2}  + v_{\sigma}^{2} \Big(2 {\lambda}_{34}  + \zeta_{34}\Big)\Big) + {\mu}_{3}\\ 
		m_{\sigma_2{\sigma}_{4}} &= \frac{1}{4} \Big(-8 v_{\eta} v_{\sigma} {\xi}_{24}  + \sqrt{2} \Big(v_{\chi}^{2} {\xi}_{34}  - v_{\rho}^{2} {\xi}_{14} \Big)\Big)\\ 
		m_{{\sigma}_{4}{\sigma}_{4}} &= \frac{1}{4} \Big(2 \Big(\Big(2 {\lambda}_4  + {\lambda}_{44}\Big)v_{\sigma}^{2}  + v_{\eta}^{2} \Big(2 {\xi}_{24}  + {\lambda}_{24}\Big)\Big) + v_{\chi}^{2} \Big(2 {\lambda}_{34}  + \zeta_{34}\Big) \nonumber\\ &\ \ \ + v_{\rho}^{2} \Big(2 {\lambda}_{14}  + \zeta_{14}\Big)\Big) + {\mu}_{4}
	\end{align} 
	The gauge fixing contributions are
	\begin{equation} 
		m^2 (\xi_{Z}) = \left( 
		\begin{array}{cccc}
			m_{\sigma_1\sigma_1} &m_{\sigma_2\sigma_1} &m_{\sigma_3\sigma_1} &m_{{\sigma}_{4}\sigma_1}\\ 
			m_{\sigma_1\sigma_2} &m_{\sigma_2\sigma_2} &m_{\sigma_3\sigma_2} &m_{{\sigma}_{4}\sigma_2}\\ 
			m_{\sigma_1\sigma_3} &m_{\sigma_2\sigma_3} &m_{\sigma_3\sigma_3} &m_{{\sigma}_{4}\sigma_3}\\ 
			m_{\sigma_1{\sigma}_{4}} &m_{\sigma_2{\sigma}_{4}} &m_{\sigma_3{\sigma}_{4}} &m_{{\sigma}_{4}{\sigma}_{4}}\end{array} 
		\right) 
	\end{equation} 
	\begin{align} 
		m_{\sigma_1\sigma_1} &= \frac{1}{3} v_{\rho}^{2} \Big(-2 \sqrt{3} g_1 g_2 R_{{1 2}}^{Z} R_{{2 2}}^{Z}  + 3 g_{1}^{2} R_{{1 2}}^{Z,2}  + g_{2}^{2} R_{{2 2}}^{Z,2} \Big)\\ 
		m_{\sigma_1\sigma_2} &= \frac{1}{6} g_2 v_{\rho} v_{\eta} \Big(g_1 R_{{1 2}}^{Z} \Big(-3 R_{{3 2}}^{Z}  + \sqrt{3} R_{{2 2}}^{Z} \Big) + g_2 R_{{2 2}}^{Z} \Big(- R_{{2 2}}^{Z}  + \sqrt{3} R_{{3 2}}^{Z} \Big)\Big)\\ 
		m_{\sigma_2\sigma_2} &= \frac{1}{12} g_{2}^{2} v_{\eta}^{2} \Big(-2 \sqrt{3} R_{{2 2}}^{Z} R_{{3 2}}^{Z}  + 3 R_{{3 2}}^{Z,2}  + R_{{2 2}}^{Z,2}\Big)\\ 
		m_{\sigma_1\sigma_3} &= -\frac{1}{6} v_{\rho} v_{\chi} \Big(-3 g_1 g_2 R_{{1 2}}^{Z} \Big(\sqrt{3} R_{{2 2}}^{Z}  + R_{{3 2}}^{Z}\Big) + 6 g_{1}^{2} R_{{1 2}}^{Z,2}  + g_{2}^{2} R_{{2 2}}^{Z} \Big(\sqrt{3} R_{{3 2}}^{Z}  + R_{{2 2}}^{Z}\Big)\Big)\\ 
		m_{\sigma_2\sigma_3} &= \frac{1}{12} g_2 v_{\eta} v_{\chi} \Big(-2 g_1 R_{{1 2}}^{Z} \Big(-3 R_{{3 2}}^{Z}  + \sqrt{3} R_{{2 2}}^{Z} \Big) + g_2 \Big(-3 R_{{3 2}}^{Z,2}  + R_{{2 2}}^{Z,2}\Big)\Big)\\ 
		m_{\sigma_3\sigma_3} &= \frac{1}{12} v_{\chi}^{2} \Big(12 g_{1}^{2} R_{{1 2}}^{Z,2}  -4 g_1 g_2 R_{{1 2}}^{Z} \Big(3 R_{{3 2}}^{Z}  + \sqrt{3} R_{{2 2}}^{Z} \Big) \nonumber\\ & \  \  \  + g_{2}^{2} \Big(2 \sqrt{3} R_{{2 2}}^{Z} R_{{3 2}}^{Z}  + 3 R_{{3 2}}^{Z,2}  + R_{{2 2}}^{Z,2}\Big)\Big)\\ 
		m_{\sigma_1{\sigma}_{4}} &= \frac{1}{6} g_2 v_{\rho} v_{\sigma} \Big(g_1 R_{{1 2}}^{Z} \Big(-3 R_{{3 2}}^{Z}  + \sqrt{3} R_{{2 2}}^{Z} \Big) + g_2 R_{{2 2}}^{Z} \Big(- R_{{2 2}}^{Z}  + \sqrt{3} R_{{3 2}}^{Z} \Big)\Big)\\ 
		m_{\sigma_2{\sigma}_{4}} &= \frac{1}{12} g_{2}^{2} v_{\eta} v_{\sigma} \Big(-2 \sqrt{3} R_{{2 2}}^{Z} R_{{3 2}}^{Z}  + 3 R_{{3 2}}^{Z,2}  + R_{{2 2}}^{Z,2}\Big)\\ 
		m_{\sigma_3{\sigma}_{4}} &= \frac{1}{12} g_2 v_{\chi} v_{\sigma} \Big(-2 g_1 R_{{1 2}}^{Z} \Big(-3 R_{{3 2}}^{Z}  + \sqrt{3} R_{{2 2}}^{Z} \Big) + g_2 \Big(-3 R_{{3 2}}^{Z,2}  + R_{{2 2}}^{Z,2}\Big)\Big)\\ 
		m_{{\sigma}_{4}{\sigma}_{4}} &= \frac{1}{12} g_{2}^{2} v_{\sigma}^{2} \Big(-2 \sqrt{3} R_{{2 2}}^{Z} R_{{3 2}}^{Z}  + 3 R_{{3 2}}^{Z,2}  + R_{{2 2}}^{Z,2}\Big)
	\end{align} 
	
	\begin{equation} 
		m^2 (\xi_{Z^\prime}) = \left( 
		\begin{array}{cccc}
			m_{\sigma_1\sigma_1} &m_{\sigma_2\sigma_1} &m_{\sigma_3\sigma_1} &m_{{\sigma}_{4}\sigma_1}\\ 
			m_{\sigma_1\sigma_2} &m_{\sigma_2\sigma_2} &m_{\sigma_3\sigma_2} &m_{{\sigma}_{4}\sigma_2}\\ 
			m_{\sigma_1\sigma_3} &m_{\sigma_2\sigma_3} &m_{\sigma_3\sigma_3} &m_{{\sigma}_{4}\sigma_3}\\ 
			m_{\sigma_1{\sigma}_{4}} &m_{\sigma_2{\sigma}_{4}} &m_{\sigma_3{\sigma}_{4}} &m_{{\sigma}_{4}{\sigma}_{4}}\end{array} 
		\right) 
	\end{equation} 
	\begin{align} 
		m_{\sigma_1\sigma_1} &= \frac{1}{3} v_{\rho}^{2} \Big(-2 \sqrt{3} g_1 g_2 R_{{1 3}}^{Z} R_{{2 3}}^{Z}  + 3 g_{1}^{2} R_{{1 3}}^{Z,2}  + g_{2}^{2} R_{{2 3}}^{Z,2} \Big)\\ 
		m_{\sigma_1\sigma_2} &= \frac{1}{6} g_2 v_{\rho} v_{\eta} \Big(g_1 R_{{1 3}}^{Z} \Big(-3 R_{{3 3}}^{Z}  + \sqrt{3} R_{{2 3}}^{Z} \Big) + g_2 R_{{2 3}}^{Z} \Big(- R_{{2 3}}^{Z}  + \sqrt{3} R_{{3 3}}^{Z} \Big)\Big)\\ 
		m_{\sigma_2\sigma_2} &= \frac{1}{12} g_{2}^{2} v_{\eta}^{2} \Big(-2 \sqrt{3} R_{{2 3}}^{Z} R_{{3 3}}^{Z}  + 3 R_{{3 3}}^{Z,2}  + R_{{2 3}}^{Z,2}\Big)\\ 
		m_{\sigma_1\sigma_3} &= -\frac{1}{6} v_{\rho} v_{\chi} \Big(-3 g_1 g_2 R_{{1 3}}^{Z} \Big(\sqrt{3} R_{{2 3}}^{Z}  + R_{{3 3}}^{Z}\Big) + 6 g_{1}^{2} R_{{1 3}}^{Z,2}  + g_{2}^{2} R_{{2 3}}^{Z} \Big(\sqrt{3} R_{{3 3}}^{Z}  + R_{{2 3}}^{Z}\Big)\Big)\\ 
		m_{\sigma_2\sigma_3} &= \frac{1}{12} g_2 v_{\eta} v_{\chi} \Big(-2 g_1 R_{{1 3}}^{Z} \Big(-3 R_{{3 3}}^{Z}  + \sqrt{3} R_{{2 3}}^{Z} \Big) + g_2 \Big(-3 R_{{3 3}}^{Z,2}  + R_{{2 3}}^{Z,2}\Big)\Big)\\ 
		m_{\sigma_3\sigma_3} &= \frac{1}{12} v_{\chi}^{2} \Big(12 g_{1}^{2} R_{{1 3}}^{Z,2}  -4 g_1 g_2 R_{{1 3}}^{Z} \Big(3 R_{{3 3}}^{Z}  + \sqrt{3} R_{{2 3}}^{Z} \Big)\nonumber\\ &\ \  \  + g_{2}^{2} \Big(2 \sqrt{3} R_{{2 3}}^{Z} R_{{3 3}}^{Z}  + 3 R_{{3 3}}^{Z,2}  + R_{{2 3}}^{Z,2}\Big)\Big)\\ 
		m_{\sigma_1{\sigma}_{4}} &= \frac{1}{6} g_2 v_{\rho} v_{\sigma} \Big(g_1 R_{{1 3}}^{Z} \Big(-3 R_{{3 3}}^{Z}  + \sqrt{3} R_{{2 3}}^{Z} \Big) + g_2 R_{{2 3}}^{Z} \Big(- R_{{2 3}}^{Z}  + \sqrt{3} R_{{3 3}}^{Z} \Big)\Big)\\ 
		m_{\sigma_2{\sigma}_{4}} &= \frac{1}{12} g_{2}^{2} v_{\eta} v_{\sigma} \Big(-2 \sqrt{3} R_{{2 3}}^{Z} R_{{3 3}}^{Z}  + 3 R_{{3 3}}^{Z,2}  + R_{{2 3}}^{Z,2}\Big)\\ 
		m_{\sigma_3{\sigma}_{4}} &= \frac{1}{12} g_2 v_{\chi} v_{\sigma} \Big(-2 g_1 R_{{1 3}}^{Z} \Big(-3 R_{{3 3}}^{Z}  + \sqrt{3} R_{{2 3}}^{Z} \Big) + g_2 \Big(-3 R_{{3 3}}^{Z,2}  + R_{{2 3}}^{Z,2}\Big)\Big)\\ 
		m_{{\sigma}_{4}{\sigma}_{4}} &= \frac{1}{12} g_{2}^{2} v_{\sigma}^{2} \Big(-2 \sqrt{3} R_{{2 3}}^{Z} R_{{3 3}}^{Z}  + 3 R_{{3 3}}^{Z,2}  + R_{{2 3}}^{Z,2}\Big)
	\end{align} 
	where $R^Z$ is the rotation matrix that diagonalize the mass of the neutral gauge boson components in the $\{W^3,W^8,X\}$ basis which is given by the following matrix
	\begin{equation}
		\left(
		\begin{array}{ccc}
			g_1^2 v_\r^2+g_1^2 v_\chi^2 & -\frac{g_1 g_2 v_\r^2}{\sqrt{3}}-\frac{g_1 g_2 v_\chi^2}{2 \sqrt{3}} & -\frac{1}{2} g_1 g_2 v_\chi^2 \\
			-\frac{g_1 g_2 v_\r^2}{\sqrt{3}}-\frac{g_1 g_2 v_\chi^2}{2 \sqrt{3}} & \frac{g_2^2 v_\r^2}{3}+\frac{g_2^2 v_\eta^2}{12}+\frac{g_2^2 v_\chi^2}{12}+\frac{g_2^2 v_\sigma^2}{12} & -\frac{g_2^2 v_\eta^2}{4 \sqrt{3}}+\frac{g_2^2 v_\chi^2}{4 \sqrt{3}}-\frac{g_2^2 v_\sigma^2}{4 \sqrt{3}} \\
			-\frac{1}{2} g_1 g_2 v_\chi^2 & -\frac{g_2^2 v_\eta^2}{4 \sqrt{3}}+\frac{g_2^2 v_\chi^2}{4 \sqrt{3}}-\frac{g_2^2 v_\sigma^2}{4 \sqrt{3}} & \frac{g_2^2 v_\eta^2}{4}+\frac{g_2^2 v_\chi^2}{4}+\frac{g_2^2 v_\sigma^2}{4} \\
		\end{array}
		\right)
	\end{equation}
	The matrix \eqref{ma0} is diagonalized by the matrix \(R^{P}\) 
	\begin{equation} 
		R^{P} m^2_{A^0} R^{P,\dagger} = m^{dia}_{2,A^0} 
	\end{equation} 
	with 
	\begin{align} 
		\sigma_1 = \sum_{j}R^{{P},*}_{j 1}A^0_{{j}}\,, \hspace{1cm} 
		\sigma_2 = \sum_{j}R^{{P},*}_{j 2}A^0_{{j}}\,, \hspace{1cm} 
		\sigma_3 = \sum_{j}R^{{P},*}_{j 3}A^0_{{j}}\\ 
		{\sigma}_{4} = \sum_{j}R^{{P},*}_{j 4}A^0_{{j}}
	\end{align} 
	\item {\bf Mass matrix for Charged Higgs}, Basis: \( \left({\rho}_{+}^*, {\eta}_1^{{+},*}, {\sigma}_2^{{p},*}, {\eta}_2^{-}, {\chi}_{-}, {\sigma}_1^{-}\right), \left({\rho}_{+}, {\eta}_1^{+}, {\sigma}_2^{p}, {\eta}_2^{{-},*}, {\chi}_{-}^*, {\sigma}_1^{{-},*}\right) \) 
	
	\begin{equation} 
		m^2_{H^-} = \left( 
		\begin{array}{cccccc}
			m_{{\rho}_{+}^*{\rho}_{+}} &0 &0 &m^*_{{\eta}_2^{-}{\rho}_{+}} &0 &m^*_{{\sigma}_1^{-}{\rho}_{+}}\\ 
			0 &m_{{\eta}_1^{{+},*}{\eta}_1^{+}} &m^*_{{\sigma}_2^{{p},*}{\eta}_1^{+}} &0 &m^*_{{\chi}_{-}{\eta}_1^{+}} &0\\ 
			0 &m_{{\eta}_1^{{+},*}{\sigma}_2^{p}} &m_{{\sigma}_2^{{p},*}{\sigma}_2^{p}} &0 &m^*_{{\chi}_{-}{\sigma}_2^{p}} &0\\ 
			m_{{\rho}_{+}^*{\eta}_2^{{-},*}} &0 &0 &m_{{\eta}_2^{-}{\eta}_2^{{-},*}} &0 &m^*_{{\sigma}_1^{-}{\eta}_2^{{-},*}}\\ 
			0 &m_{{\eta}_1^{{+},*}{\chi}_{-}^*} &m_{{\sigma}_2^{{p},*}{\chi}_{-}^*} &0 &m_{{\chi}_{-}{\chi}_{-}^*} &0\\ 
			m_{{\rho}_{+}^*{\sigma}_1^{{-},*}} &0 &0 &m_{{\eta}_2^{-}{\sigma}_1^{{-},*}} &0 &m_{{\sigma}_1^{-}{\sigma}_1^{{-},*}}\end{array} 
		\right) +  \xi_{W^+}m^2(W^+) +  \xi_{{W^\prime}^+}m^2({W^\prime}^+) 
	\end{equation} 
	\begin{align} 
		m_{{\rho}_{+}^*{\rho}_{+}} &= \frac{1}{2} \Big(2 \lambda_{1} v_{\rho}^{2}  + \Big(\lambda_{12} + \zeta_{12}\Big)v_{\eta}^{2}  + \lambda_{13} v_{\chi}^{2}  + {\lambda}_{14} v_{\sigma}^{2} \Big) + {\mu}_{1}\\ 
		m_{{\eta}_1^{{+},*}{\eta}_1^{+}} &= \frac{1}{4} \Big(2 \Big(\lambda_{12} v_{\rho}^{2}  + \Big(\lambda_{23} + \zeta_{23}\Big)v_{\chi}^{2} \Big) + 4 \lambda_{2} v_{\eta}^{2}  + v_{\sigma}^{2} \Big(2 {\lambda}_{24}  + \zeta_{24}\Big)\Big) + {\mu}_{2}\\ 
		m_{{\eta}_1^{{+},*}{\sigma}_2^{p}} &= \frac{1}{4} \Big(\sqrt{2} v_{\rho}^{2} {\xi}_{14}  + v_{\eta} v_{\sigma} \Big(4 {\xi}_{24}  + \zeta_{24}\Big)\Big)\\ 
		m_{{\sigma}_2^{{p},*}{\sigma}_2^{p}} &= \frac{1}{4} \Big(2 \Big(\Big(2 {\lambda}_4  + {\lambda}_{44}\Big)v_{\sigma}^{2}  + {\lambda}_{34} v_{\chi}^{2} \Big) + v_{\eta}^{2} \Big(2 {\lambda}_{24}  + \zeta_{24}\Big) + v_{\rho}^{2} \Big(2 {\lambda}_{14}  + \zeta_{14}\Big)\Big) + {\mu}_{4}\\ 
		m_{{\rho}_{+}^*{\eta}_2^{{-},*}} &= \frac{1}{4} \Big(-4 f v_{\chi}  + v_{\rho} \Big(2 \zeta_{12} v_{\eta}  + \sqrt{2} v_{\sigma} {\xi}_{14} \Big)\Big)\\ 
		m_{{\eta}_2^{-}{\eta}_2^{{-},*}} &= \frac{1}{4} \Big(2 \Big(\Big(\lambda_{12} + \zeta_{12}\Big)v_{\rho}^{2}  + \lambda_{23} v_{\chi}^{2} \Big) + 4 \lambda_{2} v_{\eta}^{2}  + v_{\sigma}^{2} \Big(2 {\lambda}_{24}  + \zeta_{24}\Big)\Big) + {\mu}_{2}\\ 
		m_{{\eta}_1^{{+},*}{\chi}_{-}^*} &= \frac{1}{4} \Big(-4 f v_{\rho}  + v_{\chi} \Big(2 \zeta_{23} v_{\eta}  - \sqrt{2} v_{\sigma} {\xi}_{34} \Big)\Big)\\ 
		m_{{\sigma}_2^{{p},*}{\chi}_{-}^*} &= \frac{1}{4} \Big(\sqrt{2} \Big(2 {f}_{\sigma} v_{\rho}  + v_{\eta} v_{\chi} {\xi}_{34} \Big) + v_{\chi} v_{\sigma} \zeta_{34} \Big)\\ 
		m_{{\chi}_{-}{\chi}_{-}^*} &= \frac{1}{2} \Big(2 \lambda_{3} v_{\chi}^{2}  + \lambda_{13} v_{\rho}^{2}  + \Big(\lambda_{23} + \zeta_{23}\Big)v_{\eta}^{2}  + {\lambda}_{34} v_{\sigma}^{2} \Big) + {\mu}_{3}\\ 
		m_{{\rho}_{+}^*{\sigma}_1^{{-},*}} &= \frac{1}{4} \Big(\sqrt{2} \Big(2 {f}_{\sigma} v_{\chi}  - v_{\rho} v_{\eta} {\xi}_{14} \Big) + v_{\rho} v_{\sigma} \zeta_{14} \Big)\\ 
		m_{{\eta}_2^{-}{\sigma}_1^{{-},*}} &= \frac{1}{4} \Big(- \sqrt{2} v_{\chi}^{2} {\xi}_{34}  + v_{\eta} v_{\sigma} \Big(4 {\xi}_{24}  + \zeta_{24}\Big)\Big)\\ 
		m_{{\sigma}_1^{-}{\sigma}_1^{{-},*}} &= \frac{1}{4} \Big(2 \Big(\Big(2 {\lambda}_4  + {\lambda}_{44}\Big)v_{\sigma}^{2}  + {\lambda}_{14} v_{\rho}^{2} \Big) + v_{\chi}^{2} \Big(2 {\lambda}_{34}  + \zeta_{34}\Big) + v_{\eta}^{2} \Big(2 {\lambda}_{24}  + \zeta_{24}\Big)\Big) + {\mu}_{4}
	\end{align} 
	The gauge fixing contributions are
	\begin{equation} 
		m^2 (\xi_{W^+}) = \left( 
		\begin{array}{cccccc}
			0 &0 &0 &0 &0 &0\\ 
			0 &\frac{1}{4} g_{2}^{2} v_{\eta}^{2}  &-\frac{1}{4} g_{2}^{2} v_{\eta} v_{\sigma}  &0 &-\frac{1}{4} g_{2}^{2} v_{\eta} v_{\chi}  &0\\ 
			0 &-\frac{1}{4} g_{2}^{2} v_{\eta} v_{\sigma}  &\frac{1}{4} g_{2}^{2} v_{\sigma}^{2}  &0 &\frac{1}{4} g_{2}^{2} v_{\chi} v_{\sigma}  &0\\ 
			0 &0 &0 &0 &0 &0\\ 
			0 &-\frac{1}{4} g_{2}^{2} v_{\eta} v_{\chi}  &\frac{1}{4} g_{2}^{2} v_{\chi} v_{\sigma}  &0 &\frac{1}{4} g_{2}^{2} v_{\chi}^{2}  &0\\ 
			0 &0 &0 &0 &0 &0\end{array} 
		\right) 
	\end{equation} 
	\begin{equation} 
		m^2 (\xi_{{W^\prime}^+}) = \left( 
		\begin{array}{cccccc}
			\frac{1}{4} g_{2}^{2} v_{\rho}^{2}  &0 &0 &-\frac{1}{4} g_{2}^{2} v_{\rho} v_{\eta}  &0 &\frac{1}{4} g_{2}^{2} v_{\rho} v_{\sigma} \\ 
			0 &0 &0 &0 &0 &0\\ 
			0 &0 &0 &0 &0 &0\\ 
			-\frac{1}{4} g_{2}^{2} v_{\rho} v_{\eta}  &0 &0 &\frac{1}{4} g_{2}^{2} v_{\eta}^{2}  &0 &-\frac{1}{4} g_{2}^{2} v_{\eta} v_{\sigma} \\ 
			0 &0 &0 &0 &0 &0\\ 
			\frac{1}{4} g_{2}^{2} v_{\rho} v_{\sigma}  &0 &0 &-\frac{1}{4} g_{2}^{2} v_{\eta} v_{\sigma}  &0 &\frac{1}{4} g_{2}^{2} v_{\sigma}^{2} \end{array} 
		\right) 
	\end{equation} 
	This matrix is diagonalized by \(R^{C}\)
	\begin{equation} 
		R^{C} m^2_{H^-} R^{C,\dagger} = m^{dia}_{2,H^-} 
	\end{equation} 
	with 
	\begin{align} 
		{\rho}_{+} = \sum_{j}R_{{j 1}}^{C}H^+_{{j}}\,, \hspace{1cm} 
		{\eta}_1^{+} = \sum_{j}R_{{j 2}}^{C}H^+_{{j}}\,, \hspace{1cm} 
		{\sigma}_2^{p} = \sum_{j}R_{{j 3}}^{C}H^+_{{j}}\\ 
		{\eta}_2^{-} = \sum_{j}R^{{C},*}_{j 4}H^-_{{j}}\,, \hspace{1cm} 
		{\chi}_{-} = \sum_{j}R^{{C},*}_{j 5}H^-_{{j}}\,, \hspace{1cm} 
		{\sigma}_1^{-} = \sum_{j}R^{{C},*}_{j 6}H^-_{{j}}
	\end{align} 
	\item {\bf Mass matrix for Doubly Charged Higgs} \\Basis: \( \left({\rho}^{{++},*}, {\chi}^{--}, {{\sigma}_{1}}^{--}, {{\sigma}_{2}}^{{++},*}\right), \left({\rho}^{++}, {\chi}^{{--},*}, {{\sigma}_{1}}^{{--},*}, {{\sigma}_{2}}^{++}\right) \) 
	
	\begin{equation} 
		m^2_{H^{--}} = \left( 
		\begin{array}{cccc}
			m_{{\rho}^{{++},*}{\rho}^{++}} &m^*_{{\chi}^{--}{\rho}^{++}} &m^*_{{{\sigma}_{1}}^{--}{\rho}^{++}} &m^*_{{{\sigma}_{2}}^{{++},*}{\rho}^{++}}\\ 
			m_{{\rho}^{{++},*}{\chi}^{{--},*}} &m_{{\chi}^{--}{\chi}^{{--},*}} &m^*_{{{\sigma}_{1}}^{--}{\chi}^{{--},*}} &m^*_{{{\sigma}_{2}}^{{++},*}{\chi}^{{--},*}}\\ 
			m_{{\rho}^{{++},*}{{\sigma}_{1}}^{{--},*}} &m_{{\chi}^{--}{{\sigma}_{1}}^{{--},*}} &m_{{{\sigma}_{1}}^{--}{{\sigma}_{1}}^{{--},*}} &\frac{1}{2} {\lambda}_{44} v_{\sigma}^{2}  + v_{\eta}^{2} {\xi}_{24} \\ 
			m_{{\rho}^{{++},*}{{\sigma}_{2}}^{++}} &m_{{\chi}^{--}{{\sigma}_{2}}^{++}} &\frac{1}{2} {\lambda}_{44} v_{\sigma}^{2}  + v_{\eta}^{2} {\xi}_{24}  &m_{{{\sigma}_{2}}^{{++},*}{{\sigma}_{2}}^{++}}\end{array} 
		\right) +  \xi_{Y^{++}}m^2(Y^{++}) 
	\end{equation} 
	\begin{align} 
		m_{{\rho}^{{++},*}{\rho}^{++}} &= \frac{1}{4} \Big(2 \Big(\lambda_{12} v_{\eta}^{2}  + \Big(\lambda_{13} + \zeta_{13}\Big)v_{\chi}^{2}  + \sqrt{2} v_{\eta} v_{\sigma} {\xi}_{14} \Big) + 4 \lambda_{1} v_{\rho}^{2}  \nonumber\\ &\ \ \ + v_{\sigma}^{2} \Big(2 {\lambda}_{14}  + \zeta_{14}\Big)\Big) + {\mu}_{1}\\ 
		m_{{\rho}^{{++},*}{\chi}^{{--},*}} &= \frac{1}{2} \Big(-2 f v_{\eta}  + \sqrt{2} {f}_{\sigma} v_{\sigma}  + \zeta_{13} v_{\rho} v_{\chi} \Big)\\ 
		m_{{\chi}^{--}{\chi}^{{--},*}} &= \frac{1}{4} \Big(2 \Big(\Big(\lambda_{13} + \zeta_{13}\Big)v_{\rho}^{2}  + \lambda_{23} v_{\eta}^{2} \Big) -2 \sqrt{2} v_{\eta} v_{\sigma} {\xi}_{34}  + 4 \lambda_{3} v_{\chi}^{2} \nonumber\\ &\ \ \  + v_{\sigma}^{2} \Big(2 {\lambda}_{34}  + \zeta_{34}\Big)\Big) + {\mu}_{3}\\ 
		m_{{\rho}^{{++},*}{{\sigma}_{1}}^{{--},*}} &= \frac{1}{4} \Big(4 {f}_{\sigma} v_{\chi}  + v_{\rho} \Big(-2 v_{\eta} {\xi}_{14}  + \sqrt{2} v_{\sigma} \zeta_{14} \Big)\Big)\\ 
		m_{{\chi}^{--}{{\sigma}_{1}}^{{--},*}} &= \frac{1}{4} v_{\chi} \Big(-2 v_{\eta} {\xi}_{34}  + \sqrt{2} v_{\sigma} \zeta_{34} \Big)\\ 
		m_{{{\sigma}_{1}}^{--}{{\sigma}_{1}}^{{--},*}} &= \frac{1}{2} \Big(2 \Big({\lambda}_4 + {\lambda}_{44}\Big)v_{\sigma}^{2}  + {\lambda}_{14} v_{\rho}^{2}  + {\lambda}_{24} v_{\eta}^{2}  + v_{\chi}^{2} \Big({\lambda}_{34} + \zeta_{34}\Big)\Big) + {\mu}_{4}\\ 
		m_{{\rho}^{{++},*}{{\sigma}_{2}}^{++}} &= \frac{1}{4} v_{\rho} \Big(2 v_{\eta} {\xi}_{14}  + \sqrt{2} v_{\sigma} \zeta_{14} \Big)\\ 
		m_{{\chi}^{--}{{\sigma}_{2}}^{++}} &= \frac{1}{4} \Big(4 {f}_{\sigma} v_{\rho}  + v_{\chi} \Big(2 v_{\eta} {\xi}_{34}  + \sqrt{2} v_{\sigma} \zeta_{34} \Big)\Big)\\ 
		m_{{{\sigma}_{2}}^{{++},*}{{\sigma}_{2}}^{++}} &= \frac{1}{2} \Big(2 \Big({\lambda}_4 + {\lambda}_{44}\Big)v_{\sigma}^{2}  + {\lambda}_{24} v_{\eta}^{2}  + {\lambda}_{34} v_{\chi}^{2}  + v_{\rho}^{2} \Big({\lambda}_{14} + \zeta_{14}\Big)\Big) + {\mu}_{4}
	\end{align} 
	The gauge fixing contributions are
	\begin{equation} 
		m^2 (\xi_{Y^{++}}) = \left( 
		\begin{array}{cccc}
			\frac{1}{4} g_{2}^{2} v_{\rho}^{2}  &-\frac{1}{4} g_{2}^{2} v_{\rho} v_{\chi}  &\frac{1}{2} \frac{1}{\sqrt{2}} g_{2}^{2} v_{\rho} v_{\sigma}  &-\frac{1}{2} \frac{1}{\sqrt{2}} g_{2}^{2} v_{\rho} v_{\sigma} \\ 
			-\frac{1}{4} g_{2}^{2} v_{\rho} v_{\chi}  &\frac{1}{4} g_{2}^{2} v_{\chi}^{2}  &-\frac{1}{2} \frac{1}{\sqrt{2}} g_{2}^{2} v_{\chi} v_{\sigma}  &\frac{1}{2} \frac{1}{\sqrt{2}} g_{2}^{2} v_{\chi} v_{\sigma} \\ 
			\frac{1}{2} \frac{1}{\sqrt{2}} g_{2}^{2} v_{\rho} v_{\sigma}  &-\frac{1}{2} \frac{1}{\sqrt{2}} g_{2}^{2} v_{\chi} v_{\sigma}  &\frac{1}{2} g_{2}^{2} v_{\sigma}^{2}  &-\frac{1}{2} g_{2}^{2} v_{\sigma}^{2} \\ 
			-\frac{1}{2} \frac{1}{\sqrt{2}} g_{2}^{2} v_{\rho} v_{\sigma}  &\frac{1}{2} \frac{1}{\sqrt{2}} g_{2}^{2} v_{\chi} v_{\sigma}  &-\frac{1}{2} g_{2}^{2} v_{\sigma}^{2}  &\frac{1}{2} g_{2}^{2} v_{\sigma}^{2} \end{array} 
		\right) 
	\end{equation} 
	This matrix is diagonalized by \(R_{2C}\)
	\begin{equation} 
		R_{2C} m^2_{H^{--}} R_{2C}^{\dagger} = m^{dia}_{2,H^{--}} 
	\end{equation} 
	with 
	\begin{align} 
		{\rho}^{++} = \sum_{j}R_{2C,{j 1}}H^{++}_{{j}}\,, \hspace{1cm} 
		{\chi}^{--} = \sum_{j}R^*_{{2C},{j 2}}H^{--}_{{j}}\,, \hspace{1cm} 
		{{\sigma}_{1}}^{--} = \sum_{j}R^*_{{2C},{j 3}}H^{--}_{{j}}\\ 
		{{\sigma}_{2}}^{++} = \sum_{j}R_{2C,{j 4}}H^{++}_{{j}}
	\end{align} 
\end{itemize}

\section{Renormalization group equations}
In this Appendix we present the Beta fuction of all the parameter of the minimal 331 Model in presence of the scalar sextet \cite{staub_automatic_2011,staub_sarah_2014,thomsen_rgbeta_2021}.
Here we recall the form of the scalar potential
\begin{align}
	V=&m_1\r^\dagger\r+m_2\eta^\dagger\eta+m_3\chi^\dagger\chi+\l_1(\rho^\dagger\r)^2+\l_2(\eta^\dagger\eta)^2+\l_3(\chi^\dagger\chi)^2\nonumber\\
	&+\l_{12}\r^\dagger\r\eta^\dagger\eta+\l_{13}\r^\dagger\r\chi^\dagger\chi+\l_{23}\chi^\dagger\chi\eta^\dagger\eta+\z_{12}\r^\dagger\eta\eta^\dagger\r+\z_{13}\r^\dagger\chi\chi^\dagger\r+\z_{23}\eta^\dagger\chi\chi^\dagger\eta\nonumber\\
	&+m_4\Tr(\s^\dagger\s)+\l_4(\Tr(\s^\dagger\s))^2+\l_{14}\r^\dagger\r\Tr(\s^\dagger\s)+\l_{24}\eta^\dagger\eta\Tr(\s^\dagger\s)+\l_{34}\chi^\dagger\chi\Tr(\s^\dagger\s)\nonumber\\
	&+\l_{44}\Tr(\s^\dagger\s\s^\dagger\s)+\z_{14}\r^\dagger\s\s^\dagger\rho+\z_{24}\eta^\dagger\s\s^\dagger\eta+\z_{34}\chi^\dagger\s\s^\dagger\chi\nonumber\\
	&+(\sqrt{2}f_{\r\eta\chi}\epsilon^{ijk}\r_i\eta_j\chi_k+\sqrt{2}f_{\r\s\chi}\r^T\s^\dagger\chi\nonumber\\
	&+\xi_{14}\epsilon^{ijk}\r^{*l}\s_{li}\r_j\eta_k+\xi_{24}\epsilon^{ijk}\epsilon^{lmn}\eta_i\eta_l\s_{jm}\s_{kn}+\xi_{34}\epsilon^{ijk}\chi^{*l}\s_{li}\chi_j\eta_k )+\textnormal{h.c.}\ ,
\end{align}
and the Yukawa interactions 
\begin{align}
	\mathcal{L}^{Y}_{q,\ triplet}=\ &- \overline{Q}_m\brak{Y^d_{m\a} \eta^*d_{\a R}+Y_{m\a}^u\chi^*u_{m\a}}
	-\overline{Q}_3( Y^d_{3\a}\chi d_{\a R}+Y^u_{3\a}\eta u_{m\a})+\nonumber\\
	&-\overline{Q}_{m}(Y^J_{mn}\chi J_{nR})-\overline{Q}_{3}Y^J_{3}\chi J_{3R} +\textnormal{h.c.}\ .
\end{align}

\begin{equation}
	\mathcal{L}^Y_{l,\ triplet}=G^\eta_{ab}l^i_a\cdot l^j_b\eta^{*k}\epsilon^{ijk}+h.c.
\end{equation}
considering also the sextet contribution to the leptonic Yukawa
\begin{equation}
	\mathcal{L}^Y_{l,\ sextet}=G^\s_{ab}l^i_a\cdot l^j_b\s^{*}_{ij}+\textnormal{h.c.}
\end{equation}

\subsection{Gauge Couplings}
{\allowdisplaybreaks  \begin{align} 
		\beta_{g_1}^{(1)} & =  
		22 g_{1}^{3} \\  
		\beta_{g_2}^{(1)} & =  
		-\frac{17}{3} g_{2}^{3} \\ 
		\beta_{g_3}^{(1)} & =  
		-5 g_{3}^{3} \\ 
\end{align}} 
\subsection{Quartic scalar couplings}
{\allowdisplaybreaks  \begin{align} 
		\beta_{{\xi}_{34}}^{(1)} & =  
		+2 \zeta_{13} {\xi}_{14} -6 g_{1}^{2} {\xi}_{34} -22 g_{2}^{2} {\xi}_{34} +4 \lambda_{23} {\xi}_{34} -2 \zeta_{23} {\xi}_{34} +2 {\lambda}_{24} {\xi}_{34} +4 \lambda_{3} {\xi}_{34} +4 {\lambda}_{34} {\xi}_{34} -12 {\xi}_{24} {\xi}_{34}\nonumber \\ 
		& - {\xi}_{34} \zeta_{24} +3 {\xi}_{34} \zeta_{34} +3 {\xi}_{34} \Big({Y^d_1  Y^{d *}_1}\Big) +3 {\xi}_{34} \Big({Y^d_2  Y^{d *}_2}\Big) +6 {\xi}_{34} \Big({Y^d_3  Y^{d *}_3}\Big) +6 {\xi}_{34} \Big({Y^u_1  Y^{u *}_1}\Big) \nonumber \\ 
		&+6 {\xi}_{34} \Big({Y^u_2  Y^{u *}_2}\Big) +3 {\xi}_{34} \Big({Y^u_3  Y^{u *}_3}\Big) +2 {\xi}_{34} \mbox{Tr}\Big({G^\sigma  G^{\sigma *}}\Big) \\ 
		\beta_{{\xi}_{14}}^{(1)} & =  
		-6 g_{1}^{2} {\xi}_{14} -22 g_{2}^{2} {\xi}_{14} +4 \lambda_{1} {\xi}_{14} +4 \lambda_{12} {\xi}_{14} -2 \zeta_{12} {\xi}_{14} +4 {\lambda}_{14} {\xi}_{14} +2 {\lambda}_{24} {\xi}_{14} \nonumber \\ 
		&-12 {\xi}_{14} {\xi}_{24} +2 \zeta_{13} {\xi}_{34} +3 {\xi}_{14} \zeta_{14} - {\xi}_{14} \zeta_{24} +6 {\xi}_{14} |Y^J_1|^2 +3 {\xi}_{14} \Big({Y^d_1  Y^{d *}_1}\Big) +3 {\xi}_{14} \Big({Y^d_2  Y^{d *}_2}\Big)\nonumber \\ 
		& +6 {\xi}_{14} \Big({Y^J_1  Y^{J *}_1}\Big) +6 {\xi}_{14} \Big({Y^J_2  Y^{J *}_2}\Big) +3 {\xi}_{14} \Big({Y^u_3  Y^{u *}_3}\Big) +2 {\xi}_{14} \mbox{Tr}\Big({G^\sigma  G^{\sigma *}}\Big) \\ 
		\beta_{{\xi}_{24}}^{(1)} & =  
		- {\xi}_{14}^{2} -28 g_{2}^{2} {\xi}_{24} +4 \lambda_{2} {\xi}_{24} +8 {\lambda}_{24} {\xi}_{24} +4 {\lambda}_4 {\xi}_{24} -2 {\lambda}_{44} {\xi}_{24} - {\xi}_{34}^{2} -4 {\xi}_{24} \zeta_{24} +6 {\xi}_{24} \Big({Y^d_1  Y^{d *}_1}\Big) \nonumber \\ 
		&+6 {\xi}_{24} \Big({Y^d_2  Y^{d *}_2}\Big) +6 {\xi}_{24} \Big({Y^u_3  Y^{u *}_3}\Big) +4 {\xi}_{24} \mbox{Tr}\Big({G^\sigma  G^{\sigma *}}\Big) \\ 
		\beta_{\lambda_{3}}^{(1)} & =  
		+6 g_{1}^{4} +4 g_{1}^{2} g_{2}^{2} +\frac{13}{6} g_{2}^{4} +3 \lambda_{13}^{2} +2 \lambda_{13} \zeta_{13} +\zeta_{13}^{2}+3 \lambda_{23}^{2} +2 \lambda_{23} \zeta_{23} +\zeta_{23}^{2}-12 g_{1}^{2} \lambda_{3}\nonumber \\ 
		& -16 g_{2}^{2} \lambda_{3} +28 \lambda_{3}^{2} +6 {\lambda}_{34}^{2} +2 {\xi}_{34}^{2} +4 {\lambda}_{34} \zeta_{34} +\frac{3}{2} \zeta_{34}^{2} +12 \lambda_{3} \Big({Y^d_3  Y^{d *}_3}\Big) -6 \Big(\Big({Y^d_3  Y^{d *}_3}\Big)\Big)^{2} \nonumber \\ 
		&+12 \lambda_{3} \Big({Y^u_1  Y^{u *}_1}\Big) -6 \Big(\Big({Y^u_1  Y^{u *}_1}\Big)\Big)^{2} -12 \Big({Y^u_1  Y^{u *}_2}\Big) \Big({Y^u_2  Y^{u *}_1}\Big) +12 \lambda_{3} \Big({Y^u_2  Y^{u *}_2}\Big) \nonumber \\ 
		&-6 \Big(\Big({Y^u_2  Y^{u *}_2}\Big)\Big)^{2} \\  
		\beta_{{\lambda}_{34}}^{(1)} & =  
		+\frac{13}{3} g_{2}^{4} +6 \lambda_{13} {\lambda}_{14} +2 \zeta_{13} {\lambda}_{14} +6 \lambda_{23} {\lambda}_{24} +2 \zeta_{23} {\lambda}_{24} -6 g_{1}^{2} {\lambda}_{34} -28 g_{2}^{2} {\lambda}_{34} +16 \lambda_{3} {\lambda}_{34}\nonumber \\ 
		& +4 {\lambda}_{34}^{2} +28 {\lambda}_{34} {\lambda}_4 +16 {\lambda}_{34} {\lambda}_{44} +2 {\xi}_{34}^{2} +2 \lambda_{13} \zeta_{14} +2 \lambda_{23} \zeta_{24} +4 \lambda_{3} \zeta_{34} +8 {\lambda}_4 \zeta_{34} +2 {\lambda}_{44} \zeta_{34} \nonumber \\ 
		&+\zeta_{34}^{2}+6 {\lambda}_{34} \Big({Y^d_3  Y^{d *}_3}\Big) +6 {\lambda}_{34} \Big({Y^u_1  Y^{u *}_1}\Big) +6 {\lambda}_{34} \Big({Y^u_2  Y^{u *}_2}\Big) +4 {\lambda}_{34} \mbox{Tr}\Big({G^\sigma  G^{\sigma *}}\Big) \\ 
		\beta_{\zeta_{34}}^{(1)} & =  
		+7 g_{2}^{4} +2 {\xi}_{34}^{2} +2 \zeta_{13} \zeta_{14} +2 \zeta_{23} \zeta_{24} -6 g_{1}^{2} \zeta_{34} -28 g_{2}^{2} \zeta_{34} +4 \lambda_{3} \zeta_{34} +8 {\lambda}_{34} \zeta_{34} \nonumber \\ 
		& +4 {\lambda}_4 \zeta_{34} +10 {\lambda}_{44} \zeta_{34} +5 \zeta_{34}^{2}+6 \zeta_{34} \Big({Y^d_3  Y^{d *}_3}\Big) +6 \zeta_{34} \Big({Y^u_1  Y^{u *}_1}\Big) +6 \zeta_{34} \Big({Y^u_2  Y^{u *}_2}\Big)  \nonumber \\ 
		&+4 \zeta_{34} \mbox{Tr}\Big({G^\sigma  G^{\sigma *}}\Big) \\ 
		\beta_{\zeta_{23}}^{(1)} & =  
		+\frac{5}{2} g_{2}^{4} +2 \zeta_{12} \zeta_{13} -6 g_{1}^{2} \zeta_{23} -16 g_{2}^{2} \zeta_{23} +4 \lambda_{2} \zeta_{23} +8 \lambda_{23} \zeta_{23} +6 \zeta_{23}^{2} +4 \zeta_{23} \lambda_{3} -5 {\xi}_{34}^{2} \nonumber \\ 
		&+\frac{5}{2} \zeta_{24} \zeta_{34} +6 \zeta_{23} \Big({Y^d_3  Y^{d *}_3}\Big) +6 \Big({Y^d_1  Y^{d *}_1}\Big) \Big(-2 \Big({Y^u_1  Y^{u *}_1}\Big)  + \zeta_{23}\Big)+6 \zeta_{23} \Big({Y^u_1  Y^{u *}_1}\Big)\nonumber \\ 
		& -12 \Big({Y^d_2  Y^{d *}_1}\Big) \Big({Y^u_1  Y^{u *}_2}\Big) -12 \Big({Y^d_3  Y^{d *}_1}\Big) \Big({Y^u_1  Y^{u *}_3}\Big) -12 \Big({Y^d_1  Y^{d *}_2}\Big) \Big({Y^u_2  Y^{u *}_1}\Big) \nonumber \\ 
		&+6 \Big({Y^d_2  Y^{d *}_2}\Big) \Big(-2 \Big({Y^u_2  Y^{u *}_2}\Big)  + \zeta_{23}\Big)+6 \zeta_{23} \Big({Y^u_2  Y^{u *}_2}\Big) -12 \Big({Y^d_3  Y^{d *}_2}\Big) \Big({Y^u_2  Y^{u *}_3}\Big) \nonumber \\ 
		&-12 \Big({Y^d_1  Y^{d *}_3}\Big) \Big({Y^u_3  Y^{u *}_1}\Big) -12 \Big({Y^d_2  Y^{d *}_3}\Big) \Big({Y^u_3  Y^{u *}_2}\Big) +6 \zeta_{23} \Big({Y^u_3  Y^{u *}_3}\Big) \nonumber \\ 
		&-12 \Big({Y^d_3  Y^{d *}_3}\Big) \Big({Y^u_3  Y^{u *}_3}\Big) \\ 
		\beta_{\lambda_{23}}^{(1)} & =  
		+\frac{11}{6} g_{2}^{4} +6 \lambda_{12} \lambda_{13} +2 \zeta_{12} \lambda_{13} +2 \lambda_{12} \zeta_{13} -6 g_{1}^{2} \lambda_{23} -16 g_{2}^{2} \lambda_{23} +16 \lambda_{2} \lambda_{23} +4 \lambda_{23}^{2}  \nonumber \\ 
		&+4 \lambda_{2} \zeta_{23} +2 \zeta_{23}^{2} +16 \lambda_{23} \lambda_{3}+4 \zeta_{23} \lambda_{3} +12 {\lambda}_{24} {\lambda}_{34} +7 {\xi}_{34}^{2} +4 {\lambda}_{34} \zeta_{24} +4 {\lambda}_{24} \zeta_{34} \nonumber \\ 
		&+\frac{1}{2} \zeta_{24} \zeta_{34} +6 \lambda_{23} \Big({Y^d_1  Y^{d *}_1}\Big) +6 \lambda_{23} \Big({Y^d_2  Y^{d *}_2}\Big) -12 \Big({Y^d_1  Y^{d *}_3}\Big) \Big({Y^d_3  Y^{d *}_1}\Big) \nonumber \\ 
		&+6 \lambda_{23} \Big({Y^d_3  Y^{d *}_3}\Big) +6 \lambda_{23} \Big({Y^u_1  Y^{u *}_1}\Big) +6 \lambda_{23} \Big({Y^u_2  Y^{u *}_2}\Big) -12 \Big({Y^u_1  Y^{u *}_3}\Big) \Big({Y^u_3  Y^{u *}_1}\Big)\nonumber \\ 
		& -12 \Big({Y^u_2  Y^{u *}_3}\Big) \Big({Y^u_3  Y^{u *}_2}\Big) +6 \lambda_{23} \Big({Y^u_3  Y^{u *}_3}\Big) -12 \Big({Y^d_2  Y^{d *}_3}\Big) \Big({Y^d_3  Y^{d *}_2}\Big) \\ 
		\beta_{\lambda_{2}}^{(1)} & =  
		+\frac{13}{6} g_{2}^{4} +3 \lambda_{12}^{2} +2 \lambda_{12} \zeta_{12} +\zeta_{12}^{2}-16 g_{2}^{2} \lambda_{2} +28 \lambda_{2}^{2} +3 \lambda_{23}^{2} +2 \lambda_{23} \zeta_{23} \nonumber \\ 
		&+\zeta_{23}^{2}+6 {\lambda}_{24}^{2} +12 {\xi}_{24}^{2}  +4 {\lambda}_{24} \zeta_{24} +\frac{3}{2} \zeta_{24}^{2}+12 \lambda_{2} \Big({Y^d_1  Y^{d *}_1}\Big) -6 \Big(\Big({Y^d_1  Y^{d *}_1}\Big)\Big)^{2} \nonumber \\ 
		&-12 \Big({Y^d_1  Y^{d *}_2}\Big) \Big({Y^d_2  Y^{d *}_1}\Big) +12 \lambda_{2} \Big({Y^d_2  Y^{d *}_2}\Big) -6 \Big(\Big({Y^d_2  Y^{d *}_2}\Big)\Big)^{2} +12 \lambda_{2} \Big({Y^u_3  Y^{u *}_3}\Big)\nonumber \\ 
		& -6 \Big(\Big({Y^u_3  Y^{u *}_3}\Big)\Big)^{2} \\ 
		\beta_{{\lambda}_{24}}^{(1)} & =  
		+\frac{13}{3} g_{2}^{4} +6 \lambda_{12} {\lambda}_{14} +2 \zeta_{12} {\lambda}_{14} -28 g_{2}^{2} {\lambda}_{24} +16 \lambda_{2} {\lambda}_{24} +4 {\lambda}_{24}^{2} +6 \lambda_{23} {\lambda}_{34} +2 \zeta_{23} {\lambda}_{34} \nonumber \\ 
		&+28 {\lambda}_{24} {\lambda}_4 +16 {\lambda}_{24} {\lambda}_{44} +2 {\xi}_{14}^{2} +48 {\xi}_{24}^{2} +2 {\xi}_{34}^{2} +2 \lambda_{12} \zeta_{14} +4 \lambda_{2} \zeta_{24} +8 {\lambda}_4 \zeta_{24} +2 {\lambda}_{44} \zeta_{24}\nonumber \\ 
		& +\zeta_{24}^{2}+2 \lambda_{23} \zeta_{34}  +6 {\lambda}_{24} \Big({Y^d_1  Y^{d *}_1}\Big) +6 {\lambda}_{24} \Big({Y^d_2  Y^{d *}_2}\Big)+6 {\lambda}_{24} \Big({Y^u_3  Y^{u *}_3}\Big) +4 {\lambda}_{24} \mbox{Tr}\Big({G^\sigma  G^{\sigma *}}\Big) \\ 
		\beta_{\zeta_{24}}^{(1)} & =  
		+7 g_{2}^{4} -2 {\xi}_{14}^{2} -48 {\xi}_{24}^{2} -2 {\xi}_{34}^{2} +2 \zeta_{12} \zeta_{14} -28 g_{2}^{2} \zeta_{24} +4 \lambda_{2} \zeta_{24} +8 {\lambda}_{24} \zeta_{24} +4 {\lambda}_4 \zeta_{24} \nonumber \\ 
		& +10 {\lambda}_{44} \zeta_{24} +5 \zeta_{24}^{2} +2 \zeta_{23} \zeta_{34} +6 \zeta_{24} \Big({Y^d_1  Y^{d *}_1}\Big) +6 \zeta_{24} \Big({Y^d_2  Y^{d *}_2}\Big) +6 \zeta_{24} \Big({Y^u_3  Y^{u *}_3}\Big) \nonumber \\ 
		& +4 \zeta_{24} \mbox{Tr}\Big({G^\sigma  G^{\sigma *}}\Big) \\ 
		\beta_{\zeta_{13}}^{(1)} & =  
		-12 g_{1}^{2} g_{2}^{2} +\frac{5}{2} g_{2}^{4} -12 g_{1}^{2} \zeta_{13} -16 g_{2}^{2} \zeta_{13} +4 \lambda_{1} \zeta_{13} +8 \lambda_{13} \zeta_{13} +6 \zeta_{13}^{2} +2 \zeta_{12} \zeta_{23} +4 \zeta_{13} \lambda_{3}\nonumber \\ 
		& +6 {\xi}_{14} {\xi}_{34} +\frac{5}{2} \zeta_{14} \zeta_{34} +6 |Y^J_1|^2 \Big(-2 \Big({Y^d_3  Y^{d *}_3}\Big)  + \zeta_{13}\Big)+6 \zeta_{13} \Big({Y^d_3  Y^{d *}_3}\Big) +6 \zeta_{13} \Big({Y^J_1  Y^{J *}_1}\Big)\nonumber \\ 
		& +6 \zeta_{13} \Big({Y^J_2  Y^{J *}_2}\Big)  +6 \zeta_{13} \Big({Y^u_1  Y^{u *}_1}\Big)-12 \Big({Y^J_1  Y^{J *}_1}\Big) \Big({Y^u_1  Y^{u *}_1}\Big) -12 \Big({Y^J_2  Y^{J *}_1}\Big) \Big({Y^u_1  Y^{u *}_2}\Big)  \nonumber \\ 
		&-12 \Big({Y^J_1  Y^{J *}_2}\Big) \Big({Y^u_2  Y^{u *}_1}\Big)+6 \zeta_{13} \Big({Y^u_2  Y^{u *}_2}\Big) -12 \Big({Y^J_2  Y^{J *}_2}\Big) \Big({Y^u_2  Y^{u *}_2}\Big) \\ 
		\beta_{\zeta_{12}}^{(1)} & =  
		+\frac{5}{2} g_{2}^{4} -6 g_{1}^{2} \zeta_{12} -16 g_{2}^{2} \zeta_{12} +4 \lambda_{1} \zeta_{12} +8 \lambda_{12} \zeta_{12} +6 \zeta_{12}^{2} +4 \zeta_{12} \lambda_{2} +2 \zeta_{13} \zeta_{23} -5 {\xi}_{14}^{2} \nonumber \\ 
		&+\frac{5}{2} \zeta_{14} \zeta_{24} +6 \zeta_{12} \Big({Y^d_2  Y^{d *}_2}\Big) +6 \Big({Y^d_1  Y^{d *}_1}\Big) \Big(-2 \Big({Y^J_1  Y^{J *}_1}\Big)  + \zeta_{12}\Big)+6 \zeta_{12} \Big({Y^J_1  Y^{J *}_1}\Big) \nonumber \\ 
		&-12 \Big({Y^d_2  Y^{d *}_1}\Big) \Big({Y^J_1  Y^{J *}_2}\Big) -12 \Big({Y^d_1  Y^{d *}_2}\Big) \Big({Y^J_2  Y^{J *}_1}\Big) +6 \zeta_{12} \Big({Y^J_2  Y^{J *}_2}\Big)\nonumber \\ 
		&  -12 \Big({Y^d_2  Y^{d *}_2}\Big) \Big({Y^J_2  Y^{J *}_2}\Big) +6 |Y^J_1|^2 \Big(-2 \Big({Y^u_3  Y^{u *}_3}\Big) + \zeta_{12}\Big)+6 \zeta_{12} \Big({Y^u_3  Y^{u *}_3}\Big) \\ 
		\beta_{\lambda_{13}}^{(1)} & =  
		+12 g_{1}^{4} +4 g_{1}^{2} g_{2}^{2} +\frac{11}{6} g_{2}^{4} -12 g_{1}^{2} \lambda_{13} -16 g_{2}^{2} \lambda_{13} +16 \lambda_{1} \lambda_{13} +4 \lambda_{13}^{2} +4 \lambda_{1} \zeta_{13} +2 \zeta_{13}^{2}\nonumber \\ 
		& +6 \lambda_{12} \lambda_{23}  +2 \zeta_{12} \lambda_{23} +2 \lambda_{12} \zeta_{23} +16 \lambda_{13} \lambda_{3} +4 \zeta_{13} \lambda_{3} +12 {\lambda}_{14} {\lambda}_{34} -2 {\xi}_{14} {\xi}_{34} \nonumber \\ 
		&+4 {\lambda}_{34} \zeta_{14} +4 {\lambda}_{14} \zeta_{34} +\frac{1}{2} \zeta_{14} \zeta_{34} +6 \lambda_{13} |Y^J_1|^2 +6 \lambda_{13} \Big({Y^d_3  Y^{d *}_3}\Big) +6 \lambda_{13} \Big({Y^J_1  Y^{J *}_1}\Big)\nonumber \\ 
		& +6 \lambda_{13} \Big({Y^J_2  Y^{J *}_2}\Big) +6 \lambda_{13} \Big({Y^u_1  Y^{u *}_1}\Big) +6 \lambda_{13} \Big({Y^u_2  Y^{u *}_2}\Big) \\ 
		\beta_{\lambda_{12}}^{(1)} & =  
		+\frac{11}{6} g_{2}^{4} -6 g_{1}^{2} \lambda_{12} -16 g_{2}^{2} \lambda_{12} +16 \lambda_{1} \lambda_{12} +4 \lambda_{12}^{2} +4 \lambda_{1} \zeta_{12} +2 \zeta_{12}^{2} +16 \lambda_{12} \lambda_{2} +4 \zeta_{12} \lambda_{2} \nonumber \\ 
		&+6 \lambda_{13} \lambda_{23} +2 \zeta_{13} \lambda_{23} +2 \lambda_{13} \zeta_{23} +12 {\lambda}_{14} {\lambda}_{24} +7 {\xi}_{14}^{2} +4 {\lambda}_{24} \zeta_{14} +4 {\lambda}_{14} \zeta_{24} +\frac{1}{2} \zeta_{14} \zeta_{24} \nonumber \\ 
		&+6 \lambda_{12} |Y^J_1|^2 +6 \lambda_{12} \Big({Y^d_1  Y^{d *}_1}\Big) +6 \lambda_{12} \Big({Y^d_2  Y^{d *}_2}\Big) +6 \lambda_{12} \Big({Y^J_1  Y^{J *}_1}\Big) +6 \lambda_{12} \Big({Y^J_2  Y^{J *}_2}\Big)\nonumber \\ 
		& +6 \lambda_{12} \Big({Y^u_3  Y^{u *}_3}\Big) \\ 
		\beta_{\lambda_{1}}^{(1)} & =  
		+6 g_{1}^{4} +4 g_{1}^{2} g_{2}^{2} +\frac{13}{6} g_{2}^{4} -12 g_{1}^{2} \lambda_{1} -16 g_{2}^{2} \lambda_{1} +28 \lambda_{1}^{2} +3 \lambda_{12}^{2} +2 \lambda_{12} \zeta_{12} +\zeta_{12}^{2}+3 \lambda_{13}^{2} \nonumber \\ 
		&+2 \lambda_{13} \zeta_{13} +\zeta_{13}^{2}+6 {\lambda}_{14}^{2} +2 {\xi}_{14}^{2} +4 {\lambda}_{14} \zeta_{14} +\frac{3}{2} \zeta_{14}^{2} +12 \lambda_{1} |Y^J_1|^2 -6 |Y^J_1|^4 \nonumber \\ 
		&+12 \lambda_{1} \Big({Y^J_1  Y^{J *}_1}\Big) -6 \Big(\Big({Y^J_1  Y^{J *}_1}\Big)\Big)^{2} -12 \Big({Y^J_1  Y^{J *}_2}\Big) \Big({Y^J_2  Y^{J *}_1}\Big) +12 \lambda_{1} \Big({Y^J_2  Y^{J *}_2}\Big)\nonumber \\ 
		& -6 \Big(\Big({Y^J_2  Y^{J *}_2}\Big)\Big)^{2} \\ 
		\beta_{{\lambda}_{14}}^{(1)} & =  
		+\frac{13}{3} g_{2}^{4} -6 g_{1}^{2} {\lambda}_{14} -28 g_{2}^{2} {\lambda}_{14} +16 \lambda_{1} {\lambda}_{14} +4 {\lambda}_{14}^{2} +6 \lambda_{12} {\lambda}_{24} +2 \zeta_{12} {\lambda}_{24}\nonumber \\ 
		&  +6 \lambda_{13} {\lambda}_{34} +2 \zeta_{13} {\lambda}_{34}+28 {\lambda}_{14} {\lambda}_4 +16 {\lambda}_{14} {\lambda}_{44} +2 {\xi}_{14}^{2} +4 \lambda_{1} \zeta_{14} +8 {\lambda}_4 \zeta_{14} +2 {\lambda}_{44} \zeta_{14}\nonumber \\ 
		& +\zeta_{14}^{2}+2 \lambda_{12} \zeta_{24} +2 \lambda_{13} \zeta_{34} +6 {\lambda}_{14} |Y^J_1|^2 +6 {\lambda}_{14} \Big({Y^J_1  Y^{J *}_1}\Big) +6 {\lambda}_{14} \Big({Y^J_2  Y^{J *}_2}\Big)\nonumber \\ 
		& +4 {\lambda}_{14} \mbox{Tr}\Big({G^\sigma  G^{\sigma *}}\Big) \\ 
		\beta_{\zeta_{14}}^{(1)} & =  
		+7 g_{2}^{4} +2 {\xi}_{14}^{2} -6 g_{1}^{2} \zeta_{14} -28 g_{2}^{2} \zeta_{14} +4 \lambda_{1} \zeta_{14} +8 {\lambda}_{14} \zeta_{14} +4 {\lambda}_4 \zeta_{14} +10 {\lambda}_{44} \zeta_{14} +5 \zeta_{14}^{2}\nonumber \\ 
		& +2 \zeta_{12} \zeta_{24} +2 \zeta_{13} \zeta_{34} +6 \zeta_{14} |Y^J_1|^2 +6 \zeta_{14} \Big({Y^J_1  Y^{J *}_1}\Big) +6 \zeta_{14} \Big({Y^J_2  Y^{J *}_2}\Big) \nonumber \\ 
		&+4 \zeta_{14} \mbox{Tr}\Big({G^\sigma  G^{\sigma *}}\Big) \\ 
		\beta_{{\lambda}_4}^{(1)} & =  
		+\frac{35}{3} g_{2}^{4} +3 {\lambda}_{14}^{2} +3 {\lambda}_{24}^{2} +3 {\lambda}_{34}^{2} -40 g_{2}^{2} {\lambda}_4 +40 {\lambda}_{4}^{2} +32 {\lambda}_4 {\lambda}_{44} +6 {\lambda}_{44}^{2} +8 {\xi}_{24}^{2} +2 {\lambda}_{14} \zeta_{14}\nonumber \\ 
		& +2 {\lambda}_{24} \zeta_{24} +2 {\lambda}_{34} \zeta_{34} +8 {\lambda}_4 \mbox{Tr}\Big({G^\sigma  G^{\sigma *}}\Big) \\ 
		\beta_{{\lambda}_{44}}^{(1)} & =  
		-16 \mbox{Tr}\Big({G^\sigma  G^{\sigma *}  G^\sigma  G^{\sigma *}}\Big)  + 22 {\lambda}_{44}^{2}  + 24 {\lambda}_4 {\lambda}_{44}  -40 g_{2}^{2} {\lambda}_{44}  + 5 g_{2}^{4}  + 8 {\lambda}_{44} \mbox{Tr}\Big({G^\sigma  G^{\sigma *}}\Big) \nonumber \\ 
		&   -8 {\xi}_{24}^{2}+ \zeta_{14}^{2} + \zeta_{24}^{2} + \zeta_{34}^{2}
\end{align}}

\subsection{Yukawa Couplings}
{\allowdisplaybreaks  \begin{align} 
		\beta_{G^\eta}^{(1)} & =  
		-8 {G^\sigma  G^{\sigma *}  G^\sigma}  + G^\sigma \Big(3 \Big(\Big({Y^d_1  Y^{d *}_1}\Big) \Big({Y^d_2  Y^{d *}_2}\Big)  + \Big({Y^u_3  Y^{u *}_3}\Big)\Big) -4 \mbox{Tr}\Big({G^\eta  G^{\eta *}}\Big)  -8 g_{2}^{2} \Big)\\ 
		\beta_{G^\sigma}^{(1)} & =  
		8 {G^\sigma  G^{\sigma *}  G^\sigma}  + G^\sigma \Big(2 \mbox{Tr}\Big({G^\sigma  G^{\sigma *}}\Big)  -8 g_{2}^{2} \Big)\\ 
		\beta_{Y_{1,{{i_1}}}^{d}}^{(1)} & =  
		\frac{1}{6} \Big(\Big(18 \Big({Y^d_2  Y^{d *}_2}\Big)  + 18 \Big({Y^u_3  Y^{u *}_3}\Big)  -24 g_{2}^{2}  + 30 \Big({Y^d_1  Y^{d *}_1}\Big)  + 3 \Big({Y^J_1  Y^{J *}_1}\Big)  + 3 \Big({Y^u_1  Y^{u *}_1}\Big) \nonumber \\ 
		& -48 g_{3}^{2}  -4 g_{1}^{2} \Big)Y_{1,{{i_1}}}^{d} +3 \Big(3 \Big({Y^d_1  Y^{d *}_3}\Big) Y_{3,{{i_1}}}^{d}  + 4 \Big({Y^d_1  Y^{d *}_2}\Big) Y_{2,{{i_1}}}^{d}  + 4 \Big({Y^u_1  Y^{u *}_3}\Big) Y_{3,{{i_1}}}^{d} \nonumber \\ 
		&  + \Big({Y^J_1  Y^{J *}_2}\Big) Y_{2,{{i_1}}}^{d} + \Big({Y^u_1  Y^{u *}_2}\Big) Y_{2,{{i_1}}}^{d} \Big)\Big)\\ 
		\beta_{Y_{1,{{i_1}}}^{J}}^{(1)} & =  
		\frac{1}{6} \Big(\Big(18 |Y^J_1|^2  + 18 \Big({Y^J_2  Y^{J *}_2}\Big)  -24 g_{2}^{2}  + 30 \Big({Y^J_1  Y^{J *}_1}\Big)  -34 g_{1}^{2}  + 3 \Big({Y^d_1  Y^{d *}_1}\Big)  \nonumber \\ 
		& + 3 \Big({Y^u_1  Y^{u *}_1}\Big) -48 g_{3}^{2} \Big)Y_{1,{{i_1}}}^{J} +3 \Big(4 \Big({Y^J_1  Y^{J *}_2}\Big)  + \Big({Y^d_1  Y^{d *}_2}\Big) + \Big({Y^u_1  Y^{u *}_2}\Big)\Big)Y_{2,{{i_1}}}^{J} \Big)\\ 
		\beta_{Y_{1,{{i_1}}}^{u}}^{(1)} & =  
		\frac{1}{6} \Big(\Big(-10 g_{1}^{2}  + 18 \Big({Y^d_3  Y^{d *}_3}\Big)  + 18 \Big({Y^u_2  Y^{u *}_2}\Big)  -24 g_{2}^{2}  + 30 \Big({Y^u_1  Y^{u *}_1}\Big)  + 3 \Big({Y^d_1  Y^{d *}_1}\Big)  \nonumber \\ 
		&+ 3 \Big({Y^J_1  Y^{J *}_1}\Big)  -48 g_{3}^{2} \Big)Y_{1,{{i_1}}}^{u} +3 \Big(3 \Big({Y^u_1  Y^{u *}_3}\Big) Y_{3,{{i_1}}}^{u}  + 4 \Big({Y^d_1  Y^{d *}_3}\Big) Y_{3,{{i_1}}}^{u}  + 4 \Big({Y^u_1  Y^{u *}_2}\Big) Y_{2,{{i_1}}}^{u}  \nonumber \\ 
		&+ \Big({Y^d_1  Y^{d *}_2}\Big) Y_{2,{{i_1}}}^{u}  + \Big({Y^J_1  Y^{J *}_2}\Big) Y_{2,{{i_1}}}^{u} \Big)\Big)\\ 
		\beta_{Y_{2,{{i_1}}}^{d}}^{(1)} & =  
		\frac{1}{6} \Big(12 \Big({Y^d_2  Y^{d *}_1}\Big) Y_{1,{{i_1}}}^{d} +3 \Big({Y^J_2  Y^{J *}_1}\Big) Y_{1,{{i_1}}}^{d} +3 \Big({Y^u_2  Y^{u *}_1}\Big) Y_{1,{{i_1}}}^{d} -4 g_{1}^{2} Y_{2,{{i_1}}}^{d} -24 g_{2}^{2} Y_{2,{{i_1}}}^{d} \nonumber \\ 
		&-48 g_{3}^{2} Y_{2,{{i_1}}}^{d} +18 \Big({Y^d_1  Y^{d *}_1}\Big) Y_{2,{{i_1}}}^{d} +30 \Big({Y^d_2  Y^{d *}_2}\Big) Y_{2,{{i_1}}}^{d} +3 \Big({Y^J_2  Y^{J *}_2}\Big) Y_{2,{{i_1}}}^{d} \nonumber \\ 
		&+3 \Big({Y^u_2  Y^{u *}_2}\Big) Y_{2,{{i_1}}}^{d} +18 \Big({Y^u_3  Y^{u *}_3}\Big) Y_{2,{{i_1}}}^{d} +9 \Big({Y^d_2  Y^{d *}_3}\Big) Y_{3,{{i_1}}}^{d} +12 \Big({Y^u_2  Y^{u *}_3}\Big) Y_{3,{{i_1}}}^{d} \Big)\\ 
		\beta_{Y_{2,{{i_1}}}^{J}}^{(1)} & =  
		\frac{1}{6} \Big(3 \Big({Y^d_2  Y^{d *}_1}\Big) Y_{1,{{i_1}}}^{J} +12 \Big({Y^J_2  Y^{J *}_1}\Big) Y_{1,{{i_1}}}^{J} +3 \Big({Y^u_2  Y^{u *}_1}\Big) Y_{1,{{i_1}}}^{J} -34 g_{1}^{2} Y_{2,{{i_1}}}^{J} -24 g_{2}^{2} Y_{2,{{i_1}}}^{J} \nonumber \\ 
		&-48 g_{3}^{2} Y_{2,{{i_1}}}^{J} +18 |Y^J_1|^2 Y_{2,{{i_1}}}^{J} +3 \Big({Y^d_2  Y^{d *}_2}\Big) Y_{2,{{i_1}}}^{J} +18 \Big({Y^J_1  Y^{J *}_1}\Big) Y_{2,{{i_1}}}^{J} +30 \Big({Y^J_2  Y^{J *}_2}\Big) Y_{2,{{i_1}}}^{J} \nonumber \\ 
		&+3 \Big({Y^u_2  Y^{u *}_2}\Big) Y_{2,{{i_1}}}^{J} \Big)\\ 
		\beta_{Y_{2,{{i_1}}}^{u}}^{(1)} & =  
		\frac{1}{6} \Big(3 \Big({Y^d_2  Y^{d *}_1}\Big) Y_{1,{{i_1}}}^{u} +3 \Big({Y^J_2  Y^{J *}_1}\Big) Y_{1,{{i_1}}}^{u} +12 \Big({Y^u_2  Y^{u *}_1}\Big) Y_{1,{{i_1}}}^{u} -10 g_{1}^{2} Y_{2,{{i_1}}}^{u} -24 g_{2}^{2} Y_{2,{{i_1}}}^{u} \nonumber \\ 
		& -48 g_{3}^{2} Y_{2,{{i_1}}}^{u}+3 \Big({Y^d_2  Y^{d *}_2}\Big) Y_{2,{{i_1}}}^{u} +18 \Big({Y^d_3  Y^{d *}_3}\Big) Y_{2,{{i_1}}}^{u} +3 \Big({Y^J_2  Y^{J *}_2}\Big) Y_{2,{{i_1}}}^{u}\nonumber \\ 
		& +18 \Big({Y^u_1  Y^{u *}_1}\Big) Y_{2,{{i_1}}}^{u} +30 \Big({Y^u_2  Y^{u *}_2}\Big) Y_{2,{{i_1}}}^{u} +12 \Big({Y^d_2  Y^{d *}_3}\Big) Y_{3,{{i_1}}}^{u} +9 \Big({Y^u_2  Y^{u *}_3}\Big) Y_{3,{{i_1}}}^{u} \Big)\\ 
		\beta_{Y_{3,{{i_1}}}^{d}}^{(1)} & =  
		+\frac{3}{2} \Big({Y^d_3  Y^{d *}_1}\Big) Y_{1,{{i_1}}}^{d} +2 \Big({Y^u_3  Y^{u *}_1}\Big) Y_{1,{{i_1}}}^{d} +\frac{3}{2} \Big({Y^d_3  Y^{d *}_2}\Big) Y_{2,{{i_1}}}^{d} +2 \Big({Y^u_3  Y^{u *}_2}\Big) Y_{2,{{i_1}}}^{d}\nonumber \\ 
		& -\frac{5}{3} g_{1}^{2} Y_{3,{{i_1}}}^{d} -4 g_{2}^{2} Y_{3,{{i_1}}}^{d} -8 g_{3}^{2} Y_{3,{{i_1}}}^{d} +\frac{1}{2} |Y^J_1|^2 Y_{3,{{i_1}}}^{d} +5 \Big({Y^d_3  Y^{d *}_3}\Big) Y_{3,{{i_1}}}^{d} +3 \Big({Y^u_1  Y^{u *}_1}\Big) Y_{3,{{i_1}}}^{d} \nonumber \\ 
		&+3 \Big({Y^u_2  Y^{u *}_2}\Big) Y_{3,{{i_1}}}^{d} +\frac{1}{2} \Big({Y^u_3  Y^{u *}_3}\Big) Y_{3,{{i_1}}}^{d} \\ 
		\beta_{Y^J_1}^{(1)} & =  
		\frac{1}{6} Y^J_1 \Big(18 \Big({Y^J_1  Y^{J *}_1}\Big)  + 18 \Big({Y^J_2  Y^{J *}_2}\Big)  -24 g_{2}^{2}  + 30 |Y^J_1|^2  + 3 \Big({Y^d_3  Y^{d *}_3}\Big) \nonumber \\ 
		& + 3 \Big({Y^u_3  Y^{u *}_3}\Big)  -48 g_{3}^{2}  -58 g_{1}^{2} \Big)\\ 
		\beta_{Y_{3,{{i_1}}}^{u}}^{(1)} & =  
		+2 \Big({Y^d_3  Y^{d *}_1}\Big) Y_{1,{{i_1}}}^{u} +\frac{3}{2} \Big({Y^u_3  Y^{u *}_1}\Big) Y_{1,{{i_1}}}^{u} +2 \Big({Y^d_3  Y^{d *}_2}\Big) Y_{2,{{i_1}}}^{u} +\frac{3}{2} \Big({Y^u_3  Y^{u *}_2}\Big) Y_{2,{{i_1}}}^{u} \nonumber \\ 
		&-\frac{8}{3} g_{1}^{2} Y_{3,{{i_1}}}^{u} -4 g_{2}^{2} Y_{3,{{i_1}}}^{u} -8 g_{3}^{2} Y_{3,{{i_1}}}^{u} +\frac{1}{2} |Y^J_1|^2 Y_{3,{{i_1}}}^{u} +3 \Big({Y^d_1  Y^{d *}_1}\Big) Y_{3,{{i_1}}}^{u} +3 \Big({Y^d_2  Y^{d *}_2}\Big) Y_{3,{{i_1}}}^{u} \nonumber \\ 
		&+\frac{1}{2} \Big({Y^d_3  Y^{d *}_3}\Big) Y_{3,{{i_1}}}^{u} +5 \Big({Y^u_3  Y^{u *}_3}\Big) Y_{3,{{i_1}}}^{u} 
\end{align}}

\subsection{Trilinear Scalar couplings}
{\allowdisplaybreaks  \begin{align} 
		\beta_{{f}_{\sigma}}^{(1)} & =  
		-6 {f}_{\sigma} g_{1}^{2} -18 {f}_{\sigma} g_{2}^{2} +2 {f}_{\sigma} \lambda_{13} +2 {f}_{\sigma} \zeta_{13} +2 {f}_{\sigma} {\lambda}_{14} +2 {f}_{\sigma} {\lambda}_{34} -4 f {\xi}_{14} \nonumber \\ 
		&+4 f {\xi}_{34} +4 {f}_{\sigma} \zeta_{14} +4 {f}_{\sigma} \zeta_{34} +3 {f}_{\sigma} |Y^J_1|^2 +3 {f}_{\sigma} \Big({Y^d_3  Y^{d *}_3}\Big) +3 {f}_{\sigma} \Big({Y^J_1  Y^{J *}_1}\Big) \nonumber \\ 
		&+3 {f}_{\sigma} \Big({Y^J_2  Y^{J *}_2}\Big) +3 {f}_{\sigma} \Big({Y^u_1  Y^{u *}_1}\Big) +3 {f}_{\sigma} \Big({Y^u_2  Y^{u *}_2}\Big) +2 {f}_{\sigma} \mbox{Tr}\Big({G^\sigma  G^{\sigma *}}\Big) \\ 
		\beta_{f}^{(1)} & =  
		-6 f g_{1}^{2} -12 f g_{2}^{2} +2 f \lambda_{12} -2 f \zeta_{12} +2 f \lambda_{13} -2 f \zeta_{13} +2 f \lambda_{23} -2 f \zeta_{23} \nonumber \\ 
		&-4 {f}_{\sigma} {\xi}_{14}  +4 {f}_{\sigma} {\xi}_{34} +3 f |Y^J_1|^2+3 f \Big({Y^d_1  Y^{d *}_1}\Big) +3 f \Big({Y^d_2  Y^{d *}_2}\Big) +3 f \Big({Y^d_3  Y^{d *}_3}\Big) \nonumber \\ 
		&+3 f \Big({Y^J_1  Y^{J *}_1}\Big) +3 f \Big({Y^J_2  Y^{J *}_2}\Big) +3 f \Big({Y^u_1  Y^{u *}_1}\Big) +3 f \Big({Y^u_2  Y^{u *}_2}\Big) +3 f \Big({Y^u_3  Y^{u *}_3}\Big) 
\end{align}}

\subsection{Scalar Mass Terms}
{\allowdisplaybreaks  \begin{align} 
		\beta_{{\mu}_{3}}^{(1)} & =  
		2 \Big(4 f^{2} +4 {f}_{\sigma}^{2} +3 \lambda_{13} {\mu}_{1} +\zeta_{13} {\mu}_{1} +3 \lambda_{23} {\mu}_{2} +\zeta_{23} {\mu}_{2} -3 g_{1}^{2} {\mu}_{3} -4 g_{2}^{2} {\mu}_{3} +8 \lambda_{3} {\mu}_{3}\nonumber \\ 
		& +6 {\lambda}_{34} {\mu}_{4}  +2 {\mu}_{4} \zeta_{34}+3 {\mu}_{3} \Big({Y^d_3  Y^{d *}_3}\Big) +3 {\mu}_{3} \Big({Y^u_1  Y^{u *}_1}\Big) +3 {\mu}_{3} \Big({Y^u_2  Y^{u *}_2}\Big) \Big)\\ 
		\beta_{{\mu}_{2}}^{(1)} & =  
		2 \Big(4 f^{2} +3 \lambda_{12} {\mu}_{1} +\zeta_{12} {\mu}_{1} -4 g_{2}^{2} {\mu}_{2} +8 \lambda_{2} {\mu}_{2} +3 \lambda_{23} {\mu}_{3} +\zeta_{23} {\mu}_{3} +6 {\lambda}_{24} {\mu}_{4} \nonumber \\ 
		&+2 {\mu}_{4} \zeta_{24} +3 {\mu}_{2} \Big({Y^d_1  Y^{d *}_1}\Big) +3 {\mu}_{2} \Big({Y^d_2  Y^{d *}_2}\Big) +3 {\mu}_{2} \Big({Y^u_3  Y^{u *}_3}\Big) \Big)\\ 
		\beta_{{\mu}_{1}}^{(1)} & =  
		2 \Big(4 f^{2} +4 {f}_{\sigma}^{2} -3 g_{1}^{2} {\mu}_{1} -4 g_{2}^{2} {\mu}_{1} +8 \lambda_{1} {\mu}_{1} +3 \lambda_{12} {\mu}_{2} +\zeta_{12} {\mu}_{2} +3 \lambda_{13} {\mu}_{3} \nonumber \\ 
		&+\zeta_{13} {\mu}_{3} +6 {\lambda}_{14} {\mu}_{4}  +2 {\mu}_{4} \zeta_{14}+3 {\mu}_{1} |Y^J_1|^2 +3 {\mu}_{1} \Big({Y^J_1  Y^{J *}_1}\Big) +3 {\mu}_{1} \Big({Y^J_2  Y^{J *}_2}\Big) \Big)\\ 
		\beta_{{\mu}_{4}}^{(1)} & =  
		2 \Big(-10 g_{2}^{2} {\mu}_{4}  + 14 {\lambda}_4 {\mu}_{4}  + 2 {f}_{\sigma}^{2}  + 2 {\mu}_{4} \mbox{Tr}\Big({G^\sigma  G^{\sigma *}}\Big)  + 3 {\lambda}_{14} {\mu}_{1}  + 3 {\lambda}_{24} {\mu}_{2}  + 3 {\lambda}_{34} {\mu}_{3}  \nonumber \\ 
		& + 8 {\lambda}_{44} {\mu}_{4}  + {\mu}_{1} \zeta_{14}  + {\mu}_{2} \zeta_{24}  + {\mu}_{3} \zeta_{34} \Big)
\end{align}}

\subsection{Vacuum expectation values}
{\allowdisplaybreaks  \begin{align} 
		\beta_{v_{\rho}}^{(1)} & =  
		\frac{1}{3} v_{\rho} \Big(12 g_{2}^{2}  + 3 g_{1}^{2} \xi  + 4 g_{2}^{2} \xi  + 9 g_{1}^{2}  -9 |Y^J_1|^2  -9 \Big({Y^J_1  Y^{J *}_1}\Big)  -9 \Big({Y^J_2  Y^{J *}_2}\Big) \Big)\\ 
		\beta_{v_{\eta}}^{(1)} & =  
		\frac{1}{3} v_{\eta} \Big(12 g_{2}^{2}  + 4 g_{2}^{2} \xi  -9 \Big({Y^d_1  Y^{d *}_1}\Big)  -9 \Big({Y^d_2  Y^{d *}_2}\Big)  -9 \Big({Y^u_3  Y^{u *}_3}\Big) \Big)\\ 
		\beta_{v_{\chi}}^{(1)} & =  
		\frac{1}{3} v_{\chi} \Big(12 g_{2}^{2}  + 3 g_{1}^{2} \xi  + 4 g_{2}^{2} \xi  + 9 g_{1}^{2}  -9 \Big({Y^d_3  Y^{d *}_3}\Big)  -9 \Big({Y^u_1  Y^{u *}_1}\Big)  -9 \Big({Y^u_2  Y^{u *}_2}\Big) \Big)\\ 
		\beta_{v_{\sigma}}^{(1)} & =  
		\frac{2}{3} v_{\sigma} \Big(-3 \mbox{Tr}\Big({G^\sigma  G^{\sigma *}}\Big)  + 5 g_{2}^{2} \Big(3 + \xi\Big)\Big)
\end{align}}

\end{appendices}
		
		%\listoffigures
		\addcontentsline{toc}{chapter}{Bibliography}
		\putbib[part1]
	\end{bibunit}

	\part{Anomaly-Induced Dynamics in Gauge and Gravitational Theories: Hadronic Gravitational Form Factors, Graviton Scattering Amplitudes, and Anomaly Sum Rules}
	
	\begin{bibunit}[jhep]
		\chapter{The Gravitational Form Factors of Hadrons from CFT in Momentum Space and the  Dilaton in Perturbative QCD}\label{chtjj}
\addcontentsline{toc}{section}{Synopsis}
\section*{Synopsis}

This Chapter opens the second part of the thesis, shifting the focus from the phenomenology of anomaly-driven gauge extensions of the Standard Model to the role of conformal and trace anomalies in QCD and in gravitational theories. The central object of investigation is the non-Abelian $TJJ$ correlator the three-point function of one stress-energy tensor insertion and two non-Abelian vector currents which enters at next-to-leading order in $\alpha_s$ in the gravitational form factors (GFFs) of hadrons through QCD factorization. The Chapter develops a complete off-shell parameterization of this vertex in perturbative QCD, rooted in the formalism of conformal field theory in momentum space ($\mathrm{CFT}_p$), and establishes its connection to the emergence of a dilaton-like anomaly pole associated with the QCD trace anomaly.

The Chapter begins by motivating the study of GFFs through their role in encoding the distribution of energy, momentum, angular momentum, pressure, and shear forces inside hadrons. The matrix elements of the QCD energy-momentum tensor (EMT) between hadronic states are expanded in terms of form factors $A(t)$, $B(t)$, and $D(t)$ for spin-$1/2$ hadrons, and analogous structures for spin-$0$ hadrons such as the pion. The physical interpretation of these form factors is discussed: $A(t)$ governs the momentum distribution among the hadronic constituents and satisfies $A(0) = 1$; $B(t)$ is associated with angular momentum and satisfies $B(0) = 0$; and $D(t)$, the $D$-term, characterizes the internal stress distribution and is unconstrained at zero momentum transfer. Their connection to generalized parton distributions (GPDs) through second Mellin moments is recalled, and the experimental relevance of the GFFs in the context of deeply virtual Compton scattering (DVCS), the JLab 12 GeV program, and the future Electron-Ion Collider (EIC) is outlined.

The perturbative structure of the GFFs is then analyzed within QCD factorization. At sufficiently large momentum transfer, the GFF is described by a convolution of hadronic distribution amplitudes with a hard scattering kernel, in which the $TJJ$ vertex appears at NLO as the key partonic amplitude. The Chapter emphasizes that the trace anomaly of the QCD stress-energy tensor,
\begin{equation*}
	\hat{T}^\mu_{\;\mu} = \beta(g)\, F^{a,\mu\nu}F^a_{\mu\nu} + (1 + \gamma_m)\sum_q m_q \bar{\psi}_q \psi_q,
\end{equation*}
breaks conformal symmetry at the quantum level and induces an anomalous contribution to the hard scattering, manifesting as a dilaton-like $t$-channel pole in the $TJJ$ vertex.

The formal framework is developed through the gravitational effective action of QCD, formulated in the presence of both gravitational and gauge backgrounds. The QCD partition function is expanded around a linearized metric fluctuation $h_{\mu\nu}$, and the stress-energy tensor is defined as the functional derivative of the effective action with respect to the metric. The quark sector is treated using standard conformal Ward identities (CWIs), which take the same hierarchical form as in the Abelian case and are broken only by the trace anomaly. The gluon and ghost sectors, by contrast, are governed by Slavnov-Taylor identities (STIs) due to gauge fixing, which replaces the ordinary single-derivative Ward identities with quadratic contraction conditions. This distinction between the quark and gluon sectors is a central structural feature of the analysis.

The conformal anomaly effective action, namely the Riegert action, is presented in its nonlocal form, and its linearization around flat spacetime is shown to reproduce, at the level of three-point functions, the perturbative anomaly pole through the $\Box^{-1}$ operator acting on $F^a_{\mu\nu}F^{a\,\mu\nu}$. The gauge-covariant anomaly-induced vertex is constructed from this nonlocal action, and its gauge-invariant extension to the $TJJJ$ and $TJJJJ$ correlators is presented.

The sector decomposition of the $TJJ$ vertex is then carried out systematically using transverse, transverse-traceless, and longitudinal projectors. The correlator is decomposed into a transverse-traceless (tt) sector, longitudinal sectors, and a trace sector. In the quark sector, the tt part is parameterized by four form factors $A_j^{(q)}$, $j=1,\ldots,4$, expressed in terms of scalar master integrals $B_0$ and $C_0$. The primary CWIs take the form of second-order differential equations for these form factors, which are solved analytically, and their solutions are matched against the free-field theory perturbative computation. Secondary (first-order) CWIs fix the remaining integration constants. In the gluon sector, which is treated by direct perturbative computation, additional longitudinal contributions arise that are absent in the abstract CFT solution, as well as extra terms in the trace sector proportional to the gluon equations of motion, collectively denoted $\mathcal{B}_g^{\alpha\beta ab}$.

The complete off-shell parameterization of the $TJJ$ correlator is assembled by combining the quark and gluon sectors. The final expression decomposes into a transverse-traceless sector, a longitudinal sector containing additional gluon contributions, and a trace sector of the form
\begin{equation*}
	\langle T^{\mu\nu}(q)\, J^{a\alpha}(p_1)\, J^{b\beta}(p_2)\rangle_{tr} = \frac{1}{3q^2}\hat{\pi}^{\mu\nu}(q)\left[\mathcal{A}^{\alpha\beta ab} + \mathcal{B}_g^{\alpha\beta ab}\right],
\end{equation*}
where $\hat{\pi}^{\mu\nu} \equiv q^2 g^{\mu\nu} - q^\mu q^\nu$ makes the $1/q^2$ dilaton pole explicit, and $\mathcal{A}^{\alpha\beta ab}$ is the gauge-invariant anomaly contribution proportional to the QCD $\beta$-function coefficient $(11C_A - 2n_f)$. The anomaly form factor characterizing the exchange of a dilaton state is thereby identified as $\beta(g)/q^2$.

The extension to massive quarks is then carried out. The sector decomposition remains valid, but the trace sector acquires additional structure: the anomaly form factor $\phi_1$, which projects onto the tensor structure $u^{\alpha\beta} = (p_1 \cdot p_2)g^{\alpha\beta} - p_2^\alpha p_1^\beta$, now receives mass-dependent corrections, while a second form factor $\phi_2$ accounts for terms proportional to the equations of motion for massive quarks. Explicit expressions for all scalar form factors $B_1, \ldots, B_4$ and $\chi_0$, $\chi_1$ are provided in terms of $B_0$ and $C_0$ integrals with massive arguments.

The relationship between the anomaly pole and renormalization is clarified through the $F$-basis parameterization of the $TJJ$ vertex, in which the 13 independent tensor structures are constrained by all Ward identities. In this basis, a single form factor $F_{13}$ is shown to be the unique source of ultraviolet divergence, renormalized by the $F^2$ counterterm. As $d \to 4$ in dimensional regularization, the constraint equations relating the trace form factors $F_1$ and $F_2$ to $F_{13}$ show that the anomaly pole $1/q^2$ arises solely from the failure of conformal symmetry in the renormalization limit, confirming that the dilaton pole is an unambiguous consequence of the trace anomaly and not a perturbative artifact.

The Chapter concludes by connecting the parameterization of the hard scattering to the hadronic GFFs at large $-t$, outlining how the $TJJ$ vertex enters the factorized expression for the proton and pion form factors through convolution with light-cone distribution amplitudes. The helicity decomposition of the proton matrix element is discussed, identifying the leading Fock-state contributions and the topologies through which the anomaly corrections enter at $O(\alpha_s^3)$. It is emphasized that the anomaly pole does not contribute to the form factor $A(t)$ in the massless limit due to the equations of motion, and that the anomaly form factor satisfies a dispersive sum rule whose verification extends results previously established in QED and confirms the dilaton interpretation of the anomaly pole. The sum rule, together with the pole structure, is identified as a hallmark of conformal anomalies, and its phenomenological implications for the EIC program on proton spin and tomography are highlighted as a primary motivation for future work.

The work presented in this Chapter is based primarily on, and adapted from, the publication \cite{Coriano:2024qbr}, co-authored with C.~Corianò, S.~Lionetti, and R.~Tommasi.

\section{Introduction}
Conformal symmetry imposes strong constraints on 3-point functions of scalar and tensor correlators. This symmetry allows one to establish a connection between the conventional free-field theory realization of the same correlators in Lagrangian conformal field theories (CFT), and their expected general tensorial structure. This is identified by Lorentz covariance through the use of Conformal Ward identities (CWIs) and the operator product expansion in the abstract CFT formulation. When this approach is applied in momentum space, here denoted as $CFT_p$ \cite{Coriano:2013jba,Bzowski:2013sza,Bzowski:2015pba,Coriano:2020ees}, it allows one to establish a link between the general expression of a certain correlator determined by solving the CWIs and its ordinary Feynman expansion, i.e. the corresponding amplitudes \cite{Coriano:2023hts,Coriano:2018bsy} (see \cite{Coriano:2020ees} for a review of the methods and \cite{Caloro:2022zuy,Bzowski:2020kfw}\cite{Marotta:2022jrp,Jain:2021wyn,Coriano:2024ssu} for recent extensions to higher point functions or parity-odd correlators).\\
The free-field theory expressions of the correlators are generally characterized by the presence of anomalous dimensions in their renormalization group evolution as well as by explicit violations of the conformal symmetry associated with dimensionful scales, such as mass parameters. Obviously, the latter are naturally absent in the abstract CFT solution. In the QCD case, however, the renormalization procedure is responsible for the inclusion of a renormalization scale and for the generation of a nonzero trace, the trace (scale) anomaly \cite{Capper:1975ig,Duff:1977ay,Christensen:1978gi,Duff:1993wm}, breaking conformal symmetry.\\
In a mass-independent regularization scheme such as dimensional regularization (DR), the corresponding definition of the QED/QCD $\beta$ functions are such that the anomaly contribution, which is unrelated to the mass-dependent corrections of the theories, separate. This result can be inferred from explicit computations 
in the Abelian \cite{Giannotti:2008cv,Armillis:2009pq} and non-Abelian cases \cite{Armillis:2010qk} of correlators containing the stress energy tensor.\\
In a CFT approach, the CWIs determine the solution of a generic 3-point function in terms of few constants that are matched  in free field theory by the perturbative computation of the same correlator. The analysis of such correlators in momentum space is particularly relevant in the presence of chiral and conformal (trace) anomalies, where the emergence of an anomaly contribution is directly and uniquely associated with the exchange of an anomaly pole.  
For both older and more recent perturbative analyses of trace anomalies in QCD and operatorial mixing, we refer to \cite{Freedman:1974gs, Tanaka:2022wrr, Hatta:2018sqd, Ahmed:2022adh}. \\
This procedure has been applied to various correlators, including those of both even and odd parity. In the  case of the axial-vector/vector/vector vertex, where a chiral anomaly is present, the solution to the anomalous CWIs constraining the full correlator can be obtained by introducing a chiral anomaly pole in the spectrum, highlighting the central role of this contribution \cite{Coriano:2025ceu}. A similar approach has been explored in the context of other parity-odd trace anomalies \cite{Coriano:2023hts, Coriano:2023gxa, Coriano:2023cvf}.\\
For parity even trace anomalies, the pole can be identified as a dilaton state and, as we are going to show elsewhere, is characterized by the presence of a sum rule in the anomaly form factor in which it appears.\\
At lower momentum transfers, the effects of anomalous scale breaking in hadronic matrix elements involving a $T^{\mu\nu}$ insertion must be addressed using effective models that account for chiral symmetry breaking effects \cite{Kharzeev:2021qkd} and the QCD instanton vacuum \cite{Liu:2024jno,Liu:2024vkj}. In contrast, at larger momentum transfers, the hard scattering can be analyzed using the formal approach we present.

\subsection{The TJJ and conformal symmetry in the sector decomposition}
The correlator we are going to examine in its off-shell expansion is the non-Abelian $TJJ$, which has been previously discussed in momentum space by CFT methods in \cite{Bzowski:2018fql} and in perturbative QCD (pQCD) for on-shell external gluons in \cite{Armillis:2010qk}. Here, $J$ represents a non-Abelian vector current in four dimensions ($d=4$), and $T$ is the gauge-fixed stress-energy tensor of QCD. To include this interaction in the generalized form factors (GFFs) of hadrons, we need to extend the on-shell analysis presented in \cite{Armillis:2010qk}. This extension will be achieved by closely following the formalism of CFT$_p$, as developed in \cite{Coriano:2013jba, Bzowski:2013sza}, subsequent to the initial analysis of this correlator in \cite{Armillis:2010qk}.\\
In the Abelian case, an initial parameterization of this correlator was discussed in \cite{Giannotti:2008cv, Armillis:2009pq}, where it was expressed in terms of 13 form factors (the F-basis), with only two of them containing kinematical poles. This parameterization, which we will review and apply to the quark contributions to illustrate how the anomaly emerges in dimensional regularization (DR), differs significantly from the minimal approach proposed in CFT\(_p\) for characterizing the transverse-traceless (tt) sector \cite{Bzowski:2013sza} of a tensor correlator that includes one or more stress-energy tensors.
The general structure of CFT$_p$ introduced in \cite{Bzowski:2013sza} was matched to free field theory realizations in \cite{Coriano:2018bbe, Coriano:2018zdo}. \\
The reduction in the number of form factors in this (longitudinal/transverse)  parameterization, referred to as the LT-basis, is achieved by symmetrically incorporating all three momenta of the vertex within the transverse traceless (tt) sector. 
The F-basis mentioned above turns useful in QED, but covers only the quark sector in QCD, since in QCD the WIs of QED are replaced by Slavnov-Taylor identities (STIs). 
In the case of Abelian theories, the relation between the F-basis and the LT-basis was discussed in \cite{Coriano:2018bbe}. \\
In CFT\(_p\), the correlator is constructed in the LT-basis around the  (tt) sector by reorganizing the CWIs such that they take the form of second-order differential equations (primary equations) for the form factors in this sector, and first-order (secondary equations) for the remaining sectors. The conformal anomaly is then directly linked to the inclusion of a single counterterm (\(\sim F^2\)) in the tensorial expansion, leading to the generation of a trace sector. \\
In this approach, the form factors in the (tt) sector are found explicitly, with the solution expressed in terms of a set of integration constants, which are then refined using secondary conformal constraints from the remaining ordinary Ward identities. We will clarify these aspects in the QCD context, which differs from the standard CFT$_p$ approach due to the presence of virtual gluons in quantum corrections. \\
While CFT$_p$ can handle correlators involving non-Abelian currents, this method cannot be directly applied to perturbative QCD (pQCD) due to gauge-fixing, which breaks the conformal symmetry of the QCD action. However, we will show how this limitation can be addressed by decomposing both quark and gluon contributions into the LT-basis, revealing additional form factors from the exchange of virtual gluons in the longitudinal sectors. \\
To prevent any confusion, throughout this work, {\em "sector decomposition"} refers specifically to the decomposition of a tensor into its longitudinal, transverse traceless, and trace components using projectors. This definition is unrelated to the decomposition of a Feynman integral into distinct momentum integration regions. \\
Our analysis will concentrate on the tensor structure of the hard scattering, derived from the CFT\(_p\) parameterization, guided by the standard CWIs satisfied by the quark contributions, which follows closely the Abelian case. The gluon sector, instead, is incorporated through a direct perturbative analysis. \\
In principle, also the reconstruction of the gluon sector within CFT$_p$ can be achieved by utilizing the broken CWIs of pQCD in a general framework, akin to the treatment of any conformal 3-point function. This approach should lead to the same decomposition presented in this work. However, for clarity and simplicity, we choose to organize the decomposition by employing a direct perturbative expansion, where gluon contributions are decomposed in the LT-basis and then added to the quark sector. 
The approach we propose allows for the direct isolation of all the form factors and, in particular, of the anomaly form factor and the QCD conformal anomaly present in the hard scattering process. This sets the stage for a future comprehensive analysis of the gravitational form factors (GFF) of the pion and proton at the hadronic level, within the framework of QCD factorization for exclusive processes, which will be explored in a forthcoming work.\\
From our perspective, the parameterization of the hard scattering that we present provides the optimal framework for characterizing the complete hadronic matrix elements, essential for the experimental investigation of potential anomaly effects, since anomalies, their poles and conformal symmetry are closely related.\\
The approach also allows for a straightforward formulation of higher-order perturbative corrections. Given that the longitudinal-transverse (LT) sector decomposition is always valid and minimal, it proves to be particularly valuable. In this formulation, the quark and gluon sectors combine to produce a gauge-invariant anomaly pole as a signature of the exchange of a dilaton pole due to the anomaly. \\
The introduction of mass corrections modifies the anomaly form factor, transforming 
it from a pole to a cut. However, the true significance of the anomaly lies not in the presence of the pole or the cut, but in the existence of a sum rule. This sum rule, satisfied by the integral of the spectral density of the form factor, precisely reflects the anomaly.

\subsection{Conformal Anomaly poles in Abelian and non-Abelian theories}
Dilaton-like $t$-channel exchanges in the $TJJ$ case, were originally investigated perturbatively in QED \cite{Giannotti:2008cv, Armillis:2009pq} and QCD \cite{Armillis:2010qk}, as previously mentioned, to characterize the coupling of gravity to the fields of the Standard Model via anomaly poles. A discussion of these features in the neutral current sector of the Standard Model can be found in \cite{Coriano:2011ti, 2013JHEP...06..077C}, and in supersymmetric models in \cite{Coriano:2014gja}. In the last case, for instance, these analysis provided explicit proofs of the existence of supersymmetric sum rules for the anomaly form factors of the superconformal anomaly supermultiplet, involving both chiral and conformal anomalies, as well as for the Konishi anomaly. \\
A sum rule for the $TJJ$ in QED was derived in \cite{Giannotti:2008cv}. As shown in \cite{Coriano:2023gxa}, such poles are also present in the form factor of the gravitational chiral anomaly when not only fermion currents, but also spin-1 Chern-Simons currents, are involved. This anomaly was originally discussed in \cite{Dolgov:1987yp, Dolgov:1988qx} and more recently in \cite{Agullo:2018nfv}.\\
Recent analyses of correlators involving insertions of stress-energy tensors in conformal field theory (CFT), along with the study of conformal anomaly actions, have revealed that the corresponding interaction vertices can be expanded in terms of the dimensionless parameter $R\Box^{-1}$ in a general gravitational background. In QCD, coupling the stress-energy tensor to an external metric allows us to leverage results from the formalism of the conformal anomaly effective action, which is particularly useful for addressing conformal constraints. Here, $R$ represents the scalar curvature, while the nonlocal $1/\Box$ interaction that we identify characterizes the exchange of an effective dilaton-like state in the $t$-channel of the $TJJ$ vertex.\\
Unlike the phenomenological dilaton effective action, which includes a conformal breaking scale $(\Lambda)$ and features a coupling of the form $(\chi/\Lambda) FF$, where $\chi$ is a dilaton field locally coupled to the anomaly, the nonlocal action is derived directly from the ultraviolet behavior of the diagrams we compute. These diagrams define a one-particle irreducible (1PI) effective action, from which the anomaly form factor and its associated anomaly pole can be extracted.

\subsection{The non-Abelian $TJJ$ in exclusive processes, CFT$_p$ and the anomaly effective action}
The $TJJ$  interaction enters at next-to-leading order (NLO) in the strong coupling constant \(\alpha_s\) in the gravitational form factors (GFFs) of hadrons through factorization in pQCD. This is a key component of hard scattering processes and induces a trace anomaly. 
%In this work \(T\), specifically, will denote the QCD stress-energy tensor and \(J\) the non-Abelian gluon %current.\\
We will examine this interaction in relation to the general properties of such exchanges, as investigated in pQCD, within the framework of CFT$_p$.
The phenomenological applications of the results of this analysis will be discussed in a separate work. \\
We aim to disentangle the dilaton-like effective degree which is part of the 1PI effective vertex, and show how effective nonlocal actions can be formulated for such anomalous interactions by a suitable analysis of the hard scattering.\\
Since the stress-energy tensor can be derived by varying the QCD partition function with respect to an external gravitational field, we will formulate the CWIs for this correlator using the general framework of the gravitational effective action for a non-Abelian theory, expanded around flat spacetime. The conformal anomaly interaction that we are focusing on, is a specific case within this broader formulation. To our knowledge, this approach has not been previously applied in the context of QCD, yet it proves highly effective in analyzing the constraints governing amplitudes that include a stress-energy tensor.\\
Concerning the symmetry constraints to be imposed, the correlator is governed by a less restrictive Slavnov-Taylor identity (STI) compared to an ordinary WI. Notably, the general nonperturbative CFT solutions for the \(TJJ\) correlator for both non-Abelian and Abelian currents are quite similar for non-Lagrangian theories \cite{Bzowski:2018fql}. However, in the context of QCD, these solutions are applicable only in the quark sector, as the gauge-fixing of the QCD action alters the gluon sector and breaks its conformal symmetry at $d=4$. Indeed, the hierarchical equations satisfied by the quark sector take the form of ordinary CWIs, derivable by a partition function where the gluons are treated as external fields. A similar approach in the gluon sector is more involved and requires the implementation of the broken CWIs discussed in previous works \cite{Sarkar:1974xh, Nielsen:1975ph} \cite{Braun:2018mxm}, which we will not pursue. This is not strictly necessary in our current analysis, being our work focused on the conformal limit of the interaction.

\subsection{Content of this work}
Our work is organized as follows: In \secref{gen1}, after a brief overview of the role of the GFFs of hadrons in the context of QCD factorization and their relation to the deeply virtual Compton scattering (DVCS) amplitude, we discuss the partonic \($TJJ$\) interaction as it emerges in the hard scattering of the GFFs. In \secref{three1} the correlator is discussed in full generality within the formalism of the anomaly effective action, computed in the presence of both gravitational and gauge backgrounds, of which this correlator is part. \\
In \secref{four1} we discuss the symmetries of the generating functional in the quark sector, which are essential for the formal derivation of the WIs and CWIs constraints at one loop in this sector. In this section the gluons are treated as external classical fields. In \secref{BRSTX} we turn to the analysis of the symmetries of the complete QCD partition function, deriving the relevant STIs satisfied by the correlator. 
In \secref{seven}, we describe the sector decomposition, which naturally leads to the emergence of the dilaton pole. This is followed in the next Section by an analysis of conformal constraints in this sector and their connection to perturbative results. The gluon contribution to the correlator is discussed in \secref{nine}, while \secref{ten} presents the final parameterization of the perturbative correlator for both massless and massive fermions. \\
Next we briefly review the on-shell case and explain how the dilaton anomaly pole arises in perturbation theory due to the renormalization of trace constraints in $d$ dimensions, taking the limit $d \to 4$. This is illustrated in the quark sector, along with a verification of a sum rule for the spectral density of the dilaton form factor, derived from the explicit expression of the on-shell correlator. \\
In \secref{dec22}, we conclude with a discussion of future extensions to hadronic systems, particularly the proton and pion. Several technical details are provided in the appendices, including explicit expressions for form factors in the massless case. Appendices of this Chapter A and B cover the derivation of Slavnov-Taylor identities and list the relevant Feynman rules. Appendix C summarizes the perturbative computation of the quark sector, using projectors to extract the form factors introduced in \secref{seven}. Appendices D, E, and F detail the implementation of the conformal procedure in $CFT_p$ and its modifications due to QCD gauge fixing. Appendices G and H present the transverse traceless form factors for both massless and massive cases, along with the fundamental integrals appearing in their definitions.
\section{The $TJJ$ and the gravitational form factors of hadrons }
\label{gen1}
The experimental investigation of the GFFs of the proton and the pion provide nonperturbative insights into the coupling of these hadrons to the energy-momentum tensor of QCD, revealing essential information about the distribution of their energy, spin, pressure, and shear forces. \\
GFFs are expanded in terms of the matrix elements of the EMT between hadron states. Quantum 
corrections break the conformal symmetry due to the trace anomaly of the stress energy tensor
\cite{Adler:1976zt,Nielsen:1977sy,Collins:1976yq}
\be\label{anom1}
\hat{T}_{\!\mu}^{\,\mu} \equiv 
{\beta(g)}\;F^{a,\mu\nu}{F^{a,}}_{\!\!\mu\nu}
+(1+\gamma_m)\sum_qm_q\bar\psi_q\psi_q \;,\ee
where $\beta(g)$ is the $\beta$-function of QCD and $\gamma_m$ is 
the anomalous dimension of the mass operator. The trace anomaly
has the same form in QED.\\
These expansions reveal how the internal structure of the proton is related to its energy, momentum, and stress distributions. The matrix elements of the EMT for a spin 1/2 hadron with momentum \(P\) can be expressed in terms of the GFFs as

\begin{align}
	\langle p^\prime,s^\prime| T_{\mu\nu}(0) |p,s\rangle
	= \bar u{ }^\prime\biggl[
	A(t)\,\frac{\gamma_{\{\mu} P_{\nu\}}}{2}
	+ B(t)\,\frac{i\,P_{\{\mu}\sigma_{\nu\}\rho}\Delta^\rho}{4 M}
	+ D(t)\,\frac{\Delta_\mu\Delta_\nu-g_{\mu\nu}\Delta^2}{4M}
	+ {M}\,\sum_{\hat{a}} {\bar c}^{\hat{a}}(t)\,g_{\mu\nu} \biggr]u\,
	\label{fund1}
\end{align}
where $u(p)$ and $\overline{u}(p')$ are the proton spinors, $ P = (p + p')/2$ is the average momentum, 
$\Delta = p' - p$ is the momentum transfer,  $t = \Delta^2$, and  $M$ is the mass of the proton. $\gamma^{(\mu} P^{\nu)}$denotes the symmetric combination $\gamma^\mu P^\nu + \gamma^\nu P^\mu$.\\
Using the Gordon identity, the separate components related to quarks ($\hat{a}\equiv q$) and gluons 
($\hat{a}\equiv g$)
can be expressed in the form
\bea
\la p^\prime,s^\prime| T_{\mu\nu}^{\hat{a}}(0) |p,s\rangle
= \bar u^\prime\biggl[
A^{\hat{a}}(t)\,\frac{P_\mu P_\nu}{M}
+ J^{\hat{a}}(t)\ \frac{i\,P_{\{\mu}\sigma_{\nu\}\rho}\Delta^\rho}{2 M}
+ D^{\hat{a}}(t)\,\frac{\Delta_\mu\Delta_\nu-g_{\mu\nu}\Delta^2}{4 M}
+{M}\,{\bar c}^{\hat{a}}(t)g_{\mu\nu} \biggr]u \nonumber \\
\label{Eq:EMT-FFs-spin-12-alternative} \eea
The two representations are equivalent, and the form factors are
related as $A^a(t)+B^a(t) = 2\,J^a(t)$. 
For a spin 0 hadron it takes the form 
\bea
\label{fund2}
\la p^{\,\prime\,}|\hat{T}_{\mu\nu}(0)|p\ra = 
\biggl[2 P_\mu P_\nu\, A(t) + 
\frac12(\Delta_\mu\Delta_\nu - g_{\mu\nu} \Delta^2)\,D(t)
+ {2\ m^2}\,{\bar c}(t)\,g_{\mu\nu} \biggr] \,.
\eea

The stress eenrgy tensor (see \cite{Polyakov:2018zvc} for an overview) can be investigated by studying auxiliary processes involving generalized parton distribution functions \cite{Ji:1996ek,Radyushkin:1996nd,Radyushkin:1996ru,
	Ji:1996nm} \cite{Collins:1996fb,Radyushkin:1997ki,Vanderhaeghen:1998uc} in hard exclusive reactions. 
The process provides information about  the mass and the spin of hadron 
\cite{Kobzarev:1962wt,Pagels:1966zza,Ji:1998pc,Radyushkin:2000uy,Goeke:2001tz,Diehl:2003ny,Belitsky:2005qn}. 
A perturbative analysis in the context of QCD factorization has been presented in \cite{Tong:2022zax}. \\
Together with the $D$-term \cite{Polyakov:1999gs} these form factors allow to gain information on the tomography of the proton. 
The GFFs $A(t)$, $B(t)$, and $D(t)$ in \eqref{fund1} have specific physical interpretations. For example, $A(t)$ represents the distribution of the proton's momentum among its constituents (quarks and gluons). In the forward limit (\(t = 0\)), \(A(0)\) sums to 1, indicating the total momentum of the proton.\\
$B(t)$ is associated with the distribution of the proton's angular momentum. 
The form factor $B(t)$ is present only for hadrons with $J>0$ and satisfies at zero-momentum transfer the constraint $B(0) = 0$, indicating the vanishing of the anomalous gravitomagnetic moment in the same kinematical limit. The constraints at $t=0$ derive from the fact that these form factors are related to the generators of the Poincar\`e group and henceforth to the mass and spin of the hadron. \\
On the other hand, the form factor $D(t)$ at zero-momentum transfer is unconstrained, and identifies the $D$-term, typical of any hadron. In contrast to $A(t)$ and $B(t)$ which are determined at $t=0$ by the mass and spin of the particles, the $D$-term is related to the stress tensor and internal forces.
The combination \(A(t) + B(t)\) at \(t = 0\) gives the total angular momentum carried by the quarks and gluons, as described by the sum rule  \cite{Ji:1996ek}
\beq
J_q + J_g = \frac{1}{2} [A_q(0) + B_q(0)] + \frac{1}{2} [A_g(0) + B_g(0)] = \frac{1}{2}.
\eeq

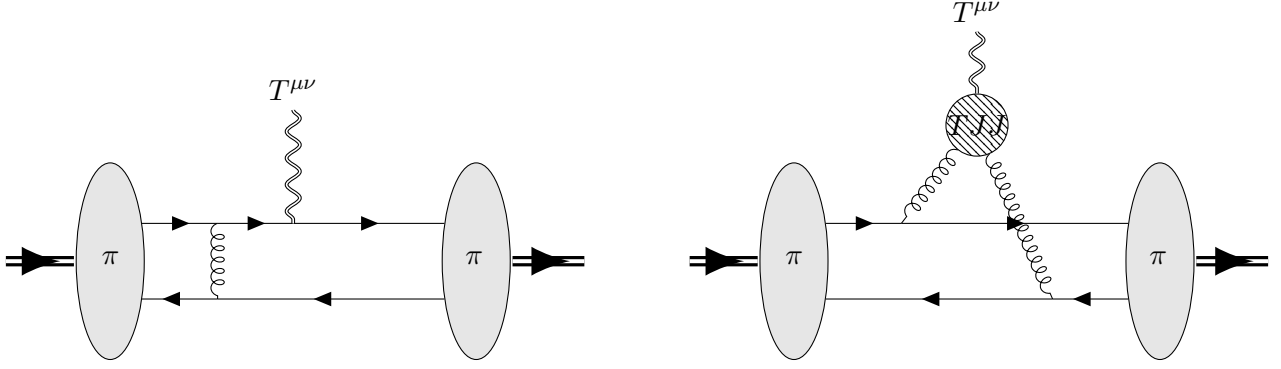
\begin{figure}
	\centering
	\begin{tikzpicture}
		\begin{feynman}
			\vertex (i1);
			\vertex[right=1cm of i1] (a1);
			\vertex[right=1cm of a1] (b1);
			\vertex[right=3cm of a1] (o1) ;
			\vertex[below=1cm of i1] (i2) ;
			\vertex[right=1cm of i2] (a2);
			\vertex[right=3cm of a2] (o2) ;
			
			%\vertex[below=2cm of i3] (t1);
			\vertex[above=1cm of a1] (t2);
			\vertex[above=2cm of a2] (t3);
			
			%\vertex[above right =1.3cm of t2] (t1);
			\vertex[above=1.5cm of b1] (i3) {$T^{\mu\nu}$};
			
			\diagram* { 
				(i1)  -- [fermion] (a1) -- [fermion] (b1)   --[fermion] (o1),
				
				(i2)  -- [anti fermion] (a2)  --[anti fermion] (o2),

				(i3)  -- [graviton] (b1),
				(a1)-- [gluon] (a2)
			};
			
			\node[] at (-0.35,-0.45) {$\pi\ $};
			\node[] at (4.46,-0.45) {$\pi\ $};

			\filldraw[fill=black, fill opacity=0.1] (-0.42,-0.5) ellipse (.45cm and 1.3cm);
			\filldraw[fill=black, fill opacity=0.1] (4.42 ,-0.5) ellipse (.45cm and 1.3cm);
			
			\draw[double, double distance=1.5,line width=1.2, postaction={decorate, decoration={
					markings,% switch on markings
					mark=at position 0.8 with  {\arrow{ latex} } } } ] (-1.8,-0.5) -- node [above] {} (-0.9,-0.5);
			
			\draw[double, double distance=1.5,line width=1.2, postaction={decorate, decoration={
					markings,% switch on markings
					mark=at position 0.8 with  {\arrow{ latex} } } } ] (4.9,-0.5) -- node [above] {} (5.85,-0.5);
		\end{feynman}
	\end{tikzpicture}
	\hspace{1cm}
	\begin{tikzpicture}
		\begin{feynman}
			\vertex (i1);
			\vertex[right=1cm of i1] (a1);
			\vertex[right=3cm of a1] (o1) ;
			\vertex[below=1cm of i1] (i2) ;
			\vertex[right=3cm of i2] (a2);
			\vertex[right=1cm of a2] (o2) ;
			
			%\vertex[below=2cm of i3] (t1);
			\vertex[above=1cm of a1] (t2);
			\vertex[above=2cm of a2] (t3);
			\vertex[blob] (t1) at (2,1.3) {{$TJJ$}};
			%\vertex[above right =1.3cm of t2] (t1);
			\vertex[above=1.5cm of t1] (i3) {$T^{\mu\nu}$};
			
			\diagram* { 
				(i1)  -- [fermion] (a1)   --[fermion] (o1),
				
				(i2)  -- [anti fermion] (a2)  --[anti fermion] (o2),

				(i3)  -- [graviton] (t1),
				(t1) -- [gluon] (a1) ,
				(t1) -- [gluon] (a2),
			};
			
			\node[] at (-0.35,-0.45) {$\pi\ $};
			\node[] at (4.46,-0.45) {$\pi\ $};

			\filldraw[fill=black, fill opacity=0.1] (-0.42,-0.5) ellipse (.45cm and 1.3cm);
			\filldraw[fill=black, fill opacity=0.1] (4.42 ,-0.5) ellipse (.45cm and 1.3cm);
			
			\draw[double, double distance=1.5,line width=1.2, postaction={decorate, decoration={
					markings,% switch on markings
					mark=at position 0.8 with  {\arrow{ latex} } } } ] (-1.8,-0.5) -- node [above] {} (-0.9,-0.5);
			
			\draw[double, double distance=1.5,line width=1.2, postaction={decorate, decoration={
					markings,% switch on markings
					mark=at position 0.8 with  {\arrow{ latex} } } } ] (4.9,-0.5) -- node [above] {} (5.85,-0.5);
		\end{feynman}
	\end{tikzpicture}
	\caption{Leading (left) and NLO contributions to the GFF of the pion.}
	\label{fig:1aaa}
\end{figure}

\subsection{The GFF-DVCS relation}
Gravitational form factors (GFFs) are related to deeply virtual Compton scattering (DVCS) of an electron (e) off a nucleon ($N$) ($eN\to e^\prime N^\prime \gamma$) with a final state photon. 
DVCS is a process where a high-energy electron scatters off a hadron (such as a proton or pion) by exchanging a virtual photon, which subsequently emits a real photon. An analysis of the process with other neutral currents is also possible \cite{Amore:2004ng}. \\
The process interpolates kinematically between the soft region, where a description in terms of QCD sum rules is possible \cite{Coriano:1993mr} and the process is dominated by the Feynman mechanism (overlap of intitial and final state hadron wavefunctions), and the inelastic region at higher energy \cite{Coriano:1998ge} \cite{Sterman:1997sx}.\\
A way to access EMT form factors is with GPDs
\cite{Ji:1996ek,Radyushkin:1996nd,Radyushkin:1996ru,
	Ji:1996nm,Collins:1996fb,Radyushkin:1997ki,Vanderhaeghen:1998uc},
which describe hard-exclusive reactions, such as deeply virtual 
Compton scattering (DVCS) $eN\to e^\prime N^\prime \gamma$ sketched 
in Fig.~\ref{dvcsx} or hard exclusive meson production 
$eN\to e^\prime N^\prime M$. In the case of the nucleon,
the second Mellin moments of unpolarized GPDs yield 
the EMT form factors $A$, $B$ and $D$
\begin{align}
	\int_{-1}^1{\rm d}x\;x\, H^a(x,\xi,t) = A^a(t) + \xi^2 D^a(t) \,, \quad \quad
	\int_{-1}^1{\rm d}x\;x\, E^a(x,\xi,t) = B^a(t) - \xi^2 D^a(t) \,.
	\label{Eq:GPD-Mellin}
\end{align}
$H$ and $E$ parameterize light-cone amplitudes with nonforward kinematics, describing the amplitude for removing from the nucleon
a parton carrying the fraction $x-\xi$ of the average momentum $P$ and 
reinserting back in the nucleon with a fraction $x+\xi$ on the light-cone.
In the process, the nucleon receives the momentum transfer $\Delta$, with $\xi$ representing a second scaling variable. 
Through their moments, this information provides insights into the GFFs.\\
The significance of GFFs in delineating hadron structure has spurred dedicated experimental endeavors, leading to the initial determinations of proton quark~\cite{Burkert:2018bqq} and gluon~\cite{Duran:2022xag} GFFs through measurements involving deeply virtual Compton scattering and $J/\psi$ photoproduction, respectively. Progress towards discerning the pion GFFs has been more constrained, with the first phenomenological constraints of the pion quark GFFs obtained from data recorded by the Belle experiment at KEKB \cite{Belle:2015oin,Savinov:2013hda,Kumano:2017lhr}. Anticipated advances in various hadron GFF determinations are expected from ongoing and forthcoming facilities such as the JLab 12 GeV program~\cite{JeffersonLabHallA:2022pnx,CLAS:2022syx} and the Electron-Ion Collider (EIC)~\cite{AbdulKhalek:2021gbh}.\\

\begin{figure}
	\centering
	\begin{tikzpicture}
		\begin{feynman}
			\vertex (i1);
			\vertex[right=1cm of i1] (a1);
			\vertex[right=1cm of a1] (b1);
			\vertex[right=3cm of a1] (o1) ;
			
			\vertex[below=1cm of i1] (i2) ;
			\vertex[right=1cm of i2] (a2);
			\vertex[right=2cm of a2] (b2);
			\vertex[right=3cm of a2] (o2) ;
			
			\vertex[below=1cm of i2] (i4) ;
			\vertex[right=1cm of i4] (a4);
			\vertex[right=2cm of a4] (b4);
			\vertex[right=3cm of a4] (o4) ;
			
			%\vertex[below=2cm of i3] (t1);
			\vertex[above=1cm of a1] (t2);
			\vertex[above=2cm of a2] (t3);
			
			%\vertex[above right =1.3cm of t2] (t1);
			\vertex[above=1.5cm of t1] (i3) {$T^{\mu\nu}$};
			
			\diagram* { 
				(i1)  -- [fermion] (a1) -- [fermion] (b1)   --[fermion] (o1),
				
				(i2)  -- [fermion] (a2)-- [fermion] (b2)  --[fermion] (o2),
				(i4)  -- [fermion] (b4)  --[fermion] (o4),

				(i3)  -- [graviton] (b1),
				(a1)-- [gluon] (a2),
				(b2)-- [gluon] (b4)
			};
			
			\node[] at (-0.4,-1.) {$P $};
			\node[] at (4.46,-1.) {$P$};

			\filldraw[fill=black, fill opacity=0.1] (-0.435,-1.) ellipse (.48cm and 1.9cm);
			\filldraw[fill=black, fill opacity=0.1] (4.435 ,-1.) ellipse (.48cm and 1.9cm);
			
			\draw[double, double distance=1.5,line width=1.2, postaction={decorate, decoration={
					markings,% switch on markings
					mark=at position 0.8 with  {\arrow{ latex} } } } ] (-1.8,-1.) -- node [above] {} (-0.9,-1.);
			
			\draw[double, double distance=1.5,line width=1.2, postaction={decorate, decoration={
					markings,% switch on markings
					mark=at position 0.8 with  {\arrow{latex} } } } ] (4.9,-1.) -- node [above] {} (5.85,-1.);
		\end{feynman}
	\end{tikzpicture}
	\hspace{1cm}
	\begin{tikzpicture}
		\begin{feynman}
			\vertex (i1);
			\vertex[right=1cm of i1] (a1);
			\vertex[right=1cm of a1] (b1);
			\vertex[right=3cm of a1] (o1) ;
			
			\vertex[below=1cm of i1] (i2) ;
			\vertex[right=1cm of i2] (a2);
			\vertex[right=2cm of a2] (b2);
			\vertex[right=1cm of a2] (c2);
			\vertex[right=3cm of a2] (o2) ;
			
			\vertex[below=1cm of i2] (i4) ;
			\vertex[right=1cm of i4] (a4);
			\vertex[right=2cm of a4] (b4);
			\vertex[right=1cm of a4] (c4);
			\vertex[right=3cm of a4] (o4) ;
			
			%\vertex[below=2cm of i3] (t1);
			\vertex[above=1cm of a1] (t2);
			\vertex[above=2cm of a2] (t3);
			
			%\vertex[above right =1.3cm of t2] (t1);
			\vertex[above=1.5cm of t1] (i3) {$T^{\mu\nu}$};
			\vertex[blob] (t1) at (2,1.3) {{$TJJ$}};
			
			\diagram* { 
				(i1)  -- [fermion] (a1)  --[fermion] (o1),
				
				(i2)  -- [fermion] (c2)-- [fermion] (b2)  --[fermion] (o2),
				(i4)  -- [fermion] (c4)  --[fermion] (o4),

				(i3)  -- [graviton] (t1),
				(a1)-- [gluon] (t1),
				(b2)-- [gluon] (t1),
				(c2)-- [gluon] (c4),
			};

			\node[] at (-0.4,-1.) {$P $};
			\node[] at (4.46,-1.) {$P$};

			\filldraw[fill=black, fill opacity=0.1] (-0.435,-1.) ellipse (.48cm and 1.9cm);
			\filldraw[fill=black, fill opacity=0.1] (4.435 ,-1.) ellipse (.48cm and 1.9cm);
			
			\draw[double, double distance=1.5,line width=1.2, postaction={decorate, decoration={
					markings,% switch on markings
					mark=at position 0.8 with  {\arrow{ latex} } } } ] (-1.8,-1.) -- node [above] {} (-0.9,-1.);
			
			\draw[double, double distance=1.5,line width=1.2, postaction={decorate, decoration={
					markings,% switch on markings
					mark=at position 0.8 with  {\arrow{latex} } } } ] (4.9,-1.) -- node [above] {} (5.85,-1.);
			
		\end{feynman}
	\end{tikzpicture}
	\caption{Typical leading (left) and NLO contributions (right) to the GFF of the proton. }
	\label{fig:2aaa}
\end{figure}
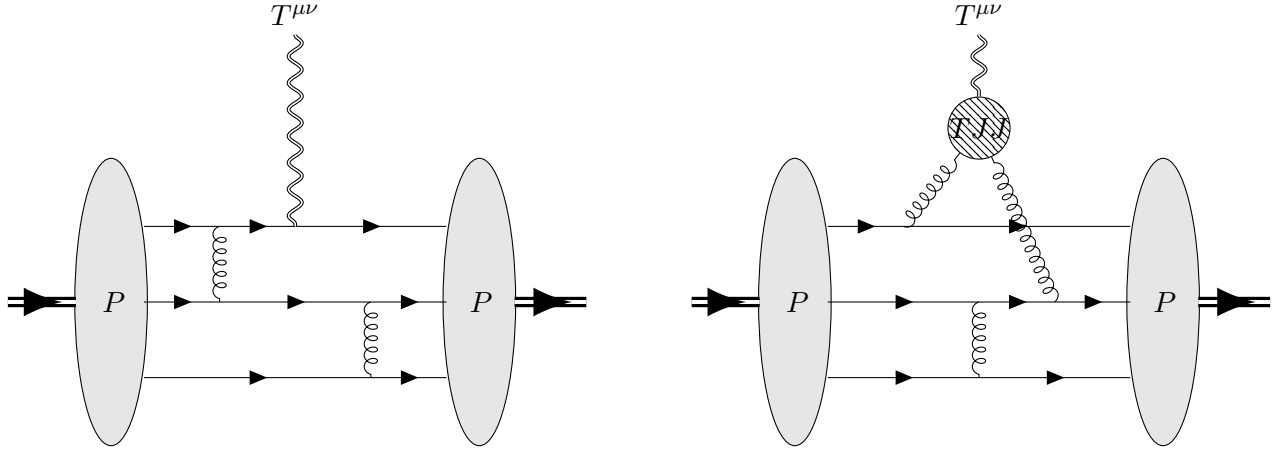

\subsection{The $TJJ$ from factorization and the conformal anomaly}
At sufficiently large momentum  transfer, the GFF is described by a factorization formula, with an insertion of the $TJJ$ vertex, which is at the center of our investigation. Also in this case one can resort to an ordinary collinear factorization, using distribution amplitudes, or to a modified factorization with the inclusion of Sudakov effects, as discussed in the case of the electromagnetic form factor of the proton \cite{Li:1992nu}. \\
The light-front Fock state expansion provides a natural representation of hadronic structures~\cite{Brodsky:1997de}. This expansion formally consists of an infinite series of Fock states, each characterized by distinct partonic configurations and their associated light-front wave functions. However, for exclusive processes involving large momentum transfers, general power counting arguments indicate that the dominant contributions arise from the lowest Fock states, those with the minimal number of partons and the least orbital angular momentum \cite{Matveev:1973ra,Brodsky:1973kr} \cite{Ji:2003fw}. 
In the factorization approach, we are going to show, before our conclusions, to briefly illustrate how the leading three-valence-quark Fock state of the proton proceeds \cite{Ji:2002xn,Ji:2003yj}, following the steps indicated in \cite{Tong:2022zax}.   \\
In general, hadronic and partonic matrix elements are connected by hadronic wave functions integrated over the fractional momenta of the hard scattering amplitude. For example, in the case of the electromagnetic form factor of a pion of initial momentum $p$ and final momentum $p'$ \cite{Sterman:1997sx}
\beq
(p' + p)_\mu F_{\pi}(q^2) = \langle \pi(p') | J_\mu(0) | \pi(p) \rangle ,
\eeq
which is the simplest case, the corresponding form factor $F_{\pi}(q^2)$ is expressed at large $q^2$ in the form 
\beq
F_{\pi}(q^2) = \int_0^1 dx \, dy \, \phi_{\pi}(y, \mu^2) \, T_H(y, x, q^2, \mu^2) \, \phi_{\pi}(x, \mu^2)
\eeq
$J^\mu$ is the electromagnetic current of the valence quarks in the pion. 
In the collinear factorization limit, the pion wave function \( \phi_{\pi}(y, \mu^2) \) describes the distribution of fractional momenta \( (x, y) \) carried by the valence \( q/\bar{q} \) pair within the pion, in its lowest Fock state. Here, \( \mu \) represents the factorization scale, whose evolution is governed by the renormalization group equation, specifically the Efremov-Radyushkin-Brodsky-Lepage (ERBL) evolution equation~\cite{Lepage:1980fj, Efremov:1979qk}.  \\
For a gravitational form factor (GFF), a similar analysis can be performed. However, identifying the anomaly contribution in the associated hard scattering process, denoted as \( T_H \), is significantly more complex. A detailed discussion of this issue will be presented in Section 11.
The process is depicted in Fig. \ref{fig:1aaa} for the pion and in Fig. \ref{fig:2aaa} for the proton. 
The amplitude is interpolated by the $TJJ$, due to the insertion at NLO of this correlator on the hard scattering amplitude. A more recent analysis of radiative corrections to the  the trace anomaly has been presented in \cite{Hatta:2018sqd}. \\
The emergence of a dilaton pole in this vertex was pointed out in QED and QCD in 
\cite{Giannotti:2008cv,Armillis:2009pq} \cite{Armillis:2010qk}. 
This point will be examined in greater detail in the next section, where we first review the result for the $TJJ$ computed in QCD with on-shell gluons before discussing its sector decomposition. Our decomposition is motivated by the $CFT_p$ approach, which effectively resolves the conformal constraints. Following the method introduced in \cite{Bzowski:2013sza}, this provides a highly economical parameterization of the vertex. The anomaly contribution arises at NLO, with the two gluons in the $TGG$ (graviton/gluon/gluon) configuration attaching to the hard scattering process in all possible ways. In contrast, the non-anomalous contributions begin at $O(\alpha_s)$ and extend to $O(\alpha_s^2)$.  
\begin{figure}[t]
	{\par\centering \resizebox*{12cm}{!}{\includegraphics{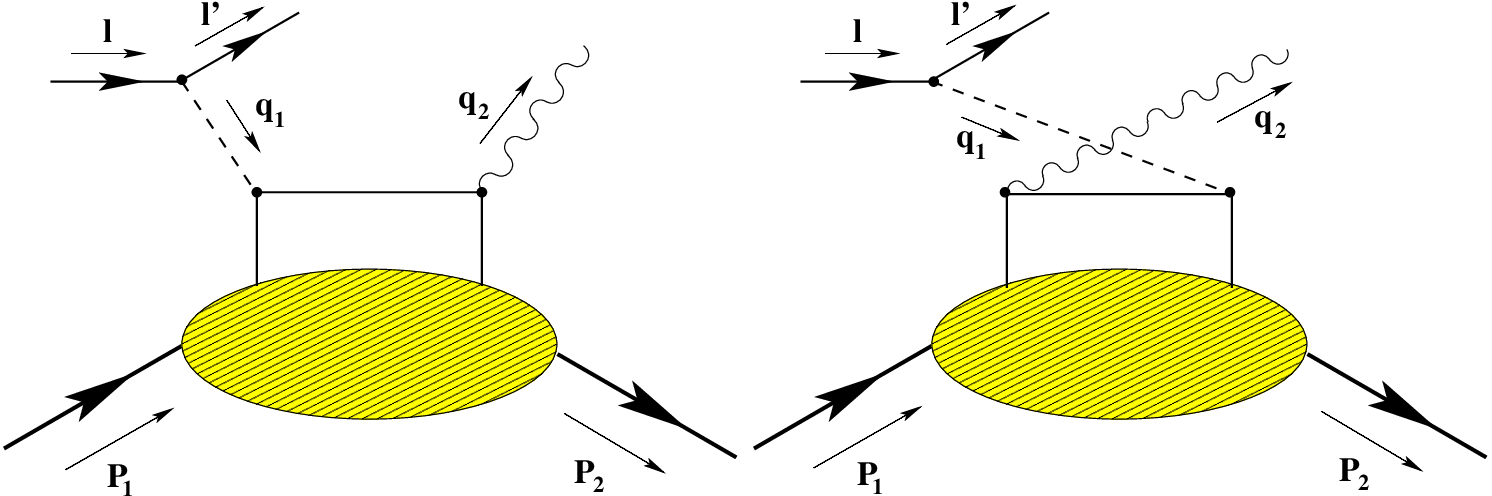}} \par}
	\caption{Leading hand-bag diagrams for the DVCS process}
	\label{DVCS_1.eps}
	\label{dvcsx}
\end{figure}

\section{The QCD gravitational effective action and the conformal limit}
\label{three1}
To derive the conformal constraints for the $TJJ$ correlation function, we employ the general formalism of the gravitational effective action, examining the partition function within gravitational and gauge field backgrounds.\\
In QCD, the presence of the gluon sector and the gauge-fixing procedure requires treating the quark and gluon sectors separately from a CFT perspective, with each sector following its own hierarchy of conformal constraints at the one-loop level. Although this separation is no longer valid at higher orders in $\alpha_s$ within the correlator's expansion, it suffices for our current analysis.\\
The anomaly in the trace of the stress-energy tensor and the appearance of the dilaton pole stem from the renormalization of the effective action, which involves distinct counterterm contributions from the quark and gluon/ghost sectors.\\
The quark sector of the correlator behaves similarly to the Abelian case, with its hierarchical identities proportional to $n_f$, the number of fermions running in the loops. Contributions from this sector comply with standard vector Ward identities on the external gluon lines. The CWIs in this sector are analogous to those in the Abelian scenario and are only broken by the anomaly.\\
Within this sector, the CWIs can be addressed both non-perturbatively by directly solving the associated differential equations and perturbatively, as we will illustrate. Reconstructing the quark contribution to the correlator can be achieved using standard CFT methods in momentum space, employing a sector decomposition. However, a similar analysis for the gluon sector is more challenging due to gauge-fixing conditions and the presence of Slavnov-Taylor identities (STIs), which replace the standard WIs. Consequently, the gluon sector is treated by us perturbatively, and its contributions are organized according to the same tensor decomposition of the quark sector, from which the dilaton state naturally emerges as a combination of the trace anomaly from both sectors.

\subsection{The action in a weak gravitational field }
For the derivation of all the relevant WIs we will be relying on the formalism of the gravitational effective action, that we are going to investigate in this and in the next sections. 
In the case of a non-Abelian gauge theory the action is given by
\bea
S_0[g,A,\psi]&=&-\frac{1}{4}\int d^4 x \sqrt{-g}F_{\mu\nu}^{{a}} F^{\mu\nu a} +\int d^4 x \sqrt{-g}\, i \bar{\psi}\gamma^\mu D_\mu \psi \nn\\
S_1[g,A]&=&\int  \sqrt{-g}J_\mu^{a} A^{\mu a}
\eea 
with
\bea
F_{\mu\nu}^{a} &=&\nabla_\mu A_{\nu}^{a} - \nabla_{\nu} A_{\mu}^{ a} + g f^{a  b  c} A_\mu^{ b} A_\nu^c\nonumber \\
&=&\partial_\mu A_{\nu}^{a} - \partial_{\nu} A_{\mu}^{ a} + g f^{a  b  c} A_\mu^{ b} A_\nu^c, \nonumber \\
\nabla_\mu A^{\nu a} &=&\partial_\mu A^{\nu a} +\Gamma_{\mu\nu}^\lambda A^{\lambda a}
\eea
with $J^{\mu  a}=g_c\bar{\psi}\gamma^\mu T^{a}\psi$ denoting the fermionic current, with $T^a$ the generators of the theory and $\nabla_\mu$ denoting the covariant derivative in the curved background on a vector field. The local Lorentz and gauge covariant derivative $(D)$ on the fermions acts via the spin connection $\omega$
\be
D_\mu \psi=\left( \partial_\mu \psi  + A_\mu^a T^a +\frac{1}{4}\omega_{\mu}^{\underline{a} \underline{b}}\sigma_{
	\underline a \underline b}\right)\psi
\ee
with $\sigma^{\ul{a}\ul{b}}=1/2 [\gamma^{\ul{a}},\gamma^{\ul{b}}]$,
having denoted with $\underline{a}\underline{b}$ the local Lorentz indices.  A local Lorentz covariant derivative $(D)$ can be similarly defined for a vector field, say $V^{\underline{a}}$, via the Vielbein $e^{\underline{a}}_{\, \,\mu}$ and its inverse  $e^{\,\,\mu}_{\underline a}$
\be
D_\mu V^{\underline a}= \partial_\mu V^{\underline a} +\omega^{\underline a}_{\mu \underline b} V^{\underline b}
\ee
with 
\be
\nabla_\mu V^\rho = e_{\underline a}^{\,\,\rho} D_\mu V^{\underline a}.
\ee
The Christoffel and the spin connection related via the holonomic relation 
\be
\Gamma^\rho_{\mu\nu}=e^{\,\,\rho}_{\ul a}\left( \partial_\mu e^{\ul a}_{\,\,\nu} +\omega^{\ul a}_{\mu \ul b} e^{\ul b}_{\,\,\nu}\right).
\ee
with
\beq
\omega_\mu^{\ul{a} \ul{b}}(x) =e_{\underline{a}}^{\,\nu}(x)e_{\underline{b}\nu;\mu}(x)\, ,
\eeq
where we have introduced the vielbein $e_{\underline{a}}^\mu(x)$,  
giving a Lagrangian of the form
\beqa \mathcal{L}_f& = & \sqrt{-g} \bigg\{\frac{i}{2}\bigg[\bar\psi\g^\mu(\mathcal{D}_\mu\psi)
- (\mathcal{D}_\mu\bar\psi)\g^\mu\psi \bigg] - m\bar\psi\psi\bigg\}       \eeqa
for the fermion sector. The gauge fixing and ghost sectors are given by 
\beq
\mathcal{L}_{\text{gf}} = -\frac{1}{2\xi} \, g^{\mu\nu} (\nabla_\mu A_\nu^a)^2\,, \qquad \mathcal{L}_{\text{ghost}} = \bar{c}^a \, \left( - g^{\mu\nu} \nabla_\mu D_\nu^{ab} \right) c^b,  \qquad \eeq
with
\beq
D_\nu^{ab} = \delta^{ab} \nabla_\nu + g f^{abc} A_\nu^c
\eeq
the gauge and diffeomorphism covariant derivative. \\
A symmetric stress energy tensor can be defined as 
\beq
T^{\mu\nu}=\frac{2}{\sqrt{-g}} \frac{\delta Z}{\delta g_{\mu\nu}}, 
\eeq
starting from the QCD partition function 
\bea
&& Z[J,\eta,\bar\eta,\chi,\bar\chi,g] =\mathcal{N}
\int \mathcal D A \, \mathcal D \psi \, \mathcal D \bar\psi \,
\mathcal D c \, \mathcal D \bar c \, \exp\bigg\{ i \int \sqrt{-g} d^4
x \left( \mathcal{L} + J_{\mu}^a A^{\mu  a} \right. \nn \\
&&\left. \hspace{6cm}   + \bar \eta \psi +
\bar \psi \eta + \bar \chi c + \bar c \chi \right)\bigg\}.
\eea
with a metric $g$ in the background. $J,\eta,\bar\eta,\chi,\bar\chi$ are the
sources of the gauge field $A$, of the fermion and antifermion fields ($\bar{\eta}, \eta$) and of the ghost and antighost fields ($\bar{\chi}$, $\chi$) respectively. \\
$\mathcal{L}$ is the standard QCD action, that a linearized level 
generates the term 
\beq
\label{Lgrav} \mathcal{L}_{grav}(x) = -\frac{\kappa}{2}T^{\mu\nu}(x)h_{\mu\nu}(x), 
\eeq 
with $h_{\mu\nu}$, describing the linear expansion of the metric around a flat spacetime.  
We use the convention $\eta_{\mu\nu}=(1,-1,-1,-1)$ for the metric in flat spacetime, parameterizing its deviations from the flat case as
\beq\label{QMM} g_{\mu\nu}(x) = \h_{\mu\nu} + \kappa \, h_{\mu\nu}(x)\,,\eeq
with the symmetric rank-2 tensor $h_{\mu\nu}(x)$ accounting for its fluctuations. We have set $\kappa^2 =16 \pi G_N$, with $G_N$ the gravitational constant.We set $\kappa\to 1$. \\

\subsection{The stress energy tensor of the gauge-fixed action}
The stress energy tensors obtained by adding the components of the field strength $(fs)$, fermionic, gauge fixing and ghost sectors of the Lagrangian  
\bea
T_{\mu\nu} = T^{f.s.}_{\mu\nu} + T^{ferm.}_{\mu\nu} + T^{g.fix.}_{\mu\nu} + T^{ghost}_{\mu\nu},
\eea
with
\beq
T^{f.s.}_{\mu\nu}
= 
\eta_{\mu\nu}\frac{1}{4}F^a_{\r\s}F^{a\,\r\s} -  F^a_{\mu\r}F^{a\,\r}_\nu  
\eeq

\bea
T^{g.f.}_{\mu\nu} &=& {1 \over \xi}\left[A_\nu^a \partial_\mu(\partial \cdot A^a) +A_\mu^a \partial_\nu(\partial \cdot A^a)\right] -{1 \over \xi}g_{\mu \nu}
\left[- \frac{1}{2} (\partial \cdot A)^2 + \partial^\rho (A_\rho^a \partial \cdot A^a)\right], \\
T^{gh}_{\mu\nu} &=& \partial_{\mu}\bar c^a D^{ab}_{\nu} c^b + \partial_{\nu}\bar c^a D^{ab}_{\mu}c^b - g_{\mu\nu} \partial^{\rho}\bar c^a D^{ab}_{\rho}c^b.
\label{gfghost}
\eea
After the inclusion of the fermion sector, the final expression takes the form

\bea
T_{\mu \nu} &=& -g_{\mu \nu} {\mathcal L}_{QCD}
-F_{\mu \rho}^a F_\nu^{a \rho} -{\frac{1} {\xi}}g_{\mu \nu}
\partial^\rho (A_\rho^a \partial^\sigma A_\sigma^a) +{\frac{1}{\xi}}(A_\nu^a \partial_\mu(\partial^\sigma A^a_\sigma)
+A_\mu^a \partial_\nu(\partial^\sigma A_\sigma^a))
\nonumber\\
&+& {\frac{i}{4}} \Big[
\overline \psi \gamma_\mu (\overrightarrow \partial_\nu
-i g T^a A_\nu^a)\psi  -\overline \psi (\overleftarrow \partial_\nu
+i g T^a A_\nu^a)\gamma_\mu\psi
+\overline \psi \gamma_\nu (\overrightarrow \partial_\mu
-i g T^a A_\mu^a)\psi
\nonumber\\
&-& \overline \psi (\overleftarrow \partial_\mu
+i g T^a A_\mu^a)\gamma_\nu\psi \Big] +\partial_\mu \overline
c^a (\partial_\nu c^a -g f^{abc} A_\nu^c  c^b)
+\partial_\nu \overline c^a (\partial_\mu c^a -g f^{abc}
A_\mu^c c^b).
\label{EMT}
\eea
The diagrams relevant in the perturbative analysis of this correlator are shown in Fig. \ref{expansionaaa}.

\subsection{The covariant expansion and the nonlocal anomaly action}
In order to derive the effective action we introduce the logarithm of the partition function that collects all the connected Green's functions 
\beq
W[J,\eta,\bar\eta,\chi,\bar\chi,h] =-i \log{Z[J,\eta,\bar\eta,\chi,\bar\chi,h]}
\eeq
(normalized to the vacuum functional) and the effective action, the generating functional $\mathcal{S}_{eff}$ of the 1-particle irreducible and truncated amplitudes. This is defined by a Legendre transform of $W$ with respect to all the non-gravitational sources
\bea
\mathcal{S}_{eff}[A_{(c)},\bar \psi_{(c)}, \psi_{(c)},
\bar c_{(c)}, c_{(c)}, h] = W[J,\eta,\bar\eta,\chi,\bar\chi,h] -
\int d^4 x \left( J_{\mu}A^{\mu}_c + \bar \eta \psi_c + \bar \psi_c \eta
+ \bar \chi c_c + \bar c_c \chi \right).
\label{1PIfunctional}
\eea
with
\bea
A^{\mu}_{(c)} = \frac{\delta W }{\delta J_{\mu}}, \qquad
\psi_{(c)} = \frac{\delta W }{\delta \bar \eta}, \qquad \bar \psi_{(c)} =
\frac{\delta W }{\delta \eta}, \qquad \omega_{(c)} = \frac{\delta W
}{\delta \bar \chi}, \qquad \bar \omega_{(c)} = \frac{\delta W }{\delta
	\chi} \label{Legendre1}
\eea
being the classical external source fields. $\mathcal{S}_{eff}$ satisfies the equations

\bea
\frac{\delta\mathcal{S}_{eff}}{\delta A^{\mu}_{(c)}} = - J_{\mu}, \qquad
\frac{\delta\mathcal{S}_{eff}    }{\delta \psi_{(c)}} = - \bar \eta, \qquad
\frac{\delta\mathcal{S}_{eff}}{\delta \bar \psi_{(c)}} = - \eta, \qquad
\frac{\delta\mathcal{S}_{eff}}{\delta \omega_{(c)}} = - \bar \chi, \qquad
\frac{\delta\mathcal{S}_{eff}}{\delta \bar \omega_{(c)}} = - \chi,
\label{Legendre2}
\eea

\bea
\frac{\delta\mathcal{S}_{eff}}{\delta h_{\mu\nu}} = \frac{\delta W}{\delta h_{\mu\nu}}. \label{Legendre3}
\eea
The vev of the stress energy tensor in a gravitational background $g_{\mu\nu}$ will be simply denoted 
as the functional derivative of $\mathcal{S}_{eff}$
\be
\langle T^{\mu\nu}\rangle=\frac{2}{\sqrt{g}}\frac{\delta \mathcal{S}_{eff}}{\delta g_{\mu\nu}}. 
\ee
The covariant expansion of $\mathcal{S}_{eff}$ around a background metric $\bar{g}$ and 
vanishing gauge field $A_{c\mu}$, with fluctuations  $\delta g$ and $\delta A_{c\,\mu}$ takes the form
\begin{eqnarray}
	\label{exps2}
	\mathcal{S}_{eff}(g,A_c) &\equiv&\sum_{n,
		k}^\infty \frac{1}{2^{n+ k} (n+k)!} \int d^d x_1\ldots d^d x_{n+k} \sqrt{g_1}\ldots \sqrt{g_n}\,\langle T^{\mu_1\nu_1}\ldots \,T^{\mu_n\nu_n} J^{a_{n+1}}_{\mu_{n+1}}\ldots J^{a_{n+k}}_{\mu_{n + k}}\rangle_{\bar{g} }\nonumber\\
	&&\times \delta g_{\mu_1\nu_1}(x_1)\ldots \delta g_{\mu_n\nu_n}(x_n)\delta A^{a_{n+1}}_{c\,\, \mu_{n+1}}(x_{n+1}) 
	\ldots \delta A^{a_{n+k}}_{c\,\, \mu_{n+k}}(x_{n+k}) +  GT \ldots
\end{eqnarray}
where the dots refer to contributions from the  pure gauge sector in the expansion, and $GT$ refers to the purely gravitational terms (i.e. with multiple insertions of only stress-energy tensors). In the future expressions we will remove the subscript "c" from the classical external source $A_{c \,\mu}^a$ for convenience. \\
The covariant normalization of the correlation functions is given by
\begin{equation}
	\label{exps1}
	\langle T^{\mu_1\nu_1}(x_1)\ldots T^{\mu_n\nu_n}(x_n)\rangle \equiv\frac{2}{\sqrt{g_1}}\ldots \frac{2}{\sqrt{g_n}}\frac{\delta^n \sm_{eff}(g,A)}{\delta g_{\mu_1\nu_1}(x_1)\delta g_{\mu_2\nu_2}(x_2)\ldots \delta g_{\mu_n\nu_n}(x_n)} 
\end{equation}
for the $n$-graviton sector, with $\sqrt{g_1}\equiv \sqrt{|\textrm{det} \, g_{{\mu_1 \nu_1}}(x_1) |} $ and so on, while mixed correlators will be defined as 
\begin{eqnarray}
	\label{exps1}
	\langle T^{\mu_1\nu_1}(x_1)\ldots T^{\mu_n\nu_n}(x_n)J^{a_{n+1} \ \mu_{n+1}}(x_{n+1})\ldots J^{a_{n+k} \, \mu_{n+k}}(x_{n+k})\rangle & \equiv & \nonumber\\
	\frac{2}{\sqrt{g_1}}\ldots \frac{2}{\sqrt{g_n}}\frac{\delta^n \mathcal{S}_{eff}(g,A)}{\delta g_{\mu_1\nu_1}(x_1)\ldots \delta g_{\mu_n\nu_n}(x_n) \, \delta  A^{a_{n+1}}_{\mu_{n+1}} (x_{n+1}) \ldots \,  \delta A^{a_{n+k}}_{\mu_{n+k}} (x_{n+k}) } && \nonumber \\
\end{eqnarray}
for the graviton/gauge sector. 
In the renormalization at one-loop of correlators like $\langle TTT\rangle_{even}$, two counterterms are required: $V_E/\epsilon$ and $V_{C^2}/\epsilon$. These are defined as 

\begin{align}
	\label{ffr}
	V_{E}(g,d) \equiv & \mu^{\epsilon} \int d^dx \, \sqrt{-g} \, E, \quad \quad \epsilon = d-4, \notag \\ 
	V_{C^2}(g,d) \equiv & \mu^{\epsilon} \int d^dx \, \sqrt{-g} \, C^2.
\end{align}
Here, $E$ is an evanescent term, the Gauss-Bonnet term, being topological at $d=4$, yet its inclusion is necessary in order for $\mathcal{S}_{eff}$  to satisfy the Wess-Zumino consistency condition. 
$C^2$ denotes the square of the Weyl tensor.\\
The anomaly arises in dimensional regularization (DR) due to the failure of these two counterterms to remain invariant under Weyl transformations  in $d$ dimensions 

\begin{equation} \label{ren}
	\begin{aligned}
		&2 g_{\mu\nu}\frac{\delta}{\delta g_{\mu\nu}}V_{E}=\frac{\delta}{\delta \phi}V_{E}=\epsilon \sqrt{g} E,\\
		&2 g_{\mu\nu}\frac{\delta}{\delta g_{\mu\nu}}V_{C^2}=\frac{\delta}{\delta \phi}V_{C^2}=\epsilon \sqrt{g} \left[ C^2 + \frac{2}{3} \square R \right]
	\end{aligned}
\end{equation}
$(\epsilon=d-4)$.
We have defined the square of the Weyl tensor at $d=4$
\beq
C^2= C_{\lambda\mu\nu\rho}C^{\lambda\mu\nu\rho} = R_{\lambda\mu\nu\rho}R^{\lambda\mu\nu\rho}
-2 R_{\mu\nu}R^{\mu\nu}  + \frac{R^2}{3},
\eeq
while the Euler (Gauss-Bonnet) term is given by
\beq
E = ^*\hskip-.2cmR_{\lambda\mu\nu\rho}\,^*\hskip-.1cm R^{\lambda\mu\nu\rho} =
R_{\lambda\mu\nu\rho}R^{\lambda\mu\nu\rho} - 4R_{\mu\nu}R^{\mu\nu}+ R^2.
\eeq
Defining, alternatively 
\begin{align}
	(C^{(d)})^2&\equiv R^{\mu \nu \rho \sigma}R_{\mu \nu \rho \sigma}-\frac{4}{d-2}R^{\mu \nu }R_{\mu \nu}+\frac{2}{(d-2)(d-1)}R^2\notag \\
	&=(C^{(4)})^2+\frac{d-4}{d-2}\left(2R^{\mu \nu }R_{\mu \nu}-\frac{d+1}{3(d-1)}R^2\right)\notag 
\end{align}
to be the Weyl tensor in $d$ dimensions, the second variation in \eqref{ren} takes the form 
\begin{equation}
	\label{ren1}
	2g_{\mu \nu}\frac{\delta}{\sqrt{-g}\delta g_{\mu \nu}}\int d^d x  \sqrt{-g} (C^{(d)})^2=(d-4)(C^{(d)})^2,
\end{equation}
showing that $\Box R$ is prescription dependent, after a comparison of the second of \eqref{ren} with \eqref{ren1}.\\
In the renormalization of the $TJJ$ we will only need the gauge counterterm 
\begin{align}
	\label{ct}
	S_{count}\equiv V_{F^2}\equiv -\frac{1}{\varepsilon}\int d^dx\,\sqrt{-g}\left(\beta\,F^2\right),
\end{align}
corresponding to the field strength $F^2$, where the coefficient $\beta$ is the ordinary QCD $\beta$ function.\\
In the $TJJ$ the naive trace of $T^{\mu\nu}$  is rather simple, since naive scale invariance gives the traceless condition
\begin{equation}
	g_{\mu\nu}T^{\m\n}=0\label{trace},	
	\eeq	
	which at linear order in the gravitational fluctuations, after renormalization,  is modified at quantum level in the form
	\begin{equation}
		g_{\mu\nu} \langle T^{\mu\nu}\rangle=\beta F^{a\, \mu\nu}F^a_{\mu\nu}.
		\label{ann}
	\end{equation}
	Additional contributions to the trace anomaly, associated with $E$ and $C^2$, are captured only by extra insertions of stress energy tensors. 
	The $TJJ$ is part of the hierarchies of correlators containing such extra insertions (see for instance \cite{Coriano:2022jkn}).
	For correlators with multiple insertions of the stress energy tensor, the renormalization of the effective action leads to the inclusion of all the three counterterms $V_E,V_{C^2}$ and $V_{F^2}$, with the generation of a parity-even trace anomaly
	\bea
	g_{\mu\nu}\langle T^{\mu\nu}\rangle =   b \, C^2 + b^{\prime} \, \left( E - \frac{2}{3} \, \square \, R\right) + b'' \, \square \, R +  c\, F^{a \, \mu \nu} F^a_{\mu \nu},
	\label{var2}
	\eea
	where  $b$ and $b'$ are related to the massless content of the virtual contributions. 
	In the non-Abelian $TJJ$ case, the expansion of the counterterm \eqref{ct} is the only one needed in order to renormalize the 2- 3- and 4- gluons vertices which are part of the gauge invariant contributions to the non-Abelian trace anomaly. In the analysis of the hierarchy of the quark sector, only the $n_f$ dependence of \eqref{ct} will be relevant, while its $C_A=N_c=3$ (colour) dependence will affect the gluon sector. \\
	The anomaly content is gauge invariant and
	can be identified by the sector decomposition that  we are going to discuss in the next sections. \\
	The anomaly part of $\mathcal{S}_{eff}$, also termed "the anomaly induced action" and denoted as $\mathcal{S}_{anom}$, can be derived by solving the functional constraint in \eqref{var2} in $d=4$ dimensions in the nonlocal form 
	\bea
	&& \hspace{-.6cm}S_{anom}[g,A] = \label{Tnonl}\\
	&&\frac {1}{8}\int d^4x\sqrt{-g}\int d^4x'\sqrt{-g'} \left(E - \frac{2}{3} \square R\right)_x
	\Delta_4^{-1} (x,x')\left[ 2b\,F
	+ b' \left(E - \frac{2}{3} \square R\right) + 2\, c\, F_{\mu\nu}F^{\mu\nu}\right]_{x'} \nonumber
	\label{var1}
	\eea
	for a general field content of massless fields in the anomaly functional, parameterized by constants $b,b'$ and $c$. We have introduced the Green's functions of the quartic (Paneitz) operator
	\beq
	\Delta_4 \equiv  \nabla_\mu\left(\nabla^\mu\nabla^\nu + 2 R^{\mu\nu} - \frac{2}{3} R g^{\mu\nu}\right)
	\nabla_\nu = \square^2 + 2 R^{\mu\nu}\nabla_\mu\nabla_\nu +\frac{1}{3} (\nabla^\mu R)
	\nabla_\mu - \frac{2}{3} R \square\,
	\label{Deldef}
	\eeq
	which is conformally covariant.  
	Around flat space, \eqref{var1}  \cite{1984PhLB..134...56R} also known as the Riegert action, can be reformulated in the form
	\beq
	S_{anom}[g,A]  \rightarrow  -\frac{c}{6}\int d^4x\sqrt{-g}\int d^4x'\sqrt{-g'}\, R^{(1)}_x
	\, \square^{-1}_{x,x'}\, [F_{\alpha\beta}^aF^{a\, \alpha\beta}]_{x'}\,,
	\label{SSimple}
	\eeq
	which holds true to the first order in the metric fluctuations around a flat background. This form is expected  to emerge in QCD and is reproduced in the perturbative picture, as shown in explicit QED \cite{Giannotti:2008cv,Armillis:2009pq}  and QCD \cite{Armillis:2010qk} computations at lowest order. \\
	Indeed, in the case of the $TJJ$ in QED, which is the lowest order,  \eqref{SSimple} acquires the specific form 
	\beq
	\label{pole}
	\mathcal{S}_{pole}= - \frac{e^2}{ 36 \pi^2}\int d^4 x \, d^4 y \left(\square h(x) - \partial_\mu\partial_\nu h^{\mu\nu}(x)\right)  \square^{-1}_{x\, y} F_{\alpha\beta}(x)F^{\alpha\beta}(y),
	\eeq
	where the Ricci scalar is expanded at linear order. In the QCD case a similar relation is valid for on-shell gluons. The $\Box^{-1}$ pole contribution to the anomaly is correctly described by such nonlocal action up to 3-point functions. The consistency of \eqref{SSimple} in the generation of the correct hierarchy of correlators, 
	constrained by the corresponding CWIs up to 3-point functions, has been verified in free field theory realizations with scalars, fermions and Abelian spin-1 fields running in the loops. This check has been performed for the $TTT$ correlator \cite{Coriano:2017mux,Coriano:2018bsy}. \\
	However, an analysis of 4-point functions reveals that, in the Abelian case, a correlator like $TTJJ$, derived from \eqref{var1}, lacks certain Weyl-invariant terms. These terms are essential for maintaining the correct hierarchical structure to which the $TTJJ$ correlator belongs. Such additional terms have been identified in the $4T$ correlator, as shown in \cite{Coriano:2021nvn}. These findings were deduced by examining the one-loop free field theory realization of the same correlator.\\
	The analysis of \cite{Armillis:2010qk}, however shows that in the case of a 3-point function such as the $TJJ$, for on-shell gluons in QCD, the nonlocal action reproduces the perturbative expansion of the anomaly form factor. \\
	In the off-shell case, we are going to show, such previous analysis can be extended, and allow to completely characterize the same anomaly form factor using a sector decomposition. This allows one to separate the anomaly contribution from those terms which are proportional to the equations of motion for the external gluons.  

	\begin{figure}[t]
		\begin{center}
			\includegraphics[scale=0.15]{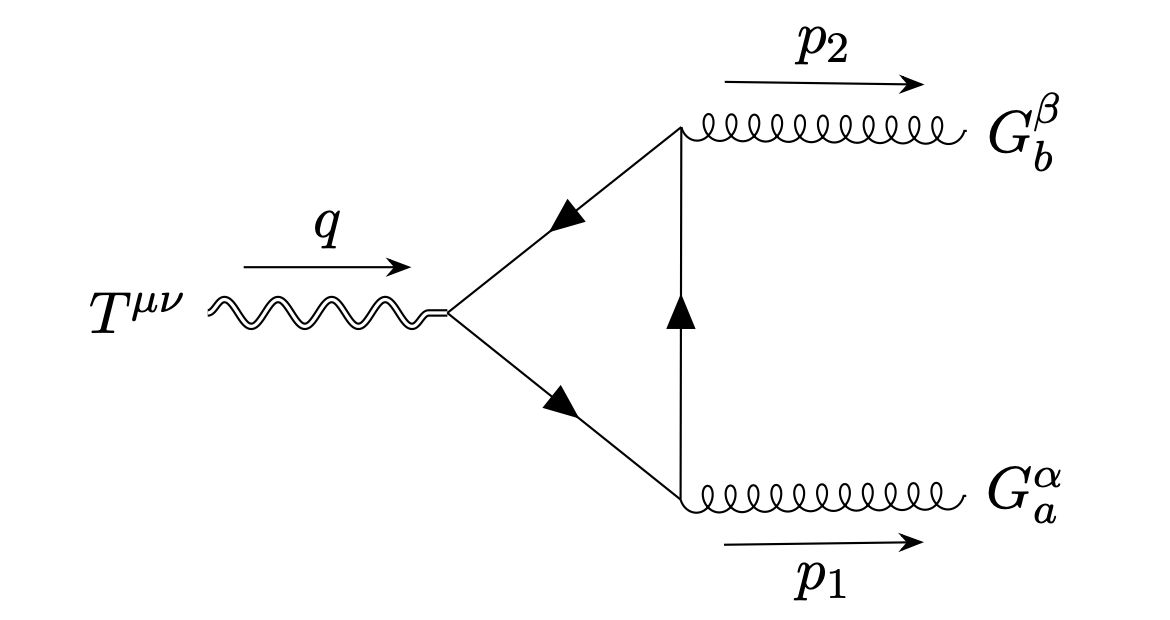}
			\includegraphics[scale=0.15]{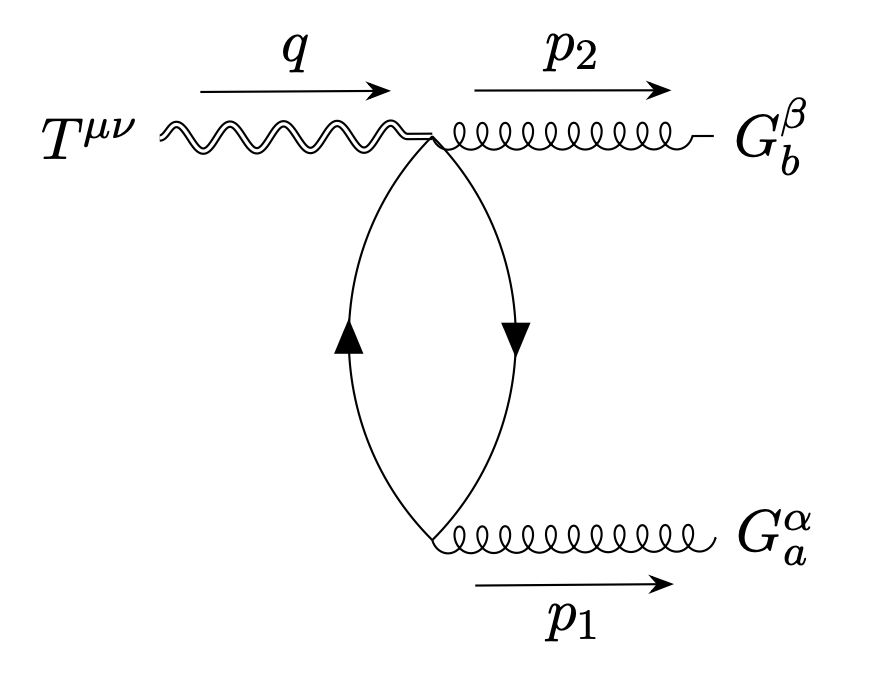}
			\includegraphics[scale=0.15]{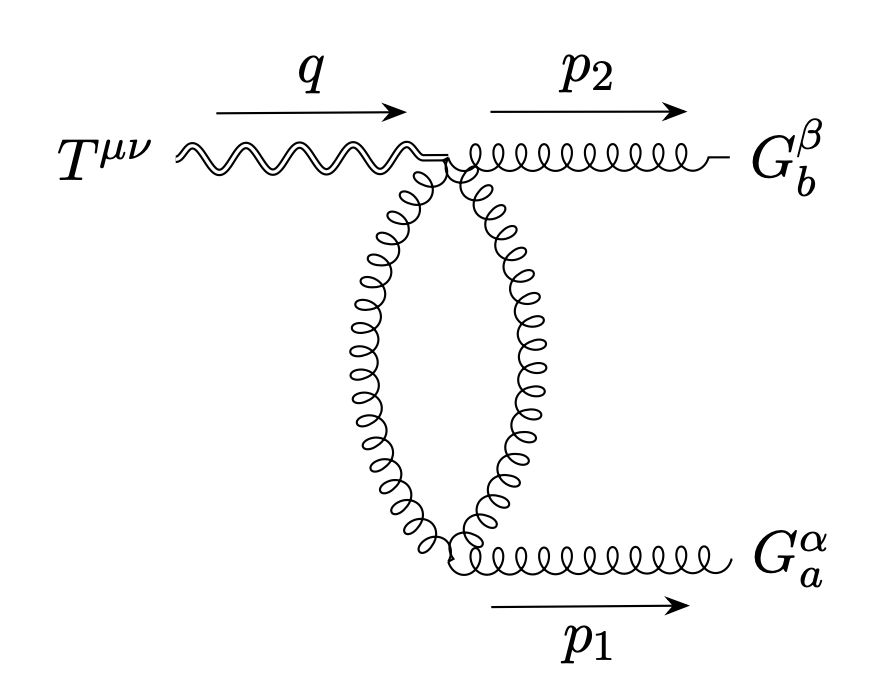}
			\includegraphics[scale=0.15]{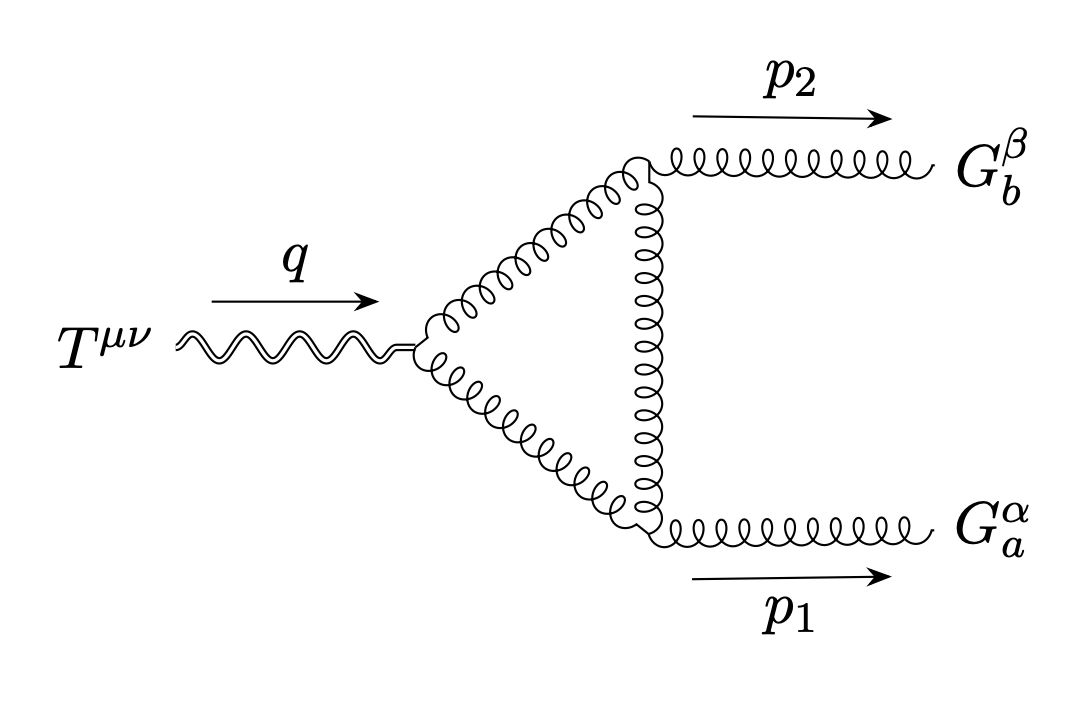}
			\includegraphics[scale=0.15]{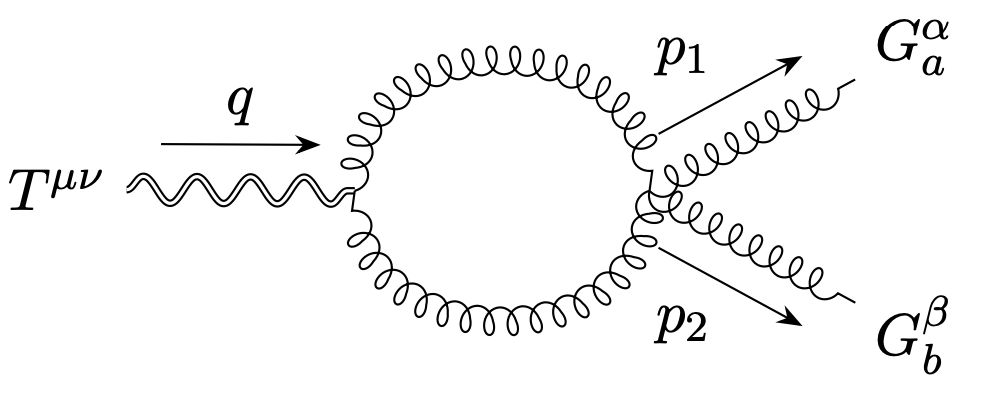}
			\includegraphics[scale=0.15]{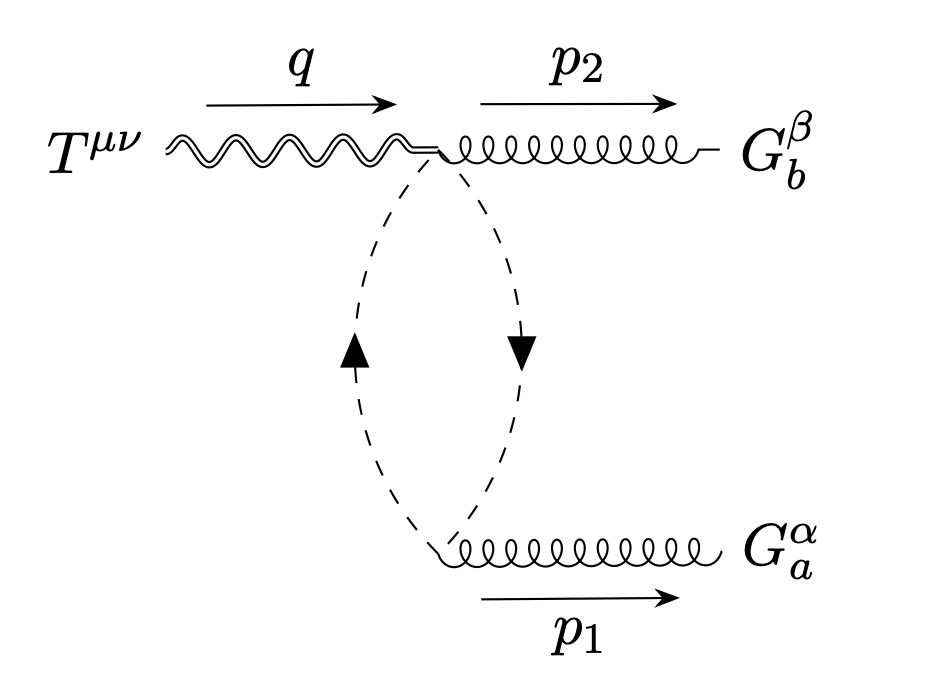}
			\includegraphics[scale=0.15]{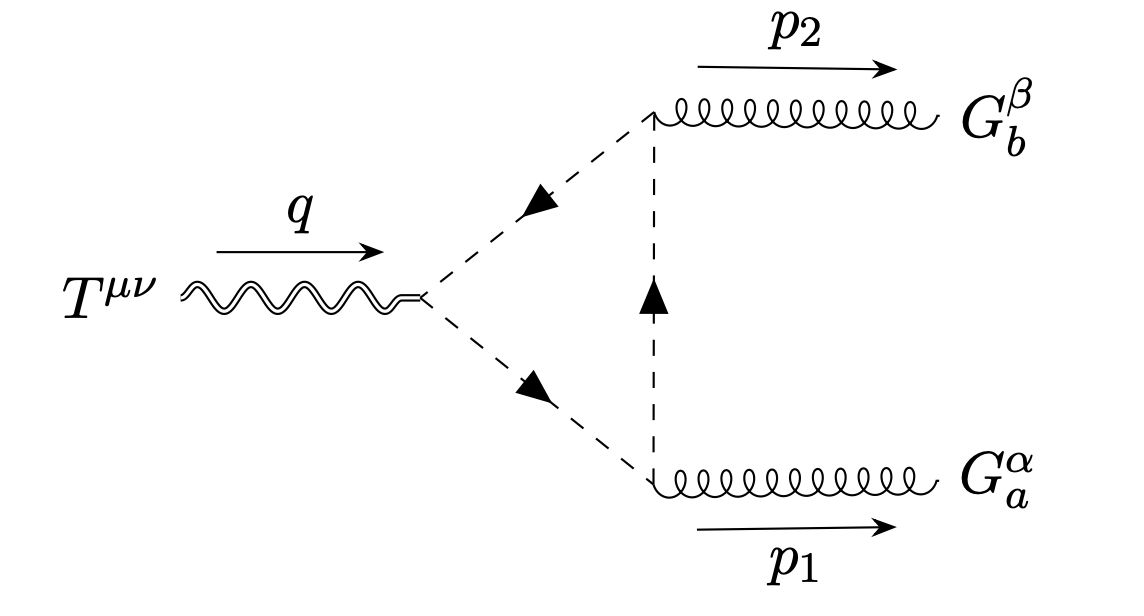}
			\caption{List of the perturbative quark, gluon, and ghost (dashed lines) sectors in the non-Abelian $TJJ$. }
			\label{expansionaaa}
		\end{center}
	\end{figure}
	
	\subsection{The anomaly sector and gauge invariance}
	In the non-Abelian case, as is customary, the anomaly computation is primarily conducted at the level of 3-point functions. The full gauge-invariant contribution is then inferred by extending the result of the 3-point function in a gauge-invariant manner. Specifically, the perturbative computation's outcome can be made gauge-covariant, at least regarding the anomaly pole part. In the non-Abelian scenario, additional contributions arise from other sectors, corresponding to extra diagrams with three and four external gluons.\\
	The $TJJ$ correlator is part of a broader set of correlation functions, including the $TJJJ$ and $TJJJJ$ correlators, all interconnected through conservation Ward identities (WIs), gauge WIs, and broken conformal WIs (CWIs) in the gluon sector. Gauge invariance permits the extraction of the gauge-invariant form of the nonlocal anomaly action from the perturbative computation of the $TJJ$. This is achieved through a standard covariantization of the results obtained from investigating this correlator. The anomaly's structure can then be recovered from the nonlocal action as in \eqref{pole}, modified by the QCD $\beta$ function 
	\bea
	S_{pole} &=&  \beta(g)  \, \int d^4 x \, d^4 y \,R^{(1)}(x)\, \square^{-1}(x,y) \, F^a_{\alpha \beta}F^{a \alpha \beta}
	\eea
	\beq
	\beta(g) = \frac{dg}{d\ln(\mu^2)} = -\beta_0 \frac{g^3}{16\pi^2}, \qquad \beta_0 = \frac{11}{3} C_A - \frac{2}{3} n_f
	\eeq
	with $n_f$ the number of massless quark flavours.
	One may define the quantum averaged stress energy tensor, with just the pole-term included  
	\be
	T^{\mu\nu}_{anom}(z) =
	\beta(g) \left(g^{\mu\nu}\sq - \partial^{\mu}\partial^{\nu}\right)_z \int\,d^4x'\, \sq_{z,x'}^{-1}
	\left[F^a_{a\alpha\beta}F^{a \alpha\beta}\right]_{x'}\,,
	\label{Tanom}
	\ee
	from which one may extract the anomaly-induced vertex at trilinear level
	
	\be
	\Gamma_{{anom}}^{\mu\nu\alpha\beta\, a b}(p_1,p_2) = \int\,d^4x\,\int\,d^4y\, e^{ip_1\cdot x + i p_2\cdot y}\,\frac{\delta^2 T^{\mu\nu}_{anom}(0)}
	{\delta A^a_{\alpha}(x) A^b_{\beta}(y)} = \beta(g)\frac{1}{3\, q^2} \left(g^{\mu\nu}q^2 - q^{\mu}q^{\nu}\right)u^{\alpha\beta}(p_1,p_2)\delta^{a b},\,
	\label{Gamanom}
	\ee
	with a trace anomaly 
	\be
	g_{\mu\nu}\Gamma^{\mu\nu\alpha\beta a b }(p,q)\big\vert_{A=0} =\beta(g)
	\,u^{\alpha\beta}(p_1,p_2)\delta^{a b},\,
	\label{Gamanomtr}
	\ee
	where 
	\beq
	u^{\a \b} (p_1,p_2)\equiv (p_1\cdot p_2) g^{\alpha\beta} - p_2^\alpha p_1^\beta
	\eeq
	This tensor structure summarizes the conformal anomaly contribution, being the Fourier transform of the anomaly term at $O(g^2)$
	\bea
	&&u^{\alpha\beta}(p_1,p_2) = -\frac{1}{4}\int\,d^4x\,\int\,d^4y\ e^{ip_1\cdot x + i p_2\cdot y}\ 
	\frac{\delta^2 \{F^a_{\mu\nu}F^{a\mu\nu}(0)\}} {\delta A_{\alpha}(x) A_{\beta}(y)}\vline_{A=0} \,.
	\label{locvarasd}\\
	\eea
	By differentiating $T^{\mu\nu}_{\text{anom}}(0)$ to higher orders with respect to the external classical gluon field, we derive analogous relations for the 4-point function ($TJJJ$)
	
	\beqa
	\Gamma_{anom}^{\mu\nu\alpha\beta\rho\, a_1 a_2 a_3}(p_1,p_2,p_3) &=& \int\,d^4x d^4y\, d^4 z \, e^{ip_1\cdot x + i p_2\cdot y + i p_3\cdot z}\,\frac{\delta^2 T^{\mu\nu}_{anom}(0)}
	{\delta A^{a_1}_{\alpha}(x) \delta A^{a_2}_{\beta}(y)\delta A^{a_3}_{\rho}(z)}\vert_{A=0} \nn\\
	&=&  \beta(g)\frac{1}{3\, q^2} \left(g^{\mu\nu}q^2 - q^{\mu}q^{\nu}\right)u^{\alpha\beta\rho\, a_1 a_2 a_3}(p_1,p_2,p_3)\,,
	\label{Gamanom}
	\eeqa
	
	\begin{equation}
		u^{\mu_1\mu_2\mu_3\,\, a_1 a_2 a_3}= g\left(f^{a_1 a_2 a_3} \left(g^{\mu_1 \mu_2} \left(p_1^{\mu_3}-p_2^{\mu_3}\right)+g^{\mu_1 \mu_3} \left(p_3^{\mu_2}-p_1^{\mu_2}\right)+g^{\mu_2 \mu_3} \left(p_2^{\mu_1}-p_3^{\mu_1}\right)\right)\right)
	\end{equation}
	
	and 5-point function $(TJJJJ)$
	
	\beqa
	\Gamma_{anom}^{\mu\nu\alpha\beta\rho\, a_1 a_2 a_3}(p_1,p_2,p_3) &=& \int\,d^4x\,d^4y\, d^4 z d^4 w \, e^{ip_1\cdot x + i p_2\cdot y + i p_3\cdot z + i p_4\cdot w}\,\frac{\delta^2 T^{\mu\nu}_{anom}(0)}
	{\delta A^{a_1}_{\alpha}(x) \delta A^{a_2}_{\beta}(y) \delta A^{a_3}_{\rho}(z) \delta A^{a_4}_{\sigma}(w)} \nn\\
	&=&  \beta(g)\frac{1}{3\, q^2} \left(g^{\mu\nu}q^2 - q^{\mu}q^{\nu}\right)u^{\alpha\beta\rho\sigma\, a_1 a_2 a_3 a_4}(p_1,p_2,p_3)\,,
	\label{Gamanom}
	\eeqa
	where 
	\begin{align}
		u^{\mu_1\mu_2\mu_3\mu_4 \,\, a_1 a_2 a_3 a_4}=g^2 \Bigl(\Bigl(&f^{a_1 a_2 b} f^{a_3 a_4 b} \left(g^{\mu_1 \mu_3} g^{\mu_2 \mu_4}-g^{\mu_1 \mu_4} g^{\mu_2 \mu_3}\right)+\nonumber\\&+f^{a_1 a_3 b} f^{a_2 a_4 b} \left(g^{\mu_1 \mu_2} g^{\mu_3 \mu_4}-g^{\mu_1 \mu_4} g^{\mu_2 \mu_3}\right)
		+f^{a_1 a_4 b} f^{a_2 a_3 b} \left(g^{\mu_1 \mu_2} g^{\mu_3 \mu_4}-g^{\mu_1 \mu_3} g^{\mu_2 \mu_4}\right)\Bigl)\Bigl).
	\end{align}
	
	\section{Symmetries of the quark partition function }
	\label{four1}
	The quark sector at one-loop satisfies ordinary anomalous CWIs, together with ordinary gauge Ward identities on the gluon vector current. In this case the treatment follows the usual approach similarly to the case of QED or any scale invariant Abelian theory, with the due generalization. Formal derivation of the equations are based on the functional integral 
	\be
	\tilde{W}[g,A] =-i \log \tilde{Z} 
	\label{GF}
	\ee
	
	where
	\be
	\tilde{Z} =\int {D\bar\psi}D\psi e^{-(S_0[g,\psi] + S_1[A,\psi ])}
	\ee
	is integrated only over the quark field $\psi$, in the background of the metric $g_{\mu\nu}$ and of the gauge field $A_\mu^a$. To fix the form of the correlator we need to impose the transverse WI on the vector lines and the conservation WI for $T^{\mu\nu}$. 
	Indeed, diffeomorphism invariance of the generating functional of the connected vertices (\ref{GF}) gives 
	\be
	\int d^d x \left(\frac{\delta\tilde{ W}}{\delta g_{\mu\nu}}\delta g_{\mu\nu}(x) +  \frac{\delta\tilde{ W}}{\delta A_{\mu}^a}\delta A_{\mu}^a(x)\right)=0,
	\label{aver}
	\ee
	where the variation of the metric and the gauge fields are the corresponding Lie derivatives, for a change of variables 
	$x^\mu\to x^\mu + \epsilon^\mu(x)$
	\bea
	\label{lie}
	\delta A_\mu^a(x) &=& -\nabla_\alpha A_\mu^a \epsilon^\alpha - A^a_\alpha \nabla_\mu \epsilon^\alpha \nonumber \\
	g_{\mu\nu}&=&-\nabla_\mu \epsilon_\nu  - \nabla_\nu \epsilon_\mu 
	\eea
	while for a gauge transformation with a parameter $\theta^a(x)$
	\be
	\delta A_\mu^a=\ul{D}_\mu \theta^a \equiv\partial_\mu \theta^a + g_c f^{a b c} A_\mu^b \theta^c.
	\ee
	The absence of propagating gluons guarantees the preservation of all the fundamental symmetries needed for the derivaton of the hierarchical WIs of this sector.\\ CWIs, conservation WIs for the stress energy tensor as well as ordinary gauge WIs can be derived by simply requiring the invariance of this functional with respect to conformal transformations, diffeomorphisms and gauge transformations.
	
	\begin{itemize}
		\item{\bf Diffeomorphism invariance}
	\end{itemize}
	Using (\ref{lie}), Eq. (\ref{aver}) becomes 
	\begin{align}
		\label{interx1}
		0=&\,\left\langle\, \int d^4 x \left( \frac{\delta (S_0 + S_1)}{\delta g_{\mu\nu}} \delta g_{\mu\nu} +\frac{\delta S_1}{\delta A_\mu^a} \delta A_{\mu}^a\right)
		\right\rangle_q\nn \\
		=&\,\left\langle \int d^4 x \sqrt{-g_x}\left[\nabla_\mu T^{\mu\nu} -(\nabla_\mu A_\nu^a -\nabla_\nu A_\mu^a) J^{\mu a} - \nabla_\mu J^{\mu a} A_\nu^a\right]\epsilon^\nu(x)\right\rangle_q
	\end{align}
	while the condition of gauge invariance gives the 
	\begin{itemize}
		\item{\bf gauge WI }
	\end{itemize}
	\be
	\int d^d x \frac{\delta \tilde{W}}{\delta A_{\mu}^a}\delta A_{\mu}^a=\left\langle \int d^4 x \sqrt{-g_x}J^\mu_a\ul{D}_\mu \theta^a \right\rangle_q=0
	\ee
	which, in turn, after an integration by parts, generates the gauge WI
	\be
	\langle \nabla_\mu J^{\mu a} \rangle_q = g_c f^{a b c} \langle J_\mu^b\rangle A^{\mu c}. 
	\ee
	Inserting this relation into \eqref{interx1} we obtain the conservation WI
	\be
	\langle \nabla^\mu T_{\mu\nu}\rangle_q -F_{\mu\nu}^a \langle J^{\mu a}\rangle_q =0.
	\ee
	In the quark sector, diffeomorphism and gauge invariance  then give the relations
	\begin{equation}
		\begin{split}
			0&=\langle \nabla^\mu T_{\mu\nu}\rangle_q -F_{\mu\nu} \langle J^{\mu}\rangle_q \\
			0&=\nabla_\n\braket{J^\nu}_q .
		\end{split}\label{transverseeee}
	\end{equation}
	
	\begin{itemize}
		\item{\bf Conformal invariance and differential equations}
	\end{itemize}
	We recall that a special conformal transformation in flat space is characterized by
	\beq
	x'^\mu=\frac{(x^\mu - b^\mu x^2)}{\Omega(x)}\qquad \textrm{with}\qquad \Omega(x)=1 - 2 b\cdot x + b^2 x^2\qquad \textrm{and}\qquad J_c\equiv \vline \frac{\partial x'}{\partial x}\vline=\Omega^{-d}.
	\label{special}
	\eeq
	On the metric and on primary scalar field $\phi(x)$ of scaling  dimension $\Delta$ it will induce the transformations
	\beq
	g'_{\mu\nu}(x')=\Omega^2 g_{\mu\nu}(x) \qquad \phi'(x')=J_c^{-\Delta/d}\phi(x)=\Omega^{\Delta}\phi(x).\qquad 
	\eeq
	and for a spin-1 field 
	\beq
	J'^\mu(x')=\Omega^{\Delta_J}\frac{\partial x'^\mu}{\partial x^\nu}J^\nu(x).
	\eeq
	
	Differential equations can be derived for all of the transformations above. Expanding these relations for $b\ll 1$
	and taking the finite part one obtains
	\beq
	\mathcal{K}^k \phi(x)=\left(- x^2\frac{\partial}{\partial x^\kappa}+ 2 x^\kappa x^\alpha \frac{\partial}
	{\partial x^\alpha} +2 \Delta_\phi x^\kappa\right) \phi(x)
	\eeq
	
	\beq
	\mathcal{K}^k J^\mu(x)= \left(- x^2\frac{\partial}{\partial x^\kappa}+ 2 x^\kappa x^\alpha \frac{\partial}
	{\partial x^\alpha} +2 \Delta_J x^\kappa\right) J^\mu(x) +2 \left(  \delta^{\mu\kappa} x_{\rho} - \delta_{\rho}^{\kappa }x^\mu  \right) J^\rho(x)
	\eeq
	for the special conformal tranformations.
	A similar transformation holds for the spinor 
	\beq
	\psi'(x')=\Omega^{\Delta_\psi} S(\Lambda(x))\psi(x)\qquad S(\Lambda(x))=e^{-\frac{i}{4}\omega_{\mu\nu}(x)\Sigma^{\mu\nu}}\qquad \Sigma^{\mu\nu}=-\frac{i}{2}[\gamma^\mu,\gamma^\nu]
	\eeq
	where $\omega^{\mu\nu}(x)$ is the infinitesimal Lorentz tranformation induced by the conformal transformation. For a special conformal transformation as in \eqref{special}
	\beq
	\delta x^\mu=\omega^{\mu}_\nu x^\nu= 1-2 b\cdot x \, x^\mu + b^\mu x^2, \qquad \omega_{\mu\nu}=-2 (b_\mu x_\nu - b_\nu x_\mu),
	\eeq
	giving 
	\beq
	\mathcal{K}^k \psi=\left(- x^2\frac{\partial}{\partial x^\kappa}+ 2 x^\kappa x^\alpha \frac{\partial}
	{\partial x^\alpha} +2 \Delta_\psi x^\kappa\right)\psi +\frac{1}{2}[\gamma^k,\gamma\cdot x] \,\psi.
	\eeq
	Due to the conformal invariance of the quark sector in the massless limit, these identities at operatorial level  act on the $TJJ$ as a derivative on each of the coordinates 
	$x_i, i=1,2,3 $ giving 
	\beq
	\mathcal{K}^\k \langle TJJ\rangle_q=0. 
	\eeq
	Defining 
	\be
	\Gamma^{\mu\nu\alpha\beta }_q(x_1,x_2,x_3)=\langle T^{\mu\nu}(x_1) J^\alpha(x_2)J^\beta(x_3)\rangle_q
	\ee
	(we omit the colour indices) their explicit expression is 
	\beqa
	\mathcal{K}^\kappa \Gamma^{\mu\nu\a\b}_q(x_1,x_2,x_3) 
	&=& \sum_{i=1}^{3} {K_i}^{ \kappa}_{scalar}(x_i) \Gamma^{\mu\nu\a\b}_q(x_1,x_2,x_3) \nonumber \\
	&&\qquad + 2 \left(  \delta^{\mu\kappa} x_{1\rho} - \delta_{\rho}^{\kappa }x_1^\mu  \right)\Gamma^{\rho \nu\alpha\beta}_q
	+ 2 \left(  \delta^{\nu\kappa} x_{1\rho} - \delta_{\rho}^{\kappa }x_1^\nu  \right)\Gamma_q^{\mu\rho \alpha\beta}\nonumber \\ 
	&&\qquad  2 \left(  \delta^{\a\kappa} x_{2\rho} - \delta_{\rho}^{\kappa }x_2^\a  \right)\Gamma_q^{\mu\nu \rho\beta}
	+  2 \left(  \delta^{\beta\kappa} x_{3\rho} - \delta_{\rho}^{\kappa }x_3^\beta  \right)\Gamma_q^{\mu\nu \alpha\rho}=0,
	\eeqa
	
	where 
	\be
	\label{ki}
	{\mathcal{K}_i}^{\kappa}_{scalar}=-x_i^2 \frac{\partial }{\partial x_\kappa} + 2 x_i^\kappa x_i^\tau\frac{\partial}{\partial x_i^\tau} + 2 \Delta_i x_i^\kappa
	\ee
	is the scalar part of the special conformal operator acting on the $i_{\textrm{th}}$ coordinate. $\Delta_i\equiv(\Delta_T,\Delta_J,\Delta_J)$ are the scaling dimensions of the operators in the correlation function ($\Delta_T=3, \Delta_J=3$ at $d=4$). \\
	Similarly, in the case of dilatations,where $\Omega(x)=\lambda^{-1}$ and 
	$x'^\mu=\lambda x^\mu$, the condition of scale invariance gives for a correlator of $n$ primary fields, scalars or tensors, 
	
	\be
	\label{scale1}
	\Phi (\lambda x_1,\lambda x_2,\ldots,\lambda x_n)=\lambda^{-\Delta} \Phi(x_1,x_2,\ldots, x_n), \qquad 
	\Delta=\Delta_1 +\Delta_2 +\ldots \Delta_n,
	\ee
	which generates the Euler equation. 
	Also in this case, the operator acts on $n-1$ of the $n$ momenta.

	\subsection{The transition to momentum space in the quark sector}
	It is convenient to resort to a symmetric relabeling of the three momenta in order to define the action of $\mathcal{K}$. The functional differentiation of \eqref{transverseeee} and \eqref{trace} allows one to derive ordinary Ward identities for the various correlators. In the $TJJ$ case we obtain, after a Fourier transformation, the  
	\begin{itemize}
		\item{\bf conservation WIs}
	\end{itemize}

	\begin{align}
		\label{trrrr}
		p_{1\n_1}\braket{T^{\mu_1\nu_1}(p_1)\,J^{\m_2}(p_2)\,J^{\m_3}(p_3)}_q&=4\,\big[\d^{\m_1\m_2}p_{2\l}\braket{J^\l(p_1+p_2)\,J^{\m_3}(p_3)}_q-p_2^{\m_1}\braket{J^{\m_2}(p_1+p_2)\,J^{\m_3}(p_3)}_q\big]\notag\\
		&+4\,\big[\d^{\m_1\m_3}p_{3\l}\braket{J^\l(p_1+p_3)\,J^{\m_2}(p_2)}_q-p_3^{\m_1}\braket{J^{\m_3}_q(p_1+p_3)\,J^{\m_2}(p_2)}_q\big].
	\end{align}
	We recall that the 2-point function of two conserved vector currents $J_i$ $(i=2,3)$ \cite{Coriano:2013jba} in any conformal field theory in $d$ dimension is given by 
	\beqa
	\langle J_2^\alpha(p)J_3^\beta(-p) \rangle =\delta_{\Delta_2\, \Delta_3}\left(c_{123} \Gamma_J \right)\pi^{\alpha\beta}(p) (p^2)^{\Delta_2-d/2},
	\qquad \Gamma_J=\frac{\pi^{d/2}}{ 4^{\Delta_2 -d/2}}\frac{\Gamma(d/2-\Delta_2)}{\Gamma(\Delta_2)},
	\eeqa
	with $c_{123}$ an overall constant and $\Delta_2=d-1$. Notice that the divergence of this 2-point function is removed by the counterterm 
	\beq
	\mathcal{L}_{\textrm{q\, count}}=-\frac{1}{(d-4)}\frac{g_s^2}{16 \pi^2}\frac{2}{3} n_f
	F^{a\, \mu\nu}F^{a}_{\mu\nu}
	\eeq
	where $n_f$ denotes the number of flavours. In the gluon sector a similar counterterm will be introduced 
	\beq
	\mathcal{L}_{\textrm{count}}=\frac{1}{(d-4)}\frac{g_s^2}{16 \pi^2}(\frac{5}{3}C_A-\frac{2}{3} n_f)
	F^{a\, \mu\nu}F^{a}_{\mu\nu}.
	\eeq
	
	In our case $\D_2=\D_3=d-1$ and Eq. (\ref{trrrr}) then takes the form 
	
	\begin{align}
		\label{2point}
		p_{1\mu_1}\braket{T^{\mu_1\nu_1}(p_1)\,J^{\m_2}(p_2)\,J^{\m_3}(p_3)}_q&=4 c_{123} \Gamma_J
		\left( \delta^{\nu_1\mu_2}\frac{p_{2\lambda}}{(p_3^2)^{d/2 -\Delta_2}} {\pi^{\lambda \mu_3}}(p_3) -
		\frac{p_2^{\nu_1}}{(p_3^2)^{d/2 -\Delta_2}}\pi^{\mu_2\mu_3}(p_3) \right.\nn\\
		& \left. + \delta^{\nu_1\mu_3}\frac{p_{3\lambda}}{(p_2^2)^{d/2-\Delta_2}}\pi^{\lambda \mu_2}(p_2) -\frac{p_3^{\nu_1}}{(p_2^2)^{d/2-\Delta_2}}\pi^{\mu_3\mu_2}(p_2)\right).
	\end{align}
	The equation can be checked perturbatively in the quark sector by the ordinary free field theory representation. In this case the constant $c_{123}$ is simply proportional to $n_f$, the number of quark flavours running in the quark loops. \\
	One can generalize this equation in the quark case to higher orders in the number of external gluons, by the inclusion of additional gluon currents on the external lines.  
	Specifically, at higher order in the hierarchy of the $TJJ$ we have the conservation WI
	\begin{equation}
		\begin{aligned}
			0=&q_{\mu } \langle T^{\mu\nu}(q)J^{\alpha a_1}(p_1)J^{\beta a_2}(p_2)J^{\gamma a_3}(p_3)\rangle_q+2
			\delta^{\alpha}_{[\mu}{p_1}_{\nu]}\langle J^{\mu a_1}(p_1-q)J^{\beta a_2}(p_2)J^{\gamma a_3}(p_3) \rangle_q\\&
			+2
			\delta^{\beta}_{[\mu}{p_2}_{\nu]}\langle J^{\alpha a_1}(p_1)J^{\mu a_2}(p_2-q)J^{\gamma a_3}(p_3) \rangle_q
			+2
			\delta^{\gamma}_{[\mu}{p_3}_{\nu]}\langle J^{\alpha a_1}(p_1)J^{\beta a_2}(p_2)J^{\mu a_3}(p_3-q) \rangle_q
			\\&-2i gf^{a_1 a_2 c} \delta^{[\alpha}_\nu \langle J^{\beta] c}(p_1+p_2-q)J^{\gamma a_3}(p_3)\rangle_q
			-2i gf^{a_1 a_3 c} \delta^{[\alpha}_\nu \langle J^{\gamma] c}(p_1+p_3-q)J^{\beta a_2}(p_2)\rangle_q
			\\&-2i gf^{a_3 a_2 c} \delta^{[\gamma}_\nu \langle J^{\beta] c}(p_3+p_2-q)J^{\alpha a_1}(p_1)\rangle_q,
		\end{aligned}
	\end{equation}
	derived from diffeomorphism invariance.
	The hierarchical equation of the $TJJJJ$ can be found in Appendix \ref{t4j}.
	\\

	\begin{itemize}
		\item{\bf gauge WIs}
	\end{itemize}
	These identities take the form
	\begin{align}
		\label{x1}
		p_{2\m_2}\braket{T^{\mu_1\nu_1}(p_1)\,J^{\m_2}(p_2)\,J^{\m_3}(p_3)}_q&=0\\
		p_{3\m_3}\braket{T^{\mu_1\nu_1}(p_1)\,J^{\m_2}(p_2)\,J^{\m_3}(p_3)}_q&=0,
	\end{align}
	These two equations are homogeneous because the $TJ$ two-point function, which is expected on their right-hand sides, is trivially zero. \\
	We also have
	\begin{equation}
		\begin{aligned}
			0=&p_{1 \alpha } \langle T^{\mu\nu}(q)J^{\alpha a_1}(p_1)J^{\beta a_2}(p_2)J^{\gamma a_3}(p_3)\rangle_q\\&+i gf^{a_1 a_2 c}  \langle T^{\mu\nu}(q)J^{\beta c}(p_2+p_1)J^{\gamma a_3}(p_3)\rangle_q+i gf^{a_1 a_3 c}  \langle T^{\mu\nu}(q)J^{\beta a_2}(p_2)J^{\gamma c}(p_3+p_1)\rangle_q
		\end{aligned}
	\end{equation}
	and 
	\begin{equation}
		\begin{aligned}
			0=&p_{1 \alpha } \langle T^{\mu\nu}(q)J^{\alpha a_1}(p_1)J^{\beta a_2}(p_2)J^{\gamma a_3}(p_3)J^{\delta a_4}(p_4)\rangle_q+i gf^{a_1 a_2 c}  \langle T^{\mu\nu}(q)J^{\beta c}(p_2+p_1)J^{\gamma a_3}(p_3)J^{\delta a_4}(p_4)\rangle_q\\ &+i gf^{a_1 a_3 c}  \langle T^{\mu\nu}(q)J^{\beta a_2}(p_2)J^{\gamma c}(p_3+p_1)J^{\delta a_4}(p_4)\rangle_q
			+i gf^{a_1 a_4 c}  \langle T^{\mu\nu}(q)J^{\beta a_2}(p_2)J^{\gamma a_3}(p_3)J^{\delta c}(p_4+p_1)\rangle_q
		\end{aligned}
	\end{equation}
	Applying functional derivatives to \eqref{ann} with respect to the gauge fields and Fourier transforming, one obtains similarly to 
	\eqref{Gamanomtr} but to higher orders in the coupling constant $g$ 
	\begin{itemize}
		\item{\bf trace WIs}
	\end{itemize}
	
	\begin{equation}
		\label{x2}
		g_{\mu\nu} \langle T^{\mu\nu}(q)J^{\alpha a_1}(p_1)J^{\beta a_2}(p_2)\rangle_q=\beta u^{\alpha \beta  \, a_1a_2}
	\end{equation}
	and those of higher order in the hierarchy
	\begin{equation}
		g_{\mu\nu} \langle T^{\mu\nu}(q)J^{\alpha a_1}(p_1)J^{\beta a_2}(p_2)J^{\gamma a_3}(p_3)\rangle_q=\beta u^{\alpha \beta \gamma \, a_1a_2a_3}
	\end{equation}
	
	\begin{equation}
		g_{\mu\nu} \langle T^{\mu\nu}(q)J^{\alpha a_1}(p_1)J^{\beta a_2}(p_2)J^{\gamma a_3}(p_3)J^{\delta a_4}(p_4)\rangle_q=\beta u^{\alpha \beta \gamma \delta \, a_1a_2a_3 a_4}.
	\end{equation}
	Finally, considering the 
	
	\begin{itemize}
		\item{\bf conformal Ward identities}
	\end{itemize}
	in the quark sector, the non-anomalous special conformal equation takes the form 
	\begin{equation}
		\begin{aligned}
			0=&\mathcal{K}^k\langle T^{\mu\nu}(q)J^{\mu_1 a_1}(p_1)\dots J^{\mu_{n-1} a_{n-1}}(p_{n-1})\rangle_q=\\&\sum_{j=1}^{n-1}\left(p_j^\kappa \frac{\partial^2}{\partial {p_j}_\alpha \partial p_j^\alpha}+2\left(\Delta_j-d\right) \frac{\partial}{\partial p_j^\kappa}-2 p_j^\alpha \frac{\partial^2}{\partial p_j^\kappa \partial p_j^\alpha}\right)\langle T^{\mu\nu}(q)J^{\mu_1 a_1}(p_1)\dots J^{\mu_{n-1} a_{n-1}}(p_{n-1})\rangle_q\\&
			+2 \sum_{j=1}^{n-1} \left[\delta^{\kappa \mu_j} \frac{\partial}{\partial p_j^{\alpha_{j}}}-\delta_{\alpha_{j}}^\kappa \frac{\partial}{\partial {p_j}_{\mu_j}}\right]\langle T^{\mu\nu}(q)J^{\mu_1 a_1}(p_1)\dots J^{\alpha_j a_j}(p_j)\dots J^{\mu_{n-1} a_{n-1}}(p_{n-1})\rangle_q
		\end{aligned}
	\end{equation}
	while the dilatation WIs are given by
	\begin{equation}
		0=\left[\sum_{j=1}^n \Delta_j-(n-1) d-\sum_{j=1}^{n-1} p_j^\lambda \frac{\partial}{\partial p_j^\lambda}\right]\langle T^{\mu\nu}(q)J^{\mu_1 a_1}(p_1)\dots J^{\mu_{n-1} a_{n-1}}(p_{n-1})\rangle_q.
	\end{equation}

	\section{BRST symmetry of the gauge-fixed action and the complete partition function}
	\label{BRSTX}
	As we include the gluon sector in the path integral representation, special conformal symmetry is broken by the gauge-fixing term in the QCD Lagrangean. For this reason, the natural approach in the investigaton of the conformal behaviour of the theory is to turn to the effective action with a Legendre transform of the complete partition function.
	The condition of diffeomorphism invariance of the generating functional $Z$ gives
	\beq\label{covW}
	Z[V,J,J^\dag,\c,\bar{\c},g] = Z[V',J',J^{'\,\dag},\c',\bar{\c}',J^{'\,\mu}, g']\, ,\eeq
	where we have allowed an arbitrary change of coordinates $x^{'\,\mu} = F^\mu(x)$ on the spacetime manifold, which can be
	parameterized locally as $x^{'\,\mu} = x^\mu + \e^\mu(x)$.
	The measure of integration is invariant under general coordinate transformations under such changes $( \mD\Phi' = \mD\Phi)$ as far as we stay in $d$ dimensions
	and we obtain to first order in $\e^\mu(x)$
	\beqa\label{preWard}
	\int \mD\Phi \, e^{i \tilde{S}}
	& = & \int \mD\Phi \quad e^{i \tilde{S}}\,\bigg( 1  + i\int d^4xd^4y\, \bigg\{- V\Theta^\mu_{\,\,\,\underline{a}}\bigg[-\d^{(4)}(x-y)\pd_\nu V^{\underline{a}}_\mu(x)
	- [\pd_\mu\d^{(4)}(x-y)]V^{\underline{a}}_\nu\bigg]\nn\\
	&& - \pd_\nu[\d^{(4)}(x-y)\bar{\c}(x)]\psi(x) - \bar{\psi}(x)\pd_\nu[\d^{(4)}(x-y)\c(x)] \nn\\
	&& - [\pd_\nu[J^\mu(x)\d^{(4)}(x-y)] + [\pd_\r\d^{(4)}(x-y)]\d^\mu_\nu J^\r(x)]A_\mu(x)\bigg\}\e^\nu(y)\bigg),
	\eeqa

	\beq \Theta^{\mu\nu} = - \frac{1}{V}\frac{\d S}{\d e^{{\underline{a}}}_\mu}e^{{\underline{a}}\nu}, \qquad 
	\Theta^{\mu\ul{a}}= \Theta^{\mu\nu} e^{\ul{a}}_{\nu} \eeq
	in terms of the determinant of the vielbein $V(x){\equiv} \left|V^{\underline{a}}_\mu(x)\right|$.
	Notice that this expression of the EMT is non-symmetric. The symmetric expression can be easily defined by the relation
	\beq T^{\mu\nu} = \frac{1}{2}(\Theta^{\mu\nu} + \Theta^{\nu\mu})\,
	\label{tmunu} \eeq
	that will be used below.\\
	\eqref{preWard}  can be brought into the form 
	\beqa\label{Ward}
	&&  \int\mD\Phi\, e^{i\tilde{S}}\, \bigg[\pd_\a T^{\a\b}(y)  -  J^\a(y)\pd^\b A_\a(y) + \pd_\a[J^\a(y) A^\b(y)]\nn\\
	&& -  \pd^\b\bar{\psi}(y)\c(y) - \bar{\c}(y)\pd^\b\psi(y)  -      \frac{1}{2}\pd_\a\bigg(\frac{\d S}{\d\psi(y)}\s^{\a\b}\psi(y)
	- \bar{\psi}(y)\s^{\a\b}\frac{\d S}{\d\bar{\psi}(y)}\bigg)\bigg] = 0\,.
	\eeqa
	where 
	\beqa
	\label{ActionToyModel}
	\tilde{S}&            =          & S + i\int d^4x\,\left( J^{\mu a} A^a_\mu  + \bar{\c}(x)\psi(x) + \textrm{h.c.}\right).\,
	\eeqa
	The conservation equation of the energy-momentum tensor takes the following form off-shell 
	\bea
	\partial^{\mu}T_{\mu\nu} &=& -\frac{\delta S}{\delta \psi} \partial_{\nu}\psi - \partial_{\nu}\bar \psi \frac{\delta S}{\delta \bar \psi} + \frac{1}{2}\partial^{\mu}\left( \frac{\delta S}{\delta \psi}\sigma_{\mu\nu}\psi - \bar \psi \sigma_{\mu\nu}\frac{\delta S}{\delta \bar \psi} \right) - \partial_{\nu}A_{\mu}^a \frac{\delta S}{\delta A_{\mu}^a} \nn\\
	&+& \partial_{\mu}\left( A_{\nu}^a \frac{\delta S}{\delta A_{\mu}^a} \right) - \frac{\delta S}{\delta c^a} \partial_{\nu}c^a - \partial_{\nu}\bar c^a\frac{\delta S}{\delta \bar c^a} \,  
	\label{EMTdivergence}
	\eea
	where $\sigma_{\mu\nu}= \frac{1}{4}[\gamma_{\mu},\gamma_{\nu}]$, where ${S}$ is the QCD action.\\
	The conservation of the energy-momentum tensor summarized in Eq.~(\ref{EMTdivergence}) in terms of classical fields, can be re-expressed
	in a functional form by a differentiation of $W$ with respect to $h_{\mu\nu}$ and the use of Eq.~\eqref{EMTdivergence}
	under the functional integral. One derives the relation \cite{Armillis:2010qk}
	\bea
	\partial_{\mu} \frac{\delta \mathcal{S}_{eff}}{\delta h_{\mu\nu}} &=& - \frac{\delta \mathcal{S}_{eff}}{\delta \psi_c}\partial^{\nu}\psi_c
	- \partial^{\nu}\bar\psi_c \frac{\delta\mathcal{S}_{eff}}{\delta \bar \psi_c} + \frac{1}{2}\partial_{\mu}\left( \frac{\delta \mathcal{S}_{eff}}{\delta \psi_c} \sigma^{\mu\nu} \psi_c
	- \bar \psi_c \sigma^{\mu\nu}\frac{\delta \mathcal{S}_{eff}}{\delta \bar \psi_c}  \right) - \partial^{\nu}A^{\mu}_c\frac{\delta \mathcal{S}_{eff}}{\delta A^{\mu}_c}
	+\partial^{\mu}\left(A^{\nu}_c\frac{\delta \mathcal{S}_{eff}}{\delta A^{\mu}_c} \right) \nn \\
	&-& \frac{\delta \mathcal{S}_{eff}}{\delta c_c}\partial^{\nu}c_c - \partial^{\nu}\bar c_c \frac{\delta \mathcal{S}_{eff}}{\delta \bar c_c}
	\label{FuncWI},
	\eea
	obtained from Eq.~\eqref{EMTdivergence} with the help of Eqs.~(\ref{Legendre1}), (\ref{Legendre2}), (\ref{Legendre3}). Some additional details on the derivation of these identities can be found in Appendix \ref{STTI} and in \cite{Coriano:2011zk}, where the analysis is extended to all the gauge sectors of the Standard Model. The relevant 
	WIs and STIs that can be used in order to fix the expression of the correlator in terms of truncated graphs are given by
	\bea
	\partial^{\mu}\langle T_{\mu\nu}(x) A_\alpha^a (x_1)A_{\beta}^b (x_2)\rangle_{trunc}  &=& - \partial_{\nu}\delta^4(x_1-x) D^{-1}_{\alpha\beta}(x_2,x)
	- \partial_{\nu}\delta^4(x_2-x) D^{-1}_{\alpha\beta}(x_1,x) \nn \\
	&+& \partial^{\mu}\left( g_{\alpha \nu} \delta^4(x_1-x) D^{-1}_{\beta\mu}(x_2,x) + g_{\beta \nu} \delta^4(x_2-x) D^{-1}_{\alpha\mu}(x_1,x)\right) \label{WIcoord} \nonumber \\
	\eea
	where $D^{-1}_{\alpha\beta}(x_1,x_2)$ is the inverse gluon propagator defined as
	\bea
	D^{-1}_{\alpha\beta }(x_1,x_2) = \langle A_{\alpha}(x_1) A_{\beta}(x_2) \rangle_{trunc}  =  \frac{\delta^2 \mathcal{S}_{eff}}{\delta A^{\alpha }_c(x_1) \delta A^{\beta }_c(x_2) }
	\eea
	and where we have omitted, for simplicity, both the colour indices and the symbol of the $T$-product.
	The first Ward identity (\ref{WIcoord}) becomes
	\bea
	k^{\mu}\langle T_{\mu\nu}(k) A_{\alpha}(p) A_{\beta}(q) \rangle_{trunc} =
	q_{\mu} D^{-1}_{\alpha\mu}(p) g_{\beta\nu} + p_{\mu} D^{-1}_{\beta\mu}(q) g_{\alpha\nu}  - q_{\nu} D^{-1}_{\alpha\beta}(p) - p_{\nu} D^{-1}_{\alpha\beta}(q) \, .
	\label{WImom}
	\eea
	with 
	\bea
	D^{-1}_{\alpha\beta}(p) = (p^2 g_{\alpha \beta} - p_{\alpha} q_{\beta}) \Pi (p^2)
	\eea
	being the gluon inverse 2-point function in momentum space and $\Pi(p^2)$ its scalar form factor. 
	A second WI can be derived using the BRST symmetry of $\mathcal{S}_{eff}$ involving two derivative. 
	For this purpose, we choose an appropriate  Green's function,  $\langle T_{\mu\nu} \partial^{\alpha} A_{\alpha}^a \bar c^b \rangle$, and then use its BRST invariance to obtain
	
	\bea
	\delta \langle T_{\mu\nu} \partial^{\alpha} A_{\alpha}^a \bar c^b \rangle = \langle \delta T_{\mu\nu} \partial^{\alpha} A_{\alpha}^a \bar c^b \rangle + \lambda \langle T_{\mu\nu} \partial^{\alpha} D^{ac}_{\alpha} c^c \bar c^b \rangle - \frac{\lambda}{\xi} \langle T_{\mu\nu} \partial^{\alpha} A_{\alpha}^a \partial^{\beta} A_{\beta}^b \rangle = 0,
	\label{BRSTward}
	\eea
	where the first two correlators, built with operators proportional to the equations of motion, contribute only with disconnected amplitudes, that are not part of the one-particle irreducible vertex function.
	From Eq.~(\ref{BRSTward}) we obtain the identity
	\bea
	\partial_{x_1}^{\alpha}\partial_{x_2}^{\beta}\langle T_{\mu\nu}(x) A_{\alpha}^a(x_1)A_{\beta}^b(x_2)\rangle_{trunc} = 0,
	\eea
	which in momentum space becomes
	\bea
	p_1^{\alpha} p_2^{\beta} \langle T_{\mu\nu}(k) A_{\alpha}^a(p_1)A_{\beta}^b(p_2)\rangle_{trunc} = 0. \label{brstwill}
	\label{pract1}
	\eea
	The broken conformal WIs of QCD in $d$ dimensions can be obtained by imposing the invariance of the generating functional 
	$W[J,\eta,\bar\eta,\chi,\bar\chi,h] $ under a conformal change of coordinates in the integrand and then rewriting the 
	constraint in the terms of $\mathcal{S}_{eff}$ using \eqref{Legendre1} and \eqref{Legendre2} \cite{Nielsen:1975ph}. They can be recast as  constraints on the 1PI vertices and truncated 2-point functions such as the $TJJ$ and the truncated $JJ$. As already mentioned in the Introduction, in the gluon sector we will not solve these equations directly using CFT$_p$ as for the quark sector, but rely on a direct one-loop computation of the gluon contributions, that will be added to the former. This is sufficient in order to obtain a consistent decomposition of the correlator in the LT-basis and proceed with an analysis of all of its sector in  the off-shell gluon case.
	
	\section{The off-shell form factors from the LT decomposition of CFT$_p$}
	\label{seven}
	%%%%%%%%%%%%%%%%%%%%%%%%%%%%%%%			%%%%%%%%%%%%%%%%%%%%%%%%%%%%%%%
	In this section, we begin by discussing the sector decomposition of the $TJJ$ to facilitate a direct analysis of its form factors. This decomposition applies to both the quark and gluon sectors, with their form factors initially treated separately before being combined.\\
	The general form of the non-Abelian $\langle TJJ \rangle$ correlator can be constructed through a decomposition into transverse,  longitudinal and trace terms \cite{Bzowski:2013sza}, exploiting its symmetries. The analysis includes additional contributions not found in the general expression for the $TJJ$ vertex derived in \cite{Bzowski:2018fql} using the CFT approach. These extra terms manifest as longitudinal components, which are naturally present in perturbative QCD (pQCD) but absent in the abstract conformal treatment of non-Abelian correlators with gauge currents.\\
	We consider the decomposition of the operators $T$ and $J$ in terms of their transverse traceless part and longitudinal (local) one, separating the quark and gluon parts, that, as we are going to see, behave differently under the application of the conformal constraints. Following \cite{Bzowski:2013sza} we define
	\begin{align}
		T^{\mu_i\nu_i}(p_i)&\equiv t^{\mu_i\nu_i}(p_i)+t_{loc}^{\mu_i\nu_i}(p_i),\label{decT}\\
		J^{a_i \, \mu_i}(p_i)&\equiv j^{a_i \, \mu_i}(p_i)+j_{loc}^{a_i \, \mu_i}(p_i),\label{decJ}
	\end{align}
	where
	\begin{align}
		\label{loct}
		t^{\mu_i\nu_i}(p_i)&=\Pi^{\mu_i\nu_i}_{\alpha_i\beta_i}(p_i)\,T^{\alpha_i \beta_i}(p_i), &&t_{loc}^{\mu_i\nu_i}(p_i)=\Sigma^{\mu_i\nu_i}_{\alpha_i\beta_i}(p)\,T^{\alpha_i \beta_i}(p_i),\\
		j^{a_i \, \mu_i}(p_i)&=\pi^{\mu_i}_{\alpha_i}(p_i)\,J^{a_i \, \alpha_i }(p_i), &&\hspace{1ex}j_{loc}^{a_i \, \mu_i}(p_i)=\frac{p_i^{\mu_i}\,p_{i\,\alpha_i}}{p_i^2}\,J^{a_i \, \alpha_i}(p_i).
	\end{align}
	having introduced the transverse-traceless ($\Pi$), transverse $(\pi)$, longitudinal ($\Sigma$) projectors, given respectively by 
	\begin{align}
		\label{prozero}
		\pi^{\mu}_{\alpha} & = \delta^{\mu}_{\alpha} - \frac{p^{\mu} p_{\alpha}}{p^2}, \\
		%%%%%%%%%%%%%%%%%%%%%%%%%%%%%%%
		\Pi^{\mu \nu}_{\alpha \beta} & = \frac{1}{2} \left( \pi^{\mu}_{\alpha} \pi^{\nu}_{\beta} + \pi^{\mu}_{\beta} \pi^{\nu}_{\alpha} \right) - \frac{1}{d - 1} \pi^{\mu \nu}\pi_{\alpha \beta}\label{TTproj}, \\
		%%%%%%%%%%%%%%%%%%%%%%%%%%%%%%%
		\Sigma^{\mu_i\nu_i}_{\alpha_i\beta_i}&=\frac{p_{i\,\beta_i}}{p_i^2}\Big[2\delta^{(\nu_i}_{\alpha_i)}p_i^{\mu_i)}-\frac{p_{i\alpha_i}}{(d-1)}\left(\delta^{\mu_i\nu_i}+(d-2)\frac{p_i^{\mu_i}p_i^{\nu_i}}{p_i^2}\right)\Big]+\frac{\pi^{\mu_i\nu_i}(p_i)}{(d-1)}\delta_{\alpha_i\beta_i}\equiv\mathcal{I}^{\mu_i\nu_i}_{\alpha_i}p_{i\,\beta_i} +\frac{\pi^{\mu_i\nu_i}(p_i)}{(d-1)}\delta_{\alpha_i\beta_i}\label{Lproj}.
		%%%%%%%%%%%%%%%%%%%%%%%%%%%%%%%
	\end{align}
	
	with
	\begin{equation} \label{a:Tttt}
		\mathscr{T}_{\mu\nu \alpha} (\bs{p}) = \frac{1}{p^2} \left[ 2 p_{(\mu} \delta_{\nu)\alpha} - \frac{p_\alpha}{d-1} \left( \delta_{\mu\nu} + (d-2) \frac{p_\mu p_\nu}{p^2} \right) \right]
	\end{equation}	
	and
	\[\label{Idecomp}
	\delta_{\mu(\alpha}\delta_{\beta)\nu}=\Pi_{\mu\nu\alpha\beta}(\bs{p})+\mathscr{T}_{\mu\nu (\alpha}(\bs{p})\,p_{\beta)}+\frac{1}{d-1}\pi_{\mu\nu}(\bs{p})\delta_{\alpha\beta}.
	\]
	Turning to the $TJJ$ case, we can divide the 3-point function into two parts: the \emph{transverse-traceless} part and the \emph{semi-local} part (indicated by subscript $loc$) expressible through the transverse and trace Ward Identities. These parts are obtained by using the projectors $\Pi$ and $\Sigma$, previously defined. 
	We can then decompose the full 3-point function as 
	
	\begin{align}
		\label{dec1111}
		\braket{T^{\m \n}\,J^{a \, \a}\,J^{b \, \b}}&=
		\braket{t^{\m \n }\,j^{a \, \a}\,j^{b \, \b}}+\braket{T^{\m \n}\,J^{a \, \a}\,j_{loc}^{b \, \b}}+\braket{T^{\m \n}\,j_{loc}^{a \, \a}\,J^{b \, \b}}+\braket{t_{loc}^{\m \n}\,J^{a \, \a}\,J^{b \, \b}}\notag\\
		&\quad-\braket{T^{\m \n }\,j_{loc}^{a \, \a}\,j_{loc}^{b \, \b}}-\braket{t_{loc}^{\m \n}\,j_{loc}^{a \, \a}\,J^{b \, \b}}-\braket{t_{loc}^{\m \n}\,J^{a \, \a}\,j_{loc}^{b \, \b}}+\braket{t_{loc}^{\m \n }\,j_{loc}^{a \, \a}\,j_{loc}^{b \, \b}}.
	\end{align}
	In a CFT approach, all the terms on the right-hand side of the decomposition, apart from the first one, may be computed by means of transverse and trace Ward Identities, with the anomaly induced by the renormalization of the hierachy by a single counterterm, proportional to the square of the field strength $(F^2)$. 
	
	\section{The sectors decomposition and the solution of the CWIs in the quark sector}
	\label{eight}
	Using the projectors $\Pi$ and $\pi$ one can write the most general form of the transverse-traceless part as
	\begin{equation}
		{\braket{t^{\m \n}(p_1)\,j^{a \, \a}(p_2)\,j^{b \, b }(p_3)}}_q =\Pi^{\m \n}_{\m_1 \n_1}(p_1)\pi^{\a}_{\a_1 }(p_2)\pi^{\b}_{\b_1}(p_3)\,\,X^{a b \, \m_1 \n_1 \a_1 \b_1}_q,
	\end{equation}
	where $X^{a b \, \m_1 \n_1 \a_1 \b_1}_q$ is a general tensor of rank four built out of the metric and momenta. We can enumerate all possible tensor that can appear in $X^{a b \, \m_1 \n_1 \a_1 \b_1}$ preserving the symmetry of the 
	correlator 
	\begin{align}
		\langle t^{\mu \nu }(p_1)j^{a \, \a}(p_2)j^{b \, \b}(p_3)\rangle_q & =
		{\Pi_1}^{\mu \nu}_{\m_1 \n_1}{\pi_2}^{\a}_{\a_1}{\pi_3}^{\b}_{\b_1}
		\left( A_{1}^{(q) a b} \ p_2^{\m_1 }p_2^{\n_1}p_3^{\a_1}p_1^{\b_1} + 
		A_{2}^{(q) a b}\ \delta^{\a_1 \b_1} p_2^{\m_1}p_2^{\n_1} + 
		A_{3}^{(q) a b}\ \delta^{\m_1\a_1}p_2^{\n_1}p_1^{\b_1}\right. \notag\\
		& \left. + 
		A_{3}^{(q) a b}(p_2\leftrightarrow p_3)\delta^{\m_1\b_1}p_2^{\n_1}p_3^{\a_1}
		+ A_{4}^{(q) a b}\  \delta^{\m_1\b_1}\delta^{\a_1\n_1}\right)\label{DecompTJJ}
	\end{align}
	with the reconstruction taking the form \cite{Bzowski:2013sza}
	\begin{framed}
		\begin{align}
			& \la T_{\mu_1 \nu_1}(\bs{p}_1) J^{\mu_2 a_2}(\bs{p}_2) J^{\mu_3 a_3}(\bs{p}_3) \ra_q = \la t_{\mu_1 \nu_1}(\bs{p}_1) j^{\mu_2 a_2}(\bs{p}_2) j^{\mu_3 a_3}(\bs{p}_3) \ra_q \nn\\[0.5ex]
			& \qquad +  2 \mathscr{T}_{\mu_1 \nu_1}^{\quad\,\,\alpha}(\bs{p}_1)\Big[\delta_{[\alpha}^{\mu_3}p_{3\beta]}  \la J^{\mu_2 a_2}(\bs{p}_2) J^{\beta a_3}(-\bs{p}_2) \ra_q
			+\delta_{[\alpha}^{\mu_2}p_{2\beta]}  \la J^{\mu_3 a_3}(\bs{p}_3) J^{\beta a_2}(-\bs{p}_3) \ra_q\Big]
			\nn\\
			& \qquad + \frac{1}{d - 1}\,\pi_{\mu_1 \nu_1}(\bs{p}_1) \mathcal{A}^{\mu_2 \mu_3 a_2 a_3}_q, \label{tjjdecccc}
		\end{align}
	\end{framed}
	where $\mathscr{T}_{\mu_1 \nu_1\alpha}$ is defined in \eqref{a:Tttt} and $\langle JJ\rangle_q$ is the 2-point function of the gluon with a virtual quark. 
	
	The transverse Ward identities are 
	\begin{align}
		& p_1^{\nu_1} \la T_{\mu_1 \nu_1}({p}_1) J^{\mu_2 a_2}({p}_2) J^{\mu_3 a_3}({p}_3) \ra_q  \nn\\[0.5ex]
		& \qquad = \: 2 \delta^{\mu_3}_{[\mu_1} p_{3\alpha]} \la J^{\mu_2 a_2}({p}_2) J^{\alpha a_3}(-{p}_2) \ra_q + 2 \delta^{\mu_2}_{[\mu_1} p_{2\alpha]} \la J^{\alpha a_2}({p}_3) J^{\mu_3 a_3}(-{p}_3) \ra_q, \\[1ex]\label{TWI_TJJ_2}
		& p_{2 \mu_2} \la T_{\mu_1 \nu_1}({p}_1) J^{\mu_2 a_2}({p}_2) J^{\mu_3 a_3}({p}_3) \ra_q = 0, \\[1ex]
	\end{align}
	while the anomalous trace WI is given by
	\begin{align}
		& \la T({p}_1) J^{\mu_2 a_2}({p}_2) J^{\mu_3 a_3}({p}_3) \ra_q = \mathcal{A}^{\mu_2 \mu_3 a_2 a_3}_q, 
	\end{align}
	where 
	\beq
	\la J^{\alpha a}({p}_1) J^{\beta b}(-{p}_1) \ra_q=
	\frac{2}{3}n_f  \,  \frac{g_s^2}{16\pi ^2} {B}_0\left({{p_1}}^2\right) {p_1}^2   \delta ^{a b}\pi^{\alpha\beta}(p_1).\eeq 
	
	They satisy the renormalized dilatation equations (diagonal in colour space) 
	\cite{Bzowski:2017poo}\cite{Coriano:2018bbe}
	
	\begin{align}
		\left(\sum_{i}^{3}p_i\sdfrac{\partial}{\partial p_i}+2\right)\,A^{(q)}_1&=0=-\m\sdfrac{\partial}{\partial \m}A^{(q)}_1\\
		\left(\sum_{i}^{3}p_i\sdfrac{\partial}{\partial p_i}\right)\,A^{(q)}_2&=\sdfrac{8\p^2\,g^2}{3}=-\m\sdfrac{\partial}{\partial \m}A^{(q)}_2\\
		\left(\sum_{i}^{3}p_i\sdfrac{\partial}{\partial p_i}\right)\,A^{(q)}_3(p_2\leftrightarrow p_3)&=\sdfrac{8\p^2\,g^2}{3}=-\m\sdfrac{\partial}{\partial \m}A^{(q)}_3\\
		\left(\sum_{i}^{3}p_i\sdfrac{\partial}{\partial p_i}-2\right)\,A^{(q)}_4&=-\sdfrac{4}{3}\p^2\,g^2(s-s_1-s_2)=-\m\sdfrac{\partial}{\partial \m}A^{(q)}_4.
	\end{align}
	while for the primary CWI's take the form 
	\begin{equation}
		\label{fst1}
		\begin{split}
			&K_{13}A^{(q)}_1=0\\
			&K_{13}A^{(q)}_2=-2A^{(q)}_1\\
			&K_{13}A^{(q)}_3=4A^{(q)}_1\\
			&K_{13}A^{(q)}_3(p_2\leftrightarrow p_3)=0\\
			&K_{13}A^{(q)}_4=2A^{(q)}_3(p_2\leftrightarrow p_3)-\sdfrac{16\,\pi^2 g^2}{3}
		\end{split}
		\hspace{1.5cm}
		\begin{split}
			&K_{23}A^{(q)}_1=0\\
			&K_{23}A^{(q)}_2=0\\
			&K_{23}A^{(q)}_3=4A^{(q)}_1\\
			&K_{23}A^{(q)}_3(q_2\leftrightarrow q_3)=-4A^{(q)}_1\\
			&K_{23}A^{(q)}_4=-2A^{(q)}_3+2A^{(q)}_3(p_2\leftrightarrow p_3).
		\end{split}
	\end{equation}
	similar to the Abelian case, discussed in \cite{Bzowski:2013sza}\cite{Coriano:2018bbe}
	
	\be
	{ K}_i\equiv \frac{\partial^2}{\partial    p_i \partial    p_i} 
	+\frac{d + 1 - 2 \Delta_i}{   p_i}\frac{\partial}{\partial   p_i} \qquad  K_{l n}\equiv K_l -K_n \qquad i,l,n=1,2,3
	\ee
	In $d=4$ the operator $K_i$ take the forms
	\begin{equation}
		K_1\equiv 4s\sdfrac{\partial^2}{\partial s^2}-4\sdfrac{\partial}{\partial s},\quad K_2\equiv 4s_1\sdfrac{\partial^2}{\partial s_1^2},\quad K_3\equiv 4s_2\sdfrac{\partial^2}{\partial s_2^2}
	\end{equation}
	The solutions of these equations in terms of ordinay master integrals $B_0$ and $C_0$, is completely equivalent to the solutions expressed in terms of 3K integrals, i.e. integrals of the product of three Bessel functions, as discussed in the appendix \ref{dd}. The differential equations for master integrals allow to reformulate the 3K solutions in terms of the ordinary free-field theory realization. \\
	It is possible, in the quark sector, to check the general CFT solution against the perturbative one. 
	For this purpose one can use the identities \cite{Coriano:2018bbe}
	
	\begin{align}
		\sdfrac{\partial}{\partial s}\,C_0&= \sdfrac{1}{s\,\s}\,\big[s(s_1+s_2-s)C_0+B_{0,R}(s_1)(s+s_1-s_2)+B_{0,R}(s_2)(s-s_1+s_2)-2s\,B_{0,R}(s)\big]\\
		\sdfrac{\partial}{\partial s_1}\,C_0&= \sdfrac{1}{s_1\,\s}\,\big[s_1(s+s_2-s_1)C_0+B_{0,R}(s)(s+s_1-s_2)+B_{0,R}(s_2)(s_1-s+s_2)-2s_1\,B_{0,R}(s_1)\big]\\
		\sdfrac{\partial}{\partial s_2}\,C_0&= \sdfrac{1}{s_2\,\s}\,\big[s_2(s+s_1-s_2)C_0+B_{0,R}(s_1)(s_2+s_1-s)+B_{0,R}(s)(s-s_1+s_2)-2s_2\,B_{0,R}(s_2)\big].
	\end{align}
	We have defined $\s=s^2-2s(s_1+s_2)+(s_1-s_2)^2$,  $B_{0,R}(s_i)\equiv B_{0,R}(s_i,0,0)=2-\log\left(-\sdfrac{s_i}{\mu^2}\right)$ and, for simplicity, $C_0\equiv C_0(s,s_1,s_2)$. \\
	The tensor nature of the correlator necessitates the imposition of additional first-order differential constraints, referred to as secondary conformal Ward identities (CWIs) in \cite{Bzowski:2013sza}. 
	These constraints can be solved at specific kinematic points as when the invariant masses of the two photons are equal (\(p_2^2 = p_3^2\)) or in the massless limit of the graviton line. We have left to appendix \ref{secc} a more detailed discussion of the procedure.
	At these points, the undetermined constants in the general solutions of the primary CWIs are constrained. The secondary CWIs are associated with the longitudinal and trace components of the correlators, and consequently, with contact terms.  Defining
	
	\begin{equation}
		A_{j }^{(q) a b}= -\,  n_f\,   \frac{g_s^2}{16\pi^2}\, \bar{A}_j^{{(q)}a b}, \qquad j=1,2\ldots 4
	\end{equation}
	where the label $(q)$ refers to the contribution from a single quark in the first diagram 
	of Fig. 4, a direct computation gives
	
	\begin{eqnarray}
		\label{Abar}
		\bar{A}_1^{(q) ab}  &=& \frac{\delta^{ab}}{48 \left(p_1^2 p_2^2-(p_1 \cdot p_2)^2 \right)^4} \Big[ A_{10} + A_{11}  B_0 (p_1^2) + A_{12} B_0(p_2^2) + A_{13} B_0(q^2) + A_{14} C_0 (p_1^2 , p_2^2 , q^2 ) \Big] \notag \\
		\bar{A}_2^{(q) ab}  &=& -\frac{\delta ^{ab}}{144 \left(p_1^2 p_2^2-(p_1 \cdot p_2)^2 \right)^3} \Big[ A^{(q)}_{20} + A^{(q)}_{21}  B_0 (p_1^2) + A^{(q)}_{22} B_0(p_2^2)  + A^{(q)}_{23} B_0(q^2) + A^{(q)}_{24} C_0 (p_1^2 , p_2^2 , q^2 ) \Big] \notag \\
		\bar{A}_3^{(q) ab}  &=& \frac{\delta ^{ab}}{72 \left(p_1^2 p_2^2-(p_1 \cdot p_2)^2 \right)^3} \Big[A^{(q)}_{31}  B_0 (p_1^2) + A^{(q)}_{32} B_0(p_2^2)  + A^{(q)}_{33} B_0(q^2) + A^{(q)}_{34} C_0 (p_1^2 , p_2^2 , q^2 ) \Big] \notag \\
		\bar{A}_4^{(q) ab}  &=& -\frac{ \delta^{ab}}{72 \left(p_1^2 p_2^2-(p_1 \cdot p_2)^2 \right)^2} \Big[A^{(q)}_{40} + A^{(q)}_{41}  B_0 (p_1^2) + A^{(q)}_{42} B_0(p_2^2) + A^{(q)}_{43} B_0(q^2) + A^{(q)}_{44} C_0 (p_1^2 , p_2^2 , q^2 ) \Big]. \notag \\
	\end{eqnarray}
	The expressions of the $A^{(q)}_i$,\, $ i=2,3,4$, are given in the Appendix \ref{ff}.  They can be isolated from the general expressions in \secref{ff} by selecting the $n_f$ dependent terms of the $A_i^{a b}$. For $i=1$ they are identical in the quark and gluon sectors, where their contributions to the final expression of the vertex differ just by $SU(3)$ colour factors.  \\
	In the explicit evaluation we can use the relations  
	
	\bea
	&  C_0 ( p_1^2,p_2^2,q^2) = \frac{ 1}{q^2} \Phi (x,y),
	\eea
	where the function $\Phi (x, y)$  is defined as
	\cite{Usyukina:1993ch}
	\bea
	\Phi( x, y) &=& \frac{1}{\lambda} \biggl\{ 2 [Li_2(-\rho  x) + Li_2(- \rho y)]  +
	\ln \frac{y}{ x}\ln \frac{1+ \rho y }{1 + \rho x}+ \ln (\rho x) \ln (\rho  y) + \frac{\pi^2}{3} \biggr\},
	\label{Phi}
	\eea
	with
	\bea
	\lambda(x,y) = \sqrt {\Delta},
	\qquad  \qquad \Delta=(1-  x- y)^2 - 4  x  y,
	\label{lambda} \\
	\rho( x,y) = 2 (1-  x-  y+\lambda)^{-1},
	\qquad  \qquad x=\frac{p_1^2}{q^2} \, ,\qquad \qquad y= \frac {p_2^2}{q^2}\, .
	\eea
	and 
	\begin{equation}
		{B}_0(p^2)=\sdfrac{1}{i\p^\frac{d}{2}}\int\,d^d l\ \frac{1}{l^2(l-p_1)^2}=\frac{ \left[\G\left(\frac{d}{2}-1\right)\right]^2\G\left(2-\frac{d}{2}\right)}{\G\left(d-2\right)(p^2)^{2-\frac{d}{2}}}\label{B0ex}.
	\end{equation}
	with 
	\beq
	B_0(p^2)=\frac{1}{\varepsilon}+\bar{B}_0(p^2)
	\eeq
	and
	\beq
	B_0^R(p^2,0,0)\equiv \bar{B}_0(p^2)= 2 + \log(\mu^2/p^2) 
	\label{BBbbbax}
	\eeq
	is the finite part in $d=4$ of the scalar integral in the $\overline{MS}$ scheme. One can check the direct cancellation of the $1/(d-4)$ poles after renormalization using the counterterm \eqref{ct}, and the $B_0$'s can be taken to be of the form $\bar{B}_0(p^2)$ in \eqref{BBbbbax}. We have left to Appendix \ref{secc} a discussion of the secondary (first order) equations for the same quark sector.

	\section{The gluon sector and the complete correlator at one-loop}
	\label{nine}
	The decomposition at one-loop of the gluon sector follows \eqref{dec1111}, 
	but its final expression is modified compared to 
	\eqref{tjjdecccc}, which is affected by new trace contributions not present in the quark sector. This sector provides a contribution of the form
	\begin{framed}
		\begin{equation}
			\begin{aligned}
				& \langle T^{\mu \nu}(q)  J^{ a\alpha}(p_1) J^{ b\beta}(p_2)\rangle_g=\langle t^{\mu \nu}(q)  j^{ a\alpha}(p_1) j^{ b\beta}(p_2)\rangle_g +\langle t^{\mu \nu }(q)j_{loc}^{a  \a}(p_1)j^{b  \b}(p_2)\rangle_g +\langle t^{\mu \nu }(q)j^{a  \a}(p_1)j_{loc}^{b  \b}(p_2)\rangle_g\\
				& \qquad+2 \mathcal{I}^{\mu \nu \rho }(q)\left[\delta_{[\rho}^{\beta} p_{2 \sigma ]}\langle J^{ a\alpha}({p}_1) J^{ b\sigma}(-{p}_1)\rangle_g +\delta_{[\rho}^{\alpha} p_{1 \sigma]}\langle J^{ b\beta}({p}_2) J^{ a\sigma}(-p_2)\rangle\right]_g  
				+\frac{1}{d-1} \pi^{\mu \nu}(q) \left[\mathcal{A}^{\alpha \beta a b}_g+\mathcal{B}^{\alpha \beta a b}_g\right]
			\end{aligned}
			\label{res}
		\end{equation}
	\end{framed} 
	There are additional local terms of the form 
	\beq
	\langle t^{\mu \nu }(q)j^{a \, \a}(p_1)j_{loc}^{b \, \b}(p_2)\rangle_g 
	\label{addsaxz}
	\eeq
	which are not part of the transverse traceless sector, but appear in the longitudinal sector of the decomposition. As we are going to discuss below, these terms are not set to zero by the Slavnov-Taylor identities, as in the quark sector, or in the general conformal solution.  
	At the same time, the trace sector is modified by the presence of extra terms which are proportional to the equations of motion of the gluons, here indicated as $\mathcal{B}^{\alpha \beta a b}_g, $ which are absent in the on-shell decomposition. Defining  
	\begin{allowdisplaybreaks}
		\begin{equation}
			A_{i }^{(g) a b}= -\,  C_A\,   \frac{g_s^2}{16\pi^2}\, \bar{A}_i^{{(g)}a b}, \qquad i=1,2\ldots 4
		\end{equation}
		$(C_A=N_c=3)$, where the functions $\bar{A}_i^{{(g)}a b}$ are extracted from  \secref{ff} by selecting the part of the 
		$A_i$ proportional to the Casimir $C_A$, their explicit expressions are given by 
		\begin{eqnarray}
			\label{Abar}
			\bar{A}_1^{(g) ab}  &=& -\frac{\delta^{ab}}{48 \left(p_1^2 p_2^2-(p_1 \cdot p_2)^2 \right)^4} \Big[ A_{10} + A_{11}  B_0 (p_1^2) + A_{12} B_0(p_2^2) + A_{13} B_0(q^2) + A_{14} C_0 (p_1^2 , p_2^2 , q^2 ) \Big] \notag \\
			\bar{A}_2^{(g) ab}  &=& -\frac{\delta ^{ab}}{144 \left(p_1^2 p_2^2-(p_1 \cdot p_2)^2 \right)^3} \Big[ A^{(g)}_{20} + A^{(g)}_{21}  B_0 (p_1^2) + A^{(g)}_{22} B_0(p_2^2)  + A^{(g)}_{23} B_0(q^2) + A^{(g)}_{24} C_0 (p_1^2 , p_2^2 , q^2 ) \Big] \notag \\
			\bar{A}_3^{(g) ab}  &=& \frac{\delta ^{ab}}{72 \left(p_1^2 p_2^2-(p_1 \cdot p_2)^2 \right)^3} \Big[A^{(g)}_{31}  B_0 (p_1^2) + A^{(g)}_{32} B_0(p_2^2)  + A^{(g)}_{33} B_0(q^2) + A^{(g)}_{34} C_0 (p_1^2 , p_2^2 , q^2 ) \Big] \notag \\
			\bar{A}_4^{(g) ab}  &=& -\frac{ \delta^{ab}}{72 \left(p_1^2 p_2^2-(p_1 \cdot p_2)^2 \right)^2} \Big[A^{(g)}_{40} + A^{(g)}_{41}  B_0 (p_1^2) + A^{(g)}_{42} B_0(p_2^2) + A^{(g)}_{43} B_0(q^2) + A^{(g)}_{44} C_0 (p_1^2 , p_2^2 , q^2 ) \Big]. \notag \\ 
		\end{eqnarray}
	\end{allowdisplaybreaks}
	The new longitudinal terms \ref{addsaxz} take the form 
	\begin{framed}
		\begin{align}
			\langle t^{\mu \nu }(q)j^{a \, \a}(p_1)j_{loc}^{b \, \b}(p_2)\rangle_g & =
			{\Pi}^{\mu \nu}_{\m_1 \n_1}(q) \, \pi^\a _{\a_1} (p_1)\, {{p_2}_{\beta_1} p_2^\beta \over p_2^2}
			\left( 	B^{ab}_1 \, p_1^{\m_1} \, p_1^{\nu_1} \, p_2^{\a_1} \, p_2^{\b_1} + B_2^{ab} \, p_1^{\m_1} \, p_2^{\b_1} \, \delta^{\a_1 \n_1}  \right) 
		\end{align}
	\end{framed}
	which is orthogonal to the trace sector. Notice that these local contributions vanish when the gluons are on-shell. The $B_i, \,\, i=1,2$ are given by  
	
	\begin{eqnarray}
		B^{ab}_1 & = &  \frac{ \,  \,  \, C_A \,  g_s^2 \,  \delta^{ab} \, p_1^2}{64\pi^2 \, p_2^2 \, \big(p_1^2 p_2^2 - (p_1 \cdot p_2)^2 \big)^2} \, \Big( -\Big[2 \, (p_1 \cdot p_2)^2 + p_1^2 p_2^2 + 3 \, p_1^2 p_1 \cdot p_2 \Big]\, B_0(p_1^2)  \notag \\
		&& -\Big[2\, (p_1 \cdot p_2)^2 + p_1^2 p_2^2 + 3 \, p_2^2 p_1 \cdot p_2 \Big] \, B_0(p_2^2) +  \Big[ 4 \, (p_1 \cdot p_2)^2 + 2 \, p_1^2 p_2^2 + 3\, (p_1^2 + p_2^2) p_1 \cdot p_2 \Big]\, B_0(q^2)  \notag \\
		&&  +\Big[  q^2 \, (p_1 \cdot p_2)^2 + 2 \, q^2 \, (p_1 \cdot p_2)^2  \Big] \,C_0(p_1^2, p_2^2, q^2) - 2 \, (p_1 \cdot p_2)^2 - 2 \, p_1^2 p_2^2 \Big) \\
		B^{ab}_2 & = & \frac{ \,  \,  \, C_A \,  g_s^2 \,  \delta^{ab} \, p_1^2}{32\pi^2 \, p_2^2 \, \big(p_1^2 p_2^2 - (p_1 \cdot p_2)^2 \big)^2} \Big( - p_1 \cdot p_2 \, B_0(p_1^2)  - p_2^2 \cdot p_2  \, B_0(p_2^2) + (p_2^2+p_1 \cdot p_2) \, B_0(q^2) \notag \\ &&
		+ \Big[p_1^2 p_2^2 + p_2^2 \, (p_1 \cdot p_2) \Big] \, C_0(p_1^2, p_2^2 , q^2) \Big).
	\end{eqnarray}
	
	The trace sector is also affected by terms that vanish for 
	on-shell gluons, indicated as $\mathcal{B}_g^{\alpha \beta a b}$, as well as the genuine anomaly term 
	$\mathcal{A}_g^{\alpha \beta a b}$. Defining 
	\begin{framed}
		\beq
		\mathcal{B}_g^{\alpha \beta a b}=\left[ C_1^{ab} \, p_1^\alpha \, p_1 ^\beta + C_2^{ab} \, p_1^\alpha \, p_2^\beta + C_3^{ab}  \, p_1^\beta \, p_2^\alpha + C_4^{ab} \, p_2^\alpha \, p_2^\beta + C_5^{ab} \, \delta^{\alpha \beta} \right]  
		\label{bbcdas}
		\eeq
	\end{framed}
	the trace sector is characterized by the anomaly contributions from the gluon sector $\mathcal{A}_g$, plus the 
	$\mathcal{B}_g$ terms proportional to the equations of motion of the gluons 
	\begin{eqnarray}
		&\mathcal{A}_g^{\alpha \beta a b}+\mathcal{B}_g^{\alpha \beta a b}=\left. g_{\mu \nu} \langle T^{\mu \nu} (q) \, J^{a \a} (p_1) \, J^{b \b} (p_2)\, \rangle_g \right| \\&
	\end{eqnarray}
	with the gluon contribution to the anomaly given by
	
	\begin{eqnarray}
		\mathcal{A}_g^{\alpha\beta ab} &=& {11 \over 3} \,  \,  \,  \, \frac{g_s^2}{16\pi^2} \, C_A  \delta^{ab}u^{\alpha\beta}(p_1,p_2)
	\end{eqnarray}
	and $u^{\alpha\beta}$ defined in \eqref{locvarasd}.
	The form factors proportional to the equations of motion in \eqref{bbcdas} take the form
	{\allowdisplaybreaks
	\begin{eqnarray}
		C_1^{ab} &=& -\frac{ \,  \,  \, C_A \, g_s^2 \, \delta^{ab} }{ 32\pi^2 \big( (p_1 \cdot p_2)^2 - p_1^2 \, p_2^2 \big)} \, \Big(
		2 \, (p_1 \cdot p_2)^2 - 2 \, p_1^2 \, p_2^2 
		- p_1^2 \, (p_1 \cdot p_2 + p_2^2 ) B_0 (p_1^2)  \notag \\ &&
		+ \big[ p_2^2 \, (p_1^2 - p_1 \cdot p_2) - 2 \, (p_1 \cdot p_2)^2 \big]  B_0(p_2^2)
		+ (p_1 \cdot p_2) \, (2 \, p_1 \cdot p_2 + p_1^2 + p_2^2) \, B_0 (q^2)  \notag \\ &&
		+ \big[ 2\, (p_1 \cdot p_2)^2 \, (p_1 \cdot p_2)^2 - p_2^2 \, \big(4 (p_1 \cdot p_2)^2   + p_1^4\big) + 5 \, p_1^2 \, p_4^4  \big] C_0(p_1^2, p_2^2 , q^2)  \Big) \\
		C_2^{ab} &=& -2  \,  \,  \, C_A \, \frac{g_s^2}{16\pi^2} \, \delta^{ab} \, (p_1 \cdot p_2) \, C_0 (p_1^2 , p_2^2 , q^2) \\
		C_3^{ab} &=& -\frac{ \,  \,  \, g_s^2 \, C_A \, (p_1^2 + p_2^2) \, \delta^{ab}}{32\pi^2 \big(p_1^2 \, p_2^2 - (p_1 \cdot p_2)^2 \big)} \Big( 
		(-p_1\cdot p_2-p_1^2)  B_0(p_1^2)+(-(p_1\cdot p_2)-p_2^2) \, B_0(p_2^2) \notag \\ &&
		+q^2 \, B_0(q^2)+\big(p_1^2 \, (p_1\cdot p_2-2 p_2^2) +(p_1\cdot p_2) \, \big(4 \, (p_1\cdot p_2)+p_2^2) \big) \, C_0(p_1^2,p_2^2, q^2)  \Big)  \\
		C_4^{ab} &=&  -\frac{ \, \,  \, g_s^2 \, C_A \, g_s^2 \,  \delta^{ab}}{32\pi^2\big((p_1 \cdot p_2)^2 - p_1^2 \, p_2^2 \big)} \Big( 
		(p_1^2 \, (p_2^2-p_1\cdot p_2)\notag \\ &&  -2 (p_1\cdot p_2)^2) \, B_0(p_1^2) -p_2^2 \, (p_1\cdot p_2+p_1^2) \, B_0(p_2^2)+(p_1\cdot p_2) \, q^2 \, B_0(q^2) \notag \\ &&  
		+ \big[-p_1^2 \, (4 \, (p_1\cdot p_2)^2+p_2^4)+2 \, (p_1\cdot p_2+p_2^2) \, (p_1\cdot p_2)^2+5 \, p_1^4 \, p_2^2 \, \big] \, C_0(p_1^2,p_2^2,q^2) \, \Big) \\
		C_5^{ab} &=& \, \,  \, \frac{g_s^2}{32\pi^2} \, C_A \,  \delta^{ab} \Big( (p_1^2 - p_2^2) \, \big[ B_0 (p_1^2) - B_0(p_2^2) \big] 
		+ \big[ p_1^4 + p_2^4 - 2 \, (p_1^2 + p_2^2) \, p_1 \cdot p_2 - 6 \, p_1^2 \, p_2^2 \big] \, C_0(p_1^2, p_2^2, q^2) \Big). \nonumber \\
	\end{eqnarray}
}

	\section{The general structure of the TJJ from the CFT decomposition}
	\label{ten}
	We may combine the parameterization of both sectors in order to derive the general expression of the hard scattering vertex. It is given by the expression 
	
	\begin{framed}
		\begin{equation}
			\begin{aligned}
				& \langle T^{\mu \nu}(q)  J^{ a\alpha}(p_1) J^{ b\beta}(p_2)\rangle=\langle t^{\mu \nu}(q)  j^{ a\alpha}(p_1) j^{ b\beta}(p_2)\rangle+\langle t^{\mu \nu }(q)j_{loc}^{a  \a}(p_1)j^{b  \b}(p_2)\rangle_g +\langle t^{\mu \nu }(q)j^{a  \a}(p_1)j_{loc}^{b  \b}(p_2)\rangle_g\\
				& \qquad+2 \mathcal{I}^{\mu \nu \rho }(q)\left[\delta_{[\rho}^{\beta} p_{2 \sigma ]}\langle J^{ a\alpha}({p}_1) J^{ b\sigma}(-{p}_1)\rangle+\delta_{[\rho}^{\alpha} p_{1 \sigma]}\langle J^{ b\beta}({p}_2) J^{ a\sigma}(-p_2)\rangle\right] 
				+\frac{1}{ 3 \, q^2} \hat\pi^{\mu \nu}(q) \left[\mathcal{A}^{\alpha \beta a b}+\mathcal{B}^{\alpha \beta a b}_g\right]
			\end{aligned}
			\label{ressa}
		\end{equation}
	\end{framed}

	where the traceless sector is
	\begin{equation}
		\begin{aligned}
			\langle T^{\mu \nu}(q)  J^{ a\alpha}(p_1) J^{ b\beta}(p_2)\rangle_{tls}=&
			\langle t^{\mu \nu}(q)  j^{ a\alpha}(p_1) j^{ b\beta}(p_2)\rangle+\langle t^{\mu \nu }(q)j_{loc}^{a  \a}(p_1)j^{b  \b}(p_2)\rangle_g +\langle t^{\mu \nu }(q)j^{a  \a}(p_1)j_{loc}^{b  \b}(p_2)\rangle_g\\
			& \qquad+2 \mathcal{I}^{\mu \nu \rho }(q)\left[\delta_{[\rho}^{\beta} p_{2 \sigma ]}\langle J^{ a\alpha}({p}_1) J^{ b\sigma}(-{p}_1)\rangle+\delta_{[\rho}^{\alpha} p_{1 \sigma]}\langle J^{ b\beta}({p}_2) J^{ a\sigma}(-p_2)\rangle\right] 
		\end{aligned}
		\label{aliqqa}
	\end{equation}
	
	and the trace part is 
	
	\beq
	\label{trc}
	\langle T^{\mu \nu}(q)  J^{ a\alpha}(p_1) J^{ b\beta}(p_2)\rangle_{tr}=\frac{1}{ 3 \, q^2} \hat\pi^{\mu \nu}(q) \left[\mathcal{A}^{\alpha \beta a b}+\mathcal{B}^{\alpha \beta a b}_g\right].
	\eeq
	%Both $\mathcal{A}$ and $\mathcal{B}$
	The trace part contains the anomaly contribution 
	$\mathcal{A}^{\alpha \beta a b}=\mathcal{A}^{\alpha \beta}\delta^{ab} $and a second term proportional to the gluon equations of motion $\mathcal{B}^{\alpha \beta a b}_g$. This is not to be considered part of the trace anomaly although it is part of the trace of the correlator since 
	\beq
	g_{\mu\nu}\langle T^{\mu \nu}(q)  J^{ a\alpha}(p_1) J^{ b\beta}(p_2)\rangle_{tr}= \left[\mathcal{A}^{\alpha \beta a b}+\mathcal{B}^{\alpha \beta a b}_g\right].
	\eeq
	The first tensor term $\mathcal{A}$ "projects into $FF$" :
	\beq
	\mathcal{A}^{\alpha\beta a b}(p_1,p_2)= A_n \delta^{a b} u^{\alpha\beta}(p_1,p_2)
	\eeq
	and it is explicitly given by  
	\beq
	\mathcal{A}^{\alpha\beta ab} = {1 \over 3} \, \,  \, \, \frac{g_s^2}{16\pi^2} \, (11C_A - 2 n_f) \delta^{ab}u^{\alpha \beta}(p_1,p_2),
	\eeq
	defining the conformal anomaly term coming from the quark and gluon sectors. $\mathcal{B}^{\alpha \beta a b}_g$ is given in \eqref{bbcdas}. The anomaly form factor characterising the exchange of a 
	dilaton pole in the $TJJ$ is then distilled in the form 
	\be
	\frac{\beta}{q^2} \delta^{ab} \subset \langle T^{\mu \nu}(q)  J^{ a\alpha}(p_1) J^{ b\beta}(p_2)\rangle
	\ee
	shown in Fig.~\ref{scvertex},
	where the $1/q^2$ anomaly pole has been explicitly extracted from the longitudinal projector $\pi^{\mu\nu}$ of the last term on the right-hand side of \eqref{ressa} by defining 
	\beq
	\hat\pi^{\mu\nu}\equiv q^2 g^{\mu\nu} - q^\mu q^\nu.
	\eeq
	Eq. \eqref{ressa} shows that the structure of the effective vertex corresponding to the $TJJ$ 
	correlator is modified compared to the ordinary CFT case, encountered in \eqref{tjjdecccc},
	with modifications affecting both the longitudinal and the trace sectors. \\
	As we have demonstrated, the decomposition is constructed using appropriate projectors, enabling a separate analysis of the different sectors of the correlator. Consequently, it remains valid at all orders in perturbation theory.
	
	\subsection{The off-shell decomposition and the anomaly form factor for massive quarks}
	Let's now consider massive quarks. The decomposition given in \eqref{aliqqa} remains valid, but the trace sector acquires a different form, that is naturally decomposed into two contributions: a trace part which is directly projected on the tensor structure $u^{\alpha\beta}$, therefore projected onto $FF$, indicated as $\phi_1$, and a second form factor 
	generalizing the $\mathcal{B}^{\alpha \beta a b}_g$ contributions given above, once we allow for massive quarks. This second term is denoted as $\phi_2$. The modifications are only present in the quark sector. This sector gives
	
	\begin{equation}
		g_{\m\n}\braket{T^{\m\n}J^{\a a}J^{\b b}}_q=\left(\phi_1^{\a\b}(p_1,p_2,q,m)+\phi_2^{\a\b}(p_1,p_2,q,m)\right)\d^{ab}
	\end{equation}
	
	where
	
	\begin{equation}
		\phi_1^{\alpha\beta a b}=\left(-\frac{2}{3}   n_f  \frac{g_s^2}{16\pi^2}+ \chi_0 (p_1,p_2,q,m)\right)\d^{ab} u^{\a\b}(p_1,p_2)
	\end{equation}
	contains the $n_f$ contribution to the anomaly pole, now with an extra term generated by the mass dependence. The $\chi_0$ function is defined as follows
	\beq
	\chi_0(p_1,p_2,q,m)\equiv \frac{B_1 - B_4 }{2\, p_1\cdot p_2}+\frac{2}{3}   n_f  \frac{g_s^2}{16\pi^2}\,,
	\eeq
	in order to make the anomaly contribution explicit.
	The trace contains extra terms that vanish upon use of the equations of motion, related to the tensor structure 
	${\phi_2}$, given by 
	
	\begin{equation}
		\phi_2^{\a\b}(p_1,p_2,q,m)= \chi_1(p_1,p_2,q,m) v^{\a\b}+ B_2 p_1^{\a}p_1^{\b}+B_2(p_1\leftrightarrow p_2) p_2^{\a}p_2^{\b}+B_3 p_1^{\a}p_2^{\b}.
	\end{equation}
	where
	\beq
	\chi_1(p_1,p_2,q,m)\equiv \frac{B_1 + B_4  }{2\, p_1\cdot p_2 }
	\eeq
	with
	\begin{equation}
		v^{\a\b}(p_1,p_2)=(p_1\cdot p_2) g^{\a\b}+p_2^\a p_1^\b
	\end{equation}
	
	and with $B_1\ldots B_4$ given below. Explicitly

	{\allowdisplaybreaks\smaller 
		\begin{align}
			\chi_0=&\frac{3 m^4 \left(p_1^4+p_2^4+q^4-2 \left(p_1^2+p_2^2\right) q^2\right) \text{C}_0\left(p_1^2,p_2^2,q^2,m^2,m^2,m^2\right) g_s^2 }{2 \pi ^2 \left(-p_1^2-p_2^2+q^2\right) \left(q^4-2 \left(p_1^2+p_2^2\right) q^2+\left(p_1^2-p_2^2\right)^2\right)}\nonumber\\&+m^2 g_s^2 \Biggl(\frac{3 \left(p_1^4+p_2^4+q^4-2 \left(p_1^2+p_2^2\right) q^2\right) }{4 \pi ^2 \left(-p_1^2-p_2^2+q^2\right) \left(q^4-2 \left(p_1^2+p_2^2\right) q^2+\left(p_1^2-p_2^2\right)^2\right)}  \nonumber\\&\quad\quad\quad\quad +\frac{3\, p_2^2\left(p_2^6+\left(p_1^2-3 q^2\right) p_2^4+\left(p_1^4+3 q^4\right) p_2^2-\left(q^2-p_1^2\right)^2 \left(3 p_1^2+q^2\right)\right)\bar B_0(p_2^2,m^2)}{4 \pi ^2 \left(-p_1^2-p_2^2+q^2\right) \left(q^4-2 \left(p_1^2+p_2^2\right) q^2+\left(p_1^2-p_2^2\right)^2\right)^2}  \nonumber\\&\quad\quad\quad\quad -\frac{3\, p_1^2\left(-3 p_2^6+\left(p_1^2+5 q^2\right) p_2^4+\left(p_1^4-q^4\right) p_2^2-\left(q^2-p_1^2\right)^3\right)\bar B_0(p_1^2,m^2)) }{4 \pi ^2 \left(p_1^2+p_2^2-q^2\right) \left(q^4-2 \left(p_1^2+p_2^2\right) q^2+\left(p_1^2-p_2^2\right)^2\right)^2}  \nonumber\\&\quad\quad\quad\quad +\frac{3 \,q^2 \left(-p_2^8+\left(2 p_1^2+3 q^2\right) p_2^6-\left(p_1^2+q^2\right) \left(2 p_1^2+3 q^2\right) p_2^4+\left(q^2-p_1^2\right) \left(-2 p_1^4+3 q^2 p_1^2+q^4\right) p_2^2+p_1^2 \left(q^2-p_1^2\right)^3\right)  \bar B_0(q^2,m^2)}{4 \pi ^2 q^2 \left(-p_1^2-p_2^2+q^2\right) \left(q^4-2 \left(p_1^2+p_2^2\right) q^2+\left(p_1^2-p_2^2\right)^2\right)^2}   \nonumber\\&\quad\quad\quad\quad -\frac{3 P_0 C_0\left(p_1^2,p_2^2,q^2,m^2,m^2,m^2\right)}{8 \pi ^2 \left(-p_1^2-p_2^2+q^2\right) \left(q^4-2 \left(p_1^2+p_2^2\right) q^2+\left(p_1^2-p_2^2\right)^2\right)^2}\Biggl)
		\end{align}
	} 
	where we have defined the function

	\begin{align}
		P_0= &-p_2^{10}+\left(p_1^2+5 q^2\right) p_2^8-2 q^2 \left(3 p_1^2+5 q^2\right) p_2^6+2 \left(5 q^2-3 p_1^2\right) q^2 \left(p_1^2+q^2\right) p_2^4\nonumber\\&-\left(q^2-p_1^2\right)^2 \left(p_1^2+q^2\right) \left(5 q^2-p_1^2\right) p_2^2+\left(q^2-p_1^2\right)^5
	\end{align}
	$\bar B_0(p^2,m^2,m^2)$ stands for the renormalized 2-point function while $C_0(p_1^2,p_2^2,q^2,m^2,m^2,m^2) $ denotes the ordinary scalar off-shell 3-point function given in Appendix \ref{scalars}.
	
	The scalar form factors related to terms in the trace which are proportional to the equations of motion are given by
	
	{\smaller\allowdisplaybreaks
		\begin{align}
			B_1=& - \frac{g_s^2}{16\p^2} m^2    \Biggl(\frac{3     \left(p_1^6-p_1^4 \left(p_2^2+3 q^2\right)-p_1^2 \left(p_2^2-3 q^2\right) \left(p_2^2+q^2\right)+\left(p_2^2-q^2\right)^3\right) C_0\left(p_1^2,p_2^2,q^2,m^2,m^2,m^2\right)}{p_1^4-2 p_1^2 \left(p_2^2+q^2\right)+\left(p_2^2-q^2\right)^2}
			\nn\\
			&\qquad \qquad \qquad +\frac{6   p_1^2 \bar B_0(p_1^2,m^2) \left(p_1^2-p_2^2-q^2\right)}{p_1^4-2 p_1^2 \left(p_2^2+q^2\right)+\left(p_2^2-q^2\right)^2} 
			+\frac{6 q^2\bar  B_0(q^2,m^2)   \left(q^2 \left(p_1^2+p_2^2\right)-\left(p_1^2-p_2^2\right)^2\right)}{q^2 \left(p_1^4-2 p_1^2 \left(p_2^2+q^2\right)+\left(p_2^2-q^2\right)^2\right)} \nn\\
			&\qquad \qquad \qquad  -\frac{6 p_2^2
				\bar B_0(p_2^2,m^2)\left(p_1^2-p_2^2+q^2\right)}{p_1^4-2 p_1^2 \left(p_2^2+q^2\right)+\left(p_2^2-q^2\right)^2} +6 \Biggl)-12   m^4 \frac{g_s^2}{16\p^2} C_0\left(p_1^2,p_2^2,q^2,m^2,m^2,m^2\right)
		\end{align}

		\begin{align}
			B_2= & - \frac{g_s^2}{16\p^2} m^2 \Biggl(\frac{12    p_2^2  \left(p_1^6-p_1^4 \left(p_2^2+q^2\right)-p_1^2 \left(p_2^4-6 p_2^2 q^2+q^4\right)+\left(p_2^2-q^2\right)^2 \left(p_2^2+q^2\right)\right) C_0\left(p_1^2,p_2^2,q^2,m^2,m^2,m^2\right)}{\left(p_1^4-2 p_1^2 \left(p_2^2+q^2\right)+\left(p_2^2-q^2\right)^2\right)^2}\nn\\ &\qquad \qquad \qquad+\frac{24    p_2^2 p_1^2\bar B_0(p_1^2,m^2)\left(2 p_1^4-p_1^2 \left(p_2^2+q^2\right)-\left(p_2^2-q^2\right)^2\right)}{p_1^2 \left(p_1^4-2 p_1^2 \left(p_2^2+q^2\right)+\left(p_2^2-q^2\right)^2\right)^2}\nn\\ &\qquad \qquad \qquad-\frac{24    p_2^2q^2\bar B_0(q^2,m^2)\left(q^2 \left(p_1^2+p_2^2\right)+\left(p_1^2-p_2^2\right)^2-2 q^4\right)}{q^2 \left(p_1^4-2 p_1^2 \left(p_2^2+q^2\right)+\left(p_2^2-q^2\right)^2\right)^2}\nn\\ &\qquad \qquad \qquad-\frac{24    p_2^2\bar B_0(p_2^2,m^2) \left(p_1^4+p_1^2 \left(p_2^2-2 q^2\right)-2 p_2^4+p_2^2 q^2+q^4\right)}{\left(p_1^4-2 p_1^2 \left(p_2^2+q^2\right)+\left(p_2^2-q^2\right)^2\right)^2}\nn\\ &\qquad \qquad \qquad+\frac{24    p_2^2  }{p_1^4-2 p_1^2 \left(p_2^2+q^2\right)+\left(p_2^2-q^2\right)^2}\Biggl)-\frac{3   m^4 p_2^2 g_s^2 C_0\left(p_1^2,p_2^2,q^2,m^2,m^2,m^2\right)}{\p^2\left(p_1^4-2 p_1^2 \left(p_2^2+q^2\right)+\left(p_2^2-q^2\right)^2\right)}
		\end{align}

		%\begin{align}
		%    A_3=& m^2 \Biggl(\frac{12 i \pi ^2 \kappa  p_1^2 g_s^2 \delta ^{a b} \left(p_1^6-p_1^4 \left(p_2^2+q^2\right)-p_1^2 \left(p_2^4-6 p_2^2 q^2+q^4\right)+\left(p_2^2-q^2\right)^2 \left(p_2^2+q^2\right)\right) C_0}{\left(p_1^4-2 p_1^2 \left(p_2^2+q^2\right)+\left(p_2^2-q^2\right)^2\right)^2}\nn\\ &+\frac{24 i \pi ^2 \kappa  g_s^2 \sqrt{p_1^2 \left(p_1^2-4 m^2\right)} \delta ^{a b} \log \left(\s(p_1,m)\right) \left(2 p_1^4-p_1^2 \left(p_2^2+q^2\right)-\left(p_2^2-q^2\right)^2\right)}{\left(p_1^4-2 p_1^2 \left(p_2^2+q^2\right)+\left(p_2^2-q^2\right)^2\right)^2}\nn\\ &-\frac{24 i \pi ^2 \kappa  p_1^2 g_s^2 \sqrt{q^2 \left(q^2-4 m^2\right)} \delta ^{a b} \log \left(\s(q,m)\right) \left(q^2 \left(p_1^2+p_2^2\right)+\left(p_1^2-p_2^2\right)^2-2 q^4\right)}{q^2 \left(p_1^4-2 p_1^2 \left(p_2^2+q^2\right)+\left(p_2^2-q^2\right)^2\right)^2}\nn\\ &-\frac{24 i \pi ^2 \kappa  p_1^2 g_s^2 \sqrt{p_2^2 \left(p_2^2-4 m^2\right)} \delta ^{a b} \log \s(p_2,m) \left(p_1^4+p_1^2 \left(p_2^2-2 q^2\right)-2 p_2^4+p_2^2 q^2+q^4\right)}{p_2^2 \left(p_1^4-2 p_1^2 \left(p_2^2+q^2\right)+\left(p_2^2-q^2\right)^2\right)^2}\nn\\ &+\frac{24 i \pi ^2 \kappa  p_1^2 g_s^2 \delta ^{a b}}{p_1^4-2 p_1^2 \left(p_2^2+q^2\right)+\left(p_2^2-q^2\right)^2}\Biggl)+\frac{48 i \pi ^2 \kappa  m^4 p_1^2 g_s^2 \delta ^{a b} C_0}{p_1^4-2 p_1^2 \left(p_2^2+q^2\right)+\left(p_2^2-q^2\right)^2}
		%\end{align}
		
		\begin{align}
			B_3=&- \frac{g_s^2}{16\p^2} m^2 \Biggl(\frac{6   \left(p_1^2+p_2^2-q^2\right) \left(p_1^6-p_1^4 \left(p_2^2+q^2\right)-p_1^2 \left(p_2^4-6 p_2^2 q^2+q^4\right)+\left(p_2^2-q^2\right)^2 \left(p_2^2+q^2\right)\right) C_0\left(p_1^2,p_2^2,q^2,m^2,m^2,m^2\right)}{\left(p_1^4-2 p_1^2 \left(p_2^2+q^2\right)+\left(p_2^2-q^2\right)^2\right)^2} \nn\\ &\qquad \qquad \qquad+\frac{12  p_1^2\bar B_0(p_1^2,m^2)\left(p_1^2+p_2^2-q^2\right) \left(2 p_1^4-p_1^2 \left(p_2^2+q^2\right)-\left(p_2^2-q^2\right)^2\right)}{p_1^2 \left(p_1^4-2 p_1^2 \left(p_2^2+q^2\right)+\left(p_2^2-q^2\right)^2\right)^2}\nn\\ &\qquad \qquad \qquad-\frac{12   q^2\bar B_0(q^2,m^2) \left(-3 q^4 \left(p_1^2+p_2^2\right)+4 p_1^2 p_2^2 q^2+\left(p_1^2-p_2^2\right)^2 \left(p_1^2+p_2^2\right)+2 q^6\right)}{q^2 \left(p_1^4-2 p_1^2 \left(p_2^2+q^2\right)+\left(p_2^2-q^2\right)^2\right)^2}\nn\\ &\qquad \qquad \qquad-\frac{12  p_2^2\bar B_0(p_2^2,m^2) \left(p_1^2+p_2^2-q^2\right) \left(p_1^4+p_1^2 \left(p_2^2-2 q^2\right)-2 p_2^4+p_2^2 q^2+q^4\right)}{p_2^2 \left(p_1^4-2 p_1^2 \left(p_2^2+q^2\right)+\left(p_2^2-q^2\right)^2\right)^2}\nn\\ &\qquad \qquad \qquad+\frac{12    \left(p_1^2+p_2^2-q^2\right)}{p_1^4-2 p_1^2 \left(p_2^2+q^2\right)+\left(p_2^2-q^2\right)^2}\Biggl)-\frac{3   m^4 g_s^2  \left(p_1^2+p_2^2\right) C_0\left(p_1^2,p_2^2,q^2,m^2,m^2,m^2\right)}{2\p^2\left(p_1^4-2 p_1^2 \left(p_2^2+q^2\right)+\left(p_2^2-q^2\right)^2\right)}
		\end{align}
		
		\begin{align}
			B_4=&- \frac{g_s^2}{16\p^2} m^2 \, p_1\cdot p_2 \Biggl(\frac{6   \left(p_1^2+p_2^2-q^2\right) \left(p_1^6-p_1^4 \left(p_2^2+3 q^2\right)+p_1^2 \left(-p_2^4+10 p_2^2 q^2+3 q^4\right)+\left(p_2^2-q^2\right)^3\right) C_0\left(p_1^2,p_2^2,q^2,m^2,m^2,m^2\right)}{\left(p_1^4-2 p_1^2 \left(p_2^2+q^2\right)+\left(p_2^2-q^2\right)^2\right)^2}\nn\\ &\qquad \qquad \qquad\qquad+\frac{12  p_2^2\bar B_0(p_2^2m^2) \left(-5 p_1^4+4 p_1^2 \left(p_2^2+q^2\right)+\left(p_2^2-q^2\right)^2\right)}{\left(p_1^4-2 p_1^2 \left(p_2^2+q^2\right)+\left(p_2^2-q^2\right)^2\right)^2}\nn\\ &\qquad \qquad \qquad\qquad+\frac{12   p_1^2\bar B_0(p_1^2,m^2)\left(p_1^4+p_1^2 \left(4 p_2^2-2 q^2\right)-5 p_2^4+4 p_2^2 q^2+q^4\right)}{\left(p_1^4-2 p_1^2 \left(p_2^2+q^2\right)+\left(p_2^2-q^2\right)^2\right)^2}\nn\\ &\qquad \qquad \qquad\qquad-\frac{12   q^2\bar B_0(q^2,m^2)\left(q^4 \left(p_1^2+p_2^2\right)+\left(p_1^2-p_2^2\right)^2 \left(p_1^2+p_2^2\right)-2 q^2 \left(p_1^4-4 p_1^2 p_2^2+p_2^4\right)\right)}{q^2 \left(p_1^4-2 p_1^2 \left(p_2^2+q^2\right)+\left(p_2^2-q^2\right)^2\right)^2}\nn\\ &\qquad \qquad \qquad\qquad+\frac{12   \left(p_1^2+p_2^2-q^2\right)}{p_1^4-2 p_1^2 \left(p_2^2+q^2\right)+\left(p_2^2-q^2\right)^2}\Biggl)-  \frac{ 3\, m^4 \,  g_s^2 \left(p_1^2+p_2^2-q^2\right)^2 C_0\left(p_1^2,p_2^2,q^2,m^2,m^2,m^2\right)}{2\left(p_1^4-2 p_1^2 \left(p_2^2+q^2\right)+\left(p_2^2-q^2\right)^2\right)}
		\end{align}
	}
	
	The corresponding form factors $A^{(q)}_i$ for the transverse traceless part of the massive case can be found in Appendix \ref{aiq}.
	
	\subsection{Comments on the structure of the decomposition}
	At this stage, we pause for some comments concerning the structure of this result also with respect to the  prediction for this correlator coming from CFT, in a non-Lagrangian formulation, as discussed in \cite{Bzowski:2013sza} both in the Abelian and non-Abelian cases. \\
	Since the gauge fixing condition breaks conformal symmetry in the non-Abelian case for a Lagrangian theory, the reconstruction of  the 
	correlator  allows extra non-conformal terms. 
	As we have discussed in the previous ections, in  the conformal case, the Ward identity on the two gauge currents are imposed by contracting the correlator with momenta either $p_1$ or $p_2$. In the case of a gauge-fixed theory, instead, the relevant STI requires a quadratic contraction with both momenta $p_1$ and $p_2$ in the form given by \eqref{brstwill} that replaces the ordinary single derivative WIs. 
	Therefore \eqref{ressa} differs in its structure compared to \eqref{tjjdecccc}, which is typical of a conformal theory. \\
	The first term, $\langle t^{\mu \nu}(q)  j^{ a\alpha}(p_1) j^{ b\beta}(p_2)\rangle$, satisfies both Ward identities (WIs) independently. In contrast, the second and third terms, such as $\langle t^{\mu \nu }(q)j_{loc}^{a \alpha}(p_1)j^{b \beta}(p_2)\rangle_g$, are constrained to vanish under the ordinary (single derivative) WI but are not constrained by the second derivative Slavnov-Taylor identity (STI).\\
	Coming to the contribution
	\beq
	\delta_{[\rho}^{\beta} p_{2 \sigma ]}\langle J^{ a\alpha}(p_1) J^{ b\sigma}(-p_1)\rangle + \delta_{[\rho}^{\alpha} p_{1 \sigma ]}\langle J^{ b\beta}(p_2) J^{ a\sigma}(-p_2)\rangle,
	\eeq
	the single derivative WIs, both in $p_1$ and $p_2$, set both terms to zero due to the transversality condition of the gauge currents in both the Abelian and non-Abelian cases. The STI also enforces this condition.\\
	Finally, the new term, $\mathcal{B}^{\alpha \beta a b}_g$, is not permitted by the single derivative WI but is allowed by the STI. The anomaly contribution and the dilaton pole, for off-shell gluons, can be extracted from the trace sector by the steps that we have outlined, taking CFT$_p$ in the quark contributions as a 
	guideline.  
	
	\begin{figure}[t]
		\begin{center}
			\includegraphics[scale=1]{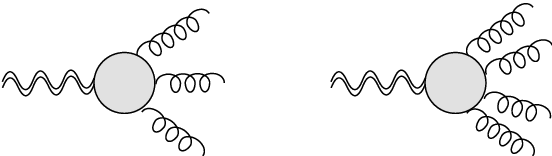}
			\caption{The $TJJJ$ and $TJJJJ$ contributions in the gauge covariant expansion of the QCD trace anomaly vertex.}
		\end{center}
	\end{figure}
	
	\begin{figure}[t]
		\begin{center}
			\includegraphics[scale=1]{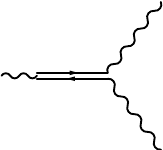}
			\caption{The dilaton intermediate state in the QCD trace anomaly vertex.}
			\label{scvertex}
		\end{center}
	\end{figure}
	
	\subsection{The anomaly in Duff's definition and the pole in DR}
	If an anomaly is understood as the failure of the trace operation to commute with a quantum average, the trace anomaly can also be defined as the difference between two trace operations on the stress-energy tensor: one performed before the quantum average and the other after. This definition was proposed by Duff \cite{Duff:1993wm,Duff:2020dqb}
	\begin{equation} \label{eq:defanomduff}
		\mathcal{A}=g^{\mu \nu}(x)\left\langle T_{\mu \nu}(x)\right\rangle-\left\langle T_\mu^\mu(x)\right\rangle,
	\end{equation}
	and is the definition considered in \cite{Bonora:2014qla} \cite{Ferrero:2023unz}, \\
	Previous computations of the $TJJ$ correlator have demonstrated the emergence of a pole, relying on a secondary decomposition first introduced in \cite{Giannotti:2008cv}, which in the case of QCD can be immeditely implemented in the quark sector
	\begin{equation}
		\Gamma_q^{\m_1\n_1\m_2\m_3 a b }(p_2,p_3)=\sum_{i=1}^{13}\,F_i(s;s_1,s_2,0)\,t_i^{\m_1\n_1\m_2\m_3}(p_2,p_3)\delta^{a b},
		\label{fs}
	\end{equation} 
	where the invariant amplitudes $F_i$ are functions of the kinematic invariants $s=p_1^2=(p_2+p_3)^2$, $s_1=p_2^2$, $s_2=p_3^2$, and the $t_i^{\m_1\n_1\m_2\m_3}$ define the basis of the independent tensor structures reproduced in Appendix \ref{genbasis1} in Table \ref{genbasistab}.
	This decomposition can be directly mapped onto the current approach, closely paralleling the analysis presented in \cite{Coriano:2018zdo}. The method is readily applicable to the quark sector, which we will illustrate here, as it is sufficient to reveal the underlying pattern. This pattern naturally extends to the gluon sector as well.\\
	\eqref{fs} is built by imposing on the $TJJ$ vertex all the Ward identities derived from diffeomorphism invariance and gauge invariance, together with Bose symmetry and conservation WIs. Additional details have been left to appendix \ref{ffdc}.
	Using the completeness of the basis in \eqref{fs} and by a direct analysis of the CWIs, we can identify the mapping between the form factors of such basis and those of the $A$-basis. In \( d=4 \), the presence of two tensor structures 
	\beq
	t_1=\left(k^2 g^{\mu\nu} - k^{\mu } k^{\nu}\right) u^{\alpha\beta}(p.q)\qquad
	t_2 =\left(k^2g^{\mu\nu} - k^{\mu} k^{\nu}\right) w^{\alpha\beta}(p.q)
	\eeq
	with nonzero trace in the \( F \)-basis initially raises questions, particularly regarding the unique relationship between anomaly poles (and associated traces) and the renormalization process. The remaining tensor structures $t_i$ are traceless. 
	The expectation is to identify a single anomaly pole originating from renormalization, while any additional poles introduced by expansions should not be linked to this process. \\
	The sets \( A_j \) presented above and \( F_i \) differ significantly, each highlighting distinct aspects of the same \( TJJ \) correlator. The \( F \)-basis, as we will show, is particularly effective in tracing the origin of the anomaly pole to a single form factor, \( F_{13} \), which diverges and thus requires renormalization.\\
	Previous analyses, such as those in \cite{Bzowski:2017poo}, indicate that the singularities in the \( A_i \)'s, specifically \( A_2 \), \( A_3 \), and \( A_4 \), align with the mapping \eqref{mapping1}, which precisely identifies the combinations involving the divergent form factor \( F_{13} \).\\
	This clear identification of the singularity's origin within the \( F \)-basis contrasts with the less direct approach in the \( A \)-basis. While the \( A_i \)'s constitute a minimal set of form factors for resolving the conformal Ward identities (CWI's) of the correlators, they obscure the origin of the singular behavior, as three out of four of these form factors exhibit UV singularities and necessitate renormalization. In contrast, the \( F \)-basis provides a straightforward method to pinpoint singularities, specifically in \( F_{13} \), which can be shown to be singular by dimensional counting. The correspondence is given in \eqref{mapping1}.\\
	To investigate the origin of the anomaly pole in the $TJJ$ correlator, we begin by considering the correlator in $d$ dimensions using the $F$-basis. Our goal is to impose that this correlator remains traceless, thereby anomaly-free, in the higher-dimensional theory. However, as we approach the physical limit $d \rightarrow 4$ using dimensional regularization, the anomaly manifests. The Ward identities associated with the trace provide crucial constraints. Specifically, imposing that the trace WI is satisfied, we derive the following conditions
	
	\begin{equation}
		\label{eq:F1}
		F_1 = \frac{(d-4)}{p_1^2(d-1)} \left[F_{13} - p_2^2 F_3 - p_3^2 F_5 - p_2 \cdot p_3 F_7\right],
	\end{equation}
	\begin{equation}
		\label{eq:F2}
		F_2 = \frac{(d-4)}{p_1^2(d-1)} \left[p_2^2 F_4 + p_3^2 F_6 + p_2 \cdot p_3 F_8\right].
	\end{equation}
	
	These equations are pivotal for understanding the renormalization process of the correlator. As $d \rightarrow 4$, it is evident from Eq. (\ref{eq:F2}) that $F_2$ vanishes:
	\begin{equation}
		F_2 = \frac{\epsilon}{(d-1) p_1^2} \left[p_2^2 F_4 + p_3^2 F_6 + p_2 \cdot p_3 F_8\right] \rightarrow 0,
	\end{equation}
	where $\epsilon \equiv d-4$. Since the form factors $F_4$, $F_6$, and $F_8$ are finite due to their dimensional scaling, $F_2$ indeed approaches zero as $d \rightarrow 4$. Consequently, in the limit $d = 4$, the $F$-basis reduces to four independent combinations of the original seven form factors (as shown in \eqref{mapping1}), which fully describe the transverse traceless sector of the theory. Additionally, one extra form factor, $F_1$, remains, which corresponds to a nonzero trace and accounts for the anomaly in four dimensions. \\
	Importantly, $F_{13}$ is the only form factor within the $F$-basis that requires renormalization. It exhibits a simple pole in $1/\epsilon$ under dimensional regularization. The fact that the singularity remains of the form $1/\epsilon$ at all perturbative orders, without higher-order poles, is a key feature of this construction. This behavior is consistent with conformal field theory, where the only available counterterm that regulates the theory is \eqref{ct} which renormalizes the two-point function $\langle JJ \rangle$ and thereby $F_{13}$. The quark sector gives (for a single fermion)
	
	\begin{equation}
		\label{eq:F13}
		F_{13} = G_0(p_1^2, p_2^2, p_3^2) - \frac{1}{2} \left[\Pi(p_2^2) + \Pi(p_3^2)\right],
	\end{equation}
	where $G_0$ can be shown to be a finite function as $d \rightarrow 4$, and the singularity is traced back to the scalar form factor $\Pi(p^2)$ of the quark contribution to the gluon 2-point function $\Pi_{\mu\nu}^{ab}(p)$. 
	\[
	\Pi_{\mu\nu}^{ab}(p) = \delta^{ab} \Pi_{\mu\nu}(p),
	\]
	where the gluon polarization tensor $\Pi_{\mu\nu}(p)$ is
	\[
	\Pi_{\mu\nu}(p) = \left(p_\mu p_\nu - g_{\mu\nu} p^2\right) \Pi(p^2),
	\]
	where $\Pi(p^2)$ is the scalar form factor that captures the momentum dependence of the quark contribution to the gluon self-energy.
	For a single quark flavor in the one-loop approximation, $\Pi(p^2)$ takes the form
	\[
	\Pi(p^2) = -\frac{g^2 T_F}{2\pi^2} \left(\frac{1}{\epsilon} - \gamma_E + \log(4\pi) + \log\left(\frac{\mu^2}{-p^2}\right) + \cdots \right),
	\]
	
	where $T_F = \frac{1}{2}$. The divergence in $F_{13}$ is then given by a single pole in $\epsilon$ is of the form
	
	\be 
	\label{renf}
	F_{13}=\frac{1}{d-4} \bar{F}_{13} + F_{13\, f},
	\eeq
	where $F_{13 f}$ is finite in the limit $d\to 4$. It is then sufficient to insert this expansion into \eqref{eq:F1} to notice the emergence of a $1/p_1^2$ dilaton pole in the limit, since all the other form factors are finite and do not contribute as $d -4 \to 0$. Therefore, this secondary parameterization shows that the anomaly is related 
	to a unique tensor structure which necessarily has to contain a pole. A similar analysis can be performed in the gluon sector. This analysis shows that the anomaly pole present in the form factor $F_1$ is the only result of the breaking of conformal symmetry as $d\to 4$ in DR. 
	\section{Connecting our parameterization of the $TJJ$ with the hadronic GFF at large $-t$}
	
	\begin{figure}
		\centering
		\begin{tikzpicture}
			\begin{feynman}
				\vertex (i1);
				\vertex[right=1cm of i1] (a1);
				\vertex[right=1cm of a1] (b1);
				\vertex[right=3cm of a1] (o1) ;
				
				\vertex[below=1cm of i1] (i2) ;
				\vertex[right=1cm of i2] (a2);
				\vertex[right=2cm of a2] (b2);
				\vertex[right=3cm of a2] (o2) ;
				
				\vertex[below=1cm of i2] (i4) ;
				\vertex[right=1cm of i4] (a4);
				\vertex[right=2cm of a4] (b4);
				\vertex[right=3cm of a4] (o4) ;
				
				%\vertex[below=2cm of i3] (t1);
				\vertex[above=1cm of a1] (t2);
				\vertex[above=2cm of a2] (t3);
				
				%\vertex[above right =1.3cm of t2] (t1);
				\vertex[above=0.6 cm of t1] (i3) {$T^{\mu\nu}_q$};
				
				\diagram* { 
					(i1)  -- [fermion] (a1) -- [fermion] (b1)   --[fermion] (o1),
					
					(i2)  -- [fermion] (a2)-- [fermion] (b2)  --[fermion] (o2),
					(i4)  -- [fermion] (b4)  --[fermion] (o4),

					(i3)  -- [graviton] (b1),
					(a1)-- [gluon] (a2),
					(b2)-- [gluon] (b4)
				};
				
			\end{feynman}
		\end{tikzpicture}
		\hspace{1cm}
		\begin{tikzpicture}
			\begin{feynman}
				\vertex (i1);
				\vertex[right=1cm of i1] (a1);
				
				\vertex[right=3cm of a1] (o1) ;
				
				\vertex[below=1cm of i1] (i2) ;
				\vertex[right=1cm of i2] (a2);
				\vertex[right=2cm of a2] (b2);
				\vertex[right=3cm of a2] (o2) ;
				
				\vertex[below=1cm of i2] (i4) ;
				\vertex[right=1cm of i4] (a4);
				\vertex[right=2cm of a4] (b4);
				\vertex[right=3cm of a4] (o4) ;
				
				%\vertex[below=2cm of i3] (t1);
				\vertex[above=1cm of a1] (t2);
				\vertex[above=2cm of a2] (t3);
				
				%\vertex[above right =1.3cm of t2] (t1);
				%\vertex[right=0.6 cm of t1] (i3);
				\vertex[above=0.5cm of a2] (b1);
				\vertex[right=2cm of b1] (i3) {$T^{\mu\nu}_g$};
				
				\diagram* { 
					(i1)  -- [fermion] (a1)   --[fermion] (o1),
					
					(i2)  -- [fermion] (a2)-- [fermion] (b2)  --[fermion] (o2),
					(i4)  -- [fermion] (b4)  --[fermion] (o4),

					(i3)  -- [graviton] (b1),
					(a1)-- [gluon] (a2),
					(b2)-- [gluon] (b4)
				};
				
			\end{feynman}
		\end{tikzpicture}
		\caption{Typical leading $O(\alpha_s^2)$ contributions to the GFF of the proton. Shown are insertions of the graviton/$f \bar{f}$ vertex contained in $T_q$ (left) and the graviton/gg vertex contained in $T_g$ (right).}
		\label{figX:2}
	\end{figure}
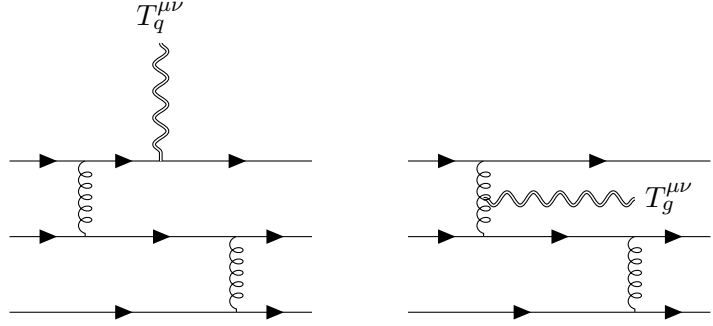

	Before presenting our conclusions, we briefly comment on the connection between our parametrization of hard scattering and \eqref{fund1} at the hadron level, focusing on the proton case. A detailed phenomenological analysis will be provided elsewhere. 
	We recall that the analysis of the form factors \( A(t) \), \( B(t) \), and \( C(t) \) at large momentum transfers, where \( t \equiv q^2 \), can be performed using a standard factorization formula. \\
	As shown in \cite{Tong:2022zax}, a convenient approach to studying the gravitational form factor (GFF) matrix element at the hadron level for large momentum transfer is to use a helicity basis. In this basis, the hadron helicities \( \Lambda \) and \( \Lambda' \) label the initial and final states, along with their respective momenta \( p \) and \( p' \)
	\begin{align}
		\langle p', \Lambda' | T^{\mu\nu}_a|p,\Lambda\rangle.
	\end{align}
	For the nucleon, the leading Fock state consists of three valence quarks, specifically \( u, u, d \), where each quark carries helicity \( \lambda_i = \pm 1/2 \). The analysis is most conveniently performed in the Breit frame, where the momenta are given by  
	
	\begin{align}
		&p^\mu = (p^+, p^-, \boldsymbol{p}_\perp) = \frac{1}{\sqrt{2}} \left(\sqrt{-t}, 0, \boldsymbol{0}_\perp \right), \\
		&p'^\mu = (p'^+, p'^-, \boldsymbol{p}'_\perp) = \frac{1}{\sqrt{2}} \left(0, \sqrt{-t}, \boldsymbol{0}_\perp \right), \\
		&\Delta^\mu = (\Delta^+, \Delta^-, \boldsymbol{\Delta}_\perp) = \left(-\sqrt{-t}, \sqrt{-t}, \boldsymbol{0}_\perp \right),
	\end{align}
	where \( t = q^2 \) is the squared momentum transfer. \\
	The total helicity of the valence quarks can take the values \( 3/2, 1/2, -1/2, -3/2 \). The relation between the hadron helicity \( \Lambda \) and the individual quark helicities \( \lambda_i \) is determined by the component of the hadron's angular momentum \( l_z \) along the scattering direction \( z \)
	
	\begin{figure}
		\centering
		\begin{tikzpicture}
			\begin{feynman}
				\vertex (i1);
				\vertex[right=1cm of i1] (a1);
				\vertex[right=1cm of a1] (b1);
				\vertex[right=3cm of a1] (o1) ;
				
				\vertex[below=1cm of i1] (i2) ;
				\vertex[right=1cm of i2] (a2);
				\vertex[right=2cm of a2] (b2);
				\vertex[right=1cm of a2] (c2);
				\vertex[right=3cm of a2] (o2) ;
				
				\vertex[below=1cm of i2] (i4) ;
				\vertex[right=1cm of i4] (a4);
				\vertex[right=2cm of a4] (b4);
				\vertex[right=1cm of a4] (c4);
				\vertex[right=3cm of a4] (o4) ;
				
				%\vertex[below=2cm of i3] (t1);
				\vertex[above=1cm of a1] (t2);
				\vertex[above=2cm of a2] (t3);
				
				%\vertex[above right =1.3cm of t2] (t1);
				\vertex[above=1.5cm of t1] (i3) {$T^{\mu\nu}$};
				\vertex[blob] (t1) at (2,1.3) {{$TJJ$}};
				
				\diagram* { 
					(i1)  -- [fermion] (a1)  --[fermion] (o1),
					
					(i2)  -- [fermion] (c2)-- [fermion] (b2)  --[fermion] (o2),
					(i4)  -- [fermion] (c4)  --[fermion] (o4),

					(i3)  -- [graviton] (t1),
					(a1)-- [gluon] (t1),
					(b2)-- [gluon] (t1),
					(c2)-- [gluon] (c4),
				};

			\end{feynman}
		\end{tikzpicture}
		\hspace{2cm}
		\begin{tikzpicture}
			\begin{feynman}
				\vertex (i1);
				\vertex[right=1cm of i1] (a1);
				\vertex[right=1cm of a1] (b1);
				\vertex[right=3cm of a1] (o1) ;
				
				\vertex[below=2cm of i1] (i2) ;
				\vertex[right=1cm of i2] (a2);
				\vertex[right=2cm of a2] (b2);
				\vertex[right=1cm of a2] (c2);
				\vertex[right=3cm of a2] (o2) ;
				
				\vertex[below=1cm of i1] (i4) ;
				\vertex[right=1cm of i4] (a4);
				\vertex[right=2cm of a4] (b4);
				\vertex[right=1cm of a4] (c4);
				\vertex[right=3cm of a4] (o4) ;
				
				%\vertex[below=2cm of i3] (t1);
				\vertex[above=1cm of a1] (t2);
				\vertex[above=2cm of a2] (t3);
				
				%\vertex[above right =1.3cm of t2] (t1);
				\vertex[above=1.5cm of t1] (i3) {$T^{\mu\nu}$};
				\vertex[blob] (t1) at (2,1.3) {{$TJJ$}};
				
				\diagram* { 
					(i1)  -- [fermion] (a1)  --[fermion] (o1),
					
					(i2)  -- [fermion] (c2)-- [fermion] (b2)  --[fermion] (o2),
					(i4)  -- [fermion] (c4)  --[fermion] (o4),

					(i3)  -- [graviton] (t1),
					(a1)-- [gluon] (t1),
					(b2)-- [gluon] (t1),
					(c2)-- [gluon] (c4),
				};

			\end{feynman}
		\end{tikzpicture}
		\caption{ Typical $O(\alpha_s^3)$ corrections appearing in the expansion of the hard scattering associated with the conformal anomaly.}
		\label{figX:3}
	\end{figure}
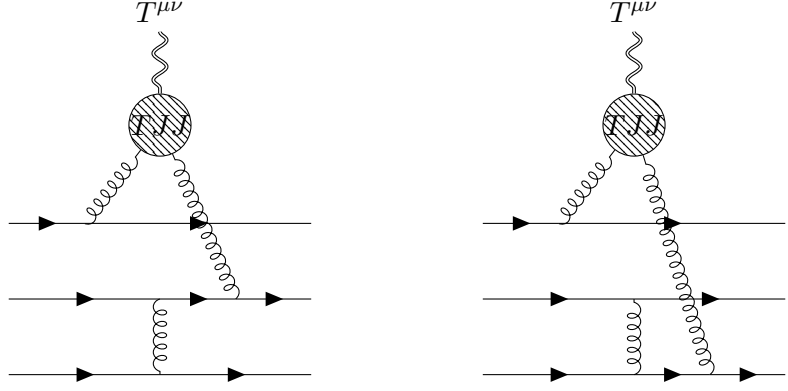
	\beq
	\Lambda=l_z - \sum_i \lambda_i.
	\eeq
	The hadronic matrix elements of definite helicities in terms of the angular momentum of the hadron in the $z$ direction
	\begin{align}
		\big| P_\uparrow  \big\rangle=\big| P_\uparrow \big\rangle^{l_z=2}_{-3/2}+\big| P_\uparrow \big\rangle^{l_z=1}_{-1/2}+\big| P_\uparrow \big\rangle^{l_z=0}_{1/2}+\big| P_\uparrow \big\rangle^{l_z=-1}_{3/2},
	\end{align}
	
	\begin{align}
		\big| P_\downarrow  \big\rangle=\big| P_\downarrow \big\rangle^{l_z=1}_{-3/2}+\big| P_\downarrow \big\rangle^{l_z=0}_{-1/2}+\big| P_\downarrow \big\rangle^{l_z=-1}_{1/2}+\big| P_\downarrow \big\rangle^{l_z=2}_{3/2}\ .
	\end{align}
	The hadronic matrix elements with definite helicities can be expressed in terms of the hadron's angular momentum along the \( z \)-direction as  
	
	\begin{align}
		\big| P_\uparrow  \big\rangle = \big| P_\uparrow \big\rangle^{l_z=2}_{-3/2} + \big| P_\uparrow \big\rangle^{l_z=1}_{-1/2} + \big| P_\uparrow \big\rangle^{l_z=0}_{1/2} + \big| P_\uparrow \big\rangle^{l_z=-1}_{3/2},
	\end{align}
	
	\begin{align}
		\big| P_\downarrow  \big\rangle = \big| P_\downarrow \big\rangle^{l_z=1}_{-3/2} + \big| P_\downarrow \big\rangle^{l_z=0}_{-1/2} + \big| P_\downarrow \big\rangle^{l_z=-1}_{1/2} + \big| P_\downarrow \big\rangle^{l_z=2}_{3/2}.
	\end{align}
	Here, the hadronic states with definite helicities are defined through either integrated or unintegrated wave functions. The integrated wave functions describe the interaction in the collinear limit, where the transverse momenta of the partons are neglected, while the unintegrated wave functions account for their transverse momentum dependence. \\
	The kinematic variables are parameterized by \( \kappa_i = (x_i, k_i) \), where \( x_i \) denotes the longitudinal momentum fraction of the parton, and \( k_i \) represents its intrinsic transverse momentum. In the unintegrated case, the wave function explicitly depends on the transverse momenta, which serve as integration variables. The hadronic initial and final states are then defined in terms of the integrated wave function \( \phi(x_1, x_2, x_3) \) and the unintegrated wave function \( \psi(\kappa_1, \kappa_2, \kappa_3) \), both of which are functions of the momentum fractions \( x_i \) and \( y_i \) corresponding to the initial and final hadron momenta \( p \) and \( p' \), respectively
	
	\begin{align}
		\big| p_\uparrow \big\rangle_{1/2}^{l_z=0}= & \int \
		\frac{[d x][d^2 \bs k]}{\sqrt{24 x_1 x_2 x_3}}
		\psi_1
		( \kappa_1,\kappa_2,\kappa_3) \ \big |\{x_i p+ \bs  k_i \}\big \rangle_{1/2},
		\label{eq_fock1} \\ 
		\big| p_\downarrow \big\rangle_{1/2}^{l_z=-1}= & \int \
		\frac{[d x][d^2 \bs k]}{\sqrt{24 x_1 x_2 x_3}}
		\big[  k_{1L} \psi_3( \kappa_2,\kappa_1,\kappa_3)+  k_{2L} \psi_4( \kappa_2,\kappa_1,\kappa_3) \big]
		\big |\{x_i p+ \bs  k_i \}\big \rangle_{1/2}.
		\label{eqf} 
	\end{align}
	
	The unintegrated wave functions \( \psi_1, \psi_2, \psi_3, \psi_4 \), classified in \cite{Ji:2002xn, Ji:2003yj}, describe the proton's internal partonic structure. The inclusion of the complex combination \( k_L = k^x - i k^y \) accounts for the presence of nonzero orbital angular momentum. However, the dominant contributions originate from the three-quark light-cone wave function with zero orbital angular momentum:
	
	\begin{align}
		\langle p'_{ \uparrow }  |
		{T^{\mu\nu }_a} | p_ \uparrow  \rangle=\langle p'_{ \uparrow }  |^{l'_z=0}_{1/2} 
		{T^{\mu\nu }_a} | p_ \uparrow  \rangle^{l_z=0}_{1/2},
	\end{align}
	
	where \( a = q, g \) denotes the quark and gluon contributions from the QCD stress-energy tensor. To illustrate how a correspondence can be established, we focus on this leading contribution, whose expression is given by  
	
	\begin{align}
		\langle p'_\uparrow |_{1/2}T^{\mu\nu}_a|p_\uparrow\rangle_{1/2}
		=&  \int \
		\frac{[d x][d y][d^2 \boldsymbol{k}'][d^2 \boldsymbol{k}]}{24(x_1 x_2 x_3 y_1 y_2 y_3)^{1/2}}
		\psi_1^*
		( \kappa'_1,\kappa'_2,\kappa'_3)\psi_1
		( \kappa_1,\kappa_2,\kappa_3) 
		\big\langle  \{y_i p'+\boldsymbol{k}'_i\}\big|_{1/2}
		{T^{\mu\nu}_a}
		\big|\{x_i p+\boldsymbol{k}_i\}\big\rangle_{1/2},
	\end{align}
	
	where the integration variables include both the momentum fractions and the intrinsic transverse momenta of the partons:  
	
	\begin{align}
		[d x ] &= d x_1 d x_2 d x_3\ \delta(1-x_1-x_2-x_3), \notag \\
		[d^2  k ] &= \frac{1}{(2\pi)^6} d^2\boldsymbol{k}_{1} d^2\boldsymbol{k}_{2} d^2\boldsymbol{k}_{3} \delta^{(2)}(\boldsymbol{k}_{1}+\boldsymbol{k}_{2}+\boldsymbol{k}_{3}).
		\label{eqm}
	\end{align}
	
	The hard scattering expansion in Fock space follows the usual approach, employing quark creation operators for plane waves with defined color and helicity. The operators \( \hat u^\dagger_{a,\lambda} \) and \( \hat d^\dagger_{a,\lambda} \) create \( u \) and \( d \) quarks, respectively, with helicity \( \lambda \) and color index \( a \) in the fundamental representation of \( SU(3) \), normalized as
	
	\begin{align}
		\{\hat u_{a,\lambda}(k), \hat u^\dagger_{b,\lambda'}(k')\} = \delta_{\lambda \lambda'}\delta_{ab}2k^+\delta(k^+-k'^+)\delta^{(2)}(\boldsymbol{k}'-\boldsymbol{k}).
	\end{align}
	
	The light-front wave functions \( \psi_1, \psi_2, \psi_3, \psi_4 \) describe the proton in terms of the kinematic variables \( \kappa_i = (x_i, \boldsymbol{k}_i) \), where \( x_i \) represents the longitudinal momentum fraction of each parton, satisfying \( \sum_i x_i = 1 \), and \( \boldsymbol{k}_i \) denotes the intrinsic transverse momentum, constrained by \( \sum_i \boldsymbol{k}_i = 0 \). 
	
	Neglecting intrinsic transverse momentum and considering only the collinear limit, we define the twist-three light-cone amplitude of the proton as~\cite{Braun:1999te}  
	
	\begin{align}
		\Phi_3(x_1,x_2,x_3)=2\int[d^2 \boldsymbol{k}] \psi_1(\kappa_1,\kappa_2,\kappa_3).
	\end{align}
	
	This expression encapsulates the leading contributions to the hadronic matrix elements in the collinear approximation, simplifying the analysis of the internal structure of the proton.
	
	The leading $A(t)$-form factor can be extracted form the amplitude $(P=(p + p')/2)$
	\begin{align}
		&\langle p'_\uparrow | T^{\mu\nu}_a|p_\uparrow\rangle
		=A_a(t)\bar u_\uparrow(p') \gamma^{(\mu} P^{\nu)} 
		u_\uparrow(p)\ ,
		\label{eq_consAmp}
	\end{align}
	while $B(t)$ is power-suppressed. One obtains 
	
	\begin{align}
		&A(t)\bar u_\uparrow(p') \gamma^{(\mu}  P^{\nu)} 
		u_\uparrow(p)
		\notag\\
		=&  \int \
		\frac{[d x][d y]}{96(x_1 x_2 x_3 y_1 y_2 y_3)^{1/2}}
		\Phi_3^*(y_1,y_2,y_3)
		\Phi_3(x_1,x_2,x_3) \big\langle  uud-udu,\{y_i p'\}\big|
		{T^{\mu\nu}}
		\big| uud-udu,\{x_i p\}\big\rangle\ ,
		\label{xs1}
	\end{align}
	where the partonic matrix element 
	\beq
	\mathcal{T}^{\mu\nu}_{1/2\,1/2}\equiv\langle  uud-udu,\{y_i p'\}\big|
	{T^{\mu\nu}}
	\big| uud-udu,\{x_i p\}\big\rangle\
	\label{tau}
	\eeq
	defines the hard scattering contribution, and we have summed over the quark and gluon contributions ($a=q,g$). 
	The dominant partonic configurations in the hard scattering process are determined by extracting the leading quark contributions from the proton's partonic wave function within the Fock space representation
	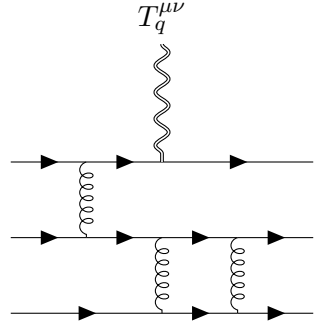
\begin{figure}
		\centering
		\begin{tikzpicture}
			\begin{feynman}
				\vertex (i1);
				\vertex[right=1cm of i1] (a1);
				\vertex[right=1cm of a1] (b1);
				\vertex[right=3cm of a1] (o1) ;
				
				\vertex[below=1cm of i1] (i2) ;
				\vertex[right=1cm of i2] (a2);
				\vertex[right=2cm of a2] (b2);
				\vertex[right=3cm of a2] (o2) ;
				
				\vertex[below=1cm of i2] (i4) ;
				\vertex[right=1cm of i4] (a4);
				\vertex[right=2cm of a4] (b4);
				\vertex[right=3cm of a4] (o4) ;
				
				%\vertex[below=2cm of i3] (t1);
				\vertex[above=1cm of a1] (t2);
				\vertex[above=2cm of a2] (t3);
				
				\vertex[right=1cm of a2] (c2);
				\vertex[below=1cm of c2] (c4);
				%\vertex[above right =1.3cm of t2] (t1);
				\vertex[above=0.6 cm of t1] (i3) {$T^{\mu\nu}_q$};
				
				\diagram* { 
					(i1)  -- [fermion] (a1) -- [fermion] (b1)   --[fermion] (o1),
					
					(i2)  -- [fermion] (a2)-- [fermion] (c2)-- [fermion] (b2)  --[fermion] (o2),
					(i4) -- [fermion] (c4) -- [fermion] (b4)  --[fermion] (o4),

					(i3)  -- [graviton] (b1),
					(a1)-- [gluon] (a2),
					(c2)-- [gluon] (c4),
					(b2)-- [gluon] (b4)
				};
				
			\end{feynman}
		\end{tikzpicture}
		\caption{Typical $O(\alpha_s^3)$ contributions from  $T_q$ not related to the conformal anomaly.}
		\label{figX:4}
	\end{figure}

	\begin{align}
		\big|%uud,
		\{p_i\} \big\rangle_{1/2}=&
		\frac{\epsilon_{abc}}{\sqrt{6}}
		\hat u_{a,\uparrow}^\dagger (p_{ 1} ) 
		\big [\hat u_{b,\downarrow}^\dagger (p_{2 })
		\hat d_{c,\uparrow}^\dagger ( p_{3 })  - \hat d_{b,\downarrow}^\dagger( p_{2 })\hat  u_{c,\uparrow}^\dagger ( p_{3 })\big]\big|0 \big\rangle
		\label{fk3}
	\end{align}
	where $p_i\equiv x_i p$ are the momenta of the initial collinear quarks and $p_i=y_i p'$ for the final state ones 
	\begin{align}
		&|uud,\{p_i\} \rangle =\frac{\epsilon_{abc}}{\sqrt{6}}
		\hat u_{a,\uparrow}^\dagger (p_{ 1} ) 
		\hat u_{b,\downarrow}^\dagger (p_{2 })
		\hat d_{c,\uparrow}^\dagger ( p_{3 })\big|0 \big\rangle
		~,\notag 
		\\
		&
		|udu,\{p_i\} \rangle =\frac{\epsilon_{abc}}{\sqrt{6}}
		\hat  u_{a,\uparrow}^\dagger (p_{ 1} ) \hat d_{b,\downarrow}^\dagger( p_{2 }) \hat u_{c,\uparrow}^\dagger ( p_{3 })\big|0 \big\rangle~.
		\label{eq_partonstate}
	\end{align} 
	We illustrate in \figref{figX:2} the typical topologies that emerge at \( O(\alpha_s^2) \) in the perturbative expansion of the hard scattering process. At tree level, the insertions of \( T_q \) into the hard scattering involve both a graviton/fermion/antifermion (\( O(g^0) \), lowest order) vertex and a graviton/gluon/fermion/antifermion (\( O(g) \)) vertex. The left diagram represents the lowest-order contribution to the hard scattering, featuring the insertion of the leading \( T_q \) vertex, while the right diagram illustrates the lowest-order insertion of \( T_g \) via the graviton/two-gluon \( O(g^0) \) vertex. \\
	Insertions of \( T_g \) include both the graviton two-gluon and graviton three-gluon vertices, corresponding to \( O(g) \) and \( O(g^2) \), respectively, though the latter is not explicitly shown in the same figure. According to the quark/gluon counting rules, hard rescattering via gluon insertions must be maintained at large \( -t \), requiring an additional gluon propagator to connect the remaining valence quark in order to ensure an elastic transition to the final state.\\
	The anomaly contributions, depicted in \figref{figX:3}, arise from replacing each gluon propagator that connects separate quark lines with the insertion of the \( O(\alpha_s) \) corrections introduced in this work. These corrections effectively replace the tree-level gluon propagator with the modified relation derived from the decomposition discussed. We have
	\beq
	S^{\alpha\beta\, a b}(l)\to S^{\alpha\alpha' \, a a'}(l)\,\Gamma^{\alpha'\beta' \, a' b'}(q, l, l+q))\,
	S^{\beta'\beta \, b' b}(l +q)
	\eeq
	where \( l^\mu \) represents a typical transverse momentum of the gluon exchanged between valence quark lines at leading order. We denote by \( \Gamma^{\alpha'\beta' \, a' b'}(q, l, l+q) \) the momentum dependence of the \( TJJ \) vertex when included in the hard scattering, where \( q \) is the incoming momentum of the graviton line. 
	
	In the pion case, such modifications affect only a single lowest-order topology, characterized by the exchange of a single gluon between the quark-antiquark pair in the hard scattering process. In contrast, for the proton, there are three relevant topologies, two of which are depicted in \figref{figX:3}. 
	
	The analysis in \cite{Tong:2022zax}, which examines the leading-order hard partonic amplitude at \( O(\alpha_s^2) \), can be systematically extended to the next perturbative order, \( O(\alpha_s^3) \), with only minor modifications. This computation requires incorporating all standard insertions into the hard scattering, including an additional gluon connecting pairs of valence quark lines, as well as self-energy corrections in the gluon propagator. The insertion of the \( TJJ \) vertex follows the same procedure as outlined earlier, though at this order, a large number of additional corrections arise that are unrelated to the anomaly. \\
	For illustration, one such correction is shown in \figref{figX:4}, where hard rescattering between two valence quark lines is mediated by the exchange of two gluons. Other relevant corrections involve graviton three-gluon and graviton/four-gluon vertices, originating from the tree-level insertions of \( T_g \) in the \( T_g ggg \) and \( T_g gggg \) interactions. These contribute at \( O(g^5) \) and \( O(g^6) \), respectively, in the proton case. \\
	By incorporating all these corrections into a matrix element that accounts for such background processes, denoted as \( \mathcal{T}^{\mu\nu}_b \), and reinstating the dependence on the strong coupling constant, the matrix element can be expressed in the form
	
	\beq
	\langle p'| T^{\mu\nu} | p\rangle
	=\alpha_s^2 \mathcal{T}^{\mu\nu} + \alpha_s^3\mathcal{T}_b^{\mu\nu} +\alpha_s^3\mathcal{T}_{\textrm{an}}^{\mu\nu}
	\eeq
	where the leading-order contributions $\mathcal{T}^{\mu\nu}$ may be decomposed in the helicity basis, of which \eqref{tau} is one of the components. \\
	In general the structure of the hard scattering, in the collinear limit will be of the form 
	\beq
	\Gamma^{\mu\nu\alpha\beta a b}(x_i p, y_j p',  t)\mathcal{N}^{\alpha\beta a b}(x_i p, y_j p',  t)
	\eeq
	to be convoluted with the ordinary hadronic wave functions as in \eqref{xs1}. We have denoted with $\mathcal{N}^{\alpha\beta a b}$ the ordinary Dirac trace appearing in the hard scattering, which can be worked out as in \cite{Tong:2022zax}. 
	The extraction of the contribution of the anomaly form factor can be obtained by tracing the hadronic matrix element. For example, one immediately obtains that $A(t)$, at large 
	$-t$, if we neglect the hadron mass, is not affected by the anomaly since the left hand side of \eqref{xs1} vanish, by use of the equations of motion on the spinors.  \\
	Notice that imposing any trace condition on the matrix element eliminates the anomaly pole from the hard scattering. This feature is explicitly revealed in the perturbative analysis, where the anomaly pole arises in the on-shell gluon case and for massless fermions.  \\
	In a related work, we will demonstrate that the same form factor obeys a general sum rule in perturbation theory. This result follows from the specific structure of the spectral density of the anomaly form factor, as given in \eqref{trc}.
	\beq
	\Phi_{an}\equiv \mathcal{A}^{\alpha \beta a b}/q^2
	\label{anf}
	\eeq
	confirming and extending previous analysis in the QED \cite{Giannotti:2008cv} and 
	QCD cases \cite{Armillis:2010qk}. The pattern, in the case of the conformal anomaly form factor \eqref{anf} is pretty similar to that discussed in the case of chiral and gravitational 
	anomalies, recently addressed in \cite{Coriano:2025ceu}. 
	
	\section{Comments and Conclusions}
	\label{dec22} 
	This work has been focused on the analysis of a specific vertex, the $TJJ$, in the non-Abelian case that, as we have illustrated, plays a key role in QCD in the GFFs of hadrons. \\
	The analysis that we have presented is directly linked with the factorization picture of exclusive processes. Such picture has brought us to consider the role of the perturbative one-loop insertion of the $TJJ$ in the hard scattering. Naturally, this insertion, in the hadron case, turns relevant at order $\alpha_s^2$, and is therefore subleading compared to the leading $O(\alpha_s)$ corrections, obtained by the direct coupling of the graviton to the collinear quarks of the hard scattering. The vertex, as pointed out, is essentially described by a nonlocal interaction that has been investigated in the past in several anomalous correlators. The extraction of this 
	interaction is rather nontrivial and requires the formalism presented in our work, which is the result of a long-term analysis of such matrix elements in $CFT_p$ using a combination of general CFT approaches and specific free field theory realizations.\\
	The interaction is described using a longitudinal/transverse/trace (sector) decomposition of such vertices, with the pole emerging in the trace channel. This factorization of the hard scattering that we have presented can be immediately extended at hadron level, in the proton and pion cases, in order to provide a possible phenomenological basis of invariant amplitudes in which we parameterize the GFF form factors.  \\
	We have aimed to bridge recent advancements in CFT$_p$ and their anomalies developed over the past decade for correlators of even and odd parity with the physics of strong interactions. This effort is particularly timely as anomalies have gained renewed attention in the context of the Electron-Ion Collider (EIC) program on the proton spin \cite{Tarasov:2020cwl,Bhattacharya:2022xxw,Castelli:2024eza}. This program is poised to play a pivotal role in the scientific agenda at BNL, contributing significantly to proton tomography and the determination of the spin and partonic content of hadrons and the anomaly contribution.\\
	We have shown that conformal anomalies are intrinsically linked to the presence of effective dilaton degrees of freedom in the hard scattering. In the perturbative framework, these anomalies are associated with the emergence of a dilaton pole in the hard scattering process. This phenomenon is accompanied by a sum rule, which we have verified in perturbation theory at the one-loop level. \\
	We have illustrated how  the standard 
	$CFT_p$ approach can be modified to account for the gauge fixing sector of QCD. 
	We have shown by an explicit computation, though limited to the on-shell gluon case, that the anomaly form factor is characterized by a dispersive part that satisfies a sum rule.
	The presence of  sum rules is a hallmark of chiral and conformal anomalies, transforming the pole into a cut, in the presence of extra scales. In other words, anomaly poles are not ordinay particle poles and their manifestation is essentially associated with sum rules. Successful verification of this character would serve as a strong indication of the exchange of a dilaton state, validating the theoretical predictions.

		\chapter{Topological Sum Rules and Spectral Flows of Chiral and Gravitational Axion-like Interactions}\label{chsr1}
\addcontentsline{toc}{section}{Synopsis}
\section*{Synopsis}

This Chapter investigates the spectral structure of chiral and gravitational anomaly form factors through a dispersive analysis of the $AVV$ ($\langle JJJ_A \rangle$) and $ATT$ ($\langle TTJ_A \rangle$) correlators, moving systematically away from the conformal limit by introducing massive fermions. The central results are the derivation of mass-independent sum rules satisfied by the anomaly form factors of both correlators in full kinematic generality, and the demonstration that these sum rules constitute the defining signature of anomaly-induced interactions and preclude a naive interpretation of the anomaly pole as an asymptotic axion-like state.

The Chapter opens by contrasting ordinary Goldstone modes from spontaneously broken non-anomalous symmetries with the anomalous case. While both lead to a $1/q^2$ pole structure in the longitudinal sector of the relevant current correlator, only the anomalous case gives rise to a sum rule constraining the spectral density of the form factor. The local interaction $a(x)F\tilde{F}$, commonly employed in axion phenomenology, is shown to lack this spectral structure: the corresponding form factor $\Phi_W$, extracted from the axial Ward identity by removing the $1/q^2$ factor, has a divergent integral over its spectral density and therefore does not obey a sum rule. In contrast, the nonlocal form factor $\Phi_0$, which retains the full $1/q^2$ structure of the 1PI effective vertex, satisfies an anomaly sum rule whose value is exactly the anomaly coefficient, independently of the fermion mass and the virtualities of the external lines.

The conformal limit of the $AVV$ correlator is briefly reviewed within the $\mathrm{CFT}_p$ framework, showing how the inclusion of a single anomaly pole in the longitudinal sector of $J_A$, combined with conformal Ward identities and Schouten identities, uniquely determines the entire correlator both its longitudinal and transverse sectors up to the anomaly coefficient $a_1$. The transverse part is expressed in terms of the master integral $I_{3\{1,0,1\}}$, related to the massless scalar triangle $C_0$. A key observation is that the anomaly coefficient acts as a structural factor throughout the correlator, not merely as a prefactor of a single tensor structure, and that the conformal solution is scale-independent and vanishes identically when $a_1 = 0$.

The analysis is then extended beyond the conformal point to massive fermions via perturbation theory. The $AVV$ vertex in the massive case is parameterized in terms of two form factors $F_1$ and $F_2$, with the anomaly residing entirely in $F_1$. The off-shell longitudinal form factor $\Phi_0$ is expressed in terms of the massive scalar triangle $C_0$ and a $1/q^2$ term reflecting the anomaly. An explicit proof is given that, for on-shell vector lines and $m \neq 0$, the apparent $1/q^2$ pole in $\Phi_0$ cancels identically against a corresponding term in the expansion of $m^2 C_0(q^2, m^2)/q^2$: the series expansion of $C_0(0,0,q^2,m^2)$ around $q^2 = 0$ shows that the pole is entirely absorbed, leaving a finite, mass-dependent form factor that decouples as $m \to \infty$. This cancellation confirms that the anomaly pole is not a particle pole in the massive theory away from the light cone.

The dispersive structure of $\Phi_0$ is then analyzed in full generality. The spectral density exhibits two localized contributions (spikes) at $s = 0$ and a continuum above the two-particle threshold $4m^2$. The first spike, of constant strength, corresponds to the anomaly coefficient; the second, of variable strength proportional to $m^2 C_0$ evaluated at $q^2 = 0$, plays the role of an effective subtraction. The sum rule is established using an inverse Laplace (Borel) transform technique: by computing $\lim_{t \to 0} \mathcal{L}^{-1}\{\Phi_0\}(t)$ via the parametric integral representation of $C_0$, the integral of the spectral density over the entire cut $[4m^2, \infty)$ is shown to equal $1/(2\pi^2)$ in full generality, for arbitrary off-shell external momenta and arbitrary fermion mass. The cancellation pattern underlying the sum rule involves a precise interplay between the two spikes at $s = 0$ and the continuum, which redistributes as the fermion mass varies and as one approaches the conformal limit $m \to 0$, where the entire spectral weight concentrates on the $\delta(s)$ spike.

The question of whether the anomaly pole can be interpreted as an asymptotic particle state is addressed by analyzing the residue of the $1/q^2$ pole in the full off-shell correlator using light-cone variables. The limit $q^2 \to 0$ is performed in a light-cone frame where $q^+$ is large while $q^- \to 0$, and it is shown that all tensor structures in the correlator vanish in this limit except those containing the $1/q^2$ pole, and only for massless vector lines and a massless fermion. For massive fermions or off-shell vector bosons, the residue $g_p = \lim_{q^2 \to 0} q^2 \langle JJJ_A \rangle$ vanishes, confirming that the anomaly pole does not correspond to a physical asymptotic state under general kinematics. The anomaly pole can be associated with a particle only when the dynamics is constrained to the light cone.

The gravitational chiral anomaly is then analyzed through the $ATT$ correlator. The conformal limit is reviewed, showing that the entire vertex including its transverse-traceless sector, parameterized by form factors $A_1$ and $A_2$ expressed in terms of $C_0$ and logarithms of momentum ratios is again fully determined by the anomaly pole in the longitudinal sector of $J_A$ via the conformal Ward identities. The perturbative computation of the massive $ATT$ is then carried out in the Breitenlohner--Maison scheme for $\gamma_5$, involving triangle, bubble, and tadpole topologies. The decomposition of the correlator in the massive case yields a longitudinal anomaly sector, a trace sector proportional to $\bar{F}_3$ generated by the breaking of conformal symmetry, and a transverse-traceless sector parameterized by $A_1$ and $A_2$ with explicit mass-dependent corrections. The anomaly form factor $\phi_{ATT} = (\bar{F}_1 - \bar{F}_2)/s$ is identified as the relevant quantity encoding the gravitational anomaly, while the combination $\bar{F}_1 + \bar{F}_2$ vanishes for on-shell gravitons and is entirely generated by explicit chiral symmetry breaking.

The spectral analysis of $\phi_{ATT}$ reveals a richer structure than the $AVV$ case, involving not only the spikes at $s = 0$ and the continuum above $4m^2$, but also a pole at $s = s_1 + s_2$ and potential anomalous thresholds at $s_\pm = (\sqrt{s_1} \pm \sqrt{s_2})^2$. The apparent anomalous threshold contributions are shown to vanish identically through explicit evaluation of $C_0$ at $s = s_\pm$, confirming that no anomalous thresholds contribute to the dispersive integral. The pole at $s = s_1 + s_2$ does contribute a localized spectral density, whose UV divergences cancel automatically in the combination of $B_0$ functions involved. The sum rule for $\phi_{ATT}$ is then derived using the same inverse Laplace transform method, yielding
\begin{equation*}
	\frac{1}{\pi} \int_0^\infty \Delta\phi_{ATT}\, ds = a_g,
\end{equation*}
where $a_g$ is the gravitational anomaly coefficient. The cancellation pattern in this case involves contributions from the continuum, the spike at $s = 0$, and the pole at $s = s_1 + s_2$, in a precise interplay that holds for arbitrary off-shell graviton virtualities and arbitrary fermion mass.

The Chapter concludes by drawing together the implications of these results. The sum rules established for both $AVV$ and $ATT$ are identified as topological properties of the anomaly vertices, encoding the conservation of spectral weight independently of the kinematic regime. The local effective action $a(x)F\tilde{F}$ fails to reproduce this sum rule and therefore cannot be derived from the 1PI vertex without additional nonperturbative input such as the dynamical generation of an axion mass at a confinement scale. In the conformal limit, the entire spectral weight concentrates on the $\delta(s)$ spike, mimicking a particle pole, but this behavior is shown to be a kinematic artifact of the light-cone limit rather than evidence for an asymptotic state. The results establish that anomaly-induced interactions are intrinsically nonlocal, and that the sum rule rather than the pole is the correct invariant characterization of the anomaly vertex across all kinematic regimes.

The work presented in this Chapter is based primarily on, and adapted from, the publication \cite{Coriano:2025ceu}, co-authored with C.~Corianò and S.~Lionetti.
\section{Introduction}
A direct approach to studying chiral and gravitational anomalies is through perturbative analysis of the 1PI effective action, particularly the axial-vector/vector/vector ($AVV$ or $\langle JJJ_A \rangle$) and axial-vector/stress-energy/stress-energy ($ATT$ or $\langle TTJ_A \rangle$) correlators. Here, \( J_A \) denotes the axial-vector current, \( J \) the vector currents, and \( T \) the stress-energy tensor, which couples to gravity. The $ATT$ vertex extends anomaly considerations into the gravitational sector.\\
Axial-vector currents may arise from chiral fermions in external fields or through interactions with gravitational waves \cite{delRio:2020cmv}. If the early Universe experienced a conformal phase, such anomaly-induced interactions could enhance fermionic chiral asymmetries. In the $ATT$ case, gravitational waves may further amplify chiral chemical potentials \cite{delRio:2021bnl}, impacting primordial magneto-hydrodynamics (MHD) and the evolution of cosmic magnetic fields \cite{Boyarsky:2011uy, Boyarsky:2015faa, Brandenburg:2017rcb}.\\
In a conformal phase, the three-point functions involving conserved currents and the stress-energy tensor are uniquely determined—up to an overall constant—by conformal symmetry. When chiral anomalies are present, these correlators are likewise uniquely fixed, again modulo an overall constant, by simultaneously imposing conformal constraints and diffeomorphism invariance. Remarkably, the full structure of these solutions can be reproduced entirely within ordinary free field theory realizations.\\
The study of the effective action associated with these interactions—manifesting in the ultraviolet (UV) regime as the exchange of intermediate states in the form of anomaly poles in the conformal limit—is characterized by the emergence of sum rules rather than of conventional asymptotic states, saturated by spectral densities which are simply proportional to $\delta$ functions. These intermediate configurations can be interpreted as ordinary particles only under specific kinematic conditions, namely when the dynamics are constrained to the light cone. In other words, these intermediate states are not ordinary particle states.\\
A related phenomenon occurs in Quantum Chromodynamics (QCD) at small values of the Bjorken variable 
$x$, where the so-called perturbative pomeron emerges. Unlike ordinary particles, the perturbative pomeron cannot be associated with a well-defined asymptotic state in the conventional sense. \\
The purpose of this work is to investigate these features for the two correlators discussed above, exploring their full kinematical structure beyond the conformal limit. This is achieved by including perturbative effects due to fermion masses and demonstrating the existence of mass-independent sum rules in all cases. In a related work, we are going to show that a similar result holds for conformal correlators such as the $TJJ$, involving the stress energy tensor and non-abelian vector currents. \\
A key result of our analysis is that the local effective action for an axion-like particle—originating in the UV through chiral anomaly interactions in a conformal setting—necessarily requires, for its extension at low energy, the inclusion of a nonperturbative intermediate scale in order to be consistently defined as a local action for an asymptotic axion. The only requirement  in the UV is described by a conformal interaction. This finding applies not only to standard chiral anomaly interactions but also to gravitational anomalies. Our detailed investigation around the conformal point demonstrates that the presence of a sum rule is essential to reach such a conclusion.  \\
Our analysis focuses on both correlators. In the conformal limit, the framework is based on momentum-space conformal field theory (CFT$_p$), where anomaly poles in the longitudinal sector of $J_A$ play a central role in solving the conformal Ward identities for both correlators \cite{Coriano:2013jba, Bzowski:2013sza, Bzowski:2015pba, Coriano:2020ees}. These poles are fundamental to understanding both parity-odd chiral anomalies and parity-even trace anomalies in the conformal limit, and they are intricately linked to the full vertex structure, including the transverse components \cite{Coriano:2023cvf, Coriano:2023hts, Coriano:2024ssu}.\\
As we move away from the conformal point, these interactions are characterized by specific sum rules satisfied by the corresponding anomaly form factors, which we examine in detail. Their spectral densities exhibit intriguing cancellation patterns between localized spectral poles and continuum contributions—structures that, to our knowledge, have not been previously analyzed with such precision. The spectral analysis is performed in full generality.  As we will demonstrate, these cancellations are closely tied to the structure of the effective action that is expected to reproduce the anomalies under broad kinematic conditions.

\subsection{Local and nonlocal effective actions}
A central question addressed in this work is whether anomaly poles appearing in the off-shell effective action—resembling axion-like interactions—can be associated with actual asymptotic axion-like states. This issue is particularly relevant given the strong interest in axion-like particles (ALPs) as potential dark matter candidates. The inclusion of such states on purely phenomenological grounds has often been proposed to explain various astrophysical observations (see the discussion in \cite{Coriano:2018uip}).\\
In standard quantum field theory, these particles are typically characterized by chiral anomaly interactions. In the conformal limit, such interactions can be analyzed using momentum-space conformal field theory ($CFT_p$), through 3-point functions governed by a nonlocal effective action. This action captures the essential features we investigate, including the emergence of anomaly poles and the associated sum rules.\\
A key part of our analysis is to demonstrate that the effective action is intrinsically nonlocal and does not, in general, support the existence of an asymptotic axion-like state. In particular, obtaining a local action from this structure would require enforcing conformal Ward identities that project out the transverse components of the anomaly vertex. However, doing so invalidates the sum rule central to our findings. The only viable path to a local effective action involves strong coupling effects that evade the UV constraints emerging from the conformal regime—mirroring the situation with standard Peccei–Quinn axions. In those cases, the modification of the functional measure (equivalent to including anomalous Ward identities) is crucial to deriving a consistent local action.\\
To clarify this further, we investigate the spectral densities of the anomaly form factors whose spectral densities manifest both anomaly poles and continuum contributions. Anomaly poles reflect the topological nature of anomaly vertices: they appear or vanish depending on the kinematic regime, governed by whether the relevant sum rule is saturated by a pole (discrete state) or by a continuum (dispersive cut). Our analysis shows that, in the perturbative regime, these interpolating states can be identified with asymptotic axion-like particles only when the interaction is constrained to the light-cone.
This unusual behavior stems from the intricate mixing between longitudinal and transverse sectors of the anomaly vertex, governed by Schouten identities. \\
These identities enable cancellations or enhancements in the residue structure of the form factors. Crucially, anomaly vertices are uniquely characterized by a sum rule always satisfied by their longitudinal form factors—a feature absent in purely local axion-like couplings of the form $a(x)\, F \tilde{F}$. While such local interactions relate to the full anomaly vertex via a Ward identity, they lack the spectral structure revealed by a dispersive analysis of both the longitudinal and transverse sectors. This distinction underlines the necessity of our nonlocal approach to fully capture the dynamics of anomaly-induced axion-like behavior.

\subsection{Anomaly Sum Rules}

Sum rules in the $AVV$ interaction have been previously studied in various kinematical configurations \cite{Dolgov:1971ri, Frishman:1980dq, Horejsi:1992tw}. A more general perspective on the chiral and conformal anomaly in the effective action is offered in \cite{Giannotti:2008cv}, with further investigations into the spectral density of anomaly form factors, including sum rules in the $\mathcal{N}=1$ superconformal anomaly multiplet and the Konishi anomaly \cite{Coriano:2014gja}.\\
Sum rules govern the absorptive amplitudes of anomaly vertices and manifest as area laws tied to the longitudinal form factors. In the $AVV$ vertex, we analyze the spectral densities of these form factors, identifying the structure of poles and cuts in the complex momentum plane ($q^2 \equiv t$), and their interplay before and after dispersion integration.
In this work, we build on these foundations by exploring different representations using Schouten identities, which can introduce or eliminate anomaly poles in vertex parameterizations. \\
We focus on their physical relevance, especially near the conformal limit, where anomaly poles are essential for satisfying conformal Ward identities \cite{Coriano:2023cvf, Coriano:2023hts, Coriano:2024ssu}. We adopt a minimal form factor basis and use $CFT_p$ methods for a sector decomposition into longitudinal and transverse components.\\
Notice that our approach leverages minimal decompositions \cite{Bzowski:2013sza}, optimized for the massless limit in parity-even trace-anomalous correlators, and recently extended to parity-odd sectors ($AVV$, $ATT$) via Schouten-based parameterizations \cite{Coriano:2023cvf, Coriano:2023hts, Coriano:2024ssu}. The transverse sectors are characterized by anomaly form factors, while the longitudinal ones encode the anomaly pole and satisfy the anomalous Ward identities. As we deviate from the conformal point, the dispersive analysis remains centered on these longitudinal sectors, where sum rules apply.
Finally, we present theoretical evidence supporting the interpretation of the interpolating states—described in \cite{Giannotti:2008cv} as involving pseudoscalar kinetic mixing—as effectively defined only on the light cone.

\section{Ordinary vs. Anomalous Couplings of Goldstone Modes}
In order to uncover the difference between ordinary (non anomalous) Goldstone modes and their coupling to the divergence of a (non anomalous) spontanously broken symmetry, and the axion case, 
we proceed with a comparison between the two cases. \\
In theories with spontaneously broken global symmetries and anomalies, Goldstone modes such as axions can couple to gauge fields through interactions like \( a(x) F \tilde{F} \). These couplings arise when a global chiral symmetry with a mixed anomaly—such as the axial-vector–vector $(AVV)$ anomaly—is spontaneously broken. The resulting Nambu–Goldstone (NG) boson couples to the anomaly through a mechanism rooted in Ward identities. This connection is often described using a local interaction, though the full anomaly structure in the currents currelator is fundamentally nonlocal.\\
To illustrate this, consider a model with a fermion \( Q \) and a complex scalar field \( \Phi \), invariant under a global chiral \( U(1)_A \) symmetry that is anomalous. When \( \Phi \) acquires a vacuum expectation value, the symmetry is spontaneously broken, and the resulting phase degree of freedom \( \pi(x) \) behaves as a Goldstone boson. The fermion gains mass via a Yukawa interaction with \( \Phi \), while the chiral current \( J_A^\mu \) acquires both fermionic and bosonic components.\\
The Goldstone mode \( \pi(x) \) couples to the divergence of the axial current via
\beq
\mathcal{L} \sim \pi(x) \partial_\mu J_A^\mu.
\eeq
If \( J_A^\mu \) is anomalous, its divergence includes a term like \( F \tilde{F} \), and this gives rise to the axion–gauge field coupling
\beq
\mathcal{L} \sim \frac{a(x)}{v} F \tilde{F}.
\eeq

However, this derivation simplifies the full structure of the anomaly. In the perturbative picture, the anomaly is encoded in the longitudinal part of a three-point function (such as the \( AVV \) vertex), which contains a \(1/q^2\) pole indicating the exchange of a massless pseudoscalar. This is not a true particle pole but rather a signal of the nonlocal nature of the anomalous interaction. Its presence arises from solving the anomalous Ward identity
\beq
q_\lambda \Delta^{\lambda \mu \nu}(q, p_1, p_2) \sim a_1 \epsilon^{\mu \nu p_1 p_2},
\eeq
which admits a solution of the form
\beq
\Delta^{\lambda \mu \nu} \sim \frac{q^\lambda}{q^2} \epsilon^{\mu \nu p_1 p_2}.
\eeq
When mass terms are included, the divergence of the axial current gains a contribution from the pseudoscalar density
\beq
\partial_\mu J_A^\mu = a_1 F \tilde{F} + 2 m_Q \bar{Q} \gamma_5 Q,
\eeq
that modifies the equation of motion for the Goldstone boson, by adding a mass correction via coupling to the pseudoscalar current. Consequently, the effective interaction becomes
\beq
\mathcal{L} \sim \frac{a(x)}{v} \left( F \tilde{F} + 2 m_Q \bar{Q} \gamma_5 Q \right).
\eeq

In momentum space, this is reflected in a vertex structure involving both the massless anomaly term and a massive correction encoded in the scalar loop integral \( C_0 \) of a three-point function (see \eqref{lcoà}). Indeed, the full axion–photon–photon vertex becomes
\beq
V^{\mu\nu} \sim \epsilon^{\mu \nu p_1 p_2} \left( 1 + 2 m^2 C_0(p_1^2, p_2^2, q^2, m^2) \right).
\eeq
While the first term corresponds to the anomaly, the second arises from explicit chiral symmetry breaking via mass terms.
Importantly, the anomaly pole (i.e., the \(1/q^2\) behavior) dominates in the conformal or massless limit. When fermion masses or off-shell momenta are included, the residue of the pole typically vanishes. This implies that the anomaly pole is not a true propagating particle, but it still defines the dominant asymptotic behavior of the vertex at large momentum transfer.\\
Furthermore, the local interaction \( a(x) F \tilde{F} \), while useful phenomenologically (e.g., in axion models like Peccei–Quinn), does not fully capture the rich tensor structure of the anomaly, especially the interplay between longitudinal and transverse components governed by conformal symmetry. The complete off-shell correlator contains both, and they are linked by the anomaly coefficient in a way that the local description obscures.\\
In contrast to NG modes from non-anomalous symmetries - which couple derivatively and decouple in the infrared - anomalous NG modes retain a distinct signature due to the anomaly pole. However, as we have mentioned above, this pole does not define a physical asymptotic state unless specific kinematic conditions are met that require a light-cone dynamics. Away from these limits, the full vertex must be retained to describe the interaction accurately.\\
Finally, while the use of Ward identities to derive local interactions is consistent at tree level or in effective field theory, the full quantum structure of anomaly vertices—especially in gravitational or mixed gauge-gravitational contexts—requires a nonlocal formulation. This ensures correct matching to perturbative computations, respects the topological origin of the anomaly, and captures sum rules and dispersion relations not visible in local couplings.

\section{Reconstructing the \(AVV\) Correlator from the Anomaly Pole and Conformal Symmetry: review}

We begin our analysis in the conformal limit, where the conformal constraints in momentum space, combined with the inclusion of a pole in the solution of these identities for the \(AVV\) correlator, play a crucial role.  
In this regime, the conformal constraints enable the complete identification of parity-odd anomaly interactions using only a single input—an anomaly pole in the longitudinal sector of each correlator. This result is inherently nonperturbative, as it does not rely on any specific Lagrangian formulation. However, once we move away from the conformal point and introduce massive fermions in virtual corrections, this picture undergoes significant modifications.  
Before delving into these corrections, which will allow for a deeper exploration of the sum rule associated with this correlator, we first provide a brief review of how the presence of a pole, together with the conformal Ward identities, enables the reconstruction of anomaly interactions in the conformal limit. This approach further facilitates the characterization of all form factors in the decomposition of these correlators into longitudinal and transverse sectors. \\
We start by imposing the external Ward identities in flat space-time for a vector and an axial-vector current respectively as quantum averages
\begin{align}\label{eq:CWIJJJ}
	\partial_\mu \braket{J^\mu}=0, \qquad\qquad \partial_\mu \braket{J^\mu_A}=a_1 \, \varepsilon^{\mu\nu\rho\sigma}F_{\mu\nu}F_{\rho\sigma}.
\end{align}
The Ward identities for the $AVV$ are extracted from \eqref{eq:CWIJJJ} by functional differentiation 
\begin{equation}\label{eq:jjjconsward3p}
	\begin{aligned}
		&p_{i\mu_i}\,\braket{J^{\mu_1}(p_1)J^{\mu_2}(p_2)J_A^{\mu_3}(p_3)}=0,\quad \quad i=1,2\\[1.2ex]
		&p_{3\mu_3}\,\braket{J^{\mu_1}(p_1)J^{\mu_2}(p_2)J_A^{\mu_3}(p_3)}=-8i  \, a_1  \, \varepsilon^{p_1p_2\mu_1\mu_2},
	\end{aligned}
\end{equation}
where $a_1$ is the anomaly coefficient. \\
Whenever we will be using a symmetric $(1,2,3)$ labeling of the momenta, useful for the conformal parameterization of the vertices, all the momenta $p_i$ will be chosen as incoming.  \\ 
In order to consider also a combined gauge and gravitational background, as in  the case for the gravitational anomaly, the quantum average of the axial vector current will be extended with the inclusion of a covariant derivative and 
curvature related terms 
\begin{equation} \label{eq:anomaliachirale}
	\nabla_\mu\langle J_A^\mu\rangle=
	a_1\, \varepsilon^{\mu \nu \rho \sigma}F_{\mu\nu}F_{\rho\sigma}+ a_2 \, \varepsilon^{\mu \nu \rho \sigma}R^{\alpha\beta}_{\hspace{0.3cm} \mu \nu} R_{\alpha\beta \rho \sigma},
\end{equation}
that will be investigated when we deal with the $ATT$ correlator. 
As already mentioned, in both the case of the $ATT$ and $AVV$, these conditions, together with the ordinary Ward identities imposed either from gauge invariance or diffeomorphism invariance, set the boundary conditions for the 
reconstruction of the corresponding correlators by the inclusion of a single anomaly pole. \\ 
In the $AVV$ case, from these relations we construct the general form of the correlator, splitting the operators into 
a transverse and a longitudinal part as
\begin{equation}
	\begin{aligned}
		\label{ex}
		&J^{\mu}(p)=j^\mu(p)+j_{loc}^\mu(p),\\
		&j^{\mu}=\pi^{\mu}_{\alpha}(p)\,J^{\alpha}(p),\quad \pi^{\mu}_\alpha(p)\equiv \delta^{\mu}_\alpha-\frac{p_\alpha\,p^\mu}{p^2},\\
		&j_{loc}^{\mu}(p)=\frac{p^\mu}{p^2}\,p\cdot J(p),
	\end{aligned}
\end{equation}
and
\begin{equation}
	\begin{aligned}
		\label{ex}
		&J_A^{\mu}(p)=j_A^\mu(p)+j_{A \,\text{loc}}^\mu(p),\\
		&j_A^{\mu}=\pi^{\mu}_{\alpha}(p)\,J_A^{\alpha}(p),\\
		&j_{ A \,\text{loc}}^{\mu}(p)=\frac{p^\mu}{p^2}\,p\cdot J_A(p).
	\end{aligned}
\end{equation}
Due to \eqref{eq:jjjconsward3p}, the correlator is purely transverse in the vector currents. We then have the following decomposition
\begin{equation}
	\braket{ J^{\mu_1 }(p_1) J^{\mu_2 } (p_2)J_A^{\mu_3}(p_3)}=\braket{ j^{\mu_1 }(p_1) j^{\mu_2 }(p_2) j_A^{\mu_3}(p_3)}+\braket{J^{\mu_1 }(p_1) J^{\mu_2 }(p_2)\, j_{A \text { loc }}^{\mu_3}(p_3)}\label{decomp}
\end{equation}
where the first term is completely transverse with respect to the momenta $p_{i\mu_i}$, $i=1,2,3$ and the second term is the longitudinal part that is proper of the anomaly contribution.
Using the anomaly constraint \eqref{eq:jjjconsward3p} on $j_{A loc}$ we obtain the expression of the longitudinal component of 
\eqref{decomp}
\begin{align}
	\label{anomc}
	\braket{J^{\mu_1 }(p_1) J^{\mu_2 }(p_2)\, j_{A \text { loc }}^{\mu_3}(p_3)}=\frac{p_3^{\mu_3}}{p_3^2}\,p_{3\,\alpha_3}\,\braket{J^{\mu_1}(p_1)J^{\mu_2}(p_2)J_A^{\alpha_3}(p_3)}=\Phi_0\,\varepsilon^{p_1p_2\mu_1\mu_2}\,p_3^{\mu_3},
\end{align}
with 
\beq
\Phi_0=-\frac{8i\,a_1}{p_3^2}
\eeq
defining the anomaly form factor in the conformal limit. The complete solution can be obtained by solving the special conformal Ward identities and using the Schouten relations in order to parameterize also the transverse sector of  
\eqref{decomp}. The procedure involves an intermediate step in which the action of the special conformal transformation on the correlator is projected onto the transverse sector and separated into second order and first order equations using the Schouten relations.Then, the general structure of the transverse part can be further simplified using the Schouten identities, resulting in the expression \cite{Coriano:2023hts}

\begin{align}
	\label{ref}
	\braket{J^{\mu_1}(p_1)J^{\mu_2}(p_2) J^{\mu_3}_A (p_3)} &=-\frac{8i\,a_1}{p_3^2}\varepsilon^{p_1p_2\mu_1\mu_2}\,p_3^{\mu_3} \nonumber \\
	&8 i a_1\,\pi^{\mu_1}_{\alpha_1}(p_1)
	\pi^{\mu_2}_{\alpha_2} (p_2) \pi^{\mu_3}_{\alpha_3}
	\left(p_3\right)\bigg[ \,p_2^2\,I_{3\{1,0,1\}}\, \varepsilon^{p_1 \alpha_1\alpha_2\alpha_3}-
	\,p_1^2\,I_{3\{0,1,1\}}\, \varepsilon^{p_2\alpha_1\alpha_2\alpha_3}  
	\bigg], \nonumber \\
\end{align}
where, remarkably, the $a_1$ the anomaly coefficient has turned into a structural factor both for the longitudinal and transverse parts.
Unlike the perturbative approach, where the anomaly coefficient is primarily determined by the prefactor in front of a Feynman amplitude, here it emerges as the residue of a pole in the longitudinal sector, as a solution of the anomaly constraint. The conformal Ward identities then propagate this pole to the other sectors of the correlator, specifically the transverse one.  \\
Notably, this parameterization of the \(AVV\) correlator is the most compact one, as it not only highlights the central role of the anomaly pole but also reveals additional features that we will discuss shortly.
The solution involves the integral
\begin{align}
	&I_{3\left\{1,0,1\right\}}=p_1\,p_3\frac{\partial^2}{\partial p_1\partial p_3}\,I_{1\{0,0,0\}},
\end{align}
where 
$I_{1\{0,0,0\}}$ is the master integral (3K integral) related to the three-point function in the theory of a free massless scalar field in $d=4$ \cite{Bzowski:2013sza}
\begin{align}
	I_{1\{0,0,0\}}=\frac{1}{4}C_{0}(p_1^2,p_2^2,p_3^2).
\end{align}
One can then express the 3K integral explicitly as follows
\begin{align}
	I_{3\left\{1,0,1\right\}}(p_1^2,p_2^2,p_3^2)=&\frac{1}{\lambda^2}\bigg\{-2 p_1^2 p_3^2 \bigg[p_1^2 \left(p_2^2-2 p_3^2\right)+p_1^4+p_2^2 p_3^2-2 p_2^4+p_3^4\bigg] C_0\left(p_1^2,p_2^2,p_3^2\right)\notag\\
	&+p_1^2 \left(\left(p_1^2-p_2^2\right)^2+4p_2^2p_3^2- p_3^4\right)\log\left(\frac{p_1^2}{p_2^2}\right)+4p_1^2 p_3^2 \left(p_1^2-p_3^2\right)\log\left(\frac{p_1^2}{p_3^2}\right)\notag\\
	&-p_3^2 \Big((p_2^2-p_3^2)^2+4p_1^2p_2^2-p_1^4\Big)\log\left(\frac{p_2^2}{p_3^2}\right)-\lambda(p_1^2-p_2^2+p_3^2)\bigg\}
\end{align}
where
\begin{equation}
	\label{lcoà}
	C_0\left(q^2,p_1^2,p_2^2\right)= \frac{1}{i \pi^2}\int d^n k\frac{1}{k^2 (k- q)^2 (k - p_1)^2},\,
\end{equation}
is the ordinary massless scalar master integral. It is given by
\beq
C_0(s,s_1,s_2) = \frac{1}{s} \Phi(x,y).
\eeq
$\Phi(x,y)$ is the special function  \cite{Usyukina:1992jd} 
\bea
\Phi( x, y) &=& \frac{1}{\lambda} \biggl\{ 2 [\textrm{Li}_2(-\rho  x) + \textrm{Li}_2(- \rho y)]  +
\ln \frac{y}{ x}\ln \frac{1+ \rho y }{1 + \rho x}+ \ln (\rho x) \ln (\rho  y) + \frac{\pi^2}{3} \biggr\},
\label{Phi}
\eea
where
\bea
\lambda(1,x,y) = \sqrt {\Delta},
\qquad  \qquad \Delta=(1-  x- y)^2 - 4  x  y,
\label{lambda} \\
\rho( x,y) = 2 (1-  x-  y+\lambda)^{-1},
\qquad  \qquad x=\frac{s_1}{s} \, ,\qquad \qquad y= \frac {s_2}{s}\qquad s\equiv p_3^2\, .
\eea
The analysis highlights the pivotal role of the anomaly pole in determining the entire anomaly interaction, particularly in the conformal limit. Remarkably, the complete off-shell correlator is fully specified by a single additional form factor, which is connected to the integral \( I_{3\{1,0,1\}} \).
Notably, the conformal solution is inherently scale-independent and entirely dictated by the anomaly coefficient. If this coefficient is set to zero, the correlator vanishes entirely.\\
At this juncture, a crucial question arises: does this $1/p_3^2\equiv 1/q^2$ interaction correspond to an asymptotic state?\\
For such an interpretation, the anomaly pole must be identified as a massless particle pole. To establish this, one must verify under general kinematic conditions whether the residue of the vertex 
\beq
\label{gpp}
g_{p} \equiv\lim_{q^2\to 0} q ^2 \braket{J^{\mu_1}(p_1)J^{\mu_2}(p_2) J^{\mu_3}_A (q)} \neq 0
\eeq
is nonvanishing.
The subtle nature of the anomaly interaction lies in the fact that the longitudinal form factor can exhibit an anomaly pole, even though \eqref{gpp} may vanish. As already pointed out in the previous sections, anomaly poles do not necessarily correspond to particle poles. As a result, the on-shell limits used to identify the anomaly coupling in the 1PI vertex are deeply connected to the properties of this vertex in its complete tensorial form rather than just with its longitudinal part. 
\\
To move away from the conformal limit, we address the massive fermion case and compute the vertex using ordinary perturbation theory. 
We can perform a decomposition similar to \eqref{ref} in the form 
\begin{align}
	\label{refm}
	\braket{J^{\mu_1}(p_1)J^{\mu_2}(p_2) J^{\mu_3}_A (p_3)} &=	\braket{J^{\mu_1 }(p_1) J^{\mu_2 }(p_2)\, j_{A \text { loc }}^{\mu_3}(p_3)}
	+ \pi^{\mu_1}_{\alpha_1}(p_1)\pi^{\mu_2}_{\alpha_2} (p_2) \pi^{\mu_3}_{\alpha_3}(p_3)
	\, \Delta_T^{\alpha_1 \alpha_2 \alpha_3 }(p_1^2,p_2^2,p_3^2,m^2) 
	\nonumber \\
\end{align}
with the longitudinal sector given by 
\begin{equation}
	\label{nonloc}
	\braket{J^{\mu_1 }(p_1) J^{\mu_2 }(p_2)\, j_{A \text { loc }}^{\mu_3}(p_3)}\equiv \Phi_0(p_1^2,p_2^2,p_3^2,m^2) \varepsilon^{p_1p_2\mu_1\mu_2}\,p_3^{\mu_3}.
\end{equation}
The expression of the transverse part is denoted by $\Delta_T$.\\
To go beyond the conformal limit, we return to perturbation theory by introducing a fermion of mass \( m \). The \( AVV \) anomaly tensor \(\Delta_{\alpha\mu\nu}(p_1, p_2)\) is defined through the triangle fermion loop
\begin{equation}
	\label{orig}
	\Delta_{\alpha\mu\nu}(p_1, p_2) = -\int \frac{d^4 k}{(2\pi)^4} \mathrm{Tr} \left( \frac{1}{\not{k} - \not{p_1} - m} \gamma_\mu \frac{1}{\not{k} - m} \gamma_\nu \frac{1}{\not{k} + \not{p_2} - m} \gamma_\alpha \gamma_5 \right) + [ (p_1, \mu) \leftrightarrow (p_2, \nu) ].
\end{equation}

The vector current remains conserved
\beq
p_1^\mu \Delta_{\alpha\mu\nu} = 0, \quad p_2^\nu \Delta_{\alpha\mu\nu} = 0,
\eeq
while the axial-vector Ward identity is anomalous
\beq
q^\alpha \Delta_{\alpha\mu\nu} = 2m \Delta_{\mu\nu}(p_1, p_2) + \frac{1}{2\pi^2} \epsilon_{\mu\nu\rho\sigma} p_1^\rho p_2^\sigma.
\eeq

The tensor structure of \(\Delta_{\alpha\mu\nu}\) is parameterized as
\begin{equation}
	\label{deff}
	\Delta_{\alpha\mu\nu} = F_1(s, m^2)\, q_\alpha \epsilon_{\mu\nu\rho\sigma} p_1^\rho p_2^\sigma 
	+ F_2(s, m^2)\, (p_{2\nu} \epsilon_{\alpha\mu\rho\sigma} - p_{1\mu} \epsilon_{\alpha\nu\rho\sigma}) p_1^\rho p_2^\sigma.
\end{equation}

The anomaly resides entirely in the \( F_1 \) form factor, which obeys
\begin{equation}
	\label{broken}
	s F_1(s, m^2) = -\frac{1}{\pi^2} \int_0^1 dx \int_0^{1-x} dy \, \frac{m^2}{m^2 - x y s - i\epsilon} + \frac{1}{2\pi^2}.
\end{equation}

The second term on the rhs of the equation above defines the anomaly, while the first accounts for explicit chiral symmetry breaking. In the massless limit, only the anomaly term survives and saturates the anomaly sum rule.

The off-shell generalization of \( F_1 \), denoted as \( \Phi_0 \), is
\begin{equation}
	\label{fstà}
	\Phi_0(p_1^2, p_2^2, p_3^2, m^2) = \frac{m^2}{\pi^2} \frac{1}{q^2} C_0(q^2, p_1^2, p_2^2, m^2) + \frac{1}{2\pi^2} \frac{1}{q^2},
\end{equation}
with \( C_0 \) the standard scalar triangle integral \eqref{lcoà} parameterized as
\begin{equation}
	C_0(q^2, p_1^2, p_2^2, m^2) = \int_0^1 dx \int_0^{1-x} dy \, \frac{1}{\mathcal{D}(x,y)},
\end{equation}
where
\beq
{D}(x,y) = m^2 + q^2 y(x + y - 1) - p_1^2 x y + p_2^2 x(x + y - 1).
\eeq

While \eqref{fstà} exhibits a \(1/q^2\) term resembling an anomaly pole, this feature is misleading, since the form factor is characterized by a branch cut rather than a true pole in the massive theory. The full cancellation of this apparent pole will be explored later, and is closely related to the anomaly sum rule. For comparison, the local form factor derived from the axial-vector Ward identity is
\begin{align}
	\label{loc}
	q_\lambda \braket{J^\mu J^\nu J_A^\lambda} &\equiv \Phi_W\, \epsilon^{\mu\nu p_1 p_2}, \\
	\Phi_W &= q^2 \Phi_0 = \frac{m^2}{\pi^2} C_0(p_1^2, p_2^2, q^2, m^2) + \frac{1}{2\pi^2}.
\end{align}
Unlike \( \Phi_0 \), the local form \( \Phi_W \) contains no \(1/q^2\) structure and does not obey an anomaly sum rule. It is commonly used in effective axion–gauge field interactions, but lacks the nonlocal and topological properties intrinsic to \( \Phi_0 \).
Thus, while \( \Phi_W \) reflects a local vertex structure, only \( \Phi_0 \) captures the full anomaly content, including its asymptotic \(1/q^2\) behavior and relation to the anomaly pole. The distinction between these two forms is central to understanding the anomaly both in conformal and massive regimes.

\subsection{Pole cancellation in $\Phi_0$ for $m\neq 0$ for on-shell vectors}
We now prove that there is no pole in $\Phi_0$ when the two vector lines are on-shell and $m\neq 0$.
One can 
show that there is a cancelation between the second and the first contribution in $\Phi_0$ in 
\eqref{fstà}.
For this purpose we consider the on shell case, $p_1^2=p_2^2=0$ and use the explicit expression of $C_0$ in the form 
\beq
\label{exp0}
C_0(0,0,q^2,m^2)\equiv C_0(q^2,m^2)=-\frac{1}{2 m^2}\,  {_3F_2} \left( 1, 1, 1; 2, \frac{3}{2}; x \right), 
\eeq
where the \(_3F_2\) hypergeometric function can be written as a series expansion for
\( q^2/4m^2 < 1\) as 
\beq
_3F_2 \left( 1, 1, 1; 2, \frac{3}{2}; x \right) = \sum_{k=0}^\infty \frac{(1)_k (1)_k (1)_k}{(2)_k \left(\frac{3}{2}\right)_k} \frac{x^k}{k!},
\eeq
with \(x = \frac{p_2^2}{4m^2}\), and where \((a)_k\) is the Pochhammer symbol, defined as
\beq
(a)_k = a(a+1)(a+2)\cdots(a+k-1), \quad (a)_0 = 1.
\eeq
Substituting the Pochhammer symbols for the given parameters
\beq
\begin{aligned}
	(1)_k &= 1 \cdot 2 \cdot 3 \cdots (k) = k!, \\
	(2)_k &= 2 \cdot 3 \cdot 4 \cdots (k+1) = (k+1)k!, \\
	\left(\frac{3}{2}\right)_k &= \frac{3}{2} \cdot \frac{5}{2} \cdot \frac{7}{2} \cdots \left(\frac{3}{2} + k - 1\right).
\end{aligned}
\eeq
one derives the expansion 
\beq
\label{exp1}
C_0(q^2, m^2) = -\frac{1}{2m^2} \left( 1 + \frac{q^2}{12m^2} + \frac{q^4}{180m^4} + \mathcal{O}\left(\frac{q^6}{m^6}\right) \right),
\eeq
giving an anomaly form factor of the form
\beq
\Phi_0 = -\frac{1}{4\pi^2} \frac{1}{12m^2} - \frac{1}{4\pi^2} \frac{q^2}{180m^4} +\ldots
\eeq
Notice that there has been a cancelation between the $1/q^2$ term in \eqref{fstà} and a similar term coming from the scalar loop $C_0$. In other words, the 1PI effective interaction is entirely mass dependent. Given the condition 
$q^2< 4 m^2$, the expansion is valid below the cut. Notice that the form factor vanishes as the fermion mass is sent to infinity, indicating the decoupling of the heavy fermion in the perturbative loop.\\ 
The area subtended by the spectral density of $\Phi_0$ (see Fig 1) over the cut plays a crucial role in the analysis of the sum rule. Therefore, we now turn to a more general examination of this form factor.
\begin{figure}[t]
	\centering
	\includegraphics[scale=0.8]{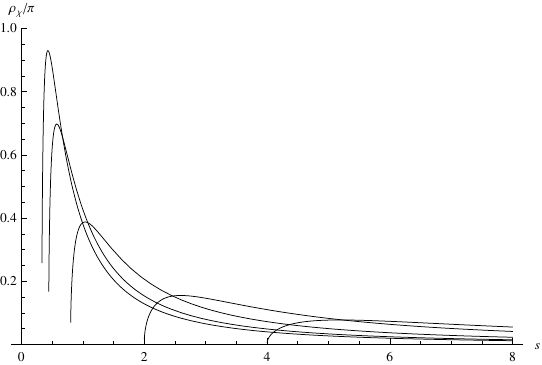}
	\caption{Spectral density flow in the $p_1^2 = p_2^2 = 0$ case as a function of the fermion mass $m$. The sum rule is an area law for this spectral flow as one moves towards the conformal point  $m\to 0$.}
	\label{dd1}
\end{figure}
\noindent

\section{Chiral sum rules }
In this section, we begin our investigation of the sum rules governing the anomaly form factors, starting with the case of the $AVV$ correlator. 
The original derivation of the sum rule for the $AVV$ had been discussed for some specific 
cases. Horejsi  \cite{Horejsi:1985qu} investigated the case $p_1^2=p_2^2=0, m\neq 0$. Previous analysis had derived the sum rule when only a single vector line is off-shell $(p_1^2=0,p_2^2\neq 0, m\neq 0)$ \cite{Dolgov:1971ri} and for 
$(p_1^2=p_2^2 <0, m=0)$ \cite{Frishman:1980dq}. The area law implicit in this specific feature of the anomaly interactions is, at least in part, a result of the property of analiticity of the vertex but also manifests a correlation between the longitudinal and transverse sectors of the interaction. This point has been observed in \cite{Coriano:2014gja}. 
\\
The analiticity conditions of $\Phi_0$ in the complex $q^2$ plane is summarised by
the following two formulas
\beqa
\oint_C \Phi_0(s,s_1,s_2,m^2)  \, ds & =& 0,
\eeqa
and
\beqa
\label{disp}
\Phi_0(q^2,s_1,s_2,m^2) &=& \frac{1}{2\pi i} 
\oint_C \frac{ \Phi_0(s,s_1,s_2,m^2)}{s - q^2} \, ds, \nn\\
\eeqa
where the contour of integration, by Jordan's lemma, is reduced to the $s\geq 0$ axis, where the function, as we are going to show, exhibits poles and a cut for $s> 4 m^2$.  
As we are going to see, the spectral density 
will contain poles in its rational factors, with a branch cut starting at $4 m^2$.

\begin{figure}
	\label{figg1}
	\centering
	\begin{tikzpicture}
		
		% Thick axes
		\draw[thick, ->] (-1,0) -- (6,0) node[right] {$\text{Re}(s)$};
		\draw[thick, ->] (0,-3) -- (0,3) node[above] {$\text{Im}(s)$};
		
		% Pole at the origin
		\filldraw[red] (0,0) circle (3pt) node[below left] {$0$};
		
		% Branch cut from 4m^2 to infinity
		\draw[thick, decorate, decoration={zigzag, amplitude=2pt, segment length=4pt}] (2,0) -- (5.9,0);
		\node at (2, -0.3) {$4m^2$};
		
		% Spikes at s=0 with different colors and closer positioning
		\draw[very thick, ->, red] (0,0) -- (0,1.5);     % Black spike with arrow
		\draw[very thick, ->, red] (0,0) -- (0,-1.5); % Red spike closer to q^2=0 with arrow
	\end{tikzpicture}
	\caption{Analyticity region of the on-shell anomaly form factor in the complex \( s\equiv q^2 \)-plane, with two bold spikes at \(s=0\) of different colour, for visual clarity. The sum rule of the anomaly form factor of the $AVV$ involves a cancelation between the two red spikes at $s=0$, leaving only the integral over the continuum for $s> 4 m^2$.} 
\end{figure}
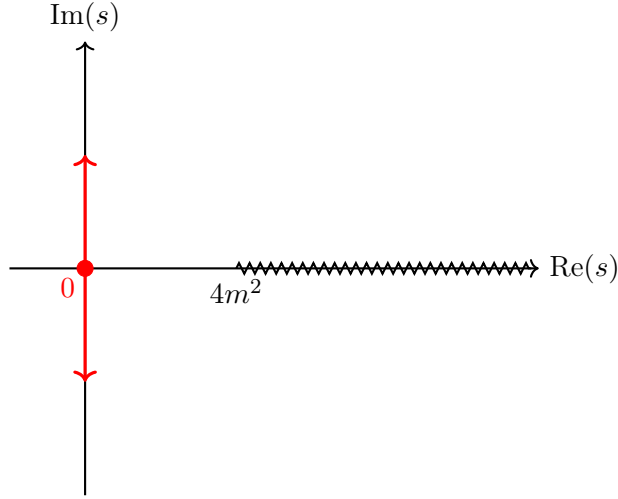
We start by defining the general discontinuity of $\Phi_0$ for $q^2 >0$, as
\beqa
\textrm{disc}\Phi_0 
&=& \Phi_0(q^2 + i \epsilon, p_1^2,p_2^2, m^2)- \Phi_0(q^2 - i \epsilon, p_1^2,p_2^2, m^2) 
\eeqa
and its imaginary part, the spectral density  $\Delta$, in the $s$ variable as
\beq
\Delta \Phi_0(s,s_1,s_2,m^2)\equiv \Im \Phi_0(s,s_1,s_2,m^2) = \frac{1}{2 i}\textrm{disc}\,\Phi_0.
\eeq
The vanishing of the boundary integral on the large radius allows one to rewrite the dispersion relation \eqref{disp} as

\beq
\Phi_0(q^2,p_1^2,p_2^2,m^2)=\frac{1}{\pi}\int_0^\infty \frac{\Delta\Phi_0(s, p_1^2,p_2^2,m^2)}{s - q^2}{ds}.
\eeq
As previously mentioned, the relation holds in the complex $q^2$ plane at fixed positive values of $s_1$ and $s_2$. Notice that the dispersion relations we derive are unsubtracted, as we enforce vector Ward identities to re-express the divergent form factors in terms of convergent ones. The computation of the spectral density of the anomaly form factors is carried out using the principal value prescription for the $1/q^2$ pole appearing in their explicit expressions
\beq
\label{ddd}
\textrm{disc}\, \frac{1}{q^2}\equiv \frac{1}{q^2 + i \epsilon}- \frac{1}{q^2  - i \epsilon}=- 2 \pi i \delta(q^2).
\eeq
Using this relation in \eqref{fstà} 
\beqa
\label{zero}
\textrm{disc}\,\Phi_0 (q^2,p_1^2,p_2^2, m^2)
&=& \frac{m^2}{\pi^2} \textrm{disc}\left( \frac{C_0(q^2,p_1^2,p_2^2,m^2)}{q^2}\right) + \frac{1}{2 \pi^2}\textrm{disc}\frac{1}{q^2}
\eeqa
with the help of \eqref{ddd}, we can separate the pole and the continuum contributions  as 
\beq
\label{first}
\textrm{disc}\,\Phi_0(q^2,p_1^2,p_2^2, m^2)= - 2 i \frac{ m^2}{\pi} \, C_0 (q^2,p_1^2,p_2^2, m^2)\, \delta(q^2)  -  \frac{i}{\pi}\, \delta(q^2)
+\frac{ m^2}{ \pi^2 q^2}\textrm{disc} \,C_0(q^2,p_1^2,p_2^2, m^2),
\eeq
showing that the spectral density exhibits two distinct localized terms at $s=0$ ($\delta$ spikes) and 
a continuum contribution for $q^2 > 4 m^2$, with an interplay among these three components, as we are going to show, which is related to the property of analiticity of $\Phi_0$. We will refer to the localized components as
\beq
\label{spk}
\begin{aligned}
	\text{spk}_1 &\equiv - \frac{i}{\pi} \, \delta(q^2), \\
	\text{spk}_2 &\equiv - 2 i\frac{ m^2}{\pi} \, C_0(q^2, p_1^2, p_2^2, m^2) \, \delta(q^2). 
\end{aligned}
\eeq
The first spike is of constant strength, while the second plays a role in equaling the integral of the density over the cut, causing a cancelation. One can think of these two components as distinct but superimposed resonances at $s=0$.
These features, as we shall demonstrate, either before or after integration, exhibit cancellation patterns tied to the external kinematics of the vertex. Such cancellations arise either within the two spikes themselves or through their interaction of  $\text{spk}_2$ with the continuum. 
The continuum corresponds to the discontinuity in the scalar loop, \(\textrm{disc} \, C_0\), which is determined by placing two intermediate lines of the triangle diagram on-shell.

\subsection{The use of the Laplace/Borel transforms }
\label{lap}
The on-shell behaviour presented above can be discussed for arbitrary kinematics. The pattern of cancelations illustrated in 
Fig. 2 takes a different form in the general case, when the vector lines are off-shell and we are away from the conformal point. 
In this case the cancelation is between the integral of $\textrm{spk}_2$ and the one over the continuum, and is illustrated in Fig. 3.  
To prove this point we introduced a method that allows to establish this result starting directly from the parametric integral. 
This method is essentially based on the use of Borel transforms.\\
The evaluation of the integral of the spectral density of $\Phi_0$ along the 
$\left[ 4 m^2,\infty)\right. $ cut
can be computed by a trick that we are going to describe, using the inverse Laplace transform of the same density (indicated by $t$), followed  by the $t\to 0$  limit on the same variable. For this, in full generality, we define the inverse Laplace transform $(f(t))$ of a generic function 
$F(s)$

\begin{equation}
	f(t) = \mathcal{L}^{-1}\{F(s)\} = \frac{1}{2\pi i} \int_{\gamma - i \infty}^{\gamma + i \infty} F(s) e^{st} \, ds,
\end{equation}
and recall that for a rational function $F(s)$ with a pole at $s={s_0}$ 	
\begin{equation}
	\mathcal{L}^{-1}\{\frac{1}{s-{s_0}}\}(t)=e^{{s_0}t}
\end{equation}
which provides a nonzero contribution to the dispersion integral as $t\to 0$. 
The method allows one to compute the integral over the continuum of the spectral density quite efficiently, but it requires some care since extra singular contributions, that can be viewed as subtractions in the reconstruction of a given form factor, are not identified by this approach. For example, double poles of the form $1/(s-{s_0})^2$ are not taken into account by the method since one obtains
\begin{equation}
	\mathcal{L}^{-1}\{\frac{1}{(s-{s_0})^2}\}(t)=t e^{{s_0}t},
\end{equation}
that vanishes in the same $t\to 0$ limit.  An example can be found in the analysis of the $ATT$ case, where 
we discuss how this point can be handled (\secref{trashà}). \\
In our case, for the anomaly form factor $\Phi_0$, given a spectral representation of the form
\beq
\Phi_0(q^2)=\frac{1}{\pi}\int_0^{\infty} \frac{\Delta \Phi_0(s)}{s- q^2}  \,ds
\eeq
the action of the inverse transform will generate the expression
\begin{equation}
	\mathcal{L}^{-1}\Bigl\{  \frac{1}{\pi} \int_0^{\infty} \frac{\Delta \Phi_0}{s-q^2}  \,ds \Bigl\}(t)  = \frac{1}{\pi} \int_0^{\infty} 
	\Delta \Phi_0(s) e^{s t}  \,ds	\end{equation}
and the integral of the spectral density will be obtained by a limit on $t$
\beq	
\lim_{t\rightarrow 0}\mathcal{L}^{-1}\{\Phi_0\}(t)  =\frac{1}{\pi}\int_0^{\infty} \Delta \Phi_0  \,ds 
\eeq
This approach allows one to derive the expression of the integral of the discontinuity along the cut and verify the sum rules of such form factors. \\
For this purpose we use the parametric representation 
\begin{equation}
	\label{parm}
	C_0\left(p_1^2,p_2^2,q^2,m^2,m^2,m^2\right)= \int_0^1 dx\, \int _0^{1-x} dy\,  \frac{1}{\D(x,y)}
\end{equation}

\begin{equation}
	\D(x,y)=m^2+q^2 y (x+y-1)-p_1^2 x y+p_2^2 x (x+y-1)
\end{equation}

that we insert in the expression of $\Phi_0$ given in \eqref{fstà}. A direct computation gives 
for the inverse Laplace transform the expression
\begin{equation}
	\mathcal{L}^{-1}\{\Phi_0\}(t)=-\int_0^1 dx \int_{0}^{1-x} d y \, \frac{g^3 \left( m^2 \exp \left(\frac{t \left(-m^2+p_1^2x y-p_2^2 x^2-p_2^2  x y+p_2^2  x\right)}{x y+y^2-y}\right)-2 m^2+p_1^2 x y-p_2^2  x^2-p_2^2  x y+p_2^2 x\right)}{\pi ^2 \left(m^2-p_1^2 x y+p_2^2  x^2+p_2^2 x y-p_2^2  x\right)}
\end{equation}
on which we perform the $t\to 0$ limit as described above, in order to derive directly the value of the integral of the discontinuity over the entire cut	
\begin{equation}
	\label{sumr1}
	\mathcal{L}^{-1}\{\Phi_0\}(t\to 0)=\frac{1}{\pi}\int_{4 m^2}^\infty  \Delta{\Phi_0}(s, p_1^2,p_2^2,m^2) ds  =\frac{1}{2\p^2}.
\end{equation}
Eq. \eqref{sumr1} holds in the most general case, for off-shell vector lines and a massive fermion. This result is in agreement with the analysis of \cite{Giannotti:2008cv}.\\	
Notice that a sum rule holds only for $\Phi_0$ and not for $\Phi_W$, the form factor extracted from the local formulation of the anomaly interaction \eqref{loc}.
A direct computation indeed shows that the corresponding parametric integral, in this case,
\beq
\int_{4 m^2}^\infty \Delta{\Phi_W}(s, p_1^2,p_2^2,m^2) ds=
\int_0^1 \left(\int_0^{1-x} \frac{g^3 m^2}{2 \pi ^2 \left(x y+y^2-y\right)} \, dy\right) \, dx 
\eeq
is indeed divergent.  \\
\begin{figure}[t]
	\centering
	\subcaptionbox{}{\includegraphics[scale=0.8]{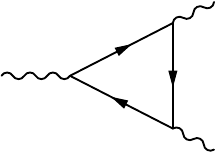}}  \hspace{2cm}
	\subcaptionbox{}{\includegraphics[scale=0.8]{figures/fermionpole.pdf}} \hspace{2cm}
	\subcaptionbox{}{\includegraphics[scale=0.8]{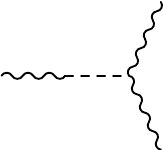}}
	\caption{The fermion loop (a); the collinear region (b); the effective pseudoscalar exchange (c). The $q^2=0$ point defining the residue of the interaction can be viewed as a massless state decaying into a massless virtual pseudoscalar state that couples to the anomaly at the last stage.}
	\label{ddiag}
\end{figure}

\section{The residue of the particle pole in the light-cone variables }  
The computation of the residue of a possible particle pole in the off-shell massive $AVV$ correlator requires the choice of special noncovariant frames where to proceed with the analysis. This allows one to test if the coupling 
$g_{p}$ in \eqref{gpp} is nonzero. The computation of the residue of the particle pole of the $AVV$ has been discussed in the past in \cite{Armillis:2009sm} and it was shown that there is no coupling of this state to 
the vertex for off-shell photons and a virtual massive fermion in the loop. The analysis was performed both for the massive and the massless cases using specific parameterizations of the correlator. We are going to elaborate on this point regarding the way the tensorial limits are performed, not discussed in \cite{Armillis:2009sm}, with the goal of extending these analysis, in the next sections, to the case of the $ATT$. \\
We recall that in order to discuss the presence of a massless particle pole interaction 
in the anomaly vertex we need to perform the limit \eqref{gpp}
and show that it is nonvanishing independently of the virtualities of the other invariants. 
We use as notations  $(s_1,s_2)\equiv (p_1^2, p_2^2)$ and $s=q^2=p_3^3$, assuming various configurations for the external invariants and the value of the fermion mass. We are going to consider two possible ways to approach the $q^2\to 0$ limit, by sending the four-momentum $q^\lambda$ on the light-cone or by performing the soft limit $q^\lambda\to 0$ in all of its components.\\
In order to perform the light-cone limit on $q$ we parameterize the four-momenta as follows 
\beq
p_3\equiv q= q^+ n^+ + q^- n^- \qquad  p_1= p_1^+n^+ + p_1^- n^- + p_\perp
\qquad p_2= p_2^+n^+ + p_2^- n^- - p_\perp
\eeq
with light-cone versors defined by ${(n^\pm)}^2=0 , \, n^+\cdot n^-=1$. In the light cone limit 
$q^+$ is large as well as $p_1^+$ and $p_2^+$, 
while the components 
\beq
q^-=\frac{q^2}{2 q^+} \qquad p_1^-=\frac{p_1^2 + p_\perp^2}{2 p_1^+}\qquad 
p_2^-=\frac{p_2^2 + p_\perp^2}{2 p_2^+}
\eeq
are small. We introduce a scale $\lambda^2\sim p_1^2\sim p_2^2$, with $\lambda \ll \bar{Q}=q^+$.   
In this limit $q_0$ and $q_3$ are both large but fixed, with  $q^-=q_0- q_3$ arbitrarily small. $\bar{Q}$ is assumed of 
any large but {\em fixed} size in the light-cone direction of $n^+$, and the limit $q^2\to 0 $ is performed, for a fixed value of $\bar{Q}$, by taking  
$q^-$, the component in the $n^-$ direction arbitrarily small. For this reason any power of $q^+$ can be dropped if multiplied by $q^-$.\\
The process for extracting the value of the residue of a tensor structure with open indices involves projecting that specific structure onto the light-cone basis. If, during this procedure, a particular component does not vanish, we can conclude that a massless $1/q^2$ pole develops a residue on the light-cone.  \\
It is straightforward to demonstrate that all tensorial structures vanish except for those that include a $1/p_3^2$ pole. This outcome is evident from any parameterization of the correlator. The only case where the residue is 
nonvanishing is for massless vector lines and a massless virtual fermion. We have left to Appendix B.1 more details concerning the technical apsects of this analysis. 
\section{The gravitational chiral anomaly vertex: review }

In this section, we begin our analysis of the \( ATT \) vertex. For simplicity, we are going to set the coupling constant to one. As for the $AVV$, our study starts in the conformal limit, revisiting earlier work that showed how the vertex can be reconstructed from the corresponding anomaly pole in \( CFT_p \) \cite{Coriano:2023gxa}.  
For the \( ATT \) vertex, the conformal realization that we examine in momentum space also encompasses the case of Chern-Simons currents \cite{Dolgov:1988qx}, which generate spin-1 anomalies at first discussed long ago and revisited in more recent studies \cite{delRio:2020cmv,Galaverni:2020xrq}. The form factor decomposition remains valid in both conformal and non-conformal cases, with the mass dependence of the correlator determined via standard perturbation theory. In \cite{Coriano:2023gxa}, it was shown that the conformal solution derived using \( CFT_p \) methods is in full agreement with the perturbative result.  Much like in the \( AVV \) case, a complete determination of the vertex in the conformal limit is also possible for the \( ATT \). Here too, the entire vertex is fixed by the residue at the pole of the longitudinal component of the axial-vector current.  
\\
We start by defining
\begin{align}
	\label{ddc}
	T^{\mu_i\nu_i}(p_i)&= t^{\mu_i\nu_i}(p_i)+t_{loc}^{\mu_i\nu_i}(p_i),\\
\end{align}
having introduced the transverse-traceless ($\Pi$), transverse $(\pi)$ and longitudinal ($\Sigma$) projectors, given respectively by 
\begin{align}
	%%%%%%%%%%%%%%%%%%%%%%%%%%%%%%%
	&\Pi^{\mu \nu}_{\alpha \beta}  = \frac{1}{2} \left( \pi^{\mu}_{\alpha} \pi^{\nu}_{\beta} + \pi^{\mu}_{\beta} \pi^{\nu}_{\alpha} \right) - \frac{1}{d - 1} \pi^{\mu \nu}\pi_{\alpha \beta}\label{TTproj}, \\&
	%%%%%%%%%%%%%%%%%%%%%%%%%%%%%%%
	\Sigma^{\mu_i\nu_i}_{\alpha_i\beta_i}=\frac{p_{i\,\beta_i}}{p_i^2}\Big[2\delta^{(\nu_i}_{\alpha_i}p_i^{\mu_i)}-\frac{p_{i\alpha_i}}{(d-1)}\left(\delta^{\mu_i\nu_i}+(d-2)\frac{p_i^{\mu_i}p_i^{\nu_i}}{p_i^2}\right)\Big]+\frac{\pi^{\mu_i\nu_i}(p_i)}{(d-1)}\delta_{\alpha_i\beta_i}\label{Lprojà}.
	%%%%%%%%%%%%%%%%%%%%%%%%%%%%%%%
\end{align}
Such decomposition allows one to split our correlation function into the following terms
\begin{equation} 
	\label{eq:splitlongttpart}
	\begin{aligned}
		\left\langle T^{\mu_{1} \n_{1}} T^{\mu_{2} \n_2} J_A^{\mu_{3}}\right\rangle=&\left\langle t^{\mu_{1} \n_{1}} t^{\mu_{2}\n_2} j_A^{\mu_{3}}\right\rangle+\left\langle T^{\mu_{1} \n_{1}} T^{\mu_{2}\n_2} j_{5\, l o c}^{\mu_{3}}\right\rangle+\left\langle T^{\mu_{1} \n_{1}} t_{l o c}^{\mu_{2}\n_2} J_A^{\mu_{3}}\right\rangle+\left\langle t_{l o c}^{\mu_{1} \n_{1}} T^{\mu_{2}\n_2} J_A^{\mu_{3}}\right\rangle \\
		&-\left\langle T^{\mu_{1} \n_{1}} t_{l o c}^{\mu_{2}\n_2} j_{A\, l o c}^{\mu_{3}}\right\rangle-\left\langle t_{l o c}^{\mu_{1} \n_{1}} t_{l o c}^{\mu_{2}\n_2} J_A^{\mu_{3}}\right\rangle-\left\langle t_{l o c}^{\mu_{1} \n_{1}} T^{\mu_{2}\n_2} j_{A\, l o c}^{\mu_{3}}\right\rangle+\left\langle t_{l o c}^{\mu_{1} \n_{1}} t_{l o c}^{\mu_{2}\n_2} j_{5\, l o c}^{\mu_{3}}\right\rangle 
	\end{aligned}
\end{equation}
which are constrained by ordinary trace Ward identities of the form 
\begin{equation} \label{ppp}
	\begin{aligned}
		&\delta_{\mu_i\nu_i }\braket{T^{\mu_1\nu_1}(p_1)T^{\mu_2\nu_2}(p_2)J_A^{\mu_3}(p_3)}=0,\qquad \quad&& i=\{1,2\}\\
		&p_{i\mu_i}\,\braket{T^{\mu_1\nu_1}(p_1)T^{\mu_2\nu_2}(p_2)J_A^{\mu_3}(p_3)}=0,\qquad \quad && i=\{1,2\}.
	\end{aligned}
\end{equation}
Notice that the trace identity (the first of the two) will be modified in the presence of mass terms, due to the breaking of conformal symmetry. The second remain invariant, being derived from the conservation of the stress energy tensor. As already mentioned the $TTA$ does not involve any renormalization and there is no trace anomaly. However, due to the presence of several diagrams of different topologies that are separately divergent, one needs a 
regularization scheme. Using the WIs above, we notice that most of the terms in Eq.~\eqref{eq:splitlongttpart} vanish, yielding
\begin{equation} 
	\begin{aligned}
		\left\langle T^{\mu_{1} \n_{1}} T^{\mu_{2} \n_2} J_A^{\mu_{3}}\right\rangle=&\left\langle t^{\mu_{1} \n_{1}} t^{\mu_{2}\n_2} j_A^{\mu_{3}}\right\rangle+\left\langle t^{\mu_{1} \n_{1}} t^{\mu_{2}\n_2} j_{A\, l o c}^{\mu_{3}}\right\rangle .
	\end{aligned}
\end{equation}
The longitudinal anomalous Ward identity is given by

\begin{equation}\label{eq:idwanomlp3}
	p_{3\mu_3}\braket{T^{\mu_1\nu_1}(p_1)T^{\mu_2\nu_2}(p_2)J_A^{\mu_3}(p_3)}= 4 i \, a_2 \, (p_1 \cdot p_2) \left\{ \left[\varepsilon^{\nu_1 \nu_2 p_1 p_2}\left(g^{\mu_1 \mu_2}- \frac{p_1^{\mu_2} p_2^{\mu_1}}{p_1 \cdot p_2}\right) +\left( \mu_1 \leftrightarrow \nu_1 \right) \right] +\left( \mu_2 \leftrightarrow \nu_2 \right) \right\},
\end{equation}
\normalsize
with the rhs of the equation above given by the Fourier transform of the second functional derivative of  the density 
$R\tilde R$.
As in the case of the $AVV$, Eq. \eqref{eq:idwanomlp3} in momentum space is solved by the insertion of a anomaly pole $1/p_3^2$ in the longitudinal sector of $J_A$
\begin{equation} \label{eq:anompolettj}
	\left\langle t^{\mu_{1} \n_{1}} t^{\mu_{2}\n_2} j_{A\, l o c}^{\mu_{3}}\right\rangle= 4i a_2 \frac{p_3^{\mu_3}}{p_3^2} \, (p_1 \cdot p_2) \left\{ \left[\varepsilon^{\nu_1 \nu_2 p_1 p_2}\left(g^{\mu_1 \mu_2}- \frac{p_1^{\mu_2} p_2^{\mu_1}}{p_1 \cdot p_2}\right) +\left( \mu_1 \leftrightarrow \nu_1 \right) \right] +\left( \mu_2 \leftrightarrow \nu_2 \right) \right\}.
\end{equation}
On the other hand, after accounting for the Bose symmetry and all the Schouten identities, the transverse-traceless component of the correlator can be expressed as \cite{Coriano:2023gxa}
\begin{equation}
	\begin{aligned}
		\langle t^{\mu_{1} \nu_{1}}\left({p}_{1}\right)& t^{\mu_{2} \nu_{2}}\left({p}_{2}\right) j_A^{\mu_{3} }\left({p_3}\right)\rangle=\Pi_{\alpha_{1} \beta_{1}}^{\mu_{1} \nu_{1}}\left({p}_{1}\right) \Pi_{\alpha_{2} \beta_{2}}^{\mu_{2} \nu_{2}}\left({p}_{2}\right) \pi_{\alpha_{3}}^{\mu_{3}}\left({p_3}\right) \bigg[
		\\
		&A_1\varepsilon^{p_1\alpha_1\alpha_2\alpha_3}p_2^{\beta_1}p_3^{\beta_2}
		-A_1(p_1\leftrightarrow p_2) \varepsilon^{p_2\alpha_1\alpha_2\alpha_3}p_2^{\beta_1}p_3^{\beta_2}\\
		&+A_2\varepsilon^{p_1\alpha_1\alpha_2\alpha_3}\delta^{\beta_1\beta_2}-
		A_2(p_1\leftrightarrow p_2)\varepsilon^{p_2\alpha_1\alpha_2\alpha_3}\delta^{\beta_1\beta_2}\\
		&+A_3\varepsilon^{p_1p_2\alpha_1\alpha_2}p_2^{\beta_1}p_3^{\beta_2}p_1^{\alpha_3}
		+A_4\varepsilon^{p_1p_2\alpha_1\alpha_2}\delta^{\beta_1\beta_2}p_1^{\alpha_3}
		\bigg],
	\end{aligned}
\end{equation}
where $A_3$ and $A_4$ are antisymmetric under the exchange $(p_1\leftrightarrow p_2)$.
The solution is \cite{Coriano:2023gxa}

\begin{allowdisplaybreaks}
	\begin{equation} \label{eq:pertresultff}
		\begin{aligned}
			A_1&=\frac{ p_2^2}{24 \pi^2 \lambda^4}
			\Bigg\{A_{11} 
			+A_{12} \log\left(\frac{p_1^2}{p_2^2}\right)+A_{13} \log\left(\frac{p_1^2}{p_3^2}\right)
			+A_{14} \, \,C_0(p_1^2,p_2^2,p_3^2)
			\Bigg\}, \\[8pt]
			A_2&=\frac{p_2^2}{48\pi^2\lambda^3}\Bigg\{
			A_{21}
			+A_{22} \log\left(\frac{p_1^2}{p_2^2}\right)+A_{23} \log\left(\frac{p_1^2}{p_3^2}\right)
			+A_{24} \, \, C_0(p_1^2,p_2^2,p_3^2)
			\Bigg\}, \\[8pt]
			A_3&=0,\\[8pt]
			A_4&=0\\[8pt],
		\end{aligned}
	\end{equation}
\end{allowdisplaybreaks}
with the K\"{a}llen $\lambda$-function given by
\begin{equation}
	\lambda \equiv \lambda(p_1,p_2,p_3)=\left(p_1-p_2-p_3\right)\left(p_1+p_2-p_3\right)\left(p_1-p_2+p_3\right).\left(p_1+p_2+p_3\right),
	\label{Kallen}
\end{equation}
expressed in terms of the magnitudes of the three momenta. 
In this case, the explicit expression aligns well with perturbation theory, which offers a free field theory realization of the anomalous conformal field theory (CFT). The coefficients \( A_{ij} \) in \eqref{eq:pertresultff} are polynomial functions of the external invariants \( q^2, p_1^2, p_2^2 \) and can be found in \cite{Coriano:2023gxa}.\\
The reconstruction of the interaction highlights the central role of the pole contribution in determining the entire vertex. Importantly, the Ward identities are sufficient to eliminate the apparent divergences in the diagram. Notably, topological contributions do not require any counterterms, provided the remaining conservation Ward identities are imposed on the vertex.
A key point here is that the anomaly pole does not manifest as a residue of the full vertex, as for the $AVV$. We will explore this phenomenon in detail.
\section{Perturbative computation of the massive $\langle TTJ_A \rangle$ correlator}
The investigation of the sum rule for the anomaly form factors of this vertex requires that we move away from the conformal point, by giving a mass to the fermion. For this reason we turn to the perturbative realization, expanding the analysis presented in \cite{Coriano:2023gxa}. 
Our calculations are carried out in the Breitenlohner-Maison scheme. To streamline the analysis, we transition to Minkowski space, where the generating functional is defined as
\begin{equation}
	e^{i \mathcal{S}[g]} = \int [\mathcal{D} \Phi]\, e^{i S_0[\Phi, g]},
\end{equation}
with the classical action $S_0$ involving a fermionic field in a gravitational and axial gauge field background
\begin{equation}
	S_0 = \int d^d x \,e  \left[\frac{i }{2}e_a^\mu\left( \bar{\psi} \gamma^a (D_\mu \psi) -  (D_\mu \bar{\psi}) \gamma^a \psi\right) -m \bar{\psi}\psi\right],
\end{equation}
where $e_a^\mu$ is the vielbein, $e$ its determinant, and $D_\mu$ the covariant derivative defined as
\begin{equation}
	\begin{aligned}
		D_\mu \psi &= \left(\partial_\mu + i g \gamma_5 B_\mu + \frac{1}{2} \omega_{\mu ab} \Sigma^{ab}\right) \psi, \\
		D_\mu \bar{\psi} &= \left(\partial_\mu - i g \gamma_5 B_\mu - \frac{1}{2} \omega_{\mu ab} \Sigma^{ab}\right) \bar{\psi}.
	\end{aligned}
\end{equation}
Here, $\Sigma^{ab}$ are the generators of the Lorentz group for spin-$\frac{1}{2}$ fields, and $\omega_{\mu ab}$ is the spin connection
\begin{equation}
	\omega_{\mu ab} = e_a^\nu \left(\partial_\mu e_{\nu b} - \Gamma_{\mu\nu}^\lambda e_{\lambda b}\right).
\end{equation}
The Greek indices refer to the curved background, while Latin indices correspond to the local flat basis.
Substituting the explicit form of $\Sigma^{ab}$, the action becomes
\begin{equation}
	S_0 = \int d^d x \, e \left[\frac{i}{2} \bar{\psi} e_a^\mu \gamma^a (\partial_\mu \psi) 
	- \frac{i}{2} (\partial_\mu \bar{\psi}) e_a^\mu \gamma^a \psi 
	- g B_\mu \bar{\psi} e_a^\mu \gamma^a \gamma_5 \psi 
	+ \frac{i}{4} \omega_{\mu ab} e_c^\mu \bar{\psi} \gamma^{abc} \psi -m \bar{\psi}\psi\right],
\end{equation}
where $\gamma^{abc} = \{\Sigma^{ab}, \gamma^c\}$.
The corresponding vertices can be computed by functional differentiation of the action with respect to the metric and the gauge field, followed by a Fourier transformation to momentum space. \\
The perturbative computation of the $\langle TTJ_A \rangle$ involves three topologically distinct Feynman diagrams, illustrated in Figure \ref{fig:feynmdiagr}.
\begin{figure}[t]
	\centering
	\includegraphics[scale=0.5]{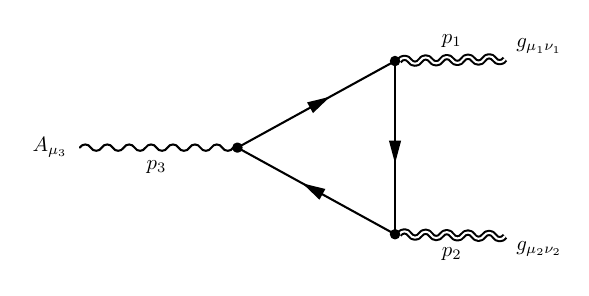}
	\includegraphics[scale=0.5]{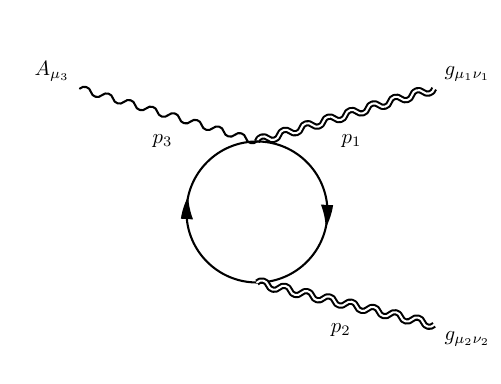}
	\includegraphics[scale=0.5]{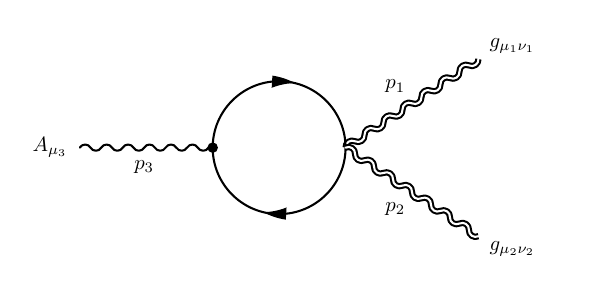}
	\includegraphics[scale=0.5]{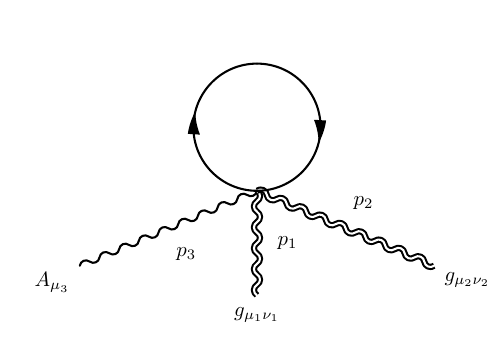}
	\caption{Feynman diagrams representing triangle, bubble, and tadpole topologies in the perturbative computation of the massive correlator.} 
	\label{fig:feynmdiagr}
\end{figure}

\subsection{The form factors for nonzero mass}\label{secffnmastta}
We present the perturbative results of the $TTA$ correlator in the massive case.
The $TTA$ can be decomposed into different sectors. In the massive case, the conformal symmetry is broken and there are additional terms that need to be considered in the decomposition. 
Indeed, the energy-momentum tensor is clearly still conserved, but the correlator exhibits trace terms due to the inclusion of a fermion mass
\begin{equation}
	p_i\left\langle T^{\mu_{1} \n_{1}} T^{\mu_{2} \n_2} J_A^{\mu_{3}}\right\rangle=0, \qquad\qquad
	g_{\mu_i\nu_i}\left\langle T^{\mu_{1} \n_{1}} T^{\mu_{2} \n_2} J_A^{\mu_{3}}\right\rangle\neq 0
	\qquad i=\{1,2\}.
\end{equation}
Therefore, one can write
\begin{equation}
	\label{ddf}
	\begin{aligned}
		\left\langle T^{\mu_{1} \n_{1}} T^{\mu_{2} \n_2} J_A^{\mu_{3}}\right\rangle=\left\langle t^{\mu_{1} \n_{1}} t^{\mu_{2}\n_2} j_A^{\mu_{3}}\right\rangle+\left\langle t^{\mu_{1} \n_{1}} t^{\mu_{2}\n_2} j_{A\,\, l o c}^{\mu_{3}}\right\rangle+\left\langle t_{loc}^{\mu_{1} \n_{1}} t^{\mu_{2}\n_2} j_A^{\mu_{3}}\right\rangle+\left\langle t^{\mu_{1} \n_{1}} t^{\mu_{2}\n_2}_{loc} j_A^{\mu_{3}}\right\rangle . 
	\end{aligned}
\end{equation}
We comment on this decomposition for clarity, noting the difference between this expansion, which includes only four terms, and the general one given in \eqref{eq:splitlongttpart}. The suffix ``$loc$'' in the current indicates the longitudinal sector, defined along the direction of momentum \( p_3 \). In the case of the stress-energy tensor, \( t^{\mu\nu}_{loc} \) denotes the sector orthogonal to the transverse-traceless one, according to the decompositions in \eqref{ddc}.  
\\
The conservation of the stress-energy tensor \( T^{\mu\nu} \) in this decomposition leaves, in the massive fermion case, only a trace component in \( t^{\mu\nu}_{loc} \). This trace cannot be accompanied, in the expansion of the correlator, by an operator such as \( j^\mu_{loc} \). The absence of contributions of the form \( \langle t_{loc} \,t\,  j_{loc}\rangle \) can also be understood from symmetry considerations: one cannot construct a tensor of odd parity in the longitudinal sector of the axial-vector current, given the symmetry of \( t \) and the presence of a trace in \( t_{loc} \), with  $j^\mu_{loc} \sim \frac{p^\mu_3}{p_3^2}$.
\\
Similar arguments show that terms of the form \( \langle t_{loc} t_{loc} j_{loc} \rangle \) cannot contribute as parity-odd components of the correlator. Indeed, a perturbative expansion confirms this result
\begin{equation}
	\begin{aligned}
		&g_{\mu_1\nu_1}\,g_{\mu_2\nu_2}\left\langle T^{\mu_{1} \n_{1}} T^{\mu_{2} \n_2} J_A^{\mu_{3}}\right\rangle=0,\\&
		{p_3}_{ \mu_3}\, g_{\mu_i\nu_i}\left\langle T^{\mu_{1} \n_{1}} T^{\mu_{2} \n_2} J_A^{\mu_{3}}\right\rangle =0, \qquad\qquad \quad i=\{1,2\}.
	\end{aligned}
\end{equation}
proving the consistency of the computation. Of all the terms contained in \eqref{ddf}, the anomaly form factor of the $ATT$ is present in the longitudinal component 
\begin{equation}
	\label{ffk}
	\left\langle t^{\mu_1 \nu_1} t^{\mu_2 \nu_2} j_{A\, \, l o c}^{\mu_3} \right\rangle = \frac{p_3^{\mu_3} }{p_3^2}\Pi^{\mu_1 \nu_1}_{\alpha_1 \beta_1}(p_1) \Pi^{\mu_2 \nu_2}_{\alpha_2 \beta_2}(p_2) \epsilon^{\alpha_1 \alpha_2 p_1 p_2} \left( \bar F_1 g^{\beta_1 \beta_2} (p_1\cdot p_2) + \bar F_2 p_1^{\beta_2} p_2^{\beta_1} \right).
\end{equation}
This term is determined by the gravitational anomaly, along with the mass corrections. The explicit expressions 
of the form factors \eqref{ffk} are given by 
\small
\begin{equation}
	\begin{aligned}
		\bar F_1&=\frac{1}{24 \pi^2 }+
		\frac{m^2}{2 \pi ^2 \lambda \left(s-s_1-s_2\right) }\bigg\{2 \big[\lambda  m^2+s {s_1} {s_2}\big] {C}_0\left(s,{s_1},{s_2},m^2\right)+Q (s) \big[s ({s_1}+{s_2})-({s_1}-{s_2})^2\big]\\&
		-{s_2} Q ({s_2}) \big[s+{s_1}-{s_2}\big]-{s_1} Q( {s_1}) \big[s-{s_1}+{s_2}\big]+\lambda \bigg\},\\
		\bar F_2&=-\frac{1}{24\pi^2 }+\frac{ m^2}{2 \pi ^2 \lambda ^2 } \bigg\{2 \big[\lambda  \left(m^2 (-s+{s_1}+{s_2})-2 {s_1} {s_2}\right)+3 {s_1} {s_2} \left(({s_1}-{s_2})^2-s ({s_1}+{s_2})\right)\big] {C}_0\left(s,{s_1},{s_2},m^2\right)\\&+{s_1} Q( {s_1}) \big[\lambda+6s_2(s+s_1-s_2)\big]+{s_2} Q ({s_2}) \big[\lambda +6 {s_1} (s-{s_1}+{s_2})\big]-Q (s) \big[12 s {s_1} {s_2}+\lambda  ({s_1}+{s_2})\big]\\&+\lambda  (-s+{s_1}+{s_2})\bigg\}.
	\end{aligned}
\end{equation}
\normalsize
The trace term of the correlator, on the other hand, can be written as
\begin{equation}
	\begin{aligned}
		\left\langle t_{loc}^{\mu_{1} \n_{1}} t^{\mu_{2}\n_2} j_A^{\mu_{3}}\right\rangle\equiv&\, \Sigma^{\mu_1 \nu_1}_{\alpha_1 \beta_1}(p_1) \Pi^{\mu_2 \nu_2}_{\alpha_2 \beta_2}(p_2)\pi^{\mu_3 }_{\alpha_3}(p_3) \left\langle T^{\mu_{1} \n_{1}} T^{\mu_{2} \n_2} J_A^{\mu_{3}}\right\rangle\\& \qquad= \frac{\pi^{\mu_1\nu_1}(p_1)}{3}\Pi^{\mu_2 \nu_2}_{\alpha_2 \beta_2}(p_2)  \pi^{\mu_3 }_{\alpha_3}(p_3)\left\langle T^{\mu}_\mu\, T^{\mu_{2} \n_2} J_A^{\mu_{3}}\right\rangle
	\end{aligned}
\end{equation}
where, in the final step, we used the explicit expression for $\Sigma^{\mu_1 \nu_1}_{\alpha_1 \beta_1}$ given in \eqref{Lprojà} along with the conservation of the energy-momentum tensor. This component is purely a trace term and it is transverse with respect to all the momenta. It can be expressed as
\begin{equation}
	\left\langle t_{loc}^{\mu_{1} \n_{1}} t^{\mu_{2}\n_2} j_A^{\mu_{3}}\right\rangle=\frac{\pi^{\mu_1\nu_1}(p_1)}{3}\Pi^{\mu_2 \nu_2}_{\alpha_2 \beta_2}(p_2)  \pi^{\mu_3 }_{\alpha_3}(p_3)\left[\bar F_3\,  \epsilon^{\alpha_2\alpha_3 p_1p_2}p_1^{\beta_{2}}\right]
\end{equation}
where
\small
\begin{equation}
	\begin{aligned}
		&\bar F_3=-\frac{  m^2  s_2 }{2\pi^2 \lambda^2 s} \bigg\{  2 \lambda  s+ 2Q(s) \Big[ 2 \lambda  s + 3  s  \Big( s_1^2 + 6  s_1  s_2 + s_2^2 \Big) + 3 (s_1+s_2)\Big( \lambda - (s_1 - s_2)^2 \Big) \Big] \\&
		- 2 Q(s_1) \Big[  s ( \lambda + 3 s_1 (s_1 + 3  s_2) ) + 3  s_1 ( \lambda - (s_1 - s_2)^2 ) \Big] 
		- 2 Q(s_2) \Big[ s ( \lambda + 3  s_2 (3  s_1 + s_2) ) + 3  s_2 ( \lambda - (s_1 - s_2)^2 ) \Big] \\
		&\qquad +\Big[
		\lambda s \left(   4m^2 + 3 (s_1 + s_2) \right) + 24  s s_1  s_2 (s_1 + s_2) 
		- \left( (s_1 - s_2)^2 - \lambda \right) \left( \lambda + 12  s_1  s_2 \right)\Big] C_0(s, s_1, s_2, m^2) 
		\bigg\}.
	\end{aligned}
\end{equation}
\normalsize
Finally, the transverse-traceless part can be expressed as
\begin{equation}
	\begin{aligned}
		\langle t^{\mu_{1} \nu_{1}}\left({p}_{1}\right)& t^{\mu_{2} \nu_{2}}\left({p}_{2}\right) j_A^{\mu_{3} }\left({p_3}\right)\rangle=\Pi_{\alpha_{1} \beta_{1}}^{\mu_{1} \nu_{1}}\left({p}_{1}\right) \Pi_{\alpha_{2} \beta_{2}}^{\mu_{2} \nu_{2}}\left({p}_{2}\right) \pi_{\alpha_{3}}^{\mu_{3}}\left({p_3}\right) \bigg[
		\\
		&A_1\varepsilon^{p_1\alpha_1\alpha_2\alpha_3}p_2^{\beta_1}p_3^{\beta_2}
		-A_1(p_1\leftrightarrow p_2) \varepsilon^{p_2\alpha_1\alpha_2\alpha_3}p_2^{\beta_1}p_3^{\beta_2}\\
		&+A_2\varepsilon^{p_1\alpha_1\alpha_2\alpha_3}\delta^{\beta_1\beta_2}-
		A_2(p_1\leftrightarrow p_2)\varepsilon^{p_2\alpha_1\alpha_2\alpha_3}\delta^{\beta_1\beta_2}\\
		&+A_3\varepsilon^{p_1p_2\alpha_1\alpha_2}p_2^{\beta_1}p_3^{\beta_2}p_1^{\alpha_3}
		+A_4\varepsilon^{p_1p_2\alpha_1\alpha_2}\delta^{\beta_1\beta_2}p_1^{\alpha_3}
		\bigg]
	\end{aligned}
\end{equation}
where
\begin{equation}
	\begin{aligned}
		A_1 &=- \frac{ s_2}{24 \pi^2 \lambda^4}\Big[\bar{A}_{11} Q(s_1)+\bar{A}_{12}Q(s_2)+\bar{A}_{13}Q(s)+\bar{A}_{14}C_0(s, s_1, s_2, m^2)+\bar{A}_{15}\Big],\\
		A_2 &=-\frac{  s_2}{48 \pi^2 \lambda^3} \Big[\bar{A}_{21} Q(s_1)+\bar{A}_{22}Q(s_2)+\bar{A}_{23}Q(s)+\bar{A}_{24}C_0(s, s_1, s_2, m^2)+\bar{A}_{25}\Big],\\
		A_3 &=0,\\
		A_4 &=0.
	\end{aligned}
\end{equation}
The complete expressions of these contributions can be found in Appendix \ref{transverse}.

\section{The sum rule of the $ATT$ for on-shell and off-shell gravitons }
Before discussing the sum rule of the gravitational anomaly form factor in full generality, we consider the case in which the two gravitons are on-shell.  
The on-shell expression of the $ATT$ takes a form close to the $AVV$.  
A direct perturbative computation shows that also in this case the only tensor structure present is the longitudinal one, proportional to $R\tilde{R}$ 
\begin{equation}
	\label{all}
	\begin{aligned}
		\langle T^{\mu_1 \nu_1} T^{\mu_2 \nu_2} J^{\mu_3}_A\rangle&=\frac{1}{96 \pi^2}\left[1+ \frac{12m^2}{p_3^2} \bigg(1+2 m^2 {C}_0\left(p_3^2,0,0, m^2\right)\bigg)\right]\\&\times \frac{p_3^{\mu_3}}{p_3^2}\bigg\{\bigg[\varepsilon^{\nu_1 \nu_2 p_1 p_2}\left(\left(p_1 \cdot p_2\right)g^{\mu_1 \mu_2}-{p_1^{\mu_2} p_2^{\mu_1}} \right)+\left(\mu_1 \leftrightarrow \nu_1\right)\bigg]+\left(\mu_2 \leftrightarrow \nu_2\right)\bigg\} .
	\end{aligned}
\end{equation}
To investigate the presence of a pole in the corresponding form factor, the \( C_0(s, 0, 0, m^2) \) integral needs to be expanded around $s=0$. We can use the expression \eqref{exp0} or, equivalently, we can rewrite it in the form 
\begin{equation}
	C_0(s, 0, 0, m^2) = \frac{\log^2 \left( \frac{\sqrt{s \left( s - 4m^2 \right)} + 2m^2 - s}{2m^2} \right)}{2s}.
\end{equation}
To better understand the behavior of \( C_0(s, 0, 0, m^2) \) near \( s = 0 \), we perform a series expansion. Expanding \( C_0(s, 0, 0, m^2) \) around \( s = 0 \), as in \eqref{exp1}, allows for a direct computation that confirms the absence of a pole at \( s = 0 \).  
Substituting this expansion into \eqref{all} and simplifying the expression reveals that the \( {1}/{s} \) and \( {1}/{s^2} \) terms cancel out. Specifically, the remaining contributions are finite and free of singularities, explicitly demonstrating the cancellation mechanism. Consequently, in the on-shell spectral function, only the cut contribution remains, in complete analogy with the \( AVV \) correlator.\\
\begin{figure}
	\centering
	\begin{tikzpicture}
		
		% Thick axes
		\draw[thick, ->] (-1,0) -- (7,0) node[right] {$\text{Re}(s)$};
		\draw[thick, ->] (0,-2) -- (0,4) node[above] {$\text{Im}(s)$};
		
		% Pole at the origin
		\filldraw[black] (0,0) circle (3pt) node[below left] {$0$};
		
		% Branch cut from 4m^2 to infinity
		\draw[thick, decorate, decoration={zigzag, amplitude=2pt, segment length=4pt}] (2,0) -- (7.0,0);
		\node at (2, -0.3) {$4m^2$};
		
		% Spikes at s=0 with different colors
		\draw[very thick, ->, red] (0,0) -- (0,1.5);     % Black spike with arrow
		\draw[very thick, ->, black] (0,0) -- (0,-2.5); % Red spike closer to q^2=0 with arrow
		
		% Simple crosses over the branch cut
		\draw[thick] (3.5,0.2) -- (3.7,-0.2); % First cross (top left to bottom right)
		\draw[thick] (3.7,0.2) -- (3.5,-0.2); % First cross (top right to bottom left)
		\node[below] at (3.6,-0.3) {$s_-$};   % Label for first cross
		
		\draw[thick] (4.5,0.2) -- (4.7,-0.2); % Second cross (top left to bottom right)
		\draw[thick] (4.7,0.2) -- (4.5,-0.2); % Second cross (top right to bottom left)
		\node[below] at (4.6,-0.3) {$s_+$};   % Label for second cross
		\draw[very thick, ->, black] (6.2,0.0) -- (6.2,-1.5); % Red spike closer to q^2=0 with arrow
		\node[below] at (6.0,-0.3) {$\bar s$}; 
	\end{tikzpicture}
	\caption{The spectral density of the off-shell gravitational anomaly form factor in the complex \( s\equiv q^2 \)-plane. Shown are the two spikes at \(s=0\), the anomalous thresholds at $s_{\pm}=(\sqrt{s_1}\pm \sqrt{s_2})^2$ and the pole at $s=\bar{s}=s_1+s_2$. In this case there is a cancelation in the sum rule between the integral of the density over the cut, the black spike at $s=0$, of variable strength, and the black spike at $s=\bar{s}$.   } 
	\label{figgXà}
\end{figure}
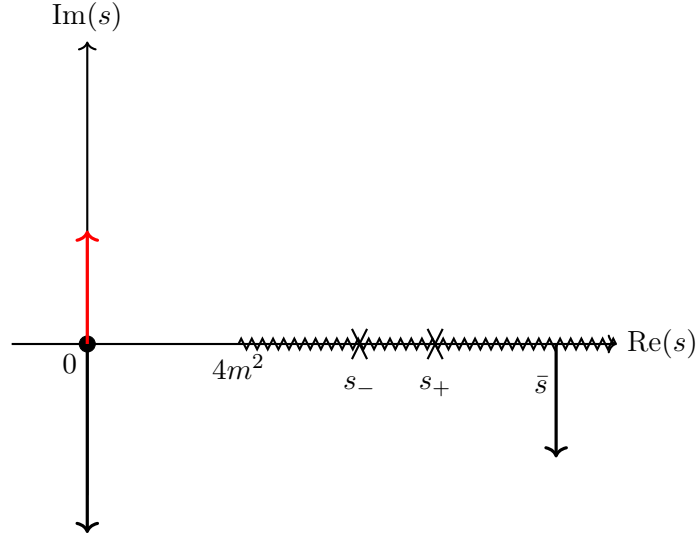
\noindent
We are going to expand on these analysis, showing the presence of a sum rule in the most general case, extending the approach presented for the $AVV$ in the former sections. 
The conformal description of the off-shell $(s_1, s_2 \neq 0, m = 0)$ $\langle TTJ_A\rangle$ or $\langle TT J_{CS}\rangle $ correlators is identical, as the only distinction between the two scenarios lies in the numerical value of the anomaly. In the conformal limit, the pole is inherently present, as demonstrated in \eqref{eq:anompolettj}. In the massive case, the same form factor can be computed perturbatively using the diagrams shown in Fig. \ref{fig:feynmdiagr}. Similarly, in this case, the perturbative computation does not necessitate any counterterms; however, an explicit regularization scheme is required to handle the $\gamma_5$ matrix within the loops. For this purpose, we have employed the Breitenlohner-Maison scheme for $\gamma_5$. We have presented the results of the computation in Section \ref{secffnmastta}. Specifically, the structure of the correlator in the longitudinal part, which defines the anomaly form factor in this context, can be written as
\begin{equation}
	\label{rr1}
	\langle t^{\mu_1 \nu_1} t^{\mu_2 \nu_2} j^{\mu_3}_{A\,loc} \rangle 
	= (\bar F_1-\bar F_2) \frac{p^{\mu_3}_3}{4\, p_3^2} 
	\left\{
	\left[
	\varepsilon^{\nu_1 \nu_2 p_1 p_2} 
	\left(
	p_1 \cdot p_2\,g^{\mu_1 \mu_2} - p_1^{\mu_2} p_2^{\mu_1}
	\right)
	+ (\mu_1 \leftrightarrow \nu_1)
	\right]
	+ (\mu_2 \leftrightarrow \nu_2)
	\right\} \nn
\end{equation}

\begin{equation}
	+ (\bar F_1+\bar F_2)\,\, \frac{p^{\mu_3}_3}{4\, p_3^2} 
	\left\{
	\left[
	\varepsilon^{\nu_1 \nu_2 p_1 p_2} 
	\left(
	p_1 \cdot p_2\,g^{\mu_1 \mu_2} + p_1^{\mu_2} p_2^{\mu_1}
	\right)
	+ (\mu_1 \leftrightarrow \nu_1)
	\right]
	+ (\mu_2 \leftrightarrow \nu_2)
	\right\}\nn
\end{equation}
\begin{equation}
	+ (\bar F_1+\bar F_2)\frac{(p_1 \cdot p_2)}{p_1^2p_2^2} \frac{p^{\mu_3}_3}{4\,p_3^2} 
	\left\{
	\left[
	\varepsilon^{\nu_1 \nu_2 p_1 p_2} 
	\left(
	(p_1 \cdot p_2)\,p_1^{\mu_1}p_2^{\mu_2} - p_2^2\,p_1^{\mu_1} p_1^{\mu_2}-p_1^2\, p_2^{\mu_1} p_2^{\mu_2}
	\right)
	+ (\mu_1 \leftrightarrow \nu_1)
	\right]
	+ (\mu_2 \leftrightarrow \nu_2)
	\right\}
\end{equation}
where the first contribution, proportional to $\bar F_1-\bar F_2$  is related to the second functional derivative of the anomaly $R\tilde{R}$ performed with respect to the metric and defines the anomaly form factor 

\begin{align}
	\label{ATT}
	\phi_{ATT}=\frac{\bar F_1-\bar F_2}{s}=\,\,&\frac{m^2 s_1 \left(s_2 \left(s_1^2-s^2\right)+s_2^2 (5 s+s_1)+(s_1-s)^3-3 s_2^3\right) B_0\left(s_1,m^2\right)}{\pi ^2 s (s-s_1-s_2) \lambda^2}\nonumber\\&+\frac{ m^2 s_2 \left(s_2 \left(3 s^2+s_1^2\right)+s_2^2 (s_1-3 s)-(s-s_1)^2 (s+3 s_1)+s_2^3\right) B_0\left(s_2,m^2\right)}{\pi ^2 s (s-s_1-s_2) \lambda^2}\nonumber\\&+\frac{ m^2 K_1(s,s_1,s_2) B_0\left(s,m^2\right)}{\pi ^2  s(s-s_1-s_2) \lambda^2}+\frac{ m^2 K_2(s,s_1,s_2) C_0\left(s,s_1,s_2,m^2\right)}{\pi ^2  s(s-s_1-s_2) \lambda^2}\nonumber\\& \frac{ m^2 \left(s^2-2 s (s_1+s_2)+s_1^2+s_2^2\right)}{\pi ^2s(s-s_1-s_2) \lambda}+ \frac{1}{12 \pi ^2 s}.
\end{align}
Notice that this form factor, at large $s$ behaves as 
\beq
\phi_{ATT} \sim \frac{1}{12 \pi ^2 s} + O(\frac{m^2}{s^2}\log\frac{s}{m^2}),
\eeq
similarly to the behaviour encountered in the case of the $AVV$, showing the dominance of the anomaly at large $q^2$. \\
The second and the third terms in \eqref{rr1}, which are related to $\bar F_1 +\bar F_2$
\begin{align}\label{eqqq}
	\bar F_1+\bar F_2=\,\,&\frac{2 m^2 s_1s_2 \left(s^2+s (s_1-2 s_2)-2 s_1^2+s_1 s_2+s_2^2\right) B_0\left(s_1,m^2\right)}{\pi ^2  (s-s_1-s_2) \lambda^2}\nonumber\\&+\frac{2  m^2  s_1s_2 \left(s_2 (s+s_1)+(s-s_1)^2-2 s_2^2\right) B_0\left(s_2,m^2\right)}{\pi ^2 s (s-s_1-s_2) \lambda^2}\nonumber\\&+\frac{2  m^2   s_1s_2\left(-2 s^2+s (s_1+s_2)+(s_1-s_2)^2\right) B_0\left(s,m^2\right)}{\pi ^2 (s-s_1-s_2) \lambda^2}\nonumber\\&+\frac{ m^2  s_1s_2
		K_3(s,s_1,s_2)\, C_0\left(s,s_1,s_2,m^2\right)}{\pi ^2  (s-s_1-s_2) \lambda^2}\nonumber\\&-\frac{2 \, m^2  s_1s_2}{\pi ^2  (s-s_1-s_2) \lambda}
\end{align}
are generated by the breaking of conformal symmetry, 
vanishing in the massless case and/or for on-shell gravitons $(s_1=s_2=0)$ and after contraction with the graviton polarizers. 
In equation \eqref{eqqq} the expression of $K_1$, $K_2$ and $K_3$ is given by
\begin{align}
	K_1(s,s_1,s_2)=\,\,&s_2 (s-s_1) \left(s^2+3 s s_1-2 s_1^2\right)+s_2^3 (3 s+2 s_1)\nonumber\\&-s_2^2 (s+s_1) (3 s+2 s_1)+s_1 (s-s_1)^3-s_2^4
\end{align}

\begin{align}
	K_2(s,s_1,s_2)=\,\,&2 m^2 \lambda \left(s^2-2 s (s_1+s_2)+s_1^2+s_2^2\right)\nonumber\\&+s_1 s_2 \left(3 s^3-5 s^2 (s_1+s_2)+s (s_1+s_2)^2+(s_1-s_2)^2 (s_1+s_2)\right)
\end{align}
\begin{align}
	K_3(s,s_1,s_2)=\,\,&s_2 \left(8 m^2 (s+s_1)+s^2-6 s s_1+s_1^2\right)+s_2^2 \left(-4 m^2+s+s_1\right)\nonumber\\&-(s-s_1)^2 \left(4 m^2+s+s_1\right)-s_2^3. 
\end{align}
We are going to show that the combination $\bar F_1-\bar F_2$ defines the correct expression of the anomaly form factor and we will verify the presence of a sum rule also in this case. Notice that the contribution proportional to $\bar F_1 +\bar F_2$, as we have mentioned, is proportional to the virtualities of the two gravitons and vanishes as we require these to be on-shell (see Eq$.$ \eqref{all}). 

\subsection{Derivation of the general sum rule for the $ATT$}
The anomaly form factor is associated with the first structure proportional to $\bar F_1 -\bar F_2$ in \eqref{rr1}. The corresponding tensor structure is the off-shell expression of the second functional derivative of $R\tilde{R}$, Fourier transformed to momentum space. The second contribution, proportional to $\bar F_1 +\bar F_2$, as we have just discussed, vanishes for on-shell-gravitons and is generated 
by the explicit breaking of the chiral symmetry for a nonzero mass parameter.  \\
Coming to the spectral function of the form factor in \eqref{ATT}, we separate the discontinuity into the pole contributions plus the continuum, in the form 
\begin{equation}
	\text{disc}(\phi_{ATT})=\text{disc}(\phi_{ATT})_{cont}+\text{disc}(\phi_{ATT})_{pole},
\end{equation}
with the discontinuity over the continuum given by 
\begin{align}
	\label{contà}
	\text{disc}(\phi_{ATT})_{cont}=\frac{ m^2 K_1(s,s_1,s_2)\, \text{disc}\, B_0\left(s,m^2\right)}{\pi ^2  s(s-s_1-s_2) \lambda^2}+\frac{g m^2 K_2(s,s_1,s_2) \,\text{disc}\,C_0\left(s,s_1,s_2,m^2\right)}{\pi ^2  s(s-s_1-s_2) \lambda^2}.
\end{align}
The discontinuity of $C_0$ is given by \eqref{d2}, while 
\begin{equation}
	\textrm{disc}\,{B_0}(s,m^2)=2\p i \sqrt{1-\frac{4m^2}{s} }\,\,\,
	\theta(s-4m^2).
\end{equation}
The pole contributions split at separate locations
\begin{align}
	\text{disc}(\phi_{ATT})_{pole}=\text{disc}(\phi_{ATT})_{0}+\text{disc}(\phi_{ATT})_{s_1+s_2}+\text{disc}(\phi_{ATT})_{s_-}+\text{disc}(\phi_{ATT})_{s_+},
\end{align}
where we have indicated the points corresponding to the zeroes of the $\lambda$ function as
\beq
s_\pm=(\sqrt{s_1} \pm \sqrt{s_2})^2.
\eeq
The pole at $s=0$ is split into two spikes 
\begin{align}
	\label{Ymà}
	\text{disc}(\phi_{ATT})_{0}=2\p i&\frac{\delta(s)}{12 \pi ^2 }-\frac{2\p i\delta(s)}{12 \pi ^2 (s_1-s_2)^3 (s_1+s_2)}\Y(s,s_1,s_2,m^2)
\end{align}
where 

\begin{align}
	\Y(s,s_1,s_2,m^2)\equiv&\biggl(12 m^2 \bigl(-\left((s_1-s_2) \left(s_1^2+s_2^2\right){B}_0\left(s,m^2\right)\right)\nonumber\\
	&+s_1 \left(s_1^2+2 s_1 s_2+3 s_2^2\right) {B}_0\left(s_1,m^2\right)-s_2 \left(3 s_1^2+2 s_1 s_2+s_2^2\right){B}_0\left(s_2,m^2\right)\nonumber\\
	&+(s_1-s_2) \left(2 m^2 \left(s_1^2+s_2^2\right)+s_1 s_2 (s_1+s_2)\right) C_0\left(s,s_1,s_2,m^2\right)\bigl)\nonumber\\ 
	& +12 m^2 \left(s_1^2+s_2^2\right) (s_1-s_2)\biggl).
\end{align}
We have outlined the various contributions in \figref{figgXà}. The "black spike"  at \( s = 0 \), which has a constant strength, corresponds to the first term on the right-hand side of \eqref{Ymà}. The second contribution, which depends on the external invariants, is represented by the "red spike" in the same figure.  \\
For \( s > 4m^2 \), across the cut, we encounter two apparent poles that, as we will demonstrate, do not contribute to the discontinuity. Additionally, there is a pole at \( s = s_1 + s_2 \), which can lie either above or below the cut (i.e., for \( s > 4m^2 \) or \( s < 4m^2 \)), depending on the values of \( s_1 \) and \( s_2 \).  
Finally, we observe that in the limit \( s \to 0 \)
\begin{align}
	\lim_{s\to 0} \Y(s,s_1,s_2)= \Y(0,s_1,s_2,m^2)\equiv&\biggl(12 m^2 \bigl(-\left((s_1-s_2) \left(s_1^2+s_2^2\right){B}^R_0\left(0,m^2\right)\right)\nonumber\\
	&+s_1 \left(s_1^2+2 s_1 s_2+3 s_2^2\right){B}_0^R\left(s_1,m^2\right)-s_2 \left(3 s_1^2+2 s_1 s_2+s_2^2\right){B}_0^R\left(s_2,m^2\right)\nonumber\\
	&+(s_1-s_2) \left(2 m^2 \left(s_1^2+s_2^2\right)+s_1 s_2 (s_1+s_2)\right){C}_0\left(0,s_1,s_2,m^2\right)\bigl)\nonumber\\ 
	& +12\, m^2 \left(s_1^2+s_2^2\right) (s_1-s_2)\biggl)  < \infty
\end{align}
\(\Y\) remains finite across the entire light-cone.

\subsection{Cancellation of the anomalous thresholds and the sum rule}
\label{trashà}
The expression for the spectral density
\begin{equation}
	\Phi_{\text{ATT}}=\frac{F_1 - F_2}{s}
\end{equation} 
can be separated into distinct components, representing contributions from regular kinematical thresholds as well as potential anomalous thresholds
\begin{equation}
	\Phi_{\text{ATT}} = \Phi_{\text{ATT}}^{\text{reg}} + \Phi_{\text{ATT}}^{(\l)},
\end{equation}
where $\Phi_{\text{ATT}}^{\text{reg}}$ corresponds to regular kinematical thresholds and $\Phi_{\text{ATT}}^{\l}$ accounts for possible anomalous thresholds. The regular part, $\Phi_{\text{ATT}}^{\text{reg}}$, includes contributions from standard kinematical thresholds located at
\begin{equation}
	q^2 = 0, \quad q^2 = s_1 + s_2, \quad \text{and} \quad q^2 = 4m^2,
\end{equation}
which manifest in the form of simple poles and cuts. These thresholds arise from well-understood physical processes. In contrast, the anomalous part, $\Phi_{\text{ATT}}^{\text{anom}}$, emerges under special conditions when the kinematical factor $\lambda = 0$ which could be present in this computation. Given that
\begin{equation}
	\lambda = (s - s_-)(s - s_+),
\end{equation}
the dispersive representation of  $ \Phi_{\text{ATT}}$ may allow terms of the form
\begin{align}
	\textrm{disc} \,\Phi_{\text{ATT}} = &c_0 \delta(s) + c_1(s,s_1, s_2) \theta(s - 4m^2) H(s) + c_2(s,s_1, s_2) \delta(s - s_1 - s_2)\nn\\&
	+ c_3(s, s_1, s_2) \delta'(s - s_-) + c_4 \delta'(s - s_+) + c_5 \delta(s - s_-) + c_6 \delta(s - s_+),
\end{align}
where the derivative of the $\delta$-function corresponds to higher order poles at $\lambda = 0$.
Indeed, the general structure of $\Phi_{\text{ATT}}$ is
\begin{align}
	\Phi_{\text{ATT}} =&  \frac{c_0(s,s_1,s_2,m^2)}{s} + \frac{c_1(s,s_1,s_2,m^2)}{s} + \text{(continuum terms)} + \frac{c_2(s,s_1,s_2,m^2)}{s - \bar{s}} + \frac{c_3(s,s_1,s_2,m^2)}{(s - s_-)^2} \nn\\&+ \frac{c_4(s,s_1,s_2,m^2)}{(s - s_+)^2} + \frac{c_5(s,s_1,s_2,m^2)}{s - s_-} + \frac{c_6(s,s_1,s_2,m^2)}{s - s_+},
\end{align}

where $\bar{s} = s_1 + s_2$ and, $c_3$ and $c_4$ define possible residues at the double poles of $\lambda^2$. These could, in principle, introduce subtractions that can be analyzed as follows. 
For example, at $s = s_-$, we have
\begin{equation}
	\Phi_{\text{ATT}}^{(\l)}(q^2 = s_-) = \frac{1}{2\pi i} \int_0^\infty \frac{c_3(s, s_1, s_2, m^2) \delta'(s - s_-)}{(s - q^2)} \, ds
	+ \frac{1}{2\pi i} \int_0^\infty \frac{c_5(s, s_1, s_2, m^2) \delta(s - s_-)}{(s - q^2)} \, ds,
\end{equation}
that simplifies to
\begin{equation}
	{2\pi i}\,\Phi_{\text{ATT}}^{(\l)}(q^2 = s_-)= -\frac{d}{ds} \left( \frac{c_3(s, s_1, s_2, m^2)}{(s - q^2)} \right) \bigg|_{s = s_-} - \frac{c_5(s_-, s_1, s_2, m^2)}{s_- - q^2}.
\end{equation}
A similar analysis gives 
\begin{equation}
	{2\pi i}\,\Phi_{\text{ATT}}^{(\lambda)}(q^2 = s_+) =  -\frac{d}{ds} \left( \frac{c_4(s, s_1, s_2, m^2)}{(s - q^2)} \right) \bigg|_{s = s_+} - \frac{c_6(s_+, s_1, s_2, m^2)}{s_+ - q^2}.
\end{equation}
For the evaluation of $\phi_{\text{ATT}}^{(\lambda)}(q^2 = s_\pm)$, we need to compute the explicit expression of the scalar integral $C_0$ at these kinematical points. We use the expressions of the scalar integral $C_0$ at these points. They are given by 
\begin{align}
	C_0(s_-,s_1,s_2,m^2)=&\frac{1}{{s_1}{s_2}\, s_-^2}\Biggl(\sqrt{{s_1}} \Biggl(\sqrt{{s_2} \left({s_2}-4 m^2\right)} \left(\sqrt{{s_1}}-\sqrt{{s_2}}\right)\times \nn\\ &
	\times \log \left(\frac{\sqrt{{s_2} \left({s_2}-4 m^2\right)}+2 m^2-{s_2}}{2 m^2}\right)\nonumber\\&+\sqrt{{s_2}} \sqrt{\left(-2 \sqrt{{s_1}} \sqrt{{s_2}}+{s_1}+{s_2}\right) \left(-4 m^2-2 \sqrt{{s_1}} \sqrt{{s_2}}+{s_1}+{s_2}\right)} \log \left(S_1\right)\Biggl)\nonumber\\&+\sqrt{{s_2}} \sqrt{{s_1} \left({s_1}-4 m^2\right)} \left(\sqrt{{s_2}}-\sqrt{{s_1}}\right) \log \left(\frac{\sqrt{{s_1} \left({s_1}-4 m^2\right)}+2 m^2-{s_1}}{2 m^2}\right)\Biggl)
\end{align}
and 
\begin{align}
	C_0( s_+,s_1,s_2,m^2)=&\frac{1}{{s_1}{s_2} s_+^2}\Biggl(\sqrt{{s_1}} \Biggl(\sqrt{{s_2} \left({s_2}-4 m^2\right)} \left(\sqrt{{s_1}}+\sqrt{{s_2}}\right)\times \nn \\&
	\times \log \left(\frac{\sqrt{{s_2} \left({s_2}-4 m^2\right)}+2 m^2-{s_2}}{2 m^2}\right)\nonumber\\&-\sqrt{{s_2}} \sqrt{\left(\sqrt{{s_1}}+\sqrt{{s_2}}\right)^4-4 m^2 \left(\sqrt{{s_1}}+\sqrt{{s_2}}\right)^2} \log \left(S_2\right)\Biggl)\nonumber\\&+\sqrt{{s_2}} \sqrt{{s_1} \left({s_1}-4 m^2\right)} \left(\sqrt{{s_1}}+\sqrt{{s_2}}\right) \log \left(\frac{\sqrt{{s_1} \left({s_1}-4 m^2\right)}+2 m^2-{s_1}}{2 m^2}\right)\Biggl), 
\end{align}

where
\begin{equation}
	S_1=\frac{\sqrt{\left(-2 \sqrt{{s_1}} \sqrt{{s_2}}+{s_1}+{s_2}\right) \left(-4 m^2-2 \sqrt{{s_1}} \sqrt{{s_2}}+{s_1}+{s_2}\right)}+2 m^2-\left(\sqrt{{s_1}}-\sqrt{{s_2}}\right)^2}{2 m^2}
\end{equation}
and 
\begin{equation}
	S_2=\frac{\sqrt{\left(\sqrt{{s_1}}+\sqrt{{s_2}}\right)^4-4 m^2 \left(\sqrt{{s_1}}+\sqrt{{s_2}}\right)^2}+2 m^2-\left(\sqrt{{s_1}}+\sqrt{{s_2}}\right)^2}{2 m^2}.
\end{equation}
Using these relations one can show by a long computation that 

\begin{equation}
	\Phi_{\text{ATT}}^{(\l)}(q^2 = s_\pm) \equiv 0,
\end{equation}
and hence 
\begin{equation}
	\Phi_{\text{ATT}}^{(\l)}=0.
\end{equation}
Therefore, the only pole to consider at finite $s$ is the one located at $s= s_1+s_2$.
This contributes to the sum rule with a localized density that is given by 
\begin{align}
	\label{boà}
	\text{disc}(\phi_{ATT})_{s_1+s_2}=&\frac{2\p i\, m^2\delta(s-s_1-s_2)}{4 \pi ^2 (s_1+s_2)}  \biggl(-2 {B}_0\left(s_1+s_2,m^2\right)+{B}_0\left(s_1,m^2\right)\nn\\&+{B}_0\left(s_2,m^2\right)+\left(4 m^2-s_1-s_2\right){C}_0\left(s_1,s_2,s_1+s_2,m^2\right)+2\biggl).
\end{align}
Notice that the UV divergences in the $B_0$'s in the expression above cancel automatically, with rule $-2 +1 +1$ in the counting of the $1/\epsilon_{UV}$ poles in each of these self-energy contributions. Therefore we can replace in \eqref{boà} each $B_0$ by $B_0^R$ with no change. No IR singularities are encountered unless we consider special values where any of the external invariants vanishes. 
\\
In order to compute the sum rule for the longitudinal part of the $ATT$ correlator we can use the same technique used before for the $AVV$. Therefore, as in the case of the anomaly form factor of the $AVV$,  we apply the inverse Laplace transform on both sides of the dispersive representation of the form factor (Cauchy's relation) to obtain
\begin{equation}
	\mathcal{L}^{-1}\Bigl\{   \int_0^{\infty} \frac{\Delta \phi_{ATT}}{s-\bar{s}}  \,ds \Bigl\} (t) = \frac{1}{\pi} \int_0^{\infty} 
	\Delta \phi_{ATT} \,\,e^{{{s}}t}  \,ds,	
\end{equation}
where 
\begin{equation}
	\Delta \phi_{ATT}\equiv \frac{1}{2 i}\text{disc}\, \phi_{ATT}
\end{equation}
and the integral of the spectral density will be obtained by performing the $t\to 0$ limit
\beq	\label{sumr}
\lim_{t\rightarrow 0}\mathcal{L}^{-1}\{\phi_{ATT}\}(t)  =\frac{1}{\pi}\int_0^{\infty} \Delta \phi_{ATT}  \,ds.  
\eeq
The techniques used to compute the inverse Laplace transform is the same described in the previous case. We have used the parametric form of the 2-points and the 3-points scalar function, exchanging the integration with the inverse Laplace transform, and once the limit on $t$ has been performed, we obtain the result of the integral in equation \eqref{sumr}
\begin{equation}
	\label{ag}
	\frac{1}{\pi}\int_0^{\infty} \Delta \Phi_{ATT}  \,ds = a_g
\end{equation}
with $a_g$ denoting the gravitational anomaly. The pattern of cancelation, in this case, is between the continuum, the pole at 
$s=\bar s=s_1 + s_2$ and the pole at $s=0$ giving the identity 
\beq
\label{eq}
\int_0^\infty \frac{1}{2 i} \left(\textrm{disc}\Phi_{ATT\, 0}(s,s_1,s_2,m^2) + \textrm{disc}\Phi_{ATT\, s_1+s_2}(s,s_1,s_2,m^2)\right)=\int_{4 m^2}^\infty 
ds \,  \frac{1}{2 i} \textrm{disc}\Phi_{ATT}(s,s_1,s_2,m^2)
\eeq 
where 
\beq
\int_0^\infty \frac{1}{2 i} \left(\textrm{disc}\Phi_{ATT\, 0}(s,s_1,s_2,m^2) + \textrm{disc}\Phi_{ATT\, s_1+s_2}\right) =
\nonumber
\eeq
\beq
\frac{1}{12 \pi} - \frac{1}{12 \pi  (s_1-s_2)^3 (s_1+s_2)}\Y(0,s_1,s_2,m^2)
\nonumber
\eeq
\beq
+ \frac{ m^2}{4 \pi  (s_1+s_2)} \biggl(-2 {B}^R_0\left(s_1+s_2,m^2\right)+{B}^R_0\left(s_1,m^2\right)
\nonumber
\eeq
\beq
+{B}^R_0\left(s_2,m^2\right)+\left(4 m^2-s_1-s_2\right) {C}_0\left(s_1,s_2,s_1+s_2,m^2\right)+2\biggr),
\label{qw1}
\eeq
while the rhs of \eqref{eq} is given by integrating \eqref{contà}. \eqref{ag} is the result of this cancelation. The 
proof of this identity can be obtained by using the parametric representation of the integrals 
and the Laplace/Borel method, followed by complete numerical checks. 

\section{Conclusions}
In this work, we have examined both local and nonlocal formulations of axion interactions, highlighting their distinctions and limitations—particularly in relation to anomaly dynamics. The local interaction, derived under specific symmetry assumptions, is well-suited to contexts such as QCD, where an axion mass emerges at the confinement scale. In such scenarios, the conceptual constraints identified in our analysis can be effectively bypassed.\\
A central challenge lies in identifying the conditions under which local and nonlocal formulations become equivalent. From the perspective of the one-particle-irreducible (1PI) effective action, the anomaly form factors associated with nonlocal interactions are constrained by perturbative sum rules. These sum rules govern the dynamics of virtual states and serve as key signatures of anomaly-induced processes. Importantly, we have demonstrated that these rules remain valid even in the presence of gravitational anomalies and for arbitrary fermion masses—despite being violated by the local action.\\
The sum rules differentiate between contributions from particle poles and branch cuts, offering a comprehensive framework for understanding the off-shell behavior of the interaction. While the local formulation captures on-shell physics, it fails to reflect this richer structure. \\
We have illustrated the nontrivial interplay between poles, branch cuts, offering a different perspective on the emergence of such interactions, relevant in the context of QCD factorization and the proton spin, which has been recently debated in several recent works \cite{Bhattacharya:2022xxw,Tarasov:2020cwl,Castelli:2024eza}. 
In the conformal limit, anomaly interactions impose nontrivial constraints on both longitudinal and transverse sectors via conformal Ward identities. Spectral density flows further reveal a redistribution of contributions between the continuum and massless poles.
This study significantly extends previous investigations by providing explicit off-shell spectral density representations of the chiral anomaly form factors for both correlators.  
For massless fermions, the anomaly and particle poles coincide only on-shell; off-shell, anomaly-induced interactions behave differently. Notably, no particle pole persists for massive fermions or vector bosons. Nonetheless, at high energies, anomaly interactions remain dominated by pole structures that can be effectively described by light-cone-localized actions involving mixed pseudoscalar fields.\\
The implications of these results extend beyond conventional axion-like particles (ALPs) in particle physics and cosmology. They also characterize emergent ALP behavior in experimental systems such as artificial gauge fields and topological materials. In these analog systems—driven by thermal gradients and chiral chemical potentials—chiral and gravitational interactions mimic those found in anomaly physics.\\
We emphasize that local ALP interactions, due to their general couplings, can only be consistently justified in the presence of nonperturbative effects. In such cases, an anomaly pole in the ultraviolet transitions into an asymptotic state below a certain scale.
		\chapter{A Dilaton Sum Rule for the Conformal Anomaly Form Factor in QCD at Order $\alpha_s$ }\label{chsr2}
\addcontentsline{toc}{section}{Synopsis}
\section*{Synopsis}

This Chapter extends the dispersive analysis of anomaly form factors, developed in the previous Chapter for chiral and gravitational anomalies, to the case of the conformal (trace) anomaly in non-Abelian gauge theory. The central object of investigation is the $TJJ$ correlator in QCD the three-point function of one stress-energy tensor and two non-Abelian vector currents whose trace sector defines a conformal anomaly form factor. The Chapter establishes that this form factor satisfies a mass-independent sum rule at $\mathcal{O}(\alpha_s)$, whose value is precisely $-\frac{1}{3}\frac{\beta(g)}{g}$, identifies the spectral structure underlying this sum rule, and analyzes the residue of the particle pole in the light-cone limit, revealing structural differences between conformal and chiral anomaly interactions.

The Chapter opens by situating the $TJJ$ correlator within the broader context of the QCD trace anomaly relation
\begin{equation*}
	\eta_{\mu\nu}\langle T^{\mu\nu}\rangle = -\frac{\beta(g)}{2g} F^a_{\mu\nu}F^{a\,\mu\nu},
\end{equation*}
and its perturbative realization through the functional expansion of the quantum average of the stress-energy tensor in an external gauge field background. The trace anomaly defines a hierarchy of correlators $TJ^n$, of which the $TJJ$ is the lowest non-trivial member, sufficient to determine the $\beta$-function coefficient. The interplay between the genuine anomaly contribution and mass-dependent explicit breaking terms is discussed through the decomposition of the trace Ward identity into an anomalous part $\mathcal{A}^{\alpha\beta ab}$ and an additional insertion of $T^\mu_\mu$ into the correlator. The relevance of this analysis for the GFFs of hadrons at the EIC, and its connection to the nucleon mass decomposition and the angular momentum sum rule, are highlighted as primary phenomenological motivations.

The conformal limit of the $TJJ$ correlator is reviewed within the $\mathrm{CFT}_p$ framework, adopting a symmetric momentum labeling with all momenta incoming. The sector decomposition into transverse-traceless, longitudinal, and trace components is performed using the projectors $\Pi$, $\pi$, and $\Sigma$. In the quark sector, standard conformal Ward identities and gauge Ward identities constrain the correlator, yielding a reconstruction formula in which the trace part is determined by the anomaly coefficient alone, projecting onto the tensor structure $u^{\alpha\beta}(p_1,p_2) = (p_1 \cdot p_2)g^{\alpha\beta} - p_2^\alpha p_1^\beta$. In the gluon sector, the gauge-fixing condition breaks conformal symmetry and ordinary Ward identities are replaced by Slavnov-Taylor identities derived from BRST symmetry, the relevant one being the quadratic contraction condition $p_2^\alpha p_3^\beta \langle T_{\mu\nu} J_\alpha^a J_\beta^b \rangle = 0$. This generates additional longitudinal contributions absent in the conformal solution, as well as extra terms $\mathcal{B}_g^{\alpha\beta ab}$ in the trace sector proportional to the gluon equations of motion, which vanish for on-shell gluons. The full renormalized correlator is assembled and the conformal anomaly form factor is identified as
\begin{equation*}
	\Phi_{TJJ} \equiv \frac{1}{3q^2}\mathcal{A}^{\alpha\beta ab}/u^{\alpha\beta}\delta^{ab},
\end{equation*}
whose $1/q^2$ pole, made explicit through the definition $\hat{\pi}^{\mu\nu} \equiv q^2 g^{\mu\nu} - q^\mu q^\nu$, signals dilaton exchange in the perturbative picture.

The analysis is then extended beyond the conformal point by including a fermion mass $m$ in the quark sector. The trace of the massive $TJJ$ correlator in the quark sector splits into two contributions: $\phi_1$, which projects onto the anomaly structure $u^{\alpha\beta}$ and contains the $n_f$ contribution to the anomaly together with a mass-dependent correction $\chi_0$; and $\phi_2$, which projects onto a second tensor structure $v^{\alpha\beta} = (p_1 \cdot p_2)g^{\alpha\beta} + p_2^\alpha p_1^\beta$ and onto terms quadratic in external momenta that vanish on-shell, representing an explicit breaking of conformal symmetry unrelated to the anomaly. The gluon sector is similarly organized, isolating the component $\chi_g$ that projects onto $u^{\alpha\beta}$ from the remaining gauge-dependent contributions. The complete anomaly form factor $\Phi_{TJJ}$ is then defined in the massive case through explicit expressions involving the massive scalar integrals $C_0(p_1^2,p_2^2,q^2,m^2)$ and $\bar{B}_0(q^2,m^2)$, together with coefficient functions $H_1,\ldots,H_6$ depending only on the external virtualities. A key observation is that $\chi_g$ introduces a gauge-dependent contribution to $\Phi_{TJJ}$, but its spectral integral vanishes, so the sum rule is gauge-invariant and equal to $-\frac{1}{3}\frac{\beta(g)}{g}$.

The dispersive analysis of $\Phi_{TJJ}$ is then carried out in full generality. The discontinuity of the form factor is decomposed into pole contributions and a continuum. The continuum arises from the cuts of $B_0(s,m^2)$ and $C_0(s,s_1,s_2,m^2)$ above the physical threshold $4m^2$, with contributions separately traceable to the $n_f$ and $C_A$ color structures. The pole structure consists of two spikes at $s=0$  one of constant strength proportional to the $\beta$-function coefficient, one of variable strength encoding the mass-dependent corrections a pole at $s = s_1 + s_2$, and potential anomalous thresholds at $s_\pm = (\sqrt{s_1} \pm \sqrt{s_2})^2$. A detailed computation shows that all anomalous threshold contributions vanish identically, $\Phi_{TJJ}^{(\lambda)} = 0$, so the only finite-$s$ pole contributing to the sum rule is at $s = s_1 + s_2$, whose residue involves $B_0$ and $C_0$ functions with UV divergences canceling automatically.

The sum rule is established by showing that the contributions of the variable-strength spike at $s=0$, the pole at $s=s_1+s_2$, and the integral of the continuum mutually cancel, leaving only the constant-strength spike at $s=0$:
\begin{equation*}
	\frac{1}{\pi}\int_0^\infty \Delta\Phi_{TJJ}(s_1,s_2,s,m^2)\,ds = -\frac{1}{3}\frac{\beta(g)}{g},
\end{equation*}
independently of the fermion mass and the gluon virtualities $s_1$ and $s_2$. The proof is established through the inverse Laplace (Borel) transform method, using the parametric representation of $\Phi_{TJJ}$ and taking the $t \to 0$ limit to extract the integral of the spectral density from the transform. This method, introduced in the previous Chapter for the $AVV$ and $ATT$ cases, is here extended to the non-Abelian setting with both quark and gluon contributions.

The on-shell behavior of the form factor is then analyzed in detail. In the fermionic sector, explicit computation of the discontinuity using distributional identities for $\mathrm{disc}(1/s)$, $\mathrm{disc}(1/s^2)$, $\mathrm{disc}(C_0/s)$, and $\mathrm{disc}(C_0/s^2)$ shows that all $\delta(s)$ and $\delta'(s)$ contributions cancel exactly, leaving only the branch cut above $4m^2$. In the gluonic sector, the form factor $\sim -11C_A/(144\pi^2 s)$ contributes a pure $\delta(s)$ spike. In the conformal limit $m \to 0$, the fermionic branch cut converges to a $\delta(s)$ distribution the spectral flow pattern familiar from the $AVV$ case and the full anomaly pole is restored with the correct QCD $\beta$-function coefficient. This spectral flow, illustrated as a sequence of spectral densities parameterized by $m^2_n = m^2/n$, is an area-law phenomenon: the integral under each spectral curve is preserved at the anomaly value as the cut migrates toward $s=0$.

The Chapter then examines the residue of the particle pole in the light-cone limit. For generic external kinematics and $m \neq 0$, the residue $\lim_{s \to 0} q^2 \langle TJJ \rangle$ vanishes, as in the chiral anomaly case. However, in the massless on-shell limit $s_1 = s_2 = m = 0$, two tensor structures survive rather than one:
\begin{equation*}
	\lim_{s \to 0} q^2 \langle T^{\mu\nu}(q) J^{\alpha a}(p_1) J^{\beta b}(p_2) \rangle 
	= -\frac{g_s^2 \delta^{ab}}{48\pi^2}\left(\frac{2n_f}{3} - \frac{11C_A}{3}\right)\tilde{\phi}_1^{\mu\nu\alpha\beta}
	- \frac{g_s^2 \delta^{ab}}{288\pi^2}(n_f - C_A)\tilde{\phi}_2^{\mu\nu\alpha\beta}.
\end{equation*}
The first structure $\tilde{\phi}_1$, proportional to $(q^2 g^{\mu\nu} - q^\mu q^\nu)u^{\alpha\beta}$, is linked to the conformal anomaly and generates the known anomaly-induced effective action $S_{\mathrm{anom}} \sim R^{(1)}\Box^{-1}F^2$. The second, traceless structure $\tilde{\phi}_2$ arises from an additional pole unrelated to the anomaly and generates a further nonlocal contribution $S_{\mathrm{extra\,pole}}$ involving $\partial^\mu\partial^\nu F_{\alpha\beta}F^{\alpha\beta}$ and related operators. This second structure vanishes only when $n_f = C_A$. This result contrasts sharply with the chiral anomaly cases ($AVV$ and $ATT$), where only the anomaly-induced structure survives on the light cone, leading to what was termed anomaly dominance. In the conformal case, the particle pole is not solely controlled by the anomaly, and the light-cone effective action requires both contributions. The full light-cone action is $\mathcal{S}_{lc} = \mathcal{S}_{\mathrm{anom}} + \mathcal{S}_{\mathrm{extra\,pole}}$, satisfying the trace anomaly constraint.

The Chapter concludes by discussing the implications for local effective actions describing dilaton interactions. The analysis confirms that, as in the axion case studied in the previous Chapter, a local $\rho FF$ interaction derived from the anomaly Ward identity fails to capture the full spectral structure of the $TJJ$ vertex, in particular the sum rule and the branch-cut continuum generated by massive fermions. A consistent local description is possible only in the presence of nonperturbative confinement effects or in a Higgs phase, where anomaly matching arguments in the UV and IR support such a local formulation. The sum rule established in this Chapter, together with those for the $AVV$ and $ATT$ correlators, forms a coherent framework for characterizing conformal, chiral, and gravitational anomaly interactions through their spectral densities, with direct implications for the study of gravitational form factors at the EIC and for gravitational scattering in the eikonal limit.

The work presented in this Chapter is based primarily on, and adapted from, the publication \cite{Coriano:2025fom}, co-authored with C.~Corianò, S.~Lionetti and L.~Torcellini.
\section{Introduction}
In the context of gauge theory, the conformal anomaly, which characterizes the breaking of scale invariance, manifests through the quantum relation \cite{Collins:1976yq,Adler:1976zt,Chanowitz:1972vd}
\begin{equation}
	\label{oper}
	\eta_{\mu\nu}\langle T^{\mu\nu}\rangle = -\frac{\beta(g)}{2 g} \mathcal A(z),  \qquad \mathcal A(z)=F_{\mu\nu}^a F^{a \,\mu\nu},
\end{equation}  
where \(T^{\mu\nu}\) is the canonical stress-energy tensor, symmetrized via the Belinfante procedure, or alternatively, defined by embedding the action of the gauge theory in a gravitational background,
by differentiation of the corresponding partition function with respect to the metric $g_{\mu\nu}$. In a non-Abelian theory such as QCD, the trace anomaly relation \eqref{oper} can be derived in perturbation theory by an analysis of the diagram in Fig. \ref{expansion} if we consider only massless quarks in the virtual corrections. In the presence of fermion mass corrections, \eqref{oper} is modified, as we are going to describe below. In order to characterize this point, one defines the partition function
\begin{equation}
	Z[g, A] = \int \, \mathcal{D}\psi \, \mathcal{D}\bar{\psi} \, e^{i S_{\text{QCD}}[A, \psi_i, \bar{\psi}, g]},
\end{equation}
with $S_{QCD}$ being the QCD action, with quarks $\psi_i$  $(i=1,\ldots n_f)$ and where $A^{\alpha a}$ is the gauge field of color index $a$. The symmetric stress energy tensor is defined as
\begin{equation}
	T^{\mu\nu}(x) = \frac{2}{\sqrt{g(x)}} \frac{\delta S_{\text{QCD}}}{\delta g_{\mu\nu}(x)},
\end{equation}
and its quantum average in the gravitational background is derived as

\begin{equation}
	\langle T^{\mu\nu}(x) \rangle = \frac{2}{\sqrt{g(x)}} \frac{\delta \ln Z[g]}{\delta g_{\mu\nu}(x)}.
	\label{average}
\end{equation}

In \eqref{oper}, \(\beta(g)\) 
\beq
\beta(g)= {1 \over 3} \, \,  \, \, \frac{g_s^3}{16\pi^2} \, (- 11C_A + 2 n_f)
\label{betagg}
\eeq
denotes the beta function of the theory, with $C_A=N_c$ for a gauge theory with $SU(N_c)$ symmetry, 
and $n_f$ fermions. The functional expansion of the quantum relation \eqref{average} in the background of an external gauge field, defines correlators of the form $TJ^n$, $(n=2,3,4)$, with multiple insertions of gauge currents $(J)$, 
the lowest nonzero contribution being the $TJJ$, with two gluon currents extracted by a functional differentiation of \eqref{average}.\\
In a non-Abelian gauge theory, identifying each individual term in the trace anomaly relation, in principle, requires computing correlators involving three and four external gluon currents. However, in practice, determining the coefficient of the anomaly—namely, the non-Abelian $\beta$-function—only necessitates evaluating the $TJJ$ correlator. Specifically, the diagrams shown in Fig. \ref{expansion} enable a direct computation of Eq. \eqref{oper} by tracing over the $\mu\nu$ indices of the following correlator
\beq
\label{eww}
\Gamma^{\mu\nu\alpha\beta \, a b}(z,x,y) \equiv \frac{\delta^2 Z}{\delta A_\alpha^a(x) \delta A_\beta^b(y)}\langle T^{\mu\nu}\rangle \Big|_{A=0},
\eeq
which lies at the heart of our analysis. As we will show, building upon results obtained in previous work, the evaluation of this correlator allows one to identify—across general non-Abelian theories, and in QCD in particular—a sum rule for a quantity we define as the conformal anomaly form factor. This quantity plays a role analogous to that of the  chiral anomaly form factors appearing in correlators of chiral currents. Much like the form factors arising in the $AVV$ (axial-vector/vector/vector) and $ATT$ (axial-vector/tensor/tensor) correlators analyzed in earlier studies \cite{Coriano:2025ceu}, the conformal anomaly form factor is extracted via a systematic decomposition of the correlator in Eq. \eqref{eww} into its independent tensor structures, extracting the trace and the traceless parts together with a longitudinal/transverse decomposition with respect to the external momenta.

\subsection{Mass corrections} 

A key feature of conformal anomalies is the clear separation between contributions arising from explicit mass terms and those that are genuinely anomalous. These second breaking terms are classified as explicit. 
This separation can be directly confirmed within perturbation theory by analyzing fundamental correlators involving a single insertion of the stress-energy tensor, using dimensional regularization with minimal subtraction. Indeed, in the case of QCD, extra contributions to the trace anomaly are generated 
by the inclusion of fermion mass corrections in the diagrams of Fig. \ref{expansion}. In the Standard Model, for example, in the spontaneously broken phase, such extra contributions are clearly proportional to the vev of the Higgs field. In this case, by taking the trace of correlators involving the stress-energy tensor and vector currents, one derives an anomalous Ward identity of the form
\begin{equation}
	\Gamma^{\alpha\beta a b}(z,x,y) 
	\equiv \eta_{\mu\nu} \left\langle T^{\mu\nu}(z) J^{ a \alpha}(x) J^{ b \beta}(y) \right\rangle 
	= \frac{\delta^2 \mathcal A(z)}{\delta A_{\alpha}^{ a }(x) \delta A_{\beta}^{ b}(y)} + \left\langle {T^\mu}_\mu(z) J^{a \alpha  }(x) J^{b\beta}(y) 
	\right\rangle,
	\label{traceid1}
\end{equation}
where $a,b$ denote color indices. 
For multiple Abelian gauge interactions in the external lines, the anomaly simplifies to
\begin{equation} \label{TraceAnomaly}
	\mathcal A(z) = -\sum_i \frac{\beta_i}{2 g_i} F^{\alpha\beta {(i)}}(z) F^{(i)}_{\alpha\beta}(z),
\end{equation}
where $\beta_i$ are the (mass-independent) beta functions for the gauge couplings $g_i$. In conformal theories where $\beta_i = 0$, the anomaly vanishes. The corresponding $\beta_i$ are just proportional to the number of fermions and scalars propagating in the virtual corrections, and coupled to the different Abelian gauge fields $A_\mu^{(i)}$. We recall that the QCD action, as any non Abelian gauge theory at $d=4$, is conformally invariant 
at the level of its classical action for massless quarks, before any gauge fixing. \\ 
We recall that a more general expression of the conformal anomaly, which involves not only the gauge-invariant background term $F^2$ but also other geometric invariants such as the square of the Weyl tensor and the Gauss--Bonnet term, is encoded in the $TTT$ (three-graviton) vertex, explicitly derived at least in the Abelian case~\cite{Coriano:2017mux,Coriano:2018bsy}. The most general conformal solution of this correlator can be obtained directly from free field theory, provided one allows for arbitrary numbers of massless scalars ($n_s$), massless fermions ($n_f$), and spin-one gauge fields ($n_V$) circulating in the loop corrections. The conformal solutions obtained in~\cite{Bzowski:2018fql,Bzowski:2013sza,Osborn:1993cr} are fully consistent with this perspective. In fact, the general solution to the conformal constraints for this correlator can be parameterized in terms of these independent constants, which correspond to the multiplicities introduced as constants of integration in the conformal Ward identities.\\
The two terms on the right-hand side of Eq.~\eqref{traceid1} can be identified explicitly. The first term comes from functional derivatives of $\mathcal A$, the genuine anomaly contribution, while the second arises from the explicit insertion of $T^\mu_\mu$ into the correlator. Their difference captures the purely anomalous contribution and allows one to isolate $\mathcal A$ through
\beq
\mathcal A^{\alpha\beta a b}(z) = \Gamma^{\alpha\beta a b}(z,x,y) - \left\langle T^\mu_\mu(z) J^{ a \alpha}(x) 
J^{ b \beta}(y) \right\rangle.
\eeq
This distinction is essential for extracting the true quantum anomaly from the stress-energy tensor's trace.\\
Beyond the anomalous term, the full dilaton–gauge–gauge vertex, defined by coupling a scalar (dilaton) field $\rho$ to 
$\Gamma^{\a\b a b}$, includes additional structure, captured by two form factors $\Sigma^{\alpha\beta a b}$ and $\Delta^{\alpha\beta a b}$ in the Standard Model, associated with loop effects from virtual fermions, gauge bosons, and scalars (including Higgs and Goldstone modes). The full momentum-space expression of $\Gamma^{\alpha\beta ab}$ can be derived by embedding the Standard Model Lagrangian in curved spacetime, performing functional variations with respect to the external metric and gauge fields, and subsequently taking the flat-space limit~\cite{2013JHEP...06..077C}. One then obtains

\begin{equation}
	\Gamma^{\alpha \beta a b}(q,p_1,p_2) = (2\pi)^4\, \delta^4(q - p_1 - p_2)\, \frac{i}{\Lambda_{cf}} 
	\left( \mathcal A^{\alpha \beta a b}(p_1,p_2) + \Sigma^{\alpha \beta a b}(p_1,p_2) + \Delta^{\alpha \beta a b}(p_1,p_2)\right),
	\label{sep}
\end{equation}
with
\begin{align}
	\mathcal A^{\alpha \beta a b}(p_1,p_2) &= \int d^4 x\, d^4 y\, e^{i p_1 \cdot x + i p_2 \cdot y}\, 
	\frac{\delta^2 \mathcal A(0)}{\delta A^{\alpha a}(x)\delta A^{\beta b}(y)}, \\
	\Sigma^{\alpha \beta a b}(p_1,p_2) + \Delta^{\alpha \beta a b}(p_1,p_2) &= \int d^4 x\, d^4 y\, e^{i p_1 \cdot x + i p_2 \cdot y}\, 
	\left\langle T^\mu_\mu(0) J^{a\alpha }(x) J^{b\beta}(y) \right\rangle.
\end{align}
$\Lambda_{cf}$ denotes a conformal breaking scale, around which the conformal anomaly interaction can be described by a local effective theory.  This defines the coupling of a possible dilaton field to the conformal anomaly. 
Detailed computations of such "$\rho FF$" vertex, with $F$ denoting the field strength of neutral gauge boson of the electroweak sector $\gamma, Z$, are discussed in \cite{2013JHEP...06..077C}, and serve to disentangle the genuine anomaly from model-dependent loop corrections. In \eqref{sep}, $\Sigma^{\alpha \beta a b}$ denote electroweak corrections which are 1-particle irreducible, while $\Delta^{\alpha \beta a b}$ contain external mixing terms with the Higgs field and can be found in \cite{2013JHEP...06..077C}. \\
Correlators of this type have been studied in QED \cite{Giannotti:2008cv,Armillis:2009pq} and in the broken electroweak sector of the Standard Model \cite{Coriano:2011zk}. From a phenomenological perspective, they are also relevant in condensed-matter contexts (for example, in experiments on topological materials), because Luttinger’s approach \cite{Luttinger:1964zz} relates thermal perturbations to fictitious gravitational fields, following the Tolman--Ehrenfest formulation \cite{Tolman:1930ona}. This correspondence enables analogue-gravity studies in condensed matter \cite{Chernodub:2021nff}.

\begin{figure}
	\centering
	\begin{tikzpicture}
		\begin{feynman}
			\vertex (i1);
			\vertex[right=1cm of i1] (a1);
			\vertex[right=1cm of a1] (b1);
			\vertex[right=3cm of a1] (o1) ;
			\vertex[below=1cm of i1] (i2) ;
			\vertex[right=1cm of i2] (a2);
			\vertex[right=3cm of a2] (o2) ;
			
			%\vertex[below=2cm of i3] (t1);
			\vertex[above=1cm of a1] (t2);
			\vertex[above=2cm of a2] (t3);
			
			%\vertex[above right =1.3cm of t2] (t1);
			\vertex[above=1.5cm of b1] (i3) {$T^{\mu\nu}$};
			
			\diagram* { 
				(i1)  -- [fermion] (a1) -- [fermion] (b1)   --[fermion] (o1),
				
				(i2)  -- [anti fermion] (a2)  --[anti fermion] (o2),

				(i3)  -- [graviton] (b1),
				(a1)-- [gluon] (a2)
			};
			
			\node[] at (-0.35,-0.45) {$\pi\ $};
			\node[] at (4.46,-0.45) {$\pi\ $};

			\filldraw[fill=black, fill opacity=0.1] (-0.42,-0.5) ellipse (.45cm and 1.3cm);
			\filldraw[fill=black, fill opacity=0.1] (4.42 ,-0.5) ellipse (.45cm and 1.3cm);
			
			\draw[double, double distance=1.5,line width=1.2, postaction={decorate, decoration={
					markings,% switch on markings
					mark=at position 0.8 with  {\arrow{ latex} } } } ] (-1.8,-0.5) -- node [above] {} (-0.9,-0.5);
			
			\draw[double, double distance=1.5,line width=1.2, postaction={decorate, decoration={
					markings,% switch on markings
					mark=at position 0.8 with  {\arrow{ latex} } } } ] (4.9,-0.5) -- node [above] {} (5.85,-0.5);
		\end{feynman}
	\end{tikzpicture}
	\hspace{1cm}
	\begin{tikzpicture}
		\begin{feynman}
			\vertex (i1);
			\vertex[right=1cm of i1] (a1);
			\vertex[right=3cm of a1] (o1) ;
			\vertex[below=1cm of i1] (i2) ;
			\vertex[right=3cm of i2] (a2);
			\vertex[right=1cm of a2] (o2) ;
			
			%\vertex[below=2cm of i3] (t1);
			\vertex[above=1cm of a1] (t2);
			\vertex[above=2cm of a2] (t3);
			\vertex[blob] (t1) at (2,1.3) {{$TJJ$}};
			%\vertex[above right =1.3cm of t2] (t1);
			\vertex[above=1.5cm of t1] (i3) {$T^{\mu\nu}$};
			
			\diagram* { 
				(i1)  -- [fermion] (a1)   --[fermion] (o1),
				
				(i2)  -- [anti fermion] (a2)  --[anti fermion] (o2),

				(i3)  -- [graviton] (t1),
				(t1) -- [gluon] (a1) ,
				(t1) -- [gluon] (a2),
			};
			
			\node[] at (-0.35,-0.45) {$\pi\ $};
			\node[] at (4.46,-0.45) {$\pi\ $};

			\filldraw[fill=black, fill opacity=0.1] (-0.42,-0.5) ellipse (.45cm and 1.3cm);
			\filldraw[fill=black, fill opacity=0.1] (4.42 ,-0.5) ellipse (.45cm and 1.3cm);
			
			\draw[double, double distance=1.5,line width=1.2, postaction={decorate, decoration={
					markings,% switch on markings
					mark=at position 0.8 with  {\arrow{ latex} } } } ] (-1.8,-0.5) -- node [above] {} (-0.9,-0.5);
			
			\draw[double, double distance=1.5,line width=1.2, postaction={decorate, decoration={
					markings,% switch on markings
					mark=at position 0.8 with  {\arrow{ latex} } } } ] (4.9,-0.5) -- node [above] {} (5.85,-0.5);
		\end{feynman}
	\end{tikzpicture}
	\caption{Leading (left) and NLO contributions to the GFF of the pion.}
	\label{fig:1}
\end{figure}
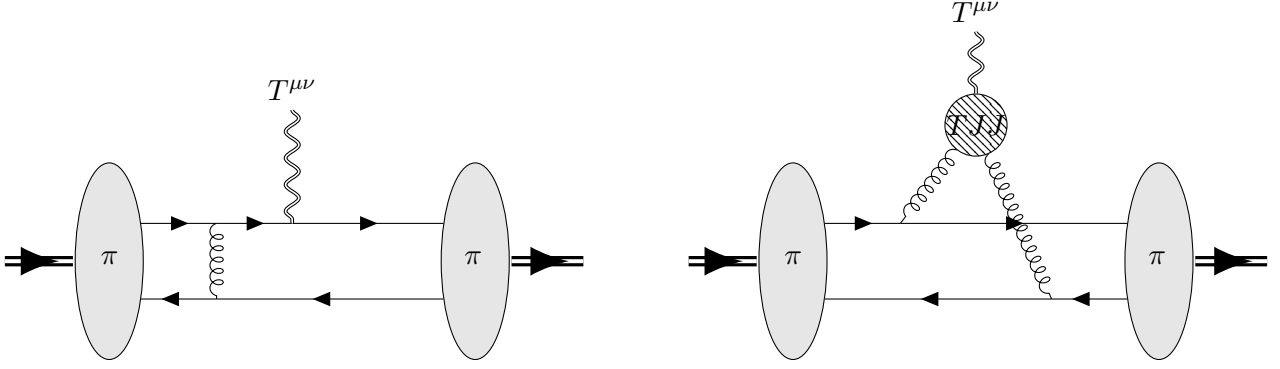

\subsection{The QCD case}
As mentioned above, in the perturbative analysis of a given anomaly vertex, the anomaly sector is characterized by the presence of a specific form factor, the anomaly form factor, which is constrained by a sum rule in its absorptive amplitude. This sum rule is associated with the emergence of an anomaly pole  - an interaction distinct from that of a standard particle pole - acting as an interpolating state in light-cone dominated kinematics.\\
This concept has been recently explored in \cite{Coriano:2025ceu}, particularly in the context of chiral and gravitational anomalies. The central aim of these studies has been to analyze the coupling and decoupling behavior of anomaly poles beyond light-cone kinematics, which defines the kinematical context in which their interaction is identical to that of an ordinary particle pole. There is, for a general kinematical setup, a clear distinction between these virtual states and conventional particle poles. To expose this behavior, explicit computations involving open indices in anomaly interactions are necessary. We will also proceed with a light-cone limit in order to illustrate 
which part of the $TJJ$ vertex dominates the interaction in this limit, as well as at small momentum transfers of the graviton 
line, showing its different behaviour compared to the case of chiral gauge and chiral gravitational interactions. In the present work, we investigate this interaction in detail for a non-Abelian theory. \\
In a related work we will show how this analysis applies to the case of the gravitational scattering of gauge fields.

The identification of the anomaly form factor within a non-Abelian gauge-fixed theory is subtle, as it involves disentangling contributions proportional to the gluon equations of motion from the genuine anomaly part. This requires a careful analysis of the interaction vertex both in the conformal limit and beyond, particularly for the $TJJ$ partonic correlator, 
which is the subject of our study, similarly to previous analysis of the chiral cases \cite{Coriano:2025ceu}. Analysis of these types of interactions in the parity-odd case - with the inclusion of anomalies - in CFT  can be found in 
\cite{Coriano:2023hts,Coriano:2023cvf,Coriano:2024ssu,Coriano:2023gxa,Coriano:2024nhv,Coriano:2024ive}.
The parity-even cases, involving the trace anomaly, had been studied in free field theory in QED in \cite{Giannotti:2008cv,Armillis:2009pq}, in QCD in \cite{Armillis:2010qk} and discussed in the general conformal approach in \cite{Bzowski:2013sza,Bzowski:2018fql}. Parity-odd trace anomaly have been addressed in \cite{Armillis:2010pa,Bonora:2014qla,Bastianelli:2019fot,Abdallah:2021eii,Larue:2023tmu,Larue:2023qxw}.\\
We recall that in Quantum Chromodynamics (QCD), the  perturbative $TJJ$ correlator plays a role in the computation of the gravitational form factors (GFFs) of the proton and the pion, albeit at subleading orders—$O(\alpha_s^2)$ and $O(\alpha_s^3)$ respectively—through standard QCD factorization of the GFFs. Indeed in  QCD, due to confinement, the quantum average in \eqref{oper} involves the computation of hadronic matrix elements, with profound implications for the analysis of the structure of hadrons at experimental level. For example, the  nucleon matrix element at zero momentum transfer $q=P'-P$, $\langle P |T^\alpha_\alpha | P \rangle$, directly relates to the nucleon mass squared. At nonzero momentum transfer, the matrix element of the uncontracted-stress energy tensor 
\beq
\langle P'| T^{\mu\nu} |P \rangle
\eeq
defining the GFF for a hadron, reveals a connection between such matrix element and the conformal anomaly \eqref{oper} for the strong interactions. \\
More generally, the interest in the GFF of the proton \cite{Tong:2022zax,Tanaka:2022wrr,Hatta:2018sqd} is part of  efforts undertaken in the analysis of such hadronic matrix elements \cite{Tong:2022zax}, which are connected with the measurements of the Deeply Virtual Compton Scattering (DVCS) cross section and of generalized parton distributions \cite{Ji:1998pc,Radyushkin:1997ki} at JLab and at the 
Electron-ion collider (EIC) at Brookhaven. The link 
between GFFS and DVCS, in the proton case, is exemplified by the angular momentum sum rule for the analysis of the quark and gluon spin of the proton \cite{Ji:1996ek}. \\
Understanding the origin of the nucleon mass has become a central objective for the future EIC \cite{Abir:2023fpo}, together with a precise determination of the quark and gluon contributions to hadron spin. Once viewed as a largely conceptual issue with limited phenomenological impact, this decomposition has now become experimentally and computationally relevant. In our analysis, we will closely track the interaction terms associated with the quark and gluon contributions, as they exhibit distinct behavior with respect to conformal symmetry and to fermion–mass effects.

\subsection{The partonic $TJJ$ and the mass corrections}
In the case of a gauge theory such as QCD, as discussed in \cite{Coriano:2024qbr}, the operatorial relation \eqref{oper} emerges at leading order in the strong coupling \(g_s\) via a similar partonic \(TJJ\) three-point function, where \(T\) couples to virtual quarks and to the gluon currents $J^{\mu a}$, using ordinary factorization for exclusive processes. Due to the conformal invariance of the classical QCD action for massless quarks, the partonic vertex 
\beq
\Gamma^{\mu\nu\alpha\beta\, a b}\equiv \langle T^{\mu\nu}(q)J^{\alpha a}(p_1) J^{\beta b}(p_2)\rangle 
\eeq
can be investigated using approaches developed from CFT in momentum space, as shown for parity-even \cite{Bzowski:2013sza} and parity-odd \cite{Coriano:2023hts,Coriano:2023cvf,Coriano:2023gxa} tensor correlators. The connection between the partonic and the hadronic result can be clearly identified within the ordinary factorization picture of exclusive processes, in terms of hadronic waves functions and of hard 
scattering contributions which are affected by the partonic $TJJ$ vertex \cite{Coriano:2024qbr}. \\
For example, 
in the analysis of the gravitational form factor of the pion, the partonic $TJJ$ appears at $O(\alpha_s^2)$, 
with the $O(\alpha_s)$ contributions characterising the direct insertion of the QCD stress energy tensor, in its  quarks and gluons contributions, on the hard scattering.  \\
Our analysis, in this work, will be focused on the perturbative $TJJ$ correlator and , in particular, on its conformal anomaly form factor. Its absorptive part is characterized by the presence of a sum rule, as pointed out in previous works \cite{Horejsi:1997yn,Giannotti:2008cv} in the Abelian case, leaving a complete analysis of these features, at hadron level, to a future study. In the conformal limit, the trace anomaly manifests itself as a massless \(1/q^{2}\) pole in such form factor. This structure is isolated by employing a momentum–space, CFT-inspired tensor decomposition of the three-point vertex \cite{Bzowski:2013sza}. In this limit, the dispersive sum rule is trivially saturated by the residue of the pole. To assess the constraint beyond conformality, we move away from the conformal limit and include finite–fermion–mass corrections. In this broader framework, the classical scale invariance of QCD is explicitly broken by terms proportional to the fermion mass, so that \eqref{oper} is modified as

\beq
\eta_{\mu\nu}\langle T^{ \mu\nu}\rangle =m \langle \bar{\psi}\psi \rangle +\frac{ {\beta(g)}}{2 g}F^{\mu\nu a}F_{\mu\nu }^a, \label{qg}
\eeq
which includes contributions from the quark mass, in agreement with \eqref{traceid1}. 
The relation discussed above holds at both the partonic and hadronic levels. At leading order (tree level), the partonic \(TJJ\) correlator is partly fixed by conformal symmetry (in the quark sector) and furnishes a free-field realization of the QCD anomaly equation. Analyses of (chiral) anomalies at the partonic level have been revisited in several recent works \cite{Tarasov:2025mvn,Tarasov:2021yll,Bhattacharya:2023wvy,Bhattacharya:2022xxw,Castelli:2024eza}; the present study adds to this literature.

\subsection{Content of this work}
In this work, we first review the CFT-based approach to determining this correlator. We then systematically characterize the interaction at the conformal point, where QCD allows a direct connection to general CFT results in the conformal limit. Beyond this limit, we extend our analysis by incorporating quark mass corrections, which are essential in orde to proceed with an explicit analysis of the sum rule, once the anomaly form factor is correctly identified. \\
We explore the nature of the particle and anomaly poles that interpolate with this vertex in both cases, showing that the anomaly form factor of this correlator satisfies a sum rule at \(\mathcal{O}(\alpha_s)\). The proof of this sum rule, as well as the characteristic coupling/decoupling behavior of the anomaly pole in the correlator, is established through a direct analysis of the spectral density. This approach follows a strategy similar to that of \cite{Coriano:2025ceu}, which investigated the chiral and gravitational anomaly vertices in QCD at the same perturbative order. We demonstrate that the emergence of a dilaton-like anomaly pole in this interaction is directly linked to the sum rule governing its spectral density. \\
Building on previous analyses of the chiral and gravitational anomaly form factors, we examine the relationship between the conformal anomaly (dilaton) pole and the corresponding particle pole that interpolates in the interaction. 
In both cases, the coupling of these intermediate states is obtained by computing the corresponding residue, as detailed in Section 5. The spectral analysis of the conformal anomaly form factor follows the methodology outlined in \cite{Coriano:2025ceu} for the chiral and gravitational anomaly cases, with necessary modifications to account for the distinct nature of the correlator. Also in this case, as in previous studies, the anomaly pole interaction is defined on the light-cone, where the gluons are on-shell and the quarks are massless, but with some differences that we are going 
to describe in the final section of this work, before our conclusions.

\section{The partonic correlator in the conformal limit }
As previously noted, our entire analysis is conducted at the partonic level. 
QCD is not a conformal theory, but at tree level the Lagrangian is conformally invariant for vanishing quark masses. The free field theory realization of the $TJJ$ in QCD is indeed conformal, except for the corrections induced by renormalization of the corresponding correlator, which at order $g_s^2$  is simply obtained by a simple counterterm action shown in \eqref{cct}. This action is naturally coupled to gravity and the counterterm vertex generated from it, renormalizes the conformal Ward identities satisfied by the correlators. These become anomalous. As mentioned, 
our goal, from the open index structure of the correlator, is to isolate the specific form factor appearing in the trace part of the 
correlator, for which we verify the presence of a sum rule in the non-Abelian case, extending previous analysis. \\ 
The identification of such (conformal anomaly) form factor is achieved by combining CFT techniques on the tensorial decomposition of such correlators \cite{Bzowski:2013sza, Coriano:2013jba} with a perturbative analysis carried out away from the conformal point, by allowing a quark mass 
$m$ in the virtual corrections. The decomposition of the $TJJ$ in a non-Abelian gauge theory is significantly more intricate than in the standard CFT approach in momentum space, owing to the presence of a gauge-fixing sector that explicitly breaks conformal symmetry at one loop  \cite{Coriano:2024qbr}.\\
This complication does not arise in QED, where the conformal solution of the $TJJ$ correlator for Abelian currents precisely matches the general result obtained by solving the conformal Ward identities using momentum-space CFT methods ($\mathrm{CFT}_p$) \cite{Bzowski:2013sza, Coriano:2018bbe}.\\
The strategy adopted in \cite{Coriano:2024qbr} for the analysis of the $TJJ$ in QCD extends the decomposition method introduced in \cite{Bzowski:2013sza} to the non-Abelian case, by isolating the quark and gluon sector contributions within this correlator. This decomposition expresses the correlator in terms of transverse-traceless, longitudinal, and trace form factors. However, in QCD, the resulting expansion deviates from the conformal non-Abelian structure discussed in \cite{Bzowski:2013sza}, due to the appearance of Slavnov–Taylor identities, which supplant the ordinary Ward identities imposed in the Abelian case \cite{Armillis:2010qk} or in the absence of a gauge fixing sector in the action.
The form factors introduced in the tensorial decomposition allow to solve for the conformal constraints in this modified setup, 
in the presence of explicit breakings, just by resorting to the perturbative analysis in free field theory.  
We are going to discuss this decomposition, reviewing the basic results of \cite{Coriano:2025ceu} before moving to the spectral analysis of the conformal anomaly form factor of the $TJJ$. 

\subsection{The tensorial decomposition of the correlator}
In $CFT$ in momentum space, $CFT_p$, correlation functions are subject to the constraints of conformal Ward identities (CWIs), which serve as tensorial conditions in momentum space. In the quark sector of the correlator, the conformal constraints, for massless quarks, are similar to those valid in QED, with the inclusion of $SU(3)$ color indices $a_i$. In the general analysis of the correlator it is convenient, for simplicity, to adopt a symmetric notation and define all the momenta to be incoming, with $p_1+p_2+p_3=0$, where $p_1\equiv q$ is the momentum of the stress energy tensor expernal (graviton) with invariant mass $s\equiv p_1^2$. Similarly we will also using $s_1$ and $s_2$ to denote the virtualities of the two external gluons $(s_1\equiv p_1^2, s_2\equiv p_2^2)$.

The special conformal Ward identities  for a generic $TJ^n$ correlator 
take the form
\begin{equation}
	\begin{aligned}
		0=&\mathcal{K}^k\langle T^{\mu\nu}(q)J^{\mu_1 a_1}(p_1)\dots J^{\mu_{n-1} a_{n-1}}(p_{n-1})\rangle_q \equiv\\&\sum_{j=1}^{n-1}\left(p_j^\kappa \frac{\partial^2}{\partial {p_j}_\alpha \partial p_j^\alpha}+2\left(\Delta_j-d\right) \frac{\partial}{\partial p_j^\kappa}-2 p_j^\alpha \frac{\partial^2}{\partial p_j^\kappa \partial p_j^\alpha}\right)\langle T^{\mu\nu}(q)J^{\mu_1 a_1}(p_1)\dots J^{\mu_{n-1} a_{n-1}}(p_{n-1})\rangle_q\\&
		+2 \sum_{j=1}^{n-1} \left[\delta^{\kappa \mu_j} \frac{\partial}{\partial p_j^{\alpha_{j}}}-\delta_{\alpha_{j}}^\kappa \frac{\partial}{\partial {p_j}_{\mu_j}}\right]\langle T^{\mu\nu}(q)J^{\mu_1 a_1}(p_1)\dots J^{\alpha_j a_j}(p_j)\dots J^{\mu_{n-1} a_{n-1}}(p_{n-1})\rangle_q
	\end{aligned}
\end{equation}
while the dilatation WIs are given by
\begin{equation}
	0=\left[\sum_{j=1}^n \Delta_j-(n-1) d-\sum_{j=1}^{n-1} p_j^\lambda \frac{\partial}{\partial p_j^\lambda}\right]\langle T^{\mu\nu}(q)J^{\mu_1 a_1}(p_1)\dots J^{\mu_{n-1} a_{n-1}}(p_{n-1})\rangle_q,
\end{equation}
where both the stress energy tensor $T^{\mu\nu}$ and the gauge currents $ J^{a_i \, \mu_i}$ are decomposed into longitudinal-trace $(loc)$ and transverse-traceless components
\cite{Bzowski:2013sza} 
\begin{align}
	T^{\mu_i\nu_i}(p_i) &\equiv t^{\mu_i\nu_i}(p_i) + t_{loc}^{\mu_i\nu_i}(p_i),\label{decT} \\
	J^{a_i \, \mu_i}(p_i) &\equiv j^{a_i \, \mu_i}(p_i) + j_{loc}^{a_i \, \mu_i}(p_i),\label{decJ}
\end{align}
where
\begin{align}
	t^{\mu_i\nu_i}(p_i) &= \Pi^{\mu_i\nu_i}_{\alpha_i\beta_i}(p_i) \,T^{\alpha_i \beta_i}(p_i), 
	&& t_{loc}^{\mu_i\nu_i}(p_i) = \Sigma^{\mu_i\nu_i}_{\alpha_i\beta_i}(p) \,T^{\alpha_i \beta_i}(p_i), \label{loct} \\
	j^{a_i \, \mu_i}(p_i) &= \pi^{\mu_i}_{\alpha_i}(p_i) \,J^{a_i \, \alpha_i }(p_i), 
	&& j_{loc}^{a_i \, \mu_i}(p_i) = \frac{p_i^{\mu_i} \, p_{i\,\alpha_i}}{p_i^2} \,J^{a_i \, \alpha_i}(p_i).
\end{align}
The decomposition is performed using the transverse-traceless $(\Pi)$, transverse $(\pi)$, and longitudinal $(\Sigma)$ projectors, defined as
\begin{align}
	\pi^{\mu}_{\alpha} &= \delta^{\mu}_{\alpha} - \frac{p^{\mu} p_{\alpha}}{p^2}, \label{prozero} \\
	\Pi^{\mu \nu}_{\alpha \beta} &= \frac{1}{2} \left( \pi^{\mu}_{\alpha} \pi^{\nu}_{\beta} + \pi^{\mu}_{\beta} \pi^{\nu}_{\alpha} \right) 
	- \frac{1}{d - 1} \pi^{\mu \nu}\pi_{\alpha \beta}, \label{TTproj} \\
	\Sigma^{\mu_i\nu_i}_{\alpha_i\beta_i} &= \frac{p_{i\,\beta_i}}{p_i^2} 
	\Bigg[ 2\delta^{(\nu_i}_{\alpha_i)} p_i^{\mu_i)} - \frac{p_{i\alpha_i}}{(d-1)} 
	\left( \delta^{\mu_i\nu_i} + (d-2) \frac{p_i^{\mu_i} p_i^{\nu_i}}{p_i^2} \right) \Bigg] 
	+ \frac{\pi^{\mu_i\nu_i}(p_i)}{(d-1)} \delta_{\alpha_i\beta_i}. \label{Lproj}
\end{align}
Additionally, we introduce
\begin{equation} \label{a:Tù}
	\mathscr{T}_{\mu\nu \alpha} (\bs{p}) = \frac{1}{p^2} \left[ 2 p_{(\mu} \delta_{\nu)\alpha} - \frac{p_\alpha}{d-1} \left( \delta_{\mu\nu} + (d-2) \frac{p_\mu p_\nu}{p^2} \right) \right].
\end{equation}
and the decomposition identity
\beq
\delta_{\mu(\alpha}\delta_{\beta)\nu} = \Pi_{\mu\nu\alpha\beta}(\bs{p}) + \mathscr{T}_{\mu\nu (\alpha}(\bs{p})\,p_{\beta)} + \frac{1}{d-1}\pi_{\mu\nu}(\bs{p})\delta_{\alpha\beta}.
\eeq
As already pointed out, in the gluon sector the conformal equations are not valid any longer, due to the presence of the gauge-fixing. However, in \cite{Coriano:2024qbr} the approach 
has been to maintain the same tensorial sector decomposition used in $CFT_p$, which is valid to all orders in perturbation theory, and carefully identify all the extra contributions connected with the presence of virtual gluons in the perturbative expansion. The contributions from both sectors are shown in Fig. 1. \\ 
Applying this framework to the $TJJ$ correlator, we distinguish two fundamental components: the \textit{transverse-traceless} part and the \textit{semi-local} part, denoted by the subscript \( loc \). These components are determined by the transverse and trace Ward identities. Using the previously defined projectors, we decompose the full 3-point function as

\begin{align}
	\label{dec1}
	\braket{T^{\m \n}\,J^{ \, \a a}\,J^{\b \, b}} &= 
	\braket{t^{\m \n }\,j^{\a \, a}\,j^{\b \, b}}
	+ \braket{T^{\m \n}\,J^{\a \, a}\,j_{loc}^{\b \, b}}
	+ \braket{T^{\m \n}\,j_{loc}^{\a \, a}\,J^{\b \, b}} 
	+ \braket{t_{loc}^{\m \n}\,J^{\a \, a}\,J^{\b \, b}} \notag \\
	&\quad - \braket{T^{\m \n }\,j_{loc}^{\a \, a}\,j_{loc}^{\b \, b}}
	- \braket{t_{loc}^{\m \n}\,j_{loc}^{\a \, a}\,J^{\b \, b}}
	- \braket{t_{loc}^{\m \n}\,J^{\a \, a}\,j_{loc}^{\b \, b}}
	+ \braket{t_{loc}^{\m \n }\,j_{loc}^{\a \, a}\,j_{loc}^{\b \, b}}.
\end{align}
Using the projectors $\Pi$ and $\pi$ one can write the most general form of the transverse-traceless part as
\begin{equation}
	{\braket{t^{\m \n}(p_1)\,j^{\a \, a}(p_2)\,j^{\b \, b }(p_3)}} =\Pi^{\m \n}_{\m_1 \n_1}(p_1)\pi^{\a}_{\a_1 }(p_2)\pi^{\b}_{\b_1}(p_3)\,\,X^{ \, \m_1 \n_1 \a_1 \b_1 \, a b},
\end{equation}
where $X^{a b \, \m_1 \n_1 \a_1 \b_1}_q$ is a general tensor of rank four built out of the metric and momenta. We can enumerate all possible tensors that can appear in $X^{a b \, \m_1 \n_1 \a_1 \b_1}$ preserving the symmetry of the 
correlator, that in the quark case are represented by the functions  $A_{i}^{(q) a b}$ taking the form
\begin{align}
	\langle t^{\mu \nu }(p_1)j^{a \, \a}(p_2)j^{b \, \b}(p_3)\rangle_q & =
	{\Pi_1}^{\mu \nu}_{\m_1 \n_1}{\pi_2}^{\a}_{\a_1}{\pi_3}^{\b}_{\b_1}
	\left( A_{1}^{(q) a b} \ p_2^{\m_1 }p_2^{\n_1}p_3^{\a_1}p_1^{\b_1} + 
	A_{2}^{(q) a b}\ \delta^{\a_1 \b_1} p_2^{\m_1}p_2^{\n_1} + 
	A_{3}^{(q) a b}\ \delta^{\m_1\a_1}p_2^{\n_1}p_1^{\b_1}\right. \notag\\
	& \left. + 
	A_{3}^{(q) a b}(p_2\leftrightarrow p_3)\delta^{\m_1\b_1}p_2^{\n_1}p_3^{\a_1}
	+ A_{4}^{(q) a b}\  \delta^{\m_1\b_1}\delta^{\a_1\n_1}\right)\label{DecompTJJx}.
\end{align}
A similar expression for the transverse-traceless sector is generated by the virtual gluons 
holds, with form factors  $A_{i}^{(g) a b}$. The expression of the form factors identified by the expansion can be found in \cite{Coriano:2024qbr}.

\subsection {The reconstruction of the quark and gluon sectors in QCD}
The perturbative analysis requires the derivation of the QCD stress-energy tensor. 
It is obtained by adding the components of the field strength $(fs)$, fermionic, gauge fixing 
$(g. f.)$ and ghost $(gh)$ sectors of the Lagrangian  
\bea
T_{\mu\nu} = T^{f.s.}_{\mu\nu} + T^{ferm.}_{\mu\nu} + T^{g.fix.}_{\mu\nu} + T^{ghost}_{\mu\nu},
\eea
with
\beq
T^{f.s.}_{\mu\nu}
= 
-\eta_{\mu\nu}\frac{1}{4}F^a_{\r\s}F^{a\,\r\s}  + F^a_{\mu\r}F^{a\,\r}_\nu  
\eeq

\bea
T^{g.f.}_{\mu\nu} &=& -{1 \over \xi}\left[A_\nu^a \partial_\mu(\partial \cdot A^a) +A_\mu^a \partial_\nu(\partial \cdot A^a)\right] +{1 \over \xi}g_{\mu \nu}
\left[- \frac{1}{2} (\partial \cdot A)^2 + \partial^\rho (A_\rho^a \partial \cdot A^a)\right], \\
T^{gh}_{\mu\nu} &=&- \partial_{\mu}\bar c^a D^{ab}_{\nu} c^b - \partial_{\nu}\bar c^a D^{ab}_{\mu}c^b + g_{\mu\nu} \partial^{\rho}\bar c^a D^{ab}_{\rho}c^b.
\label{gfghost}
\eea
After the inclusion of the fermion sector, its final expression takes the form
\bea
T_{\mu \nu} &=& - g_{\mu \nu} {\mathcal L}_{QCD}
-F_{\mu \rho}^a F_\nu^{a \rho} -{\frac{1} {\xi}}g_{\mu \nu}
\partial^\rho (A_\rho^a \partial^\sigma A_\sigma^a) +{\frac{1}{\xi}}(A_\nu^a \partial_\mu(\partial^\sigma A^a_\sigma)
+A_\mu^a \partial_\nu(\partial^\sigma A_\sigma^a))
\nonumber\\
&+& {\frac{i}{4}} \Big[
\overline \psi \gamma_\mu (\overrightarrow \partial_\nu
-i g T^a A_\nu^a)\psi  -\overline \psi (\overleftarrow \partial_\nu
+i g T^a A_\nu^a)\gamma_\mu\psi
+\overline \psi \gamma_\nu (\overrightarrow \partial_\mu
-i g T^a A_\mu^a)\psi
\nonumber\\
&-& \overline \psi (\overleftarrow \partial_\mu
+i g T^a A_\mu^a)\gamma_\nu\psi \Big] +\partial_\mu \overline
c^a (\partial_\nu c^a -g f^{abc} A_\nu^c  c^b)
+\partial_\nu \overline c^a (\partial_\mu c^a -g f^{abc}
A_\mu^c c^b).
\label{EMT}
\eea
$\xi$ is an arbitrary gauge-fixing parameter and 
($c,\bar{c}$) are the ghost/antighost fields.
Note that the stress energy tensor is normalized in the non-Abelian case in agreement with \cite{Armillis:2010qk}.  
The derivation is performed by embedding the gauge-fixed QCD action into a gravitational background as described in 
\cite{Coriano:2011zk} for the Standard Model with 
\beq
T^{\mu\nu }_{QCD} \equiv \frac{2}{\sqrt{-g}}\frac{\delta\mathcal{S}_{QCD}}{\delta g_{\mu\nu}}.
\eeq

\begin{figure}[t]
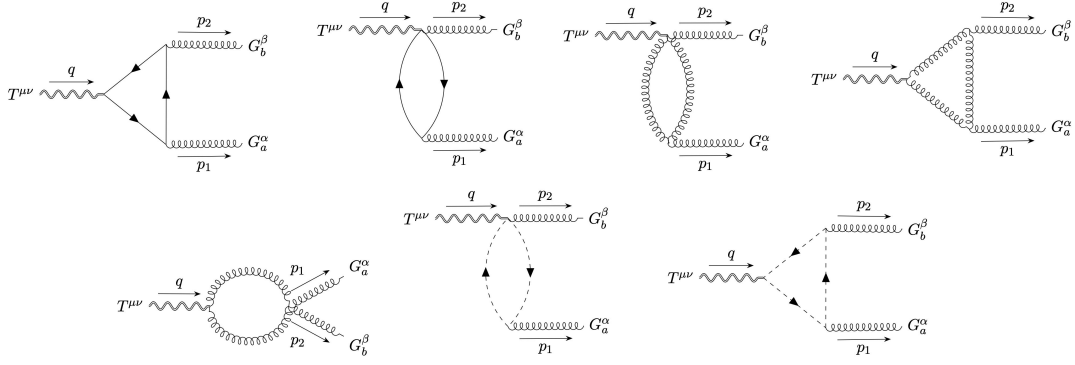

	\begin{center}
		\includegraphics[scale=0.10]{figures/oneone.jpeg}
		\includegraphics[scale=0.10]{figures/onetwo.jpeg}
		\includegraphics[scale=0.10]{figures/onethree.jpeg}
		\includegraphics[scale=0.10]{figures/onefour.jpeg}
		\includegraphics[scale=0.10]{figures/onefive.jpeg}
		\includegraphics[scale=0.10]{figures/onesix.jpeg}
		\includegraphics[scale=0.10]{figures/oneseven.jpeg}
		\caption{List of the perturbative quark, gluon, and ghost (dashed lines) sectors in the non-Abelian $TJJ$. }. 
		\label{expansion}
	\end{center}
\end{figure}

\subsection{Ward identities in the quark sector}
In the context of conformal field theory (CFT), the decomposition of \eqref{dec1} reveals that all terms on the right-hand side - except for the transverse-traceless leading contribution -  arise as specific solutions to the transverse and trace Ward identities. These identities encode the effects of quantum anomalies, which originate from renormalization and are dictated by a single counterterm proportional to the squared field strength \( F^2 \), given by \eqref{cct}. The conservation Ward identities take the form 

\begin{align}
	\label{trù}
	p_{1\n_1}\braket{T^{\mu_1\nu_1}(p_1)\,J^{\m_2 a}(p_2)\,J^{\m_3 b}(p_3)}_q&=4\,\big[\d^{\m_1\m_2}p_{2\l}\braket{J^{\l a}(p_1+p_2)\,J^{\m_3 b}(p_3)}_q-p_2^{\m_1}\braket{J^{\m_2 a}(p_1+p_2)\,J^{\m_3 b}(p_3)}_q\big]\notag\\
	&+4\,\big[\d^{\m_1\m_3}p_{3\l}\braket{J^{\l a}(p_1+p_3)\,J^{\m_2 b}(p_2)}_q-p_3^{\m_1}\braket{J^{\m_3 a }_q(p_1+p_3)\,J^{\m_2 b}(p_2)}_q\big],
\end{align}
while the gauge Ward identities are given by
\begin{align}
	\label{x1}
	p_{2\m_2}\braket{T^{\mu_1\nu_1}(p_1)\,J^{\m_2 a }(p_2)\,J^{\m_3 b}(p_3)}_q&=0\\
	p_{3\m_3}\braket{T^{\mu_1\nu_1}(p_1)\,J^{\m_2 a }(p_2)\,J^{\m_3 b}(p_3)}_q&=0.
\end{align}
These two equations are homogeneous because the $TJ$ two-point function, which is expected on their right-hand sides, is trivially zero. The trace Ward identity, instead, includes the anomaly contribution 
\begin{equation}
	\label{x2}
	g_{\mu\nu} \langle T^{\mu\nu}(q)J^{\alpha a_1}(p_1)J^{\beta a_2}(p_2)\rangle_q=\beta u^{\alpha \beta  \, a_1a_2},
\end{equation}
with 
\bea
&&u^{\alpha\beta a_1 a_2}(p_1,p_2, a_1, a_2) = -\frac{1}{4}\int\,d^4x\,\int\,d^4y\ e^{ip_1\cdot x + i p_2\cdot y}\ 
\frac{\delta^2 \{F^a_{\mu\nu}F^{a\mu\nu}(0)\}} {\delta A_{\alpha}^{a_1}(x) A_{\beta}^{a_2}(y)}\vline_{A^a=0} \,
\label{locvar}\\
\eea
explicitly given by
\beq
u^{\a \b a b} (p_1,p_2)\equiv \delta^{a b}\left((p_1\cdot p_2) g^{\alpha\beta} - p_2^\alpha p_1^\beta\right).
\eeq
This tensor structure summarizes the conformal anomaly contribution, being the Fourier transform of the anomaly term at $O(g^2)$. Additional terms are present at the level of three and four external gluons, which are also part of the hierarchy of these equations, as discussed in \cite{Coriano:2024qbr}, but are not part of this analysis.\\
The 2-point function of two conserved vector currents $J_i$ $(i=2,3)$ shown above in \eqref{trù}, in any CFT in $d$ dimensions is given by 
\beqa
\langle J_2^{\alpha a}(p)J_3^{\beta b}(-p) \rangle =\delta^{a b}\delta_{\Delta_2\, \Delta_3}\left(c_{123} \Gamma_J \right)\pi^{\alpha\beta}(p) (p^2)^{\Delta_2-d/2},
\qquad \Gamma_J=\frac{\pi^{d/2}}{ 4^{\Delta_2 -d/2}}\frac{\Gamma(d/2-\Delta_2)}{\Gamma(\Delta_2)},
\eeqa
with $c_{123}$ an overall constant and $\Delta_2=d-1$ is the scaling dimension of the vector current. 
The divergence of this 2-point function is removed by the counterterm vertex $\Gamma_{\text{count}}$ introduced below in \eqref{cct1}, as we will explain. In our case $\D_2=\D_3=d-1$ and Eq. (\ref{trù}) then takes the form 
\begin{align}
	\label{2point}
	p_{1\mu_1}\braket{T^{\mu_1\nu_1}(p_1)\,J^{\m_2 a }(p_2)\,J^{\m_3 b}(p_3)}_q&=4\delta^{a b} c_{123} \Gamma_J
	\left( \delta^{\nu_1\mu_2}\frac{p_{2\lambda}}{(p_3^2)^{d/2 -\Delta_2}} {\pi^{\lambda \mu_3}}(p_3) -
	\frac{p_2^{\nu_1}}{(p_3^2)^{d/2 -\Delta_2}}\pi^{\mu_2\mu_3}(p_3) \right.\nn\\
	& \left. + \delta^{\nu_1\mu_3}\frac{p_{3\lambda}}{(p_2^2)^{d/2-\Delta_2}}\pi^{\lambda \mu_2}(p_2) -\frac{p_3^{\nu_1}}{(p_2^2)^{d/2-\Delta_2}}\pi^{\mu_3\mu_2}(p_2)\right).
\end{align}\\

\subsection{The tensorial sectors decomposition for quarks }
\label{eight}

Using the projectors $\Pi$ and $\pi$ one can write the most general form of the transverse-traceless part as
\begin{equation}
	{\braket{t^{\m \n}(p_1)\,j^{ \a \, a }(p_2)\,j^{\b \, b }(p_3)}}_q =\Pi^{\m \n}_{\m_1 \n_1}(p_1)\pi^{\a}_{\a_1 }(p_2)\pi^{\b}_{\b_1}(p_3)\,\,X^{ \, \m_1 \n_1 \a_1 \b_1 a b}_q,
\end{equation}
where $X^{a b \, \m_1 \n_1 \a_1 \b_1}_q$ is a general tensor of rank four built out of the metric and momenta. We can enumerate all possible tensor that can appear in $X^{a b \, \m_1 \n_1 \a_1 \b_1}$ preserving the symmetry of the 
correlator 
\begin{align}
	\langle t^{\mu \nu }(p_1)j^{a \, \a}(p_2)j^{b \, \b}(p_3)\rangle_q & =
	{\Pi_1}^{\mu \nu}_{\m_1 \n_1}{\pi_2}^{\a}_{\a_1}{\pi_3}^{\b}_{\b_1}
	\left( A_{1}^{(q) a b} \ p_2^{\m_1 }p_2^{\n_1}p_3^{\a_1}p_1^{\b_1} + 
	A_{2}^{(q) a b}\ \delta^{\a_1 \b_1} p_2^{\m_1}p_2^{\n_1} + 
	A_{3}^{(q) a b}\ \delta^{\m_1\a_1}p_2^{\n_1}p_1^{\b_1}\right. \notag\\
	& \left. + 
	A_{3}^{(q) a b}(p_2\leftrightarrow p_3)\delta^{\m_1\b_1}p_2^{\n_1}p_3^{\a_1}
	+ A_{4}^{(q) a b}\  \delta^{\m_1\b_1}\delta^{\a_1\n_1}\right)\label{DecompTJJ}
\end{align}
with the reconstruction taking the form 
\begin{align}
	& \la T_{\mu_1 \nu_1}(\bs{p}_1) J^{\mu_2 a_2}(\bs{p}_2) J^{\mu_3 a_3}(\bs{p}_3) \ra_q = \la t_{\mu_1 \nu_1}(\bs{p}_1) j^{\mu_2 a_2}(\bs{p}_2) j^{\mu_3 a_3}(\bs{p}_3) \ra_q \nn\\[0.5ex]
	& \qquad +  2 \mathscr{T}_{\mu_1 \nu_1}^{\quad\,\,\alpha}(\bs{p}_1)\Big[\delta_{[\alpha}^{\mu_3}p_{3\beta]}  \la J^{\mu_2 a_2}(\bs{p}_2) J^{\beta a_3}(-\bs{p}_2) \ra_q
	+\delta_{[\alpha}^{\mu_2}p_{2\beta]}  \la J^{\mu_3 a_3}(\bs{p}_3) J^{\beta a_2}(-\bs{p}_3) \ra_q\Big]
	\nn\\
	& \qquad + \frac{1}{d - 1}\,\pi_{\mu_1 \nu_1}(\bs{p}_1) \mathcal{A}^{\mu_2 \mu_3 a_2 a_3}_q, \label{tjjdec}
\end{align}
where $\mathscr{T}_{\mu_1 \nu_1\alpha}$ is defined in \eqref{a:Tù} and $\langle JJ\rangle_q$ is the 2-point function of the gluon with a virtual quark
and
\beq
\mathcal{A}_q^{\alpha\beta ab} = -{1 \over 3} \, \,  \, \, \frac{g_s^2}{16\pi^2} \,  2n_f \delta^{ab}u^{\alpha \beta}(p_1,p_2),
\eeq
is the contribution of the anomaly to the expansion of the correlator. 

\subsection{Renormalization of the Ward Identities and the gluon sector}

The decomposition at one-loop of the gluon sector follows \eqref{dec1}, 
but its final expression is modified compared to 
\eqref{tjjdec}, which is affected by new trace contributions not present in the quark sector. The main difference come from the external identities constraining the correlator in the presence of a gauge-fixing sector in the QCD action. The Ward identities do not hold any longer and re replaced by Slavnov-Taylor identities, derived from the BRST symmetry. As shown in \cite{Coriano:2024qbr}, the relevant 
WIs and STIs that can be used in order to fix the expression 
The first Ward identity is similar to \eqref{trù}.
A second identity can be derived using the BRST symmetry of the gauge fixed action. For this purpose, it is sufficient to choose an appropriate Green's function with one stress energy tensor, a derivative of the gluon field and a ghost field,  $\langle T_{\mu\nu} \partial^{\alpha} A_{\alpha}^a \bar c^b \rangle$, and then use its BRST invariance to obtain
\cite{Coriano:2024qbr,Armillis:2010qk}
the identity
\beq
\partial_{x_1}^{\alpha}\partial_{x_2}^{\beta}\langle T_{\mu\nu}(x) J_{\alpha}^a(x_1)J_{\beta}^b(x_2)\rangle = 0,
\eeq
which in momentum space becomes
\begin{equation}
	p_2^{\alpha} p_3^{\beta} \langle T_{\mu\nu}(p_1) J_{\alpha}^a(p_2)J_{\beta}^b(p_3)\rangle = 0. \label{brstwi}
\end{equation}
Notice that this constraint is weaker compared to an ordinary Ward identity, since we require a vanishing constraint after contractions with both vector momenta. This allows additional tensor structures to be nonzero in the sector decomposition of the correlator, compared to the quark sector. This sets a distinction between the ordinary CFT analysis of this correlator, which imposes just ordinary non-Abelian Ward identities on the vector lines, compared to the free-field theory realization that we discuss in our work.  
The renormalization of the Ward/Slavnov-Taylor identities for the entire vertex is obtained by the counterterm  action
\beq
\label{cct}
\mathcal{S}_{\textrm{count}}=\frac{1}{(d-4)} \frac{g_s^2}{16 \pi^2}(\frac{5}{3}C_A-\frac{2}{3} n_f)
\int d^4 x g \sqrt{g} F^{a\, \mu\nu}F^{a}_{\mu\nu}.
\eeq
From this expression we extract the associated counterterm vertex $\Gamma_{\textrm{count}} (q,p_1,p_2)$ by functional differentiation with respect to the external fields and then Fourier transforming to momentum space
\beq
\label{cct1}
\Gamma^{\mu\nu\alpha\beta\, ab}_{\text{count}}(z,x,y) =  \frac{2}{\sqrt{-g}}\frac{\delta^3 S_{\textrm{count}}}{\delta{g_{\mu\nu}}\delta A^a_\alpha(x)\, \delta A^b_\beta(y)}\vline_{A = 0}.
\eeq
Using this counterterm all the hierarchical WIs (conservation, trace, etc. ) of $\Gamma +\Gamma_{\text{count}}$.\\
As shown in \cite{Coriano:2024qbr}, in free field theory the tensorial sector expansion is given by
\begin{equation}
	\begin{aligned}
		& \langle T^{\mu \nu}(q)  J^{ a\alpha}(p_1) J^{ b\beta}(p_2)\rangle_g=\langle t^{\mu \nu}(q)  j^{ a\alpha}(p_1) j^{ b\beta}(p_2)\rangle_g +\langle t^{\mu \nu }(q)j_{loc}^{a  \a}(p_1)j^{b  \b}(p_2)\rangle_g +\langle t^{\mu \nu }(q)j^{a  \a}(p_1)j_{loc}^{b  \b}(p_2)\rangle_g\\
		& \qquad+2 \mathcal{I}^{\mu \nu \rho }(q)\left[\delta_{[\rho}^{\beta} p_{2 \sigma ]}\langle J^{ a\alpha}({p}_1) J^{ b\sigma}(-{p}_1)\rangle_g +\delta_{[\rho}^{\alpha} p_{1 \sigma]}\langle J^{ b\beta}({p}_2) J^{ a\sigma}(-p_2)\rangle\right]_g  
		+\frac{1}{d-1} \pi^{\mu \nu}(q) \left[\mathcal{A}^{\alpha \beta a b}_g+\mathcal{B}^{\alpha \beta a b}_g\right]
	\end{aligned}
	\label{res}
\end{equation}
with additional local terms of the form 
\beq
\langle t^{\mu \nu }(q)j^{\a \, a}(p_1)j_{loc}^{\b \, b}(p_2)\rangle_g. 
\label{adds}
\eeq
The contribution from the gluons to the anomaly at $O(\alpha_s)$ is given by
\begin{eqnarray}
	\mathcal{A}_g^{\alpha\beta ab} &=& {11 \over 3} \,  \frac{g_s^2}{16\pi^2} \, C_A  \delta^{ab}u^{\alpha\beta}(p_1,p_2).
\end{eqnarray}
As mentioned above, unlike the transverse traceless sector, the longitudinal sector of the decomposition retains additional contributions that do not vanish under Slavnov-Taylor identities, as they do in the quark sector or the general conformal solution discussed in 
\cite{Bzowski:2013sza}. These conditions, being homogeneous second-order differential constraints in coordinate space, are less restrictive than conventional Ward identities. Notice, however, that such terms are not part of the trace anomaly, since they are traceless by construction. Simultaneously, the trace sector undergoes further modifications due to additional terms linked to the gluon equations of motion, denoted here as \( \mathcal{B}^{\alpha \beta a b}_g \). These terms, absent in the on-shell decomposition, introduce corrections that alter the structure of the trace sector that define additional explicit breakings of the conformal symmetry, even for a vanishing quark mass. These are expected since, as already mentioned, the conformal symmetry of the one-loop expansion of the correlator is also violated by the gauge fixing condition.    
The new longitudinal terms \eqref{adds} take the form 

\begin{align}
	\langle t^{\mu \nu }(q)j^{\a \, a}(p_1)j_{loc}^{\b \, b}(p_2)\rangle_g & =
	{\Pi}^{\mu \nu}_{\m_1 \n_1}(q) \, \pi^\a _{\a_1} (p_1)\, {{p_2}_{\beta_1} p_2^\beta \over p_2^2}
	\left( 	B^{ab}_1 \, p_1^{\m_1} \, p_1^{\nu_1} \, p_2^{\a_1} \, p_2^{\b_1} + B_2^{ab} \, p_1^{\m_1} \, p_2^{\b_1} \, \delta^{\a_1 \n_1}  \right), 
\end{align}

which is orthogonal to the trace sector. Notice that these local (longitudinal) contributions vanish when the gluons are on-shell. 
\subsection{The complete correlator}
The complete expansion of the correlator is obtained by combining the two sectors. 
In the massless fermion limit, the perturbative $TJJ$ is given by the expression

\begin{equation}
	\begin{aligned}
		\Gamma^{\mu\nu\alpha\beta a b}(q,p_1,p_2) & =\langle t^{\mu \nu}(q)  j^{ \a a}(p_1) j^{ \b b}(p_2)\rangle+\langle t^{\mu \nu }(q)j_{loc}^{\a  a}(p_1)j^{\b  b}(p_2)\rangle_g +\langle t^{\mu \nu }(q)j^{\a  a}(p_1)j_{loc}^{\b  b}(p_2)\rangle_g\\
		& \qquad+2 \mathcal{I}^{\mu \nu \rho }(q)\left[\delta_{[\rho}^{\beta} p_{2 \sigma ]}\langle J^{ a\alpha}({p}_1) J^{ b\sigma}(-{p}_1)\rangle+\delta_{[\rho}^{\alpha} p_{1 \sigma]}\langle J^{ b\beta}({p}_2) J^{ a\sigma}(-p_2)\rangle\right] \\
		&	+\frac{1}{ 3 \, q^2} \hat\pi^{\mu \nu}(q) \left[\mathcal{A}^{\alpha \beta a b}+\mathcal{B}^{\alpha \beta a b}_g\right]
	\end{aligned}
	\label{resù}
\end{equation}
with 
\beq
\mathcal{A}^{\alpha \beta a b}=\mathcal{A}^{\alpha \beta a b}_q + \mathcal{A}^{\alpha \beta a b}_g,
\eeq
and the trace anomaly given by the expression
\beq
\mathcal{A}^{\alpha\beta ab} = {1 \over 3} \, \,  \, \, \frac{g_s^2}{16\pi^2} \, (11C_A - 2 n_f) \delta^{ab}u^{\alpha \beta}(p_1,p_2).
\eeq
Eq. \eqref{resù} shows that the structure of the effective vertex corresponding to the $TJJ$ 
correlator is modified compared to the ordinary CFT case, encountered in \eqref{tjjdec},
with modifications affecting both the longitudinal and the trace sectors. 
The traceless sector in $\Gamma$ is given by
\begin{equation}
	\begin{aligned}
		\langle T^{\mu \nu}(q)  J^{ a\alpha}(p_1) J^{ b\beta}(p_2)\rangle_{tls}=&
		\langle t^{\mu \nu}(q)  j^{ a\alpha}(p_1) j^{ b\beta}(p_2)\rangle+\langle t^{\mu \nu }(q)j_{loc}^{a  \a}(p_1)j^{b  \b}(p_2)\rangle_g +\langle t^{\mu \nu }(q)j^{a  \a}(p_1)j_{loc}^{b  \b}(p_2)\rangle_g\\
		& \qquad+2 \mathcal{I}^{\mu \nu \rho }(q)\left[\delta_{[\rho}^{\beta} p_{2 \sigma ]}\langle J^{ a\alpha}({p}_1) J^{ b\sigma}(-{p}_1)\rangle+\delta_{[\rho}^{\alpha} p_{1 \sigma]}\langle J^{ b\beta}({p}_2) J^{ a\sigma}(-p_2)\rangle\right], 
	\end{aligned}
	\label{ali}
\end{equation}

and the trace part by

\beq
\label{anommX}
\langle T^{\mu \nu}(q)  J^{ a\alpha}(p_1) J^{ b\beta}(p_2)\rangle_{tr}=\frac{1}{ 3 \, q^2} \hat\pi^{\mu \nu}(q) \left[\mathcal{A}^{\alpha \beta a b}+\mathcal{B}^{\alpha \beta a b}_g\right],
\eeq
with
\beq
\mathcal{B}_g^{\alpha \beta\, a b }\equiv \delta^{a b} \mathcal{B}_g^{\alpha \beta }. 
\eeq
Therefore, in the conformal limit, from \eqref{anommX} we derive the trace anomaly condition

\beq
\label{vg}
g_{\mu\nu}\langle T^{\mu \nu}(q)  J^{ \a a}(p_1) J^{ \b b}(p_2)\rangle_{tr}= \left[\mathcal{A}^{\alpha \beta a b}+\mathcal{B}^{\alpha \beta a b}_g\right].
\eeq
The contraction with $g_{\mu\nu}$ has removed the anomaly poles from the trace sector, while the $\mathcal{B}_g$ in \eqref{vg} is uniquely introduced by the gauge-dependent contributions introduced by the gluon sector in the vertex. \\
These terms vanish for 
on-shell gluons and take the form
\beq
\mathcal{B}_g^{\alpha \beta}(q^2, p_1^2,p_2^2)=\left[ C_1 \, p_1^\alpha \, p_1 ^\beta + C_2 \, p_1^\alpha \, p_2^\beta + C_3  \, p_1^\beta \, p_2^\alpha + C_1(p_1\leftrightarrow p_2) \, p_2^\alpha \, p_2^\beta + C_4\, g^{\alpha \beta} \right].  
\label{bb}
\eeq
They are generated by the explicit breaking of the classical conformal symmetry of the QCD action induced by the gauge fixing procedure.\\
The coefficient functions in \eqref{bb} $C_1,\ldots C_4$ are explicitly given in Appendix \ref{gg} and depend only on 
the virtualities of the external momenta. Notice that in \eqref{resù} and \eqref{anommX}, the \({1}/{q^2} \) anomaly pole has been explicitly extracted from the longitudinal projector \( \pi^{\mu\nu} \) - the last term on the right-hand side of \eqref{resù} - by defining
\beq
\hat\pi^{\mu\nu} \equiv q^2 g^{\mu\nu} - q^\mu q^\nu, \qquad  \pi^{\mu\nu}=\frac{1}{q^2}\hat\pi^{\mu\nu}. 
\eeq
In the tensorial decomposition of \eqref{resù}, the pole emerges directly as a result of algebraic manipulations, at this stage without any particular physical interpretation. Its role becomes evident through an analysis of the sum rule associated with a specific tensor component of the trace sector and signals a dilaton exchange in the perturbative picture. \\ 
In the presence of fermion masses, the pole is smoothed into a momentum–dependent form factor, ${\Phi}_{\text{TJJ}}(q^{2},p_1^2,p_2^2,m^{2})$, which we analyze dispersively and which reproduces the \(1/q^{2}\) anomaly pole as \(m\to0\) and for 
on-shell gluons.
This form factor, here shown in the conformal limit, can then be defined in such a limit as 
\beq
\label{holds}
\Phi_{TJJ}\equiv \frac{\mathcal{A}^{\alpha\beta ab}}{q^2}. 
\eeq
The inclusion of the $1/q^2$ factor is essential for the saturation of the sum rule in the conformal limit, a point that we will address in the next sections.
In the conformal limit in which \eqref{holds} holds, this and other types of anomaly interactions manifest identical features. This occurs for chiral, conformal, and gravitational anomalies \cite{Giannotti:2008cv, Armillis:2010qk, Armillis:2009pq}, including the study of parity-violating trace anomalies (CFTs)~\cite{Coriano:2023hts}.\\
The term { anomaly pole}, clearly identified in the conformal limit, has been introduced to emphasize the subordinate nature of this interaction compared to that of a conventional particle pole, which corresponds to an asymptotic state. Its physical relevance becomes tied to the saturation of a sum rule that governs the absorptive part of the vertex. \\
This sum rule may be saturated by a pole, a branch cut, or both, depending on the external kinematical conditions and the presence or absence of mass corrections.
As we shall see, the identification of which terms contribute to the anomaly form factor in the general off-shell and massive case, as opposed to those that don't, is a subtle task that we are going to address. In fact, we will retain, in the definition of the anomaly form factor, a specific component of $\mathcal{B}_g$ which is proportional to $u^{\alpha\beta}$, the image of the anomaly in momentum space. This clearly implies that the anomaly form factor is gauge-dependent, but the contributions of these extra terms to the sum rule, as we are going to show, vanish. The sum rule is gauge-invariant and numerically related to $\beta(g)$.
As we move to the massive fermion case, the corresponding expression of \eqref{anommX} includes extra mass-dependent terms. These characterize the explicit breaking of conformal symmetry due to mass corrections coming from the quark sector, while the contribution of the gluon sector is already contained in \eqref{anommX} and remains unmodified.  

\section{Identifying the anomaly form factor for the massive correlator}
In order to move away from the conformal point and separate and identify the conformal anomaly form factor, we re-investigate the correlator 
by including a fermion mass $m$ in the perturbative expansion. \\
We proceed from the quark sector. The decomposition given in \eqref{resù} remains valid; however, the trace sector takes a modified form, naturally splitting into two distinct contributions. We obtain the trace contribution
\begin{equation}
	\label{sept}
	g_{\mu\nu} \braket{T^{\mu\nu}J^{\alpha a}J^{\beta b}}_q = \left(\phi_1^{\alpha\beta}(p_1,p_2,q,m) + \phi_2^{\alpha\beta}(p_1,p_2,q,m)\right) \delta^{ab},
\end{equation}  
separated into two terms.
The first, denoted as \(\phi_1\), is directly projected onto the tensor structure \(u^{\alpha\beta}\), while the second one \(\phi_2\), as we are going to show, doesn't.  In $\phi_1$ we identify two contributions
\begin{equation}
	\phi_1^{\alpha\beta a b} = \left(-\frac{2}{3} n_f \frac{g_s^2}{16\pi^2} + \chi_0(p_1,p_2,q,m)\right) \delta^{ab} u^{\alpha\beta}(p_1,p_2),
\end{equation}  
containing the quark flavour number \(n_f\) contribution to the anomaly pole in the conformal limit, now supplemented by an additional mass-dependent term, $\chi_0$. The function \(\chi_0\) is given by
\begin{equation}
	\chi_0(p_1,p_2,q,m) \equiv \frac{B_1 - B_4 }{2\, p_1\cdot p_2}, 
\end{equation}
where the scalar functions $B_1,\ldots B_4$ are given in the appendix. Explicitly

\begin{align}
	\label{parm}
	\chi_0=&g_s^2m^4  H_1\, C_0 \left(p_1^2,p_2^2,q^2,m^2\right) +g_s^2m^2 \Bigl(H_2  \,C_0 \left(p_1^2,p_2^2,q^2,m^2\right)  \nn\\&+ H_3 \,\bar{B}_0(q^2,m^2)+ H_4 \,\bar{B}_0(p_1^2,m^2)+ H_5 \,\bar{B}_0(p_2^2,m^2) +H_6\,\Bigl).
\end{align}
The coefficients functions $H_1\ldots H_6$ ($H_i\equiv H_i(p_1^2,p_2^2,q^2))$
depend only on the virtualities of the external momenta and not on $m$.  $C_0 \left(p_1^2,p_2^2,q^2,m^2\right)$ denotes the massive scalar 3-point function and $\bar{B}_0(q^2,m^2)$ is the renormalized scalar massive 2-point function. Their expressions are given in Appendix \ref{Co}.\\
Concerning the tensor structure $\phi_2$ in \eqref{sept}, as just mentioned, it appears in the trace but does not generate terms proportional to the tensor structure $u^{\alpha\beta}$ and as such it is not included in the anomaly form factor. Indeed we obtain
\begin{equation}
	\label{p2}
	\phi_2^{\alpha\beta}(p_1,p_2,q,m) = \chi_1(p_1,p_2,q,m) v^{\alpha\beta} + \left( B_2 p_1^{\alpha}p_1^{\beta} + B_2(p_1 \leftrightarrow p_2) p_2^{\alpha}p_2^{\beta} + B_3 p_1^{\alpha}p_2^{\beta} \right), 
\end{equation}  
with
\begin{equation}
	v^{\alpha\beta}(p_1,p_2) = (p_1\cdot p_2) g^{\alpha\beta} + p_2^\alpha p_1^\beta
\end{equation} 
and
\begin{equation}
	\chi_1(p_1,p_2,q,m) \equiv \frac{B_1 + B_4 }{2\, p_1\cdot p_2}.
\end{equation} 
$B_1\ldots B_4$ are given in Appendix \ref{appfunc}.
Explicitly 
\allowdisplaybreaks{
	\begin{align}
		\chi_1=s_1\, s_2\,m^2\,g_s^2 n_f\,\delta ^{a b} \Biggl(&\frac{ \left(-s_2 (s+s_1)-(s-s_1)^2+2 s_2^2\right) {B}_0\left(s_2,m^2\right)}{6 \pi ^2 s (s-s_1-s_2) \left(s^2-2 s (s_1+s_2)+(s_1-s_2)^2\right)^2}
		\nn\\&+\frac{  \left(2 s^2-s (s_1+s_2)-(s_1-s_2)^2\right) {B}_0\left(s,m^2\right)}{6 \pi ^2 s (s-s_1-s_2) \left(s^2-2 s (s_1+s_2)+(s_1-s_2)^2\right)^2}
		\nn\\&+\frac{ \left(-s^2-s (s_1-2 s_2)+(s_1-s_2) (2 s_1+s_2)\right) {B}_0\left(s_1,m^2\right)}{6 \pi ^2 s (s-s_1-s_2) \left(s^2-2 s (s_1+s_2)+(s_1-s_2)^2\right)^2}
		\nn\\&+\frac{ \left(s^3-s^2 (s_1+s_2)-s \left(s_1^2-6 s_1 s_2+s_2^2\right)+(s_1-s_2)^2 (s_1+s_2)\right) {C}_0\left(s_1,s_2,s,m^2\right)}{12 \pi ^2 s (s-s_1-s_2) \left(s^2-2 s (s_1+s_2)+(s_1-s_2)^2\right)^2}
		\nn\\&+\frac{1}{6 \pi ^2 s (s-s_1-s_2) \left(s^2-2 s (s_1+s_2)+(s_1-s_2)^2\right)}	\nn\\&+\frac{m^2  {C}_0\left(s_1,s_2,s,m^2\right)}{3 \pi ^2 s (s-s_1-s_2) \left(s^2-2 s (s_1+s_2)+(s_1-s_2)^2\right)}\Biggl).
	\end{align}
}
In summary, the decomposition in \eqref{sept} is globally organized in terms of a projector onto the tensor structure $u^{\alpha\beta}$—which captures the genuine anomaly contribution—and a second structure, $v^{\alpha\beta}$, which is unrelated to the anomaly, identifying an explicit breaking. The scalar function $\chi_0$ accounts for the mass corrections to the anomaly in this sector.
The remaining terms in brackets in \eqref{p2} vanish if multiplied by transverse polarization vectors of the two gluons.
This separation is conceptually natural: in the quark sector, the anomaly condition admits a decomposition into the pure anomaly term and a mass-dependent contribution, the latter arising from the insertion of the scalar operator $\sim m \bar\psi \psi$ into the $JJ$ two-point function. \\
\subsection{Extension to the gluon sector}
This approach is naturally extended to the gluon sector. This sector is unaffected by the mass corrections. For this reason,  we organize the gluonic part of the trace $\mathcal{B}^{\alpha\beta}_g$ in a similar way, following the same template. Obviously, this second sector does not carry any mass term, but we require that the tensor decomposition in terms of $v^{\alpha\beta}$ and the remaining tensor structures quadratic in $p_i^{\alpha} p_j^{\beta}\, i,j=1,2$ be identical to that of the quark sector.  
Therefore the contribution to the trace of the correlator from the virtual gluon sector is organised in the form
\begin{equation}
	g_{\m\n}\langle {T^{\m\n}J^{\a a}J^{\b b}}\rangle_g=\left( \frac{g_s^2}{16\pi^2}\, \frac{11}{3}C_A  \right) u^{\a\b}\delta^{ab} +\mathcal{B}^{\alpha \beta}_g \delta^{ab},
\end{equation}
as extracted from \eqref{vg}. Also in this case, the contribution proportional to $u^{\alpha \beta}$ in $\mathcal{B}^{\alpha \beta}_g$ is identified by the decomposition
\begin{equation}
	\mathcal{B}^{\alpha \beta a b}_g=\chi_g(p_1,p_2,q)u^{\a\b}\d^{ab}+\phi_g^{\alpha \beta a b}
\end{equation}
where only
\beq
\chi_g(p_1,p_2,q)\equiv \frac{C_4- (p_1\cdot p_2)C_3 }{2\, p_1\cdot p_2}
\label{chig}
\eeq
projects into the anomaly structure $u^{\a\b}$.
The remaining tensor component 
\begin{equation}
	\phi_g^{\alpha \beta a b}=\chi_g'(p_1,p_2,q)v^{\a\b}\d^{ab}+C_1 \, p_1^\alpha \, p_1 ^\beta \d^{ab}+ C_2 \, p_1^\alpha \, p_2^\beta  \d^{ab}+ C_1(p_1\leftrightarrow p_2) \, p_2^\alpha \, p_2^\beta\d^{ab}
\end{equation}
weither contains terms that do not project on $u^{\a\b}$  
\begin{equation}
	\chi_g'(p_1,p_2,q)\equiv \frac{C_4+ (p_1\cdot p_2)C_3 }{2\, p_1\cdot p_2}.
\end{equation}
or vanish by contracting with the polarization vectors of the gluons. The functions $C_i$, are given in Appendix \ref{gg}. \\
The final decomposition of the trace of the $TJJ$ correlator then can be written as
{\small
\begin{align}
	\langle T^{\mu \nu}(q)  J^{ a\alpha}(p_1) J^{ b\beta}(p_2)\rangle_{tr}=\frac{1}{ 3 \, q^2}\hat\pi^{\mu \nu}(q) \biggl[& \, \,  \, \,  \Biggl(\frac{g_s^2}{16\pi^2} \, \bigg(\frac{11}{3}C_A - \frac{2}{3} n_f \bigg)+ \chi_0(p_1,p_2,q,m)+\chi_g(p_1,p_2,q)  \Biggl)\delta^{ab}u^{\alpha \beta}(p_1,p_2) \nn\\&+ \left(\phi_2^{\alpha \beta a b} +\phi_g^{\alpha \beta a b}\right) \biggl],
\end{align}}
that allows one to define the anomaly form factor as 
\beq
\Phi_{TJJ} (p_1^2,p_2^2,q^2,m^2)\equiv \frac{1}{3q^2}\mathcal{A}=\frac{1}{3 q^2}\Big(\frac{g_s^2}{48\pi^2}(11C_A - 2 n_f) + \chi_0(p_1,p_2,q,m)+\chi_g(p_1,p_2,q)\Big).
\eeq
Combining this result with \eqref{parm}, its expression can be rewritten in the form 
\begin{align}
	\Phi_{TJJ}=\Bigl(&\frac{g_s^2}{144 \pi^2q^2}(11C_A - 2 n_f)  +\frac{1}{3q^2} g_s^2m^4 H_1\, C_0 \left(p_1^2,p_2^2,q^2,m^2\right) +\frac{1}{3q^2}g_s^2m^2 \Bigl(H_2\,C_0 \left(p_1^2,p_2^2,q^2,m^2\right)  \nn\\&+ H_3\, \bar{B}_0(q^2,m^2)+ H_4\, \bar{B}_0(p_1^2,m^2)+ H_5 \,\bar{B}_0(p_2^2,m^2) +H_6\,+\chi_g(p_1,p_2,q)\Bigl).
	\label{tjj}
\end{align}
This defines the function that we are going to investigate in our spectral analysis. 
Notice that the first $1/q^2$ contribution in \eqref{tjj} has a residue given by the QCD $\beta$ function up to a sign $\sim 1/3 (11C_A - 2 n_f)$. With the definition above, the trace sector can be expressed in terms of $\Phi_{TJJ}$ as 
\begin{align}
	\label{last}
	\langle T^{\mu \nu}(q)  J^{ a\alpha}(p_1) J^{ b\beta}(p_2)\rangle_{tr}=&\hat\pi^{\mu \nu}(q)\delta^{ab}\Bigl[ \Phi_{TJJ}(q)\, u^{\alpha\beta}(p_1,p_2) \nn \\
	& +\frac{1}{3}\left( \chi_1(p_1,p_2,q,m) 
	+ \chi'_g(p_1,p_2) \right)  v^{\alpha\beta}(p_1,p_2) \nn\\
	& +\frac{1}{3} \left((B_2 + C_1) p_1^{\alpha}p_1^{\beta} + \left[ B_2(p_1 \leftrightarrow p_2) + C_1(p_1 \leftrightarrow p_2)\right] p_2^{\alpha}p_2^{\beta} + ( B_3 + C_2) p_1^{\alpha}p_2^{\beta}\right) \Bigl],
\end{align}
where we have separated the contribution projecting onto the anomaly term $F^2$ from the remaining terms
that are part of the trace but not of the anomaly. The remaining terms, not included in the definition of the form factor, are those proportional to the tensor structure
$v^{\alpha b}$ and—in the last bracket of \eqref{last}—terms that vanish once contracted with the external polarizations of the gluons. We will show, through a detailed analysis of its spectral density, that the spectral density of this form factor is characterized by a sum rule valid for generic virtualities of the external gluons.\\
Notice that $\chi_g$ in \eqref{chig}, originating from the gluon sector, is a gauge-dependent contribution that must necessarily be included in the definition of the anomaly form factor when performing a dispersive analysis for off-shell gluons.

\section{Derivation of the sum rule}
Having identified the anomaly form factor, we can proceed with the analysis of the spectral function, proving the sum rule satified by its spectral density.\\  
The analyticity properties of the function \(\Phi_0\equiv \Phi_{TJJ}\) in the complex \(q^2\)-plane  \(\Phi_0\) satisfies relation of the form
\beq
\Phi_0(q^2, s_1, s_2, m^2) = \frac{1}{2\pi i} \oint_C \frac{\Phi_0(s, s_1, s_2, m^2)}{s - q^2}\, ds,
\eeq
with $s_1$ and $s_2$ real, where \(C\) is a closed contour encircling the singularities of \(\Phi_0\). The integration can be deformed to run over the $s=q^2 \geq 0$ on the real axis. 
As we will demonstrate, \(\Phi_0\) possesses both poles and a branch cut on the real axis, with the cut beginning at the physical threshold \(s = 4m^2\). Correspondingly, the associated spectral density will reflect this analytic structure, featuring poles from rational terms and a discontinuity across the cut for \(s > 4m^2\).\\
We begin by characterizing the discontinuity of the function \(\Phi_0\) across the branch cut for time-like momenta \(q^2 > 0\), defined as:
\beq
\label{discù}
\mathrm{Disc}\,\Phi_0(q^2) = \Phi_0(q^2 + i\epsilon, p_1^2, p_2^2, m^2) - \Phi_0(q^2 - i\epsilon, p_1^2, p_2^2, m^2),
\eeq
which encodes the non-analytic behavior of \(\Phi_0\) in the complex \(q^2\)-plane. 
\\
The spectral density of $\Phi_0$, denoted by \(\Delta \Phi_0\), is then introduced via the relation
\beq
\label{delta}
\Delta\Phi_0(s, s_1, s_2, m^2) \equiv \mathrm{Im}\,\Phi_0(s, s_1, s_2, m^2) = \frac{1}{2i} \mathrm{disc}\,\Phi_0(s).
\eeq
The contribution from the boundary at infinity vanishes and the dispersion relation simplifies to
\beq
\Phi_0(q^2, p_1^2, p_2^2, m^2) = \frac{1}{\pi} \int_0^\infty \frac{\Delta\Phi_0(s, p_1^2, p_2^2, m^2)}{s - q^2} \, ds.
\eeq
Coming to the spectral function of this form factor, we separate the discontinuity into the pole contributions plus the continuum, in the form 
\begin{equation}
	\text{disc}(\Phi_{TJJ})=\text{disc}(\Phi_{TJJ})_{cont}+\text{disc}(\Phi_{TJJ})_{pole},
\end{equation}
with the discontinuity of \eqref{tjj} over the branch cut given by 
\begin{align}
	\label{cont}
	\text{disc}(\Phi_{TJJ})_{cont}=&\frac{n_F\,g_s^2\,  m^2 K_1(s,s_1,s_2)\, \text{disc}\, {B}_0\left(s,m^2\right)}{12\pi ^2  s(s-s_1-s_2) \lambda^2}+\frac{n_F\,g_s^2\,  m^2  K_2(s,s_1,s_2) \,\text{disc}\,{C}_0\left(s,s_1,s_2,m^2\right)}{24\pi ^2  s(s-s_1-s_2) \lambda^2}\nn\\&\frac{C_A g_s^2 (s_1+s_2)\,\text{disc}\,  {B}_0(s)}{48 \pi ^2 \l}+\frac{C_A g_s^2\, K_3\, \text{disc}\, {C}_0\left(s,s_1,s_2\right)}{96 \pi ^2 s (s-s_1-s_2) \l},
\end{align}
with
\begin{align}
	K_1=-s_2 (s-s_1) \left(s^2+3 s s_1-2 s_1^2\right)-s_2^3 (3 s+2 s_1)+s_2^2 (s+s_1) (3 s+2 s_1)-s_1 (s-s_1)^3+s_2^4
\end{align}
\begin{align}
	K_2=&-4 m^2 \left(s^2-2 s (s_1+s_2)+(s_1-s_2)^2\right) \left(s^2-2 s (s_1+s_2)+s_1^2+s_2^2\right)\nn\\&+s_2^4 (5 s+s_1)-2 s s_2^3 (5 s+3 s_1)+2 s s_2^2 (s+s_1) (5 s-3 s_1)\nn\\&-s_2 (5 s-s_1) (s+s_1) (s-s_1)^2+(s-s_1)^5-s_2^5
\end{align}
\begin{align}
	K_3=&\,\,s_2^2 (-9 s^2+3 s s_1-10 s_1^2)+s_2 (3 s^3-10 s^2 s_1+3 s s_1^2+8 s_1^3)\nn\\&+s_2^3 (9 s+8 s_1)+3 s_1 (s-s_1)^3-3 s_2^4\, .
\end{align}
Notice that the contributions to the cuts or poles in the spectral density originating from the quark and gluon sectors can be traced separately by examining the coefficients of $n_f$ and $C_A$ in \eqref{cont}, respectively.\\
The general discontinuity of the scalar 3-point function $C_0$, given in Appendix \ref{Co}, is given by
\begin{equation}
	\label{d2}
	\text{disc}\,{C_0}(s,s_1,s_2,m^2)=-\frac{2i  \pi  \log \left(\frac{m^2-s (w-\bar z) (\bar w-z)}{m^2-s (w-z) (\bar w-\bar z)}\right)}{s (z-\bar z)}\,\,\, \theta(s-4m^2),
\end{equation}
where	

\beq
z=\frac{1}{2 s } \left(\sqrt{\lambda} + s_1 - s_2 +s\right) \qquad \bar z=z=\frac{1}{2 s } \left(- \sqrt{\lambda} + s_1 - s_2 +s\right)
\eeq
with
\beq
w=\frac{1}{2} \left(1+\sqrt{1-\frac{4 m^2}{s}}\right)	\qquad \bar w=\frac{1}{2} \left(1-\sqrt{1-\frac{4 m^2}{s}}\right),
\eeq	
and 
\beq
\text{disc}\, {C}_0\left(s,s_1,s_2\right)
= - \frac{2i\pi}{\sqrt{\lambda(s, s_1, s_2)}}
\log \left(
\frac{(2s - s_1 + s_2 + \sqrt{\lambda})(s_1 - s_2 + s + \sqrt{\lambda})}
{(2s - s_1 + s_2 - \sqrt{\lambda})(s_1 - s_2 + s - \sqrt{\lambda})}
\right),
\eeq

while 
\begin{equation}
	\label{bos}
	\textrm{disc}\,{B_0}(s,m^2)=2\p i \sqrt{1-\frac{4m^2}{s} }\,\,\,
	\theta(s-4m^2).
\end{equation}
The pole contributions split at separate locations $(s=0, s=s_1+s_2, s=s_\pm)$
\begin{align}
	\text{disc}(\Phi_{TJJ})_{pole}=\text{disc}(\Phi_{TJJ})_{0}+\text{disc}(\Phi_{TJJ})_{s_1+s_2}+\text{disc}(\Phi_{TJJ})_{s_-}+\text{disc}(\Phi_{TJJ})_{s_+},
\end{align}
where we have indicated the points corresponding to the zeroes of the $\lambda$ function as

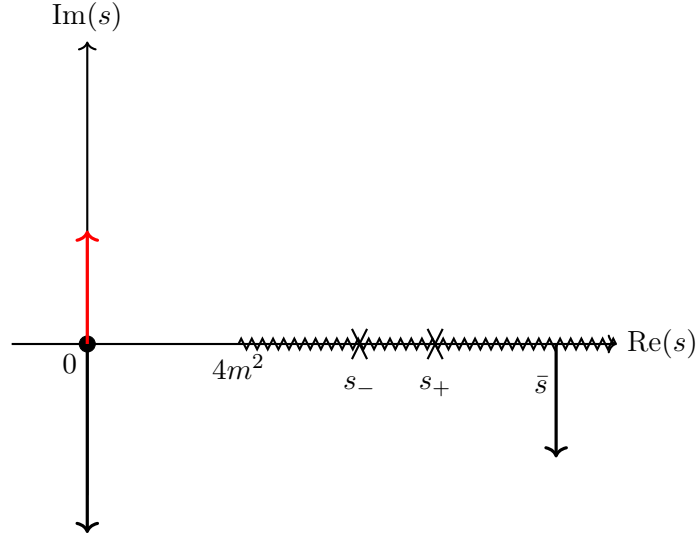
\begin{figure}
	\centering
	\begin{tikzpicture}
		
		% Thick axes
		\draw[thick, ->] (-1,0) -- (7,0) node[right] {$\text{Re}(s)$};
		\draw[thick, ->] (0,-2) -- (0,4) node[above] {$\text{Im}(s)$};
		
		% Pole at the origin
		\filldraw[black] (0,0) circle (3pt) node[below left] {$0$};
		
		% Branch cut from 4m^2 to infinity
		\draw[thick, decorate, decoration={zigzag, amplitude=2pt, segment length=4pt}] (2,0) -- (7.0,0);
		\node at (2, -0.3) {$4m^2$};
		
		% Spikes at s=0 with different colors
		\draw[very thick, ->, red] (0,0) -- (0,1.5);     % Black spike with arrow
		\draw[very thick, ->, black] (0,0) -- (0,-2.5); % Red spike closer to q^2=0 with arrow
		
		% Simple crosses over the branch cut
		\draw[thick] (3.5,0.2) -- (3.7,-0.2); % First cross (top left to bottom right)
		\draw[thick] (3.7,0.2) -- (3.5,-0.2); % First cross (top right to bottom left)
		\node[below] at (3.6,-0.3) {$s_-$};   % Label for first cross
		
		\draw[thick] (4.5,0.2) -- (4.7,-0.2); % Second cross (top left to bottom right)
		\draw[thick] (4.7,0.2) -- (4.5,-0.2); % Second cross (top right to bottom left)
		\node[below] at (4.6,-0.3) {$s_+$};   % Label for second cross
		\draw[very thick, ->, black] (6.2,0.0) -- (6.2,-1.5); % Red spike closer to q^2=0 with arrow
		\node[below] at (6.0,-0.3) {$\bar s$}; 
	\end{tikzpicture}
	\caption{The spectral density of the off-shell conformal anomaly form factor in the complex \( s\equiv q^2 \)-plane. Shown are the two spikes at \(s=0\), the anomalous thresholds at $s_{\pm}=(\sqrt{s_1}\pm \sqrt{s_2})^2$ and the pole at $s=\bar{s}=s_1+s_2$. In this case there is a cancelation in the sum rule between the integral of the density over the cut, the black spike at $s=0$, of variable strength, and the black spike at $s=\bar{s}$.    } 
	\label{figgX}
\end{figure}
\beq
s_\pm=(\sqrt{s_1} \pm \sqrt{s_2})^2.
\eeq
One extra pole is present at $s= s_1 + s_2$, beside the pole at $s=0$.
The poles at $s=0$ are split into two spikes, one of constant and the other of variable strength, represented in 
Fig. 3 
\begin{align}
	\label{Ym}
	\text{disc}(\Phi_{TJJ})_{0}=2\p i&\delta(s){ \, \,   \frac{g_s^2}{144\pi^2} \, (11C_A - 2 n_f)}-2\p i\delta(s)\Y(s_1,s_2,s, m^2)
\end{align}
where 
{\allowdisplaybreaks\small
	\begin{align}
		\Psi(s_1,s_2,s=0, m^2)=&g_s^2\,\,C_A\,  \left(-\frac{s_1 s_2 {B}_0^R(s_1,0)}{24 \pi ^2 \left(s_1^2-s_2^2\right)}+\frac{s_1 s_2 {B}_0^R(s_2,0)}{24 \pi ^2 \left(s_1^2-s_2^2\right)}+\frac{\left(3 s_1^2-2 s_1 s_2+3 s_2^2\right) {C}_0(s_1,s_2,0)}{96 \pi ^2 (s_1+s_2)}+\frac{1}{48 \pi ^2}\right)\nn\\&
		+g_s^2\,\,n_F \, \Biggl(m^2 \biggl(-\frac{\left(s_1^2+s_2^2\right) {B}_0^R\left(0,m^2\right)}{12 \pi ^2 (s_1-s_2)^2 (s_1+s_2)}+\frac{s_1 \left(s_1^2+2 s_1 s_2+3 s_2^2\right) \text{B}_0^R\left(s_1,m^2\right)}{12 \pi ^2 (s_1-s_2)^3 (s_1+s_2)}\nn\\&
		+\frac{s_2 \left(3 s_1^2+2 s_1 s_2+s_2^2\right) {B}_0^R\left(s_2,m^2\right)}{12 \pi ^2 (s_2-s_1)^3 (s_1+s_2)}+\frac{\left(s_1^2+s_2^2\right) {C}_0\left(s_1,s_2,0,m^2\right)}{24 \pi ^2 (s_1-s_2)^2}+\frac{s_1^2+s_2^2}{12 \pi ^2 (s_1-s_2)^2 (s_1+s_2)}\biggl)\nn\\&
		+\frac{m^4 \left(s_1^2+s_2^2\right) {C}_0\left(s_1,s_2,0,m^2\right)}{6 \pi ^2 (s_1-s_2)^2 (s_1+s_2)}\Biggl).
		\label{psi}
	\end{align}}
	
	where  $C_0(s_1,s_2,0)$ (in three variables) is the massless $C_0(s_1,s_2,s)$ integral evaluated at $s=0$, and 
	${C}_0(s_1,s_2,0,m^2)$ is the massive $C_0(s_1,s_2, s,m^2)$ integral evaluated at the same point. We have denoted with 
	We have outlined the various contributions in \figref{figgX}. The "black spike" at \( s = 0 \), which maintains a constant strength, corresponds to the first term on the right-hand side of \eqref{Ym}. The second contribution, which is dependent on the external invariants, is represented by the "red spike" in the same figure. \\
	For \( s > 4m^2 \), across the cut, we encounter two apparent poles. As we will demonstrate, these poles do not contribute to the discontinuity. Additionally, there is a pole at \( s = s_1 + s_2 \), which can appear either above or below the cut (i.e., for \( s > 4m^2 \) or \( s < 4m^2 \)), depending on the specific values of \( s_1 \) and \( s_2 \).
	Finally, we observe that in the limit \( s \to 0 \)
	\begin{align}
		\lim_{s\to 0} \Y(s,s_1,s_2)=&g_s^2\,\,C_A\,  \left(-\frac{s_1 s_2 {B}_0^R(s_1,0)}{24 \pi ^2 \left(s_1^2-s_2^2\right)}+\frac{s_1 s_2{B}_0^R(s_2,0)}{24 \pi ^2 \left(s_1^2-s_2^2\right)}+\frac{\left(3 s_1^2-2 s_1 s_2+3 s_2^2\right){C}_0(s_1,s_2,0)}{96 \pi ^2 (s_1+s_2)}+\frac{1}{48 \pi ^2}\right)\nn\\&
		+g_s^2\,\,n_F \, \Biggl(m^2 \biggl(-\frac{\left(s_1^2+s_2^2\right){B}_0^R\left(0,m^2\right)}{12 \pi ^2 (s_1-s_2)^2 (s_1+s_2)}+\frac{s_1 \left(s_1^2+2 s_1 s_2+3 s_2^2\right){B}_0^R\left(s_1,m^2\right)}{12 \pi ^2 (s_1-s_2)^3 (s_1+s_2)}\nn\\&
		+\frac{s_2 \left(3 s_1^2+2 s_1 s_2+s_2^2\right) {B}_0^R\left(s_2,m^2\right)}{12 \pi ^2 (s_2-s_1)^3 (s_1+s_2)}+\frac{\left(s_1^2+s_2^2\right) {C}_0\left(s_1,s_2,0,m^2\right)}{24 \pi ^2 (s_1-s_2)^2}+\frac{s_1^2+s_2^2}{12 \pi ^2 (s_1-s_2)^2 (s_1+s_2)}\biggl)\nn\\&
		+\frac{m^4 \left(s_1^2+s_2^2\right) {C}_0\left(s_1,s_2,0,m^2\right)}{6 \pi ^2 (s_1-s_2)^2 (s_1+s_2)}\Biggl)  < \infty.
	\end{align}
	Notice that \(\Y\) remains finite across the entire light-cone for fixed $s_1$ and $s_2$.

	\subsection{Cancellation of the anomalous thresholds and the sum rule}
	\label{trash}
	The expression for the spectral density of $\Phi_{TJJ}$ can be separated into distinct components, representing contributions from regular kinematical thresholds as well as potential anomalous thresholds. Therefore we split the form factor as
	\begin{equation}
		\Phi_{{TJJ}} = \Phi_{{TJJ}}^{\text{reg}} + \Phi_{{TJJ}}^{(\l)},
	\end{equation}
	where $\Phi_{{TJJ}}^{\text{reg}}$ corresponds to regular kinematical thresholds and $\Phi_{{TJJ}}^{\l}$ accounts for possible anomalous thresholds in $\Phi_{{TJJ}}$. The regular part, $\Phi_{{TJJ}}^{\text{reg}}$, includes contributions from standard kinematical thresholds located at
	\begin{equation}
		q^2 = 0, \quad q^2 = s_1 + s_2, \quad \text{and} \quad q^2 = 4m^2,
	\end{equation}
	which manifest in the form of simple poles and cuts. These thresholds arise from well-understood physical processes. In contrast, the anomalous part, $\Phi_{{TJJ}}^{\text{anom}}$, emerges under special conditions when the kinematical factor $\lambda = 0$ which could be present in this computation. Given that
	\begin{equation}
		\lambda = (s - s_-)(s - s_+),
	\end{equation}
	the dispersive representation of  $ \Phi_{\text{TJJ}}$ may allow terms of the form
	\begin{align}
		\textrm{disc} \,\Phi_{{TJJ}} = &c_0 \delta(s) + c_1(s,s_1, s_2) \theta(s - 4m^2) H(s) + c_2(s,s_1, s_2) \delta(s - s_1 - s_2)\nn\\&
		+ c_3(s, s_1, s_2) \delta'(s - s_-) + c_4 \delta'(s - s_+) + c_5 \delta(s - s_-) + c_6 \delta(s - s_+),
	\end{align}
	where the derivative of the $\delta$-function corresponds to higher order poles at $\lambda = 0$.
	Indeed, the general structure of $\Phi_{\text{TJJ}}$ is
	\begin{align}
		\Phi_{{TJJ}} =&  \frac{c_0(s,s_1,s_2,m^2)}{s} + \frac{c_1(s,s_1,s_2,m^2)}{s} + \text{(continuum terms)} + \frac{c_2(s,s_1,s_2,m^2)}{s - \bar{s}} + \frac{c_3(s,s_1,s_2,m^2)}{(s - s_-)^2} \nn\\&+ \frac{c_4(s,s_1,s_2,m^2)}{(s - s_+)^2} + \frac{c_5(s,s_1,s_2,m^2)}{s - s_-} + \frac{c_6(s,s_1,s_2,m^2)}{s - s_+},
	\end{align}
	where $\bar{s} = s_1 + s_2$ and, $c_3$ and $c_4$ define possible residues at the double poles of $\lambda^2$. These could, in principle, introduce subtractions that can be analyzed as follows. \\
	For example, at $s = s_-$, we have
	\begin{equation}
		\Phi_{{TJJ}}^{(\l)}(q^2 = s_-) = \frac{1}{2\pi i} \int_0^\infty \frac{c_3(s, s_1, s_2, m^2) \delta'(s - s_-)}{(s - q^2)} \, ds
		+ \frac{1}{2\pi i} \int_0^\infty \frac{c_5(s, s_1, s_2, m^2) \delta(s - s_-)}{(s - q^2)} \, ds.
	\end{equation}
	This simplifies to:
	\begin{equation}
		{2\pi i}\,\Phi_{{TJJ}}^{(\l)}(q^2 = s_-)= -\frac{d}{ds} \left( \frac{c_3(s, s_1, s_2, m^2)}{(s - q^2)} \right) \bigg|_{s = s_-} - \frac{c_5(s_-, s_1, s_2, m^2)}{s_- - q^2}.
	\end{equation}
	A similar analysis gives 
	\begin{equation}
		{2\pi i}\,\Phi_{{TJJ}}^{(\lambda)}(q^2 = s_+) =  -\frac{d}{ds} \left( \frac{c_4(s, s_1, s_2, m^2)}{(s - q^2)} \right) \bigg|_{s = s_+} - \frac{c_6(s_+, s_1, s_2, m^2)}{s_+ - q^2}.
	\end{equation}
	For the evaluation of $\Phi_{{TJJ}}^{(\lambda)}(q^2 = s_\pm)$, we need to compute the explicit expression of the scalar integral $C_0$ at these kinematical points. We use the expressions of the scalar integral $C_0$ at these points 
	($C_0(s_-,s_1,s_2,m^2)$ and $C_0( s_+,s_1,s_2,m^2)$). Their expressions can be found in appendix \ref{Co}.
	
	Using these relations one can show by a long computation that 
	
	\begin{equation}
		\Phi_{{TJJ}}^{(\l)}(q^2 = s_\pm) \equiv 0,
	\end{equation}
	and hence 
	\begin{equation}
		\Phi_{{TJJ}}^{(\l)}=0.
	\end{equation}
	Therefore, the only pole to consider at finite $s$ is the one located at $s= s_1+s_2$.
	This contributes to the sum rule with a localized density that is given by 
	{\small\begin{align}
			\text{disc}(\Phi_{TJJ})_{s_1+s_2}=&g_s^2\,\,C_A  \left(\frac{(s_2-s_1) \text{B}_0^R(s_1)}{96 \pi ^2 (s_1+s_2)}+\frac{(s_1-s_2) {B}_0^R(s_2)}{96 \pi ^2 (s_1+s_2)}-\frac{\left(s_1^2-6 s_1 s_2+s_2^2\right) {C}_0(s_1,s_2,s_1+s_2)}{96 \pi ^2 (s_1+s_2)}-\frac{1}{48 \pi ^2}\right)\nn\\&
			+g_s^2 \,\, n_F \Biggl(m^2 \biggl(-\frac{{B}_0^R\left(s_1,m^2\right)}{48 \pi ^2 (s_1+s_2)}+\frac{{B}_0^R\left(s_1+s_2,m^2\right)}{24 \pi ^2 (s_1+s_2)}-\frac{{B}_0^R\left(s_2,m^2\right)}{48 \pi ^2 (s_1+s_2)}+\frac{\text{C}_0\left(s_1,s_2,s_1+s_2,m^2\right)}{48 \pi ^2}\nn\\&-\frac{1}{24 \pi ^2 (s_1+s_2)}\biggl)-\frac{m^4 {C}_0\left(s_1,s_2,s_1+s_2,m^2\right)}{12 \pi ^2 (s_1+s_2)}\Biggl).
			\label{bo}
	\end{align}}
	Notice that the UV divergences in the $B_0$'s in the expression above cancel automatically, in the counting of the $1/\epsilon_{UV}$ poles in each of these self-energy contributions. Therefore we have replaced in \eqref{bo} each $B_0$ by $B_0^R$ with no change. No IR singularities are encountered unless we consider special values where any of the external invariants vanishes. 
	\\
	The mutual cancellation of some of the discrete contributions with the continuum, separately for the $n_f$ and $C_A$ terms, after integration over $s$, can be shown by a direct computation. One can verify the nontrivial relation 
	\beq
	\int_{4 m^2}^\infty d s\, \text{disc}(\Phi_{TJJ})_{cont}- 2 \pi i\psi(s_1,s_2,s=0,m^2) + \text{disc}(\Phi_{TJJ})_{s_1+s_2}=0,
	\eeq
	where we have used \eqref{Ym}, the expression of $\psi(s_1,s_2,s=0,m^2)$ defined in \eqref{psi}, \eqref{bo} and the continuum contribution given in \eqref{cont}. Using \eqref{Ym} one is clearly left only with the anomaly pole contribution at $s=0$ (the first term of \eqref{Ym}), giving 
	
	\beq
	\int _{0}^{\infty} ds \,\Delta \Phi_{TJJ}(s_1,s_2,s,m^2)=-\frac{1}{3}\frac{\beta(g)}{g},
	\eeq
	which is mass-independent and gauge invariant. This equation is the sum rule satisfied by the absorptive part of the conformal anomaly form factor in QCD at $O(\alpha_s)$.

	\section{Sum rules of anomaly form factors in the on-shell limit }
	In the on-shell case, the sum rules take a simpler form, and both the integral of the spectral density and the relation between the continuum and the anomaly pole exhibit a clearer pattern. While the chiral and conformal anomaly sum rules share structural similarities, they are not identical. To highlight the distinction between the two, we first summarize the main results of the chiral case before turning to the conformal case \cite{Coriano:2025ceu}.
	
	\subsection{The chiral case: review}
	We consider the $AVV$ correlator (axial-vector/vector/vector) and illustrate how the branch cut of the spectral density of its anomaly form factor evolves into a pole as the fermion mass approaches zero.
	To explore deviations from conformal invariance in the structure of this form factor one considers the case of massive fermions and computes the correlator using standard perturbative technique. The $AVV$ correlator is constructed with two vector currents $J$ and one axial-vector current $J_A$, all Abelian, and can be parameterized in the form 
	\begin{align}
		\label{refm}
		\braket{J^{\mu_1}(p_1)J^{\mu_2}(p_2) J^{\mu_3}_A (p_3)} &=	\braket{J^{\mu_1 }(p_1) J^{\mu_2 }(p_2)\, j_{A \text { loc }}^{\mu_3}(p_3)}
		+ \pi^{\mu_1}_{\alpha_1}(p_1)\pi^{\mu_2}_{\alpha_2} (p_2) \pi^{\mu_3}_{\alpha_3}(p_3)
		\, \Delta_T^{\alpha_1 \alpha_2 \alpha_3 }(p_1^2,p_2^2,p_3^2,m^2) 
		\nonumber \\
	\end{align}
	with the longitudinal sector given by 
	\begin{equation}
		\label{nonloc}
		\braket{J^{\mu_1 }(p_1) J^{\mu_2 }(p_2)\, j_{A \text { loc }}^{\mu_3}(p_3)}\equiv \bar \Phi_0(p_1^2,p_2^2,p_3^2,m^2) \varepsilon^{p_1p_2\mu_1\mu_2}\,p_3^{\mu_3}.
	\end{equation}
	We have used a decomposition based on the inclusion of projectors as in \eqref{loct}.	
	The expression of the transverse part is denoted by $\Delta_T$ and will not be relevant for our current analysis. A direct computation of the longitudinal part gives
	\beq
	\label{fst}
	\bar \Phi_0(p_1^2,p_2^2,p_3^2,m^2)=\frac{ m^2}{ \pi ^2}\,\frac{1}{q^2}C_0\left(q^2,p_1^2 \,,p_2^2\,, m^2\right)+\frac{1 }{2 \pi ^2}\,\frac{1}{q^2},
	\eeq	
	whose discontinuity defined as in \eqref{discù}, for $s_1\equiv p_1^2=0,\, s_2=p_2^2=0$ - defining the on-shell case - takes the form 
	\beqa
	\label{zero}
	\textrm{disc}\,\bar \Phi_0 (q^2, m^2) 
	&=&  \frac{m^2}{ \pi^2} \textrm{disc}\left( \frac{C_0(q^2,m^2)}{q^2}\right) +\frac{1}{2 \pi^2} \textrm{disc}\frac{1}{q^2},
	\eeqa
	and can be re-expressed as
	\begin{equation}
		\mathrm{disc} \, \bar \Phi_0(q^2,m^2) =-\frac{1}{2\pi^2} \left(2 \pi i \, \delta(q^2)
		+ 4 i \pi m^2 \, C_0(q^2 \to 0,m^2) \delta(q^2) - \frac{2 m^2}{q^2} \mathrm{disc} \, C_0(q^2, m^2)\right).
		\label{second}
	\end{equation}
	There are three contributions to this function:
	two spikes at $q^2\equiv s=0$, represented by two uparrows, one of constant and the other of varaible strength, and a cut covering the $q^2>4 m^2$ 
	region.
	Introducing the limit 
	\begin{equation}
		C_0(q^2 \to 0,m^2) = -\frac{1}{2 m^2}
	\end{equation}
	one notices the cancellation of the two localized contributions at $s = 0$. 
	The anomaly form factor does not exhibit any pole at $p_3^2=0$, but has only a branch cut  for $q^2\equiv p_3^2 > 4\, m^2$. A similar result can be obtained form the explicit expression of $\Phi_0$, that can be derived in the form 
	\begin{equation}
		\bar \Phi_0= \frac{1}{2 \pi^2 q^2} \left( 1 + \frac{m^2}{q^2} \log^2 \left( \frac{\sqrt{\tau(q^2, m^2)} + 1}{\sqrt{\tau(q^2, m^2)} - 1} \right) \right), \qquad q^2 > 0,
		\label{a6}
	\end{equation}
	with $\tau(q^2, m^2) = 1 - \frac{4m^2}{q^2}$ with
	\begin{align}
		\mathrm{disc} \, {C}_0(q^2, m^2) &= \frac{1}{i \pi^2} 
		\int d^4 l \frac{2 \pi i \delta_+(l^2 - m^2) 2 \pi i \delta_+((l - q)^2 - m^2)}{(l - p_1)^2 - m^2 + i \epsilon} \notag \\
		&= \frac{2 \pi}{i q^2} \log \left( \frac{1 + \sqrt{\tau(q^2, m^2)}}{1 - \sqrt{\tau(q^2, m^2)}} \right) \theta(q^2 - 4 m^2)
		\label{cut}
	\end{align}
	that gives
	\beq
	\label{twist}
	\textrm{disc} \, \bar \Phi_0(q^2, m^2) =-2 i  \frac{m^2}{\p (q^2)^2}\log \frac{1 + \sqrt{\tau(q^2,m^2)}}{1 - \sqrt{\tau(q^2,m^2)} }\theta(q^2- 4 m^2).
	\eeq 
	
	Then, the final spectral density takes the form
	\begin{equation}
		\Delta {\bar \Phi_0}(q^2,m^2) = \frac{ m^2}{\pi s^2} \log \left( \frac{1 + \sqrt{\tau(s, m^2)}}{1 - \sqrt{\tau(s, m^2)}} \right) \theta(s - 4 m^2). 
		\label{exp}
	\end{equation}
	and is only given by the contribution coming from the cut. 
	A direct integration of this function gives 
	\begin{equation}
		\label{sumrr}
		\frac{1}{\pi}\int_{4 m^2}^\infty ds\, \Delta{\bar \Phi_0}(s, 0, 0, m^2) = \frac{1}{2\p^2},
	\end{equation}
	proving the validity of a sum rule. 
	This demonstrates the exchange of a branch-cut in the on-shell case $(s_1=s_2=0)$ for general massive fermion in the loop rather than poles. The change of the cut into a pole can be illustrated by moving towards the conformal limit. 
	For this purpose we set $m^2_n\equiv (1/n) m^2$ and reduce the value of the mass parameters for $n\to \infty$. \\
	The sequence $\Delta^{(n)}{\Phi_0}$, characterising the discontinuity with parameter $m^2_n$, exhibits an interesting behavior, ultimately converging to a Dirac delta function in the limit \(m \to 0\). Specifically, the spectral flow is given by the sequence of functions 
	\beq
	\lim_{m^2_n\to 0}\Delta^{(n)}{\bar \Phi_0} (s, m_n^2) = \lim_{m_n\to 0} \frac{ m_n^2}{\pi s^2} \log\left(\frac{1 + \sqrt{\tau(s, m_n^2)}}{1 - \sqrt{\tau(s, m_n^2)}}\right) \theta(s - 4m_n^2) = \frac{1}{2\pi} \delta(s),
	\eeq
	where the convergence to $\delta(s)$ is understood in the distributional sense. The sequence of functions are strongly suppressed at large $s$. This expression encapsulates the essence of how the spectral densities evolve as the mass parameter \(m_n\) approaches zero. 
	The pattern is shown in Fig \ref{dd1}. In the figure, the area under the curve $\Delta^{(n)}{\Phi_0}$ is unrelated to \(m_n\), due to the sum rule \eqref{sumrr} and equals the anomaly. We will refer to this phenomenon as to an "area law" or 
	a spectral flow of the sequence of spectral densities $\Delta^{(n)}{\Phi_0}$ towards $s=0$, where the sum rule is indeed saturated by the exchange of an anomaly pole in the conformal limit, while the continuum disappears. We are going to describe the behaviour in the conformal case.

	\subsection{On-shell analysis of the conformal anomaly form factor}
	At this point, we extend the analysis presented above turning to a description of the on-shell behavior of the $\Phi_{TJJ}$ form factor, that in the presence of a massive fermion loop and pure gauge contributions is given by
	\begin{equation}
		\Phi_{TJJ}(s) = \frac{n_F g_s^2 \delta^{ab} \left(3 m^2 (s - 4m^2) \,{C}_0(s,m^2) - 6m^2 + s\right)}{72 \pi^2 s^2} - \frac{11 C_A g_s^2 \delta^{ab}}{144 \pi^2 s},
	\end{equation}
	where ${C}_0(s,m^2) \equiv C_0(s,0,0,m^2)$ is the standard scalar three-point function (see Appendix \ref{Co}). In the on-shell limit, it takes the form:
	\begin{equation}
		{C}_0(s,m^2) = \frac{1}{2s} \log^2\left( \frac{\sqrt{s(s-4m^2)} + 2m^2 - s}{2m^2} \right).
	\end{equation}
	
	To study the spectral behavior of $\Phi_{TJJ}$, we compute its discontinuity across the real axis $(s\geq 0)$ using
	\begin{equation}\label{lleq1}
		\text{disc}\left(\frac{1}{s}\right)=-2i\pi\delta(s), \qquad \text{disc} \left( \frac{1}{s^2} \right) = 2 i \pi \delta'(s), 
	\end{equation}
	\begin{equation}\label{lleq2}
		\text{disc} \left( \frac{C_0(s, m^2)}{s} \right)
		= -2i \frac{\pi}{s^2} \log \left( \frac{1 + \sqrt{\tau(s, m^2)}}{1 - \sqrt{\tau(s, m^2)}} \right) \theta(s - 4m^2) - i \pi\,A(0) \delta(s),
	\end{equation}
	
	and
	\begin{align}\label{lleq4}
		\text{disc} \left( \frac{C_0(s,m^2)}{s^2} \right) = -\frac{2 i \pi}{s^3} \log\left( \frac{1 + \sqrt{\tau(s,m^2)}}{1 - \sqrt{\tau(s,m^2)}} \right) \theta(s - 4m^2) + i \pi \delta'(s) A(s), 
	\end{align}
	where we defined
	\begin{align}
		\tau(s,m^2) = 1 - \frac{4m^2}{s}, 
	\end{align}
	\begin{align}
		A(s) = {C}_0(s + i\epsilon, m^2) + {C}_0(s - i\epsilon, m^2).
	\end{align}
	We use the general identity for the distributional difference of powers of propagators gives:
	\begin{equation}
		\left( \frac{1}{x + i\epsilon} \right)^n - \left( \frac{1}{x - i\epsilon} \right)^n = (-1)^n \frac{2\pi i}{(n-1)!} \delta^{(n-1)}(x),
	\end{equation}
	which implies the presence of $\delta(s)$ and $\delta'(s)$ terms in the discontinuities of rational functions like $1/s^n$.
	
	The contribution proportional to $\delta'(s)$ in equation \eqref{lleq4} can be expressed in terms of $\delta$ derivatives acting on $A(s)$:
	\begin{equation}
		\delta'(s) A(s) = \delta'(s) A(0) - \delta(s) A'(0), \quad \text{with} \quad A(0) = -\frac{1}{m^2}, \quad A'(0) = -\frac{1}{12 m^4}.
	\end{equation}
	
	As a result, the full discontinuity of the fermionic part of $\Phi_{TJJ}$ takes the form
	\begin{align}
		\text{disc} \left( \Phi_{TJJ}(s) \right)_{\text{fermion}}
		=\,& \frac{n_F g_s^2 \delta^{ab}}{72 \pi^2} \, \text{disc} \left( \frac{3 m^2 (s - 4m^2) C_0(s,m^2) - 6m^2 + s}{s^2} \right) \\
		=\,& \frac{n_F g_s^2 \delta^{ab}}{72 \pi^2} \biggl[ 3 m^2  \left(\text{disc} \left( \frac{C_0(s,m^2)}{s} \right)+4m^2\text{disc} \left( \frac{C_0(s,m^2)}{s^2} \right)\right) \nonumber\\  &-6m^2  \text{disc} \left( \frac{1}{s^2} \right)  + \text{disc} \left( \frac{1}{s} \right)\biggl]
	\end{align}
	It is now straightforward to use Eqs.~\eqref{lleq1}, \eqref{lleq2} and \eqref{lleq4} to verify that the $\delta'(s)$ and $\delta(s)$ terms cancel exactly. 
	In other words, there is no pole contribution to the spectral density of $\Phi_{TJJ}$ in the fermionic sector, only the continuous cut survives in the on-shell limit. Explicitly, we find
	\begin{equation}
		\text{disc}[\Phi_{TJJ}(s)]_{\text{fermion}} = -2i \, \frac{n_F g_s^2 \delta^{ab} \left(3 m^2 (s - 4m^2) \right)}{72 \pi s^3} \log \left( \frac{1 + \sqrt{\tau(s, m^2)}}{1 - \sqrt{\tau(s, m^2)}} \right) \theta(s - 4m^2).
	\end{equation}
	This result reflects the cancellation of the conformal anomaly pole in the fermionic sector, consistent with the ``area law'' holding for the spectral function.
	In contrast, the gluonic contribution to the form factor is given by
	\begin{equation}
		\Phi_{TJJ}^{\text{gluon}}(s) = - \frac{11 C_A g_s^2 \delta^{ab}}{144 \pi^2 s},
	\end{equation}
	whose discontinuity is entirely localized at $s=0$:
	\begin{equation}
		\text{disc}[\Phi_{TJJ}(s)]_{\text{pole}} = \frac{11 C_A g_s^2 \delta^{ab}}{144 \pi^2} \cdot 2\pi i \delta(s).
	\end{equation}
	Therefore, the anomaly pole is absent in the fermionic sector for $m \neq 0$, but remains present in the gluonic sector. In the conformal limit, the branch cut generated by the fermionic contribution over the real axis shrinks to the origin and turns into a pole, while the gluon contributions remain anchored at s=0, being this sector massless. Their combinations restores the full structure of the anomaly pole, whose residue bears the entire conformal anomaly.  The complete anomaly pole re-emerges as
	\begin{equation}
		\text{disc}[\Phi_{TJJ}(s)] = - \frac{(2n_F - 11 C_A)\, i g_s^2 \delta^{ab}}{72 \pi} \delta(s).
	\end{equation}
	Introducing the factor $1/2 i$ coming from the definition of the spectral density we obtain 
	\beq
	\Delta \Phi_{TJJ}(s)=-\frac{1}{3}\frac{\beta(g)}{g} \delta(s)
	\eeq
	that reproduces the anomaly sum rule under integration for $s\geq 0$. Notice that the factor $1/3$ comes from 
	the definition of the conformal anomaly form factor, due to the transverse projector  ${1}/{ (3 \, q^2)} \hat\pi^{\mu \nu}(q)$ in 
	\eqref{anommX}, that is erased once we perform the trace on the correlator. Notice that, in our conventions, the quarks contribute negatively and the gluons positively to the anomaly. 
	
	\subsection{Alternative tests using of the Laplace/Borel transforms }
	\label{lap}
	We are now going to show that the integral of the spectral density of $\Phi_0$ along the cut from $\left[4 m^2, \infty \right)$ can be 
	evaluated as follows. We use the inverse Laplace transform of the density, denoted as $t$, followed by the $t \to 0$ limit. 
	
	In the most general context, we define the inverse Laplace transform, $\mathcal{L}^{-1}\{F(s)\}$, for a given function $F(s)$ as
	\beq
	f(t) = \mathcal{L}^{-1}\{F(s)\} = \frac{1}{2\pi i} \int_{\gamma - i \infty}^{\gamma + i \infty} F(s) e^{st} \, ds.
	\eeq
	Additionally, for a rational function $F(s)$ possessing a pole at $s = s_0$, the inverse Laplace transform is
	\beq
	\mathcal{L}^{-1}\left\{\frac{1}{s - s_0}\right\} = e^{s_0 t}.
	\eeq
	This contribution is nonzero as $t \to 0$, making it relevant for the dispersion integral. While this method efficiently computes integrals over the continuous spectrum of the spectral density, caution is required since it does not account for singular contributions, which can appear as subtractions when reconstructing a specific form factor. Notably, double poles, such as $\frac{1}{(s - s_0)^2}$, are not captured in this method, where one obtains
	\beq
	\mathcal{L}^{-1}\left\{\frac{1}{(s - s_0)^2}\right\} = t e^{s_0 t},
	\eeq
	which vanishes in the limit $t \to 0$.\\
	For the anomaly form factor $\Phi_0$, which has the spectral representation
	\beq
	\Phi_0(q^2) = \frac{1}{\pi} \int_0^\infty \frac{\Delta \Phi_0(s)}{s - q^2} \, ds,
	\eeq
	applying the inverse Laplace transform results in
	\beq
	\mathcal{L}^{-1}\left\{\int_0^\infty \frac{\Delta \Phi_0(s)}{s - q^2} \, ds \right\}(t) = \frac{1}{\pi} \int_0^\infty \Delta \Phi_0(s) e^{st} \, ds.
	\eeq
	Thus, the integral of the spectral density is evaluated in the $t \to 0$ limit
	\beq
	\lim_{t \to 0} \mathcal{L}^{-1}\{\Phi_0\}(t) = \frac{1}{\pi} \int_0^\infty \Delta \Phi_0(s) \, ds.
	\eeq
	This technique provides a means of deriving the integral along the cut and verifying sum rules associated with form factors. 
	To apply this method, we utilize the parametric representation of the scalar integrals in $\Phi_0\equiv \Phi_{TJJ}$ from \eqref{tjj} and proceed by direct computation of the inverse Laplace transform. This yields
	\small
	\begin{equation}
		\mathcal{L}^{-1}\{\Phi_0\}(t) = -\int_0^1 dx \int_0^{1-x} dy \, \frac{g_s^2 \delta^{ab} \left(2 n_f \left(-6 (4 y (x+y-1)+1) e^{-\frac{m^2 t}{y (x+y-1)}} - 30 m^2 t + 24 y (x+y-1) + 7\right) - 11 C_A \right)}{72 \pi^2}.
	\end{equation}
	\normalsize
	This result is valid for the on-shell case, though the extension to more general situations is straightforward, albeit lengthy. Taking the limit as $t \to 0$ leads directly to the value of the integral of the discontinuity across the entire cut
	\beq
	\mathcal{L}^{-1}\{\Phi_0\}(t \to 0) = \int_{4 m^2}^\infty \Delta \Phi_0(s, p_1^2, p_2^2, m^2) \, ds = \frac{g_s^2}{144 \pi^2}(11 C_A - 2 n_f)=-\frac{1}{3}\frac{\beta(g)}{g}
	\eeq
	This expression holds in the most general case, including for off-shell vector lines and a massive fermion, and is consistent with the findings in \cite{Giannotti:2008cv}. 
	\begin{figure}[t]
		\centering
		{\includegraphics[scale=1]{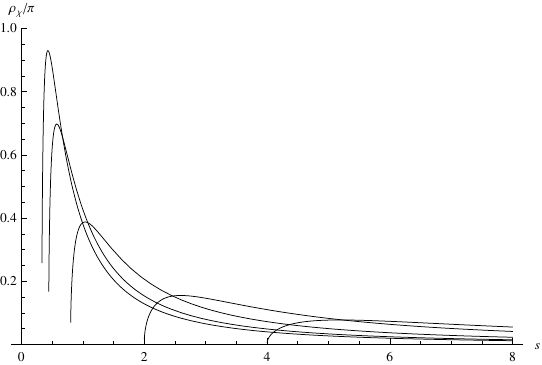}}  \hspace{2cm}
		\caption{Chiral spectral density flow in the $p_1^2 = p_2^2 = 0$ case as a function of the fermion mass $m^2_n\equiv m^2/n$ for the $AVV$ for $n\to \infty$. The sum rule is an "area law" for this spectral flow as one moves towards the conformal point  $m\to 0$ and modifies the cuts at $4 m_n^2$ into a pole 
			($\sim \delta(s),$ $s\equiv q^2$). The area is subtended by the spectral curves and is preserved by the sequence as $n$ grows and the densities get increasingly peaked at $s=0$. 
			In the conformal case, this identical pattern is reproduced only by the fermionic contributions, which are mass-dependent, while the contribution to the anomaly coming from the gluons (not shown) stays always localized at $s=0$, 
			proportional to $\delta(s)$. }
		\label{dd1}
	\end{figure}

	\noindent
	
	\section{The residue of the particle pole  of the conformal anomaly form factor in the light-cone variables }  
	\label{vert}
	
	In order to investigate the possible presence of a massless interpolating state---a particle pole---in the renormalized $TJJ$ interaction (the full vertex), we analyze in this section the residue of the correlator for generic values of the invariants $s_1$ and $s_2$. \\
	For the chiral anomaly vertex, the corresponding residue at $q^2=0$ vanishes unless $s_1$, $s_2$, and the fermion mass $m$ are simultaneously set to zero (we adopt the notation $(s_1, s_2) \equiv (p_1^2, p_2^2)$ and $s = q^2 = p_3^2$). This distinctive feature of the interaction~\cite{Coriano:2025ceu} will also appear in the conformal anomaly case. The cancellation arises from delicate interferences among the various tensor components of the correlators. \\
	As shown in~\cite{Coriano:2025ceu}, the only tensor structure that survives the limit $s \to 0$ at $s_1 = s_2 = m = 0$ is proportional to the Pontryagin density $F\tilde{F}$ for the chiral gauge anomaly, or to $R\tilde{R}$ in the case of the chiral gravitational anomaly. Thus, when all invariants vanish, the residue reduces to these anomaly-induced structures. This can be interpreted as an \emph{anomaly dominance} of the interaction on the light-cone, since the transverse sectors of the corresponding vertices are subleading. In other words, the light-cone behavior of the interaction is controlled by the anomaly in both cases. A similar result holds in the soft limit, when all four components of $q^\lambda$ are taken to zero. \\
	To highlight the difference between the chiral and conformal cases, we now repeat the analysis for the $TJJ$ correlator. Also here it is essential to consider the limit
	\begin{equation}
		\lim_{s \to 0} \, q^2 \,
		\big\langle T^{\mu\nu}(q)\, J^{\alpha a}(p_1)\, J^{\beta b}(p_2) \big\rangle,
	\end{equation}
	and, for a particle pole to exist, one must show that this expression remains nonzero, regardless of the virtualities of the other invariants. We therefore examine different configurations of the external invariants together with the fermion mass. Two prescriptions for the $q^2 \to 0$ limit will be considered:  
	(i) sending the four-momentum $q^\lambda$ along a lightlike direction (light-cone limit), and  
	(ii) taking $q^\lambda \to 0$ in all components (soft limit). \\
	For the light-cone limit, the four-momenta are parametrized as
	\begin{equation}
		p_3 \equiv q = q^+ n^+ + q^- n^-, \qquad  
		p_1 = p_1^+ n^+ + p_1^- n^- + p_\perp, \qquad  
		p_2 = p_2^+ n^+ + p_2^- n^- - p_\perp,
	\end{equation}
	with light-cone unit vectors satisfying $(n^\pm)^2 = 0$ and $n^+ \cdot n^- = 1$. In this regime, $q^+$, $p_1^+$, and $p_2^+$ are large, while $q^-$, $p_1^-$, and $p_2^-$ are small. The components are given by
	\begin{equation}
		q^- = \frac{q^2}{2 q^+}, \qquad 
		p_1^- = \frac{p_1^2 + p_\perp^2}{2 p_1^+}, \qquad 
		p_2^- = \frac{p_2^2 + p_\perp^2}{2 p_2^+}.
	\end{equation}
	We introduce a scale $\lambda^2 \sim p_1^2 \sim p_2^2$ with $\lambda \ll \bar{Q} = q^+$, where $\bar{Q}$ is large but fixed in the $n^+$ direction. In this scenario, $q_0$ and $q_3$ are large but fixed, while $q^- = q_0 - q_3$ becomes arbitrarily small. The limit $q^2 \to 0$ is then taken at fixed $\bar{Q}$, with $q^- \to 0$, so that any term proportional to $q^+ q^-$ can be neglected. \\
	To extract the residue of a tensor structure with open indices, we project onto the light-cone basis. If a component survives this projection, it indicates the presence of a massless $1/q^2$ pole in the residue along the light-cone. The actual behavior of the residue depends on the kinematic configuration of the external momenta and the fermion masses, and is determined by a one-loop computation. \\
	For the $AVV$ and $TTJ_5$ correlators, as discussed above, the anomaly form factor contributes with a nonzero residue in the limit $q^2 \to 0$ only when all other invariants vanish. In both cases, the residues match the anomaly coefficients. By contrast, the $TJJ$ correlator behaves differently: the particle pole does not coincide with the anomaly, because an additional pole at zero momentum contributes in the massless, on-shell limit. We will clarify this point in what follows. \\
	In particular, if any external momentum is off-shell or if the internal fermion mass is nonzero, the residue of the $TJJ$ correlator vanishes:
	\begin{equation}
		\lim_{s \to 0} \, q^2 \braket{T^{\mu\nu}(q) J^{\alpha a}(p_1) J^{\beta b}(p_2)} = 0 \qquad  ( \textrm{for } s_1\neq 0 \, \textrm{or}\, s_2\neq 0 \,\textrm{or} \, m\neq 0).
	\end{equation}
	This mirrors the behavior found in the chiral anomaly case. However, in the conformal case the interaction exhibits a massless pole only in the light-cone configuration
	\begin{equation}
		\label{mod}
		\lim_{s \to 0} \, q^2 \braket{T^{\mu\nu}(q) J^{\alpha a}(p_1) J^{\beta b}(p_2)} \neq 0 \qquad  (  \textrm{for } s_1=0, s_2=0, m= 0).
	\end{equation}
	The nonzero residue arises upon contracting the correlator with the polarization vectors of the external spin-1 states and imposing transversality. This setup corresponds to a light-cone process, namely a virtual graviton decaying into two on-shell photons or gluons. \\
	As discussed above, in the case of the chiral anomalies only the longitudinal tensor structures of the 
	$AVV$ and $TTJ_5$ correlators survive, where the longitudinal direction is defined by the momentum of the chiral current,  
	showing that the interaction is anomaly dominated. In the case of the $TJJ$, on the other hand, one can verify the validity of 
	\eqref{mod}, but two tensor structures survive: one related to the anomaly ($FF$) and a second, traceless one. Specifically, we obtain
	\beq\label{tjjpolelim}
	\lim_{s \to 0} \, q^2 \braket{T^{\mu\nu}(q) J^{\alpha a}(p_1) J^{\beta b}(p_2)} 
	= -\frac{g_s^2 \, \delta^{ab}}{48 \pi^2} \left(\frac{2}{3} n_f - \frac{11}{3} C_A \right) \tilde\phi_1^{\mu\nu\alpha\beta} 
	- \frac{g_s^2 \, \delta^{ab}}{288 \pi^2} \left(n_f - C_A \right) \tilde\phi_2^{\mu\nu\alpha\beta}.
	\eeq
	The analysis in the soft limit is similar, preserving the same structures. 
	This result contains two contributions: the first, proportional to $\tilde\phi_1^{\mu\nu\alpha\beta}$, is linked to the conformal anomaly, while the second, associated with $\tilde\phi_2^{\mu\nu\alpha\beta}$, corresponds to an additional pole and is related to a traceless tensor structure. 
	The two tensor structures are
	\begin{align}
		\tilde\phi_1^{\, \mu \nu \alpha \beta} (p_1,p_2) &= 
		\left(s \, g^{\mu\nu} - q^{\mu} q^{\nu}\right) \, u^{\alpha \beta} (p_1,p_2),
		\label{widetilde1ù}\\
		\tilde\phi_2^{\, \mu \nu \alpha \beta} (p_1,p_2) &= 
		- 2 \, u^{\alpha \beta} (p_1,p_2) \left[ s \, g^{\mu \nu} + 2 \left(p_1^\mu p_1^\nu + p_2^\mu p_2^\nu \right)
		- 4 \left(p_1^\mu p_2^\nu + p_2^\mu p_1^\nu \right) \right],
		\label{widetilde2ù}
	\end{align}
	from which it is straightforward to see that $\tilde\phi_2^{\mu\nu\alpha\beta}$ is traceless if the two gluons are taken to be transverse. 
	This second traceless structure allows us to identify an additional contribution to the anomaly effective action describing the interaction around the light-cone, of the form
	\begin{align}\label{expoleac}
		S_{\text{extra pole}} = \frac{g_s^2}{72 \pi^2} \left(n_f - C_A \right) \int &d^4x \sqrt{-g} \int d^4x' \sqrt{-g'}\, h_{\mu\nu}(x) \Box^{-1}_{x,x'} 
		\nonumber\\&
		\times \left[ 
		3 (\partial^\mu \partial^\nu F^a_{\alpha\beta}) F^{\alpha\beta a}
		+ \frac{1}{4} \left( g^{\mu\nu} \Box - 4 \partial^\mu \partial^\nu \right) F^a_{\alpha\beta} F^{\alpha\beta a}
		\right]_{x'},
	\end{align}
	while the first tensor structures is associated with the well-known effective action of the form \cite{Armillis:2009pq}
	\bea
	S_{\text{anom}} &=&\frac{1}{3} \, \frac{g^3}{16 \pi^2} \left (  - \frac{11}{3} \, C_A + \frac{2}{3} \, n_f \right)  \, \int d^4 x \, d^4 y \,R^{(1)}(x)\, \square^{-1}(x,y) \, F^a_{\alpha \beta}F^{\alpha \beta a},
	\eea
	that generalizes the expression derived in \cite{Giannotti:2008cv} for QED. $R^{(1)}(x)$ is the linearized scalar curvature.
	Therefore, in order extract the effective action describing the light-cone behaviour of the correlator, we need to combine the contributions of two functionals generating both $\tilde{\phi_1}$ and 
	$\tilde{\phi}_2$ of \eqref{widetilde1ù} and \eqref{widetilde2ù}, which are obtained by varying the general light-cone $(\text{lc})$ action 
	\beq
	\mathcal{S}_{lc}=\mathcal{S}_{\text{anom}} + \mathcal{S}_{\text{extra pole}},
	\eeq
	with respect to the external metric fluctuations 
	$h^{\mu\nu}$ and non-Abelian gauge fields $A^{\alpha a}$ as 
	
	\begin{equation}
		\left\langle T^{\mu_1\nu_1}(x) J^{\mu_2 a}(y) J^{\mu_3 b}(z) \right\rangle_{\text{lc}}
		= \left. \frac{\delta^3 \mathcal{S}_{\text{lc}}}{\delta h_{\mu_1\nu_1}(x) \delta A^a_{\mu_2}(y) \delta A^b_{\mu_3}(z)} \right|_{h=0,\,A=0}, 
	\end{equation}
	followed by a Fourier transform to momentum space 
	\beq
	\Gamma^{\mu_1\nu_1 \mu_2\mu_3 a b }_{lc}=\int d^4 x d^4 y e^{i p_1 x + i p_2 y}\left\langle T^{\mu_1\nu_1}(0) J^{\mu_2 a}(y) J^{\mu_3 b}(z) \right\rangle_{\text{lc}}.
	\eeq
	The effective action obviously satisfies the trace anomaly constraints 
	\bea
	g_{\mu\nu}\,  \Gamma_{lc}^{\mu\nu\alpha\beta a b}(p,q) =  - 2 \, \frac{\b (g)}{g}\, u^{\a \b}(p,q) \delta^{a b},
	\eea
	with the QCD $\b$ function given by \eqref{betagg}. \\
	It should be emphasized that this extra contribution does not correspond to an additional effective energy--momentum tensor that is separately conserved, since $\mathcal{S}_{\text{extra pole}}$ is not diffeomorphism invariant.  
	Recall that the conservation of the stress--energy tensor is equivalent to the requirement of diffeomorphism invariance of the underlying action.  
	In this case, the conservation of the energy--momentum tensor associated with $\mathcal{S}_{\text{extra pole}}$ can be verified only for on-shell vectors, and the absence of diffeomorphism invariance in this action does not contradict the general 
	Noether theorem, which holds true for any 
	off-shell action which is diffeomorphism invariant.  
	This observation aligns with the findings of \cite{Giannotti:2008cv}, which analyzed similar phenomena within the framework of QED. The specific tensor structure associated with $\mathcal{S}_{\text{extra pole}}$ excludes the interpretation of the amplitude as arising from scalar particle exchange. As observed in \cite{Giannotti:2008cv}, the emergence of a pole featuring a nontrivial tensor character can be traced to non-vanishing contributions from terms such as $\partial^\mu \partial^\nu F_{\alpha\beta}$ and $\partial^\mu \partial^\nu (F_{\alpha\beta}F^{\alpha\beta})$, evaluated in the presence of external gauge field backgrounds. These background configurations inherently violate Lorentz symmetry, and the resulting massless excitations 
	are not scalar in nature.  \\
	In summary, our analysis shows that, in restricted light-cone kinematics, the residue at the particle pole of the \(TJJ\) correlator differs from its \(AVV\) and \(TTJ_5\) counterparts~\cite{Coriano:2025ceu}. In general it is not solely tied to the trace anomaly; only in the special case \(n_f = N_c = C_A\) does \(\mathcal{S}_{\text{extra pole}}\) vanish. This suggests that the chiral and gravitational anomalies impose stricter constraints on the relation between the particle-pole residue and light-cone behaviour, making the anomaly--correlator interplay more intricate than in the other two cases.
	
	\section{Comments: the local effective actions for axions and dilatons}
	The use of a local action to describe an asymptotic state raises significant questions, particularly regarding the treatment of axion and dilaton interactions at the phenomenological level. Specifically, if an anomaly is associated with an anomaly pole, and this pole does not necessarily represent an asymptotic state (i.e., a particle pole), one might argue that a local action transforming a pole into an asymptotic state has limited applicability.\\
	Typically, the local interaction is derived by coupling the divergence of an anomaly current— in this case, the dilaton current— to the dilaton field, thus generating the ordinary local $\rho FF$ interaction. Interactions of this form, such as the dilaton case or the $\alpha(x) F\tilde{F}$ form in the axion case, are naturally obtained by applying Ward identities to the underlying anomaly vertex. However, when moving away from the conformal limit, at least in a perturbative framework, these interactions appear not to be consistently motivated.\\
	In the context of confining gauge theories, this issue can be effectively resolved by recognizing the breakdown of the perturbative picture when describing certain anomaly interactions below the nonperturbative confinement scale. For example, in the case of axions, the local action and its interactions are derived through manipulations of the chiral fermion measure in the path integral. These manipulations are equivalent to applying Ward identities, and the off-shell effective action derived from perturbative expansion takes a local form.\\
	The framework we have explored is closely tied to the conformal properties of such vertices, which are clearly broken by nonperturbative effects. A second possibility to endorse a local effective description emerges from the analysis of similar interactions in the presence of a Higgs phase. Both the axion and dilaton cases are expected to be consistent with the results derived in the UV through the analysis of all such anomalies, supported by anomaly matching arguments in both the UV and the IR, only in these two cases.

	\section{Conclusions}
	We have shown that sum rules associated with interactions linked to the conformal anomaly impose crucial constraints on the absorptive parts of the corresponding anomaly form factor. Our findings reveal that, under a wide range of kinematic conditions, the spectral density of this anomaly can exhibit unexpected behaviors, including the emergence of dilaton-like interpolating states in the correlators. Depending on the kinematic configuration, contributions may cancel, or appear as poles or cuts in the complex plane of the stress-energy tensor's momentum. Notably, even when mass effects convert a pole into a cut in the spectral densities, the sum rule remains effectively satisfied due to the emergence of a pole. This behavior reflects the analytic structure of the amplitudes and the intrinsic characteristics of the interactions.\\
	The breaking of conformal symmetry is thus intimately tied to the appearance of massless intermediate states. Combining this analysis with our earlier study of chiral and gravitational anomaly form factors, we observe that in the conformal and on-shell limits, anomaly poles behave as true propagating degrees of freedom. Conversely, the introduction of mass effects and renormalization corrections complicates the spectral structure but preserves the sum rule. These phenomena have been explored through both perturbative methods and momentum-space conformal field theory ($CFT_p$), which decomposes correlation functions into transverse and longitudinal components.\\
	It would be interesting to investigate how the specific constraints on the partonic correlator, as we have explored, manifest at the hadronic level, particularly in the analysis of gravitational form factors for the pion and proton. Another noteworthy aspect of this analysis is its potential implications for general gravitational scatterings in the eikonal limit involving photons and gravitons, which will be explored in forthcoming works.
		\chapter{The Gravitational Form Factor of the Pion in Perturbative QCD with a Dilaton Interaction}\label{chpionff}
\addcontentsline{toc}{section}{Synopsis}
\section*{Synopsis}

This Chapter applies the field-theoretic framework developed in the preceding Chapters to a concrete phenomenological problem: the computation of the pion gravitational form factors (GFFs) at large momentum transfer, with explicit inclusion of the trace anomaly through the insertion of the non-Abelian $TJJ$ correlator into the hard-scattering kernel. The Chapter establishes a direct connection between the QCD conformal anomaly controlled by the beta-function coefficient $\beta_0 = (11C_A - 2n_f)/3$ and observable mechanical properties of the pion, particularly the $D$-term form factor and the trace form factor.

The Chapter begins by reviewing the kinematics of the pion GFFs and their definition through the matrix element of the QCD energy-momentum tensor between pion states,
\begin{equation*}
	\langle \pi(P_2)|\hat{T}^{\mu\nu}|\pi(P_1)\rangle = 2P^\mu P^\nu A_\pi(q^2) + \frac{1}{2}(q^\mu q^\nu - g^{\mu\nu}q^2)D_\pi(q^2),
\end{equation*}
where $A_\pi(t)$ encodes the momentum distribution of the pion's constituents and $D_\pi(t)$ characterizes the internal pressure and shear force distribution. The role of generalized parton distributions (GPDs) in providing experimental access to these form factors through DVCS and related exclusive processes is recalled, and the motivation from the JLab 12 GeV program and the future EIC is emphasized.

The factorization framework is set up in the Breit frame with light-cone kinematics, working at large spacelike momentum transfer $Q^2 = -q^2 \to \infty$ in the chiral limit. The pion is described by its leading-twist distribution amplitude (DA) $\tilde{\phi}_\pi(\alpha)$, expanded in Gegenbauer polynomials with moments $a_n(\mu)$ obeying ERBL evolution. The tree-level hard kernel, obtained from seven Feynman diagrams involving one hard gluon exchange and one EMT insertion, is computed and expressed in terms of the scalar form factors $F_i^{\rm tree}$, which are used to extract $A_\pi$ and $D_\pi$ through the convolution formula
\begin{equation*}
	\langle \pi(P_2)|T^{\mu\nu}|\pi(P_1)\rangle = \frac{if_\pi^2 C_F g_s^2}{N_C} \int_0^1 d\alpha_1 \int_0^1 d\alpha_2\, \tilde{\phi}_\pi(\alpha_1)\, K_\pi^{\mu\nu}(\alpha_1,\alpha_2,q)\, \tilde{\phi}_\pi(\alpha_2).
\end{equation*}
The gauge independence of the tree-level result in the chiral limit is verified explicitly.

The QCD energy-momentum tensor is then reviewed, including its gauge-fixed form with contributions from the field strength, fermion, gauge-fixing, and ghost sectors. The trace anomaly relation in the chiral limit,
\begin{equation*}
	T^\mu_{\;\mu} = \frac{\beta(g_s)}{2g_s} F^a_{\rho\sigma}F^{a\,\rho\sigma} + \text{BRST-exact and equation-of-motion terms},
\end{equation*}
is identified as the source of the anomaly-induced hard-scattering correction, controlled by the QCD $\beta$-function.

The tensor-sector decomposition of the non-Abelian $TJJ$ correlator is reviewed following the $\mathrm{CFT}_p$-inspired framework of the preceding Chapters. The correlator is decomposed into transverse-traceless, longitudinal, and trace components using the projectors $\Pi$, $\pi$, and $\Sigma$. In the quark sector, four form factors $A_i^{(q)ab}$ parameterize the transverse-traceless part, expressed in terms of scalar integrals $B_0$ and $C_0$, and the trace sector is controlled by the anomaly coefficient $\mathcal{A}_q^{\alpha\beta ab} \propto n_f u^{\alpha\beta}\delta^{ab}$. In the gluon sector, Slavnov-Taylor identities replace ordinary Ward identities and generate additional longitudinal contributions $\langle t^{\mu\nu} j^a j^b_{\rm loc}\rangle_g$ as well as extra trace terms $\mathcal{B}_g^{\alpha\beta ab}$ proportional to the gluon equations of motion. The gluon anomaly contribution $\mathcal{A}_g^{\alpha\beta ab} \propto (11/3)C_A u^{\alpha\beta}\delta^{ab}$ combines with the quark sector to reproduce the full QCD trace anomaly $(11C_A - 2n_f)/3$.

The insertion of the $TJJ$ correlator into the hard-scattering factorization formula is then carried out explicitly. The hard kernel receives a one-loop correction from the $TJJ$ vertex, and explicit expressions for the five scalar coefficients $F_i^{TJJ}$ are derived. A central observation is that the anomaly coefficient
\begin{equation*}
	\mathcal{A} = \frac{g_s^2}{16\pi^2}\beta_0
\end{equation*}
appears directly in $F_5^{TJJ}$, the coefficient controlling the scalar ($g^{\mu\nu}$) component of the hard kernel, which projects onto the $D$-term. In contrast, the isolated anomaly contribution cancels in the $A_\pi$ projection due to the tensor structure of the spin-zero EMT matrix element in the chiral limit: taking the trace of the spin-zero decomposition gives $g_{\mu\nu}\langle \pi|T^{\mu\nu}|\pi\rangle \simeq (Q^2/2)[A_\pi + 3D_\pi]$, showing that the scalar sector affects the combination $A_\pi + 3D_\pi$ directly. The full $TJJ$ insertion still modifies $A_\pi$ through non-anomalous tensor structures. The gauge independence of the final physical form factors is verified: contributions from the Slavnov-Taylor-constrained sectors of the gluon decomposition, including $\mathcal{B}_g$ and the local longitudinal terms, either cancel algebraically upon contraction with the pion state projectors or vanish upon integration over the momentum fractions, confirming that the extracted $D_\pi$ isolates a genuine gauge-invariant physical quantity.

To extend the analysis beyond the strict collinear approximation and to regularize endpoint-enhanced contributions, Sudakov resummation is incorporated through the modified perturbative approach of Li and Sterman. The pion distribution amplitude is promoted to a transverse-momentum dependent wave function in impact-parameter space,
\begin{equation*}
	\tilde{\Psi}_\pi(\alpha,b,Q,\mu) = \phi_\pi(\alpha,\mu_0)\exp[-S(\alpha,b,Q,\mu)]\exp\left[-\beta_\pi^2\frac{m_q^2}{\alpha(1-\alpha)}\right]\exp\left[-\frac{\alpha(1-\alpha)b^2}{4\beta_\pi^2}\right],
\end{equation*}
where $S(\alpha,b,Q,\mu)$ is the Sudakov exponent resumming double and single logarithms to next-to-leading logarithmic accuracy, and the Gaussian profile in $b$ encodes the intrinsic transverse size of the pion parameterized by $\beta_\pi$ and a constituent quark mass $m_q$. The Sudakov factor exponentially suppresses both large transverse separations $b \gtrsim 1/\Lambda_{\rm QCD}$ and endpoint configurations $\alpha \to 0,1$, ensuring that the convolution integral is dominated by the perturbatively calculable short-distance region. The factorization formula is then promoted to a threefold convolution in $\alpha_1$, $\alpha_2$, and $b$, with the hard kernel Fourier-transformed to impact-parameter space. The transverse momentum dependence is retained exclusively in the exchanged gluon propagator, allowing the Fourier transform to be performed analytically in terms of modified Bessel functions.

The numerical analysis is carried out using phenomenologically motivated parameters consistent with the electromagnetic pion form factor: $a_2(\mu_0) = 0.2$ at $\mu_0 = 1$~GeV, $m_q = 0.33$~GeV, $\beta_\pi = 0.24$~GeV$^{-1}$, and $\Lambda_{\rm QCD} = 0.25$~GeV. Results are presented for the three observables $A_\pi(Q^2)$, $D_\pi(Q^2)$, and the trace form factor $\Theta_\pi(Q^2) = g_{\mu\nu}\langle\pi|T^{\mu\nu}|\pi\rangle$, with comparison to lattice QCD data from Hackett \textit{et al.} The key findings are as follows. In $A_\pi(Q^2)$, the isolated anomaly cancels but the full $TJJ$ insertion lowers the leading Sudakov curve at small $Q^2$, improving agreement with the lattice data in the low-momentum window. In $D_\pi(Q^2)$, the anomaly gives an important contribution through the scalar sector, generating a visible correction at moderate $Q^2$; a complete order-$\alpha_s^2$ description would require additional non-anomalous radiative terms. In $\Theta_\pi(Q^2)$, the trace form factor most directly isolates the anomaly contribution, which dominates the $TJJ$ correction over the entire momentum-transfer range considered. 

The Chapter concludes by noting that the analysis establishes a concrete perturbative realization of the conformal anomaly inside the hard-scattering description of a hadronic exclusive process, and that the qualitative improvement in the comparison with lattice data after including the anomaly-related corrections, even in a kinematic region not strictly within the asymptotic perturbative regime, supports the physical relevance of the trace anomaly as a structural ingredient in the pion's gravitational form factors. Extensions to the proton GFFs, to graviton scattering in the eikonal limit, and to a complete order-$\alpha_s^2$ hard kernel incorporating all non-anomalous radiative contributions are identified as natural directions for future work.

The work presented in this Chapter is based primarily on, and adapted from, two preprints: a shorter Letter \cite{Coriano:2026gyl}, co-authored with C.~Corianò and H.~Li, which outlines the primary results, and a more comprehensive, complete work co-authored with C.~Corianò, H.~Li, and L.~Torcellini containing the full theoretical framework and detailed derivations \cite{Coriano:2026ssa}.

\section{Introduction}

Gravitational form factors (GFFs) provide a fundamental characterization of hadron structure, encoding the distributions of energy, momentum, pressure, shear forces, and angular momentum carried by quarks and gluons inside hadrons. Although they are defined through matrix elements of the QCD energy--momentum tensor (EMT) and therefore do not correspond to direct gravitational probes, they furnish one of the most comprehensive field-theoretic descriptions of the internal dynamics of hadrons. In particular, they offer access to the mechanical structure of strongly interacting bound states, complementing the information obtained from electromagnetic and axial form factors and extending the partonic description of hadrons beyond charge and spin observables.

Experimentally, GFFs can be accessed through hard exclusive scattering processes. A primary channel in this context is deeply virtual Compton scattering (DVCS), in which a high--energy electron scatters off a hadron by exchanging a highly virtual photon that subsequently converts into a real photon,
\begin{equation}
	eN \to e^\prime N^\prime \gamma .
	\label{chza:eq:DVCSprocess}
\end{equation}
Over the past decade, the central role of GFFs in elucidating hadron structure has motivated an extensive experimental program. First extractions of the proton quark~\cite{Burkert:2018bqq} and gluon~\cite{Duran:2022xag} GFFs have been achieved through DVCS measurements and exclusive \(J/\psi\) photoproduction, respectively. In comparison, the determination of pion GFFs remains more challenging, although initial phenomenological constraints on the pion quark GFFs have been obtained from high--precision data collected at the Belle experiment at KEKB~\cite{Belle:2015oin,Savinov:2013hda,Kumano:2017lhr}. Significant progress is expected in the near future from ongoing and planned experimental programs, including the upgraded \(12\;\mathrm{GeV}\) Jefferson Lab~\cite{JeffersonLabHallA:2022pnx,CLAS:2022syx} and the forthcoming Electron--Ion Collider~\cite{AbdulKhalek:2021gbh}, which will substantially extend the kinematic reach and precision of DVCS and related exclusive measurements. These developments provide strong motivation for refined theoretical analyses of GFFs, especially in the hard-scattering regime where QCD factorization and symmetry-based methods can be applied systematically. In this context, related neutral-current processes can also be investigated and further enlarge the phenomenological reach of the framework~\cite{Amore:2004ng}.

Perturbative studies of hadronic GFFs at large momentum transfer have also been developed directly within QCD factorization.  In particular, the gluon GFFs of the pion and the nucleon were analyzed in~\cite{Tong:2021ctu}, and a broader perturbative calculation of GFFs at large momentum transfer was presented in~\cite{Tong:2022zax}.  These works provide useful benchmarks for the large-\(Q^2\) behavior of quark and gluon contributions and form part of the theoretical background for the factorized treatment considered below.

DVCS occupies a special position in the kinematic landscape of strong-interaction processes. At moderate momentum transfer, it interpolates between a soft regime, where descriptions based on QCD sum rules are applicable and the reaction is largely governed by the Feynman mechanism, namely by the overlap of initial- and final-state hadronic wave functions~\cite{Coriano:1993mr}, and a genuinely hard regime at larger virtualities, where factorization theorems isolate short-distance kernels from long-distance hadronic matrix elements. This dual character makes DVCS an especially powerful probe of hadron structure across different distance scales, allowing one to connect nonperturbative physics with perturbatively calculable amplitudes in a controlled manner.

A systematic and model-independent connection between DVCS observables and EMT form factors is provided by generalized parton distributions (GPDs)~\cite{Ji:1996ek,Radyushkin:1996nd,Radyushkin:1996ru,Ji:1996nm,Collins:1996fb,Radyushkin:1997ki,Vanderhaeghen:1998uc}. GPDs parameterize nonforward light-cone matrix elements of quark and gluon operators and thus interpolate between ordinary parton distributions and elastic form factors. Their Mellin moments encode local operator matrix elements, among which the EMT plays a central role. In the nucleon case, the second Mellin moments of the unpolarized GPDs \(H^a(x,\xi,t)\) and \(E^a(x,\xi,t)\) are directly related to the EMT form factors \(A^a(t)\), \(B^a(t)\), and \(D^a(t)\),
\begin{align}
	\int_{-1}^{1} dx \, x\, H^a(x,\xi,t)
	&=
	A^a(t)+\xi^2 D^a(t),
	\\
	\int_{-1}^{1} dx \, x\, E^a(x,\xi,t)
	&=
	B^a(t)-\xi^2 D^a(t).
	\label{chza:Eq:GPD-Mellin}
\end{align}
Here \(t=q^2\) denotes the invariant momentum transfer and \(\xi\) is the longitudinal momentum asymmetry, or skewness, parameter. Physically, the GPDs \(H\) and \(E\) describe the amplitude for removing from the nucleon a parton carrying a light-cone momentum fraction \(x-\xi\) of the average nucleon momentum \(P\) and reinserting it with fraction \(x+\xi\), while the hadron absorbs the momentum transfer \(q\). Through their Mellin moments, GPDs therefore provide direct access to the GFFs and to the internal mechanical and angular-momentum structure of hadrons.

While much of the phenomenological discussion of GFFs has been developed in terms of GPDs and related exclusive observables, the perturbative analysis of EMT matrix elements at large momentum transfer raises a more specific field-theoretic question: how the EMT couples to colored constituents in QCD in the hard regime. This problem naturally leads to the study of the non-Abelian \(TJJ\) correlator, which governs the coupling of the EMT to two gauge currents, or equivalently to two external gluon fields in the background-field formulation. The importance of this correlator is twofold. First, it controls the short-distance structure entering hard exclusive amplitudes in which the EMT couples to partonic constituents. Second, it encodes the quantum breaking of scale invariance through the trace anomaly and thus provides the perturbative origin of anomaly-mediated contributions.

In the conformal limit, the trace of the EMT vanishes at the classical level in massless QCD, but acquires a nonvanishing expectation value after renormalization, proportional to the QCD beta function. In momentum space this effect is captured by a specific form factor whose spectral density satisfies a nontrivial sum rule. In suitable kinematic limits, and in particular for massless quarks, this structure reduces to an effective pole contribution that can be interpreted as the exchange of a scalar state induced by the anomaly. This effective scalar exchange, often described as dilaton-like, provides a natural bridge between the perturbative description of the trace anomaly and the factorized treatment of hadronic GFFs.

A systematic analysis of this mechanism requires a tensor decomposition of the non-Abelian \(TJJ\) correlator into transverse--traceless, longitudinal, and trace sectors. While analogous decompositions are standard in conformal field theory, their implementation in gauge theories is more subtle because gauge fixing and BRST symmetry modify the off-shell structure of the correlator. As a consequence, the quark and gluon sectors must be treated separately at one loop. The quark contribution is structurally close to the Abelian case and obeys ordinary vector Ward identities on the external gluon lines. The gluon sector, by contrast, is constrained by Slavnov--Taylor identities and contains additional local terms that are absent in a purely conformal treatment. The anomaly emerges only after these distinct sectors are combined consistently.

%In this work we investigate the role of these anomaly-induced contributions in the perturbative description of the pion GFFs at large momentum transfer. Our analysis combines the hard-scattering framework for pion EMT matrix elements with the structure of the non-Abelian \(TJJ\) correlator in QCD, allowing us to isolate the dilaton-like interaction generated by the trace anomaly and to study its contribution to the form factors in the factorized regime. In this sense, the present analysis provides a concrete realization, in the pion case, of the general framework developed in~\cite{Coriano:2024qbr}.

In this work, we investigate the role of these anomaly-induced contributions in the perturbative description of the pion GFFs at large momentum transfer. Specifically, we embed the non-Abelian $TJJ$ correlator, which corrects, at one-loop level, the vertex coupling of the local energy-momentum tensor to two external gluons, into the standard hard-scattering framework for exclusive hadronic processes. By factorizing the pion EMT matrix elements into a convolution of nonperturbative pion distribution amplitudes (DAs) and perturbatively calculable hard-scattering kernels, we isolate the dynamical effects of the trace anomaly. This approach allows us to explicitly track how the anomaly-induced, dilaton-like effective scalar exchange propagates into the physical form factors in the asymptotic, large-$Q^2$ regime. In this sense, the present analysis provides a concrete realization, in the pion case, of the general field-theoretic framework developed in~\cite{Coriano:2024qbr}, establishing a direct link between a fundamental field-theoretic anomaly and observable mechanical properties of a pseudo-Goldstone boson. The remainder of this paper is structured as follows. 

In Section~\ref{chza:aaaasec2}, we define the kinematics of the hard exclusive process and review the definition of the pion GFFs $A_\pi(t)$ and $D_\pi(t)$. After a brief review in Section~\ref{chza:sec:emt_conformal} of the QCD energy-momentum tensor,  Section~\ref{chza:seven} is dedicated to the field-theoretic structure of the off-shell non-Abelian $TJJ$ correlator at one-loop order; here, we detail the separation between the quark and gluon loop contributions and highlight how gauge-fixing terms and Slavnov-Taylor identities affect the longitudinal and trace sectors. In Section~\ref{chza:contjjtra}, we compute the hard-scattering kernels, explicitly incorporating the $TJJ$ anomaly vertex into the pion's partonic subprocesses. 
In Section~\ref{chza:phenomesec1}, we analyze the Sudakov resummation framework for the pion form factor at large momentum transfer. This resummation is essential to handle the soft-gluon divergences that arise in the endpoint regions of the pion DAs, providing a reliable and stable perturbative expansion by incorporating transverse-momentum effects. In Section~\ref{chza:phenomesec2}, we introduce a concrete phenomenological model for the pion wave function based on a Gaussian distribution in transverse momentum. This specific ansatz allows us to construct a realistic nonperturbative input and establish a complete phenomenological framework to quantify our anomaly-induced corrections and evaluate their numerical impact on the GFFs.
Finally, in Section~\ref{chza:numericalsec}, we present the analysis of our results, discussing the numerical impact of the trace anomaly on the $Q^2$ evolution of the pion form factors. The paper also includes three appendices. Appendix~\ref{chza:rules} collects 
the Feynman rules and vertex expressions employed in the diagrammatic 
evaluation of the amplitudes. Appendix~\ref{chza:ff} provides the explicit 
form factors of the off-shell non-Abelian $TJJ$ correlator 
in the transverse-traceless sector, including the full dependence on the 
loop integral coefficients for both the quark and gluon contributions. 
Appendix~\ref{chza:kernelexpb} contains the complete hard-scattering kernel 
expressions in impact-parameter space, covering both the tree-level 
contributions and the corrections induced by the $TJJ$ insertion for 
each GFF.

We find that the trace anomaly introduces a distinct behavior in the trace sector of the form factors, which significantly alters the behavior of the pion $D$-term at large momentum transfer compared to purely traditional, anomaly-free hard-scattering predictions. In particular, the effective scalar exchange induced by the anomaly leaves a unique imprint on the internal pressure and shear force distributions within the pion, demonstrating that quantum scale violation plays an indispensable structural role even in the deeply perturbative, high-energy domain.

\section{The pion form factor in the limit of large momentum transfer}\label{chza:aaaasec2}

A systematic and controlled description of hadronic form factors within QCD becomes possible in the asymptotic regime of large spacelike momentum transfer,
\begin{equation}
	Q^2 = -q^2 \to \infty, \qquad q^\mu = P_2^\mu - P_1^\mu,
\end{equation}
where short-distance dynamics dominates and long-distance nonperturbative effects can be factorized into universal hadronic quantities. In this regime, the interaction resolves the hadron at partonic distances of order $1/Q$, allowing for a perturbative treatment of the hard scattering subprocess.

As a prototypical example, we consider the GFFs of the pion, defined through the matrix element of the QCD energy-momentum tensor $ T^{\mu\nu}$,
\begin{equation}\label{chza:eq:GFF_definition_improved}
	\langle \pi(P_2)| {T}^{\mu\nu} | \pi(P_1) \rangle
	= 2 P^\mu P^\nu A^\pi(q^2)
	+ \frac{1}{2} \left( q^\mu q^\nu - g^{\mu\nu}q^2 \right) D^\pi(q^2),
\end{equation}
where
\begin{equation}
	P = \frac{P_1 + P_2}{2},
\end{equation}
and $A^\pi(q^2)$ and $D^\pi(q^2)$ encode, respectively, the momentum and mechanical (pressure and shear) distributions inside the pion.

To make the scaling properties manifest, it is convenient to work in the Breit frame, where the momentum transfer is purely spatial. Neglecting the pion mass in the large-$Q^2$ limit, one may choose the kinematics
\begin{equation}
	P_1^\mu = \left( \frac{Q}{2}, 0, 0, -\frac{Q}{2} \right), \qquad 
	P_2^\mu = \left( \frac{Q}{2}, 0, 0, \frac{Q}{2} \right), \qquad 
	q^\mu = (0, 0, 0, Q),
\end{equation}
so that $P_1^2 = P_2^2 \simeq 0$. Introducing light-cone vectors,
\begin{equation}
	n^\mu = \frac{1}{\sqrt{2}}(1,0,0,-1), \qquad 
	\bar n^\mu = \frac{1}{\sqrt{2}}(1,0,0,1), \qquad 
	n^2 = \bar n^2 = 0, \qquad n \cdot \bar n = 1,
\end{equation}
the external momenta can be written as
\begin{equation}
	P_1^\mu = \frac{Q}{2} n^\mu, \qquad 
	P_2^\mu = \frac{Q}{2} \bar n^\mu,
	\label{chza:eq:lightcone_p_improved}
\end{equation}
making explicit the collinear nature of the incoming and outgoing states.

In this frame, the pion is probed in a highly boosted configuration, and its dominant Fock component consists of a collinear quark--antiquark pair carrying longitudinal momentum fractions of order unity. Denoting by $k_1, k_2$ ($k_1', k_2'$) the parton momenta in the initial (final) state, one has the scaling behavior
\begin{equation}
	k_i^\mu \sim \alpha_i P_1^\mu + k_{i\perp}^\mu, \qquad 
	k_i^{\prime \mu} \sim \bar\alpha_i P_2^\mu + k_{i\perp}^\mu,
\end{equation}
with
\begin{equation}
	\alpha_i \sim \mathcal{O}(1), \qquad 
	|k_\perp| \ll Q.
\end{equation}
The transverse momenta remain of order $\Lambda_{\mathrm{QCD}}$ and are therefore power suppressed in the hard scattering subprocess.

This hierarchy of scales underlies the factorization of the matrix element into a short-distance kernel and universal nonperturbative DAs. At asymptotically large $Q^2$, higher Fock components involving additional soft gluons or sea quarks are suppressed by powers of $\Lambda_{\mathrm{QCD}}/Q$, and the dominant contribution arises from the valence quark--antiquark configuration.

Physically, the insertion of the energy-momentum tensor corresponds to a localized interaction occurring over a time scale $\Delta t \sim 1/Q$. During this short interval, the partons behave as quasi-free and collinear, and the graviton probes a frozen configuration of the pion. This justifies the use of perturbative QCD for the hard subprocess, while the long-distance structure is encoded in the pion DA.

Within this framework, the GFF can be computed using the hard-scattering mechanism. The momentum transfer injected by the graviton must be redistributed among the constituents through the exchange of a hard gluon carrying virtuality of order $Q^2$, ensuring the recombination of the partons into the final pion state. This mechanism is directly analogous to the perturbative treatment of the electromagnetic form factor, with the crucial difference that the energy-momentum tensor probes the full energy flow inside the hadron.

Neglecting transverse momenta at leading power, one adopts the collinear approximation
\begin{equation}
	k_1 = \alpha_1 P_1, \qquad k'_1 = \bar{\alpha}_1 P_2, \qquad
	k_2 = \alpha_2 P_1, \qquad k'_2 = \bar{\alpha}_2 P_2,
	\label{chza:eq:collinear_approx_improved}
\end{equation}
with $\bar{\alpha}_i = 1 - \alpha_i$. The pion states can then be expanded in terms of valence Fock components as
\begin{align}
	\label{chza:wf}
	|\pi(P_1)\rangle &= \int_0^1 d\alpha_1\, \tilde{\phi}_\pi(\alpha_1)\,
	| u(\alpha_1 P_1)\, \bar{d}(\bar{\alpha}_1 P_1) \rangle, \\
	\langle \pi(P_2)| &= \int_0^1 d\alpha_2\, \tilde{\phi}_\pi(\alpha_2)\,
	\langle u(\alpha_2 P_2)\, \bar{d}(\bar{\alpha}_2 P_2) |,
\end{align}
The light-cone momentum fractions \(\alpha_{1,2}\in[0,1]\) label the momentum carried by the quark in each pion DA.
where $\tilde{\phi}_\pi(\alpha)$ is the leading-twist pion DA, normalized as
\begin{equation}
	\int_0^1 d\alpha\, \tilde{\phi}_\pi(\alpha) = 1.
\end{equation}

The matrix element can then be expressed, within collinear factorization, as a convolution of the DAs with a perturbatively calculable hard kernel,
\begin{equation}
	\langle \pi(P_2)| {T}^{\mu\nu} | \pi(P_1) \rangle
	=  \frac{i\,f_\pi^2 C_F g_s^2}{N_C}\int_0^1 d\alpha_1 \int_0^1 d\alpha_2 \,
	\tilde{\phi}_\pi(\alpha_1)\,
	T_H^{\mu\nu}(\alpha_1,\alpha_2,Q)\,
	\tilde{\phi}_\pi(\alpha_2).
	\label{chza:eq:factorization_master}
\end{equation}

At leading order in $\alpha_s(Q)$, the hard kernel $T_H^{\mu\nu}$ is determined by the tree-level diagrams involving a single hard gluon exchange. The perturbative expansion is controlled by the running coupling evaluated at the scale $Q$, which is small in the asymptotic regime.

The explicit computation proceeds by inserting the QCD interaction vertices and performing Wick contractions of quark and gluon fields. The resulting amplitude can be reduced to a set of tensor structures consistent with Lorentz symmetry,
\begin{equation}
	T_H^{\mu\nu} =
	F_1 P_1^\mu P_2^\nu
	+ F_2 P_2^\mu P_1^\nu
	+ F_3 P_1^\mu P_1^\nu
	+ F_4 P_2^\mu P_2^\nu
	+ F_5 g^{\mu\nu},
\end{equation}
where the scalar functions $F_i(\alpha_1,\alpha_2,Q^2)$ are obtained from the evaluation of the diagrams.

Matching this decomposition onto Eq.~\eqref{chza:eq:GFF_definition_improved} allows one to extract the GFFs as convolution integrals over the pion DA,
\begin{align}
	A_\pi(q^2) &= \int_0^1 d\alpha_1 \int_0^1 d\alpha_2\,
	\tilde{\phi}_\pi(\alpha_1)\,
	\mathcal{A}(\alpha_1,\alpha_2,Q^2)\,
	\tilde{\phi}_\pi(\alpha_2), \\
	D_\pi(q^2) &= \int_0^1 d\alpha_1 \int_0^1 d\alpha_2\,
	\tilde{\phi}_\pi(\alpha_1)\,
	\mathcal{D}(\alpha_1,\alpha_2,Q^2)\,
	\tilde{\phi}_\pi(\alpha_2),
\end{align}
where $\mathcal{A}$ and $\mathcal{D}$ are determined by appropriate linear combinations of the $F_i$.

This result establishes the leading-power factorization of the pion GFFs at large momentum transfer, showing explicitly that their asymptotic behavior is governed by short-distance QCD dynamics convoluted with universal nonperturbative DAs.

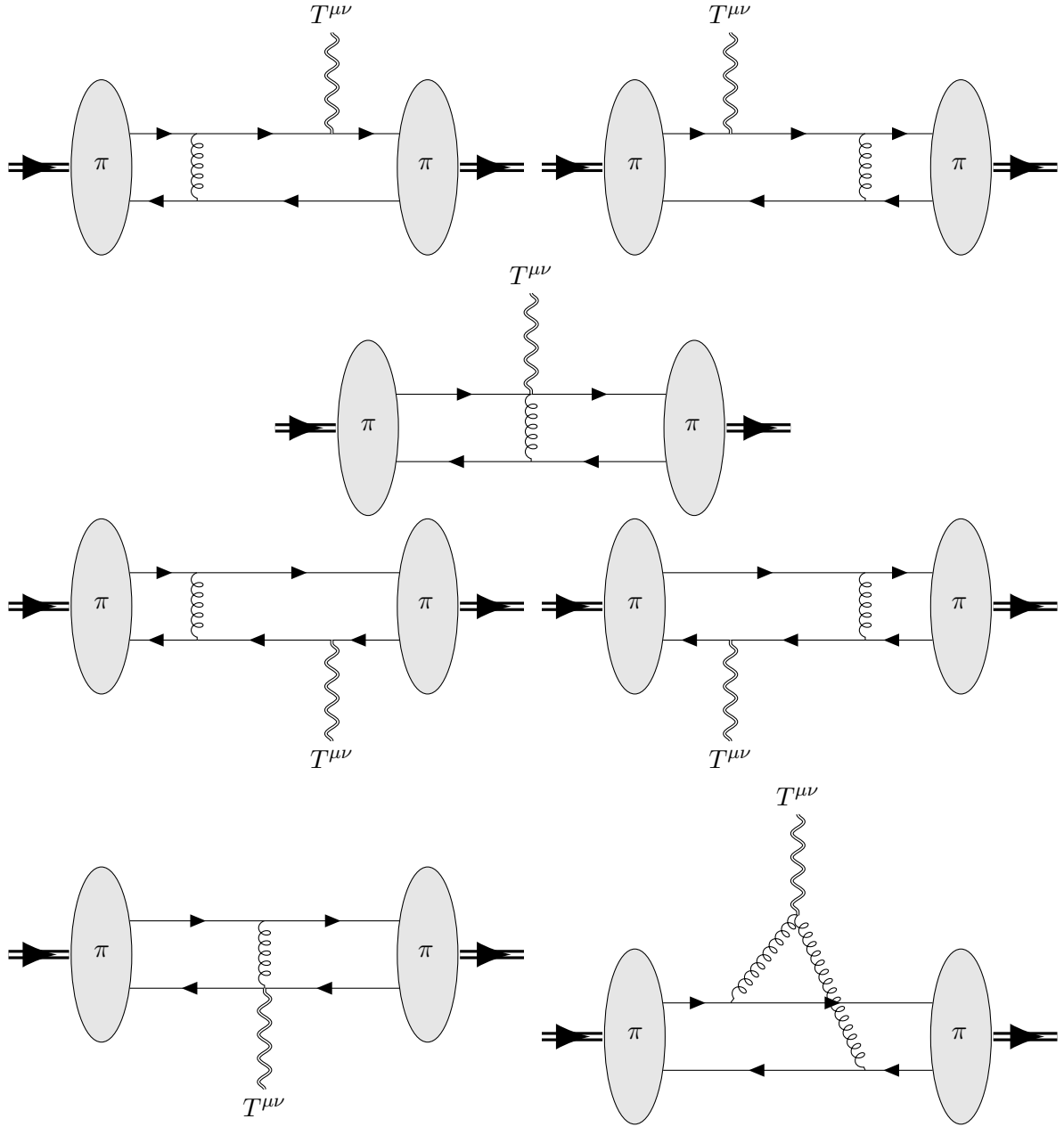
\begin{figure}
	\centering
	\begin{tikzpicture}
		\begin{feynman}
			\vertex (i1);
			\vertex[right=1cm of i1] (a1);
			\vertex[right=2cm of a1] (b1);
			\vertex[right=3cm of a1] (o1) ;
			\vertex[below=1cm of i1] (i2) ;
			\vertex[right=1cm of i2] (a2);
			\vertex[right=3cm of a2] (o2) ;
			
			%\vertex[below=2cm of i3] (t1);
			\vertex[above=1cm of a1] (t2);
			\vertex[above=2cm of a2] (t3);
			
			%\vertex[above right =1.3cm of t2] (t1);
			\vertex[above=1.5cm of b1] (i3) {$T^{\mu\nu}$};
			
			\diagram* { 
				(i1)  -- [fermion] (a1) -- [fermion] (b1)   --[fermion] (o1),
				
				(i2)  -- [anti fermion] (a2)  --[anti fermion] (o2),

				(i3)  -- [graviton] (b1),
				(a1)-- [gluon] (a2)
			};
			
			\node[] at (-0.35,-0.45) {$\pi\ $};
			\node[] at (4.46,-0.45) {$\pi\ $};

			\filldraw[fill=black, fill opacity=0.1] (-0.42,-0.5) ellipse (.45cm and 1.3cm);
			\filldraw[fill=black, fill opacity=0.1] (4.42 ,-0.5) ellipse (.45cm and 1.3cm);
			
			\draw[double, double distance=1.5,line width=1.2, postaction={decorate, decoration={
					markings,% switch on markings
					mark=at position 0.8 with  {\arrow{ latex} } } } ] (-1.8,-0.5) -- node [above] {} (-0.9,-0.5);
			
			\draw[double, double distance=1.5,line width=1.2, postaction={decorate, decoration={
					markings,% switch on markings
					mark=at position 0.8 with  {\arrow{ latex} } } } ] (4.9,-0.5) -- node [above] {} (5.85,-0.5);
		\end{feynman}
	\end{tikzpicture}
	\begin{tikzpicture}
		\begin{feynman}
			\vertex (i1);
			\vertex[right=1cm of i1] (a1);
			\vertex[right=2cm of a1] (b1);
			\vertex[right=3cm of a1] (o1) ;
			\vertex[below=1cm of i1] (i2) ;
			\vertex[right=3cm of i2] (a2);
			\vertex[right=1cm of a2] (o2) ;
			
			%\vertex[below=2cm of i3] (t1);
			\vertex[above=1cm of a1] (t2);
			\vertex[above=2cm of a2] (t3);
			
			%\vertex[above right =1.3cm of t2] (t1);
			\vertex[above=1.5cm of a1] (i3) {$T^{\mu\nu}$};
			
			\diagram* { 
				(i1)  -- [fermion] (a1) -- [fermion] (b1)   --[fermion] (o1),
				
				(i2)  -- [anti fermion] (a2)  --[anti fermion] (o2),

				(i3)  -- [graviton] (a1),
				(b1)-- [gluon] (a2)
			};
			
			\node[] at (-0.35,-0.45) {$\pi\ $};
			\node[] at (4.46,-0.45) {$\pi\ $};

			\filldraw[fill=black, fill opacity=0.1] (-0.42,-0.5) ellipse (.45cm and 1.3cm);
			\filldraw[fill=black, fill opacity=0.1] (4.42 ,-0.5) ellipse (.45cm and 1.3cm);
			
			\draw[double, double distance=1.5,line width=1.2, postaction={decorate, decoration={
					markings,% switch on markings
					mark=at position 0.8 with  {\arrow{ latex} } } } ] (-1.8,-0.5) -- node [above] {} (-0.9,-0.5);
			
			\draw[double, double distance=1.5,line width=1.2, postaction={decorate, decoration={
					markings,% switch on markings
					mark=at position 0.8 with  {\arrow{ latex} } } } ] (4.9,-0.5) -- node [above] {} (5.85,-0.5);
		\end{feynman}
	\end{tikzpicture}
	\begin{tikzpicture}
		\begin{feynman}
			\vertex (i1);
			\vertex[right=2cm of i1] (a1);
			\vertex[right=2cm of a1] (b1);
			\vertex[right=3cm of a1] (o1) ;
			\vertex[below=1cm of i1] (i2) ;
			\vertex[right=2cm of i2] (a2);
			\vertex[right=2cm of a2] (o2) ;
			
			%\vertex[below=2cm of i3] (t1);
			\vertex[above=1cm of a1] (t2);
			\vertex[above=2cm of a2] (t3);
			
			%\vertex[above right =1.3cm of t2] (t1);
			\vertex[above=1.5cm of a1] (i3) {$T^{\mu\nu}$};
			
			\diagram* { 
				(i1)  -- [fermion] (a1) -- [fermion] (b1)  ,
				
				(i2)  -- [anti fermion] (a2)  --[anti fermion] (o2),

				(i3)  -- [graviton] (a1),
				(a1)-- [gluon] (a2)
			};
			
			\node[] at (-0.35,-0.45) {$\pi\ $};
			\node[] at (4.46,-0.45) {$\pi\ $};

			\filldraw[fill=black, fill opacity=0.1] (-0.42,-0.5) ellipse (.45cm and 1.3cm);
			\filldraw[fill=black, fill opacity=0.1] (4.42 ,-0.5) ellipse (.45cm and 1.3cm);
			
			\draw[double, double distance=1.5,line width=1.2, postaction={decorate, decoration={
					markings,% switch on markings
					mark=at position 0.8 with  {\arrow{ latex} } } } ] (-1.8,-0.5) -- node [above] {} (-0.9,-0.5);
			
			\draw[double, double distance=1.5,line width=1.2, postaction={decorate, decoration={
					markings,% switch on markings
					mark=at position 0.8 with  {\arrow{ latex} } } } ] (4.9,-0.5) -- node [above] {} (5.85,-0.5);
		\end{feynman}
	\end{tikzpicture}

	\begin{tikzpicture}
		\begin{feynman}
			\vertex (i1);
			\vertex[right=1cm of i1] (a1);
			\vertex[right=1cm of a2] (b2);
			\vertex[right=3cm of a1] (o1) ;
			\vertex[below=1cm of i1] (i2) ;
			\vertex[right=1cm of i2] (a2);
			\vertex[right=3cm of a2] (o2) ;
			
			%\vertex[below=2cm of i3] (t1);
			\vertex[above=1cm of a1] (t2);
			\vertex[above=2cm of a2] (t3);
			
			%\vertex[above right =1.3cm of t2] (t1);
			\vertex[below=1.5cm of b2] (i3) {$T^{\mu\nu}$};
			
			\diagram* { 
				(i1)  -- [fermion] (a1)    --[fermion] (o1),
				
				(i2)  -- [anti fermion] (a2) -- [anti fermion] (b2) --[anti fermion] (o2),

				(i3)  -- [graviton] (b2),
				(a1)-- [gluon] (a2)
			};
			
			\node[] at (-0.35,-0.45) {$\pi\ $};
			\node[] at (4.46,-0.45) {$\pi\ $};

			\filldraw[fill=black, fill opacity=0.1] (-0.42,-0.5) ellipse (.45cm and 1.3cm);
			\filldraw[fill=black, fill opacity=0.1] (4.42 ,-0.5) ellipse (.45cm and 1.3cm);
			
			\draw[double, double distance=1.5,line width=1.2, postaction={decorate, decoration={
					markings,% switch on markings
					mark=at position 0.8 with  {\arrow{ latex} } } } ] (-1.8,-0.5) -- node [above] {} (-0.9,-0.5);
			
			\draw[double, double distance=1.5,line width=1.2, postaction={decorate, decoration={
					markings,% switch on markings
					mark=at position 0.8 with  {\arrow{ latex} } } } ] (4.9,-0.5) -- node [above] {} (5.85,-0.5);
		\end{feynman}
	\end{tikzpicture}
	\begin{tikzpicture}
		\begin{feynman}
			\vertex (i1);
			\vertex[right=3cm of i1] (a1);
			\vertex[right=2cm of a2] (b2);
			\vertex[right=1cm of a1] (o1) ;
			\vertex[below=1cm of i1] (i2) ;
			\vertex[right=1cm of i2] (a2);
			\vertex[right=3cm of a2] (o2) ;
			
			%\vertex[below=2cm of i3] (t1);
			\vertex[above=1cm of a1] (t2);
			\vertex[above=2cm of a2] (t3);
			
			%\vertex[above right =1.3cm of t2] (t1);
			\vertex[below=1.5cm of a2] (i3) {$T^{\mu\nu}$};
			
			\diagram* { 
				(i1)  -- [fermion] (a1)    --[fermion] (o1),
				
				(i2)  -- [anti fermion] (a2) -- [anti fermion] (b2) --[anti fermion] (o2),

				(i3)  -- [graviton] (a2),
				(a1)-- [gluon] (b2)
			};
			
			\node[] at (-0.35,-0.45) {$\pi\ $};
			\node[] at (4.46,-0.45) {$\pi\ $};

			\filldraw[fill=black, fill opacity=0.1] (-0.42,-0.5) ellipse (.45cm and 1.3cm);
			\filldraw[fill=black, fill opacity=0.1] (4.42 ,-0.5) ellipse (.45cm and 1.3cm);
			
			\draw[double, double distance=1.5,line width=1.2, postaction={decorate, decoration={
					markings,% switch on markings
					mark=at position 0.8 with  {\arrow{ latex} } } } ] (-1.8,-0.5) -- node [above] {} (-0.9,-0.5);
			
			\draw[double, double distance=1.5,line width=1.2, postaction={decorate, decoration={
					markings,% switch on markings
					mark=at position 0.8 with  {\arrow{ latex} } } } ] (4.9,-0.5) -- node [above] {} (5.85,-0.5);
		\end{feynman}
	\end{tikzpicture}
	\begin{tikzpicture}
		\begin{feynman}
			\vertex (i1);
			\vertex[right=2cm of i1] (a1);
			\vertex[right=2cm of a1] (b1);
			\vertex[right=3cm of a1] (o1) ;
			\vertex[below=1cm of i1] (i2) ;
			\vertex[right=2cm of i2] (a2);
			\vertex[right=2cm of a2] (o2) ;
			
			%\vertex[below=2cm of i3] (t1);
			\vertex[above=1cm of a1] (t2);
			\vertex[above=2cm of a2] (t3);
			
			%\vertex[above right =1.3cm of t2] (t1);
			\vertex[below=1.5cm of a2] (i3) {$T^{\mu\nu}$};
			
			\diagram* { 
				(i1)  -- [fermion] (a1) -- [fermion] (b1)  ,
				
				(i2)  -- [anti fermion] (a2)  --[anti fermion] (o2),

				(i3)  -- [graviton] (a2),
				(a1)-- [gluon] (a2)
			};
			
			\node[] at (-0.35,-0.45) {$\pi\ $};
			\node[] at (4.46,-0.45) {$\pi\ $};

			\filldraw[fill=black, fill opacity=0.1] (-0.42,-0.5) ellipse (.45cm and 1.3cm);
			\filldraw[fill=black, fill opacity=0.1] (4.42 ,-0.5) ellipse (.45cm and 1.3cm);
			
			\draw[double, double distance=1.5,line width=1.2, postaction={decorate, decoration={
					markings,% switch on markings
					mark=at position 0.8 with  {\arrow{ latex} } } } ] (-1.8,-0.5) -- node [above] {} (-0.9,-0.5);
			
			\draw[double, double distance=1.5,line width=1.2, postaction={decorate, decoration={
					markings,% switch on markings
					mark=at position 0.8 with  {\arrow{ latex} } } } ] (4.9,-0.5) -- node [above] {} (5.85,-0.5);
		\end{feynman}
	\end{tikzpicture}
	\begin{tikzpicture}
		\begin{feynman}
			\vertex (i1);
			\vertex[right=1cm of i1] (a1);
			\vertex[right=3cm of a1] (o1) ;
			\vertex[below=1cm of i1] (i2) ;
			\vertex[right=3cm of i2] (a2);
			\vertex[right=1cm of a2] (o2) ;
			
			%\vertex[below=2cm of i3] (t1);
			\vertex[above=1cm of a1] (t2);
			\vertex[above=2cm of a2] (t3);
			\vertex[] (t1) at (2,1.3) ;
			%\vertex[above right =1.3cm of t2] (t1);
			\vertex[above=1.5cm of t1] (i3) {$T^{\mu\nu}$};
			
			\diagram* { 
				(i1)  -- [fermion] (a1)   --[fermion] (o1),
				
				(i2)  -- [anti fermion] (a2)  --[anti fermion] (o2),

				(i3)  -- [graviton] (t1),
				(t1) -- [gluon] (a1) ,
				(t1) -- [gluon] (a2),
			};
			
			\node[] at (-0.35,-0.45) {$\pi\ $};
			\node[] at (4.46,-0.45) {$\pi\ $};

			\filldraw[fill=black, fill opacity=0.1] (-0.42,-0.5) ellipse (.45cm and 1.3cm);
			\filldraw[fill=black, fill opacity=0.1] (4.42 ,-0.5) ellipse (.45cm and 1.3cm);
			
			\draw[double, double distance=1.5,line width=1.2, postaction={decorate, decoration={
					markings,% switch on markings
					mark=at position 0.8 with  {\arrow{ latex} } } } ] (-1.8,-0.5) -- node [above] {} (-0.9,-0.5);
			
			\draw[double, double distance=1.5,line width=1.2, postaction={decorate, decoration={
					markings,% switch on markings
					mark=at position 0.8 with  {\arrow{ latex} } } } ] (4.9,-0.5) -- node [above] {} (5.85,-0.5);
		\end{feynman}
	\end{tikzpicture}
	\caption{Diagrams of the hard scattering contribution to the pion GFF corresponding to the order $\mathcal{O}(\alpha_s)$.}
	\label{chza:GFFdiagramsTree}
\end{figure}
To compute the hadronic matrix element, we insert the energy-momentum tensor and the gluon exchanges at second order in $g_s$, therefore
\begin{align}
	T^{\mu\nu} &= \int d^4x\, d^4y\, \langle \pi(P_2) |
	T \left\{
	ig_s\, \bar{u}(x) \slashed{A}^a(x) \frac{\lambda^a}{2} u(x)\, 
	\hat{T}^{\mu\nu}(0)\,
	ig_s\, \bar{d}(y) \slashed{A}^b(y) \frac{\lambda^b}{2} d(y)
	\right\}
	| \pi(P_1) \rangle,
\end{align}
where $\hat{T}^{\mu\nu}(0)$ is the local energy-momentum tensor insertion point, and the gluon fields mediate the hard scattering between quarks.

To proceed with the evaluation of the hadronic matrix element, we perform Wick contractions of the gluon and quark fields. 
As for the $u$-quark fields appearing in the matrix element involving the energy-momentum tensor $T^{\mu\nu}$, there are two types of contractions corresponding to the two hard-scattering diagrams involving
\begin{align}
	\langle 0 | T \{ u^j_\alpha(x) \bar{u}^i_\beta(0) \} | 0 \rangle &\\
	\langle 0 | T \{ u^j_\alpha(0) \bar{u}^i_\beta(x) \} | 0 \rangle &
	\label{chza:eq:quark_props}
\end{align}
corresponding respectively to diagrams in Fig. \ref{chza:GFFdiagramsTree}, and similarly for the $d$ quark. We can insert the valence Fock states of the pion in the collinear approximation, using \eqref{chza:wf},
and evaluating matrix elements of field operators on the pion state, we obtain
\begin{align}
	\langle 0 | \bar{d}_n(y)\, u_l(0) | \pi(P_1) \rangle 
	&= \frac{f_\pi}{4} \int_0^1 d\alpha_1\, e^{-i\bar{\alpha}_1 P_1 \cdot y}\, \tilde{\phi}_\pi(\alpha_1)\, (\slashed{P}_1 \gamma_5)_{l n} , \\
	\langle \pi(P_2)| \bar{u}_i(x)\, d_m(y) | 0 \rangle 
	&= \frac{f_\pi}{4} \int_0^1 d\alpha_2\, e^{i\alpha_2 P_2 \cdot x + i\bar{\alpha}_2 P_2 \cdot y}\, \tilde{\phi}_\pi(\alpha_2)\, (\slashed{P}_2 \gamma_5)_{i m},
\end{align}
here \(f_\pi\) is the pion decay constant.

These expressions originate from the Fourier transform of the light-cone expansion of quark fields, with momenta \( \alpha_i P_i \) and \( \bar{\alpha}_i P_i \), and reproduce the correct Dirac structure for a pseudoscalar meson. Higher-order terms in $y^\mu$ are neglected because the typical distance between quarks is $\mathcal{O}(1/Q)$ at large $Q$.

These results allow us to carry out the full diagrammatic calculation of the GFFs $A^\pi(q^2)$  and  $ D^\pi(q^2)$ within the collinear factorization framework at the tree level. 
The evaluation of the diagrams in Fig. \ref{chza:GFFdiagramsTree} is straightforward once the previously introduced identities are employed. In momentum space, the result can be expressed as
\begin{equation}\label{kerfact}
	\mel{\pi(P_2)}{T^{\mu\nu}}{\pi(P_1)} = \frac{i\,f_\pi^2 C_F g_s^2}{N_C}\int_0^1 d\alpha_1 \int_0^1 d\alpha_2 \, 
	\tilde{\phi}_\pi(\alpha_1)\, K_\pi^{\mu\nu}(\alpha_1,\alpha_2)\, \tilde{\phi}_\pi(\alpha_2),
\end{equation}
with \(C_F=(N_C^2-1)/(2N_C)\).
The kernel $K_\pi^{\mu\nu}(\alpha_1,\alpha_2)$ can be decomposed in terms of form factors as
\begin{equation}\label{decompkpi}
	K_\pi^{\mu\nu}(\alpha_1,\alpha_2) = F_1\, P_1^{\mu } P_2^{\nu } 
	+ F_2\, P_2^{\mu } P_1^{\nu } 
	+ F_3\, P_1^{\mu } P_1^{\nu } 
	+ F_4\, P_2^{\mu } P_2^{\nu } 
	+ F_5\, g^{\mu\nu},
\end{equation}
with
\begin{equation}
	F_1=F_2=-4\frac{{\alpha_1} (2 \alpha_2-1)-\alpha_2+1}{(\alpha_1-1) \alpha_1 (\alpha_2-1) \alpha_2 \,q^2},
\end{equation}

\begin{equation}
	F_3=F_4=-2\frac{{\alpha_1} (2 \alpha_2-1)-\alpha_2+1}{(\alpha_1-1) \alpha_1 (\alpha_2-1) \alpha_2 \,q^2},
\end{equation}
\begin{equation}
	F_5=\frac{{\alpha_1} (2 \alpha_2-1)-\alpha_2+1}{(\alpha_1-1) \alpha_1 (\alpha_2-1) \alpha_2}.
	%\frac{\alpha_1 (2 \alpha_2-3)-3 \alpha_2+3 }{16 (\alpha_1-1) \alpha_1 (\alpha_2-1) \alpha_2}.  
\end{equation}

The direct computation of the form factors $F_i$ immediately reveals the symmetry under the exchange $\mu \leftrightarrow \nu$ as well as $P_1 \leftrightarrow P_2$, which is associated with the conditions
\begin{equation}
	F_1 = F_2 \ \ \ \ \ \ \ \ F_3 = F_4(\alpha_1\leftrightarrow\alpha_2).
\end{equation}
It also reveals that in the limit of large momentum transfer ($Q^2 \gg m_\pi^2$), which corresponds to the chiral (massless pion) limit, the calculation exhibits explicit gauge invariance at the amplitude level. In this regime, the unphysical gauge-dependent terms proportional to the longitudinal components of the gauge propagators are naturally canceled out. This tree-level gauge independence ensures a robust baseline for the factorization framework before radiative corrections are taken into account.

Finally, the relations connecting the form factors $F_i$ with the GFFs $A_\pi$ and $D_\pi$ are
\begin{equation}\label{chza:atree}
	A_\pi= \frac{f_\pi^2 C_F g_s^2}{N_C}\int_0^1 d\alpha_1 \int_0^1 d\alpha_2\, \tilde{\phi}_\pi(\alpha_1)\,
	(F_1+F_3)\,
	\tilde{\phi}_\pi(\alpha_2),
\end{equation}
\begin{equation}\label{chza:dtree}
	D_\pi=-\frac{f_\pi^2 C_F g_s^2}{N_C}\int_0^1 d\alpha_1 \int_0^1 d\alpha_2\, \tilde{\phi}_\pi(\alpha_1)\,
	\frac{2 F_5}{q^2}\,
	\tilde{\phi}_\pi(\alpha_2).
\end{equation}

One important question concerns the convergence of the factorization formula already at leading order. In performing the integrals over the DAs in the full interval $0 < \alpha_{1,2} < 1$, we tacitly assume that, at the endpoint $\alpha_i \to 1$, the function $\tilde{\phi}_\pi(\alpha)$ vanishes at least linearly in $\bar{\alpha} = 1 - \alpha$, ensuring the finiteness of the integrals. In fact, the DA at finite factorization scale can be written as a Gegenbauer expansion,
\begin{equation}
	\tilde\phi_{\pi}(\a,\mu) = 6\a(1-\a) \left[
	1 + \sum_{n=2,4,\dots} a_n(\mu)\, C^{3/2}_n(2\a-1)
	\right],
\end{equation}
with the Gegenbauer moments
\begin{equation}
	a_n(\mu)
	=
	a_n(\mu_0)
	\left(
	\frac{\alpha_s(\mu)}{\alpha_s(\mu_0)}
	\right)^{\gamma_n},
\end{equation}
where the anomalous dimensions $\gamma_n$ are
\begin{equation}
	\gamma_n
	=C_F	\left[	-3+ 4 \sum_{j=1}^{n+1} \frac{1}{j}-\frac{2}{(n+1)(n+2)}\right].
\end{equation}
The running coupling 
\begin{equation}\label{chza:evolalph}
	\alpha_s(\mu) =\frac{4\pi}{\beta_0\log\frac{\mu^2}{\Lambda_{\rm QCD}^2}},
\end{equation}  
with the QCD scale $\Lambda_{\rm QCD}=0.25$ GeV and the QCD beta function at one-loop level
\begin{equation}\label{chza:eq:beta0def}
	\beta_0=\frac{11C_A-2n_f}{3},
\end{equation}
where \(C_A=N_C=3\) and \(n_f\) is the number of active quark flavors.
Eq. \eqref{chza:evolalph} is evaluated at the largest hard scale in the convolution, and reduces to
\begin{equation}
	\tilde\phi^{\text{as}}_{\pi}(\a) = 6\a(1-\a),
\end{equation}
at asymptotically large scales. Therefore, the convergence of the integral in Eqs \eqref{chza:atree} and \eqref{chza:dtree} is  straightforward.

However, if we include power-suppressed corrections, for example effects of order $\mathcal{O}(k_\perp/Q^2)$ that were neglected in the collinear approximation, we encounter divergent integrals in the factorization approach. To regularize these divergences, various extensions to the factorization framework have been proposed, such as incorporating the transverse-momentum dependence of the pion DAs
\begin{equation}
	\tilde{\phi}_\pi(\alpha) \rightarrow \tilde{\phi}_\pi(\alpha, {k}_\perp).
\end{equation}

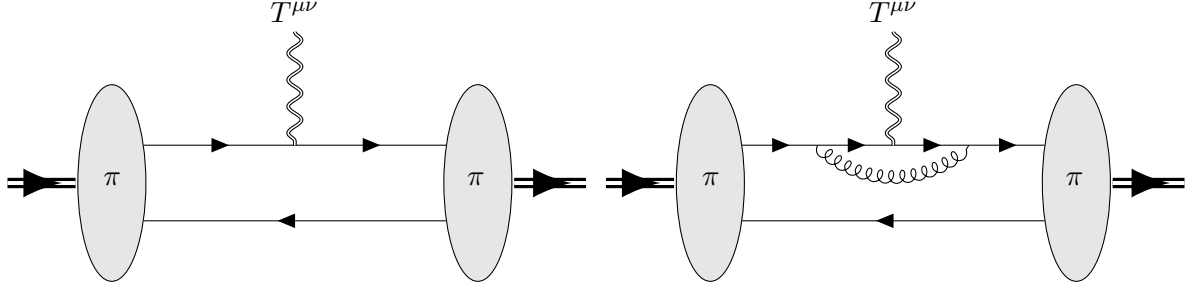
\begin{figure}
	\centering
	\begin{tikzpicture}
		\begin{feynman}
			\vertex (i1);
			\vertex[right=1cm of i1] (ax2);
			\vertex[right=3cm of i1] (ax3);
			\vertex[right=2cm of i1] (a1);
			\vertex[right=2cm of a1] (b1);
			\vertex[right=3cm of a1] (o1) ;
			\vertex[below=1cm of i1] (i2) ;
			
			\vertex[below=1cm of b1] (o2) ;
			
			%\vertex[below=2cm of i3] (t1);
			\vertex[above=1cm of a1] (t2);
			\vertex[above=2cm of a2] (t3);
			
			%\vertex[above right =1.3cm of t2] (t1);
			\vertex[above=1.5cm of a1] (i3) {$T^{\mu\nu}$};
			
			\diagram* { 
				(i1)  -- [fermion] (a1) -- [fermion]  (b1)  ,
				
				(i2)    --[anti fermion] (o2),

				(i3)  -- [graviton] (a1),
				
			};
			
			\node[] at (-0.35,-0.45) {$\pi\ $};
			\node[] at (4.46,-0.45) {$\pi\ $};

			\filldraw[fill=black, fill opacity=0.1] (-0.42,-0.5) ellipse (.45cm and 1.3cm);
			\filldraw[fill=black, fill opacity=0.1] (4.42 ,-0.5) ellipse (.45cm and 1.3cm);
			
			\draw[double, double distance=1.5,line width=1.2, postaction={decorate, decoration={
					markings,% switch on markings
					mark=at position 0.8 with  {\arrow{ latex} } } } ] (-1.8,-0.5) -- node [above] {} (-0.9,-0.5);
			
			\draw[double, double distance=1.5,line width=1.2, postaction={decorate, decoration={
					markings,% switch on markings
					mark=at position 0.8 with  {\arrow{ latex} } } } ] (4.9,-0.5) -- node [above] {} (5.85,-0.5);
		\end{feynman}
	\end{tikzpicture}
	\begin{tikzpicture}
		\begin{feynman}
			\vertex (i1);
			\vertex[right=1cm of i1] (ax2);
			\vertex[right=3cm of i1] (ax3);
			\vertex[right=2cm of i1] (a1);
			\vertex[right=2cm of a1] (b1);
			\vertex[right=3cm of a1] (o1) ;
			\vertex[below=1cm of i1] (i2) ;
			
			\vertex[below=1cm of b1] (o2) ;
			
			%\vertex[below=2cm of i3] (t1);
			\vertex[above=1cm of a1] (t2);
			\vertex[above=2cm of a2] (t3);
			
			%\vertex[above right =1.3cm of t2] (t1);
			\vertex[above=1.5cm of a1] (i3) {$T^{\mu\nu}$};
			
			\diagram* { 
				(i1)   -- [fermion] (ax2) -- [fermion] (a1) -- [fermion] (ax3) -- [fermion] (b1)  ,
				
				(i2)    --[anti fermion] (o2),

				(i3)  -- [graviton] (a1),
				(ax2)-- [gluon, out=-45, in=-135] (ax3)
			};
			
			\node[] at (-0.35,-0.45) {$\pi\ $};
			\node[] at (4.46,-0.45) {$\pi\ $};

			\filldraw[fill=black, fill opacity=0.1] (-0.42,-0.5) ellipse (.45cm and 1.3cm);
			\filldraw[fill=black, fill opacity=0.1] (4.42 ,-0.5) ellipse (.45cm and 1.3cm);
			
			\draw[double, double distance=1.5,line width=1.2, postaction={decorate, decoration={
					markings,% switch on markings
					mark=at position 0.8 with  {\arrow{ latex} } } } ] (-1.8,-0.5) -- node [above] {} (-0.9,-0.5);
			
			\draw[double, double distance=1.5,line width=1.2, postaction={decorate, decoration={
					markings,% switch on markings
					mark=at position 0.8 with  {\arrow{ latex} } } } ] (4.9,-0.5) -- node [above] {} (5.85,-0.5);
		\end{feynman}
	\end{tikzpicture}
	\caption{Soft-overlap contribution diagrams to the pion form factor. The left panel shows the lowest-order contribution, while the right panel shows one of the diagrams contributing to the $\mathcal{O}(\a_s)$ correction.}
	\label{chza:softover}
\end{figure}

\section{Energy--momentum tensor and conformal properties}
\label{chza:sec:emt_conformal}
The stress-tensor operator is defined by varying the gauge-fixed QCD action with respect to the metric,
\begin{equation}
	T_{\mu\nu}(x)
	=
	-\left.
	\frac{2}{\sqrt{-g}}
	\frac{\delta S_{\mathrm{QCD}}}{\delta g^{\mu\nu}(x)}
	\right|_{g=\eta}.
	\label{chza:eq:emt_def_short}
\end{equation}
Equivalently, in the weak-field expansion \(g_{\mu\nu}=\eta_{\mu\nu}+\kappa h_{\mu\nu}\),
\begin{equation}
	S_{\mathrm{QCD}}[g]
	=
	S_{\mathrm{QCD}}[\eta]
	-\frac{\kappa}{2}\int d^4x\,h_{\mu\nu}(x)T^{\mu\nu}(x)
	+\mathcal O(h^2).
	\label{chza:eq:linear_emt_short}
\end{equation}
In flat space we use
\begin{equation}
	\mathcal L_{\mathrm{QCD}}
	=
	-\frac{1}{4}F_{\rho\sigma}^aF^{a\,\rho\sigma}
	+
	\frac{i}{2}\bar\psi\gamma^\rho\overleftrightarrow{D}_\rho\psi
	-
	m\bar\psi\psi
	-
	\frac{1}{2\xi}(\partial^\rho A_\rho^a)^2
	+
	\partial^\rho\bar c^{\,a}D_\rho^{ab}c^b,
	\label{chza:eq:qcd_lagrangian_short}
\end{equation}
with
\begin{equation}
	F_{\mu\nu}^a
	=
	\partial_\mu A_\nu^a-\partial_\nu A_\mu^a+g_s f^{abc}A_\mu^bA_\nu^c,
	\qquad
	D_\mu^{ab}c^b=\partial_\mu c^a+g_s f^{acb}A_\mu^c c^b .
	\label{chza:eq:field_strength_short}
\end{equation}
The explicit Hilbert tensor entering the perturbative \(TJJ\) vertex is then
\begin{align}
	T_{\mu\nu}
	&=
	-\eta_{\mu\nu}\mathcal L_{\mathrm{QCD}}
	-
	F_{\mu\rho}^aF_{\nu}^{a\,\rho}
	-
	\frac{1}{\xi}\eta_{\mu\nu}\,\partial^\rho\!\left(A_\rho^a\,\partial^\sigma A_\sigma^a\right)
	\nonumber\\
	&\quad
	+
	\frac{1}{\xi}
	\left[
	A_\nu^a\,\partial_\mu(\partial^\sigma A_\sigma^a)
	+
	A_\mu^a\,\partial_\nu(\partial^\sigma A_\sigma^a)
	\right]
	\nonumber\\
	&\quad
	+
	\frac{i}{4}
	\Big[
	\bar\psi\gamma_\mu(\overrightarrow{\partial}_\nu-i g_s A_\nu^a t^a)\psi
	-
	\bar\psi(\overleftarrow{\partial}_\nu+i g_s A_\nu^a t^a)\gamma_\mu\psi
	\nonumber\\
	&\hspace{2.4cm}
	+
	\bar\psi\gamma_\nu(\overrightarrow{\partial}_\mu-i g_s A_\mu^a t^a)\psi
	-
	\bar\psi(\overleftarrow{\partial}_\mu+i g_s A_\mu^a t^a)\gamma_\nu\psi
	\Big]
	\nonumber\\
	&\quad
	+
	\partial_\mu \bar c^{\,a}\,D_\nu^{ab}c^b
	+
	\partial_\nu \bar c^{\,a}\,D_\mu^{ab}c^b .
	\label{chza:eq:emt_explicit_short}
\end{align}

In the classical massless limit the gauge-invariant part of the tensor is conserved and traceless, up to equations of motion and gauge-fixing terms.  Renormalization breaks this Weyl invariance.  In physical matrix elements the massless trace is governed by the gauge-invariant anomaly term
\begin{equation}
	T^\mu_{\ \mu}
	=
	\frac{\beta(g_s)}{2g_s}\,
	F_{\rho\sigma}^aF^{a\,\rho\sigma}
	\,+\,
	\sum_f (1+\gamma_m)m_f\bar\psi_f\psi_f
	\,+\,
	\hbox{BRST-exact and equation-of-motion terms}.
	\label{chza:eq:qcd_trace_anomaly_short}
\end{equation}
Thus, in the chiral limit, the conformal breaking relevant for the pion hard kernel is controlled by the QCD beta function.  In the form-factor calculation below the non-Abelian \(TJJ\) correlator is used as a one-loop stress-tensor insertion in the two-gluon hard subgraph.  The quark, gluon and ghost loop sectors combine into a renormalized off-shell \(TJJ\) vertex; its trace part contains precisely the beta-function coefficient that is later projected onto the pion GFFs.

\section{Tensor-sector decomposition of the non-Abelian $TJJ$ correlator}
\label{chza:seven}
%%%%%%%%%%%%%%%%%%%%%%%%%%%%%%%			%%%%%%%%%%%%%%%%%%%%%%%%%%%%%%%
The second ingredient is the sector decomposition of the off-shell \(TJJ\) vertex.  We separate the correlator into transverse--traceless, longitudinal or semi-local, and trace sectors, treating the quark and gluon loop contributions separately before combining them in the renormalized result.  This is the minimal organization needed for the subsequent matching onto the pion form factors.

The general form of the non-Abelian $\langle TJJ \rangle$ correlator in $d$ dimensions can be constructed through a decomposition into transverse, longitudinal and trace terms \cite{Bzowski:2013sza}, exploiting its symmetries. The perturbative QCD correlator contains additional longitudinal terms with respect to the abstract conformal field theory (CFT) solution of \cite{Bzowski:2018fql}. These terms are tied to gauge fixing, BRST invariance and the Slavnov--Taylor identities, and they vanish or reduce in the appropriate on-shell limits.

We consider the decomposition of the operators $T$ and $J$ in terms of their transverse-traceless part and longitudinal (local) one. Following \cite{Bzowski:2013sza} we define
\begin{align}
	T^{\mu_i\nu_i}(p_i)&\equiv t^{\mu_i\nu_i}(p_i)+t_{loc}^{\mu_i\nu_i}(p_i),\label{chza:decT}\\
	J^{a_i \, \mu_i}(p_i)&\equiv j^{a_i \, \mu_i}(p_i)+j_{loc}^{a_i \, \mu_i}(p_i),\label{chza:decJ}
\end{align}
where
\begin{align}
	\label{chza:loct}
	t^{\mu_i\nu_i}(p_i)&=\Pi^{\mu_i\nu_i}_{\alpha_i\beta_i}(p_i)\,T^{\alpha_i \beta_i}(p_i), &&t_{loc}^{\mu_i\nu_i}(p_i)=\Sigma^{\mu_i\nu_i}_{\alpha_i\beta_i}(p_i)\,T^{\alpha_i \beta_i}(p_i),\\
	j^{a_i \, \mu_i}(p_i)&=\pi^{\mu_i}_{\alpha_i}(p_i)\,J^{a_i \, \alpha_i }(p_i), &&\hspace{1ex}j_{loc}^{a_i \, \mu_i}(p_i)=\frac{p_i^{\mu_i}\,p_{i\,\alpha_i}}{p_i^2}\,J^{a_i \, \alpha_i}(p_i).
\end{align}
Having introduced the transverse-traceless ($\Pi$), transverse $(\pi)$, longitudinal ($\Sigma$) projectors, given respectively by 
\begin{align}
	\label{chza:prozero}
	&\pi^{\mu}_{\alpha}  = \delta^{\mu}_{\alpha} - \frac{p^{\mu} p_{\alpha}}{p^2}, \\
	%%%%%%%%%%%%%%%%%%%%%%%%%%%%%%%
	&\Pi^{\mu \nu}_{\alpha \beta}  = \frac{1}{2} \left( \pi^{\mu}_{\alpha} \pi^{\nu}_{\beta} + \pi^{\mu}_{\beta} \pi^{\nu}_{\alpha} \right) - \frac{1}{d - 1} \pi^{\mu \nu}\pi_{\alpha \beta}\label{chza:TTproj}, \\
	%%%%%%%%%%%%%%%%%%%%%%%%%%%%%%%
	&\Sigma^{\mu_i\nu_i}_{\alpha_i\beta_i}=\frac{p_{i\,\beta_i}}{p_i^2}\Big[2\delta^{(\nu_i}_{\alpha_i}p_i^{\mu_i)}-\frac{p_{i\alpha_i}}{(d-1)}\left(\delta^{\mu_i\nu_i}+(d-2)\frac{p_i^{\mu_i}p_i^{\nu_i}}{p_i^2}\right)\Big]+\frac{\pi^{\mu_i\nu_i}(p_i)}{(d-1)}\delta_{\alpha_i\beta_i}\equiv\mathscr{T}^{\mu_i\nu_i}_{\alpha_i}p_{i\,\beta_i} +\frac{\pi^{\mu_i\nu_i}(p_i)}{(d-1)}\delta_{\alpha_i\beta_i}\label{chza:Lproj}.
	%%%%%%%%%%%%%%%%%%%%%%%%%%%%%%%
\end{align}
with
\begin{equation} \label{chza:a:T}
	\mathscr{T}_{\mu\nu \alpha} ({p}) = \frac{1}{p^2} \left[ 2 p_{(\mu} \delta_{\nu)\alpha} - \frac{p_\alpha}{d-1} \left( \delta_{\mu\nu} + (d-2) \frac{p_\mu p_\nu}{p^2} \right) \right]
\end{equation}	
and
\begin{equation}\label{chza:Idecomp}
	\delta_{\mu(\alpha}\delta_{\beta)\nu}=\Pi_{\mu\nu\alpha\beta}({p})+\mathscr{T}_{\mu\nu (\alpha}({p})\,p_{\beta)}+\frac{1}{d-1}\pi_{\mu\nu}({p})\delta_{\alpha\beta}.
\end{equation}
Turning to the $TJJ$ case, we can divide the three-point function into two parts: the \emph{transverse-traceless} part and the \emph{semi-local} part (indicated by subscript $loc$) expressible through the transverse and trace Ward identities. These parts are obtained by using the projectors $\Pi$ and $\Sigma$, previously defined. 
We can then decompose the full three-point function as 

\begin{align}
	\label{chza:dec1}
	\braket{T^{\m \n}\,J^{a \, \a}\,J^{b \, \b}}&=
	\braket{t^{\m \n }\,j^{a \, \a}\,j^{b \, \b}}+\braket{T^{\m \n}\,J^{a \, \a}\,j_{loc}^{b \, \b}}+\braket{T^{\m \n}\,j_{loc}^{a \, \a}\,J^{b \, \b}}+\braket{t_{loc}^{\m \n}\,J^{a \, \a}\,J^{b \, \b}}\notag\\
	&\quad-\braket{T^{\m \n }\,j_{loc}^{a \, \a}\,j_{loc}^{b \, \b}}-\braket{t_{loc}^{\m \n}\,j_{loc}^{a \, \a}\,J^{b \, \b}}-\braket{t_{loc}^{\m \n}\,J^{a \, \a}\,j_{loc}^{b \, \b}}+\braket{t_{loc}^{\m \n }\,j_{loc}^{a \, \a}\,j_{loc}^{b \, \b}}.
\end{align}
In a CFT approach, all the terms on the right-hand side of the decomposition, apart from the first one, may be computed by means of transverse and trace Ward identities, with the anomaly induced by the renormalization of the theory by a single counterterm, proportional to the square of the field strength $(F^2)$. 

\subsection{The decomposition of the correlator in the quark sector}
\label{chza:eight}
Using the projectors $\Pi$ and $\pi$ one can write the most general form of the transverse-traceless part as
\begin{equation}
	{\braket{t^{\m \n}(q)\,j^{a \, \a}(p_1)\,j^{b \, b }(p_2)}}_q =\Pi^{\m \n}_{\m_1 \n_1}(q)\pi^{\a}_{\a_1 }(p_1)\pi^{\b}_{\b_1}(p_2)\,\,X^{a b \, \m_1 \n_1 \a_1 \b_1}_q,
\end{equation}
where $X^{a b \, \m_1 \n_1 \a_1 \b_1}_q$ is a general tensor of rank four built out of the metric and momenta. We can enumerate all possible tensors that can appear in $X^{a b \, \m_1 \n_1 \a_1 \b_1}$ while preserving the symmetry of the 
correlator 
\begin{align}
	\langle t^{\mu \nu }(q)j^{a \, \a}(p_1)j^{b \, \b}(p_2)\rangle_q & =
	{\Pi_1}^{\mu \nu}_{\m_1 \n_1}{\pi_2}^{\a}_{\a_1}{\pi_3}^{\b}_{\b_1}
	\left( A_{1}^{(q) a b} \ p_1^{\m_1 }p_1^{\n_1}p_2^{\a_1}q^{\b_1} + 
	A_{2}^{(q) a b}\ \delta^{\a_1 \b_1} p_1^{\m_1}p_1^{\n_1} + 
	A_{3}^{(q) a b}\ \delta^{\m_1\a_1}p_1^{\n_1}q^{\b_1}\right. \notag\\
	& \left. + 
	A_{3}^{(q) a b}(p_1\leftrightarrow p_2)\delta^{\m_1\b_1}p_1^{\n_1}p_2^{\a_1}
	+ A_{4}^{(q) a b}\  \delta^{\m_1\b_1}\delta^{\a_1\n_1}\right)\label{chza:DecompTJJ}
\end{align}
with the reconstruction taking the form \cite{Bzowski:2013sza}
\begin{align}
	& \la T_{\mu_1 \nu_1}(q) J^{\mu_2 a_2}({p}_1) J^{\mu_3 a_3}({p}_2) \ra_q = \la t_{\mu_1 \nu_1}(q) j^{\mu_2 a_2}({p}_1) j^{\mu_3 a_3}({p}_2) \ra_q \nn\\[0.5ex]
	& \qquad +  2 \mathscr{T}_{\mu_1 \nu_1}^{\quad\,\,\alpha}(q)\Big[\delta_{[\alpha}^{\mu_3}p_{2\beta]}  \la J^{\mu_2 a_2}({p}_1) J^{\beta a_3}(-{p}_1) \ra_q
	+\delta_{[\alpha}^{\mu_2}p_{1\beta]}  \la J^{\mu_3 a_3}({p}_2) J^{\beta a_2}(-{p}_2) \ra_q\Big]
	\nn\\
	& \qquad + \frac{1}{d - 1}\,\pi_{\mu_1 \nu_1}(q) \mathcal{A}^{\mu_2 \mu_3 a_2 a_3}_q, \label{chza:tjjdec}
\end{align}
where $\mathscr{T}_{\mu_1 \nu_1\alpha}$ is defined in \eqref{chza:a:T} and $\langle JJ\rangle_q$ is the 2-point function of the gluon with a virtual quark, namely
\beq
\la J^{\alpha a}({p}_1) J^{\beta b}(-{p}_1) \ra_q=
\frac{2}{3}n_f  \,  \frac{g_s^2}{16\pi ^2} {B}_0\left({{p_1}}^2\right) {p_1}^2   \delta ^{a b}\pi^{\alpha\beta}(p_1),
\eeq 
where \(B_0(p^2)\) is the standard scalar two-point Passarino--Veltman function in dimensional regularization.
The transverse Ward identities are 
\begin{align}
	& q^{\nu_1} \la T_{\mu_1 \nu_1}(q) J^{\mu_2 a_2}({p}_1) J^{\mu_3 a_3}({p}_2) \ra_q  \nn\\[0.5ex]
	& \qquad = \: 2 \delta^{\mu_3}_{[\mu_1} p_{2\alpha]} \la J^{\mu_2 a_2}({p}_1) J^{\alpha a_3}(-{p}_1) \ra_q + 2 \delta^{\mu_2}_{[\mu_1} p_{1\alpha]} \la J^{\alpha a_2}({p}_2) J^{\mu_3 a_3}(-{p}_2) \ra_q, \\[1ex]\label{chza:TWI_TJJ_2}
	& p_{1 \mu_2} \la T_{\mu_1 \nu_1}(q) J^{\mu_2 a_2}({p}_1) J^{\mu_3 a_3}({p}_2) \ra_q = 0, \\[1ex]
\end{align}
while the anomalous trace Ward identity is given by
\begin{align}
	& \la T(q) J^{\mu_2 a_2}({p}_1) J^{\mu_3 a_3}({p}_2) \ra_q = \mathcal{A}^{\mu_2 \mu_3 a_2 a_3}_q=-{2 \over 3} \, \,  \, \, \frac{g_s^2}{16\pi^2} \,   n_f \delta^{a_2a_3}u^{\mu_2 \mu_3}(p_1,p_2),
\end{align}
where $u$ stands for the second derivative of the field strength squared $F^2$ which in momentum space is
\begin{equation}\label{chza:defu}
	u^{\a \b}(p_1,p_2)= (p_1\cdot p_2) g^{\a\b}-p_2^\a p_1^\b
\end{equation}
The form factors of the transverse part of the correlator can be explicitly defined as
\begin{equation}\label{chza:aiq}
	A_{j }^{(q) a b}= -\,   \frac{g_s^2}{16\pi^2}\, \bar{A}_j^{{(q)}a b}, \qquad j=1,2,3, 4
\end{equation}
where the label $(q)$ refers to the contribution from a single quark loop in the perturbative expansion; a direct computation gives
\begin{eqnarray}
	\label{chza:Abar}
	\bar{A}_1^{(q) ab}  &=& \frac{n_f\delta^{ab}}{48 \left(p_1^2 p_2^2-(p_1 \cdot p_2)^2 \right)^4} \Big[ A_{10} + A_{11}  B_0 (p_1^2) + A_{12} B_0(p_2^2) + A_{13} B_0(q^2) + A_{14} C_0 (p_1^2 , p_2^2 , q^2 ) \Big] \notag \\
	\bar{A}_2^{(q) ab}  &=& -\frac{\delta ^{ab}}{144 \left(p_1^2 p_2^2-(p_1 \cdot p_2)^2 \right)^3} \Big[ A^{(q)}_{20} + A^{(q)}_{21}  B_0 (p_1^2) + A^{(q)}_{22} B_0(p_2^2)  + A^{(q)}_{23} B_0(q^2) + A^{(q)}_{24} C_0 (p_1^2 , p_2^2 , q^2 ) \Big] \notag \\
	\bar{A}_3^{(q) ab}  &=& \frac{\delta ^{ab}}{72 \left(p_1^2 p_2^2-(p_1 \cdot p_2)^2 \right)^3} \Big[A^{(q)}_{31}  B_0 (p_1^2) + A^{(q)}_{32} B_0(p_2^2)  + A^{(q)}_{33} B_0(q^2) + A^{(q)}_{34} C_0 (p_1^2 , p_2^2 , q^2 ) \Big] \notag \\
	\bar{A}_4^{(q) ab}  &=& -\frac{ \delta^{ab}}{72 \left(p_1^2 p_2^2-(p_1 \cdot p_2)^2 \right)^2} \Big[A^{(q)}_{40} + A^{(q)}_{41}  B_0 (p_1^2) + A^{(q)}_{42} B_0(p_2^2) + A^{(q)}_{43} B_0(q^2) + A^{(q)}_{44} C_0 (p_1^2 , p_2^2 , q^2 ) \Big]. \notag \\
\end{eqnarray}
The expressions of the $A_{ij}$ are given in Appendix \ref{chza:ff}, while $C_0(p_1^2,p_2^2,q^2)$ is the massless three-point scalar integral whose explicit expression is given in Eq.~\eqref{chza:c0formula}.  The explicit formulas of $\bar A_i^{(q)ab}$ can be isolated from the general 
expressions for $A_{ij}$ in Appendix \ref{chza:ff}  by selecting the $n_f$ dependent terms in $A_{ij}$.

\subsection{The decomposition of the correlator in the gluon sector}
\label{chza:nine}
The one-loop decomposition of the gluon sector follows \eqref{chza:dec1}, 
but its final expression is modified compared to 
\eqref{chza:tjjdec}, which is affected by new trace contributions not present in the quark sector. This sector provides a contribution of the form
\begin{equation}
	\begin{aligned}
		& \langle T^{\mu \nu}(q)  J^{ a\alpha}(p_1) J^{ b\beta}(p_2)\rangle_g=\langle t^{\mu \nu}(q)  j^{ a\alpha}(p_1) j^{ b\beta}(p_2)\rangle_g +\langle t^{\mu \nu }(q)j_{loc}^{a  \a}(p_1)j^{b  \b}(p_2)\rangle_g +\langle t^{\mu \nu }(q)j^{a  \a}(p_1)j_{loc}^{b  \b}(p_2)\rangle_g\\
		& \qquad+2 \mathscr{T}^{\mu \nu \rho }(q)\left[\delta_{[\rho}^{\beta} p_{2 \sigma ]}\langle J^{ a\alpha}({p}_1) J^{ b\sigma}(-{p}_1)\rangle_g +\delta_{[\rho}^{\alpha} p_{1 \sigma]}\langle J^{ b\beta}({p}_2) J^{ a\sigma}(-p_2)\rangle\right]_g  
		+\frac{1}{d-1} \pi^{\mu \nu}(q) \left[\mathcal{A}^{\alpha \beta a b}_g+\mathcal{B}^{\alpha \beta a b}_g\right]
	\end{aligned}
	\label{chza:res}
\end{equation}
There are additional local terms of the form 
\beq
\langle t^{\mu \nu }(q)j^{a \, \a}(p_1)j_{loc}^{b \, \b}(p_2)\rangle_g 
\label{chza:adds}
\eeq
which are not part of the transverse-traceless sector, but appear in the longitudinal sector of the decomposition. These terms are not set to zero by the Slavnov-Taylor identities, as in the quark sector, or in the general conformal solution.  
At the same time, the trace sector is modified by the presence of extra terms which are proportional to the equations of motion of the gluons, here indicated as $\mathcal{B}^{\alpha \beta a b}_g, $ which are absent in the on-shell decomposition. Defining  
\begin{allowdisplaybreaks}
	\begin{equation}\label{chza:aig}
		A_{i }^{(g) a b}= -\,   \frac{g_s^2}{16\pi^2}\, \bar{A}_i^{{(g)}a b}, \qquad i=1,2\ldots 4
	\end{equation}
	their explicit expressions are given by 
	\begin{eqnarray}
		\label{chza:Abar_g}
		\bar{A}_1^{(g) ab}  &=& -\frac{C_A\delta^{ab}}{48 \left(p_1^2 p_2^2-(p_1 \cdot p_2)^2 \right)^4} \Big[ A_{10} + A_{11}  B_0 (p_1^2) + A_{12} B_0(p_2^2) + A_{13} B_0(q^2) + A_{14} C_0 (p_1^2 , p_2^2 , q^2 ) \Big] \notag \\
		\bar{A}_2^{(g) ab}  &=& -\frac{\delta ^{ab}}{144 \left(p_1^2 p_2^2-(p_1 \cdot p_2)^2 \right)^3} \Big[ A^{(g)}_{20} + A^{(g)}_{21}  B_0 (p_1^2) + A^{(g)}_{22} B_0(p_2^2)  + A^{(g)}_{23} B_0(q^2) + A^{(g)}_{24} C_0 (p_1^2 , p_2^2 , q^2 ) \Big] \notag \\
		\bar{A}_3^{(g) ab}  &=& \frac{\delta ^{ab}}{72 \left(p_1^2 p_2^2-(p_1 \cdot p_2)^2 \right)^3} \Big[A^{(g)}_{31}  B_0 (p_1^2) + A^{(g)}_{32} B_0(p_2^2)  + A^{(g)}_{33} B_0(q^2) + A^{(g)}_{34} C_0 (p_1^2 , p_2^2 , q^2 ) \Big] \notag \\
		\bar{A}_4^{(g) ab}  &=& -\frac{ \delta^{ab}}{72 \left(p_1^2 p_2^2-(p_1 \cdot p_2)^2 \right)^2} \Big[A^{(g)}_{40} + A^{(g)}_{41}  B_0 (p_1^2) + A^{(g)}_{42} B_0(p_2^2) + A^{(g)}_{43} B_0(q^2) + A^{(g)}_{44} C_0 (p_1^2 , p_2^2 , q^2 ) \Big], \notag \\ 
	\end{eqnarray}
\end{allowdisplaybreaks}
where the functions $A_{ij}^{(g)}$ are extracted from  \secref{chza:ff} by selecting the part of the $A_{ij}$ proportional to the Casimir $C_A$.
The new longitudinal terms \eqref{chza:adds} take the form 
\begin{align}
	\langle t^{\mu \nu }(q)j^{a \, \a}(p_1)j_{loc}^{b \, \b}(p_2)\rangle_g & =
	{\Pi}^{\mu \nu}_{\m_1 \n_1}(q) \, \pi^\a _{\a_1} (p_1)\, {{p_2}_{\beta_1} p_2^\beta \over p_2^2}
	\left( 	B^{ab}_1 \, p_1^{\m_1} \, p_1^{\nu_1} \, p_2^{\a_1} \, p_2^{\b_1} + B_2^{ab} \, p_1^{\m_1} \, p_2^{\b_1} \, \delta^{\a_1 \n_1}  \right) 
\end{align}
which is orthogonal to the trace sector. Notice that these local contributions vanish when the gluons are on-shell. The $B_i, \,\, i=1,2$ are given by  

\begin{eqnarray}
	B^{ab}_1 & = &  \frac{ \,  \,  \, C_A \,  g_s^2 \,  \delta^{ab} \, p_1^2}{64\pi^2 \, p_2^2 \, \big(p_1^2 p_2^2 - (p_1 \cdot p_2)^2 \big)^2} \, \Big( -\Big[2 \, (p_1 \cdot p_2)^2 + p_1^2 p_2^2 + 3 \, p_1^2 p_1 \cdot p_2 \Big]\, B_0(p_1^2)  \notag \\
	&& -\Big[2\, (p_1 \cdot p_2)^2 + p_1^2 p_2^2 + 3 \, p_2^2 p_1 \cdot p_2 \Big] \, B_0(p_2^2) +  \Big[ 4 \, (p_1 \cdot p_2)^2 + 2 \, p_1^2 p_2^2 + 3\, (p_1^2 + p_2^2) p_1 \cdot p_2 \Big]\, B_0(q^2)  \notag \\
	&&  +\Big[  q^2 \, (p_1 \cdot p_2)^2 + 2 \, q^2 \, (p_1 \cdot p_2)^2  \Big] \,C_0(p_1^2, p_2^2, q^2) - 2 \, (p_1 \cdot p_2)^2 - 2 \, p_1^2 p_2^2 \Big) \\
	B^{ab}_2 & = & \frac{ \,  \,  \, C_A \,  g_s^2 \,  \delta^{ab} \, p_1^2}{32\pi^2 \, p_2^2 \, \big(p_1^2 p_2^2 - (p_1 \cdot p_2)^2 \big)^2} \Big( - p_1 \cdot p_2 \, B_0(p_1^2)  - p_2^2 \cdot p_2  \, B_0(p_2^2) + (p_2^2+p_1 \cdot p_2) \, B_0(q^2) \notag \\ &&
	+ \Big[p_1^2 p_2^2 + p_2^2 \, (p_1 \cdot p_2) \Big] \, C_0(p_1^2, p_2^2 , q^2) \Big).
\end{eqnarray}

The trace sector is also affected by terms that vanish for 
on-shell gluons, indicated as $\mathcal{B}_g^{\alpha \beta a b}$, as well as the genuine anomaly term 
$\mathcal{A}_g^{\alpha \beta a b}$. Defining 
\beq
\mathcal{B}_g^{\alpha \beta a b}=\left[ C_1^{ab} \, p_1^\alpha \, p_1 ^\beta + C_2^{ab} \, p_1^\alpha \, p_2^\beta + C_3^{ab}  \, p_1^\beta \, p_2^\alpha + C_4^{ab} \, p_2^\alpha \, p_2^\beta + C_5^{ab} \, \delta^{\alpha \beta} \right]  
\label{chza:bb}
\eeq
the trace sector is characterized by the anomaly contributions from the gluon sector $\mathcal{A}_g$, plus the 
$\mathcal{B}_g$ terms proportional to the equations of motion of the gluons 
\begin{equation}
	\mathcal{A}_g^{\alpha \beta a b}+\mathcal{B}_g^{\alpha \beta a b}= g_{\mu \nu} \langle T^{\mu \nu} (q) \, J^{a \a} (p_1) \, J^{b \b} (p_2)\, \rangle_g 
\end{equation}
with the gluon contribution to the anomaly given by
\begin{eqnarray}
	\mathcal{A}_g^{\alpha\beta ab} &=& {11 \over 3} \,  \,  \,  \, \frac{g_s^2}{16\pi^2} \, C_A  \delta^{ab}u^{\alpha\beta}(p_1,p_2)
\end{eqnarray}
and $u^{\alpha\beta}$ defined in \eqref{chza:defu}.
The form factors proportional to the equations of motion in \eqref{chza:bb} take the form
{\allowdisplaybreaks
	\begin{eqnarray}
		C_1^{ab} &=& -\frac{ \,  \,  \, C_A \, g_s^2 \, \delta^{ab} }{ 32\pi^2 \big( (p_1 \cdot p_2)^2 - p_1^2 \, p_2^2 \big)} \, \Big(
		2 \, (p_1 \cdot p_2)^2 - 2 \, p_1^2 \, p_2^2 
		- p_1^2 \, (p_1 \cdot p_2 + p_2^2 ) B_0 (p_1^2)  \notag \\ &&
		+ \big[ p_2^2 \, (p_1^2 - p_1 \cdot p_2) - 2 \, (p_1 \cdot p_2)^2 \big]  B_0(p_2^2)
		+ (p_1 \cdot p_2) \, (2 \, p_1 \cdot p_2 + p_1^2 + p_2^2) \, B_0 (q^2)  \notag \\ &&
		+ \big[ 2\, (p_1 \cdot p_2)^2 \, (p_1 \cdot p_2)^2 - p_2^2 \, \big(4 (p_1 \cdot p_2)^2   + p_1^4\big) + 5 \, p_1^2 \, p_4^4  \big] C_0(p_1^2, p_2^2 , q^2)  \Big) \\
		C_2^{ab} &=& -2  \,  \,  \, C_A \, \frac{g_s^2}{16\pi^2} \, \delta^{ab} \, (p_1 \cdot p_2) \, C_0 (p_1^2 , p_2^2 , q^2) \\
		C_3^{ab} &=& -\frac{ \,  \,  \, g_s^2 \, C_A \, (p_1^2 + p_2^2) \, \delta^{ab}}{32\pi^2 \big(p_1^2 \, p_2^2 - (p_1 \cdot p_2)^2 \big)} \Big( 
		(-p_1\cdot p_2-p_1^2)  B_0(p_1^2)+(-(p_1\cdot p_2)-p_2^2) \, B_0(p_2^2) \notag \\ &&
		+q^2 \, B_0(q^2)+\big(p_1^2 \, (p_1\cdot p_2-2 p_2^2) +(p_1\cdot p_2) \, \big(4 \, (p_1\cdot p_2)+p_2^2) \big) \, C_0(p_1^2,p_2^2, q^2)  \Big)  \\
		C_4^{ab} &=&  -\frac{ \, \,  \, g_s^2 \, C_A \, g_s^2 \,  \delta^{ab}}{32\pi^2\big((p_1 \cdot p_2)^2 - p_1^2 \, p_2^2 \big)} \Big( 
		(p_1^2 \, (p_2^2-p_1\cdot p_2)\notag \\ &&  -2 (p_1\cdot p_2)^2) \, B_0(p_1^2) -p_2^2 \, (p_1\cdot p_2+p_1^2) \, B_0(p_2^2)+(p_1\cdot p_2) \, q^2 \, B_0(q^2) \notag \\ &&  
		+ \big[-p_1^2 \, (4 \, (p_1\cdot p_2)^2+p_2^4)+2 \, (p_1\cdot p_2+p_2^2) \, (p_1\cdot p_2)^2+5 \, p_1^4 \, p_2^2 \, \big] \, C_0(p_1^2,p_2^2,q^2) \, \Big) \\
		C_5^{ab} &=& \, \,  \, \frac{g_s^2}{32\pi^2} \, C_A \,  \delta^{ab} \Big( (p_1^2 - p_2^2) \, \big[ B_0 (p_1^2) - B_0(p_2^2) \big] 
		+ \big[ p_1^4 + p_2^4 - 2 \, (p_1^2 + p_2^2) \, p_1 \cdot p_2 - 6 \, p_1^2 \, p_2^2 \big] \, C_0(p_1^2, p_2^2, q^2) \Big). \nonumber \\\label{chza:cig}
	\end{eqnarray}
}

\section{The contribution of the $TJJ$ and of the trace anomaly in the hard kernel}\label{chza:contjjtra}

In this section we discuss the insertion of the non-Abelian \(TJJ\) correlator into the hard-scattering factorization formula for the pion GFFs, with particular emphasis on the relation between the off-shell trace sector of the correlator and the anomaly-induced contribution to the scalar gravitational structure of the pion. We also clarify the scaling of the various terms entering the hard kernel, the role of the two-point functions generated by the loop expansion, and the behavior of the convolution integrals with the pion DA \(\tilde{\phi}_\pi(\alpha)\).

\subsection{Factorization structure and notation}

Working in momentum space and in the hard regime
\begin{equation}\label{chza:eq:factorisation_regime}
	q^2\equiv -Q^2,\qquad Q^2\gg \Lambda_{\rm QCD}^2,\; m_\pi^2,
\end{equation}
%the matrix element of the energy-momentum tensor between pion states can be written in the collinear factorisation framework as
the correction to the tree-level calculation in Eq. \eqref{kerfact} of the matrix element of the energy-momentum tensor between pion states in the collinear factorization framework due to the $TJJ$ insertion is given by
\begin{equation}\label{chza:eq:factorisation}
	%\langle\pi(P_2)|T^{\mu\nu}|\pi(P_1)\rangle= 
	\frac{i\,f_\pi^2 C_F g_s^2}{N_C}\int_0^1 d\alpha_1 \int_0^1 d\alpha_2 \;
	\tilde{\phi}_\pi(\alpha_1)\, K_{TJJ}^{\mu\nu}(\alpha_1,\alpha_2,q)\, \tilde{\phi}_\pi(\alpha_2).
\end{equation}
Here \(K_{TJJ}^{\mu\nu}\) is the perturbative hard kernel obtained by inserting the correlator \(TJJ\) at leading nontrivial order in the strong coupling. The light-cone momentum fractions \(\alpha_{1,2}\in[0,1]\) label the momentum carried by the quark in each pion DA.

We decompose the kernel in the tensor basis constructed from the pion momenta \(P_1\), \(P_2\) and the metric,
\begin{equation}\label{chza:eq:Kdecomp}
	K_\pi^{\mu\nu}(\alpha_1,\alpha_2,q)
	= \sum_{i=1}^5 F_i^{TJJ}(\alpha_1,\alpha_2,q)\, T_i^{\mu\nu},
\end{equation}
following the same decomposition previously used for the tree-level kernel in Eq. \eqref{decompkpi}
\begin{equation}\label{chza:eq:tensorbasis}
	T_1^{\mu\nu}=P_1^\mu P_2^\nu,\qquad
	T_2^{\mu\nu}=P_2^\mu P_1^\nu,\qquad
	T_3^{\mu\nu}=P_1^\mu P_1^\nu,\qquad
	T_4^{\mu\nu}=P_2^\mu P_2^\nu,\qquad
	T_5^{\mu\nu}=g^{\mu\nu}.
\end{equation}
The form-factor coefficients \(F_i^{TJJ}\) encode the full dependence on the partonic fractions and the hard scale \(q^2\). In the expressions below we keep the superscript \(TJJ\) to emphasize that these are the pieces originating from the non-Abelian \(TJJ\) insertion, while the tree-level kernel pieces will be denoted by \(F_i^{\rm tree}\).

\subsection{Explicit form of the kernel coefficients and their interpretation}

From the perturbative calculation one obtains the following compact form for the anomaly-induced pieces. For the momentum-space form factors one finds\allowdisplaybreaks
\begin{align}\label{chza:eq:F12}
	F_1^{TJJ}=F_2^{TJJ}=\frac{1}{3(\alpha_1-1)\alpha_1(\alpha_2-1)\alpha_2 q^2}\  \Biggl[&{q}^2 \bigg(4  \left(6 \text{$\alpha_1$} \text{$\alpha_2$}+(2 \text{$\alpha_1 $}-5) \text{$\alpha_1 $}+2 \text{$\alpha_2 $}^2-5 \text{$\alpha_2$}+2\right){A_2}\nn\\&+3  (3 \text{$\alpha_1$}+\text{$\alpha_2$}-2){A_3}+2  (-2 \text{$\alpha_1$} \text{$\alpha_2$}+\text{$\alpha_1$}+\text{$\alpha_2$}-2)\mathcal{A}\bigg)\nn\\&+4 \bigg((8 \text{$\alpha_1$}+8 \text{$\alpha_2$}-4) \mathcal{JJ}\left[(\text{$\alpha_2$}-1) {{P_2}}-(\text{$\alpha_1$}-1){{P_1}}\right]\nn\\&
	-4 (2 \text{$\alpha_1$}+2 \text{$\alpha_2$}-3) \mathcal{JJ}\left[\text{$\alpha_1$} {{P_1}}-\text{$\alpha_2$} P_2\right]+4 {A_4}-{C_5}\bigg)\Biggl]
\end{align}
\begin{align}\label{chza:eq:F3}
	F_3^{TJJ}
	=-\frac{2}{3(\alpha_1-1)\alpha_1(\alpha_2-1)\alpha_2 q^2}\  \Biggl[& {q}^2 \left( \left(13 \text{$\alpha_1$}^2+2 \text{$\alpha_1$} (3 \text{$\alpha_2$}-8)+(\text{$\alpha_2$}-2)^2\right){A_2}+(-2 \text{$\alpha_1$} \text{$\alpha_2$}+\text{$\alpha_1$}+\text{$\alpha_2$}-2)\mathcal{A} \right)\nn\\&+2 \bigg(-2 (7 \text{$\alpha_1$}+\text{$\alpha_2$}-6) \mathcal{JJ} \left[\text{$\alpha_1$} {P_1}-\text{$\alpha_2 $} {P_2}\right]\nn\\&+2 (7 \text{$\alpha_1$}+\text{$\alpha_2$}-2) \left(\mathcal{JJ} \left[(\text{$\alpha_2$}-1) {P_2}-(\text{$\alpha_1$}-1) {P_1}\right]\right)+{A_4}-{C_5}\bigg) \Biggl],
\end{align}
\begin{equation}\label{chza:eq:F4}
	F_4^{TJJ}
	=F_3^{TJJ}(\a_1\leftrightarrow\a_2,P_1\leftrightarrow P_2),
\end{equation}
and the scalar, trace-sector coefficient
\begin{align}\label{chza:eq:F5}
	F_5^{TJJ}=&-\frac{1}{3(\alpha_1-1)\alpha_1(\alpha_2-1)\alpha_2 q^2}\Biggl[
	q^2\Bigl(\bigl((\alpha_2-2)^2+\alpha_1^2+(6\alpha_2-4)\alpha_1\bigr)A_2 \nonumber\\
	&\qquad\qquad\qquad\qquad\qquad\qquad
	+2\bigl((2\alpha_2-1)\alpha_1-\alpha_2+2\bigr)\mathcal{A}\Bigr)\nonumber\\
	&\qquad\qquad\qquad\qquad\qquad\qquad
	+2\Bigl(2(\alpha_1+\alpha_2-2)\,\mathcal{JJ}[(\alpha_2-1)P_2-(\alpha_1-1)P_1]\nonumber\\
	&\qquad\qquad\qquad\qquad\qquad\qquad
	-2(\alpha_1+\alpha_2)\,\mathcal{JJ}[\alpha_1 P_1-\alpha_2 P_2]+A_4+2C_5\Bigr)\Biggr].
\end{align}
The coefficient \(\mathcal{A}\) is proportional to the one-loop QCD trace-anomaly coefficient, and it can be written as
\begin{equation}\label{chza:eq:Adef}
	\mathcal{A}=\frac{g_s^2}{16\pi^2}\,\beta_0
\end{equation}
in terms of the leading coefficient of the QCD beta function in Eq. \eqref{chza:eq:beta0def}.
Its appearance in \(F_i^{TJJ}\) signals that the anomaly insertion reweights the trace structure and generates a contribution proportional to the running of the coupling.

The remaining quantities \(A_2\), \(A_3\), \(A_4\), and \(C_5\) are scalar coefficients originating from the loop integrals and from the tensor reduction of the triangle and bubble diagrams contributing to the \(TJJ\) correlator. They are finite functions of the renormalization scale \(\mu\) and of the gluon momenta \(p_1\) and \(p_2\). Their relation to the pion external momenta is fixed by the collinear kinematics,
\begin{equation}\label{chza:eq:p1p2def}
	p_1=\alpha_2 P_2-\alpha_1 P_1,\qquad
	p_2=\bar{\alpha}_2 P_2-\bar{\alpha}_1 P_1,
\end{equation}
with \(\bar{\alpha}_i=1-\alpha_i\). The coefficients \(A_2\), \(A_3\), \(A_4\), and \(C_5\) are understood here as the scalar functions obtained from the off-shell decomposition of the \(TJJ\) correlator after combining the quark and gluon sectors.

The symbol \(\mathcal{JJ}[p]\) denotes the scalar two-point function generated by the vacuum-polarisation substructure of the correlator. In the present normalization it reads
\begin{equation}\label{chza:eq:JJdef}
	\mathcal{JJ}[p]=\frac{g_s^2}{16\pi^2}\left(\frac{2}{3}n_f-\frac{5}{3}C_A\right)p^2 B_0(p^2),
\end{equation}
%where \(B_0(p^2)\) is the standard scalar two-point Passarino--Veltman function in dimensional regularisation. 
At large Euclidean momentum one has the expected logarithmic behavior
\begin{equation}\label{chza:eq:B0asympt}
	B_0(p^2)\sim \ln\!\left(\frac{-p^2}{\mu^2}\right)+\text{const.},
\end{equation}
so that \(\mathcal{JJ}[p]\sim g_s^2 p^2 \ln(-p^2/\mu^2)\).

For the large-\(Q^2\) analysis it is important to observe that \(A_2\) and \(A_3\) do not introduce additional parametric powers of \(Q^2\) beyond those displayed explicitly in Eqs.~\eqref{chza:eq:F12}--\eqref{chza:eq:F5}, whereas \(A_4\), \(C_5\), and \(\mathcal{JJ}[p]\) scale as \(Q^2\), but this growth is compensated by the overall factor \(1/q^2\) in the hard coefficients. As a consequence, the anomaly-induced kernel contributes at order \(1/Q^2\), up to logarithmic corrections.

\subsection{From the kernel to the GFFs \(A_\pi\) and \(D_\pi\)}

Taking the trace of Eq.~\eqref{chza:eq:GFF_definition_improved} gives
\begin{equation}\label{chza:eq:trace_spinzero}
	g_{\mu\nu}\langle \pi(P_2)|T^{\mu\nu}|\pi(P_1)\rangle
	=2P^2 A_\pi(q^2)-\frac{3}{2}q^2D_\pi(q^2)
	=2\left(m_\pi^2-\frac{q^2}{4}\right)A_\pi(q^2)-\frac{3}{2}q^2D_\pi(q^2),
\end{equation}
where in the last equality we used \(P_1^2=P_2^2=m_\pi^2\). In the hard, chiral limit used in the factorized calculation, \(q^2=-Q^2\) and \(m_\pi^2\ll Q^2\), so
\begin{equation}\label{chza:eq:trace_hard_chiral}
	g_{\mu\nu}\langle \pi(P_2)|T^{\mu\nu}|\pi(P_1)\rangle
	\simeq -\frac{q^2}{2}\left[A_\pi(q^2)+3D_\pi(q^2)\right]
	=\frac{Q^2}{2}\left[A_\pi(-Q^2)+3D_\pi(-Q^2)\right].
\end{equation}
This relation makes explicit why the anomaly insertion is naturally tied to the scalar projection of the pion matrix element. Since the full QCD energy-momentum tensor is conserved, no independent form factor multiplying \(g^{\mu\nu}\) is needed in Eq.~\eqref{chza:eq:GFF_definition_improved}. Terms of this type may appear in separated quark and gluon sectors, or in the off-shell decomposition of the \(TJJ\) correlator, but after imposing the Ward or Slavnov--Taylor identities and matching onto the on-shell pion matrix element they are absorbed into the two conserved form factors \(A_\pi\) and \(D_\pi\). In the present leading-power matching the anomaly coefficient \(\mathcal{A}\) enters through the trace-sector coefficient \(F_5^{TJJ}\), and therefore gives its most direct correction to \(D_\pi\).

For the numerical interpretation it is important to keep the full \(TJJ\) insertion distinct from its beta-function part.  In the \(A_\pi\) projection the term proportional to the anomaly coefficient cancels, so the isolated anomaly does not contribute to the momentum form factor.  The full \(TJJ\) correction can nevertheless shift the leading Sudakov curve downward at low \(Q^2\), because it also contains non-anomalous tensor structures.  In the \(D_\pi\) projection the anomaly survives and gives an important scalar contribution.  The plotted \(TJJ\) term, however, is not the complete order-\(\alpha_s^2\) hard kernel; additional non-anomalous radiative terms may still modify the comparison with the lattice data.

Using this decomposition, one projects onto the two pion GFFs as
\begin{align}
	A_\pi(q^2)
	&= \frac{f_\pi^2 C_F g_s^2}{N_C} \int_0^1 d\alpha_1 \int_0^1 d\alpha_2\, \tilde{\phi}_\pi(\alpha_1)\,
	\big(F_1^{\rm tree}+F_1^{TJJ}+F_3^{\rm tree}+F_3^{TJJ}\big)\,
	\tilde{\phi}_\pi(\alpha_2),\label{chza:eq:Api}\\[6pt]
	D_\pi(q^2)
	&= - \frac{f_\pi^2 C_F g_s^2}{N_C}\int_0^1 d\alpha_1 \int_0^1 d\alpha_2\, \tilde{\phi}_\pi(\alpha_1)\,
	\frac{2\big(F_5^{\rm tree}+F_5^{TJJ}\big)}{q^2}\,
	\tilde{\phi}_\pi(\alpha_2).\label{chza:eq:Dpi}
\end{align}
Here the \(F_i^{\rm tree}\) denote the tree-level contributions of the hard kernel, while the \(F_i^{TJJ}\) encode the one-loop \(TJJ\) corrections, including the anomaly-controlled trace sector of the non-Abelian correlator.

These expressions make the role of the trace anomaly particularly transparent. First, the \(D\)-term is directly sensitive to the coefficient \(\mathcal{A}\), and therefore to the QCD beta-function coefficient \(\beta_0\). This is natural, since the scalar part of the matrix element probes the trace sector of the stress tensor, whose anomalous contribution is proportional to \(\beta(g_s)\). The result is a correction to \(D_\pi\) proportional to \(\beta_0 g_s^2\), which distinguishes it from the non-anomalous contributions already present at tree level.

Second, the dependence on the two-point function \(B_0(p^2)\), appearing through \(\mathcal{JJ}\) and implicitly in the remaining scalar coefficients, generates the expected logarithmic dependence on the hard scale. After convolution with the pion DA, these logarithms translate into scale-dependent terms of the form \(\ln(Q^2/\mu^2)\), superimposed on the overall \(1/Q^2\) suppression. The renormalization scale \(\mu\) therefore enters both through the hard kernel and through the evolution of the pion DA \(\tilde{\phi}_\pi(\alpha;\mu)\).

Finally, one must examine the endpoint behavior of the convolution integrals. The denominators \((\alpha_i-1)\alpha_i\) appearing in the coefficients suggest possible singularities as \(\alpha_i\to 0,1\). In practice, these are regulated by the pion DA, which vanishes linearly at the endpoints. 
Therefore the convolution integrals in Eqs.~\eqref{chza:eq:Api} and \eqref{chza:eq:Dpi} are finite at leading twist. 
Only when one goes beyond the strict collinear approximation, for instance by including transverse-momentum effects or power-suppressed endpoint-enhanced terms, does one expect a stronger sensitivity to the endpoint region and the need for an extended factorization framework.

To verify the gauge independence of the physical pion GFFs, let us examine the structure of the three-point Green's function embedded within the hard-scattering kernel. The off-shell Slavnov-Taylor identities dictate that all gauge-fixing dependent modifications are strictly confined to two isolated structures: the non-local longitudinal sector $\mathcal{B}_g$ and the local operator sector denoted by $\langle t^{\mu\nu}(q)j^{a\,\alpha}(p_1)j_{\text{loc}}^{b\,\beta}(p_2)\rangle$. At the diagrammatic level, any unphysical variation contributing to this Green's function yields structures that scale explicitly with the inverse gluon propagators, and are thus proportional to the virtualities $p_1^2$ and $p_2^2$. In collinear kinematics, as stated previously, these off-shell lines correspond to $p_1^2 = -\alpha_1 \alpha_2 Q^2, \quad \text{and} \quad p_2^2 = -(1-\alpha_1)(1-\alpha_2)Q^2.$ Crucially, every form factor residing within the $\langle t^{\mu\nu}(q)j^{a\,\alpha}(p_1)j_{\text{loc}}^{b\,\beta}(p_2)\rangle$ sector is automatically cancelled out upon algebraic contraction with the valence quark projector of the pion state. The only remaining contribution from this gauge-dependent sector is the coefficient $C_5$ which only appears in $D_\pi$. It tracks the remnant off-shell equations of motion and vanishes identically once the full phase-space integration over the hard kernel is performed against the pion DAs. In other words, while individual off-shell lines inside the unintegrated loop parameterize gauge variations, the collinear projection mapping acts as a natural gauge filter. Since these unphysical terms vanish automatically upon convolution, complete gauge independence is fully restored to the hard-scattering kernel. This ensures that our final extraction of the $D_\pi$ form factor safely isolates the physical contribution.

In summary, the insertion of the one-loop non-Abelian \(TJJ\) correlator into the hard kernel generates a calculable correction to the pion GFFs. Its trace sector is proportional to the QCD beta function and therefore represents the perturbative realization of the conformal anomaly inside the hard-scattering description. After convolution with the pion DA, this produces a contribution of order \(1/Q^2\), modulated by logarithmic dependence on the hard scale.

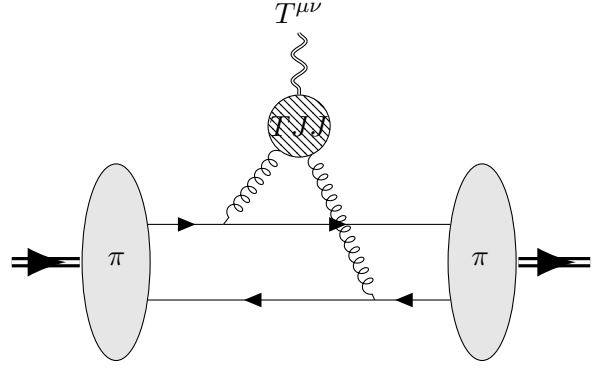
\begin{figure}
	\centering
	\begin{tikzpicture}
		\begin{feynman}
			\vertex (i1);
			\vertex[right=1cm of i1] (a1);
			\vertex[right=3cm of a1] (o1) ;
			\vertex[below=1cm of i1] (i2) ;
			\vertex[right=3cm of i2] (a2);
			\vertex[right=1cm of a2] (o2) ;
			
			%\vertex[below=2cm of i3] (t1);
			\vertex[above=1cm of a1] (t2);
			\vertex[above=2cm of a2] (t3);
			\vertex[blob] (t1) at (2,1.3) {{$TJJ$}};
			%\vertex[above right =1.3cm of t2] (t1);
			\vertex[above=1.5cm of t1] (i3) {$T^{\mu\nu}$};
			
			\diagram* { 
				(i1)  -- [fermion] (a1)   --[fermion] (o1),
				
				(i2)  -- [anti fermion] (a2)  --[anti fermion] (o2),

				(i3)  -- [graviton] (t1),
				(t1) -- [gluon] (a1) ,
				(t1) -- [gluon] (a2),
			};
			
			\node[] at (-0.35,-0.45) {$\pi\ $};
			\node[] at (4.46,-0.45) {$\pi\ $};

			\filldraw[fill=black, fill opacity=0.1] (-0.42,-0.5) ellipse (.45cm and 1.3cm);
			\filldraw[fill=black, fill opacity=0.1] (4.42 ,-0.5) ellipse (.45cm and 1.3cm);
			
			\draw[double, double distance=1.5,line width=1.2, postaction={decorate, decoration={
					markings,% switch on markings
					mark=at position 0.8 with  {\arrow{ latex} } } } ] (-1.8,-0.5) -- node [above] {} (-0.9,-0.5);
			
			\draw[double, double distance=1.5,line width=1.2, postaction={decorate, decoration={
					markings,% switch on markings
					mark=at position 0.8 with  {\arrow{ latex} } } } ] (4.9,-0.5) -- node [above] {} (5.85,-0.5);
		\end{feynman}
	\end{tikzpicture}
	\caption{TJJ contribution to the GFF of the pion.}
	\label{chza:fig:1}
\end{figure}

\section{Sudakov corrections and modification of the pion wave function}\label{chza:phenomesec1}

In the previous sections the factorized description of the pion GFFs has been formulated in the collinear approximation, where the transverse momenta of the partons are neglected and the nonperturbative dynamics is entirely encoded in the DA \(\tilde{\phi}_\pi(\alpha)\). While this framework correctly captures the leading-power structure at large momentum transfer, it is well known that radiative corrections generate large logarithms associated with soft and collinear gluon emission. These logarithms must be resummed in order to obtain a reliable description of the hard amplitude, and their resummation leads to the emergence of Sudakov suppression factors.

The inclusion of Sudakov effects requires a refinement of the factorization scheme, in which transverse degrees of freedom are retained and the pion DA is promoted from a purely collinear DA to a transverse-momentum-dependent object. In this section we describe how Sudakov resummation modifies the structure of the pion DA and how it affects the convolution formula for the GFFs.

The perturbative corrections to the hard kernel involve loop integrals in which virtual gluons become soft or collinear with respect to the external partons. In these regions the amplitude develops double logarithmic enhancements of the form \cite{Sterman:1986aj} \cite{Botts:1989kf}\cite{Li:1992nu,Li:1994cka}
\begin{equation}\label{chza:eq:sudakov_logs}
	\alpha_s \ln^2\!\left(\frac{Q^2}{k_\perp^2}\right),
\end{equation}
where \(k_\perp\) is the transverse momentum of the parton inside the pion. Such contributions arise from the overlap of soft and collinear divergences and appear to all orders in perturbation theory. In the collinear approximation these logarithms are absent.

The resummation of these logarithms leads to an exponential suppression factor, which damps the contributions from regions of phase space where soft-gluon emission would otherwise enhance the amplitude. This suppression plays a crucial role in regulating endpoint configurations and ensuring the self-consistency of the perturbative expansion.

To incorporate Sudakov effects, one introduces a transverse-momentum-dependent (TMD) pion wave function
\begin{equation}\label{chza:eq:TMDwf}
	\Psi_\pi(\alpha,k_\perp,\mu),
\end{equation}
which generalizes the collinear DA according to
\begin{equation}\label{chza:eq:LCDA_reduction}
	\tilde{\phi}_\pi(\alpha,\mu)=\int d^2 k_\perp\, \Psi_\pi(\alpha,k_\perp,\mu).
\end{equation}
It is convenient to work in impact-parameter space, defined by the Fourier transform
\begin{equation}\label{chza:eq:bspace}
	\tilde{\Psi}_\pi(\alpha,b,\mu)=\int \frac{d^2 k_\perp}{(2\pi)^2}\,
	e^{-i \vec{k}_\perp\cdot \vec{b}}\,
	\Psi_\pi(\alpha,k_\perp,\mu),
\end{equation}
where \(b\) is conjugate to \(k_\perp\). In this representation, the Sudakov factor exponentiates in a simple form.

The factorized expression for the matrix element \eqref{chza:eq:factorisation} is then modified into
\begin{align}\label{chza:eq:factorisation_sudakov}
	\langle\pi(P_2)|T^{\mu\nu}|\pi(P_1)\rangle
	= \frac{if_\pi^2 C_Fg_s^2}{N_C}
	\int_0^1 d\alpha_1 \int_0^1 d\alpha_2
	\int d^2 b\;
	\tilde{\Psi}_\pi(\alpha_1,b,\mu)\,
	K_\pi^{\mu\nu}(\alpha_1,\alpha_2,b,Q,\mu)\,
	\tilde{\Psi}_\pi(\alpha_2,b,\mu),
\end{align}
where the hard kernel now depends explicitly on the transverse separation \(b\), and the convolution includes the Sudakov suppression factor.

The resummation of soft and collinear logarithms leads to a universal exponential factor modifying the wave function,
\begin{equation}\label{chza:eq:sudakov_factor}
	\tilde{\Psi}_\pi(\alpha,b,Q,\mu)
	=
	\exp\!\left[-S(\alpha,b,Q,\mu)\right]\,
	\tilde{\Psi}_\pi^{(0)}(\alpha,b,\mu),
\end{equation}
where \(\tilde{\Psi}_\pi^{(0)}\) is a nonperturbative input function and \(S(\alpha,b,Q,\mu)\) is the Sudakov exponent.

At next-to-leading logarithmic accuracy, the Sudakov exponent can be written as
\begin{equation}\label{chza:eq:sudakov_explicit}
	S(\alpha,b,Q,\mu)
	=
	s(\alpha, Q,b)+s(1-\alpha,Q,b)
	+2\int_{1/b}^{\mu}\frac{d\mu'}{\mu'}\gamma_q(\alpha_s(\mu')),
\end{equation}
where \(\gamma_q\) is the quark anomalous dimension and the function \(s(\alpha,Q,b)\) resums the double logarithms,
\begin{equation}\label{chza:eq:s_function}
	s(\alpha,Q,b)
	=
	\int_{1/b}^{\alpha\,Q}\frac{d\mu'}{\mu'}\left[
	\ln\!\left(\frac{\alpha\,Q}{\mu'}\right) A(\alpha_s(\mu'))
	+ B(\alpha_s(\mu'))
	\right].
\end{equation}
The coefficient \(A(\alpha_s)\) is proportional to the cusp anomalous dimension and controls the double logarithmic behavior, while \(B(\alpha_s)\) encodes single logarithmic corrections.
The Sudakov suppression factor can be analytically computed and expressed as 
\begin{equation}\label{chza:eq:sudakov_explicit_results}
	S(\alpha,b,Q,\mu)
	= s(\alpha, Q,b)+s(1-\alpha,Q,b)-\frac{1}{\b_1}\log \frac{ \log (t/\Lambda_{ \rm QCD})}{\log (1/b\Lambda_{\rm QCD})}
	%+2\int_{1/b}^{\mu}\frac{d\mu'}{\mu'}\gamma_q(\alpha_s(\mu')),
\end{equation}
with
\begin{align}
	s(\alpha, Q,b) = &\frac{A^{(1)}}{2\beta_1} \hat{q} \ln \left( \frac{\hat{q}}{\hat{b}} \right) + \frac{A^{(2)}}{4\beta_1^2} \left( \frac{\hat{q}}{\hat{b}} - 1 \right) - \frac{A^{(1)}}{2\beta_1} (\hat{q} - \hat{b})-\frac{A^{(1)}\beta_2}{16\beta_1^3}\hat{q}\left[\frac{\ln(2\hat{b})+1}{\hat{b}} - \frac{\ln(2\hat{q})+1}{\hat{q}}\right]\nn\\
	&- \left[\frac{A^{(2)}}{4\beta_1^2} - \frac{A^{(1)}}{4\beta_1}\ln\left(\frac{e^{2\gamma-1}}{2}\right)\right]\ln\left(\frac{\hat{q}}{\hat{b}}\right)- \frac{A^{(1)}\beta_2}{32\beta_1^3}\left[\ln^2(2\hat{q}) - \ln^2(2\hat{b})\right]
\end{align}
and
\begin{equation}
	\hat{q} \equiv \ln \left(\alpha Q / (\sqrt{2} \Lambda_{ \rm QCD}) \right)
\end{equation}
\begin{equation}
	\hat{b} \equiv \ln (1 / b \Lambda_{\rm QCD})
\end{equation}
\begin{equation}
	t=\text{max}\left(Q\sqrt{\a_1\a_2},{1/b}\right)
\end{equation}
The coefficients $A^{(i)}$ and $\beta_i$ are
\begin{equation}
	\beta_1 = \frac{33 - 2n_f}{12}\,, \quad \beta_2 = \frac{153 - 19n_f}{24}\,,
\end{equation}
\begin{equation}
	A^{(1)} = \frac{4}{3}\,, \quad A^{(2)} = \frac{67}{9} - \frac{\pi^2}{3} - \frac{10}{27}n_f + \frac{8}{3}\beta_1 \ln\left(\frac{e^\gamma}{2}\right)\,,
\end{equation}
where $n_f = 3$ in this case is the number of quark flavors and $\gamma$ is the Euler constant.

The exponential factor \(\exp[-S]\) suppresses configurations with large transverse separation \(b\), corresponding to soft gluon exchange and long-distance dynamics. As a result, the dominant contribution to the convolution integral comes from the perturbative region \(b\lesssim 1/\Lambda_{\rm QCD}\), where the use of the hard kernel remains justified.

\subsection{Impact on the anomaly-induced amplitude}

The inclusion of Sudakov effects modifies the convolution structure of the anomaly-induced contributions. In particular, the form factors \(F_i^{TJJ}\) are now convoluted with the Sudakov-suppressed wave functions rather than with the purely collinear DA. Symbolically, one has
\begin{equation}\label{chza:eq:F_sudakov}
	F_i^{TJJ} \;\longrightarrow\;
	\int d^2 b \;
	\tilde{\Psi}_\pi(\alpha_1,b,Q,\mu)\,
	F_i^{TJJ}(\alpha_1,\alpha_2,q,b)\,
	\tilde{\Psi}_\pi(\alpha_2,b,Q,\mu).
\end{equation}
The most important consequence of this modification is the suppression of endpoint regions \(\alpha\to 0,1\), where the denominators in Eqs.~\eqref{chza:eq:F12}--\eqref{chza:eq:F5} would otherwise enhance the integrand. The Sudakov factor therefore stabilizes the perturbative expansion and reduces the sensitivity to nonperturbative configurations.

Furthermore, the logarithmic dependence introduced by the two-point function \(B_0(p^2)\), which generates terms of the form \(\ln(Q^2/\mu^2)\), is now consistently combined with the Sudakov logarithms arising from soft-gluon resummation. The resulting structure reflects the interplay between ultraviolet renormalization, governed by the QCD beta function and the anomaly coefficient \(\mathcal{A}\), and infrared dynamics, encoded in the Sudakov exponent.

The Sudakov suppression provides a dynamical mechanism that enforces the dominance of short-distance configurations in exclusive processes at large momentum transfer. In the present context, it ensures that the anomaly-induced contribution to the GFFs remains under perturbative control, even in the presence of potentially dangerous endpoint configurations.

At the same time, the modification of the pion wave function reflects the fact that the hadronic structure probed in the hard process is not purely collinear, but involves a nontrivial interplay between longitudinal momentum fractions and transverse spatial distributions. The Sudakov factor effectively reshapes the wave function, suppressing large transverse separations and enhancing the contribution from compact configurations.

In summary, the inclusion of Sudakov corrections leads to a consistent extension of the factorization framework in which both perturbative and nonperturbative effects are properly organized. This refinement is essential for a quantitatively reliable description of the anomaly-induced contributions to the pion GFFs at large but finite momentum transfer.

\section{A phenomenological model for the Sudakov-improved pion wave function and its impact on the anomaly contribution}\label{chza:phenomesec2}

In order to make the Sudakov-resummation-improved factorization formula operational, one needs a concrete model for the pion wave function in transverse coordinate space. The purpose of this section is to introduce a simple but widely used phenomenological parametrization, to combine it with the Sudakov exponential, and to show how the anomaly-induced contribution to the pion GFFs is modified once the transverse degrees of freedom are taken into account explicitly. The discussion also makes clear how the endpoint suppression induced by Sudakov resummation stabilizes the convolution integrals that define the hard amplitude.

The starting point is the transverse-momentum-dependent pion wave function \(\Psi_\pi(\alpha,k_\perp,\mu)\), whose Fourier transform to impact-parameter space has been introduced in Eq.~\eqref{chza:eq:bspace}. A natural phenomenological ansatz consists in factorizing the longitudinal and transverse dependence of the nonperturbative input at a low scale \(\mu_0\),
\begin{equation}\label{chza:eq:psi_model_k}
	\Psi_\pi^{(0)}(\alpha,k_\perp,\mu_0)
	=
	\phi_\pi(\alpha,\mu_0)\,\Sigma_\pi(\alpha,k_\perp),
\end{equation}
where \(\phi_\pi(\alpha,\mu_0)\) is the usual leading-twist DA and \(\Sigma_\pi\) describes the intrinsic transverse momentum profile \cite{Lepage:1980fj,Efremov:1979qk,Chernyak:1983ej}\cite{Braun:1989iv,Ball:2006wn,Jakob:1993iw}\cite{Huang:1994dy}

A commonly adopted Gaussian form is
\begin{equation}\label{chza:eq:gaussian_k}
	\Sigma_\pi(\alpha,k_\perp)
	=
	\frac{\beta_\pi^2}{\pi\,\alpha(1-\alpha)}
	\exp\!\left[-\beta_\pi^2\frac{k_\perp^2+m_q^2}{\alpha(1-\alpha)}\right],
\end{equation}
where \(\beta_\pi\) is a hadronic size parameter and \(m_q\) is an effective constituent-like quark mass introduced to regulate the infrared behavior of the model. The normalization is chosen so that
\begin{equation}\label{chza:eq:sigma_norm}
	\int d^2k_\perp\,\Sigma_\pi(\alpha,k_\perp)
	=
	\exp\!\left[-\beta_\pi^2\frac{m_q^2}{\alpha(1-\alpha)}\right].
\end{equation}
When \(m_q\to 0\), the transverse profile integrates to unity, and the collinear DA is recovered directly from the transverse-momentum-dependent wave function. For finite \(m_q\), the exponential factor produces an additional suppression near the endpoints, which is phenomenologically useful in the modeling of exclusive amplitudes.

Passing to impact-parameter space, the Fourier transform of Eq.~\eqref{chza:eq:gaussian_k} gives
\begin{equation}\label{chza:eq:gaussian_b}
	\tilde{\Sigma}_\pi(\alpha,b)
	=
	\exp\!\left[-\beta_\pi^2\frac{m_q^2}{\alpha(1-\alpha)}\right]
	\exp\!\left[-\frac{\alpha(1-\alpha)b^2}{4\beta_\pi^2}\right],
\end{equation}
up to the conventional overall normalization associated with the definition of the Fourier transform. The corresponding pion wave function in \(b\)-space may then be written as
\begin{equation}\label{chza:eq:psi0_b}
	\tilde{\Psi}_\pi^{(0)}(\alpha,b,\mu_0)
	=
	\phi_\pi(\alpha,\mu_0)\,
	\exp\!\left[-\beta_\pi^2\frac{m_q^2}{\alpha(1-\alpha)}\right]
	\exp\!\left[-\frac{\alpha(1-\alpha)b^2}{4\beta_\pi^2}\right].
\end{equation}
The Sudakov-improved wave function is obtained by multiplying this nonperturbative input by the resummed exponential factor introduced in Eq.~\eqref{chza:eq:sudakov_factor},
\begin{equation}\label{chza:eq:psi_b_sud}
	\tilde{\Psi}_\pi(\alpha,b,Q,\mu)
	=
	\exp\!\left[-S(\alpha,b,Q,\mu)\right]\,
	\tilde{\Psi}_\pi^{(0)}(\alpha,b,\mu_0).
\end{equation}
Substituting Eq.~\eqref{chza:eq:psi0_b} into Eq.~\eqref{chza:eq:psi_b_sud}, one finds the explicit form
\begin{equation}\label{chza:eq:psi_b_full}
	\tilde{\Psi}_\pi(\alpha,b,Q,\mu)
	=
	\phi_\pi(\alpha,\mu_0)\,
	\exp\!\left[-S(\alpha,b,Q,\mu)\right]
	\exp\!\left[-\beta_\pi^2\frac{m_q^2}{\alpha(1-\alpha)}\right]
	\exp\!\left[-\frac{\alpha(1-\alpha)b^2}{4\beta_\pi^2}\right].
\end{equation}
This expression makes the structure of the improved wave function transparent. The first exponential in Eq.~\eqref{chza:eq:psi_b_full} is perturbative and universal, encoding the resummation of soft and collinear logarithms. The second and third exponentials represent the intrinsic nonperturbative transverse structure of the pion. Their combined effect is to suppress both large transverse separations and endpoint regions, thus improving the convergence of the hard convolution.

The DA entering Eq.~\eqref{chza:eq:psi_b_full} may itself be modelled at the input scale \(\mu\) by a truncated Gegenbauer expansion keeping only the first nontrivial term,
\begin{equation}\label{chza:eq:phi_model_simple}
	\phi_\pi(\alpha,\mu)
	=
	6\alpha(1-\alpha)
	\left[
	1+a_2(\mu)\,C_2^{3/2}(2\alpha-1)
	\right],
\end{equation}
which already captures the main departure from the asymptotic shape and is sufficient for a first numerical analysis.
The Gegenbauer moment $a_2(\mu)$ is given by
\begin{equation}
	a_2(\mu)
	=
	a_2(\mu_0)
	\left(
	\frac{\alpha_s(\mu)}{\alpha_s(\mu_0)}
	\right)^{\frac{25\,C_F}{6}},
\end{equation}
and the running coupling is evaluated using Eq.~\eqref{chza:evolalph} at the largest hard scale in the convolution. The parameter $a_2(\mu_0)=0.2$ is chosen at the scale $\mu_0=1$ GeV following \cite{BAKULEV2001279}.

The hard amplitude for the pion GFFs is then rewritten as a threefold convolution in the longitudinal fractions and in the transverse separation,
\begin{align}\label{chza:eq:full_conv_sud}
	\langle\pi(P_2)|T^{\mu\nu}|\pi(P_1)\rangle
	=
	\frac{if_\pi^2 C_Fg_s^2}{N_C}
	\int_0^1 d\alpha_1
	\int_0^1 d\alpha_2
	\int_0^\infty 2\pi b\,db\;
	\tilde{\Psi}_\pi(\alpha_1,b,Q,\mu)\,
	K_\pi^{\mu\nu}(\alpha_1,\alpha_2,b,Q,\mu)\,
	\tilde{\Psi}_\pi(\alpha_2,b,Q,\mu).
\end{align}
Here the hard kernel includes both the tree-level and the anomaly-induced components,
\begin{equation}\label{chza:eq:kernel_split_sud}
	K_\pi^{\mu\nu}(\alpha_1,\alpha_2,b,Q,\mu)
	=
	K_{{\rm tree}}^{\mu\nu}(\alpha_1,\alpha_2,b,Q,\mu)
	+
	K_{TJJ}^{\mu\nu}(\alpha_1,\alpha_2,b,Q,\mu).
\end{equation}
The kernels appearing in Eq.~\eqref{chza:eq:kernel_split_sud} are obtained from the corresponding momentum-space amplitudes through a Fourier transformation with respect to the transverse momentum exchanged by the hard gluon,
\begin{equation}
	K_\pi^{\mu\nu}(\alpha_1,\alpha_2,b,Q,\mu)
	=
	\int \frac{d^2{k}_T}{(2\pi)^2}\,
	e^{-i{k}_T\cdot{b}}\,
	K_\pi^{\mu\nu}(\alpha_1,\alpha_2,{k}_T,Q,\mu),
\end{equation}
where ${b}$ denotes the transverse separation between the valence quarks in impact-parameter space. The same definition applies separately to the tree-level and to the $TJJ$ contributions.

The transverse momentum dependence is retained exclusively in the propagator of the exchanged gluon, while the intrinsic transverse momenta carried by the external constituent quarks are neglected in the quark propagators and in the Dirac numerators.
This approximation isolates the dominant non-collinear contribution associated with the exchanged hard gluon and allows the Fourier transform to be performed analytically in terms of modified Bessel functions.

The simplification becomes particularly important for the anomaly-induced $TJJ$ contribution. Indeed, the insertion generates non-local momentum structures and logarithmic terms in the gluon virtuality, making the exact two-dimensional Fourier transform analytically intractable if the full transverse dependence of all propagators is retained. Restricting the transverse dynamics to the exchanged gluon propagator preserves the factorized structure of the kernel and leads to a computable representation in impact-parameter space. In this way, both the tree-level and the anomaly-corrected amplitudes can be expressed in terms of one-dimensional Bessel transforms, which are suitable for stable numerical evaluation.

The explicit expressions of the tree-level kernel, the modified propagator structure induced by the $TJJ$ insertion, and the corresponding impact-parameter-space kernels are collected in Appendix~\ref{chza:kernelexpb}, together with a detailed discussion of the approximations employed in the treatment of the transverse momentum dependence.

Projecting again onto the two independent form factors, one obtains the Sudakov-improved expressions
\begin{align}
	A_\pi(q^2)
	&= \frac{f_\pi^2 C_F g_s^2}{N_C}
	\int_0^1 d\alpha_1
	\int_0^1 d\alpha_2
	\int_0^\infty 2\pi b\,db\;
	\tilde{\Psi}_\pi(\alpha_1,b,Q,\mu)
	\Bigl[
	F_1^{\rm tree}+F_1^{TJJ}+F_3^{\rm tree}+F_3^{TJJ}
	\Bigr]
	\tilde{\Psi}_\pi(\alpha_2,b,Q,\mu),
	\label{chza:eq:Api_sud}
	\\[6pt]
	D_\pi(q^2)
	&=
	- \frac{f_\pi^2 C_F g_s^2}{N_C}
	\int_0^1 d\alpha_1
	\int_0^1 d\alpha_2
	\int_0^\infty 2\pi b\,db\;
	\tilde{\Psi}_\pi(\alpha_1,b,Q,\mu)
	\frac{2\bigl(F_5^{\rm tree}+F_5^{TJJ}\bigr)}{q^2}
	\tilde{\Psi}_\pi(\alpha_2,b,Q,\mu),
	\label{chza:eq:Dpi_sud}
\end{align}
where the \(F_i\) now depend implicitly on \(b\) through the transverse-momentum dependence retained in the hard scattering kernel, and are explicitly given in Appendix~\ref{chza:kernelexpb}. The anomaly-induced correction to the \(D\)-term is therefore given by
\begin{equation}\label{chza:eq:Dpi_anom_sud}
	D_\pi^{\rm anom}(q^2)
	=
	-\frac{f_\pi^2 C_F g_s^2}{N_C}
	\int_0^1 d\alpha_1
	\int_0^1 d\alpha_2
	\int_0^\infty 2\pi b\,db\;
	\tilde{\Psi}_\pi(\alpha_1,b,Q,\mu)
	\frac{2F_5^{TJJ}(\alpha_1,\alpha_2,b,q,\mu)}{q^2}
	\tilde{\Psi}_\pi(\alpha_2,b,Q,\mu).
\end{equation}

At this stage it is useful to examine the effect of the Sudakov improvement on the anomaly contribution. In the purely collinear treatment, the form factor \(F_5^{TJJ}\) contains denominators proportional to
\begin{equation}\label{chza:eq:endpoint_den}
	\alpha_1(1-\alpha_1)\alpha_2(1-\alpha_2),
\end{equation}
which enhance the integrand near the endpoints \(\alpha_i\to 0,1\). Although the asymptotic DA \(\phi_\pi^{\rm as}(\alpha)=6\alpha(1-\alpha)\) already renders the convolution finite, the endpoint region may still contribute significantly once logarithmic corrections and non-asymptotic models are included. The Sudakov factor cures this problem dynamically. Indeed, for small \(\alpha\) or \(1-\alpha\), the exponent \(S(\alpha,b,Q,\mu)\) grows, so that
\begin{equation}\label{chza:eq:endpoint_supp}
	\exp[-S(\alpha,b,Q,\mu)]\to 0,
	\qquad
	\alpha\to 0,1.
\end{equation}
This implies a strong suppression of the endpoint configurations that would otherwise endanger the perturbative stability of the result.

A second important effect concerns the \(b\)-integration. Without Sudakov suppression, the convolution in Eq.~\eqref{chza:eq:full_conv_sud} would receive substantial contributions from large \(b\), corresponding to long-distance transverse separations where the perturbative hard kernel cannot be trusted. The Sudakov exponent suppresses precisely this region,
\begin{equation}\label{chza:eq:large_b_supp}
	\exp[-S(\alpha,b,Q,\mu)]\to 0,
	\qquad
	b\to \infty,
\end{equation}
and therefore ensures that the dominant contribution arises from a finite window of transverse separations. In this way the improved factorization formula becomes internally consistent: the nonperturbative information remains encoded in the model wave function, while the hard kernel is sampled mainly in the region where perturbation theory is applicable.

For practical numerical applications, one needs to specify the scales entering the Sudakov exponent. A standard choice is to set the factorization and renormalization scale to the largest perturbative scale in the problem,
\begin{equation}\label{chza:eq:hard_scale_choice}
	t=\max\!\left(\sqrt{\alpha_1\alpha_2}\,Q,\frac{1}{b}\right),
\end{equation}
and to evaluate the running coupling \(\alpha_s(t)\) at this scale. The Sudakov exponent is then written in terms of \(t\), while the DA is evolved from the input scale \(\mu_0\) to the same hard scale by the usual ERBL evolution. In a first phenomenological analysis, however, one may keep the evolution of \(\phi_\pi\) fixed at a representative scale and focus on the dominant effect of the Sudakov exponential itself.

Within such a setup, the principal effect of the Sudakov improvement on the anomaly-induced amplitude can be summarized schematically as
\begin{equation}\label{chza:eq:supp_scheme}
	D_\pi^{\rm anom}(Q^2)
	=
	D_{\pi,\,{\rm coll}}^{\rm anom}(Q^2)\,
	\mathcal{S}_D(Q^2),
\end{equation}
where \(D_{\pi,\,{\rm coll}}^{\rm anom}\) is the result obtained in the collinear approximation and \(\mathcal{S}_D(Q^2)\) is an effective suppression factor generated by the \(b\)-space convolution with the Sudakov-improved wave functions. Although \(\mathcal{S}_D(Q^2)\) is not universal and depends on the chosen model parameters, it is typically smaller than unity and decreases the relative importance of the endpoint regions. In the \(A_\pi(Q^2)\) projection the anomalous part cancels, while the full \(TJJ\) insertion still affects the curve through its non-anomalous tensor components.  The effect is more directly controlled in the \(D\)-term and in the trace form factor because the trace-sector coefficient \(F_5^{TJJ}\) is sensitive to the scalar and logarithmic pieces of the kernel.

The phenomenological parameters entering the model, such as \(\beta_\pi\), \(m_q\), and the Gegenbauer moments \(a_n(\mu_0)\), may be fixed by comparison with other exclusive observables, for instance the electromagnetic pion form factor or the \(\gamma\gamma^*\to\pi\) transition form factor. Once these parameters are chosen, the same Sudakov-improved wave function can be used consistently in the computation of the GFFs. This makes the framework predictive and allows one to assess whether the anomaly-induced correction survives as a numerically appreciable effect once the realistic suppression of long-distance configurations is taken into account.

In summary, the introduction of a phenomenological \(b\)-space model for the pion wave function provides the missing ingredient needed to turn the Sudakov-improved factorization formula into a calculable scheme. The wave function is modified in two essential ways: first, by the intrinsic transverse Gaussian profile, which encodes the finite transverse size of the pion, and second, by the Sudakov exponent, which suppresses large transverse separations and endpoint regions. When these effects are incorporated into the convolution for \(D_\pi(q^2)\), the anomaly-induced contribution remains present but is reshaped and typically reduced relative to the purely collinear estimate. This mechanism is essential for a realistic perturbative analysis of the pion GFFs at finite momentum transfer.

\section{Numerical Results}\label{chza:numericalsec}

\begin{figure}
	\centering
	\includegraphics[scale=0.5]{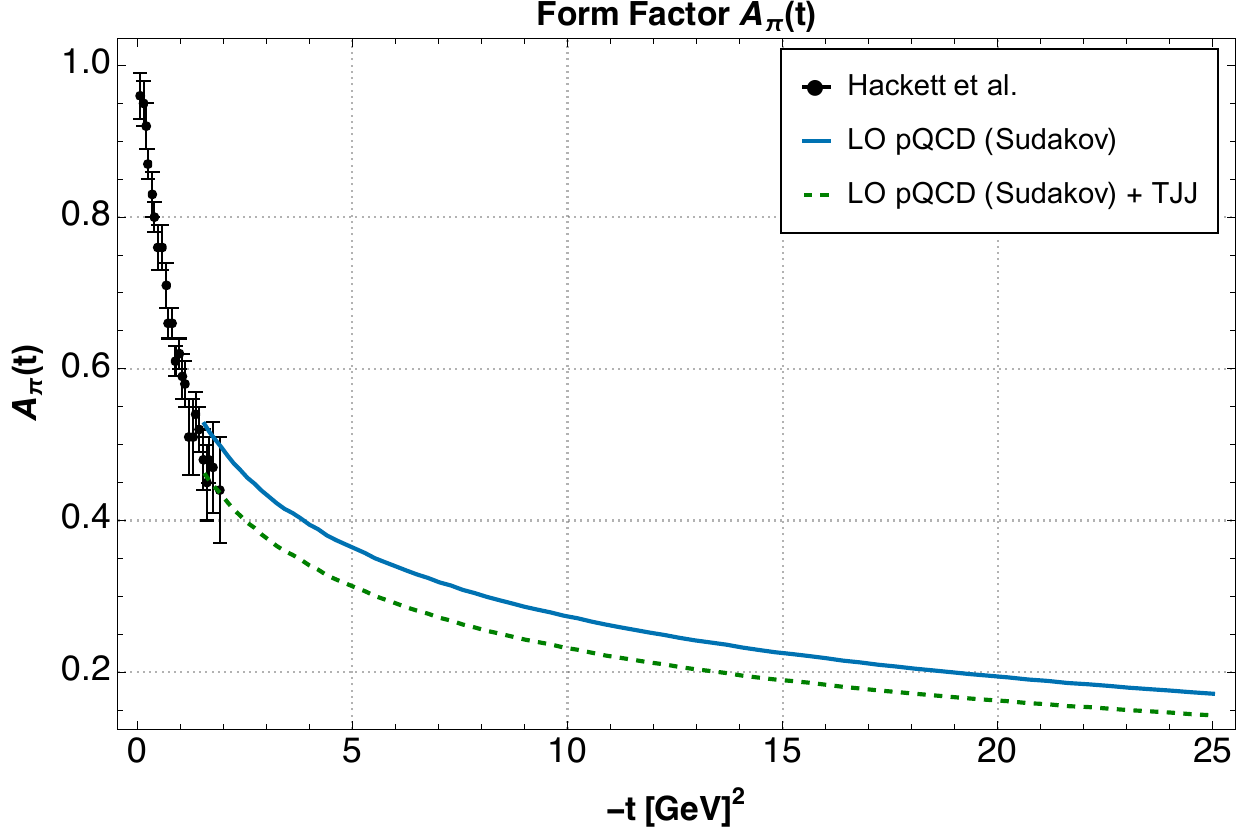}
	\caption{$Q^2=-t$. In the momentum form factor \(A_\pi\), the isolated anomaly cancels and does not generate a separate contribution.  The full \(TJJ\) insertion nevertheless lowers the leading Sudakov curve at small \(Q^2\), improving the local overlap with the lattice data in the first kinematic window. The curves are obtained for $a_2(\mu_0)=0.2$, $\mu_0=1$ GeV, $m_q=0.33\,\, \text{GeV}$ and $\beta_\pi=0.24 \,\,\text{GeV}^{-1}$.}
	\label{chza:ap}
\end{figure}
\begin{figure}
	\centering
	\includegraphics[scale=0.5]{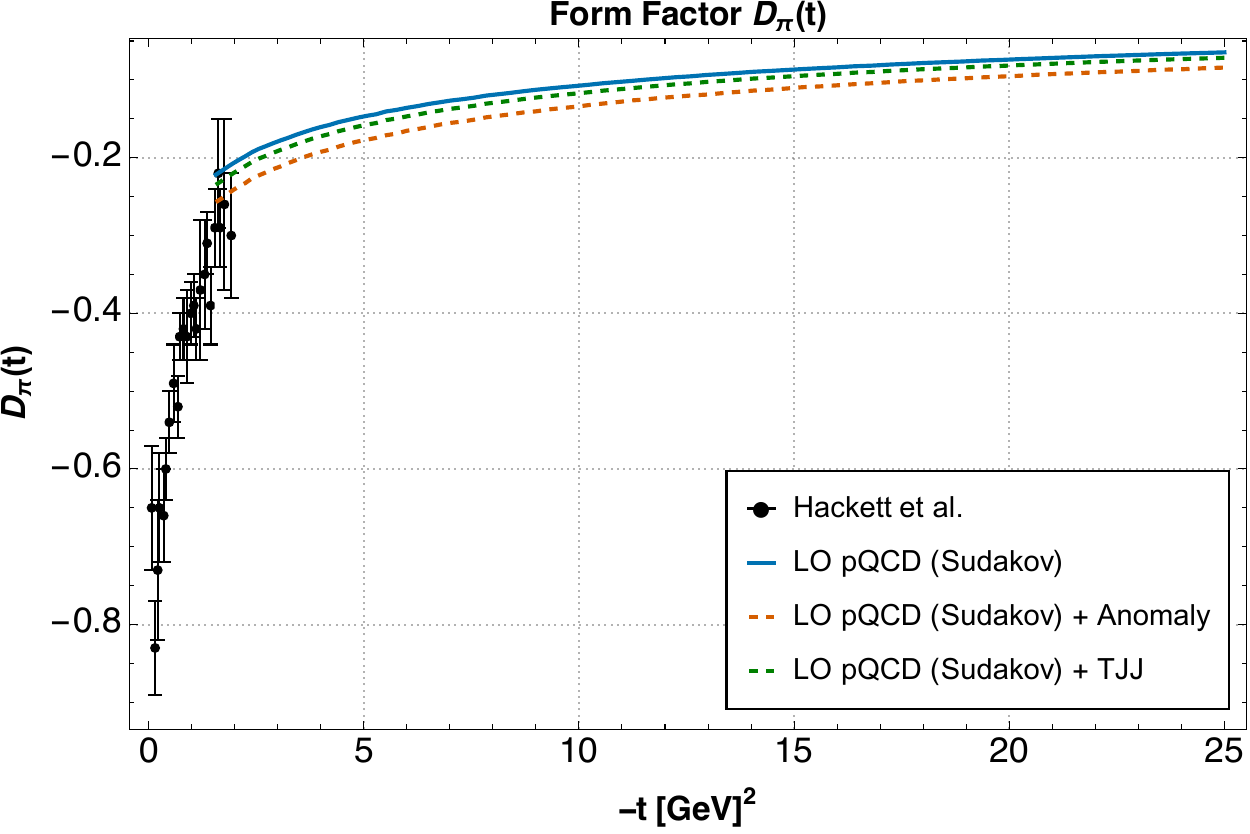}
	\caption{In the \(D_\pi\) form factor the anomaly gives an important contribution.  The displayed full \(TJJ\) correction is more modest and should not be read as the complete order-\(\alpha_s^2\) hard kernel; additional non-anomalous terms may improve the comparison with the data. The curves are obtained for $a_2(\mu_0)=0.2$, $\mu_0=1$ GeV, $m_q=0.33\,\, \text{GeV}$ and $\beta_\pi=0.24 \,\,\text{GeV}^{-1}$.}
	\label{chza:dp}
\end{figure}

In this section we discuss the numerical behavior of the pion GFFs obtained from the Sudakov-improved factorization framework described above, including the contribution generated by the non-Abelian $TJJ$ insertion.

A crucial aspect of the numerical calculation concerns the treatment of the impact-parameter integration.  The transverse separation $b$ is restricted to the perturbative region,
\begin{equation}
	0 \le b \lesssim \frac{1}{\Lambda_{\rm QCD}},
\end{equation}
since larger transverse distances correspond to configurations dominated by genuinely nonperturbative soft-gluon dynamics. In practice, the Sudakov exponent suppresses the integrand strongly as $b$ approaches $1/\Lambda_{\rm QCD}$, leading to a natural dynamical cutoff of the convolution integral. As emphasized in Ref.~\cite{Li:1992nu}, the perturbative contribution becomes increasingly dominated by small transverse separations as the hard scale grows.

In the present analysis we implement this prescription explicitly by truncating the numerical integration over $b$ at
\begin{equation}
	b_{\rm max}\sim \frac{1}{\Lambda_{\rm QCD}},
\end{equation}
which corresponds approximately to the onset of the nonperturbative region. This cutoff is particularly important for the stability of the anomaly-induced $TJJ$ contribution. Indeed, the $TJJ$ insertion generates additional logarithmic dependence on the gluon virtuality, enhancing the sensitivity of the Fourier transform to large transverse separations. Without Sudakov suppression and the restriction to the perturbative domain, the convolution would receive sizable contributions from regions where the perturbative treatment of the hard kernel is no longer reliable.

The numerical results for the two GFFs are shown in Fig.~\ref{chza:ap} and in Fig.~\ref{chza:dp}, while the trace form factor is shown in Fig.~\ref{chza:trp}, where the leading-order Sudakov-improved perturbative prediction is compared with the corresponding curves including the isolated anomaly contribution and the full $TJJ$ insertion. The calculations are also compared with the lattice-QCD determination of Hackett \textit{et al.}  in the low- and intermediate-$Q^2$ region \cite{Hackett:2023pyn}.

The resulting form factors exhibit a characteristic projection hierarchy. The momentum form factor $A_\pi(Q^2)$ receives only a moderate correction from the complete $TJJ$ insertion, while the isolated anomaly contribution cancels in the leading-twist projection, consistently with the tensor decomposition of the stress-tensor correlator. The cancellation is straightforward from Eqs \eqref{chza:eq:F12}  and \eqref{chza:eq:F3}. Nevertheless, the non-anomalous tensor structures contained in the full $TJJ$ kernel lower the leading-order prediction at small momentum transfer, improving the qualitative agreement with the lattice-QCD data.

A different behavior is observed for the D-term form factor $D_\pi(Q^2)$. In this channel the scalar structures associated with the trace anomaly project much more efficiently, producing a sizable correction already at moderate values of $Q^2$. This enhancement reflects the sensitivity of the D-term to the internal stress distribution and to the scalar sector of the energy-momentum tensor.

The clearest signal of the anomaly emerges in the trace form factor,
\begin{equation}
	\Theta_\pi(Q^2)
	=
	g_{\mu\nu}
	\langle \pi(P_2)|T^{\mu\nu}|\pi(P_1)\rangle ,
\end{equation}
where the scalar projection directly selects the trace component of the $TJJ$ correlator. In this case the anomaly-induced contribution dominates the full $TJJ$ correction over the entire momentum-transfer region considered in the analysis. This behavior confirms that the trace form factor is the most direct observable sensitive to the beta-function-controlled scalar channel generated by the conformal anomaly.

\begin{figure}
	\centering
	\includegraphics[scale=0.65]{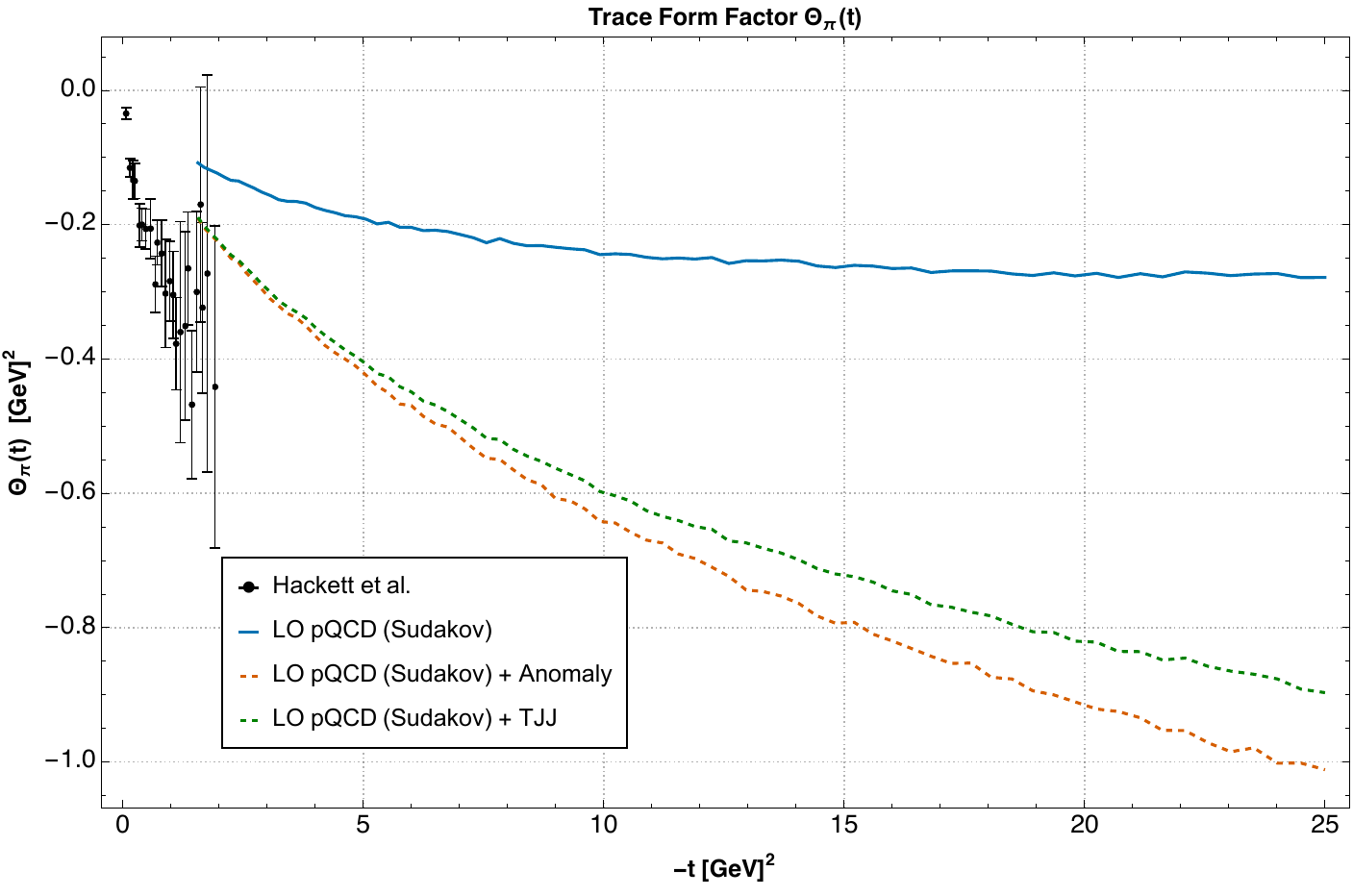}
	\caption{The trace form factor \(\Theta_\pi\) is the channel in which the anomaly is most significant: it dominates the \(TJJ\) contribution at both low and high momentum transfer.  Treating the anomaly as a constrained contribution in the hard part improves the trace prediction and helps the interpretation of the \(D_\pi\) form factor. The curves are obtained for $a_2(\mu_0)=0.2$, $\mu_0=1$ GeV, $m_q=0.33\,\, \text{GeV}$ and $\beta_\pi=0.24 \,\,\text{GeV}^{-1}$.}
	\label{chza:trp}
\end{figure}

The hierarchy observed among the three form factors therefore provides a nontrivial consistency check of the tensor decomposition of the $TJJ$ vertex. A universal shift of all form factors would simply correspond to a global normalization effect, whereas the selective enhancement of the scalar observables demonstrates that the anomaly contribution is correctly projected onto the trace sector of the pion energy-momentum tensor.

Finally, it should be emphasized that the present calculation is not intended as a complete low-energy description of the pion GFFs. At low momentum transfer, additional contributions associated with soft overlap, higher-twist effects, chiral dynamics, and quark-mass corrections to the trace become relevant. The purpose of the present analysis is instead to isolate the perturbative trace-anomaly contribution within a controlled Sudakov-improved factorization framework and to study its projection onto the various GFFs.

Although the momentum-transfer range explored by the lattice data is not strictly within the asymptotic perturbative regime, the comparison is still useful in order to investigate the effect of the trace-anomaly contribution on the various GFFs. In particular, both the isolated anomaly contribution and the complete $TJJ$ correction tend to lower the pure leading-order perturbative prediction in the low/intermediate-$Q^2$ region, producing a visibly improved overlap with the available lattice points.

This effect is especially evident in the trace form factor $\Theta_\pi(Q^2)$ and, to a lesser extent, in the D-term form factor $D_\pi(Q^2)$, where the scalar component associated with the conformal anomaly contributes more efficiently through the tensor decomposition of the $TJJ$ correlator. In the case of $A_\pi(Q^2)$, the isolated anomaly contribution cancels at leading twist, but the full $TJJ$ insertion still generates a moderate downward shift of the perturbative curve at low momentum transfer.

At the same time, the comparison should be interpreted with some caution. In the low-$Q^2$ region the pion dynamics is expected to receive important contributions from soft overlap, chiral effects, higher-twist corrections, and genuinely nonperturbative physics \cite{Liu:2024vkj}, which are not fully captured within a perturbative factorization framework. Consequently, the present calculation should not be viewed as a precision description of the low-energy pion GFFs. Rather, its purpose is to isolate the perturbative trace-anomaly contribution and to study how the non-Abelian $TJJ$ insertion modifies the hierarchy of the various tensor projections.

From this perspective, the qualitative agreement obtained after including the anomaly-related corrections is encouraging. It indicates that the scalar channel associated with the QCD beta function contributes in the expected direction and improves the phenomenological behavior of the perturbative prediction even in a kinematic region where strict collinear factorization is only partially applicable.

\section{Dilaton Exchange as a Correlated Pair Channel}
\label{sec:dilaton-correlated-pair}

The perturbative dilaton discussed in the analysis of GFFs should not be interpreted, in the first instance, as an elementary scalar added to QCD.  It is a scalar projection of the renormalized \(TJJ\) vertex.  More precisely, it is the part of the stress-tensor insertion which is selected by the trace sector and whose normalization is fixed by the beta function.  This interpretation follows the perturbative anomaly-pole analysis of the \(TJJ\) correlator and its gravitational effective action representation developed in Refs.~\cite{Armillis:2009pq,Armillis:2010qk,Coriano:2018zdo}. The corresponding scalar form factor in the \(TJJ\) correlator is therefore the perturbative realization of a dilaton-like exchange.  Its role in hard exclusive processes is to provide a definite trace-channel component of the hard kernel, rather than a universal rescaling of all GFFs.  The same mechanism underlies the recent formulation of hadron GFFs from momentum-space CFT and perturbative QCD~\cite{Coriano:2024qbr}.

The point can be made precise by using the momentum-space decomposition of the \(TJJ\) vertex developed by Bzowski, McFadden and Skenderis and later applied to anomaly form factors in QCD~\cite{Bzowski:2013sza,Bzowski:2018fql,Coriano:2018bbe,Coriano:2020ees,Coriano:2024qbr,Coriano:2025fom}.  In the conformal limit the trace sector contains a form factor of the form
\begin{equation}
	\Phi_{\rm anom}(q^2)\sim \frac{\mathcal A}{q^2},
	\qquad
	\mathcal A =
	\frac{g_s^2}{16\pi^2}\,\beta_0 ,
\end{equation}
with \(q\) the momentum carried by the stress tensor.  This pole is the momentum-space image of the nonlocal anomaly action. In a local phenomenological language one may introduce a dilaton field coupled to \(F^2\).  In the perturbative construction, however, the same object is already present in the one-particle-irreducible \(TJJ\) correlator as the scalar anomaly pole.  The connection between this pole and the scalar component of the gravitational exchange has also been emphasized in the anomaly-mediated description of conformal sectors~\cite{Coriano:2026vya}.

\begin{figure}[t]
	\centering
	\includegraphics[width=0.22\textwidth]{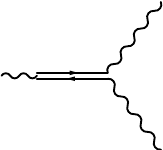}
	\caption{Effective pole representation of the anomalous \(TJJ\) interaction. The diagram gives a graphical representation of the nonlocal \(R^{(1)}\Box^{-1}F^2\) exchange generated by the anomaly form factor, as in the Abelian and non-Abelian analyses of Refs.~\cite{Armillis:2009pq,Armillis:2010qk,Coriano:2025fom}.}
	\label{fig:dilaton-pole-exchange}
\end{figure}

This interpretation is sharpened by Duff's definition of the anomaly, namely the failure of the trace operation to commute with the quantum average,
\begin{equation}
	\mathcal A
	=
	g_{\mu\nu}\langle T^{\mu\nu}\rangle
	-
	\langle T^\mu_{\ \mu}\rangle .
\end{equation}
In dimensional regularization the \(TJJ\) tensor basis shows that the anomaly pole is not produced by an arbitrary pole term inserted by hand.  The singularity is traced to a unique renormalized form factor, originally introduced in \cite{Giannotti:2008cv}.  The trace Ward identities imply that, when \(d\to4\), the \(1/(d-4)\) singularity of this form factor feeds the traced tensor structure and generates the \(1/q^2\) pole \cite{Coriano:2018zdo}.  All other form factors are finite in this limit.  Thus the dilaton pole is the residual effect of renormalization in the trace sector, consistently with the CFT reconstruction of anomalous graviton vertices and with the anomaly-action analysis of higher stress-tensor correlators~\cite{Coriano:2017mux,Coriano:2018bbe,Coriano:2020ees}.

The same conclusion has a dispersive formulation.  For general virtualities and nonzero fermion mass the scalar form factor can be represented as
\begin{equation}
	\Phi_{\rm anom}(q^2;p_1^2,p_2^2,m^2)
	=
	\frac{1}{\pi}
	\int_0^\infty ds\,
	\frac{\rho_{\rm anom}(s;p_1^2,p_2^2,m^2)}
	{s-q^2-i0},
	\label{eq:anom-dispersion}
\end{equation}
with a spectral density obeying the anomaly sum rule~\cite{Coriano:2014gja,Coriano:2025fom}
\begin{equation}
	\frac{1}{\pi}
	\int_0^\infty ds\,
	\rho_{\rm anom}(s;p_1^2,p_2^2,m^2)
	=
	\mathcal A .
	\label{eq:anom-area-law}
\end{equation}
The right-hand side is independent of the mass and of the external virtualities.  The spectral distribution may change, but the total area is fixed.  This is the sense in which the dilaton exchange can be viewed as a correlated exchange of intermediate pairs.  The spectral density is generated by the intermediate two-particle states of the \(TJJ\) subgraph, such as quark-antiquark pairs in the fermion loop and gluon pairs in the non-Abelian sector.  These states are not exchanged independently; they are projected onto the color-singlet scalar channel selected by the trace of the stress tensor.  The dilaton is therefore the coherent scalar channel carried by these correlated partonic pairs.  This spectral-flow interpretation extends the earlier anomaly sum-rule analyses of dilaton and superconformal channels~\cite{Coriano:2014gja,Coriano:2025fom}.

After the \(TJJ\) correlator is embedded in the pion matrix element, the same sum rule should be interpreted as a matching constraint on the scalar part of the hard kernel.  It does not become a separate universal sum rule for each individual pion GFF.  Rather, the fixed anomaly strength in Eq.~\eqref{eq:anom-area-law} is projected onto the hadronic amplitude by the pion DA and, at finite momentum transfer, by the Sudakov-improved wave function.  The convolution redistributes this scalar strength among the observable projections: it may cancel, or be numerically suppressed, in the anomalous part of \(A_\pi\), while it remains visible in \(D_\pi\) and is selected most directly by the trace form factor.

At lower momentum transfer the same statement admits a hadronic spectral interpretation.  The partonic two-particle cut of the \(TJJ\) subgraph is then replaced by scalar hadronic intermediate states, including pion continua, scalar resonances, and gluonic components.  The detailed saturation of the sum rule is nonperturbative, but the total anomaly strength remains fixed by the trace identity.  In this sense the perturbative dilaton pole is the short-distance representative of a scalar spectral channel which, at hadron level, is redistributed over physical intermediate states.

In the massive theory the same strength is distributed over the cut, beginning at the physical threshold \(s=4m^2\) in the fermion sector.  As the conformal limit is approached, the sequence of spectral densities is squeezed toward \(s=0\).  In distributional form one obtains a spectral flow of the type
\begin{equation}
	\rho_{\rm anom}(s;m^2)
	\longrightarrow
	\pi\,\mathcal A\,\delta(s),
	\qquad
	m\to0,
\end{equation}
so that the continuum representation collapses into the anomaly pole.  The pole is therefore not a separate assumption; it is the limiting form of a correlated two-particle spectral distribution whose area is fixed by the anomaly.  In the gluonic part of the non-Abelian vertex the localization at \(s=0\) is already present in the massless sector, while fermion mass effects smear the pole into a continuum before the conformal limit is taken.

There is, however, an important qualification.  The analysis of the particle-pole residue in light-cone variables shows that the anomaly pole and a genuine particle pole are not identical notions in the \(TJJ\) case~\cite{Giannotti:2008cv,Armillis:2009pq,Armillis:2010qk,Coriano:2025fom}.  If the external vector legs are off shell, or if the fermions in the loop are massive, the residue of \(q^2\langle TJJ\rangle\) vanishes as \(q^2\to0\).  A nonzero residue appears only in the massless, on-shell, light-cone configuration corresponding to a virtual graviton decaying into two transverse on-shell gluons or photons.  In the on-shell basis of the Abelian \(TJJ\) correlator, later extended to the non-Abelian case, the surviving residue contains two independent tensor structures.  With \(k=p_1+p_2\), \(s=k^2\), \(p_1^2=p_2^2=0\) and Eq. \eqref{chza:defu},
one may write the light-cone residue by separating the scalar anomaly pole from the transverse-traceless pole contribution as
\begin{equation}
	\lim_{s\to0}
	s\,\langle T^{\mu\nu}(k)J^{a \alpha}(p_1)J^{b \beta}(p_2)\rangle
	=
	\frac{g_s^2}{48\pi^2}\delta^{a b}\left(\frac{11}{3}C_A-\frac{2}{3}n_f\right)
	\tilde\phi_1^{\mu\nu\alpha\beta}
	-\frac{g_s^2}{288\pi^2}\delta^{a b}(n_f - C_A)\,
	\tilde\phi_2^{\mu\nu\alpha\beta}.
	\label{eq:tjj-lightcone-two-poles}
\end{equation}
The second term denotes a traceless contribution, already present in the Abelian case.\\
The first tensor is the scalar anomaly structure,
\begin{equation}
	\tilde\phi_1^{\mu\nu\alpha\beta}
	=
	\left(sg^{\mu\nu}-k^\mu k^\nu\right)
	u^{\alpha\beta}(p_1,p_2),
\end{equation}
and is the residue associated with the \(R^{(1)}\Box^{-1}F^2\) exchange represented in Fig.~\ref{fig:dilaton-pole-exchange}.\\  The second tensor is
\begin{equation}
	\tilde\phi_2^{\mu\nu\alpha\beta}
	=
	-2\,u^{\alpha\beta}(p_1,p_2)
	\left[
	sg^{\mu\nu}
	+2\left(p_1^\mu p_1^\nu+p_2^\mu p_2^\nu\right)
	-4\left(p_1^\mu p_2^\nu+p_2^\mu p_1^\nu\right)
	\right],
	\label{eq:traceless-pole-tensor}
\end{equation}
which obeys \(g_{\mu\nu}\tilde\phi_2^{\mu\nu\alpha\beta}=0\) in four dimensions and is therefore not part of the trace anomaly. In the Abelian case (QED), discussed in \cite{Giannotti:2008cv}, the additional massless pole associated to this structure was interpreted as a plasmon-like mode of the anomaly vertex, proportional to the number of fermion flavors $n_f$.  As discussed in \cite{Coriano:2025fom}, it is a light-cone pole which in the context of a gravitational exchange is mediated by the spin--2 part of the graviton propagator rather than by the scalar trace projection \cite{Coriano:2026vya}. It plays a role in the scattering of two gluons $(g g  \to gg)$ once two $TJJ$ vertices are coupled by the exchange of an intermediate graviton. The factor $(n_f - C_A)$ in front of $\tilde\phi_2$ in \eqref{eq:tjj-lightcone-two-poles} accounts for both the quark and the gluon/ghost contributions.\\
This second traceless structure contributes to the anomaly effective action describing the interaction around the light-cone, of the form
\begin{align}\label{expoleac}
	S_{\text{extra pole}} = \frac{g_s^2}{72 \pi^2} \left(n_f - C_A \right) \int &d^4x \sqrt{-g} \int d^4x' \sqrt{-g'}\, h_{\mu\nu}(x) \Box^{-1}_{x,x'} 
	\nonumber\\&
	\times \left[ 
	3 (\partial^\mu \partial^\nu F^a_{\alpha\beta}) F^{\alpha\beta a}
	+ \frac{1}{4} \left( g^{\mu\nu} \Box - 4 \partial^\mu \partial^\nu \right) F^a_{\alpha\beta} F^{\alpha\beta a}
	\right]_{x'},
\end{align}
while the first tensor structures is associated with the well-known effective action of the form \cite{Armillis:2009pq}
\bea
S_{\text{anom}} &=&\frac{1}{3} \, \frac{g^3}{16 \pi^2} \left (  - \frac{11}{3} \, C_A + \frac{2}{3} \, n_f \right)  \, \int d^4 x \, d^4 y \,R^{(1)}(x)\, \square^{-1}(x,y) \, F^a_{\alpha \beta}F^{\alpha \beta a},
\eea
that generalizes the expression derived in \cite{Giannotti:2008cv} for QED. $R^{(1)}(x)$ is the linearized scalar curvature.

This sector has the same massless propagator as the scalar anomaly exchange, but its residue is traceless.  It should therefore be kept conceptually separate from the dilaton channel: the dilaton is the scalar projection fixed by the trace Ward identity and by the anomaly sum rule, while the plasmon-like pole is a light-cone excitation of the transverse-traceless sector.  This separation is precisely the reason why the CFT momentum-space decomposition and the explicit perturbative \(TJJ\) calculation have to be used together when the hard kernel is projected onto pion GFFs.

For the pion GFFs this distinction is essential.  The hard kernel contains the one-loop \(TJJ\) insertion, and the pion DAs attach to the external partonic lines.  The trace-anomaly contribution is then the scalar projection of the correlated intermediate pair exchanged inside this hard subgraph.  It contributes most directly to the trace form factor and to the scalar part of the \(D_\pi\)-term, while it cancels in the anomalous part of the \(A_\pi\) projection.  This is precisely what one expects if the dilaton is not an elementary external state, but the coherent trace-channel component of the two-parton intermediate spectrum.

\section{Conclusions}

We have analyzed the pion GFFs in the hard-scattering regime by combining the standard collinear factorization picture with the one-loop non-Abelian \(TJJ\) correlator. The calculation isolates the trace-anomaly part of the correlator and shows how it enters the short-distance kernel through the coefficient \(\mathcal{A}=g_s^2\beta_0/(16\pi^2)\). This coefficient is fixed by the QCD beta function and therefore provides a perturbative realization of the scale anomaly inside the hard kernel.

For a spin-zero external state the trace projection relates the scalar matrix element to the combination \(A_\pi+3D_\pi\) in the chiral large-\(Q^2\) limit. In the explicit matching performed here, the anomaly contribution appears most directly in the coefficient \(F_5^{TJJ}\), and hence in the projection onto the \(D\)-term. The isolated anomaly cancels in \(A_\pi\), while the full \(TJJ\) insertion still lowers the leading curve at small \(Q^2\) through its non-anomalous tensor components.  In \(D_\pi\) the anomaly gives an important contribution, although a complete order-\(\alpha_s^2\) hard kernel would also require the remaining non-anomalous radiative terms.

The Sudakov-improved formulation extends the collinear result by including transverse separation and the resummation of soft-collinear logarithms. The resulting suppression of large \(b\) and endpoint configurations stabilizes the anomaly-induced convolution and makes the perturbative description more reliable at finite momentum transfer. The numerical analysis therefore supports a picture in which the trace anomaly gives a distinct, beta-function-controlled correction to the mechanical structure of the pion, with its clearest impact in the trace form factor and an important contribution to \(D_\pi\).

The comparison with the instanton-vacuum analysis of Ref. \cite{Liu:2024vkj} clarifies the domain of validity of this result. Their calculation supplies a nonperturbative soft and semi-hard realization of chiral and scale breaking, including twist-3 pion DAs and pressure/shear distributions. The present calculation supplies the short-distance $TJJ$
anomaly kernel. A complete phenomenological description of pion GFFs should combine these ingredients, matching the
nonperturbative instanton or chiral component to the perturbative TJJ-improved kernel while avoiding double counting
in the scalar trace sector.

		\chapter{Anomaly-mediated Scalar Gravitational  Interactions and the Coupling of Conformal Sectors}\label{chgsampl}
\addcontentsline{toc}{section}{Synopsis}
\section*{Synopsis}

This final Chapter investigates the role of the conformal (trace) anomaly in activating a scalar gravitational channel in four-point scattering amplitudes, connecting the field-theoretic framework developed throughout the thesis to the broader context of gravitational wave physics, induced gravity, and the structure of graviton exchange amplitudes. The central question addressed is whether and how the trace anomaly, which generates anomaly poles in three-point correlators such as the $TJJ$ and $TTT$, can produce a genuine scalar gravitational response in four-point functions obtained by sewing these vertices through single-graviton exchange in the Einstein--Hilbert (EH) framework.

The Chapter opens by situating the problem within the current landscape of gravitational physics. The persistent tensions in cosmological parameters motivate the consideration of extensions of GR involving additional propagating degrees of freedom. A broad class of such extensions introduces scalar gravitational waves, which can arise either from higher-derivative corrections to the EH action as in the Starobinsky model where $R^2$ introduces a propagating scalaron or, in the context examined here, from the quantum breaking of classical conformal symmetry through the trace anomaly. The Chapter emphasizes that these two mechanisms are conceptually distinct: in the Starobinsky case the scalar arises from a local modification of the gravitational action, while in the anomaly-induced case it emerges from a nonlocal effective interaction of the form $R\Box^{-1}F^2$ generated by integrating out conformal matter. The analysis is carried out entirely in the perturbative EH framework around flat spacetime, where both the conformal correlators and the graviton propagator are under full perturbative control.

The induced-gravity perspective of Sakharov is reviewed, showing that integrating out conformal matter fields whether visible or dark generates an effective gravitational action whose dominant quantum contribution, in the conformal case, is the nonlocal Riegert (anomaly-induced) action
\begin{equation*}
	\mathcal{S}_{\rm anom}^{NL}[g] = \frac{b'}{8}\int d^4x\sqrt{-g}\int d^4x'\sqrt{-g'}\left[\left(E - \tfrac{2}{3}\Box R\right) + \tfrac{b}{b'}C^2\right]_x \Delta_4^{-1}(x,x') \left[\left(E - \tfrac{2}{3}\Box R\right) + \tfrac{b}{b'}C^2\right]_{x'},
\end{equation*}
which is intrinsically nonlocal through the inverse Paneitz operator $\Delta_4^{-1}$. Around flat spacetime this action is characterized by $\Box^{-1}R$ interactions that select a specific scalar component of the metric, the gauge-invariant combination $\Phi_{\rm cov} = h - \Box w$, through the projector $\pi^{\mu\nu}$.

The $TJJ$ correlator is then analyzed in its conformal and perturbative realizations. In the conformal limit, the full correlator is reconstructed from the anomaly pole in the longitudinal sector via the conformal Ward identities, expressing it in terms of transverse-traceless, longitudinal, and trace components. The Gram-matrix method is used to extract the $tt$ form factors $\tilde{A}_i$ from the basis of independent tensor structures, providing a clean algebraic extraction free from longitudinal or trace contamination. In the on-shell photon limit the correlator reduces to three structures: one associated with the anomaly pole, one traceless pole related to a secondary nonlocal effective action $S_{\rm extra\,pole}$, and one finite renormalized term $F_{13,R}$. The residue of the $1/q^2$ pole is shown to be nonvanishing only in the strict massless on-shell light-cone configuration, confirming the kinematic selectivity of anomaly-pole couplings established in previous Chapters.

The spin decomposition of the graviton propagator in the de Donder gauge is carried out using the Barnes--Rivers projectors $P^{(2)}$, $P^{(1)}$, $P^{(0\text{-}s)}$, and $P^{(0\text{-}w)}$. The spin-2 and spin-0 components of the propagator are identified as
\begin{equation*}
	D^{(2)}_{\mu\nu,\alpha\beta}(k) = -\frac{1}{2}\frac{\Pi_{\mu\nu,\alpha\beta}(k)}{k^2}, \qquad D^{(0\text{-}s)}_{\mu\nu,\alpha\beta}(k) = -\frac{1}{6}\frac{\pi_{\mu\nu}(k)\pi_{\alpha\beta}(k)}{k^2},
\end{equation*}
and their gauge independence (with respect to the de Donder parameter $\xi_D$) is established. The covariant decomposition of $h_{\mu\nu}$ into transverse-traceless, transverse vector, and scalar sectors is performed, identifying $\Phi_{\rm cov}$ as the unique gauge-invariant scalar combination. A key result is that $\Phi_{\rm cov}$ does not appear in the quadratic EH action and thus does not propagate as an independent degree of freedom in pure Einstein gravity. Only the two helicity $\pm 2$ tensor polarizations are physical in the EH framework.

The question of whether explicit conformal-symmetry breaking through a non-minimal coupling $\xi \neq 1/6$ or a scalar mass $m$ can activate genuine scalar graviton exchange is then examined through tree-level $2 \to 2$ scalar scattering. The full amplitude is shown to take the form
\begin{equation*}
	i\mathcal{M}(\xi,m) = i\mathcal{M}_0(m) - \frac{i\kappa^2}{4}(2\xi - 6\xi^2)(s+t+u),
\end{equation*}
where the $\xi$-dependent part is polynomial and proportional to $s+t+u$. For massless scalars on-shell, $s+t+u=0$ and the amplitude is $\xi$-independent; for massive scalars, the amplitude retains $\xi$-dependence on-shell through $s+t+u=4m^2$. This demonstrates that the presence of a nonvanishing trace in the stress-energy tensor and an apparent scalar contribution to the exchange do not constitute evidence for propagating scalar graviton exchange; explicit breaking does not activate an independent scalar gravitational mode.

The anomaly-induced scalar activation is then analyzed in the Fierz--Pauli Lagrangian. Starting from the covariant decomposition of $h_{\mu\nu}$, the scalar sector of the quadratic EH action is shown to produce a kinetic term for $\Phi_{\rm cov}$,
\begin{equation*}
	S_{EH,S}^{(2)} = \frac{3}{32\kappa^2}\int d^4x\, \Phi_{\rm cov}\Box\Phi_{\rm cov},
\end{equation*}
combined with which the anomaly pole action $S_{\rm pole} \sim \int R^{(1)}\Box^{-1}F^2$ generates, using $R^{(1)} = -\frac{3}{4}\Box\Phi_{\rm cov}$, a coupling of the form $\Phi_{\rm cov} F^2$. The resulting action
\begin{equation*}
	S_\Phi = \int d^4x\left[\frac{3}{32\kappa^2}\Phi_{\rm cov}\Box\Phi_{\rm cov} - \frac{3b}{32\pi^2}\Phi_{\rm cov}F^2\right]
\end{equation*}
yields the modified equation of motion $\Box\Phi_{\rm cov} = (\kappa^2 b/\pi^2)F^2$, showing that $\Phi_{\rm cov}$ is sourced by the anomaly. This is the key result: while $\Phi_{\rm cov}$ is non-dynamical in pure EH gravity, the conformal anomaly activates it as a distinct response channel.

The four-point amplitude for photon--photon scattering mediated by the anomaly-induced scalar exchange is then computed. Sewing two $TJJ$ vertices through the spin-0 component of the graviton propagator and using $\pi_{\mu\nu}P^{(0\text{-}s)\,\mu\nu}{}_{\rho\sigma}\pi^{\rho\sigma} = 9/q^2$, the scalar exchange amplitudes in the $s$, $t$, and $u$ channels take the compact form
\begin{equation*}
	\mathcal{A}_{s,t,u}^{(0)} \propto u^{\alpha\gamma}(p_1,p_2)\frac{1}{s,t,u}u^{\beta\delta}(p_3,p_4),
\end{equation*}
where $u^{\alpha\beta}(p,q) = (p\cdot q)g^{\alpha\beta} - q^\alpha p^\beta$ is the momentum-space image of the $F^2$ operator. A complete table of helicity amplitudes is presented. A crucial observation is that for on-shell massless photons, $\mathcal{U}_{ij}\mathcal{U}_{kl}/s_{\rm ch} \sim s_{\rm ch}$, so each channel contributes a polynomial contact term $\sim s_{\rm ch}/M_P^2$ rather than a physical pole. The anomaly pole is therefore manifest at the level of the effective action but does not generate an observable long-range scalar exchange in this process. This is consistent with the residue analysis: the pole residue in the $TJJ$ vertex is nonvanishing only in the strict massless on-shell light-cone configuration, and the kinematic condition for a genuine particle-like scalar exchange is not satisfied in the four-point process considered.

The extension to massive fermions is carried out through the spectral analysis of the conformal anomaly form factor $\Phi_0(s,m)$ in the on-shell limit. The discontinuity of the fermionic sector is computed using distributional identities for $\mathrm{disc}(1/s)$, $\mathrm{disc}(C_0/s)$, and $\mathrm{disc}(C_0/s^2)$, showing that all $\delta(s)$ and $\delta'(s)$ contributions cancel identically for $m \neq 0$: only the branch cut above $4m^2$ survives in the fermionic sector. In the conformal limit $m \to 0$ the cut converges to a $\delta(s)$ distribution, restoring the anomaly pole and reproducing the sum rule $\int \Delta\Phi_0\,ds = -\frac{1}{3}\beta(e)/e$.

Graviton--graviton scattering with anomaly corrections is then analyzed using the $TTT$ correlator from the nonlocal anomaly effective action. The anomaly-induced vertex $\mathcal{A}_2^{\mu_1\nu_1\mu_2\nu_2}$ is shown to factorize as
\begin{equation*}
	\mathcal{A}_2^{\alpha\beta\rho\sigma}(p,q) = 4(b+b')\left[u^{\alpha\rho}(p,q)u^{\beta\sigma}(p,q) + u^{\alpha\sigma}(p,q)u^{\beta\rho}(p,q)\right],
\end{equation*}
revealing a double-copy-like structure in which the graviton amplitude is expressed as a square of the photon amplitude $\mathcal{U}_{ij}^2$. Complete helicity tables for the $2 \to 2$ graviton scattering amplitude are presented. For scalar emission ($g + g \to \phi + \phi$), the $s$-channel spin-0 exchange survives as a contact term $\sim s^3$, while the $t$- and $u$-channel spin-0 components vanish on-shell. The total cross section is computed for equal and opposite polarizations, scaling as $s^5/M_P^{12}$, quartic in the conformal multiplicities $b$ and $b'$.

The Chapter concludes by discussing the extension to dark sectors, where a visible conformal sector (e.g.\ QED) and a dark conformal sector interact only through gravity. In this case the anomaly coefficients $b$ and $b'$ receive separate contributions from both sectors, enhancing the effective scalar coupling and imprinting the internal structure of the hidden sector onto the amplitude. An important distinction is drawn between the perturbative EH framework used throughout the Chapter, where all anomaly-mediated effects are Planck-suppressed contact terms, and the full nonlocal Riegert action, where the conformalon $\varphi$ can mediate scale-independent infrared effects potentially relevant in strongly curved backgrounds such as neutron-star mergers. The analysis establishes that within the EH framework the anomaly activates a distinguished scalar response channel through $\Phi_{\rm cov}$ without introducing a new fundamental scalar particle, and that the scalar gravitational effects identified here are best understood as a coherent signature of the conformal anomaly in gravitational scattering rather than as evidence for new propagating graviscalar degrees of freedom.

The work presented in this Chapter is based primarily on, and adapted from, the preprint currently currently under revision for publication \cite{Coriano:2026vya} co-authored with C.~Corianò S.~Lionetti and L.~Torcellini.
\section{Introduction}

The standard cosmological model, $\Lambda$CDM, combines a cosmological constant and cold dark matter with Einstein's General Relativity (GR), and provides an excellent description of the Cosmic Microwave Background, large-scale structure and many other observations. At the same time, persistent tensions in parameters such as the Hubble constant $H_0$, the clustering amplitude $S_8$, and the baryonic acoustic oscillations (BAO)-inferred sound horizon suggest that the standard picture may be incomplete. If these discrepancies are not entirely due to unresolved systematics, they may point either to extensions of the cosmological sector or to modifications of gravity itself. This motivates a closer examination of theoretically consistent alternatives to GR and of new observational probes capable of testing them \cite{CosmoVerseNetwork:2025alb}.

A broad class of extensions of GR introduces additional propagating degrees of freedom beyond the massless spin--2 graviton. Such fields can affect the expansion history, the growth of structure, and the properties of gravitational radiation. In particular, scalar gravitational waves may arise whenever the gravitational sector contains a dynamical scalar mode in addition to the two tensor polarizations of GR \cite{Eardley1973,Lee1974,Will1977DipoleRadiationBinaryPulsar,Gong2018,Alves2010,Hyun2019}.
In Einstein gravity, diffeomorphism invariance together with the Einstein--Hilbert action imply that only transverse and traceless tensor perturbations propagate, corresponding to the familiar plus and cross polarizations.

Gravitational-wave astronomy has made these issues directly testable. Observations by LIGO and Virgo \cite{Aasi:2015cqb,Acernese:2014hva,Abbott:2016blz,Abbott:2016nmj,Abbott:2016pxj,Abbott:2017vtc,Abbott:2017oio} have already yielded stringent tests of GR in the strong-field regime, including bounds on the graviton mass and constraints on deviations from the post-Newtonian predictions of Einstein gravity. In this context, the polarization content of gravitational radiation provides a particularly clean probe of additional propagating modes \cite{Callister:2017ocg}.\\
Observationally, scalar gravitational waves are constrained both by binary pulsar timing and by interferometric measurements of gravitational-wave polarizations \cite{DamourEspositoFarese1996,AntoniadisEtAl2013,ShaoEtAl2017PRX,ZhuEtAl2019J1713Symmetries,Callister:2017ocg}. Present data are compatible with purely tensorial radiation, but they still leave room for subleading scalar components. A detection of such modes would provide direct evidence for additional gravitational degrees of freedom, whereas increasingly strong null results further sharpen the empirical success of GR.\\
Kinematically, scalar waves differ sharply from tensor waves: a massless scalar produces an isotropic transverse \emph{breathing} deformation, while a massive scalar can also excite a longitudinal component. These signatures provide, at least in principle, a way to distinguish scalar radiation from the purely tensorial signal of GR.
In a generic metric theory of gravity, gravitational waves can indeed excite several independent components of the spacetime curvature compatible with wave propagation \cite{Eardley1973,Lee1974,Will1977DipoleRadiationBinaryPulsar,Gong2018,Alves2010,Hyun2019,Kausar2016,Isi2023,Capozziello2021,Shankaranarayanan2019}. Their effect on freely falling test particles is governed by the geodesic deviation equation
\begin{equation}
	\frac{d^{2}\xi^{i}}{dt^{2}} = - R_{0i0j}\,\xi^{j},
	\label{geodev}
\end{equation}
so that the polarization content of a plane wave can be classified in terms of the independent components of $R_{0i0j}$. The $E(2)$ classification of null momenta shows that a general metric theory can contain up to six independent polarizations \cite{Eardley1973}: two tensor modes, two vector modes, and two scalar modes. For propagation along the $z$ direction, the tensor modes are the usual $h_+$ and $h_\times$ of GR; the vector modes, often denoted $h_x$ and $h_y$, generate transverse shear distortions; and the scalar sector contains a transverse breathing mode $h_b$ together with a longitudinal mode $h_L$. GR is the special case in which only the two tensor polarizations are dynamical.

Extra degrees of freedom arise naturally once higher--derivative corrections are included. Such terms have been studied both as possible ingredients of a perturbatively renormalizable extension of gravity \cite{tHooftVeltman1974,DeserNieuwenhuizen1974,DeserVanNieuwenhuizenTsao1976,Stelle1977,Stelle1978} and as an effective framework for early-universe cosmology. In the latter context, higher-curvature interactions can generate an inflationary phase without introducing elementary scalar inflaton fields by hand, and they also provide a mechanism for its termination. This addresses the so-called graceful-exit problem: an inflationary model is physically viable only if the quasi-de Sitter accelerated expansion is not eternal, but evolves smoothly into the standard decelerated Friedmann--Robertson--Walker evolution, allowing reheating and the onset of the radiation-dominated era. A paradigmatic example is the Starobinsky model,

\begin{equation}
	S = M_P^2 \int d^4x \sqrt{-g}\, f(R) + S^{(m)}(g,\psi),
	\label{eq:starobinsky_action}
\end{equation}
with
\begin{equation}
	f(R)=R+\frac{R^2}{M^2},
	\label{eq:fR_starobinsky}
\end{equation}
where \(\psi\) collectively denotes the matter fields and \(S^{(m)}\) is their action. The Einstein--Hilbert term,
\begin{equation}
	S_{EH}\equiv M_P^2 \int d^4x \sqrt{-g}\,R,
	\label{eq:EH_term}
\end{equation}
is therefore supplemented by a quadratic curvature correction capable of driving inflation \cite{Starobinsky1980}. In the metric formulation, this theory can be recast, via a field redefinition and a conformal transformation to the Einstein frame, as general relativity plus an independent massive scalar degree of freedom (the scalaron). By contrast, in Palatini formulations the \(R^2\) term does not by itself introduce the same propagating scalar degree of freedom, and inflation typically requires additional scalar fields  \cite{Antoniadis:2018ywb}. Despite their phenomenological appeal, higher--derivative corrections raise well-known issues of unitarity and ghost propagation, and are best interpreted within an effective-field-theory framework. In ordinary GR this program has been proposed, for instance, in \cite{Donoghue1994GRasEFT}.

\subsection{Scalar dilaton-like couplings to the anomaly}
In the present work, however, our emphasis is different: we focus on scalar modes generated by the conformal anomaly, namely by the quantum breaking of classical conformal symmetry in matter sectors coupled to gravity.
Classically conformal matter does not source a scalar mode at tree level, but at the quantum level the trace anomaly generates an effective scalar coupling even for classically conformal fields \cite{Capper:1974ic,Duff:1977ay,Mottola:2016mpl}. In many theories this scalar couples to the trace of the energy--momentum tensor,
\begin{equation}
	\mathcal{L}_{\text{int}} \sim \varphi \, T^\mu_{\ \mu},
\end{equation}
so that sources with nonvanishing trace can emit scalar radiation. In the anomaly-induced case, by contrast, the scalar degree of freedom is more naturally interpreted as an effective excitation associated with the intrinsically nonlocal structure \(R\,\Box^{-1}\), which singles out and activates a particular scalar component of the metric. In this respect it is conceptually distinct from an ordinary local propagating field. We recall that a genuine asymptotic mode is associated with a field which, in the absence of external sources, obeys an independent hyperbolic equation of motion and is therefore able to propagate freely to infinity. Our discussion is precisely aimed at clarifying this distinction and at analyzing the subtle behaviour of the anomaly-triggered scalar excitation generated by the conformal anomaly.\\
This specific coupling emerges in perturbative analyses of conformal matter coupled to gravity through anomalous vertices such as $TJJ$ and $TTT$, where $T$ denotes the stress energy tensor and $J$ are Abelian currents. The analysis can be extended to non-Abelian currents but it will be discussed elsewhere. \\
For the \(TJJ\) vertex, first analyzed in free-field theories such as QED and QCD \cite{Giannotti:2008cv,Armillis:2009pq,Armillis:2010qk}, the anomaly manifests as a single massless exchange (a massless pole) in the conformal limit, when all the fermion masses are set to vanish. The residue of this interaction is nonzero on the light cone and affects, in our construction, both conformal sectors. The role of linearized gravity is to transfer this interaction.\\
The anomaly-induced action  can be recast in a symmetric form involving a single dynamical scalar field, the conformalon $\varphi$ by the inclusion of a Weyl invariant term \cite{Mottola:2016mpl}, which is allowed by the solution. From this perspective, anomaly-induced gravity may be regarded as a particular realization of Sakharov's induced-gravity idea \cite{Sakharov1967,Visser:2002ew}, in which the scalar sector emerges from the nonlocal effective action generated by the anomaly \cite{Mottola:2016mpl,Mottola:2010gp}. 
\\
While the framework discussed in \cite{Mottola:2016mpl} is based on the nonlocal anomaly-induced action with gravity treated as a background field, our analysis is carried out directly within the dynamical framework of GR. The discussion is largely formal, since in our setting the Planck-scale suppression of anomaly-mediated effective interactions is entirely natural within a perturbative treatment. Our aim is therefore not immediate phenomenological applicability, but rather a detailed investigation of how trace-anomaly contributions to three-point functions interface with tree-level graviton exchange in quantum gravity.

\subsection{The activation of extra modes in gravity}
A static source with $T\neq 0$ can generate only a near-zone scalar field; genuine radiation requires time-dependent scalar multipoles, $\dot{M}_\phi\neq 0$, $\ddot{\mathbf D}_\phi\neq 0$, $\dddot{Q}^{\,\phi}_{ij}\neq 0$, $\ldots$. In ordinary scalar--tensor systems, dipole radiation may dominate over the tensor quadrupole channel \cite{Will2014LRR,Eardley1975}, although it vanishes when the scalar charges of the two bodies are equal. By contrast, the anomaly-induced scalar ($\varphi$) is not sourced by body-dependent scalar charges but by the anomaly functional itself. In the local form of the anomaly action,
\begin{equation}
	S_{\rm anom} \sim 
	\int d^4x \sqrt{-g}\,
	\varphi \left(E-\frac{2}{3}\Box R\right)
	+\ldots ,
\end{equation}
the field $\varphi$ couples universally to curvature invariants, and its equation of motion is sourced directly by the terms appearing in the trace anomaly \cite{1984PhLB..134...56R,Fradkin:1983tg,Mottola:2006ew,Deser:1976yx}. The corresponding scalar excitation is therefore tied to metric perturbations and spacetime curvature rather than to independent scalar charges carried by compact objects.\\
For these reasons, more broadly, anomalies and the breaking of conformal symmetry have also been invoked in early-universe physics, including inflationary dynamics, matter-antimatter asymmetry, and alternative cosmological scenarios \cite{Starobinsky1980,Mavromatos:2024szb,Basilakos:2019acj,Alexander:2004us,McFaddenSkenderis2010,McFaddenSkenderis2011,HinterbichlerKhoury2012,Rubakov2014,HinterbichlerKhouryLevyMatas2013,BarsSteinhardtTurok2014,BarsSteinhardtTurok2013,Shaposhnikov:2018xkv,Shaposhnikov:2022zhj}. They can be  envisioned as transient phenomena, before that other sources of symmetry breakings set in, through other Higgs-like scalars. \\
The activation of additional degrees of freedom, which in this case may become dynamical, can only be discussed at a qualitative level within a quantum-gravity setting, given the well-known limitations of ordinary quantum-field-theoretic methods even in the analysis of the simplest curved spacetimes.\\
In effective field theory, scalar interactions can modify amplitudes and produce infrared enhancements even when the ordinary tensor contribution from the metric is suppressed. This is especially relevant in processes controlled by anomalous couplings, where a scalar channel can dominate the low-energy behavior despite being absent in classical Einstein gravity.\\
We recall that infrared corrections to GR of the type discussed in \cite{Mottola:2016mpl} are induced by conformal quantum sectors, while the gravitational dynamics itself remains classical, as already emphasized. In this context, the anomaly-induced action plays a central role, since it resums these effects into an effective description and is related to stress-energy-tensor correlators with an arbitrary number of insertions, generated after integrating out the conformal matter sector.

Quantum effects related to the anomaly are described by effective actions which are characterized by the presence of specific invariants of mass dimension four.   
Quadratic invariants such as $R^2$, the Gauss--Bonnet density $E_4$, and the Weyl invariant $C^2$ arise generically in effective actions and as counterterms in the renormalization of conformal quantum field theories coupled to gravity \cite{tHooftVeltman1974,Duff:1993wm,Capper:1975ig,Buchbinder:1992rb,Birrell:1982ix}. In four dimensions, $E_4$ is topological at the classical level, but it reappears naturally in the anomaly-induced effective action \cite{Coriano:2021nvn,Coriano:2020ees}. Within this framework, the anomaly-associated scalar mode has a special status and should be distinguished from the more familiar scalar degree of freedom encountered, for instance, in $f(R)$ gravity.\\
Its simplest realization occurs in conformal free-field theories such as QED and QCD coupled to external gravity, where the relevant anomalous interaction is encoded in the $TJJ$ and $TTT$ correlators, for example, that will be the subject of our investigation \cite{Coriano:2011ti,Giannotti:2008cv}.  
Related parity-odd correlators, that have been studied extensively \cite{Bonora:2015nqa,Armillis:2010pa,Abdallah:2023cdw,Bastianelli:2019zrq,Larue:2023qxw} may also play a role in early universe cosmology, enhancing the generation of chiral currents and generating possible coupling to gravitational waves \cite{Coriano:2023cvf,delRio:2021bnl}.  The conformal anomaly sector may also provide a bridge to dark conformal sectors coupled to ordinary matter only through gravity, a point that we will also address in our analysis. \\ 
In this work, the nonlocal conformal-anomaly action enters only in the analysis of \(2\to2\) amplitudes with four external gravitons, where it determines the structure of the two gravitationally coupled \(TTT\) vertices. In this case, the resulting amplitude exhibits an interesting double-copy-like structure, expressed in terms of local \(F^2\)-type interactions.

\subsection{Content of this work}
The aim of this work is to clarify the role of scalar contributions in virtual gravitational exchange and to distinguish ordinary gauge-dependent scalar pieces from the genuinely anomaly-induced scalar channel. This issue is especially subtle in gravity, since tree-level amplitudes in covariant gauges may contain spin--0 exchanges that do not correspond to physical propagating modes, while the trace anomaly introduces a nonlocal scalar contribution of a different origin.\\
The paper is organized as follows. In Sections 2 and 3 we review induced gravity in the presence of visible or dark conformal sectors and summarize the structure of the nonlocal anomaly action \cite{Sakharov1967,Visser:2002ew,Mottola:2016mpl,Mottola:2010gp}. Section 4 is devoted to the \(TJJ\) correlator, both in its general conformal form and in its perturbative realization \cite{Bzowski:2013sza,Armillis:2009pq}, with special emphasis on its anomaly pole and on the associated spin decomposition. In Sections 5 and 6 we move to four-point functions and analyze the graviton propagator and tree-level graviton exchange in the de~Donder gauge, isolating the spin--0 contribution. Section 7 then discusses scalar gravitational scattering in the case of explicit, non-anomalous breaking, showing the absence of a genuine scalar exchange in ordinary four-point functions.\\
The central part of the paper begins in Section 8, where we study the Fierz--Pauli Lagrangian in the presence of the anomaly and show how the anomaly activates a distinguished scalar response channel while leaving the TT and vector sectors unaffected. This analysis is closely related to the role played by anomaly poles in the solution of anomalous conformal Ward identities \cite{Coriano:2023hts,Coriano:2023cvf,Coriano:2023gxa}. Section 9 applies this result to the conformal limit of the photon four-point function, where the anomaly-induced interaction can be represented in terms of an effective scalar exchange. Section 10 extends the analysis to the massive fermion case and discusses how explicit mass corrections modify the anomaly form factor and its discontinuity structure.\\
In Sections 11 and 12 we turn to graviton scattering. We first analyze anomaly corrections to graviton--graviton scattering at tree level, using the anomalous part of the \(TTT\) vertex as studied both perturbatively and through the anomaly action \cite{Coriano:2017mux}, and then discuss the corresponding double-copy-like structure and scalar emission channels. Section 13 is devoted to the anomaly correction to the eikonal phase, while Section 14 briefly comments on the extension of the analysis to dark sectors. Our conclusions are presented in Section 15.

\section{Sakharov's induced gravity for conformal and visible/dark sectors}
The role of conformal symmetry and its anomalous breaking cannot be understated. A spacetime which is conformal before inflation allows a wider evolution compared to an ordinary De Sitter space, due to the presence of a dynamic conformal factor.\\
The integration of quantum matter fields, either visible or dark, generically induces an effective gravitational action. 
If the degrees of freedom that are integrated out belong to a dark sector, gravity mediates the interaction between the visible and dark sectors through corrections to the Einstein--Hilbert (EH) action. 
When the integrated-out matter sector is non-conformal, the induced corrections typically include local terms—possibly reproducing the EH action itself—together with higher-curvature operators accompanied by dimensionful coefficients. 
This mechanism realizes Sakharov’s idea of induced gravity~\cite{Sakharov1967} (see~\cite{Visser:2002ew} for a review). \\
At the opposite extreme, when conformal matter (visible or dark) is integrated out, the dominant contribution originates from the breaking of Weyl invariance through the trace anomaly, inducing a scaleless and intrinsically nonlocal effective action $\mathcal{S}_{anom}$, defined in \eqref{Snonlsq}. 
Around flat spacetime, this action can be formally expanded in terms of nonlocal structures of the form $\Box^{-1} R\, $, where $R$ is the scalar curvature \cite{Giannotti:2008cv,Coriano:2022ftl}. As we are going to clarify, this interaction appearing on each external graviton leg 
of a correlator, identifies a transverse projector $\pi^{\mu \nu}$, function of the graviton momentum, which is sandwiched by the scalar component $P^{(0\text{-}s)}_{\mu\nu,\alpha\beta}$ of the graviton propagator in the De Donder gauge. The same projector 
\beq
P^{(0\text{-}s)}_{\mu\nu,\alpha\beta}
=
\frac13\,\pi_{\mu\nu}\pi_{\alpha\beta},
\eeq
acting on the covariant decomposition of the metric 
will extract the only gauge invariant scalar component of the metric, denoted as $\equiv h - \Box w$ in \cite{Mottola:2016mpl}, and denoted as $\Phi_{cov}\equiv \varphi$ in our notations, extracted in a covariant decomposition of its fluctuations around flat space, that plays 
a key role in the analysis of the anomaly induced action.  
$\mathcal{S}_{anom}$ defines the backreaction on the metric, bringing in additional contributions to the gravitational dynamics, supplementing the local EH term 
\beq
\mathcal{S}=\mathcal{S}_{EH} +\mathcal{S}_{anom}.
\eeq
These aspects will be discussed in detail with reference to a diagrammatic expansion presented in the sections below that should, as we hope, clarify the way in which anomaly interactions are mediated 
in GR at quadratic level.
If metric fluctuations are quantized and normalized as fields of mass dimension one,
\begin{equation}
	g_{\mu\nu} = \eta_{\mu\nu} + \kappa\, h_{\mu\nu} \, ,
\end{equation}
with $\kappa^{2} = 16\pi G_{N}$, the resulting gravitational interaction mediating the anomaly is mediated by $\Phi_{cov}$ and strongly suppressed below the Planck scale.\\
The two sectors are interpolated by the ordinary two-point function of gravitons as determined from the EH action. 
This analysis, however, is clearly not exhaustive. 
If the full gravitational action is assumed to combine the Einstein--Hilbert (EH) action with the anomaly--induced (Riegert) action, one may write
\begin{equation}
	\label{total}
	\mathcal{S}_g \equiv \mathcal{S}_{\mathrm{EH}} + \mathcal{S}_{\mathrm{anom}} \, .
\end{equation}
The anomaly--induced contribution $\mathcal{S}_{\mathrm{anom}}$ admits a local formulation through the introduction of an auxiliary scalar degree of freedom, the conformalon $\varphi$ \cite{Mottola:2016mpl}. In this representation the interaction generated by $\mathcal{S}_{\mathrm{anom}}$ is scaleless and resums infrared contributions that are not contained in the Einstein--Hilbert sector. The presence of both terms in \eqref{total} therefore corresponds to two distinct dynamical regimes. The first is governed by $\mathcal{S}_{\mathrm{EH}}$, whose effects are suppressed by the Planck scale but become relevant as the characteristic energy approaches that scale. The second originates from $\mathcal{S}_{\mathrm{anom}}$, which encodes the resummation of infrared effects and can become important in backgrounds with sufficiently large curvature.

In the present analysis we restrict our attention to the dynamics generated by $\mathcal{S}_{\mathrm{EH}}$. This sector can be consistently studied in the vicinity of flat space within a framework entirely based on free--field realizations of the corresponding interactions. In particular, the four--point functions constructed in this work arise from the coupling of ordinary conformal sectors---characterized by well--defined three--point correlators, such as those appearing in QED---with the interactions mediated by $\mathcal{S}_{\mathrm{EH}}$ at leading order in the gravitational coupling, as illustrated in Fig.~1. In this regime the relevant dynamics is described by the (massless) Fierz-Pauli action obtained by expanding the Einstein--Hilbert action around flat spacetime.
The gravitational coupling of vertices affected by the conformal anomaly allows for a clean disentanglement of the standard spin--2 graviton exchange from the anomaly-induced scalar channel, which turns physical.  

\begin{figure}[t]
	\begin{center}
		\includegraphics[scale=1,angle=0]{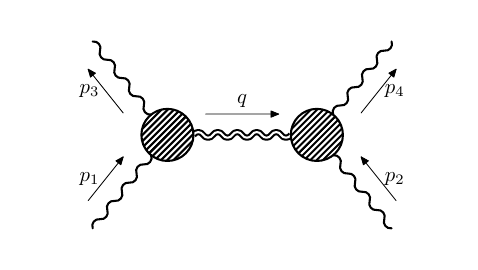}
		\includegraphics[scale=1,angle=0]{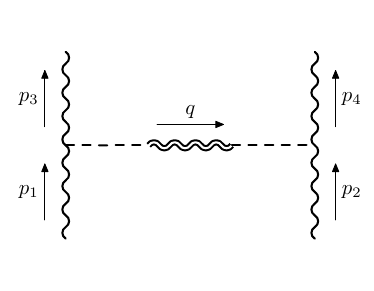}
		\caption{$t$-channel scattering of two spin-1 states $p_1+p_2 \to p_3 + p_4$, mediated by the exchange of a single graviton (left). The two conformal sectors, represented by the darker blobs, may both belong to the visible sector or alternatively to one visible and one dark sector. The anomaly-mediated interaction between the two conformal sectors, responsible for the exchange of a scalar gravitational component, is shown on the right. This contribution arises from the combination of the external $R\Box^{-1}$ mixing with the spin-0 component of the graviton propagator.}
	\end{center}
	\label{fig1x}
\end{figure}

\section{The anomaly and the nonlocal action: review} 
In order to set the stage for our discussion in a consistent manner, we briefly review the construction of anomaly--induced actions as solutions of a variational problem, before turning to their simplest realizations at the level of three--point functions.\\
The fundamental identity characterising the anomaly is that of the one-point function, evaluated in a generic metric background, which takes the general form
\be
g_{\m\n} \big\langle T^{\m\n}(x)\big\rangle_g \equiv 
\mathcal{A}(x)\ee
with
\be
\mathcal{A}(x)= T_{cl} +  b\, C^2 +
b' \big(E - \tfrac{2}{3}\sq R\big) + b'' \sq R + \sum_i \beta_i \,\cL_i .
\label{trace1}
\ee

The symmetric energy-momentum-stress tensor $T^{\m\n}(x)$ coupling to the gravitational metric field 
$g_{\m\n}(x)$ is defined by the metric variation
\be
T^{\m\n}(x)\equiv \frac{2}{\sqrt{-g(x)}}\frac{\ \del{\rm S_{cl}}[\bar\Phi, g] }{\del g_{\m\n}(x)}
\ee
where ${\rm S_{cl}} [\bar\Phi, g]$ is the classical action of the fields, denoted generically by $\Phi = \{\Phi_i\}$. Here ${\rm S_{cl}}[\bar\Phi, g]$ is viewed 
as a functional also of the general curved spacetime background metric $g$. From a 
phenomenological perspective, $\Phi$ can be either visible or dark. If visible, in the sense that they interact ordinarily with fields of the Standard Model, their interactions with gravity will be taken as mass suppressed, but this assumption is not the only one. For example, if the interaction is present before the inflationary epoch, they may be simply the result of a sector coupled to gravity via a portal.

Here $T_{cl}$ is the trace expected from the non-invariance of the classical action itself
\beq
T_{cl} = \frac{2\,g_{\m\n}(x)}{\sqrt{-g(x)}}\left\langle\frac{\ \delta{\rm S_{cl}}[\bar \Phi, g] }{\delta g_{\m\n}(x)}\right\rangle_{\!g}
\label{cltrace}
\eeq
absent any anomalies, and the additional finite terms in (\ref{trace1}) are the quantum anomalous contributions. 
These are given in terms of the dimension-four curvature invariants 
\bes
\bea
&&E \equiv\, ^*\hskip-.5mm R_{\m\a\n\b}\,^*\hskip-.5mm R^{\m\a\n\b} =
R_{\m\a\n\b}R^{\m\a\n\b}-4R_{\m\n}R^{\m\n} + R^2 \\
&&C^2\equiv C_{\m\a\n\b}C^{\m\a\n\b} = R_{\m\a\n\b}R^{\m\a\n\b} -2 R_{\m\n}R^{\m\n}  + \tfrac{1}{3}R^2 \label{EFdef}
\eea\ees
which are the Euler-Gauss-Bonnet invariant and square of the Weyl conformal tensor respectively (with $R_{\m\a\n\b}$ the Riemann curvature tensor,
$^*\hskip-.5mm R_{\m\a\n\b}$ its dual, together with the Ricci tensor $R_{\mu\nu}$ and Ricci scalar $R$).
\(\mathcal{L}_i\) include, in general, several background (classical) operators 
\begin{equation}
	\mathcal{L}_i \in
	\left\{
	F_{\mu\nu}F^{\mu\nu},
	\;
	R\,\phi^2,
	\;
	(\nabla \phi)^2,
	\;
	\phi^4,
	\;
	\dots
	\right\},
	\label{eq:Li-examples}
\end{equation}
as well as Yukawa terms and gauge-scalar mixing terms, 
depending on the matter content of the theory. They arise when the quantum theory is embedded in more general backgrounds, with extra classical gauge and diff-invariant terms of dimension-four other than pure curvature invariants. Their coefficients \(\beta_i\) are the corresponding beta functions of the renormalized couplings. Thus, $\beta_i \mathcal{L}_i$
encodes running couplings and explicit scale breaking, rather than new gravitational dynamics.\\
The coefficients $b$, $b'$, and $\beta_i$ depend on the number of light fields in the theory and contain no suppression by the Planck scale (see \cite{Coriano:2020ees} for a review), while $b''$ is 
prescription-dependent. As already mentioned above, unlike local curvature corrections arising in conventional effective field theory expansions, the anomaly gives rise to a non-local effective action with logarithmic scaling at large distances, making it \emph{marginally relevant} in the infrared \cite{Mottola:2016mpl}. Its effects are cumulative on macroscopic scales, even in spacetimes of small curvature. In the presence of a dark sector, gravitationally coupled to a visible one, $b$, $b'$, get separate contribution from the multiplicities of massless conformal fields present in both sectors. \\
One expects such contributions to manifest themselves in the form of scalar gravitational waves directly generated by the anomaly, which could have been present in the early Universe at energy scales that may be regarded as pre--inflationary and could therefore leave an imprint on the stochastic gravitational--wave background. 
\subsection{The nonlocal action}
No local covariant action exists whose variation reproduces \eqref{trace1}. The anomaly can be integrated to construct a nonlocal effective action, a procedure that relies on the existence of conformally covariant operators, most notably the Paneitz operator \cite{Paneitz2008}, whose Green’s functions exhibit the characteristic nonlocal nature of conformal symmetry breaking by the anomaly. 
The anomaly contribution can be generated by the nonlocal Riegert action, which in its compact form reads \cite{1984PhLB..134...56R} \cite{AntoniadisMottola1992}

\bea
&&\hspace{-8mm}\cS_{\rm anom}^{^{NL}}[g] =\sdfrac {b'}{8\,}\!\int \!d^4x\,\sqrt{-g_x}\int\! d^4x'\,\sqrt{-g_{x'}}\, 
\bigg[\big(E - \tfrac{2}{3}\sq R\big) +  \sdfrac{b}{\,b'}\,C^2\bigg]_{x}\!
D_4(x,x')\bigg[\big(E - \tfrac{2}{3}\sq R\big) +  \sdfrac{b}{\,b'}\,C^2\bigg]_{x'}\nn\\
&&\hspace{2cm} = \sdfrac {1}{8b'}\int d^4x \int d^4x'\, \cA(x)\, D_4(x,x')\, \cA(x')
\label{Snonlsq}
\eea
where
\begin{equation}
	\Delta_{4}
	= 
	\Box^{2} 
	+ 2 R^{\mu\nu}\nabla_{\mu}\nabla_{\nu}
	-\frac{2}{3} R\, \Box
	+\frac{1}{3} (\nabla^{\mu} R)\nabla_{\mu}.
	\label{eq:PaneitzOp}
\end{equation}
The inverse Paneitz operator \(G_{4}(x)\equiv \Delta_{4}^{-1}(x)\) \cite{Paneitz2008} renders the action \eqref{Snonlsq} intrinsically nonlocal and is responsible for the massless scalar pole characterizing anomalous correlators. The combination $1/4(E-2/3 \Box R)$ defines the first realization of the 
$Q$-curvature operators whose siblings in higher dimensions have been investigated in the context of conformal geometry. 
The anomaly interaction mediated by this action has been explicitly verified at the level of three-point functions, in particular for the \(TTT\) correlator \cite{Coriano:2017mux} against the perturbative description. A comparison with perturbative flat-space results for four-point functions, such as \(TTJJ\), however indicates that additional Weyl-invariant terms are required in order to ensure consistency with the hierarchical conformal Ward identities (CWIs) \cite{Coriano:2022jkn}.

Since the analysis of conformal interactions on curved backgrounds is technically challenging—owing to the absence of a conventional S-matrix— here we focus instead on flat spacetime, where both perturbative and non-perturbative formulations are available and the conformal constraints are fully under control \cite{Bzowski:2013sza,Bzowski:2018fql,Bzowski:2017poo,Coriano:2017mux,Coriano:2018bbe,Coriano:2023gxa,Coriano:2023hts}. This approach allows one to trace, in a controlled way, the activation of a virtual scalar mode in gravitational interactions induced by the trace anomaly. The construction is rigorously defined, since each intermediate step in the derivation of the four–point functions can be reproduced within a standard perturbative framework. In particular, the structure of the relevant conformal three–point functions in momentum space can be reconstructed explicitly using free field theories. This ensures that the correlators entering the analysis admit a fully perturbative realization, while still capturing the anomaly–induced scalar dynamics. Therefore, our analysis combines three-point functions obtained from the solution of the CWIs with single-graviton exchange in the \(s\), \(t\), and \(u\) channels, revealing anomaly-driven corrections that excite a scalar state and modify the standard spin-2 exchange. This analysis will be presented in Section~11, where we first briefly review the derivation of the anomaly-induced part of the $TTT$ correlator and then turn to the computation of the four-graviton amplitude in the presence of anomaly-mediated interactions.

\section{The $TJJ$ correlator and anomaly mediation}

In this section we start investigating the role of trace anomaly interactions and their coupling via a gravitational exchange. The construction of the coupling of two conformal sectors starts from each 3-point vertex, assuming that this is determined from conformal invariance. We start from the $TJJ$ correlator, which allows the description of the photon-photon-graviton vertex in flat space. The vertex will be used in the later sections in the study of 4-point functions.

The structure of these correlators is fully constrained by conformal symmetry and can be obtained by solving the corresponding conformal Ward identities. The resulting solutions are expressed in terms of Appell hypergeometric functions \cite{Coriano:2013jba}, which enter the independent form factors appearing in the tensorial decomposition of the correlator in the form of 3K integrals \cite{Bzowski:2013sza}. These are parametric integrals of three Bessel functions. The general solution is determined up to a finite set of constants and is most naturally analyzed in momentum space. In this representation, the dynamics of the anomaly interaction are encoded in massless exchanges \cite{Giannotti:2008cv,Coriano:2025fom,Armillis:2009pq}, which exhibit a nonvanishing residue on the light-cone. In the present analysis we restrict our attention to the Abelian case, relevant for spin-1 gauge fields, whether belonging to the visible or to a dark sector. The conformal constraints take the form of a dilatation Ward identity

\begin{equation}
	0=\left[\sum_{j=1}^3\Delta_j-(n-1)d-\sum_{j=1}^2\,p_j^\a\sdfrac{\partial}{\partial p_j^\alpha}\right]\braket{{T^{\mu_1\nu_1}(p_1)\,J^{\m_2}(p_2)\,J^{\mu_3}(\bar p_3)}},
	\label{DilatationTJJ}
\end{equation}
where $\Delta_j$ are the scaling dimensions of $T$ ($\Delta_1=4$)  and $J$ $(\Delta_2=\Delta_3=3)$ as for conserved currents.\\
Other constraints are the Lorentz ones 
\begin{align}
	0&=\sum_{j=1}^{2}\left[p_j^\nu\sdfrac{\partial}{\partial p_{j\mu}}-p_j^{\mu}\sdfrac{\partial}{\partial p_{j\nu}}\right]\braket{{T^{\mu_1\nu_1}(p_1)\,J^{\m_2}(p_2)\,J^{\mu_3}(\bar p_3)}}\notag\\
	&\qquad+2\left(\delta^{\nu}_{\a_1}\d^{\mu(\mu_1}-\delta^{\mu}_{\alpha_1}\delta^{\nu(\mu_1}\right)\braket{{T^{\nu_1)\alpha_1}(p_1)\,J^{\m_2}(p_2)\,J^{\mu_3}(\bar p_3)}}\notag\\
	&\qquad+\left(\delta^{\nu}_{\a_2}\d^{\mu\mu_2}-\delta^{\mu}_{\alpha_2}\delta^{\nu\mu_2}\right)\braket{{T^{\mu_1\nu_1}(p_1)\,J^{\alpha_2}(p_2)\,J^{\mu_3}(\bar p_3)}}\notag\\
	&\qquad+\left(\delta^{\nu}_{\a_3}\d^{\mu\mu_3}-\delta^{\mu}_{\alpha_3}\delta^{\nu\mu_3}\right)\braket{{T^{\mu_1\nu_1}(p_1)\,J^{\mu_2}(p_2)\,J^{\alpha_3}(\bar p_3)}}\label{RotationTJJ},
\end{align}
where, in our convention 
\begin{equation}
	\d^{\n(\m_1}\,\braket{{T^{\nu_1)\alpha_1}\,J^{\m_2}\,J^{\mu_3}}}\equiv\sdfrac{1}{2}\big(\d^{\n\m_1}\,\braket{{T^{\nu_1\alpha_1}\,J^{\m_2}\,J^{\mu_3}}}+\d^{\n\n_1}\,\braket{{T^{\mu_1\alpha_1}\,J^{\m_2}\,J^{\mu_3}}}\big),
\end{equation}
and the special conformal Ward identities
\begin{align}
	0&=\sum_{j=1}^{2}\left[2(\Delta_j-d)\sdfrac{\partial}{\partial p_j^\k}-2p_j^\a\sdfrac{\partial}{\partial p_j^\a}\sdfrac{\partial}{\partial p_j^\k}+(p_j)_\k\sdfrac{\partial}{\partial p_j^\a}\sdfrac{\partial}{\partial p_{j\a}}\right]\braket{{T^{\mu_1\nu_1}(p_1)\,J^{\m_2}(p_2)\,J^{\mu_3}(\bar p_3)}}\notag\\
	&\qquad+4\left(\d^{\k(\mu_1}\sdfrac{\partial}{\partial p_1^{\a_1}}-\delta^{\k}_{\alpha_1}\delta^{\l(\mu_1}\sdfrac{\partial}{\partial p_1^\l}\right)\braket{{T^{\nu_1)\alpha_1}(p_1)\,J^{\m_2}(p_2)\,J^{\mu_3}(\bar p_3)}}\notag\\
	&\qquad+2\left(\d^{\k\mu_2}\sdfrac{\partial}{\partial p_2^{\a_2}}-\delta^{\k}_{\alpha_2}\delta^{\l\mu_2}\sdfrac{\partial}{\partial p_2^\l}\right)\braket{T^{\mu_1\nu_1}(p_1)\,J^{\alpha_2}(p_2)\,J^{\mu_3}(\bar p_3)}.
	\label{SCWTJJ}
\end{align} 
We have denoted with $\bar p_3$ the momentum of one of the current, taken as a function of the other two incoming momenta $p_1$ and $p_2$ as  $\bar p_3=-p_1 -p_2$. Momentum differentiation should be performed only with respect to $p_1$ and $p_2$ and implicitly on $\bar p_3$ as a function of the first two. 
The former constraints are accompanied by conservation Ward identities
\begin{align}
	\label{tr}
	p_{1\n_1}\braket{T^{\mu_1\nu_1}(p_1)\,J^{\m_2}(p_2)\,J^{\m_3}(p_3)}_q&=4\,\big[\d^{\m_1\m_2}p_{2\l}\braket{J^\l(p_1+p_2)\,J^{\m_3}(p_3)}_q-p_2^{\m_1}\braket{J^{\m_2}(p_1+p_2)\,J^{\m_3}(p_3)}_q\big]\notag\\
	&+4\,\big[\d^{\m_1\m_3}p_{3\l}\braket{J^\l(p_1+p_3)\,J^{\m_2}(p_2)}_q-p_3^{\m_1}\braket{J^{\m_3}_q(p_1+p_3)\,J^{\m_2}(p_2)}_q\big].
\end{align}
with the 2-point function of two conserved vector currents $J_i$ $(i=2,3)$  in any conformal field theory in $d$ dimensions given by 
\beqa
\langle J_2^\alpha(p)J_3^\beta(-p) \rangle =\delta_{\Delta_2\, \Delta_3}\left(c_{123} \Gamma_J \right)\pi^{\alpha\beta}(p) (p^2)^{\Delta_2-d/2},
\qquad \Gamma_J=\frac{\pi^{d/2}}{ 4^{\Delta_2 -d/2}}\frac{\Gamma(d/2-\Delta_2)}{\Gamma(\Delta_2)},
\eeqa
with $c_{123}$ an overall constant and $\Delta_2=d-1$ are the ordinary scaling dimensions of conserved currents. In the coupling of the two conformal sectors, once we take the on-shell limit of the spin-1 external states, then one finds that the spin-1 part of the propagator decouples in the amplitude. In other terms the conservation Ward identity can be taken as a homogeneous equation. 
Concerning the $JJ$, this is affected by a divergence that can be identified, for example, just by a simple free field theory analysis with $n_f$ fermions and requires the introduction of the counterterm
\beq
\mathcal{L}_{\textrm{q\, count}}=-\frac{1}{(d-4)}\frac{g_s^2}{16 \pi^2}\frac{2}{3} n_f
F^{a\, \mu\nu}F^{a}_{\mu\nu}.
\eeq
$n_f$ denotes the number of flavours running in the loop of the $JJ$ and $d$ the spacetime dimensions. Free field theory parallels in every single step the reconstruction of this correlator, otherwise performed by the general approach developed in \cite{Bzowski:2013sza}.
Indeed, in the case of the $TJJ$, using a free-field theory realization with $n_f$ fermion flavours, the general solution of the conformal Ward identities in the Abelian case takes the form 
\begin{align}
	\!\big\langle
	T_{\mu_1\nu_1}(p_1)\,
	J_{\mu_2}(p_2)\,
	J_{\mu_3}(p_3)
	\big\rangle
	&=
	\langle
	t_{\mu_1\nu_1}(p_1)\,
	j_{\mu_2}(p_2)\,
	j_{\mu_3}(p_3)
	\rangle
	\nonumber\\
	&\quad
	+
	2\,\mathscr{T}_{\mu_1\nu_1}{}^{\alpha}(p_1)
	\Big[
	\delta^{\mu_3}{}_{[\alpha}\,p_{3\beta]}\,
	\langle
	J_{\mu_2}(p_2)\,
	J^{\beta}(-p_2)
	\rangle
	\nonumber\\
	&\qquad\qquad\qquad
	+
	\delta^{\mu_2}{}_{[\alpha}\,p_{2\beta]}\,
	\big\langle
	J_{\mu_3}(p_3)\,
	J^{\beta}(-p_3)
	\big\rangle\!
	\Big]
	\nonumber\\
	&\quad
	+
	\frac{1}{d-1}\,
	\pi_{\mu_1\nu_1}(p_1)\,
	\mathcal A^{\mu_2\mu_3},
	\label{eq:TJJreconstructed2}
\end{align}
with
\begin{equation} \label{a:T}
	\mathscr{T}_{\mu\nu \alpha} (\bs{p}) = \frac{1}{p^2} \left[ 2 p_{(\mu} \delta_{\nu)\alpha} - \frac{p_\alpha}{d-1} \left( \delta_{\mu\nu} + (d-2) \frac{p_\mu p_\nu}{p^2} \right) \right]
\end{equation}	
and
\[\label{Idecomp}
\delta_{\mu(\alpha}\delta_{\beta)\nu}=\Pi_{\mu\nu\alpha\beta}(\bs{p})+\mathscr{T}_{\mu\nu (\alpha}(\bs{p})\,p_{\beta)}+\frac{1}{d-1}\pi_{\mu\nu}(\bs{p})\delta_{\alpha\beta}.
\]
The contribution of the anomaly to the expansion of the correlator is given by
\beq
\mathcal{A}_q^{\alpha\beta} = -{1 \over 3} \, \,  \, \, \frac{g_s^2}{16\pi^2} \,  2n_f u^{\alpha \beta}(p_1,p_2),
\label{anoms}
\eeq 
The correlator is decomposed into a transverse traceless part $\langle
t_{\mu_1\nu_1}(p_1)\,
j_{\mu_2}(p_2)\,
j_{\mu_3}(p_3)
\rangle$, a longitudinal one (proportional to $JJ$) and a trace part.
The trace Ward identity includes the anomaly contribution 
\begin{equation}
	\label{x2}
	g_{\mu\nu} \langle T^{\mu\nu}(p_1)J^{\alpha }(p_2)J^{\beta }(p_3)\rangle_q=\beta(e) u^{\alpha \beta  },
\end{equation}
with 
\bea
&&u^{\alpha\beta }(p,q) = -\frac{1}{4}\int\,d^4x\,\int\,d^4y\ e^{ip\cdot x + i q\cdot y}\ 
\frac{\delta^2 \{F_{a\mu\nu}F^{a\mu\nu}(0)\}} {\delta A_{\alpha}(x) \delta A_{\beta}(y)}\vline_{A=0} \,
\label{locvar}
\eea
explicitly given by
\beq\label{uab}
u^{\a \b } (p,q)\equiv (p\cdot q) g^{\alpha\beta} - p^\beta q^\alpha.
\eeq
The pole in the trace sector is part of the $\pi^{\mu\nu}$ projector. 
This tensor structure summarizes the conformal anomaly contribution, being the Fourier transform of the anomaly term at $O(e^2)$. Notice that the spin decomposition in \eqref{eq:TJJreconstructed2} is 
characterized by a spin-2 part, coupling to the $P^{(2)}$ component of the graviton propagator, and a spin-0 
part associated to the anomaly, with an interpolating $P^{(0)}$ projector. 
The transverse traceless (spin-2) gravitational sector, is related to the $tjj$ part of the correlator, the spin-1 $(jj)$ part decouples due to energy momentum conservation once we couple two $TJJ$ vertices with a virtual graviton. These are the three orthogonal sectors that are matched with a similar decomposition of the correlators.

The transverse traceless 
part has been discussed in \cite{Bzowski:2013sza,Bzowski:2018fql} using a regularization of the conformal integrals (3K integrals) useful for the analysis of general conformal theories. The identification of the form factors appearing in the decomposition of this sector, the only relevant for the exchange of a spin-2 graviton in a BIM-like diagram can be conveniently done using a Gram decomposition. 
The $tt$ sector is decomposed, with a minor modification of the notation of \cite{Bzowski:2018fql}, as  

\begin{equation}
	\langle t jj\rangle_{\mathrm{TT}}
	=
	\sum_{i=1}^{5}
	\tilde{A}_i(s,s_1,s_2)\,
	{T}_i^{\mu_1\nu_1\mu_2\mu_3}.
	\label{eq:TTdecomp}
\end{equation}

\begin{align}
	T_1^{\mu_1\nu_1\mu_2\mu_3}
	&=
	\Pi_{\mu_1\nu_1\alpha_1\beta_1}(p_1)\,
	\pi^{\mu_2}_{\alpha_2}(p_2)\,
	\pi^{\mu_3}_{\alpha_3}(p_3)\,
	p_2^{\alpha_1} p_2^{\beta_1} p_3^{\alpha_2} p_1^{\alpha_3},
	\\[2mm]
	T_2^{\mu_1\nu_1\mu_2\mu_3}
	&=
	\Pi_{\mu_1\nu_1\alpha_1\beta_1}(p_1)\,
	\pi^{\mu_2}_{\alpha_2}(p_2)\,
	\pi^{\mu_3}_{\alpha_3}(p_3)\,
	\delta^{\alpha_2\alpha_3}\,
	p_2^{\alpha_1} p_2^{\beta_1},
	\\[2mm]
	T_3^{\mu_1\nu_1\mu_2\mu_3}
	&=
	\Pi_{\mu_1\nu_1\alpha_1\beta_1}(p_1)\,
	\pi^{\mu_2}_{\alpha_2}(p_2)\,
	\pi^{\mu_3}_{\alpha_3}(p_3)\,
	\delta^{\alpha_1\alpha_2}\,
	p_2^{\beta_1} p_1^{\alpha_3},
	\\[2mm]
	T_4^{\mu_1\nu_1\mu_2\mu_3}
	&=
	\Pi_{\mu_1\nu_1\alpha_1\beta_1}(p_1)\,
	\pi^{\mu_2}_{\alpha_2}(p_2)\,
	\pi^{\mu_3}_{\alpha_3}(p_3)\,
	\delta^{\alpha_1\alpha_3}\,
	p_2^{\beta_1} p_3^{\alpha_2},
	\\[2mm]
	T_5^{\mu_1\nu_1\mu_2\mu_3}
	&=
	\Pi_{\mu_1\nu_1\alpha_1\beta_1}(p_1)\,
	\pi^{\mu_2}_{\alpha_2}(p_2)\,
	\pi^{\mu_3}_{\alpha_3}(p_3)\,
	\delta^{\alpha_1\alpha_3}\,
	\delta^{\alpha_2\beta_1}.
\end{align}
Notice that $T_3$ and $T_4$ are related by the exchange of the momenta and indices of the two $J$ currents. The scalar functions $\tilde{A_i}$ are the $tt$ form factors. By construction, they are free from kinematic ambiguities and encode the genuinely dynamical content of the three--point function. 
We equip the $tt$ tensor space with the natural bilinear form
\begin{equation}
	(X,Y)
	\equiv
	X^{\mu_1\nu_1\mu_2\mu_3}
	Y_{\mu_1\nu_1\mu_2\mu_3}.
	\label{eq:innerprod}
\end{equation}
Using this pairing, we define the Gram matrix associated with the basis \(\{{T}_i\}\) as
\begin{equation}
	G_{ij}
	\equiv
	({T}_i,{T}_j)
	=
	{T}_i^{\mu_1\nu_1\mu_2\mu_3}
	{T}_{j\,\mu_1\nu_1\mu_2\mu_3}.
	\label{eq:Gram}
\end{equation}
For generic values of \((s,s_1,s_2)\), the tensors \({T}_i\) are linearly independent and the Gram matrix is non--degenerate. Its inverse \(G^{-1}\) therefore exists. Contracting the decomposition \eqref{eq:TTdecomp} with \({T}_j\) and using \eqref{eq:Gram}, one obtains the linear system
\begin{equation}
	({T}_j,\langle TJJ\rangle)
	=
	\sum_{k=1}^{5}
	G_{jk}\,\tilde{A}_k.
	\label{eq:linearsystem}
\end{equation}
Multiplication by the inverse Gram matrix immediately yields the projection formula
\begin{equation}
	\tilde A_i(s,s_1,s_2)
	=
	\sum_{j=1}^{5}
	(G^{-1})_{ij}\,
	{T}_j^{\mu_1\nu_1\mu_2\mu_3}
	\,
	\langle T_{\mu_1\nu_1} J_{\mu_2} J_{\mu_3} \rangle.
	\label{eq:projection}
\end{equation}
This relation is purely algebraic and holds independently of any perturbative realization. In particular, the full correlator may be used on the right--hand side, since the $tt$ tensors are orthogonal to the non--$tt$ sector.
The regularization of the correlator follows the use of the counterterm action from which one extracts the corresponding vertex $\Delta_{ct} TJJ$ which is then projected onto all the components $\tilde{A}_i$ using the same Gram projection shown above, where the vertex extracted from the counterterm by functional differentiation is projected over all the $tt$ form factors

\begin{equation}
	A_{j }^{}= -\,  n_f\,   \frac{g_s^2}{16\pi^2}\, \bar{A}_j, \qquad j=1,2\ldots 4
\end{equation}

\begin{eqnarray}
	\label{Abar}
	\bar{A}_1  &=& \frac{1}{48 \left(p_1^2 p_2^2-(p_1 \cdot p_2)^2 \right)^4} \Big[ A_{10} + A _{11} B_0 (p_1^2) + A_{12} B_0(p_2^2) + A_{13} B_0(q^2) + A_{14} C_0 (p_1^2 , p_2^2 , q^2 ) \Big] \notag \\
	\bar{A}_2  &=& -\frac{1}{144 \left(p_1^2 p_2^2-(p_1 \cdot p_2)^2 \right)^3} \Big[ A_{20} + A_{21}  B_0 (p_1^2) + A_{22} B_0(p_2^2)  + A_{23} B_0(q^2) + A_{24} C_0 (p_1^2 , p_2^2 , q^2 ) \Big] \notag \\
	\bar{A}_3  &=& \frac{1}{72 \left(p_1^2 p_2^2-(p_1 \cdot p_2)^2 \right)^3} \Big[A_{31}  B_0 (p_1^2) + A_{32} B_0(p_2^2)  + A_{33} B_0(q^2) + A_{34} C_0 (p_1^2 , p_2^2 , q^2 ) \Big] \notag \\
	\bar{A}_5  &=& -\frac{1}{72 \left(p_1^2 p_2^2-(p_1 \cdot p_2)^2 \right)^2} \Big[A_{40} + A_{41}  B_0 (p_1^2) + A_{42} B_0(p_2^2) + A_{43} B_0(q^2) + A_{44} C_0 (p_1^2 , p_2^2 , q^2 ) \Big]. \notag \\
\end{eqnarray}
and $A_4$ obtained from $A_3$ with the exchange of $p_1$ with $p_2$.
We give here, as an example, the first polynomial appearing in $\bar{A}_1$, the other exhibiting a similar structure
\begin{eqnarray}
	A_{10} &=& 16\, (p_1\cdot p_2)^7+36 \, p_1^2 \, (p_1\cdot p_2)^6+36 \, p_2^2 \, (p_1\cdot p_2)^6+12 \, p_1^4 \, (p_1\cdot p_2)^5+12 \, p_2^4 \, (p_1\cdot p_2)^5+240 \, p_1^2 \, p_2^2 \, (p_1\cdot p_2)^5 \notag \\ &&
	+188 \, p_1^2 \, p_2^4 \, (p_1\cdot p_2)^4+188 \, p_1^4 \, p_2^2 \, (p_1\cdot p_2)^4+46 \, p_1^2 \, p_2^6 \, (p_1\cdot p_2)^3-108 \, p_1^4 \, p_2^4 \, (p_1\cdot p_2)^3+46 \, p_1^6 \, p_2^2 \, (p_1\cdot p_2)^3 \notag \\ &&
	-204 \, p_1^4 \, p_2^6 \, (p_1\cdot p_2)^2-204 \, p_1^6 \, p_2^4 \, (p_1\cdot p_2)^2-20 \, p_1^6 \, p_2^8 \,-20 \, p_1^8 \, p_2^6 \,-58 \, p_1^4 \, p_2^8 \, (p_1\cdot p_2) \notag \\ &&
	-148 \, p_1^6 \, p_2^6 \, (p_1\cdot p_2)-58 \, p_1^8 \, p_2^4 \, (p_1\cdot p_2) 
\end{eqnarray}
Their explicit expressions can be found in Appendix G of \cite{Coriano:2024qbr}, by sending the Casimir 
$C_A\to 0$.\\
The Gram--matrix method ensures that this extraction is unique and free from ambiguities related to trace or longitudinal terms. The tensors structures are identical to those introduced in 
\cite{Bzowski:2018fql}, which are four,  since $\tilde{A_3}$ and $\tilde{A_4}$  and $T_3$ and $T_4$ differ just by the exchange of the momenta/polarization of the two photons.  
As we have mentioned, this scheme can be perfectly matched by the perturbative analysis of two combined free field theory with multiplicities $n_s$ and $n_f$ 
describing the general structure allowed by the correlator in a CFT. Such correspondences have been investigated both within the abstract CFT approach and the perturbative realizations, and extended to parity odd correlators as well \cite{Coriano:2023hts,Coriano:2023gxa,Coriano:2023cvf}. 
\subsection{The general conformal solution and its on-shell limit} \label{sksubseq}
The general conformal solution is expressed in terms of an overall normalization constant $C_1$ and the normalization of the $JJ$ 2-point function $C_{JJ}$. Another coefficient $D_{JJ}$ is introduced by the renormalization scheme implemented in \cite{Bzowski:2018fql}.

\begin{equation}
	\big\langle
	J^{\mu}(p)\,J^{\nu}(-p)
	\rangle
	=
	\pi^{\mu\nu}(p)\,
	p^{2N-2}
	\left[
	C_{JJ}\,
	\ln\!\left(\frac{p^{2}}{\mu^{2}}\right)
	+
	D_{JJ}
	\right],
	\label{eq:JJ2pt}
\end{equation}
One can check that the solution presented in free field theory or given in \cite{Bzowski:2018fql} 
is infrared safe for on-shell spin-1. Indeed, in the on-shell photon limit the expression of the transverse traceless form factors  simplify drastically and take the form

\begin{align}
	A_1 &= \frac{8 C_1}{p_1^2} \nn \\
	A_2 &= 2 C_1 + 2 C_{JJ} \left(\frac{4}{3} - \log p_1^2\right) - 2 D_{JJ} \nn \\
	A_3 &=A_4= -2 C_1 + 2 C_{JJ} \left(\frac{4}{3} - \log p_1^2\right) - 2 D_{JJ}\nn \\
	A_5 &=  2p_1^2 \left[ C_1 - C_{JJ} \left(\frac{2}{3}  - \log p_1^2\right) + D_{JJ} \right].
	\label{bmcs}
\end{align}
The complete expression of the correlator in the notations of \cite{Bzowski:2018fql} is given in Appendix 
\eqref{bms}. The finiteness of the result can be directly checked by an explicit computation and deserves few comments. 
In the on-shell limit of the two photons in which $p_2^2, p_3^2\to 0$, before using the transversality condition on the two external vector lines, the computation manifests a double pole $1/(d-4)^2$ as well 
as a single pole, which is attributed to the scalar $C_0$ integral. The double pole characterizes a soft and collinear singularity in the variable $q^2\equiv p_1^2$ describing the invariant mass of the off-shell graviton decaying into two on-shell spin 1. Once we use the transversality conditions on the photons, the amplitude for an off-shell graviton 
to decay into two on-shell photons is finite and nonzero.   

\subsection{The perturbative implementation of the $TJJ$ correlator}
Here we briefly review the perturbative computation of the correlator, in order to summarize its structure in the case of on-shell photons. \\

We consider the standard coupling in an Abelian theory, which for simplicity we take to be QED. The same analysis applies to a generic conformal sector. In that case, the result presented below is simply multiplied by a multiplicity factor \(n_f\), accounting for a generic number of fermions, while \(e\) denotes the Abelian gauge coupling. In QED, the energy--momentum tensor (EMT) of a fermion of mass \(m\) is
\begin{equation}
	T_{\mu\nu}
	=
	\frac{i}{4}\,\bar{\psi}
	\left(
	\gamma_\mu \overleftrightarrow{D}_\nu
	+
	\gamma_\nu \overleftrightarrow{D}_\mu
	\right)\psi
	-
	g_{\mu\nu}\mathcal{L},
\end{equation}
where \(\mathcal{L}\) is the QED Lagrangian. When computing the \(TJJ\) correlator
\begin{equation}
	\langle T_{\mu\nu}(k) J_\alpha(p) J_\beta(q)\rangle,
	\qquad k=p+q,
\end{equation}
one must include both the triangle diagrams and the contact terms generated by the covariant derivative
\begin{equation}
	D_\mu=\partial_\mu+i e A_\mu.
\end{equation}
The fermionic part of the energy--momentum tensor is
\begin{equation}
	T^{\mu\nu}_\psi
	=
	\frac{i}{4}
	\bar\psi
	\left(
	\gamma^\mu \overleftrightarrow{\partial}^{\nu}
	+
	\gamma^\nu \overleftrightarrow{\partial}^{\mu}
	\right)\psi
	-\eta^{\mu\nu}\,
	\bar\psi (i\gamma^\rho \partial_\rho - m)\psi,
	\label{Tmunu_fermion}
\end{equation}
with
\begin{equation}
	\bar\psi \overleftrightarrow{\partial}^\mu \psi
	=
	\bar\psi (\partial^\mu \psi)
	-
	(\partial^\mu \bar\psi)\psi.
	\label{leftrightarrow_def}
\end{equation}

In momentum space, with incoming fermion momentum \(p\) and outgoing momentum \(p'\), the graviton--fermion--fermion vertex is
\begin{equation}
	V^{\mu\nu}_{h\bar\psi\psi}(p',p)
	=
	-\,\frac{i\kappa}{8}
	\left[
	\gamma^\mu (p'+p)^\nu
	+
	\gamma^\nu (p'+p)^\mu
	-2 \eta^{\mu\nu}\left(\gamma^\rho (p'+p)_\rho-2m\right)
	\right].
	\label{vertex_hpsipsi}
\end{equation}
The Maxwell energy--momentum tensor is
\begin{equation}
	T^{\mu\nu}_{\text{Maxwell}}
	=
	F^{\mu\lambda}F^\nu_{\ \lambda}
	-\frac14 \eta^{\mu\nu}F^{\rho\sigma}F_{\rho\sigma},
	\label{Tmunu_Maxwell}
\end{equation}
and, for photon momenta \(p\) and \(q\), the \(hAA\) vertex reads
\begin{equation}
	\begin{aligned}
		V^{\mu\nu\alpha\beta}_{hAA}(p,q)
		=
		-\,i\kappa
		\Big[
		& \eta^{\alpha\beta} p^{(\mu} q^{\nu)}
		-\eta^{\mu\nu} p^{(\alpha} q^{\beta)}
		+ \eta^{\mu(\alpha} q^{\beta)} p^{\nu}
		+ \eta^{\nu(\alpha} p^{\beta)} q^{\mu}
		- \eta^{\mu(\alpha}\eta^{\beta)\nu} (p\!\cdot\! q)
		\Big],
	\end{aligned}
	\label{vertex_hAA}
\end{equation}
where symmetrization is defined by
\begin{equation}
	a^{(\mu} b^{\nu)}
	=
	\frac12 \left(a^\mu b^\nu + a^\nu b^\mu\right).
\end{equation}
Expanding the Dirac action in curved space also generates the mixed contact interaction
\begin{equation}
	h_{\mu\nu} A_\alpha \bar\psi \psi,
	\label{mixed_structure}
\end{equation}
whose momentum-space vertex is
\begin{equation}
	V^{\mu\nu\alpha}_{hA\bar\psi\psi}
	=
	\frac{i\kappa e}{4}
	\left[
	\eta^{\mu\alpha}\gamma^\nu
	+
	\eta^{\nu\alpha}\gamma^\mu
	-
	\eta^{\mu\nu}\gamma^\alpha
	\right].
	\label{vertex_hApsipsi}
\end{equation}
The full Abelian \(TJJ\) amplitude is then
\begin{equation}
	\Gamma^{\mu\nu\alpha\beta}
	=
	\Gamma^{\mu\nu\alpha\beta}_{\text{triangle}}(p,q)
	+
	\Gamma^{\mu\nu\beta\alpha}_{\text{triangle}}(q,p)
	+
	\Gamma^{\mu\nu\alpha\beta}_{\text{contact}}(p,q)
	+
	\Gamma^{\mu\nu\beta\alpha}_{\text{contact}}(q,p),
\end{equation}
where \(p\) and \(q\) are the photon momenta. The two contributions are given by the following loop integrals:
\begin{equation}
	\Gamma^{\mu\nu\alpha\beta}_{\text{triangle}}(p,q)
	=
	-\int\frac{d^4 l}{(2\pi)^4}
	\, {\rm tr}\left\{
	V^{\mu\nu}_{h\bar\psi\psi}(l-q,l+p)
	\frac{1}{\lsl-\qsl-m}
	\gamma^{\beta}
	\frac{1}{\lsl-m}
	\gamma^{\alpha}
	\frac{1}{\lsl+\psl-m}
	\right\},
\end{equation}
and
\begin{equation}
	\Gamma^{\mu\nu\alpha\beta}_{\text{contact}}(p,q)
	=
	-\int\frac{d^4 l}{(2\pi)^4}
	\, {\rm tr}\left\{
	V_{hA\bar\psi\psi}^{\mu\nu\alpha}
	\frac{1}{\lsl-\qsl-m}
	\gamma^{\beta}
	\frac{1}{\lsl-m}
	\right\}.
	\label{loop}
\end{equation}
The full amplitude satisfies the Ward identities
\begin{equation}
	p_\alpha \Gamma^{\mu\nu\alpha\beta}=0,
	\qquad
	q_\beta \Gamma^{\mu\nu\alpha\beta}=0.
	\label{Ward_identities}
\end{equation}

\subsection{The anomaly}
\label{reff2}
At the operator level, the trace of the EMT is
\begin{equation}
	T^\mu_{\ \mu}
	=
	\mathcal{A}(x),
\end{equation}
where \(\mathcal{A}(x)\) denotes the trace anomaly. For a single massless fermion, the electromagnetic contribution 
\begin{equation}
	\mathcal{A}
	=
	\frac{\beta(e)}{2e}
	F_{\mu\nu}F^{\mu\nu}
	=
	\frac{\alpha}{3\pi}
	F_{\mu\nu}F^{\mu\nu}.
\end{equation}

\begin{figure}[t]
	\begin{center}
		\includegraphics[scale=0.7,angle=0]{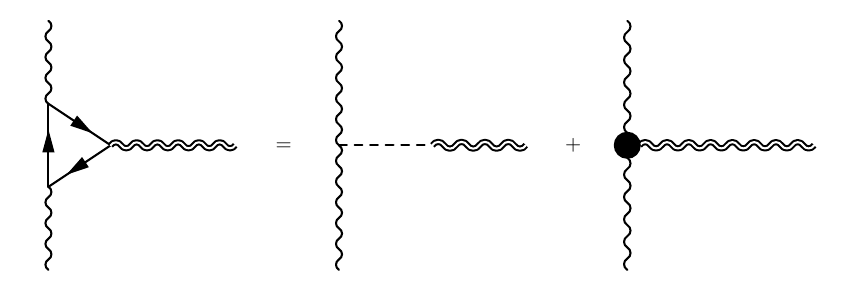}
		\caption{Decomposition of the \(TJJ\) vertex into an anomalous contribution and a transverse--traceless sector (black blob).}
		\label{expansionX}
	\end{center}
\end{figure}

The Abelian correlator has been studied in the past in a basis of thirteen tensor structures \(t_1,\ldots,t_{13}\), reported in Appendix \ref{reff}, with corresponding form factors \(F_1,\ldots,F_{13}\). For on-shell photons and massless fermions, the nonvanishing form factors are ($s\equiv k^2$)
\bea
F_{1} (s, 0, 0, 0) &=& - \frac{e^2}{18 \pi^2  s}, \\
F_{3} (s, 0, 0, 0) &=&  F_{5} (s, 0, 0, 0) = - \frac{e^2}{144 \pi^2 \, s}, \\
F_{7} (s, 0, 0, 0) &=& -4 \, F_{3} (s, 0, 0, 0), \\
F_{13, R} (s, 0, 0, 0) &=& - \frac{e^2}{144 \pi^2} \, \left[ 12 \log \left(-\frac{s}{\mu^2}\right) - 35\right],
\qquad s <0 \eea
Accordingly, the correlator with two on-shell photons can be written as \cite{Armillis:2009pq}
\bea
\Gamma^{\mu\nu\alpha\beta}  (s; 0,0,0) &=&
F_{1} (s, 0, 0, 0) \, \widetilde{t}^{\, \mu \nu \alpha \beta}_{1}
+ F_{3} (s, 0, 0, 0) \,
\left(
\widetilde{t}^{\, \mu \nu \alpha \beta}_{3}
+ \widetilde{t}^{\, \mu \nu \alpha \beta}_{5}
- 4 \widetilde{t}^{\, \mu \nu \alpha \beta}_{7}
\right)
+ F_{13, R} \, \widetilde{t}^{\, \mu \nu \alpha \beta}_{13},
\label{gamma00}
\eea
with
\bea
\widetilde t_1^{\, \mu \nu \a \b} &=&
(s \, g^{\mu\nu} - k^{\mu}k^{\nu}) \, u^{\a \b} (p,q)
=\hat{\pi}^{\mu\nu}(s) \, u^{\a \b} (p,q),
\label{widetilde1}\\
\widetilde t_3^{\, \mu \nu \a \b}
+ \widetilde t_5^{\, \mu \nu \a \b}
- 4 \widetilde t_7^{\, \mu \nu \a \b} &=&
- 2 \, u^{\a \b} (p,q)
\left(
s \, g^{\mu \nu}
+ 2 (p^\mu p^\nu + q^\mu q^\nu)
- 4 (p^\mu q^\nu + q^\mu p^\nu)
\right),
\eea
and
\bea
\widetilde{t}^{\, \mu \nu \alpha \beta}_{13} &=&
\big(p^{\mu} q^{\nu} + p^{\nu} q^{\mu}\big)g^{\alpha\beta}
+ \frac{s}{2} \big(g^{\alpha\nu} g^{\beta\mu} + g^{\alpha\mu} g^{\beta\nu}\big) \nn \\
&&
- g^{\mu\nu} \left(\frac{s}{2} g^{\alpha \beta}- q^{\alpha} p^{\beta}\right)
-\big(g^{\beta\nu} p^{\mu}
+ g^{\beta\mu} p^{\nu}\big)q^{\alpha}
-\big(g^{\alpha\nu} q^{\mu}
+ g^{\alpha\mu} q^{\nu }\big)p^{\beta}.
\label{widetilde13}
\eea
Each of these three tensor structures is separately conserved and gauge invariant on the photon lines in the massless limit. This solution can be mapped to the basis introduced in Section \ref{sksubseq} through
\bea
F_3(s,0,0,0)&=& -\frac{A_1 s^2-2 A_3 s-4 A_5}{24 s^2},\nn\\
F_{13,R}(s,0,0,0)&=& -\frac{A_3}{2}.
\eea
Two characteristic features of this expression should be emphasized. The first is associated with the tensor structure \(\tilde t_1\) and the corresponding form factor \(F_1\), which contains the anomaly pole and contributes to the traced part of the correlator (see Fig.~2). This term controls the exchange of the scalar component of the graviton and originates from the nonlocal nature of the effective interaction. A second pole appears in the traceless sector and is encoded in the form factor \(F_3\). This contribution has been pointed out in earlier analyses and interpreted as a possible ``plasmon'' mode \cite{Giannotti:2008cv}; in the present framework it is generated by exchange of the spin--2 component of the virtual graviton. While the scalar mode is interpolated by a  projector, as we are going to show in the next sections, the two plasmon modes couple through the spin--2 sector of the exchanged graviton.\\
The decomposition of this correlator derived in \cite{Armillis:2009pq} in the on-shell limit has been extended to the non-Abelian case in \cite{Coriano:2025fom}, in the analysis of dilaton sum rules in QCD. In particular, the residue at the \(k^2=0\) pole was computed in \cite{Coriano:2025fom}. The result contains two contributions: the first, proportional to \(\tilde\phi_1^{\mu\nu\alpha\beta}\), is associated with the conformal anomaly, while the second, proportional to \(\tilde\phi_2^{\mu\nu\alpha\beta}\), corresponds to an additional pole in a traceless tensor structure. The corresponding pole relation is
\beq\label{tjjpolelim}
\lim_{s \to 0} \, k^2 \braket{T^{\mu\nu}(k) J^{\alpha }(p) J^{\beta }(q)}
= -\frac{g_s^2}{48 \pi^2} \left(\frac{2}{3} n_f\right) \tilde\phi_1^{\mu\nu\alpha\beta}
- \frac{g_s^2}{288 \pi^2} n_f \, \tilde\phi_2^{\mu\nu\alpha\beta},
\eeq
which is in agreement with \eqref{gamma00}. Indeed, one finds
\begin{align}
	\tilde\phi_1^{\, \mu \nu \alpha \beta} (p,q) &= \tilde{t}_1^{\mu\nu\a\b},
	\label{widetilde1phi} \\
	\tilde\phi_2^{\, \mu \nu \alpha \beta} (p,q) &=
	\widetilde t_3^{\, \mu \nu \a \b}
	+ \widetilde t_5^{\, \mu \nu \a \b}
	- 4 \widetilde t_7^{\, \mu \nu \a \b}.
	\label{widetilde2}
\end{align}
These identify massless exchanges in distinct channels. The anomaly pole contribution behaves as
\(\sim s^{-1}\tilde{\phi}_1^{\mu\nu \a\b}\),
while the additional pole in the traceless sector,
\(\sim s^{-1}\tilde{\phi}_2^{\mu\nu \a\b}\),
is reproduced by the nonlocal effective action
\begin{equation}
	S_{\text{extra pole}}
	=
	\frac{g_s^2}{72\pi^2} n_f
	\int d^4x \sqrt{-g}
	\int d^4x' \sqrt{-g'}
	\, h_{\mu\nu}(x)
	\, \Box^{-1}_{x,x'}
	\, \mathcal{O}^{\mu\nu}(x'),
	\label{extra_pole_action}
\end{equation}
where
\begin{equation}
	\mathcal{O}^{\mu\nu}
	=
	3(\partial^\mu\partial^\nu F_{\alpha\beta}^a)F^{\alpha\beta a}
	+\frac14 (g^{\mu\nu}\Box-4\partial^\mu\partial^\nu)
	F_{\alpha\beta}^aF^{\alpha\beta a}.
	\label{operator_def}
\end{equation}
This is in agreement with the analysis of Giannotti and Mottola \cite{Giannotti:2008cv}. The second traceless structure therefore identifies an additional nonlocal contribution to the effective action, describing an interaction localized around the light cone,
\begin{align}\label{expoleac}
	S_{\text{extra pole}}
	=
	\frac{g_s^2}{72 \pi^2} n_f
	\int d^4x \sqrt{-g}
	\int d^4x' \sqrt{-g'}\, h_{\mu\nu}(x) \Box^{-1}_{x,x'}
	\left[
	3 (\partial^\mu \partial^\nu F^a_{\alpha\beta}) F^{\alpha\beta a}
	+ \frac{1}{4} \left( g^{\mu\nu} \Box - 4 \partial^\mu \partial^\nu \right) F^a_{\alpha\beta} F^{\alpha\beta a}
	\right]_{x'},
\end{align}
whereas the first tensor structure is associated with the familiar anomaly action \cite{Armillis:2009pq}
\bea
S_{\text{anom}} &=&
\frac{1}{3} \, \frac{g^3}{16 \pi^2}
\left(  \frac{2}{3} \, n_f \right)
\int d^4 x \, d^4 y \,
R^{(1)}(x)\, \square^{-1}(x,y) \,
F^a_{\alpha \beta}F^{\alpha \beta a}.
\eea

In the next sections we are going to show how the nonlocal structure of the anomaly effective action with the inclusion of $S_{\text{anom}}$ is responsible for the activation of a scalar degree of freedom in the case of 4-point functions, discussing the gravitational scattering of photons. For this, we need to turn to 
the massless Fierz-Pauli Lagrangian, extracted from the EH action at quadratic level in the metric fluctuations and address the gauge dependence of the virtual graviton.

\section{Towards 4-point functions: spin projectors and the scalar sector in the de~Donder gauge}

In order to fix our conventions, we start at linearized level, with the metric in the  Einstein--Hilbert action in mostly minus signature
\begin{equation}
	S_{\rm EH}=\frac{1}{2\kappa^{2}}\int d^{4}x\,\sqrt{-g}\,R,
	\qquad
	\kappa^{2}=16\pi G_N 
	\label{EHaction}
\end{equation}
and
\begin{equation}
	\kappa^2 = \frac{1}{M_P^2},
\end{equation}
where $M_P$ denotes the reduced Planck mass.
We linearize the metric around flat spacetime
\begin{equation}
	g_{\mu\nu}=\eta_{\mu\nu}+\kappa h_{\mu\nu},
	\qquad
	|h_{\mu\nu}|\ll1 ,
	\label{metricexpand}
\end{equation}
using $\eta_{\mu\nu}$ to raise and lower indices. The linearized Christoffel symbols are
\begin{equation}
	\Gamma^{\rho}{}_{\mu\nu}
	=\frac12\,\eta^{\rho\sigma}\left(
	\partial_\mu h_{\nu\sigma}
	+\partial_\nu h_{\mu\sigma}
	-\partial_\sigma h_{\mu\nu}
	\right),
\end{equation}
from which one obtains the linearized Ricci tensor and scalar
\begin{equation}
	R^{(1)}_{\mu\nu}
	=\frac12\left(
	\partial_{\rho}\partial_{\mu}h^{\rho}{}_{\nu}
	+\partial_{\rho}\partial_{\nu}h^{\rho}{}_{\mu}
	-\Box h_{\mu\nu}
	-\partial_{\mu}\partial_{\nu}h
	\right),
\end{equation}
\begin{equation}
	R^{(1)}=\partial_{\mu}\partial_{\nu}h^{\mu\nu}-\Box h ,
\end{equation}
with $\Box=\eta^{\mu\nu}\partial_\mu\partial_\nu$. The linearized Einstein tensor is
\begin{equation}
	G^{(1)}_{\mu\nu}
	=
	\frac12\left(
	\Box h_{\mu\nu}
	-\partial_{\mu}\partial^{\rho}h_{\rho\nu}
	-\partial_{\nu}\partial^{\rho}h_{\rho\mu}
	+\partial_{\mu}\partial_{\nu}h
	+\eta_{\mu\nu}\partial_{\rho}\partial_{\sigma}h^{\rho\sigma}
	-\eta_{\mu\nu}\Box h
	\right)
	\label{FierzPauli}
\end{equation}
and the linearization of Einstein’s equations gives the Fierz--Pauli equations
\begin{equation}
	G^{(1)}_{\mu\nu}=0 .
	\label{FPeq}
\end{equation}
It is convenient to introduce the trace--reversed field
\begin{equation}
	\bar h_{\mu\nu}=h_{\mu\nu}-\frac12\eta_{\mu\nu}h,
	\qquad
	\bar h=-h ,
\end{equation}
and impose the de~Donder (harmonic) gauge condition
\begin{equation}
	\partial^\mu\bar h_{\mu\nu}=0 .
	\label{dedonder}
\end{equation}
This condition implies
\begin{equation}
	\partial^\mu h_{\mu\nu}=\frac12\,\partial_\nu h 
\end{equation}
and reduces the equations of motion to the diagonal form
\begin{equation}
	\Box\,\bar h_{\mu\nu}=0 .
	\label{waveeq}
\end{equation}
At the level of the action, the quadratic Einstein--Hilbert (massless Fierz-Pauli, FP) Lagrangian before gauge fixing is
\begin{equation}
	\mathcal L^{(2)} = M_P^2\left(
	\frac12\,\partial_\lambda h_{\mu\nu}\partial^\lambda h^{\mu\nu}
	- \partial_\mu h^{\mu\nu}\partial^\lambda h_{\lambda\nu}
	+ \partial_\mu h^{\mu\nu}\partial_\nu h
	- \frac12\,\partial_\lambda h \partial^\lambda h
	\right),
	\label{FP}
\end{equation}
with $h=h^\mu{}_\mu$. Implementing the de~Donder gauge via the gauge--fixing function
\begin{equation}
	\mathcal F_\nu=\partial^\mu h_{\mu\nu}-\frac12\partial_\nu h ,
\end{equation}
one adds the term
\begin{equation}
	\mathcal L_{\rm GF}
	=
	-\frac{M_P^2}{2\xi_D}\,\mathcal F_\nu\mathcal F^\nu ,
	\label{DD}
\end{equation}
where $\xi_D$ is a gauge parameter. Even after this gauge fixing, a residual gauge symmetry remains, corresponding to transformations with parameters $\xi_\mu$ satisfying
\begin{equation}
	\Box\xi_\mu=0 .
	\label{residual}
\end{equation}
This residual symmetry plays a crucial role in eliminating nonphysical components of \(h_{\mu\nu}\). To make the covariant structure of the field explicit, we introduce the decomposition
\begin{equation}
	h_{\mu\nu}
	=
	h_{\mu\nu}^{TT}
	+\partial_\mu V_\nu+\partial_\nu V_\mu
	+\left(\partial_\mu\partial_\nu-\frac14\eta_{\mu\nu}\Box\right)w
	+\frac14\eta_{\mu\nu}h,
	\label{eq:cov_decomp_ped}
\end{equation}
with
\begin{equation}
	\partial^\mu h_{\mu\nu}^{TT}=0,
	\qquad
	h^{TT\,\mu}{}_{\mu}=0,
	\qquad
	\partial^\mu V_\mu=0.
	\label{eq:cov_constraints_ped}
\end{equation}

This decomposition is purely kinematical. The original symmetric tensor \(h_{\mu\nu}\) has \(10\) independent components, which are reorganized covariantly as
\begin{equation}
	10 = 5 + 3 + 1 + 1,
	\label{eq:10_5311_ped}
\end{equation}
namely the covariant transverse--traceless tensor \(h_{\mu\nu}^{TT}\), carrying \(5\) components;
the transverse vector \(V_\mu\), carrying \(3\) components; the two scalars \(w\) and \(h\), carrying \(1+1\) components.\\
Under linearized diffeomorphisms
\begin{equation}
	\delta h_{\mu\nu}=\partial_\mu \xi_\nu+\partial_\nu \xi_\mu,
	\label{eq:lin_diff_ped}
\end{equation}
with
\begin{equation}
	\xi_\mu=\xi_\mu^T+\partial_\mu \sigma,
	\qquad
	\partial^\mu \xi_\mu^T=0,
	\label{eq:xi_split_ped}
\end{equation}
one finds
\begin{equation}
	\delta h_{\mu\nu}^{TT}=0,
	\qquad
	\delta V_\mu=\xi_\mu^T,
	\qquad
	\delta w=2\sigma,
	\qquad
	\delta h=2\Box\sigma.
	\label{eq:cov_gauge_transform_ped}
\end{equation}

Hence the \(3\) transverse components of \(\xi_\mu^T\) gauge away the \(3\) components of \(V_\mu\), while the scalar parameter \(\sigma\) removes one scalar combination of \(w\) and \(h\). Therefore the covariant gauge-invariant content is
\begin{equation}
	6 = 5 + 1,
	\label{eq:6_51_ped}
\end{equation}
where the unique scalar gauge-invariant combination is
\begin{equation}
	\Phi_{cov}\equiv h-\Box w.
	\label{eq:Phi_cov_ped}
\end{equation}

Neither \(w\) nor \(h\) is gauge invariant individually. However,
\begin{equation}
	\delta(h-\Box w)=2\Box\sigma-\Box(2\sigma)=0,
\end{equation}
so \(\Phi_{cov}\) is gauge invariant. This is the component investigated by Mottola in \cite{Mottola:2016mpl}.
It is the unique gauge-invariant scalar built from \(h_{\mu\nu}\) at linear order in a covariant decomposition.\\
Notice that the reduction from 5 to 2 degrees propagating of freedom in $h_{\mu\nu}^{TT}$ is associated 
with the residual gauge freedom \eqref{residual}. Indeed, in the covariant de Donder gauge

At this point it is important to stress the distinction between gauge-invariant variables and propagating degrees of freedom. The counting \eqref{eq:6_51_ped} is kinematical: it tells us how many independent gauge-invariant covariant combinations can be built from \(h_{\mu\nu}\). It does \emph{not} mean that all these variables propagate.\\
Indeed, the linearized curvature scalar is related to \(\Phi_{cov}\) by
\begin{equation}
	\delta R
	=
	\partial_\mu\partial_\nu h^{\mu\nu}
	-\Box h
	=
	-\frac34\,\Box \Phi_{cov}.
	\label{eq:R_Phicov_ped}
\end{equation}
Substituting the decomposition \eqref{eq:cov_decomp_ped} into the quadratic Einstein--Hilbert action, using the transversality conditions and integrating by parts, all contributions involving \(V_\mu\), \(w\), \(h\), and therefore \(\Phi_{cov}\), cancel identically up to boundary terms. The action reduces to
\begin{equation}
	S_{EH}^{(2)}
	=
	\frac{1}{8}
	\int d^4x\,
	\partial_\lambda h_{\mu\nu}^{TT}\partial^\lambda h^{TT\,\mu\nu}.
	\label{eq:SEH_TT_only_ped}
\end{equation}

Thus the gauge-invariant scalar \(\Phi_{cov}\) does not appear in the quadratic Einstein--Hilbert action. In particular,
\begin{equation}
	\frac{\delta S_{EH}^{(2)}}{\delta \Phi_{cov}}=0,
	\label{eq:no_Phicov_variation_ped}
\end{equation}
so there is no Euler--Lagrange equation associated with \(\Phi_{cov}\) and no independent scalar phase-space degree of freedom generated by the quadratic Einstein--Hilbert dynamics.\\
Therefore, covariantly, one can construct \(5+1\) gauge-invariant variables, namely \(h_{\mu\nu}^{TT}\) and \(\Phi_{cov}\). However, only the spin--2 sector contributes to the quadratic Einstein--Hilbert action, while the scalar invariant \(\Phi_{cov}\) is non-dynamical. Hence the covariant decomposition should be interpreted as a kinematical organization of the field variables, not as a direct counting of propagating modes. The physical massless graviton still carries only the two helicity-\(\pm 2\) propagating degrees of freedom.\\
In summary, while the de~Donder gauge gives rise to wave equations for the trace-reversed field and for its trace, the covariant decomposition of the quadratic Einstein--Hilbert action makes clear that the gauge-invariant scalar combination \(\Phi_{cov} = h - \Box w\) does not enter the action. The constraint structure of general relativity thus guarantees that only the two transverse--traceless tensor polarizations propagate as physical degrees of freedom. In the following sections, we revisit this issue in detail after including the anomaly contribution in the effective action.

\subsection{Spin decomposition of the graviton propagator}

To make contact between virtual graviton exchange in Einstein--Hilbert gravity using the FP Lagrangian, it is useful to work in momentum space and decompose the graviton propagator into its spin components. This makes explicit which sectors are diffeomorphis invariant (gauge independent), which are gauge independent, and how the gauge-invariant scalar combination of the metric, introduced previously, propagate in this basis. \\For this reason, following Barnes and Rivers \cite{Barnes1965,Rivers1964}, define the transverse and longitudinal projectors
\begin{equation}
	\pi_{\mu\nu}\equiv \eta_{\mu\nu}-\frac{k_\mu k_\nu}{k^2},
	\qquad
	\omega_{\mu\nu}\equiv \frac{k_\mu k_\nu}{k^2}.
	\label{eq:piomega}
\end{equation}
Here \(k^2\neq 0\), and poles are treated in the propagator, not in the projector algebra. In four dimensions, the Barnes--Rivers projectors for symmetric rank-two tensors are
\begin{align}
	P^{(2)}_{\mu\nu,\alpha\beta}
	&=
	\frac12\!\left(\pi_{\mu\alpha}\pi_{\nu\beta}+\pi_{\mu\beta}\pi_{\nu\alpha}\right)
	-\frac13\,\pi_{\mu\nu}\pi_{\alpha\beta},
	\label{eq:P2}
	\\
	P^{(1)}_{\mu\nu,\alpha\beta}
	&=
	\frac12\!\left(
	\pi_{\mu\alpha}\omega_{\nu\beta}
	+\pi_{\mu\beta}\omega_{\nu\alpha}
	+\pi_{\nu\alpha}\omega_{\mu\beta}
	+\pi_{\nu\beta}\omega_{\mu\alpha}
	\right),
	\label{eq:P1}
	\\
	P^{(0\text{-}s)}_{\mu\nu,\alpha\beta}
	&=\frac13\,\pi_{\mu\nu}\pi_{\alpha\beta},
	\qquad\qquad
	P^{(0\text{-}w)}_{\mu\nu,\alpha\beta}
	=\omega_{\mu\nu}\omega_{\alpha\beta}.
	\label{eq:P0}
\end{align}
They satisfy
\begin{equation}
	\left(P^{(2)}+P^{(1)}+P^{(0\text{-}s)}+P^{(0\text{-}w)}\right)_{\mu\nu,\alpha\beta}
	=
	I_{\mu\nu,\alpha\beta},
	\qquad\qquad
	I_{\mu\nu,\alpha\beta}\equiv
	\frac12\!\left(\eta_{\mu\alpha}\eta_{\nu\beta}+\eta_{\mu\beta}\eta_{\nu\alpha}\right),
	\label{eq:completeness}
\end{equation}
together with orthogonality and idempotency:
\(P^{(i)}P^{(j)}=\delta^{ij}P^{(i)}\). Contracting \(h_{\mu\nu}\) with \(\pi^{\mu\nu}\) isolates the transverse scalar component,
\begin{equation}
	\pi^{\mu\nu}h_{\mu\nu}(k)=\Phi_{cov}(k),
	\label{eq:PhiProj}
\end{equation}
which is the momentum-space representative of the gauge-invariant scalar combination discussed in the previous section. With a de Donder gauge-fixing term parametrized by \(\xi_D\) \eqref{DD}, the quadratic kinetic operator is
\begin{equation}
	\mathcal O_{\mu\nu,\alpha\beta}(k)=
	k^2\!\left[
	P^{(2)}
	-2\,P^{(0\text{-}s)}
	+\frac{1}{\xi_D}\!\left(P^{(1)}+\frac12 P^{(0\text{-}w)}\right)
	\right]_{\mu\nu,\alpha\beta}.
	\label{eq:Ok}
\end{equation}
Using projector orthogonality, its inverse is decomposed as 
\begin{equation}
	D_{\mu\nu,\alpha\beta}(k)=
	\frac{1}{k^2}\!\left[
	P^{(2)}
	-\frac12 P^{(0\text{-}s)}
	+\xi_D\!\left(P^{(1)}+\frac12 P^{(0\text{-}w)}\right)
	\right]_{\mu\nu,\alpha\beta}.
	\label{eq:Dk}
\end{equation}
We denote by $D^{(2)}$ and $D^{(0-s)}$ the spin-2 and spin-0 part of the propagator (Figure \ref{fig:graviton_propagator_decomposition}) 
\begin{equation}\label{eq:propagator_spin_decomposition}
	D^{(0-s)}_{\mu\nu,\alpha\beta}(k) = -\frac{1}{6}\frac{\pi^{\mu\nu}(k)\pi^{\alpha\beta}(k)}{k^2}, \qquad \qquad D^{(2)}_{\mu\nu,\alpha\beta}(k) = -\frac{1}{2}\frac{\Pi^{\mu\nu,\alpha\beta}(k)}{k^2}.
\end{equation}
All \(\xi_D\)-dependence is confined to \(P^{(1)}\) and \(P^{(0\text{-}w)}\), i.e. the pure-gauge sectors. The spin-2 part \(P^{(2)}\) and the scalar projector \(P^{(0\text{-}s)}\) are gauge-parameter independent (see Fig. 3). In Einstein gravity, this scalar sector does not represent an independent propagating physical degree of freedom. For \(\xi_D=1\) (de Donder in this convention),
\begin{equation}
	D_{\mu\nu,\alpha\beta}(k)\big|_{\xi_D=1}
	=
	\frac{1}{k^2}
	\left(
	P^{(2)}-\frac12 P^{(0\text{-}s)}+P^{(1)}+\frac12 P^{(0\text{-}w)}
	\right)_{\mu\nu,\alpha\beta}.
	\label{eq:DkDeDonder}
\end{equation}
For conserved external sources (\(k_\mu T^{\mu\nu}=k_\alpha T'^{\alpha\beta}=0\)), the longitudinal projectors do not contribute, and one obtains the exact identity
\begin{equation}
	T^{\mu\nu}D_{\mu\nu,\alpha\beta}(k)T'^{\alpha\beta}
	=
	\frac{1}{k^2}\,
	T^{\mu\nu}
	\left[
	\frac12\!\left(
	\eta_{\mu\alpha}\eta_{\nu\beta}
	+\eta_{\mu\beta}\eta_{\nu\alpha}
	-\eta_{\mu\nu}\eta_{\alpha\beta}
	\right)
	\right]
	T'^{\alpha\beta}.
	\label{eq:FPonConserved}
\end{equation}
Thus the exchange kernel, for conserved stres energy tensors, is exactly the Fierz--Pauli tensor structure on conserved sources.

\begin{figure}
	\centering
	\includegraphics[scale=1.2,angle=0]{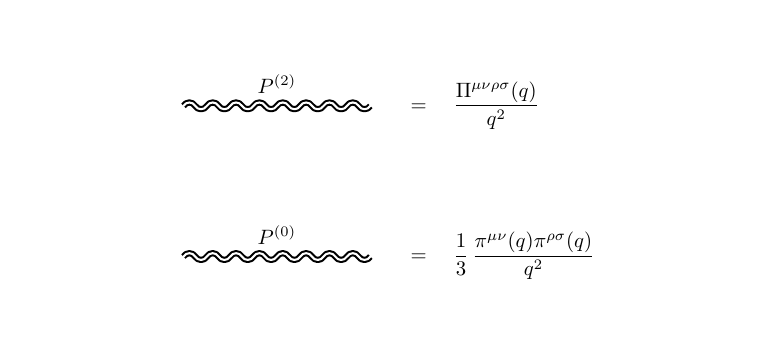}
	\caption{Spin-2 and spin-0 part of the graviton propagator.}
	\label{fig:graviton_propagator_decomposition}
\end{figure}

\section{Tree--level graviton exchange and the spin--0 channel in the De~Donder gauge }
To characterize the role of the scalar exchange in the De Donder gauge, and extend \eqref{eq:FPonConserved}, we consider two independent matter sectors, redefined as $L$ and $R$, interacting only through gravity. A similar analysis was presented in \cite{Giannotti:2008cv}.
One can think of $L$ as a visible and of $R$ as a dark sector, with gravity mediating their interaction. At tree level, the exchange of a graviton is described by the amplitude
\begin{equation}
	\mathcal A_t
	=
	\frac{\kappa^2}{4}\,
	T^{\mu\nu}_L(k)\,
	D_{\mu\nu,\rho\sigma}(k)\,
	T^{\rho\sigma}_R(-k),
	\label{Atree_def}
\end{equation}
where $k$ is the momentum transfer. Throughout this section the stress--energy tensors are assumed to be conserved
\begin{equation}
	k_\mu T^{\mu\nu}_{L}(k)=0,
	\qquad
	k_\mu T^{\mu\nu}_{R}(k)=0,
	\label{T_conservation}
\end{equation}
but not necessarily traceless
\begin{equation}
	T^\mu{}_{\mu\,L}\neq0,
	\qquad
	T^\mu{}_{\mu\,R}\neq0 .
	\label{T_trace}
\end{equation}
Using the covariant spin decomposition of the graviton propagator discussed in the previous section, the de~Donder propagator can be written as
\begin{equation}
	D_{\mu\nu,\rho\sigma}(k)
	=
	\frac{i}{k^2+i\epsilon}
	\left[
	P^{(2)}_{\mu\nu,\rho\sigma}
	-
	\frac12\,P^{(0\text{-}s)}_{\mu\nu,\rho\sigma}
	\right],
	\label{prop_spin}
\end{equation}
where only the transverse spin--2 and transverse scalar projectors appear. Substituting \eqref{prop_spin} into \eqref{Atree_def}, the amplitude splits into a spin--2 and a spin--0 contribution,
\begin{equation}
	\mathcal A_t
	=
	\frac{i\kappa^2}{4k^2}
	\left[
	T^{\mu\nu}_L
	P^{(2)}_{\mu\nu,\rho\sigma}
	T^{\rho\sigma}_R
	-
	\frac12\,
	T^{\mu\nu}_L
	P^{(0\text{-}s)}_{\mu\nu,\rho\sigma}
	T^{\rho\sigma}_R
	\right].
	\label{Atree_split}
\end{equation}
Because of energy--momentum conservation, all longitudinal components drop out and the projectors act effectively as algebraic operators on the stress--energy tensors. The spin--2 contribution evaluates to
\begin{equation}
	T^{\mu\nu}_L
	P^{(2)}_{\mu\nu,\rho\sigma}
	T^{\rho\sigma}_R
	=
	T^{\mu\nu}_L T_{R\,\mu\nu}
	-
	\frac13\,
	T^\mu{}_{\mu\,L}\,
	T^\nu{}_{\nu\,R},
	\label{spin2}
\end{equation}
where the trace subtraction reflects the irreducibility of the spin--2 representation. The spin--0 contribution follows from
\begin{equation}
	T^{\mu\nu}_L
	P^{(0\text{-}s)}_{\mu\nu,\rho\sigma}
	T^{\rho\sigma}_R
	=
	\frac13\,
	\left(\pi_{\mu\nu}T_L^{\mu\nu}\right)
	\left(\pi_{\rho\sigma}T_R^{\rho\sigma}\right).
\end{equation}
Using conservation once more, $\pi_{\mu\nu}T^{\mu\nu}_{L,R}=T^\mu{}_{\mu\,L,R}$, one finds
\begin{equation}
	T^{\mu\nu}_L
	P^{(0\text{-}s)}_{\mu\nu,\rho\sigma}
	T^{\rho\sigma}_R
	=
	\frac13\,
	T^\mu{}_{\mu\,L}\,
	T^\nu{}_{\nu\,R}.
	\label{spin0}
\end{equation}
Inserting \eqref{spin2} and \eqref{spin0} into \eqref{Atree_split}, the full tree--level amplitude becomes
\begin{equation}
	\mathcal A_t
	=
	\frac{i\kappa^2}{4k^2}
	\left[
	T^{\mu\nu}_L T_{R\,\mu\nu}
	-
	\frac12\,
	T^\mu{}_{\mu\,L}\,
	T^\nu{}_{\nu\,R}
	\right].
	\label{Atree_final}
\end{equation}
Equation \eqref{Atree_final} shows that, whenever the stress--energy tensors are not traceless, an apparent scalar contribution proportional to the product of the traces remains alongside the spin--2 exchange. This scalar term is associated with the same massless pole, $1/k^2$, as the spin--2 part and matches precisely the scalar projector $P^{(0\text{-}s)}$ identified previously. Its interpretation, however, requires some care, since the mere appearance of a trace contribution does not necessarily imply that the trace contributes to a physical scattering amplitude through a genuinely propagating scalar degree of freedom. To make this point precise, it is useful to turn to the analysis of four--point functions and examine anomaly--mediated interactions in that framework, where the physical content of the exchange can be extracted more clearly. We shall first consider the case of explicit breaking and then move to anomalous breaking.

\section{Scalar gravitational scattering and explicit breaking:  the absence of scalar exchange in 4-point functions}
\label{sec:scalar_xi_dependence_summary}
As just mentioned, the appearance of a nonvanishing trace in the scalar-graviton vertex is not, by itself, sufficient to conclude that a scalar component of the metric is being exchanged. This statement is best understood by analyzing the same process in two complementary descriptions: first in the standard covariant formulation, where the full graviton propagator is used explicitly, and then in the transverse-traceless gauge, where only the propagating spin-2 part is manifest and the remaining metric components appear through constraint equations. The two derivations lead to the same physical amplitude. In particular, they show that in ordinary Einstein gravity the presence of explicit conformal-symmetry-breaking terms in the matter sector does not activate an independent propagating scalar mode.

The scalar theory considered below allows for two distinct sources of explicit breaking of conformal symmetry. The first is the non-minimal coupling \(\xi\neq 1/6\), which already breaks conformal invariance in the massless theory. The second is the scalar mass \(m\), which provides an additional and independent breaking. These two effects behave differently at the level of the amplitude. In the massless case with generic \(\xi\), the tree-level amplitude is \(\xi\)-dependent off-shell, but the \(\xi\)-dependence disappears on-shell. In the massive case, by contrast, the amplitude remains \(\xi\)-dependent even on-shell. One may therefore wonder whether the simultaneous presence of a trace contribution and of a mass term could signal the exchange of a scalar component of the metric. The answer is negative. The explicit computations below show that the trace of the vertex and the existence of a genuine scalar exchange are logically distinct issues.

We begin with the action in \(d=4\) for a real scalar field coupled non-minimally to gravity,
\begin{equation}
	S_\phi=\int d^4x\,\sqrt{-g}\left[
	\frac12 g^{\mu\nu}\nabla_\mu\phi\nabla_\nu\phi
	-\frac12 m^2\phi^2
	-\frac12 \xi R\phi^2
	\right],
\end{equation}
using the mostly-minus convention
\begin{equation}
	\eta_{\mu\nu}=\mathrm{diag}(+,-,-,-).
\end{equation}
At linear order in the metric fluctuation \(h_{\mu\nu}\), the scalar-graviton vertex is
\begin{equation}
	V_{\mu\nu}(p_1,p_2)=
	p_{1\mu}p_{2\nu}+p_{1\nu}p_{2\mu}
	-\eta_{\mu\nu}(p_1\!\cdot\! p_2-m^2)
	+2\xi\big(\eta_{\mu\nu}q^2-q_\mu q_\nu\big),
	\qquad q=p_1+p_2.
\end{equation}
Its trace is
\begin{equation}
	V^\mu{}_\mu=-2\,p_1\!\cdot\! p_2+4m^2+6\xi q^2.
	\label{eq:traceV_standalone_rewritten}
\end{equation}
The fact that this trace does not vanish for generic \(\xi\) and \(m\) will be central in what follows. However, as we shall show, its presence does not imply the propagation of a scalar metric degree of freedom.

We consider the tree-level \(2\to2\) scattering of scalar particles without initially imposing any on-shell condition. The kinematics is conveniently described in terms of the Mandelstam invariants
\begin{equation}
	s=s_{12}=(p_1+p_2)^2,
	\qquad
	t=s_{13}=(p_1-p_3)^2,
	\qquad
	u=s_{14}=(p_1-p_4)^2,
\end{equation}
which satisfy the off-shell identity
\begin{equation}
	s+t+u=\sum_{i=1}^4 p_i^2.
	\label{eq:stu_identity_standalone_rewritten}
\end{equation}
For each channel $s$, $t$, and $u$, the exchange amplitude can be written as
\begin{equation}
	i\mathcal M_\mathrm{ch}=
	-\frac{i\kappa^2}{4s_\mathrm{ch}}
	\left[
	V^{(\mathrm{ch})}_{\mu\nu}V^{(\mathrm{ch})\mu\nu}
	-\frac12 V^{(\mathrm{ch})}V^{(\mathrm{ch})}
	\right],
	\label{eq:Mx_standalone_rewritten}
\end{equation}
where \(s_\mathrm{ch}\in\{s,t,u\}\) denotes the corresponding channel invariant.

To isolate the effect of the non-minimal coupling, it is convenient to split the vertex into a minimal part and a purely \(\xi\)-dependent contribution ($p$ and $q$ are generic momenta of the scalars in the graviton/scalar/scalar vertex)
\begin{equation}
	V_{\mu\nu}^{(\mathrm{ch})}(p,q)=
	V_{0\,\mu\nu}^{(\mathrm{ch})}(p,q)+2\xi\,X_{\mu\nu}^{(\mathrm{ch})}(p,q),
\end{equation}
with
\begin{equation}
	X_{\mu\nu}^{(\mathrm{ch})}(p,q)=
	\eta_{\mu\nu}(p+q)^2-(p+q)_\mu(p+q)_\nu,
\end{equation}
and
\begin{equation}
	V_{0\,\mu\nu}^{(\mathrm{ch})}(p,q)=
	p_\mu q_\nu+p_\nu q_\mu-\eta_{\mu\nu}(p\!\cdot\! q-m^2).
\end{equation}
Defining
\begin{equation}
	N^{(\mathrm{ch})}\equiv
	V^{(\mathrm{ch})}_{\mu\nu}V^{(\mathrm{ch})\mu\nu}
	-\frac12 V^{(\mathrm{ch})}V^{(\mathrm{ch})},
\end{equation}
one finds that the \(\xi\)-dependence reorganizes in the simple form
\begin{equation}
	N^{(\mathrm{ch})}
	=
	N_0^{(\mathrm{ch})}+(2\xi-6\xi^2)s_\mathrm{ch},
	\label{eq:N_decomposition_rewritten}
\end{equation}
where \(N_0^{(\mathrm{ch})}\) denotes the $\xi$-independent contribution. Substituting this decomposition into \eqref{eq:Mx_standalone_rewritten}, the single-channel amplitude becomes
\begin{equation}
	i\mathcal M_\mathrm{ch}
	=
	-\frac{i\kappa^2}{4s_\mathrm{ch}}N_0^{(\mathrm{ch})}
	-\frac{i\kappa^2}{4}(2\xi-6\xi^2)s_\mathrm{ch}.
\end{equation}
Summing over the three channels gives
\begin{equation}
	i\mathcal M(\xi,m)=
	i\mathcal M_0(m)-\frac{i\kappa^2}{4}(2\xi-6\xi^2)(s+t+u),
	\qquad
	i\mathcal M_0(m)\equiv
	-\frac{i\kappa^2}{4}
	\left(
	\frac{N_0^{(\mathrm{s})}}{s}+\frac{N_0^{(\mathrm{t})}}{t}+\frac{N_0^{(\mathrm{u})}}{u}
	\right).
	\label{eq:master_amp_standalone_rewritten}
\end{equation}
This is the basic formula from which the subsequent discussion follows. The minimal numerators can be written entirely in terms of the invariants \(s,t,u\) and the external virtualities \(p_i^2\):
{\smaller
\begin{align}
	N_0^{(\mathrm{s})}&=
	\frac12\!\left[
	(t-p_1^2-p_3^2)(t-p_2^2-p_4^2)
	+(u-p_1^2-p_4^2)(u-p_2^2-p_3^2)
	-(s-p_1^2-p_2^2)(s-p_3^2-p_4^2)
	\right] +\left(2sm^2-6m^4\right),\\[1mm]
	N_0^{(\mathrm{t})}&=
	\frac12\!\left[
	(s-p_1^2-p_2^2)(s-p_3^2-p_4^2)
	+(u-p_1^2-p_4^2)(u-p_2^2-p_3^2)
	-(t-p_1^2-p_3^2)(t-p_2^2-p_4^2)
	\right]
	+\left(2tm^2-6m^4\right),\\[1mm]
	N_0^{(\mathrm{u})}&=
	\frac12\!\left[
	(s-p_1^2-p_2^2)(s-p_3^2-p_4^2)
	+(t-p_1^2-p_3^2)(t-p_2^2-p_4^2)
	-(u-p_1^2-p_4^2)(u-p_2^2-p_3^2)
	\right]
	+\left(2um^2-6m^4\right).
\end{align}}

\subsection{Off-shell and on-shell \texorpdfstring{$\xi$}{xi} dependence}

The structure of \eqref{eq:master_amp_standalone_rewritten} makes the dependence on \(\xi\) particularly transparent
\begin{equation}
	i\mathcal M(\xi,m)=
	i\mathcal M_0(m)-\frac{i\kappa^2}{4}(2\xi-6\xi^2)(s+t+u).
	\label{eq:summary_master_rewritten}
\end{equation}
This immediately shows that the \(\xi\)-dependent part is polynomial and proportional to the sum of the channel invariants.

In the massless case, \(m=0\), the off-shell amplitude is
\begin{equation}
	i\mathcal M^{\rm off}(\xi,0)=
	i\mathcal M_0^{\rm off}(0)-\frac{i\kappa^2}{4}(2\xi-6\xi^2)(s+t+u),
\end{equation}
and is therefore \(\xi\)-dependent. If one goes on-shell, \(p_i^2=0\), then \(s+t+u=0\), so that
\begin{equation}
	i\mathcal M^{\rm on}(\xi,0)=i\mathcal M_0^{\rm on}(0)
	=-\frac{i\kappa^2}{4}
	\left(
	\frac{tu}{s}+\frac{su}{t}+\frac{st}{u}
	\right).
\end{equation}
Thus the massless on-shell amplitude is independent of \(\xi\), even though the local vertex has a nonvanishing trace for generic \(\xi\).

The situation is different in the massive case. Off-shell, the amplitude is again \(\xi\)-dependent by \eqref{eq:summary_master_rewritten}. For equal external masses on-shell, \(p_i^2=m^2\), one has
\begin{equation}
	s+t+u=4m^2,
\end{equation}
and therefore
\begin{equation}
	i\mathcal M^{\rm on}(\xi,m)=
	i\mathcal M_0^{\rm on}(m)-i\kappa^2m^2(2\xi-6\xi^2).
	\label{eq:summary_massive_on_rewritten}
\end{equation}
Hence, unlike the massless case, the massive on-shell amplitude retains a nontrivial dependence on \(\xi\). In particular, at the conformal value \(\xi=1/6\),
\begin{equation}
	i\mathcal M^{\rm on}(\xi=1/6,m)=
	i\mathcal M_0^{\rm on}(m)-\frac{i\kappa^2m^2}{6}.
\end{equation}
The conclusion is therefore clear. In the massless theory, the amplitude is \(\xi\)-dependent off-shell but \(\xi\)-independent on-shell. In the massive theory, by contrast, the amplitude depends on \(\xi\) both off-shell and on-shell. This difference reflects the distinct role played by the two explicit breakings of conformal symmetry, namely the non-minimal coupling and the scalar mass. However, this distinction should not be confused with the propagation of an independent scalar gravitational degree of freedom. Even when the trace of the vertex is nonzero, and even when the amplitude retains a nontrivial dependence on \(\xi\), the full exchange still need not correspond to scalar propagation.

\section{The activation of the scalar sector in the Fierz-Pauli Lagrangian with the anomaly}
\label{closing}
To illustrate the unusual behaviour of the scalar interaction mediated by virtual gravitational exchanges 
we start from the quadratic Einstein--Hilbert action in Fierz--Pauli form, and isolate the scalar longitudinal sector in \(d=4\) from \eqref{eq:cov_decomp_ped}  
\begin{equation}
	h_{\mu\nu}^{(S)}
	=
	\left(\partial_\mu\partial_\nu-\frac14\eta_{\mu\nu}\Box\right)w
	+\frac14\eta_{\mu\nu}h,
	\qquad
	\Phi_{cov}\equiv h-\Box w,
	\label{eq:scalar_decomp}
\end{equation}
having used the covariant decomposition.
Introducing \(U\equiv \Box w\), one has
\begin{equation}
	\partial_\mu h^{\mu\nu}=\frac14\,\partial^\nu(h+3U),\qquad
	h=\eta^{\mu\nu}h_{\mu\nu}^{(S)}.
	\label{eq:divh}
\end{equation}
Substituting \eqref{eq:scalar_decomp} into the Fierz-Pauli Lagrangian \eqref{FP} we obtain
\begin{equation}
	\mathcal L_{S}^{(2)}
	=
	-\frac38\Big[(\partial h)^2-2\,\partial h\!\cdot\!\partial U+(\partial U)^2\Big]
	=
	-\frac38\big(\partial(h-U)\big)^2
	=
	-\frac38(\partial\Phi_{cov})^2.
	\label{eq:Lsc}
\end{equation}
Therefore the scalar contribution generates the action
\begin{equation}
	S_{EH,S}^{(2)}
	=
	-\frac{3}{32\kappa^2}\int d^4x\,(\partial\Phi_{cov})^2
	=
	\frac{3}{32\kappa^2}\int d^4x\,\Phi_{cov}\,\Box\Phi_{cov},
	\label{eq:SEHscalar}
\end{equation}
up to boundary terms, which is the scalar kinetic term quoted in the main text. Using the 
linearized expression of the Ricci scalar 
\begin{equation}
	R^{(1)}=\partial_\mu\partial_\nu h^{\mu\nu}-\Box h
	\label{eq:R1def}
\end{equation}
from \eqref{eq:scalar_decomp},
\begin{equation}
	\partial_\mu\partial_\nu h^{\mu\nu}
	=\frac14\Box h+\frac34\Box^2 w,
\end{equation}
one derives the relation
\begin{equation}
	R^{(1)}
	=
	-\frac34\Box(h-\Box w)
	=
	-\frac34\Box\Phi_{cov}.
	\label{eq:R1Phi}
\end{equation}
At this point, if we include the anomaly-pole interaction from the $TJJ$ vertex we rewrite the pole action as
\begin{equation}
	S_{\mathrm{pole}}
	=
	\frac{b}{8\pi^2}\int d^4x\,
	R^{(1)}\frac1\Box F^2
	=
	\frac{b}{8\pi^2}\int d^4x
	\left(-\frac34\Box\Phi_{cov}\right)\frac1\Box F^2
	=
	-\frac{3b}{32\pi^2}\int d^4x\,\Phi_{cov} F^2.
	\label{scalar1}
\end{equation}
Combining \eqref{eq:SEHscalar} and \eqref{scalar1} the action accounting for the anomaly becomes
\begin{equation}
	S_\Phi
	=
	\int d^4x\left[
	\frac{3}{32\kappa^2}\Phi_{cov}\Box\Phi_{cov}
	-\frac{3b}{32\pi^2}\Phi_{cov} F^2
	\right].
	\label{eq:SphiTotal}
\end{equation}
that once varies in the presence of an anomaly, eq. \eqref{eq:no_Phicov_variation_ped}, gives the
modified equation
\begin{equation}
	\label{cov}
	\frac{3}{32\kappa^2}\Box\Phi_{cov}-\frac{3b}{32\pi^2}F^2=0
	\quad\Longrightarrow\quad
	\Box\Phi_{cov}=\frac{\kappa^2 b}{\pi^2}F^2.
\end{equation}
with $b$ denoting the multiplicity factor (the number of massless conformal states) of the conformal states which have been integrated over in the quantum corrections (in the $TJJ$).
Clearly, $\Phi_{cov}$ is sourced by the anomaly.\\
At this point it is useful to distinguish between two related but conceptually different effective descriptions of the anomaly-induced interaction.

If one integrates out the matter fields while keeping both the metric and the gauge field as explicit variables, the anomaly appears in the mixed 1PI effective action \(\Gamma[g,A]\) through the nonlocal \(TJJ\) vertex. To linear order in the metric fluctuation and quadratic order in the gauge field, this gives a contribution of the schematic form
\begin{equation}
	\Gamma_{\rm anom}[g,A]\supset \int d^4x\, R^{(1)}\,\Box^{-1}F^2,
	\label{eq:mixed_anom_action}
\end{equation}
or, equivalently, in terms of the covariant scalar sector,
\begin{equation}
	\Gamma_{\rm anom}[g,A]\supset \int d^4x\, \Phi_{cov}\,F^2,
	\label{eq:phicov_f2_action}
\end{equation}
using \eqref{eq:R1Phi}. In this representation the anomaly-induced term is part of the nonlocal effective action of the coupled gravity-matter system; it is not an external source inserted by hand. Most importantly, the anomaly pole selects a unique scalar direction in the metric sector, namely \(\Phi_{cov}\), when the latter is described in the covariant decomposition, as emphasized in \cite{Mottola:2016mpl}.

A different effective description is obtained if one eliminates the graviton as well. In that case one sews two \(TJJ\) vertices through an internal graviton line and obtains a reduced effective action for the gauge field alone. As we will show below, this generates schematically a nonlocal four-point interaction of the form, setting \(\kappa=1\),
\begin{equation}
	\Gamma_{\rm red}[A]\supset \int d^4x\, d^4y\,
	F^2(x)\,G(x-y)\,F^2(y),
	\label{eq:reduced_gauge_action}
\end{equation}
where \(G\) denotes the kernel induced by the exchanged gravitational channel and is schematically of \(\Box^{-1}\) type. Restoring the Planck scale, the corresponding momentum-space structure is
\begin{equation}
	\Gamma_{\rm red}[A]\sim \frac{1}{M_P^2}\int d^4 q\,
	F^2(q)\,\frac{1}{q^2}\,F^2(-q)
	\label{eq:F2boxF2}
\end{equation}
up to numerical coefficients, with 
\begin{equation}
	F^2(q)=
	2\int \frac{d^4p}{(2\pi)^4}
	\left[
	p_\mu (q-p)_\nu-\eta_{\mu\nu}\,p\!\cdot\!(q-p)
	\right]
	A^\mu(p)\,A^\nu(q-p).
\end{equation}

It is important to stress that \eqref{eq:F2boxF2} is not a new anomaly structure independent of the original \(TJJ\) vertex. Rather, it is the four-point exchange amplitude generated by the same anomaly-selected scalar channel. For this reason, the appearance of an \(F^2\Box^{-1}F^2\)-type term should be interpreted as the manifestation of a single exchanged scalar contribution in the effective interaction, not as evidence for two independent scalar gravitational degrees of freedom. In covariant language this exchanged channel is again the unique invariant \(\Phi_{cov}\); in the non-covariant decomposition it corresponds to one specific linear combination of two gauge-invariant variables, while the orthogonal scalar combination remains non-propagating. The non covariant decomposition is discussed in Appendix B.\\
A final remark is in order. Since this four-point amplitude is obtained by combining two conformal \(TJJ\) vertices with an intermediate gravitational propagator governed by the Einstein--Hilbert action, which is not conformally invariant, the pole structure of the full sewn amplitude is not guaranteed to remain manifest in a completely factorized form once one passes to momentum space and includes the explicit expressions of the two $F^2$ invariant in \eqref{eq:F2boxF2}. Thus the existence of the anomaly pole is clearest at the level of the mixed \(TJJ\) effective action, while its realization in the reduced four-point amplitude requires a separate analysis.
\section{The anomaly interaction in the conformal limit and the photon 4-point function}
\label{section9}
The structure of the $TJJ$ correlator, as we have seen, is quite simple in the on-shell limit, being characterized by one pole in the trace sector, and one extra pole and a log 
in the traceless sector. The two poles are interpolated by the virtual graviton, coupled to its spin-2 and spin-0 components. Both components can be discussed separately. In each $L/R$ sector the exchange of such poles is interpreted as a light-cone dominated process. This is the kinematical situation in which the residue of the pole at each of the two TJJ vertices is nonvanishing. This point is rather technical and has been discussed  in detail in \cite{Coriano:2025fom}. In this section we are going to 
discuss the structure of this decomposition, extending our former analysis and isolating the contributions related to the spin-2 and spin-0 gravitational components in a rigorous way. We consider the scattering of two on-shell photons mediated by a graviton.  
In this specific kinematic configuration each of the vertices
is characterized by an anomaly form factor satisfying a
sum rule.
We have the amplitudes in the $t$, $s$ and $u$ channels.

We introduce the effective coupling
\begin{equation}
	\frac{\bar C_0}{M_P^2}
	\equiv
	\kappa^2 \frac{3}{2}
	\left(\frac{e^2}{48\pi^2}\right)^2 ,
\end{equation}
Hence
\begin{equation}
	\label{C0}
	\bar C_0 = \frac{3}{2}
	\left(\frac{e^2}{48\pi^2}\right)^2 .
\end{equation}
Notice that the effect of the inclusion of an intermediate scalar projector, once we carry out the contractions 
in the $s , t$ and $u$ channels  is just to produce a factor $9/q^2$ in the exchange.   
Since we have the contractions
\begin{align}
	\pi_{\mu_2\nu_2}(q)\,
	{P^{(0)\,\mu_2\nu_2}{}_{\rho_2\sigma_2}(q)}\frac{1}{q^2}\,
	\pi^{\rho_2\sigma_2}(q) 
	=
	\left(
	\pi_{\mu_2\nu_2}(q)\,
	\pi^{\mu_2\nu_2}(q)
	\right)
	\left(
	\pi_{\rho_2\sigma_2}(q)\,
	\pi^{\rho_2\sigma_2}(q)
	\right)
	\frac{1}{q^2} 
	= \frac{9}{q^2}
\end{align}
on each internal scalar line.  The result is that the interaction between the two sectors is mediated by a single scalar exchange, with the two anomaly component of each vertex appearing symmetrically. 
A similar amplitude can be computed in the other $s$ and $u$ channels. The scalar–exchange amplitudes then take the compact form

\begin{equation}
	\mathcal{A}_s^{(0)}
	=
	-\,\frac{\bar C_0}{M_P^2}
	\, u^{\alpha\gamma}(p_1,p_2)
	\frac{1}{s}
	u^{\beta\delta}(p_3,p_4) ,
\end{equation}

\begin{equation}
	\mathcal{A}_t^{(0)}
	=
	-\,\frac{\bar C_0}{M_P^2}
	\, u^{\alpha\beta}(-p_1,p_3)
	\frac{1}{t}
	u^{\gamma\delta}(-p_2,p_4) ,
\end{equation}

\begin{equation}
	\mathcal{A}_u^{(0)}
	=
	-\,\frac{\bar C_0}{M_P^2}
	\, u^{\alpha\delta}(-p_1,p_4)
	\frac{1}{u}
	u^{\gamma\beta}(-p_2,p_3) .
\end{equation}

It is useful to stress the meaning of this scalar-exchange representation. The
spin--0 contribution isolated in this way should be understood as an effective
anomaly-induced scalar channel, equivalently described covariantly by
\(\Phi_{cov}\), rather than as a new elementary asymptotic graviton degree of
freedom. In other words, the scalar exchange appearing in the factorized
amplitudes is a convenient parametrization of the nonlocal pole structure of the
full \(TJJ\)-mediated interaction. It reproduces the same \(1/\Box\) behaviour of
the complete amplitude, but it should not be interpreted as the LSZ exchange of
an independent fundamental scalar particle. This interpretation is consistent
with the analysis of Section~\ref{app:TT_anomaly_reorganized}, where the scalar
sector of the Einstein--Hilbert/Fierz--Pauli action remains constrained and the
anomaly activates a distinguished nonlocal scalar response channel without
introducing an additional local propagating graviscalar in the spectrum.

\begin{table}[t]
	\centering
	\small
	\renewcommand{\arraystretch}{1.12}
	\setlength{\tabcolsep}{4.5pt}
	\begin{tabular}{cc cc c @{\hspace{1em} \vrule width 0.75pt\hspace{1em} } cc cc c}
		\toprule
		$p_i$ & $\epsilon_i$ & $p_j$ & $\epsilon_j$ & $\epsilon(p_i)_\alpha \epsilon(p_j)_\beta u^{\alpha\beta}(p_i,p_j)$ &
		$p_i$ & $\epsilon_i$ & $p_j$ & $\epsilon_j$ & $\epsilon(p_i)_\alpha \epsilon(p_j)_\beta u^{\alpha\beta}(p_i,p_j)$ \\
		\midrule
		$p_1$ & $+$ & $p_1$ & $+$ & $0$ &
		$p_2$ & $+$ & $p_2$ & $+$ & $0$ \\
		
		$p_1$ & $+$ & $p_1$ & $-$ & $0$ &
		$p_2$ & $+$ & $p_2$ & $-$ & $0$ \\
		
		$p_1$ & $+$ & $p_2$ & $+$ & $-\frac{s}{2}$ &
		$p_2$ & $+$ & $p_3$ & $+$ & $\frac{u(t-u)}{2s}$ \\
		
		$p_1$ & $+$ & $p_2$ & $-$ & $0$ &
		$p_2$ & $+$ & $p_3$ & $-$ & $\frac{tu}{s}$ \\
		
		$p_1$ & $+$ & $p_3$ & $+$ & $\frac{t(s+2t)}{2s}$ &
		$p_2$ & $+$ & $p_4$ & $+$ & $\frac{t(u-t)}{2s}$ \\
		
		$p_1$ & $+$ & $p_3$ & $-$ & $-\frac{tu}{s}$ &
		$p_2$ & $+$ & $p_4$ & $-$ & $\frac{tu}{s}$ \\
		
		$p_1$ & $+$ & $p_4$ & $+$ & $\frac{u(u-t)}{2s}$ &
		$p_2$ & $-$ & $p_2$ & $-$ & $0$ \\
		
		$p_1$ & $+$ & $p_4$ & $-$ & $-\frac{tu}{s}$ &
		$p_2$ & $-$ & $p_3$ & $+$ & $\frac{tu}{s}$ \\
		
		$p_1$ & $-$ & $p_1$ & $-$ & $0$ &
		$p_2$ & $-$ & $p_3$ & $-$ & $\frac{u(t-u)}{2s}$ \\
		
		$p_1$ & $-$ & $p_2$ & $+$ & $0$ &
		$p_2$ & $-$ & $p_4$ & $+$ & $\frac{tu}{s}$ \\
		
		$p_1$ & $-$ & $p_2$ & $-$ & $-\frac{s}{2}$ &
		$p_2$ & $-$ & $p_4$ & $-$ & $\frac{t(u-t)}{2s}$ \\
		
		$p_1$ & $-$ & $p_3$ & $+$ & $-\frac{tu}{s}$ &
		$p_3$ & $+$ & $p_3$ & $+$ & $0$ \\
		
		$p_1$ & $-$ & $p_3$ & $-$ & $\frac{t(s+2t)}{2s}$ &
		$p_3$ & $+$ & $p_3$ & $-$ & $0$ \\
		
		$p_1$ & $-$ & $p_4$ & $+$ & $-\frac{tu}{s}$ &
		$p_3$ & $+$ & $p_4$ & $+$ & $\frac{s}{2}$ \\
		
		$p_1$ & $-$ & $p_4$ & $-$ & $\frac{u(u-t)}{2s}$ &
		$p_3$ & $+$ & $p_4$ & $-$ & $0$ \\
		
		$p_3$ & $-$ & $p_3$ & $-$ & $0$ &
		$p_4$ & $+$ & $p_4$ & $+$ & $0$ \\
		
		$p_3$ & $-$ & $p_4$ & $+$ & $0$ &
		$p_4$ & $+$ & $p_4$ & $-$ & $0$ \\
		
		$p_3$ & $-$ & $p_4$ & $-$ & $\frac{s}{2}$ &
		$p_4$ & $-$ & $p_4$ & $-$ & $0$ \\
		\bottomrule
	\end{tabular}
	\caption{All possible contractions of polarization vectors with $u^{\alpha\beta}$.}
	\label{tab:u_eps_eps}
\end{table}
Introducing the polarization four-vectors in the center of mass frame of the scattering photons
\begin{align}
	\varepsilon^\mu(p_1,\pm)
	&= \frac{1}{\sqrt{2}}(0,1,\pm i,0),
	&
	\varepsilon^\mu(p_2,\pm)
	&= \frac{1}{\sqrt{2}}(0,1,\mp i,0),
	\\[6pt]
	\varepsilon^\mu(p_3,\pm)
	&= \frac{1}{\sqrt{2}}(0,\cos\theta,\pm i,-\sin\theta),
	&
	\varepsilon^\mu(p_4,\pm)
	&= \frac{1}{\sqrt{2}}(0,-\cos\theta,\pm i,\sin\theta).
\end{align}
we compute the contraction
\begin{equation}
	\mathcal U_{p,q}(\lambda_1,\lambda_2)
	=
	\varepsilon_\alpha(p,\lambda_1)\,
	\varepsilon_\beta(q,\lambda_2)\,
	u^{\alpha\beta}(p,q).
\end{equation}
for the various helicity configurations. The results are collected in Table \ref{tab:u_eps_eps}. Notice that $\mathcal{U}_{p,q}$ satisfies the following property
\begin{equation}\label{eq:U_property}
	\mathcal U_{-p,q}(\lambda_1,\lambda_2)
	=
	-\,\mathcal U_{p,q}(\lambda_1,\lambda_2),
\end{equation}
indeed, using transversality, the contraction on  the left hand side reduces to
\begin{equation}
	\mathcal U_{p,q}
	=
	(p\!\cdot\!q)\,
	\varepsilon(p)\!\cdot\!\varepsilon(q);
\end{equation}
and using the identity
\begin{equation}
	u^{\alpha\beta}(-p,q) = -\,u^{\alpha\beta}(p,q),
\end{equation}
eq. \eqref{eq:U_property} follows immediately.

Upon contraction with the polaritazion vectors the amplitudes for the three channels read:
\begin{align}\label{eq:channels_TJJ}
	\mathcal{M}_s(\lambda_1,\lambda_2,\lambda_3,\lambda_4) &= \frac{\bar C_0}{M_P^2}\mathcal{U}_{p_1,p_2}(\lambda_1,\lambda_2)\frac{1}{s}\mathcal{U}_{p_1,p_2}(\lambda_3,\lambda_4),\\
	\mathcal{M}_t(\lambda_1,\lambda_2,\lambda_3,\lambda_4) &= \frac{\bar C_0}{M_P^2}\mathcal{U}_{-p_1,p_3}(\lambda_1,\lambda_3)\frac{1}{t}\mathcal{U}_{-p_2,p_4}(\lambda_2,\lambda_4),\\
	\mathcal{M}_u(\lambda_1,\lambda_2,\lambda_3,\lambda_4) &= \frac{\bar C_0}{M_P^2}\mathcal{U}_{-p_1,p_4}(\lambda_1,\lambda_4)\frac{1}{u}\mathcal{U}_{-p_2,p_3}(\lambda_2,\lambda_3),
\end{align}
\begin{table}[t]
	\centering
	\small
	\renewcommand{\arraystretch}{1.12}
	\setlength{\tabcolsep}{4.2pt}
	\resizebox{\textwidth}{!}{%
		\begin{tabular}{cccc c c c c  @{\hspace{1em} \vrule width 0.75pt\hspace{1em} }cccc c c c c}
			\toprule
			$p_1$&$p_2$&$p_3$&$p_4$&
			$\bar{ \mathcal{M}}_s$&$\bar{ \mathcal{M}}_t$&$\bar{ \mathcal{M}}_u$&$\bar{ \mathcal{M}}$&
			$p_1$&$p_2$&$p_3$&$p_4$&
			$\bar{ \mathcal{M}}_s$&$\bar{ \mathcal{M}}_t$&$\bar{ \mathcal{M}}_u$&$\bar{ \mathcal{M}}$
			\\
			\midrule
			$+$&$+$&$+$&$+$&$-\frac{s}{4}$&$-\frac{t (t-u)^2}{4 s^2}$&$-\frac{u (t-u)^2}{4 s^2}$&$-\frac{t u}{s}$&
			$-$&$+$&$+$&$+$&$0$&$\frac{t u (t-u)}{2 s^2}$&$\frac{t u (u-t)}{2 s^2}$&$0$ \\
			
			$+$&$+$&$+$&$-$&$0$&$\frac{t u (t-u)}{2 s^2}$&$\frac{t u (u-t)}{2 s^2}$&$0$&
			$-$&$+$&$+$&$-$&$0$&$-\frac{t u^2}{s^2}$&$-\frac{u (t-u)^2}{4 s^2}$&$-\frac{u}{4}$ \\
			
			$+$&$+$&$-$&$+$&$0$&$\frac{t u (t-u)}{2 s^2}$&$\frac{t u (u-t)}{2 s^2}$&$0$&
			$-$&$+$&$-$&$+$&$0$&$-\frac{t (t-u)^2}{4 s^2}$&$-\frac{t^2 u}{s^2}$&$-\frac{t}{4}$ \\
			
			$+$&$+$&$-$&$-$&$-\frac{s}{4}$&$-\frac{t u^2}{s^2}$&$-\frac{t^2 u}{s^2}$&$-\frac{(t-u)^2}{4 s}$&
			$-$&$+$&$-$&$-$&$0$&$\frac{t u (t-u)}{2 s^2}$&$\frac{t u (u-t)}{2 s^2}$&$0$ \\
			
			$+$&$-$&$+$&$+$&$0$&$\frac{t u (t-u)}{2 s^2}$&$\frac{t u (u-t)}{2 s^2}$&$0$&
			$-$&$-$&$+$&$+$&$-\frac{s}{4}$&$-\frac{t u^2}{s^2}$&$-\frac{t^2 u}{s^2}$&$-\frac{(t-u)^2}{4 s}$ \\
			
			$+$&$-$&$+$&$-$&$0$&$-\frac{t (t-u)^2}{4 s^2}$&$-\frac{t^2 u}{s^2}$&$-\frac{t}{4}$&
			$-$&$-$&$+$&$-$&$0$&$\frac{t u (t-u)}{2 s^2}$&$\frac{t u (u-t)}{2 s^2}$&$0$ \\
			
			$+$&$-$&$-$&$+$&$0$&$-\frac{t u^2}{s^2}$&$-\frac{u (t-u)^2}{4 s^2}$&$-\frac{u}{4}$&
			$-$&$-$&$-$&$+$&$0$&$\frac{t u (t-u)}{2 s^2}$&$\frac{t u (u-t)}{2 s^2}$&$0$ \\
			
			$+$&$-$&$-$&$-$&$0$&$\frac{t u (t-u)}{2 s^2}$&$\frac{t u (u-t)}{2 s^2}$&$0$&
			$-$&$-$&$-$&$-$&$-\frac{s}{4}$&$-\frac{t (t-u)^2}{4 s^2}$&$-\frac{u (t-u)^2}{4 s^2}$&$-\frac{t u}{s}$ \\
			\bottomrule
	\end{tabular}}
	\caption{Four-particle configurations and corresponding values. Here $\bar{ \mathcal{M}}_s=\mathcal U(p_1,p_2)\frac1s\mathcal U(p_3,p_4)$, $\bar{\mathcal M}_t=\mathcal U(-p_1,p_3)\frac1t\mathcal U(-p_2,p_4)$, $\bar{\mathcal M}_u=\mathcal U(-p_1,p_4)\frac1u\mathcal U(-p_2,p_3)$, and $\bar{\mathcal M}$ is the total amplitude. In $\bar{ \mathcal{M}}$ we have omitted an overall constant $C/M_P^2$.}
	\label{tab:TJJ-TTJ}
\end{table}
The amplitudes can be generated from the action 
\begin{equation}
	S_{\text{eff}}
	=
	\frac{\bar C_0}{M_P^2}
	\int d^4x \, d^4y \,
	\big(F_{\mu\nu}F^{\mu\nu}\big)(x)\,
	G(x-y)\,
	\big(F_{\rho\sigma}F^{\rho\sigma}\big)(y),
	\tag{6.81}
\end{equation}
where the Green's function satisfies
\begin{equation}
	\Box_x G(x-y)=\delta^{(4)}(x-y).
\end{equation}
Equivalently,
\begin{equation}
	\mathcal S_{\rm eff}
	=
	\frac{\bar C_0^2}{M_P^2}
	\int d^4x \,
	\big(F_{\mu\nu}F^{\mu\nu}\big)\,
	\frac{1}{\Box}\,
	\big(F_{\rho\sigma}F^{\rho\sigma}\big).
\end{equation}
Although the overall mass dimension of the Lagrangian density is four,
\begin{equation}
	\left[
	\frac{1}{M_P^2}
	(F^2)\frac{1}{\Box}(F^2)
	\right]=4,
\end{equation}
the operator is intrinsically nonlocal because
\begin{equation}
	\left(\frac{1}{\Box} f\right)(x)
	=
	\int d^4y\, G(x-y)\, f(y),
\end{equation}
so the action couples field strengths at different spacetime points. 
This nonlocality, however, does not imply that the corresponding momentum-space amplitude exhibits a genuine propagating massless scalar exchange. As we show below, in the present kinematic setting it reduces instead to a contact contribution, namely to a term polynomial in the external momenta. This should not be surprising: contact interactions commonly arise when an underlying interaction is probed at energies much smaller than the characteristic mass scale of its mediator, although in the anomaly-induced case the interpretation is more subtle.\\
For instance, in Fermi theory of weak interactions one encounters the four-fermion contact term
\begin{equation}
	\mathcal L_F
	=
	-\frac{G_F}{\sqrt{2}}
	\left(\bar\psi \gamma^\mu (1-\gamma^5)\psi\right)
	\left(\bar\chi \gamma_\mu (1-\gamma^5)\chi\right),
	\label{fermitheory}
\end{equation}
which is a local dimension-six operator suppressed by a heavy mass scale,
\begin{equation}
	G_F \sim \frac{1}{M_{W/Z}^2}.
\end{equation}
In this case the heavy mediators have been integrated out, producing a genuine local contact interaction valid at energies \(E \ll M_{W/Z}\). By contrast, the term
\begin{equation}
	\frac{1}{M_P^2}\,F^2\,\frac{1}{\Box}\,F^2
\end{equation}
does not originate from integrating out a heavy field. Rather, it is associated with the exchange of a massless channel encoded in a nonlocal effective interaction, with the overall strength suppressed by the Planck scale. For this reason, the inverse d'Alembertian cannot be expanded into a finite tower of local operators, and the effective action remains intrinsically nonlocal even at low energies.\\
`
\subsection{Contact interactions mediated by the EH action}
It is interesting to investigate the implications of this result more closely. 
In momentum space the amplitude is of the form 
\begin{equation}
	\mathcal M_\Phi(q)\propto \frac{V_{F^2}(q)\,V'_{F^2}(q)}{q^2},
	\qquad
	\operatorname{Res}_{q^2=0}\mathcal M_\Phi
	\propto
	\lim_{q^2\to0}V_{F^2}(q)V'_{F^2}(q),
	\label{eq:ResidueCriterion}
\end{equation}
more details will we given in the next sections.
For photon scattering, the anomaly-induced spin--0 exchange can be represented
in terms of an effective vertex using \eqref{uab},
\begin{equation}
	iV^{\alpha\beta}(p_i,p_j)
	=
	ig_\Phi\,u^{\alpha\beta}(p_i,p_j),
	\qquad
	u^{\alpha\beta}(p_i,p_j)
	=
	(p_i\!\cdot p_j)\eta^{\alpha\beta}-p_j^\alpha p_i^\beta.
	\label{eq:Vertexu}
\end{equation}
where \(\Phi\) denotes the effective scalar channel associated with the
anomaly-induced spin--0 exchange. \\
A detailed analysis of these 2-to-2 amplitudes will be discussed in the next sections, but here we anticipate a result in order to characterize their specific features. \\
Recalling that by definition 
\begin{equation}
	\mathcal{U}_{ij} = 
	u^{\alpha\beta}(p_i,p_j)\,\varepsilon_{i\alpha}\varepsilon_{j\beta}
	=
	(p_i\!\cdot p_j)(\varepsilon_i\!\cdot\varepsilon_j)
	-(p_j\!\cdot\varepsilon_i)(p_i\!\cdot\varepsilon_j),
	\label{eq:Uij}
\end{equation}
with photon polarization 4-vectors $\varepsilon_j$ and scalar propagator \(i/(q^2+i0)\), the three channels give \begin{equation}
	i\mathcal M_s=i g_\Phi^2\frac{\mathcal{U}_{12}\,\mathcal{U}_{34}}{s+i0},\qquad
	i\mathcal M_t=i g_\Phi^2\frac{\mathcal{U}_{13}\,\mathcal{U}_{24}}{t+i0},\qquad
	i\mathcal M_u=i g_\Phi^2\frac{\mathcal{U}_{14}\,U_{23}}{u+i0},
\end{equation}
thus
\begin{equation}
	i\mathcal M_{\gamma\gamma\to\gamma\gamma}^{(\Phi)}
	=
	i g_\Phi^2\left(
	\frac{\mathcal{U}_{12}\,\mathcal{U}_{34}}{s+i0}
	+\frac{\mathcal{U}_{13}\,\mathcal{U}_{24}}{t+i0}
	+\frac{\mathcal{U}_{14}\,\mathcal{U}_{23}}{u+i0}
	\right).
	\label{eq:FullAmpPhi}
\end{equation}
and for on-shell photons (with $\kappa\to 1$)
\begin{equation}
	\frac{\mathcal{U}\,\mathcal{U}'}{s_{\mathrm{ch}}}\sim s_{\mathrm{ch}},
\end{equation}
where $s_{\mathrm{ch}}\in\{s,t,u\}$,
i.e. each channel is polynomial as \(s_{\mathrm{ch}}\to0\) and the pole residue vanishes. Correcting by the Planck mass, the amplitude takes the form of a contact term 
\beq
\mathcal M_{\mathrm{ch}}\sim \frac{s_{\mathrm{ch}}}{M_P^2}.
\label{eq:OffshellPole}
\end{equation}
Therefore the \(1/\Box\) structure is present at the level of the effective action \eqref{eq:ResidueCriterion}, while the visibility of a physical scalar pole in a given observable is controlled by the channel-dependent residue criterion in the same equation.  

One should not be surprised of the unusual behaviour associated with anomaly-mediated interactions. One can show by a careful analysis of the interaction, that the coupling of massless poles and their residues are strictly connected with the off-shell/on-shell kinematics of each 3-point vertex.\\
For example, one can show that the actual residue of a single pole in each anomaly vertex is nonzero only if  the graviton line is off-shell and the two spin-1 are on-shell. The interaction has to take place strictly on the light cone in order for each residue in each 3-point vertex to be nonzero \cite{Giannotti:2008cv,Coriano:2025fom}. \\
Notice that the existence of a massless pole in the \(TJJ\) correlator must be tested through the residue
\begin{equation}
\mathcal R^{\mu\nu\alpha\beta}(p_1,p_2)
\equiv
\lim_{s\to 0}
s\,\langle T^{\mu\nu}(q)\,J^\alpha(p_1)\,J^\beta(p_2)\rangle,
\qquad s\equiv q^2,\quad q=p_1+p_2.
\label{eq:Rdef_lightcone}
\end{equation}
A genuine particle-like massless exchange requires \(\mathcal R^{\mu\nu\alpha\beta}\neq0\). The key result  is that this condition is highly restrictive: the residue survives only in the massless, on-shell light-cone configuration.
To make this explicit, one compares two limiting prescriptions for \(q^2\to0\): (i) a light-cone scaling at fixed large \(q^+\), and (ii) a soft limit \(q^\lambda\to0\), as discussed in \cite{Coriano:2025fom}. In both cases, the same selection rule emerges. If at least one external current is off-shell, or if the loop fermion is massive, the residue vanishes:
\begin{equation}
\lim_{s\to0}
s\,\langle T^{\mu\nu}(q)\,J^\alpha(p_1)\,J^\beta(p_2)\rangle
=0,
\qquad
(s_1\neq0\ \text{or}\ s_2\neq0\ \text{or}\ m\neq0),
\label{eq:residue_zero_generic}
\end{equation}
where \(s_1=p_1^2\), \(s_2=p_2^2\). Conversely, in the strict on-shell massless configuration,
\begin{equation}
\lim_{s\to0}
s\,\langle T^{\mu\nu}(q)\,J^\alpha(p_1)\,J^\beta(p_2)\rangle
\neq0,
\qquad
(s_1=s_2=0,\ m=0).
\label{eq:residue_nonzero_strict}
\end{equation}
Therefore, the anomaly-coupled pole is not a generic off-shell feature of the correlator; it is activated only on a specific kinematic support.
As discussed in Section \ref{reff2}, in this limit the residue contains two tensor structures,
\begin{equation}
\mathcal R^{\mu\nu\alpha\beta}
=
c_1\,\tilde\phi_1^{\mu\nu\alpha\beta}
+
c_2\,\tilde\phi_2^{\mu\nu\alpha\beta},
\label{eq:residue_two_structures}
\end{equation}
defined in \eqref{widetilde1} and \eqref{widetilde2}. The first term is the anomaly-related trace contribution, while the second is traceless and corresponds to the additional pole term identified in \eqref{expoleac}.\\
Thus, even when the residue is nonvanishing, the conformal case is more restrictive than the naive picture of an ordinary particle pole mediating the interaction.\\
This clarifies why anomaly mediation need not generate an observable pole in a given four-point process and provides the appropriate conceptual bridge: anomaly poles couple only under sharply constrained kinematical conditions, and a nonzero residue is a light-cone/on-shell property rather than a generic off-shell feature of an anomaly vertex. This explains why one may consistently find no anomaly-induced scalar pole in a specific four-point amplitude, despite the existence of a nonlocal anomaly action. In such situations, the surviving contributions should instead be interpreted as contact terms, which are a genuine and characteristic manifestation of anomaly-mediated interactions.

\begin{figure}[t]
\begin{center}
	\includegraphics[scale=0.7,angle=0]{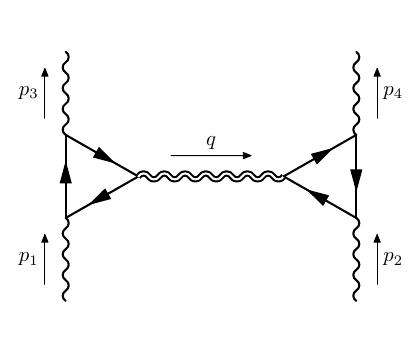}
	\includegraphics[scale=0.7,angle=0]{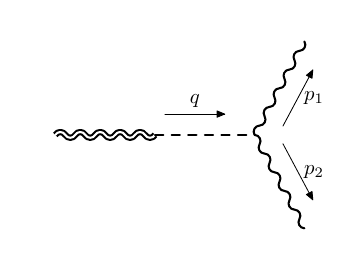}
	\caption{The free field theory realization of the $TJJ$ vertex, here represented for an Abelian theory such as QED. The $J$ are coupled to photon lines (left). The mixing interaction induced by the graviton and the, corresponding in momentum space of a projector $\pi^{\mu\nu}$ on the external graviton line of any anomaly vertex (right).} 
	\label{expansionX1}
\end{center}
\end{figure}

\section{Anomaly mediation with a massive fermion}

We now turn to the on-shell regime and analyze the spectral properties of the conformal-anomaly form factor \(\Phi_{TJJ}\) in the presence of $n_F$ massive fermions. The spectral analysis of this form factor has been presented in \cite{Coriano:2025fom} for QCD, but it can be easily extrapolated to the Abelian 
case. We are going to briefly summarize it.  \\
The anomaly contribution 
\beq
\frac{1}{ 3 \, q^2} \hat\pi^{\mu \nu}(q) \mathcal{A}^{\alpha \beta}
\eeq
appearing in \eqref{eq:TJJreconstructed2}, with the anomaly $\mathcal{A}^{\alpha \beta}$ given by\eqref{anoms} is replaced by 

\begin{equation}\label{PhiTJJ_onshell}
\Phi_{0}(s,m)
=
\frac{n_F g^2}{72\pi^2 s^2}
\Bigl[3m^2(s-4m^2)\,C_0(s,m^2)-6m^2+s\Bigr],
\end{equation}
where \(s=q^2\) and \(C_0(s,m^2)\equiv C_0(s,0,0,m^2)\) denotes the standard scalar three-point function with two onshell external lines
\begin{equation}\label{C0_onshell}
C_0(s,m^2)
=
\frac{1}{2s}
\log^2\!\left(
\frac{\sqrt{s(s-4m^2)}+2m^2-s}{2m^2}
\right).
\end{equation}

The spectral content of \(\Phi_0\) is extracted from the discontinuity across the physical cut along the real axis, \(s\ge 0\). We define
\begin{equation}
\text{disc}\,F(s)\equiv F(s+i0)-F(s-i0),
\end{equation}
for any function admitting boundary values on the cut. 
Introducing
\begin{equation}
\tau(s,m^2)=1-\frac{4m^2}{s},
\qquad
A(s)=C_0(s+i0,m^2)+C_0(s-i0,m^2),
\end{equation}
one finds
\begin{align}\label{disc_C0_over_s_clean}
\text{disc}\!\left(\frac{C_0(s,m^2)}{s}\right)
&=
-2\pi i\,\frac{1}{s^2}
\log\!\left(
\frac{1+\sqrt{\tau(s,m^2)}}{1-\sqrt{\tau(s,m^2)}}
\right)\theta(s-4m^2)
-\pi i\,A(0)\,\delta(s),
\\[2mm]
\text{disc}\!\left(\frac{C_0(s,m^2)}{s^2}\right)
&=
-2\pi i\,\frac{1}{s^3}
\log\!\left(
\frac{1+\sqrt{\tau(s,m^2)}}{1-\sqrt{\tau(s,m^2)}}
\right)\theta(s-4m^2)
+\pi i\,\delta'(s)\,A(s).
\end{align}
The term proportional to \(\delta'(s)\) is understood as
\begin{equation}\label{disc_powers_clean}
\delta'(s)\,A(s)=\delta'(s)\,A(0)-\delta(s)\,A'(0),
\qquad
A(0)=-\frac{1}{m^2},
\qquad
A'(0)=-\frac{1}{12m^4}.
\end{equation}

Using these relations, the discontinuity of the fermionic contribution to \(\Phi_{TJJ}\) can be organized as
\begin{equation}
\text{disc}\bigl[\Phi_0(s,m)\bigr]_{\text{fermion}}
=
\frac{n_F g^2}{72\pi^2}
\text{disc}\!\left(
\frac{3m^2(s-4m^2)C_0(s,m^2)-6m^2+s}{s^2}
\right).
\end{equation}
Substituting \eqref{disc_powers_clean} and \eqref{disc_C0_over_s_clean}, one finds that all terms proportional to \(\delta'(s)\) and \(\delta(s)\) cancel identically for \(m\neq0\). Therefore the fermionic sector does not generate any pole contribution to the spectral density; only the continuous cut above threshold survives:
\begin{equation}\label{disc_Phi_fermion_final_clean}
\text{disc}\bigl[\Phi_{0}(s,m)\bigr]_{\text{fermion}}
=
-2\pi i\,
\frac{n_F e^2\,\delta^{ab}}{72\pi}
\frac{3m^2(s-4m^2)}{s^3}
\log\!\left(
\frac{1+\sqrt{\tau(s,m^2)}}{1-\sqrt{\tau(s,m^2)}}
\right)
\theta(s-4m^2).
\end{equation}
This explicitly shows that the anomaly pole is absent in the fermionic sector at finite mass: the massless pole dissolves into a branch cut. In the conformal limit \(m\to0\), however, the cut collapses to the origin,
\begin{equation}
\text{disc}\bigl[\Phi_0(s)\bigr]
=
-\frac{(2n_F)\,i e^2}{72\pi}\,\delta(s),
\end{equation}
Introducing the standard factor \((2i)^{-1}\) in the definition of the spectral density, one recovers
\begin{equation}
\Delta\Phi_0(s)
\equiv\frac{1}{2 i}\text{disc}\bigl[\Phi_0(s)\bigr]=
-\frac{1}{3}\,\frac{\beta(e)}{e}\,\delta(s),
\end{equation}
which reproduces the anomaly sum rule upon integration over \(s\ge0\),
\beq
\int_{4m^2}^{\infty}\Delta\Phi_0(s,m^2)\,ds
=
\frac{g_s^2}{144\pi^2}\,(-2n_f)
=
-\frac{1}{3}\frac{\beta(e)}{e}.
\eeq
In this case the anomaly interaction is characterized by the presence of branch cuts rather than poles on each vertex. The scalar amplitude is modified to
\begin{equation}
\widetilde{\mathcal{A}}^{(0)}
=
\widetilde{\mathcal{A}}_s^{(0)}
+
\widetilde{\mathcal{A}}_t^{(0)}
+
\widetilde{\mathcal{A}}_u^{(0)}.
\end{equation}
In components this takes the form
\begin{equation}
\begin{aligned}
	\widetilde{\mathcal{A}}^{(0)\alpha\beta\gamma\delta}
	=
	\frac{\bar C_0}{M_P^2}
	\Bigg[&
	\Phi_0^2(s,m)\,
	u^{\alpha\gamma}(p_1,p_2)\,
	\frac{1}{s}\,
	u^{\beta\delta}(p_3,p_4)
	\\
	&\quad
	+
	\Phi_0^2(t,m)\,
	u^{\alpha\beta}(p_1,p_3)\,
	\frac{1}{t}\,
	u^{\gamma\delta}(p_2,p_4)
	+
	\Phi_0^2(u,m)\,
	u^{\alpha\delta}(p_1,p_4)\,
	\frac{1}{u}\,
	u^{\gamma\beta}(p_2,p_3)
	\Bigg].
\end{aligned}
\end{equation}
A direct computation gives
\begin{equation}
\widetilde{\mathcal{A}}_{\lambda_1\lambda_2\lambda_3\lambda_4}
=
Y
\begin{dcases}
	-\frac{4 (t+u)^6}{s^3} \Phi_0(s,m)^2 - t (s-t+3 u)^2 \Phi_0(t,m)^2 - u (s+3 t-u)^2 \Phi_0(u,m)^2   \hfill\text{for } \pm\pm\pm\pm \\
	-\frac{4 (t+u)^6}{s^3} \Phi_0(s,m)^2 - \frac{t^2 (s+t-3 u)^2}{u} \Phi_0(u,m)^2 - \frac{u^2 (s-3 t+u)^2}{t} \Phi_0(t,m)^2 \quad \text{for } \pm\pm\mp\mp \\
	-t (s-t+3 u)^2 \Phi_0(t,m)^2 - \frac{t^2 (s+t-3 u)^2}{u} \Phi_0(u,m)^2 \hfill \text{for } \pm\mp\pm\mp \\
	-\frac{u^2 (s-3 t+u)^2}{t} \Phi_0(t,m)^2 - u (s+3 t-u)^2 \Phi_0(u,m)^2
	\hfill  \text{for } \pm\mp\mp\pm \\
	u (s-3 t+u) (s-t+3 u) \Phi_0(t,m)^2 + t (s+3 t-u) (s+t-3 u) \Phi_0(u,m)^2
	\hfill \text{for odd flips}
\end{dcases}
\end{equation}
with
\beq
Y\equiv \frac{\bar C_0}{16M_P^2(t+u)^2},
\eeq
and \(\bar C_0\) given by \eqref{C0}.

\begin{figure}[t]
\begin{center}
	\includegraphics[scale=0.6,angle=0]{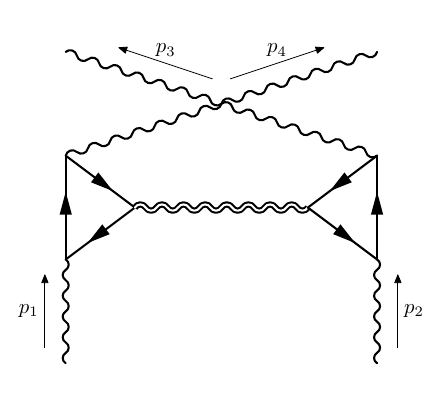}
	\includegraphics[scale=0.6,angle=0]{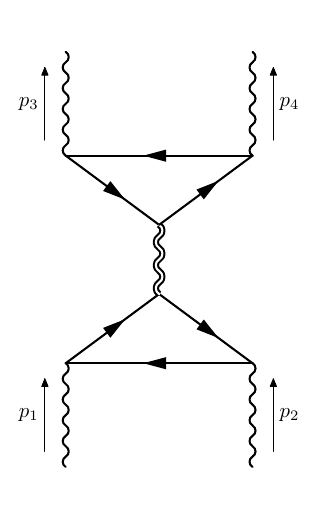}
	\caption{$t$ and $s$ channel diagrams in the scattering of spin-1 with anomaly mediation. } 
	\label{expansionX2}
\end{center}
\end{figure}

\section{ Graviton-Graviton scattering at tree level and the anomaly corrections}
The extension of our previous analysis to graviton--graviton scattering can be carried out using the explicit expression of the \(TTT\) correlator. This correlator has been analysed within the general framework based on the solution of the conformal Ward identities developed in \cite{Bzowski:2013sza}, and subsequently re-investigated perturbatively in \cite{Coriano:2018bsy}. In the latter work, the integration constants appearing in the general conformal solution of \cite{Osborn:1993cr}, reconstructed in momentum space in \cite{Bzowski:2013sza}, were matched to three independent multiplicity factors in arbitrary spacetime dimensions.\\
The analysis of \cite{Coriano:2018bsy} is, to our knowledge, the only one that yields infrared-safe results for the vertex when two of the three external momenta are taken on shell, a feature that we have already shown to occur in the case of the \(TJJ\) correlator. The difference with respect to the dimensional-regularization scheme adopted in \cite{Bzowski:2013sza} lies in the specific subtraction prescription used in the renormalization of the triple-\(K\) integrals, which generates additional logarithmic terms depending on the renormalization scale and which fail to remain finite in the double on-shell limit.\\
While this issue requires a dedicated discussion that we defer to future work, our present interest is focused on the anomaly contribution to the vertex, for which the approaches of \cite{Bzowski:2013sza} and \cite{Coriano:2018bsy} are in agreement. The same anomaly-induced effective vertex has also been shown to coincide with the result obtained from the nonlocal anomaly action, a method that we now briefly review.\\
It becomes possible to recast the generally covariant non-local effective action (\ref{Snonlsq}) in local form by the introduction of  
only a single new scalar field $\vf$, the conformalon field \cite{Mottola:2016mpl}. 
Thus, if the anomaly effective action is recast in local form
\bea
&&\hspace{-1.5cm} \cS_{\rm anom}[g;\vf] \equiv -\sdfrac{b'}{2} \int d^4x\,\sqrt{-g}\, \Big[ (\sq \vf)^2 - 2 \big(R^{\m\n} - \tfrac{1}{3} R g^{\m\n}\big)
(\na_\m\vf)(\na_\n \vf)\Big]\nn \\
&& \hspace{1.5cm} +\, \sdfrac{1}{2}\,\int d^4x\,\sqrt{-g}\  \Big[b'\big(E - \tfrac{2}{3}\sq R\big) + b\,C^2 \Big]\,\vf
\label{Sanom}
\eea
and varied with respect to $\varphi$, the linear eq. of motion takes the form 
\be
\sqrt{-g}\,\D_4\, \vf = \sqrt{-g}\left[\sdfrac{E}{2}- \sdfrac{\!\sq R\!}{3} + \sdfrac{b}{2b'}\, C^2\right] = \frac{1}{2b'} \,\cA
\label{phieom}
\ee

\begin{figure}[t]
\centering
\includegraphics[scale=0.7]{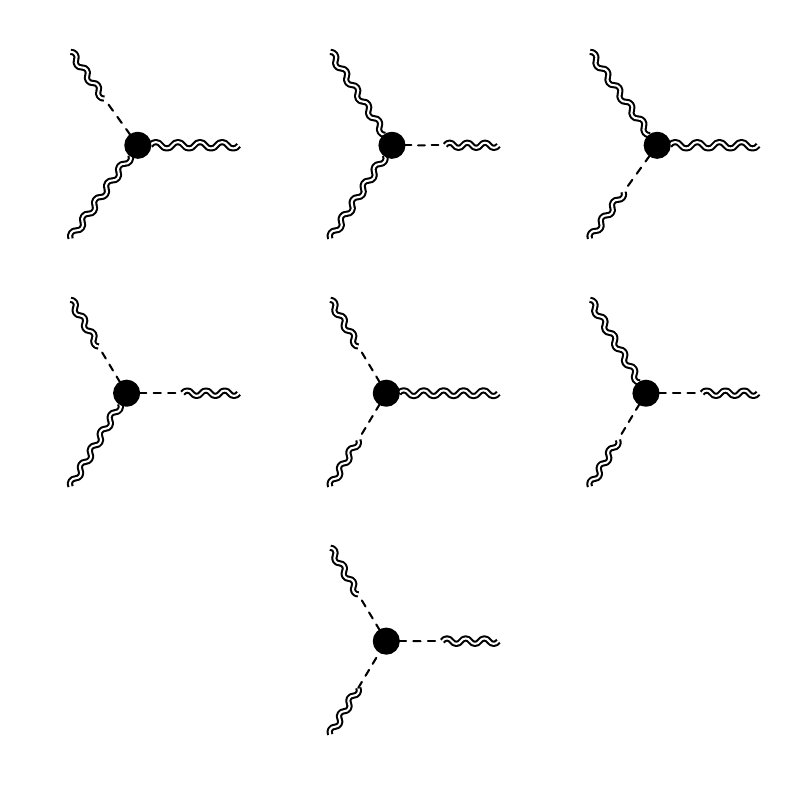}
\caption{Anomaly interactions mediated by the exchange of one, two or three poles. The poles are generated by the renormalization of the longitudinal sector of the $TTT$.}
\label{dec}
\end{figure}

In order to obtain the contributions of the anomaly effective action (\ref{Sanom}) to the three-point function
$\cS_3^{\m_1\n_1\m_2\n_2\m_3\n_3}$ we require the expansion of $\cS_{\rm anom}[g;\vf]$ to third order
in deviations from flat space. The consistent expansion  of $\cS_{\rm anom}[g;\vf]$ around flat space is 
defined by the simultaneous expansion of the metric $g_{\m\n}$ and conformalon scalar $\vf$
\bes
\bea
g_{\m\n} &=& g_{\m\n}^{(0)} + g_{\m\n}^{(1)} + g_{\m\n}^{(2)} + \dots \equiv \eta_{\m\n} + h_{\m\n} + h_{\m\n}^{(2)} + \dots\\
\vf &=& \vf^{(0)} +  \vf^{(1)} +  \vf^{(2)}  + \dots
\eea
\ees
substitution of these expansions into (\ref{phieom}) and identification of terms of the same order. Thus to the lowest three orders of the expansion we have
\bes
\bea
&&\hspace{4cm}\sqb^2 \vf^{(0)} = 0 \label{eom0}\\
&&\hspace{-1.5cm}(\sqrt{-g} \D_4)^{(1)} \vf^{(0)} + \sqb^2 \vf^{(1)} = \left[\sqrt{-g}
\left( \sdfrac{E}{2}- \sdfrac{\!\sq R\!}{3} + \sdfrac{b}{2b'}\, C^2 \right)\right]^{(1)}
= - \sdfrac{\!1\!}{3}\, \sqb R^{(1)} \label{eom1}\\
&&\hspace{-2cm}(\sqrt{-g} \D_4)^{(2)} \vf^{(0)} + (\sqrt{-g} \D_4)^{(1)} \vf^{(1)} + \sqb^2 \vf^{(2)} =
\left[\sqrt{-g}\left(\sdfrac{E}{2}- \sdfrac{\!\sq R\!}{3} + \sdfrac{b}{2b'}\, C^2 \right)\right]^{(2)} \nn \\
&&\hspace{2cm}= \sdfrac{1}{2}E^{(2)} - \sdfrac{1}{3}\, [\sqrt{-g}\sq R]^{(2)} + \sdfrac{b}{2b'}\, [C^2]^{(2)} \label{eom2}
\eea
\ees
where $\sqb$ is the d'Alembert wave operator in flat Minkowski spacetime, and we have used the fact that $E$ and $C^2$ are second order in curvature invariants while the Ricci scalar $R$ starts at first order. 

\be
\vf^{(1)} = - \sdfrac{\!1 \!}{3\sqb} \, R^{(1)}
\label{vf1}
\ee
and hence the solution of (\ref{eom2}) is
\be
\vf^{(2)} = \sdfrac{1}{\sqb^2} \left\{ (\sqrt{-g} \D_4)^{(1)}\sdfrac{\!1 \!}{3\sqb} \, R^{(1)} 
+  \sdfrac{1}{2}E^{(2)} - \sdfrac{1}{3}\, [\sqrt{-g}\sq R]^{(2)} + \sdfrac{b}{2b'}\, [C^2]^{(2)}\right\}
\label{vf2}
\ee
giving at cubic order the effective action
\bea
&&\hspace{-5mm} \cS_{\rm anom}^{(3)} =
\sdfrac{b'}{9} \int\! d^4x \int\!d^4x'\!\int\!d^4x''\!\left\{\big(\pa_{\m} R^{(1)})_x\left(\sdfrac{1}{\sqb}\right)_{\!xx'}  
\!\left(R^{(1)\m\n}\! - \!\sdfrac{1}{3} \eta^{\m\n} R^{(1)}\right)_{x'}\!
\left(\sdfrac{1}{\sqb}\right)_{\!x'x''}\!\big(\pa_{\n} R^{(1)})_{x''}\right\}\nn \\
&&\hspace{-6mm}- \sdfrac{1}{6}\! \int\! d^4x\! \int\!d^4x'\! \left(b'\, E^{\!(2)} + b\,  [C^2]^{(2)}\right)_{\!x}\! \left(\sdfrac{1}{\sqb}\right)_{\!xx'} \!R^{(1)}_{x'}
+ \sdfrac{b'}{18} \! \int\! d^4x\, R^{(1)}\left(2\, R^{\!(2)} + (\sqrt{-g})^{(1)} R^{(1)}\right)
\label{S3anom3}
\eea
which in momentum space takes the form 
\begin{align}
^{(\cA)}S_3&^{\m_1\n_1\m_2\n_2\m_3\n_3} = \sdfrac{1}{3}\, \pi^{\m_1\n_1}(p_1)\,\eta_{\a_1\b_1}\,^{(\cA)}S_3^{\a_1\b_1\m_2\n_2\m_3\n_3}
+\sdfrac{1}{3}\, \pi^{\m_2\n_2}(p_2)\,\eta_{\a_2\b_2}\,^{(\cA)}S_3^{\m_1\n_1\a_2\b_2\m_3\n_3} \nonumber\\
& +\sdfrac{1}{3}\, \pi^{\m_3\n_3}(p_3)\,\eta_{\a_3\b_3}\,^{(\cA)}S_3^{\m_1\n_1\m_2\n_2\a_3\b_3} 
-\sdfrac{1}{9}\,\pi^{\m_1\n_1}(p_1)\,\pi^{\m_3\n_3}(p_3)\,\eta_{\a_1\b_1}\eta_{\a_3\b_3}\,^{(\cA)}S_3^{\a_1\b_1\m_2\n_2\a_3\b_3} \nonumber\\
&-\sdfrac{1}{9}\,\pi^{\m_2\n_2}(p_2)\pi^{\m_3\n_3}(p_3)\,\eta_{\a_2\b_2}\eta_{\a_3\b_3}\,^{(\cA)}S_3^{\m_1\n_1\a_2\b_2\a_3\b_3}
-\sdfrac{1}{9}\,\pi^{\m_1\n_1}(p_1)\pi^{\m_2\n_2}(p_2)\eta_{\a_1\b_1}\, \eta_{\a_2\b_2}\,^{(\cA)}S_3^{\a_1\b_1\a_2\b_2\m_3\n_3} \nonumber\\
& +\sdfrac{1}{27}\,\pi^{\m_1\n_1}(p_1)\pi_2^{\m_2\n_2} (p_2)\pi^{\m_3\n_3}(p_3)\,\eta_{\a_1\b_1}\eta_{\a_2\b_2}\eta_{\a_3\b_3}\,^{(\cA)}S_3^{\a_1\b_1\a_2\b_2\a_3\b_3} \,. \label{fin1}
\end{align}
where 
\bea
&&\hspace{-1.5cm}\eta_{\a_1\b_1}\,^{\!(\cA)}\cS_3^{\a_1\b_1\m_2\n_2\m_3\n_3}(p_1,p_2,p_3)\Big\vert_{p_3 = -(p_1 + p_2)} =\nn \\
&& 8 b\, \big[(C^2)^{(2)}\big]^{\m_2\n_2\m_3\n_3} (p_2, p_3) 
+ \,8b'\, \big[E^{(2)}\big]^{\m_2\n_2\m_3\n_3} (p_2, p_3) = 4 \, \cA_2^{\m_2\n_2\m_3\n_3} (p_2, p_3)
\label{traceS3}
\eea
denotes the trace of the third functional derivative of anomaly functional around flat space. the other structures, describing contact interactions in coordinate space are obtained by double tracing

\bea
&&\eta_{\a_1\b_1}\eta_{\a_3\b_3}\,^{\!(\cA)}\cS_3^{\a_1\b_1\m_2\n_2\a_3\b_3}(p_1,p_2,p_3)\big\vert_{p_3 = -(p_1 + p_2)} 
= 8b' \, \eta_{\a_3\b_3}\big[E^{(2)}\big]^{\m_2\n_2\a_3\b_3} (p_2, p_3)  \nn \\
&& \qquad = 16b'\, Q^{\m_2\n_2}(p_1,p_2,p_3)\big\vert_{p_3 = -(p_1 + p_2)} \, + \ 8b'\,p_2^2\, \left( p_1^2  + p_1\cdot p_2\right)\pi^{\m_2\n_2}(p_2) 
\label{dbltrace3}
\eea
with 
\bea
&&Q^{\m_2\nu_2}(p_1,p_2,p_3) \equiv p_{1\m}\, [R^{\m\nu}]^{\m_2\nu_2}(p_2)\,p_{3\n} \nn \\
&&= \sdfrac{1}{2} \,\Big\{(p_1\cdot p_2)(p_2\cdot p_3)\, \eta^{\m_2\n_2} 
+ p_2^2 \ p_1^{(\m_2}\, p_3^{\n_2)} - (p_2\cdot p_3) \, p_1^{(\m_2}\, p_2^{\n_2)} - (p_1\cdot p_2) \, p_2^{(\m_2}\, p_3^{\n_2)}\Big\}
\eea

and triple traces of the same vertex
\be
\eta_{\a_1\b_1}\eta_{\a_2\b_2}\eta_{\a_3\b_3}\,^{\!(\cA)}\cS_3^{\a_1\b_1\a_2\b_2\a_3\b_3}(p_1,p_2,p_3)\big\vert_{p_3 = -(p_1 + p_2)} =
16b' \left[ p_1^2\,p_2^2 - (p_1\cdot p_2)^2\right].
\label{triptrace3}
\ee

Second functional derivatives of the anomaly functional in flat space have a specific local structure 

\bea
&&\hspace{-1.5cm}\big[E^{(2)}\big]^{\m_i\nu_i\m_j\nu_j} =\big[R_{\m\a\n\b}^{(1)}R^{(1)\m\a \n\b}\big]^{\m_i\nu_i\m_j\nu_j}
-4\,\big[R_{\m\n}^{(1)}R^{(1)\m\n}\big]^{\m_i\nu_i\m_j\nu_j}
+\big[ \big(R^{(1)}\big)^2\big]^{\m_i\nu_i\m_j\nu_j}\\
&&\hspace{-1.5cm} \big[(C^2)^{(2)}\big]^{\m_i\nu_i\m_j\nu_j}= \big[R_{\m\a\n\b}^{(1)}R^{(1)\m\a \n\b}\big]^{\m_i\nu_i\m_j\nu_j}
-2\,\big[R_{\m\n}^{(1)}R^{(1)\m\n}\big]^{\m_i\nu_i\m_j\nu_j}
+ \sdfrac{1}{3}\,\big[(R^{(1)})^2\big]^{\m_i\nu_i\m_j\nu_j}
\eea
Notice that $\big[(C^2)^{(2)}\big]^{\m_i\nu_i\m_j\nu_j}$ is proportional to the transverse traceless projector 
$P^{(2)}$ in flat space.

\begin{figure}[t]
\begin{center}
	\includegraphics[scale=0.5,angle=0]{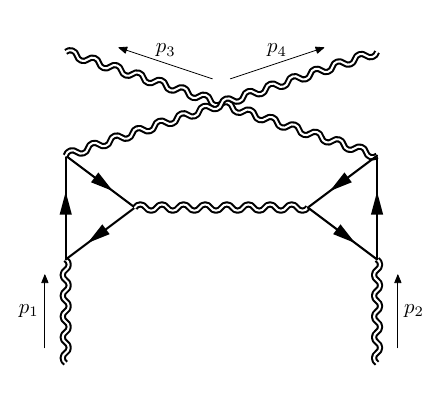}
	\includegraphics[scale=0.5,angle=0]{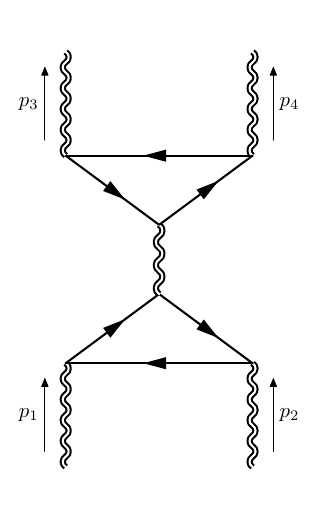}
	\includegraphics[scale=0.5,angle=0]{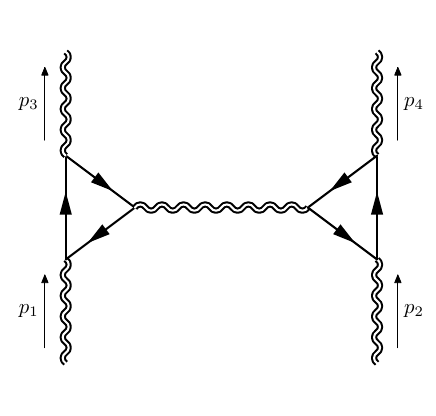}
	\caption{Anomaly mediation of graviton-graviton scattering. Shown are the $s, t$ and $u$ channels. } 
	\label{expansionX3}
\end{center}
\end{figure}

\subsection{The scalar contribution to a 2-to-2  scattering and the double copy}
In order to discuss the impact of anomaly mediation we can send on-shell two of the three graviton lines and consider only the interpolating graviton in its scalar component, in order to build the scalar exchange shown in Fig. 1 (right). The amplitude is built by sewing together two $TTT$ 3-point functions using the scalar component of the De Donder propagator (see Fig. 7). 
Since we are only interested in this specific component of the exchange, together with small momentum transfers which 
are relevant for the study of the eikonal limit, we need just to select a single term from \eqref{fin1}, which in this case is given by (for the left vertex)
\beq
\Gamma^{\mu_1\nu_1,\mu_2\nu_2,\mu_3\nu_3}(p_1,p_2,p_3)=\sdfrac{1}{3}\, \pi^{\mu_2\nu_2}(p_2)\,\eta_{\sigma_1\sigma_2}\,^{(\cA)}S_3^{\mu_1\nu_1\sigma_1\sigma_2\mu_3\nu_3}(p_1,p_2,p_3)
\eeq
The anomaly appears via a single trace contraction of $S_3$ which is proportional to  $\mathcal{A}_2(p_1,q)$ as specified by \eqref{traceS3}
\begin{equation}
\Gamma^{\mu_1\nu_1,\mu_2\nu_2,\mu_3\nu_3}(p_1,p_2,p_3)
=
\frac{1}{3}\mathcal{A}_2^{\mu_1\nu_1,\mu_3\nu_3}(p_1,p_3)\,
\pi^{\mu_2\nu_2}(p_2),
\end{equation}
where
\begin{equation}
\pi^{\mu_2\nu_2}(p_2)
\equiv
\left(
{R}^{(1)}(p_2)\,\Box^{-1}
\right)^{\mu_2\nu_2}
\end{equation}
is the scalar projector appearing on the exchanged gravitational leg (see Fig. 1).
The intermediate propagator is given by the scalar component of the De Donder propagator \eqref{eq:propagator_spin_decomposition}, which is proportional to the spin-0 projector in flat space
\beq
D^{(0-s)}(k)_{\rho\sigma\alpha\beta}  = \frac{1}{2} \frac{P^{(0){}}_{\rho\sigma\alpha\beta}(k)}{k^2} = -\frac{1}{6}\frac{\pi_{\rho\sigma}\pi_{\alpha\beta}}{k^2} .
\end{equation}
We contract the two vertices to obtain for the channel $t$
\begin{align}
\Gamma_4^{\mu_1\nu_1,\mu_2\nu_2,\mu_3\nu_3,\mu_4\nu_4}(p_1,p_2,p_3,p_4) & \equiv \Gamma^{\mu_1\nu_1,\rho\sigma,\mu_2\nu_2}(p_1,-q,p_3)\,D^{(0-s)}_{\rho\sigma\alpha\beta}(q)\,\Gamma^{\mu_3\nu_3,\alpha\beta,\mu_4\nu_4}(p_3,q,p_4) \nonumber\\
&= -\frac{1}{6}\,\tilde{\mathcal{A}}_2^{\mu_1\nu_1,\mu_3\nu_3}(p_1,p_3)\,
\frac{1}{q^2}\tilde{\mathcal{A}}_2^{\mu_2\nu_2,\mu_4\nu_4}(p_2,p_4)\,
,
\end{align}
where $q$ is the exchanged momentum. Adding the contributions of all the three channels we obtain 
\begin{equation}\label{eq:tot_ampl_gg-D0s-gg}
\begin{aligned}
\mathcal{M}^{(0)}_{\mu_1\nu_1\,\mu_2\nu_2\,\mu_3\nu_3\,\mu_4\nu_4}
=
-\frac{1}{6M_P^6}
\Big[
&\mathcal{A}_2^{\mu_1\nu_1\,\mu_2\nu_2}(p_1,p_2)
\frac{1}{s}
\mathcal{A}_2^{\mu_3\nu_3\,\mu_4\nu_4}(p_3,p_4) + 
\mathcal{A}_2^{\mu_1\nu_1\,\mu_3\nu_3}(-p_1,p_3)
\frac{1}{t}
\mathcal{A}_2^{\mu_2\nu_2\,\mu_4\nu_4}(-p_2,p_4)
\\
&+
\mathcal{A}_2^{\mu_1\nu_1\,\mu_4\nu_4}(-p_1,p_4)
\frac{1}{u}
\mathcal{A}_2^{\mu_2\nu_2\,\mu_3\nu_3}(-p_2,p_3)
\Big] .
\end{aligned}
\end{equation}
with $\cA_2^{\mu_1\nu_1,\mu_2,\nu_2}$ given by \eqref{traceS3}. For the generic channel we have 
\begin{equation}\label{eq:A2_simplified}
\begin{aligned}
\mathcal{A}^{\mu_1\nu_1,\mu_2,\nu_2}_2(p_i,p_j)&=\frac{1}{4}\,(b+b')\Big(
-2s_\mathrm{ch}\,p_j^{\mu_1}p_i^{\mu_2}g^{\nu_1\nu_2}
-2s\,p_i^{\mu_2}p_j^{\nu_1}g^{\mu_1\nu_2}
-2s\,p_j^{\mu_1}p_i^{\nu_2}g^{\mu_2\nu_1}
-2s\,p_j^{\nu_1}p_i^{\nu_2}g^{\mu_1\mu_2}
\\
&\qquad\qquad
+s^{2}g^{\mu_1\nu_2}g^{\mu_2\nu_1}
+s^{2}g^{\mu_1\mu_2}g^{\nu_1\nu_2}
+8\,p_j^{\mu_1}p_i^{\mu_2}p_j^{\nu_1}p_i^{\nu_2}
\Big).
\end{aligned}
\end{equation}
with $s_\mathrm{ch} \in\{ s, t, u\}$. Observe that on-shell $\mathcal{A}_2^{\mu_1\nu_1\mu_2\nu_2}$ can be rewritten in the following way:
\begin{align}\label{eq:A2_u_decomposition}
\mathcal{A}^{\alpha\beta\rho\sigma}_2(p,q)&\equiv \prescript{(\mathcal{A})}{}{S_3}^{\alpha\beta\rho\sigma\mu\nu}(p,q,-p-q)\eta_{\mu\nu}\nonumber \\
&= 4(b+b')\left[u^{\alpha\rho}(p,q)u^{\beta\sigma}(p,q)+u^{\alpha\sigma}(p,q)u^{\beta\rho}(p,q)\right],
\end{align}
where $u^{\alpha\beta}(p,q)$ defined in \eqref{uab} represent exactly the anomaly part of the $TJJ$. Notice that as one should expect the rhs of \eqref{eq:A2_u_decomposition} is invariant under the exchange of the two gravitons and is symmetric under $\{\alpha, \beta\} \leftrightarrow \{\rho, \sigma\}$. The relationship:
\begin{equation}
\mathcal{A}_2^{\alpha\beta\rho\sigma}=4(b+b')\left[u^{\alpha\rho}u^{\beta\sigma}+u^{\alpha\sigma}u^{\beta\rho}\right]	
\end{equation}
can be interpreted as a manifestation of the double-copy.

The polarization tensor of the graviton can be written as product of two polarization vectors:
\beq
\varepsilon_{\mu\nu}(p,\lambda) = \varepsilon_{\mu}(p,\lambda)\varepsilon_{\nu}(p,\lambda)
\eeq
If we require it to be transverse-traceless then:
\begin{align}
\varepsilon^{TT}_{\mu\nu}(p,\lambda)& = \Pi_{\mu\nu}^{\;\;\;\;\alpha\beta}(p)\varepsilon_{\alpha\beta}(p,\lambda)\nonumber\\
&= \left[\frac{1}{2}\Big(\pi_{\mu}^{\;\;\alpha}(p) \pi_{\nu}^{\;\;\beta}(p)+ \pi_{\nu}^{\;\;\alpha}(p) \pi_{\mu}^{\;\;\beta}(p)\Big) - \frac{1}{3} \pi_{\mu\nu}(p) \pi^{\alpha\beta}(p)\right]\varepsilon_{\alpha}(p)\varepsilon_{\beta}(p).
\end{align}
If $\varepsilon_{\alpha}(p)$ and $\varepsilon_{\beta}(p)$ are transverse this expression simplifies:
\begin{align}
\varepsilon^{TT}_{\mu\nu}(p,\lambda)& = \varepsilon^T_\mu(p) \varepsilon^T_\nu(p)- \frac{1}{3}\pi_{\mu\nu}\, \varepsilon^T(p)\cdot\varepsilon^T(p).
\end{align}
Notice that helicity projectors are both transverse and lightlike $p\cdot \varepsilon(p,\pm) = \varepsilon(p,\pm)\cdot\varepsilon(p,\pm)$ = 0, so that:
\begin{equation}
\varepsilon^{TT}_{\mu\nu}(p,\pm) = \varepsilon_{\mu}(p,\pm)\varepsilon_{\nu}(p,\pm).
\end{equation}
The contraction of the $\mathcal{A}_2$ vertex with the polarization tensors is given by:
\begin{align}\label{eq:mathsfA2}
\varepsilon_{\mu_1\nu_1}(p_1,\lambda_1) \varepsilon_{\mu_2\nu_2}(p_2,\lambda_2) \mathcal{A}^{\mu_1\nu_1\mu_2\nu_2}_2(p_1,p_2) =8(b+b')\mathcal{U}^2_{p_1,p_2}(\lambda_1,\lambda_2). 
\end{align}
This relation can be inserted into \eqref{eq:tot_ampl_gg-D0s-gg} to compute the total scattering amplitude. Here we report the results for the three channels separately:
\begin{align}
\mathcal{M}_s(\lambda_1,\lambda_2,\lambda_3,\lambda_4) &= -\frac{32\,(b+b')^2}{3 M_P^6} \mathcal{U}^2_{p_1,p_2}(\lambda_1,\lambda_2)^2\frac{1}{s}\,\mathcal{U}^2_{p_3,p_4}(\lambda_3,\lambda_4),\\
\mathcal{M}_t (\lambda_1,\lambda_2,\lambda_3,\lambda_4) &= -\frac{32\,(b+b')^2}{3M_P^6} \mathcal{U}^2_{-p_1,p_3}(\lambda_1,\lambda_3)\frac{1}{t}\,\mathcal{U}^2_{-p_2,p_4}(\lambda_2,\lambda_4),\\
\mathcal{M}_u (\lambda_1,\lambda_2,\lambda_3,\lambda_4) &=-\frac{32\,(b+b')^2}{3 M_P^6}  \mathcal{U}^2_{-p_1,p_4}(\lambda_1,\lambda_4)\frac{1}{u}\,\mathcal{U}^2_{-p_2,p_3}(\lambda_2,\lambda_3).
\end{align}
A comparison with \eqref{eq:channels_TJJ} shows the presence of a "double copy relation" for these amplitudes. The explicit results for the scattering amplitudes corresponding to various helicity configurations are collected in Table \ref{table:graviton_DC1}.

\begin{table}[t]
\centering
\small
\renewcommand{\arraystretch}{1.12}
\setlength{\tabcolsep}{4.2pt}
\resizebox{\textwidth}{!}{%
\begin{tabular}{cccc c c c c @{\hspace{1.0em}  \vrule width 0.75pt\hspace{1em} } cccc c c c c}
	\toprule
	$\lambda_1$&$\lambda_2$&$\lambda_3$&$\lambda_4$&
	$\mathcal{M}_s$&$\mathcal{M}_t$&$\mathcal{M}_u$&$\mathcal{M}_{\text{tot}}$&
	$\lambda_1$&$\lambda_2$&$\lambda_3$&$\lambda_4$&
	$\mathcal{M}_s$&$\mathcal{M}_t$&$\mathcal{M}_u$&$\mathcal{M}_{\text{tot}}$
	\\
	\midrule
	$+$&$+$&$+$&$+$&$-\frac{s}{4}$&$-\frac{t (t-u)^2}{4 s^2}$&$-\frac{u (t-u)^2}{4 s^2}$&$\frac{s^7+(t-u)^4 \left(t^3+u^3\right)}{16 s^4}$&
	$-$&$+$&$+$&$+$&$0$&$\frac{t u (t-u)}{2 s^2}$&$\frac{t u (u-t)}{2 s^2}$&$-\frac{t^2 u^2 (t-u)^2}{4 s^3}$ \\[0.3em]
	
	$+$&$+$&$+$&$-$&$0$&$\frac{t u (t-u)}{2 s^2}$&$\frac{t u (u-t)}{2 s^2}$&$-\frac{t^2 u^2 (t-u)^2}{4 s^3}$&
	$-$&$+$&$+$&$-$&$0$&$-\frac{t u^2}{s^2}$&$-\frac{u (t-u)^2}{4 s^2}$&$\frac{u^3 \left(16 t^3 u+(t-u)^4\right)}{16 s^4}$ \\[0.3em]
	
	$+$&$+$&$-$&$+$&$0$&$\frac{t u (t-u)}{2 s^2}$&$\frac{t u (u-t)}{2 s^2}$&$-\frac{t^2 u^2 (t-u)^2}{4 s^3}$&
	$-$&$+$&$-$&$+$&$0$&$-\frac{t (t-u)^2}{4 s^2}$&$-\frac{t^2 u}{s^2}$&$\frac{t^3 \left(16 t u^3+(t-u)^4\right)}{16 s^4}$ \\[0.3em]
	
	$+$&$+$&$-$&$-$&$-\frac{s}{4}$&$-\frac{t u^2}{s^2}$&$-\frac{t^2 u}{s^2}$&$\frac{s^3}{16}-\frac{t^3 u^3}{s^3}$&
	$-$&$+$&$-$&$-$&$0$&$\frac{t u (t-u)}{2 s^2}$&$\frac{t u (u-t)}{2 s^2}$&$-\frac{t^2 u^2 (t-u)^2}{4 s^3}$ \\[0.3em]
	
	$+$&$-$&$+$&$+$&$0$&$\frac{t u (t-u)}{2 s^2}$&$\frac{t u (u-t)}{2 s^2}$&$-\frac{t^2 u^2 (t-u)^2}{4 s^3}$&
	$-$&$-$&$+$&$+$&$-\frac{s}{4}$&$-\frac{t u^2}{s^2}$&$-\frac{t^2 u}{s^2}$&$\frac{s^3}{16}-\frac{t^3 u^3}{s^3}$ \\[0.3em]
	
	$+$&$-$&$+$&$-$&$0$&$-\frac{t (t-u)^2}{4 s^2}$&$-\frac{t^2 u}{s^2}$&$\frac{t^3 \left(16 t u^3+(t-u)^4\right)}{16 s^4}$&
	$-$&$-$&$+$&$-$&$0$&$\frac{t u (t-u)}{2 s^2}$&$\frac{t u (u-t)}{2 s^2}$&$-\frac{t^2 u^2 (t-u)^2}{4 s^3}$ \\[0.3em]
	
	$+$&$-$&$-$&$+$&$0$&$-\frac{t u^2}{s^2}$&$-\frac{u (t-u)^2}{4 s^2}$&$\frac{u^3 \left(16 t^3 u+(t-u)^4\right)}{16 s^4}$&
	$-$&$-$&$-$&$+$&$0$&$\frac{t u (t-u)}{2 s^2}$&$\frac{t u (u-t)}{2 s^2}$&$-\frac{t^2 u^2 (t-u)^2}{4 s^3}$ \\[0.3em]
	
	$+$&$-$&$-$&$-$&$0$&$\frac{t u (t-u)}{2 s^2}$&$\frac{t u (u-t)}{2 s^2}$&$-\frac{t^2 u^2 (t-u)^2}{4 s^3}$&
	$-$&$-$&$-$&$-$&$-\frac{s}{4}$&$-\frac{t (t-u)^2}{4 s^2}$&$-\frac{u (t-u)^2}{4 s^2}$&$\frac{s^7+(t-u)^4 \left(t^3+u^3\right)}{16 s^4}$ \\[0.3em]
	\bottomrule
	\end{tabular}}
	\caption{Four-particle configurations and corresponding scattering amplitude values for four gravitons. Here $\bar{\mathcal{M}}_s$, $\bar{\mathcal{M}}_t$, and $\bar{\mathcal{M}}_u$ represent the respective channels, and $\bar{\mathcal{M}}_{\text{tot}}$ is the total amplitude. As in Table \ref{tab:TJJ-TTJ} we have omitted the common factor $-32\,(b+b')^2/(3 M_P^6)$ for simplicity.}
	\label{table:graviton_DC1}
\end{table}

\section{Scalar emission in graviton-graviton scattering}

\begin{figure}[t]
\begin{center}
\includegraphics[scale=0.4,angle=0]{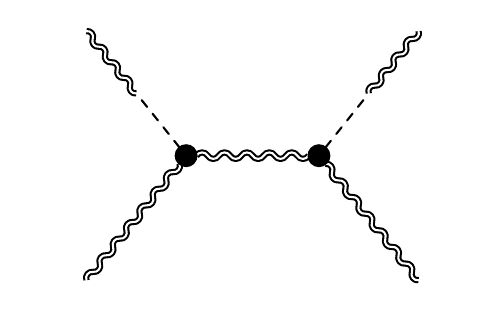}
\includegraphics[scale=0.4,angle=0]{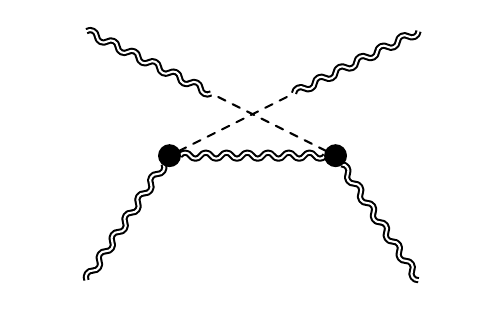}
\includegraphics[scale=0.4,angle=0]{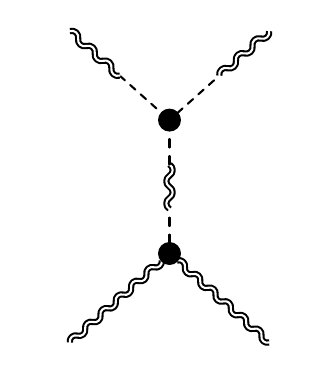}
\includegraphics[scale=0.4,angle=0]{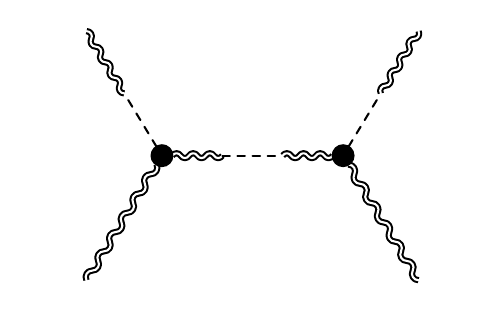}
\includegraphics[scale=0.4,angle=0]{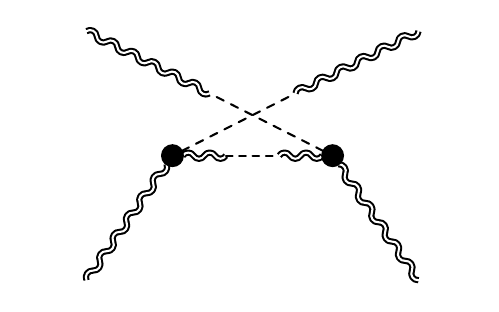}
\caption{ Scalar emission in graviton-graviton scattering with anomaly mediation.} 
\label{fig:scalar_emission}
\end{center}
\end{figure}

We now study the three channels for the process
\[
g_{\lambda_1}(p_1)\,g_{\lambda_2}(p_2)\to \phi(p_3)\,\phi(p_4),
\]
where \(\phi\) is the spin--0 component of the graviton (Figure \ref{fig:scalar_emission}). We define
\begin{equation}
\mathcal{A}_0(p_1,p_2) \equiv 4 \eta_{\mu_1\nu_1}\eta_{\mu_2\nu_2}\mathcal{A}_2^{\mu_1\nu_1\mu_2\nu_2}(p_1,p_2) = 16\,b'(p_1^2 p_2^2-p_1\cdot p_2)
\end{equation}
The polarization tensor can be decomposed in a spin-2 and spin-0 part as:
\begin{equation}
\varepsilon_{\mu\nu}(p) = \varepsilon^{(0)}_{\mu\nu}(p) + \varepsilon^{(2)})_{\mu\nu}(p)
\end{equation}
where:
\begin{align}
\varepsilon^{(2)}_{\mu\nu}(p) &= \Pi_{\mu\nu}^{\;\;\;\alpha\beta}(p)\epsilon_{\alpha\beta}(p) = \varepsilon_{\mu\nu}(p,\pm), \label{eq:epsilon2_def}\\
\varepsilon^{(0)}_{\mu\nu}(p) &= \frac{1}{3}\pi_{\mu\nu}(p).\label{eq:epsilon0_def}
\end{align}
We will use \eqref{eq:epsilon2_def} for the gravitons of helicities $\pm 2$ and \eqref{eq:epsilon0_def} for the scalars. Notice also that $\varepsilon^{(0)}$ satisfies:
\begin{equation} \label{eq:epsilon0_times_pi}
\pi^{\mu\nu}(p)\varepsilon^{(0)}_{\mu\nu}(p) = 1.
\end{equation}
The $s$-channel open indices amplitude for spin-0 polarization vector reads:
\begin{align}
\mathcal{M}_s^{\mu_1\nu_1\mu_2\nu_2\mu_3\nu_3\mu_4\nu_4} (p_1,p_2,p_3,p_4) = & \,\frac{1}{M_P^6}\left[\frac{1}{3}\,\mathcal{A}_2^{\mu_1\nu_1\mu_2\nu_2}(p_1,p_2)\pi^{\rho \sigma}(p_1+p_2)\right]D^{(0-s)}_{\rho\sigma\alpha\beta}(p)\nonumber\\
& \times \left[\frac{1}{27}\,\pi^{\alpha \beta}(p)\mathcal{A}_0(-p_3,-p_4)\pi^{\mu_3\nu_3}(-p_3)\pi^{\mu_4\nu_4}(-p_4)\right]
\end{align}
where the expression for the verteces contained inside the square brakets can be extrapolated from \eqref{fin1}. Notice that only the $D^{(0-s)}$ part of the propagator contributes, since
\begin{equation}
\pi^{\rho\sigma}(p)D^{(2)}_{\rho\sigma\alpha\beta}(p)=0.
\end{equation}
Using \eqref{eq:epsilon0_times_pi} and 
\begin{equation}
\pi^{\rho\sigma}(p)D^{(0-s)}_{\rho\sigma\alpha\beta}(p)\pi_{\alpha\beta}(p) = -\frac{3}{2}\frac{1}{p^2} 
\end{equation}
we can compute the scattering amplitude:
\begin{equation}
\mathcal{M}_s(\lambda_1,\lambda_2) \equiv \varepsilon^{(2)}_{\mu_1\nu_1}(p_1,\lambda_1) \varepsilon^{(2)}_{\mu_2\nu_2}(p_2,\lambda_2) \varepsilon^{(0)}_{\mu_3\nu_3}(p_3) \varepsilon^{(0)}_{\mu_4\nu_4}(p_4) \,\mathcal{M}_s^{\mu_1\nu_1\mu_2\nu_2\mu_3\nu_3\mu_4\nu_4} (p_1,p_2,p_3,p_4);
\end{equation}
for different polarizations. The result is:
\begin{equation}
\mathcal{M}_s(\lambda_1,\lambda_2)= \begin{dcases}
\;\frac{1}{27 M_P^6}\, b'(b'+b) \, s^3 \qquad &\text{if} \qquad \lambda_1 = \lambda_2,
\\
\;0 \qquad &\text{if} \qquad \lambda_1 \neq \lambda_2.
\end{dcases}\label{eq:amp_ggss_sChannel}
\end{equation}
In the $t$-channel both $D^{(0-s)}$ and $D^{(2)}$ contribute: 
\begin{align}
\mathcal{M}_t^{\mu_1\nu_1\mu_2\nu_2\mu_3\nu_3\mu_4\nu_4} (p_1,p_2,p_3,p_4) = & \frac{1}{M_P^6}\left[\frac{1}{3}\,\mathcal{A}_2^{\mu_1\nu_1\rho\sigma}(p_1,p_3-p_1)\pi^{\mu_3\nu_3}(p_3)\right]\Big(D^{(0-s)}_{\rho\sigma\alpha\beta}+D^{(2)}_{\rho\sigma\alpha\beta}\Big)(p_1-p_3)\nonumber\\
& \times \left[\frac{1}{3}\,\mathcal{A}_2^{\mu_2\nu_2\alpha\beta}(p_2,p_4-p_2)\pi^{\mu_4\nu_4}(-p_4)\right]
\end{align}
hence given the amplitude
\begin{align}
\mathcal{M}_t(\lambda_1,\lambda_2) \equiv \varepsilon^{(2)}_{\mu_1\nu_1}(p_1,\lambda_1) \varepsilon^{(1)}_{\mu_2\nu_2}(p_2,\lambda_2) \varepsilon^{(0)}_{\mu_3\nu_3}(p_3) \varepsilon^{(0)}_{\mu_4\nu_4}(p_4) \mathcal{M}_t^{\mu_1\nu_1\mu_2\nu_2\mu_3\nu_3\mu_4\nu_4} (p_1,p_2,p_3,p_4) \nonumber 
\end{align}
we can separate it into two components: the first one where the scattering is mediated by a spin-0 propagator and the second one where it is mediated by a spin-2 propagator \eqref{expansionX2}
\begin{equation}
\mathcal{M}_t = \mathcal{M}_t^{(0-s)} + \mathcal{M}_t^{(2)}.
\end{equation}
On shell one finds:
\begin{align}
\mathcal{M}_t^{(0-s)} &= 0\\
\mathcal{M}_t^{(2)}(\lambda_1,\lambda_2) &= \begin{dcases}
-\frac{(b+b')^2}{36} \, \left[t^3\left(1 + \frac{4\, u^2}{s^2}\right)^2\right]\qquad &\text{if} \qquad \lambda_1 = \lambda_2,\\
-\frac{4(b+b')^2}{9M_P^6}\, \left(t^3\,\frac{u^4}{s^4}\right)\qquad &\text{if} \qquad \lambda_1 \neq \lambda_2.
\end{dcases}\label{eq:amp_ggss_tChannel}
\end{align}
Following the same line of reasoning for the $u$-channel:
\begin{align}
\mathcal{M}_u^{\mu_1\nu_1\mu_2\nu_2\mu_3\nu_3\mu_4\nu_4} (p_1,p_2,p_3,p_4) = & \frac{1}{M_P^6}\left[\frac{1}{3}\mathcal{A}_2^{\mu_1\nu_1\rho\sigma}(p_1,p_4-p_1)\pi^{\mu_4\nu_4}(p_4)\right]\Big(D^{(0-s)}_{\rho\sigma\alpha\beta}+D^{(2)}_{\rho\sigma\alpha\beta}\Big)(p_1-p_4)\nonumber\\
& \times \left[\mathcal{A}_2^{\mu_2\nu_2\alpha\beta}(p_2,p_3-p_2)\pi^{\mu_4\nu_4}(-p_3)\right]
\end{align}
and separating the amplitude based on the spin of the propagator:
\begin{equation}
\mathcal{M}_u = \mathcal{M}_u^{(0-s)} + \mathcal{M}_u^{(2)}
\end{equation}
going on shell one finds:
\begin{align}
\mathcal{M}_u^{(0-s)} &= 0\\
\mathcal{M}_u^{(2)}(\lambda_1,\lambda_2) &= \begin{dcases}
- \frac{(b+b')^2}{36M_P^6} \, \left[u^3\left(1 + \frac{4\,t^2}{s^2}\right)^2\right]\qquad &\text{if} \qquad \lambda_1 = \lambda_2,\\
-\frac{4(b+b')^2}{9M_P^6}\, \left(u^3\,\frac{t^4}{s^4}\right)\qquad &\text{if} \qquad \lambda_1 \neq \lambda_2.
\end{dcases}\label{eq:amp_ggss_uChannel}
\end{align}
The result of this analysis is that the interaction manifests only as contact terms.
\subsection{Total amplitude cross section}
Using equations \eqref{eq:amp_ggss_sChannel},\eqref{eq:amp_ggss_tChannel}and \eqref{eq:amp_ggss_uChannel} we can compute the total scattering amplitude for equal and opposite polarizations
\begin{align}
\mathcal{M}_\text{eq} &\equiv  \mathcal{M}^{(2)}_t(\pm,\pm) + \mathcal{M}^{(2)}_u(\pm,\pm)= + \frac{4}{9 M_P^6}   \left[\frac{b'(b'+b)}{108}\,s^3 + (b+b')^2\left(\frac{t^3 u^3}{s^3} + \frac{t^2 u^2}{2s} -\frac{t^3 + u^3}{16}\right) \right], \\
\mathcal{M}_\text{opp} &\equiv \mathcal{M}^{(2)}_t(\pm,\mp) + \mathcal{M}^{(2)}_u(\pm,\mp) =  \frac{4(b+b')^2}{9M_P^6} \,\left(\frac{t^3 u^3}{s^3} \right),
\end{align}
and derive the differential cross section which is given by
\begin{equation}
\frac{d\sigma}{d\Omega} = \frac{1}{2}\frac{1}{64\pi^2 s}\,|\mathcal M|^2 = \frac{1}{128\pi^2 s}\left[|\mathcal M_\text{eq}|^2 + |\mathcal M_\text{opp}|^2 + 2\,\text{Re}(\mathcal M_\text{eq} \mathcal M_\text{opp}^*)\right]
\end{equation}
where the factor of $1/2$ accounts for the identical final states. Integrating over the solid angle we find the total cross sections
\begin{equation}
\sigma_\text{eq} = \frac{1}{64\pi s}\int_{-1}^{1} dx\,|\mathcal M_\text{eq}|^2 
\qquad \qquad 
\sigma_\text{opp} = \frac{1}{64\pi s}\int_{-1}^{1} dx\,|\mathcal M_\text{opp}|^2 \qquad \qquad 
\sigma_\text{int} = \frac{1}{64\pi s}\int_{-1}^{1} dx\,\text{Re}(\mathcal M_\text{eq} \mathcal M_\text{opp}^*),
\end{equation}
where we have defined $x \equiv \cos\theta$. Explicitly it is given by 
\begin{align}
\sigma_\text{eq} &= \frac{\tilde{C}^2 }{M_P^{12}}\,s^5 (b+b')^2 \left[c_{\text{eq}_1}b^2 + c_{\text{eq}_2} b\,b' + c_{\text{eq}_3} b'^2\right], \label{eq:tot_sigma_eq}\\
\sigma_\text{opp} &= \frac{\tilde{C}^2 }{M_P^{12}}\,s^5 (b+b')^4 c_{\text{opp}}, \label{eq:tot_sigma_opp}\\
\sigma_\text{int} &= \frac{\tilde{C}^2 }{M_P^{12}}\,s^5 (b+b')^3 \left[c_{\text{int}_1}b+c_{\text{int}_2}b'\right],\label{eq:tot_sigma_int}
\end{align}
where the numerical coefficients are
\begin{gather}
c_{\text{eq}_1} = \frac{2609}{138378240\pi}, \qquad
c_{\text{eq}_2} = \frac{739807}{16812956160 \pi}, \qquad
c_{\text{eq}_3} = \frac{7770803}{302633210880\pi}, \nonumber\\
c_{\text{opp}_1} = \frac{1}{1945944\pi},  \nonumber \\
c_{\text{int}_1} = \frac{1189}{467026560\pi}, \qquad
c_{\text{int}_2} = \frac{4139}{1401079680\pi}.
\end{gather}
Classical GR conventionally employs a linear representation of the excitations. The two are related by a simple change of basis
\beq
|+\rangle_{\rm lin}=\frac{|+2\rangle+|-2\rangle}{\sqrt2},
\qquad
|\times\rangle_{\rm lin}=\frac{|+2\rangle-|-2\rangle}{\sqrt2},
\eeq
and express the scattering amplitudes in the linear basis as
\beq
\mathcal M_{[+][+]}=\mathcal M_{\rm eq}+\mathcal M_{\rm opp},
\qquad
\mathcal M_{[\times][\times]}=\mathcal M_{\rm eq}-\mathcal M_{\rm opp},
\qquad
\mathcal M_{[+][\times]}=\mathcal M_{[\times][+]}=0.
\eeq
The corresponding cross sections are given by
\begin{align}\label{eq:sigma_plus_cross_def}
\sigma_{[+][+]}=\sigma_{\text{eq}}+\sigma_{\text{opp}}+\sigma_{\text{int}},\qquad \qquad \sigma_{[\times][\times]}=\sigma_{\text{eq}}+\sigma_{\text{opp}}-\sigma_{\text{int}}.
\end{align}
Substituting the expressions for the three total cross sections from eq.~\eqref{eq:tot_sigma_eq}-\eqref{eq:tot_sigma_int} we find
\begin{align}
\sigma_{[+][+]}&=\frac{\tilde{C}^2 }{M_P^{12}}\,s^5 (b+b')^2 \left[(c_{\text{eq}_1} +c_{\text{opp}} + c_{\text{int}_1})b^2 + (c_{\text{eq}_2} + 2c_{\text{opp}} + c_{\text{int}_1} + c_{\text{int}_2})bb' + (c_{\text{eq}_3} +c_{\text{opp}}+c_{\text{int}_2}){b'}^2\right],\\
\sigma_{[\times][\times]}&=\frac{\tilde{C}^2 }{M_P^{12}}\,s^5 (b+b')^2 \left[(c_{\text{eq}_1} +c_{\text{opp}} - c_{\text{int}_1})b^2 + (c_{\text{eq}_2} + 2c_{\text{opp}} - c_{\text{int}_1} - c_{\text{int}_2})bb' + (c_{\text{eq}_3} +c_{\text{opp}} - c_{\text{int}_2}){b'}^2\right],\\
\sigma_{[+][\times]}&=\sigma_{[\times][+]}=0.
\end{align}
The unpolarized cross section is obtained by averaging over all possible initial helicity configurations, and can be expressed in terms of the helicity cross sections as
\beq
\sigma_{\rm unpol}
=
\frac14\left(
\sigma_{++}+\sigma_{+-}+\sigma_{-+}+\sigma_{--}
\right)
=
\frac12(\sigma_{\rm eq}+\sigma_{\rm opp}).
\eeq
Using \eqref{eq:sigma_plus_cross_def} we can also express the result in terms of the linear polarization cross sections as
\begin{equation}
\sigma_{\rm unpol} = \frac{1}{4}\left(\sigma_{[+][+]}+\sigma_{[\times][\times]}\right).
\end{equation}
Substituting the explicit expressions either for \(\sigma_{\rm eq}\) and \(\sigma_{\rm opp}\) or for \(\sigma_{[+][+]}\) and \(\sigma_{[\times][\times]}\),  the unpolarized cross section takes the form
\begin{equation}
\sigma_{\rm unpol} = \frac{\tilde{C}^2 }{2 M_P^{12}}\,s^5 (b+b')^2 \left[(c_{\text{eq}_1} +c_{\text{opp}})b^2 + (c_{\text{eq}_2} + 2c_{\text{opp}})bb' + (c_{\text{eq}_3} +c_{\text{opp}}){b'}^2\right].
\end{equation}
The cross section has a behaviour of $\mathcal{O}(s^5/{M_P}^{12})$ and it is quartic in the multiplicites of the conformal states appearing in the virtual corrections.

\section{Anomaly correction to the eikonal phase}

In this section we analyze how anomaly-mediated corrections modify high-energy photon scattering in impact-parameter space. Our goal is to separate long-distance effects, controlled by non-analytic \(t\)-channel singularities, from short-distance contributions, which are analytic in \(t\) and therefore local in impact-parameter space. We focus on a helicity-preserving channel and suppress explicit helicity labels whenever no confusion can arise. We use the standard eikonal representation
\begin{equation}
\mathcal{M}(s,t)
=
2s\int d^2\mathbf{b}\;e^{-i\mathbf{q}\cdot\mathbf{b}}
\left(e^{i\chi(s,\mathbf{b})}-1\right),
\qquad
t=-\mathbf{q}^2,
\label{eq:eikonal-convention}
\end{equation}
which at leading eikonal order reduces to
\begin{equation}
e^{i\chi}-1\simeq i\chi,
\qquad
\chi(s,\mathbf{b})
=
\frac{1}{2si}\int\frac{d^2\mathbf{q}}{(2\pi)^2}\,
e^{i\mathbf{q}\cdot\mathbf{b}}\,
\mathcal{M}(s,-\mathbf{q}^2).
\label{eq:chi-linear}
\end{equation}
At tree level, the Einstein--Hilbert contribution is
\begin{equation}
\mathcal{M}^{(0)}(s,t,u)
=
\mathcal{M}^{(0)}_{s}
+\mathcal{M}^{(0)}_{t}
+\mathcal{M}^{(0)}_{u}.
\label{eq:tree-full-stu-clean}
\end{equation}
In the Regge/eikonal limit,
\begin{equation}
s\gg |t|,
\qquad
t=-\mathbf{q}^{2},
\label{eq:regge-limit}
\end{equation}
the only term that generates a long-range phase is the \(t\)-channel pole,
\begin{equation}
\mathcal{M}^{(0)}_{t}(s,t)\simeq i\,\frac{\kappa^2 s^2}{4t},
\label{eq:t-pole}
\end{equation}
while \(\mathcal{M}^{(0)}_{s}\), \(\mathcal{M}^{(0)}_{u}\), and the regular part of \(\mathcal{M}^{(0)}_{t}\) are analytic in \(t\), and therefore correspond to contact terms in \(\mathbf b\)-space. For \(b\neq 0\), one then finds
\begin{equation}
\chi^{(0)}(s,b)
=
\frac{1}{2si}\int\frac{d^2\mathbf{q}}{(2\pi)^2}\,
e^{i\mathbf{q}\cdot\mathbf{b}}
\left(i\,\frac{\kappa^2 s^2}{4t}\right)
=
-\frac{\kappa^2 s}{8}
\int\frac{d^2\mathbf{q}}{(2\pi)^2}\,
\frac{e^{i\mathbf{q}\cdot\mathbf{b}}}{\mathbf{q}^2}.
\label{eq:chi0-pre}
\end{equation}
Using
\begin{equation}
\int\frac{d^2\mathbf{q}}{(2\pi)^2}\,
\frac{e^{i\mathbf{q}\cdot\mathbf{b}}}{\mathbf{q}^2}
=
\frac{1}{2\pi}\log\frac{L}{b},
\label{eq:2d-log-ft}
\end{equation}
we obtain
\begin{equation}
\chi^{(0)}(s,b)
=
-\frac{\kappa^2 s}{16\pi}\log\frac{L^2}{b^2},
\qquad (b\neq0),
\label{eq:chi-tree-final-clean}
\end{equation}
up to the overall sign convention for \(\mathcal M\). The physical meaning of \(\chi(s,b)\) follows from the stationary-phase approximation applied to \eqref{eq:eikonal-convention}. The phase of the Fourier transform is
\begin{equation}
\Phi(\mathbf b)=\mathbf q\cdot \mathbf b+\chi(s,b),
\label{eq:phase-def}
\end{equation}
and the saddle-point condition reads
\begin{equation}
\nabla_b \Phi(\mathbf b)=0
\qquad\Longrightarrow\qquad
\mathbf q_\perp=-\nabla_b \chi(s,b).
\label{eq:saddle-condition}
\end{equation}
Thus the transverse momentum transfer is determined directly by the gradient of the eikonal phase. For two massless particles in the center-of-mass frame,
\begin{equation}
p=\frac{\sqrt{s}}{2},
\qquad
t=-2p^2(1-\cos\theta),
\label{eq:t-angle}
\end{equation}
and for small-angle scattering,
\begin{equation}
q_\perp^2=-t\simeq p^2\theta^2.
\label{eq:small-angle-relation}
\end{equation}
Therefore the deflection angle is
\begin{equation}
\theta(s,b)
=
-\frac{1}{p}\,\partial_b \delta(s,b),
\qquad
\delta(s,b)\equiv \Re \chi(s,b),
\label{eq:theta-from-eikonal}
\end{equation}
or, equivalently, in absolute value,
\begin{equation}
|\theta(s,b)|=\frac{1}{p}\,|\partial_b\delta(s,b)|.
\label{eq:theta-abs}
\end{equation}
Since in the present tree-level channel \(\chi^{(0)}(s,b)\) is real up to conventions, we may identify \(\delta=\chi^{(0)}\), and \eqref{eq:chi-tree-final-clean} gives
\begin{equation}
\theta^{(0)}(s,b)
=
-\frac{1}{p}\,\partial_b \chi^{(0)}(s,b)
=
\frac{\kappa^2\sqrt{s}}{4\pi b},
\label{eq:theta-tree}
\end{equation}
again up to the overall sign convention, while the magnitude of the deflection is unambiguous.

\subsection{Anomaly correction}

From the channel decomposition obtained in the previous section,
\begin{equation}
\mathcal{M}^{\mathrm{anom}}_{s}
=
-\frac{s}{4},
\qquad
\mathcal{M}^{\mathrm{anom}}_{t}
=
-\frac{t(t-u)^2}{4s^2},
\qquad
\mathcal{M}^{\mathrm{anom}}_{u}
=
-\frac{u(t-u)^2}{4s^2},
\end{equation}
hence
\begin{equation}
\mathcal{M}^{\mathrm{anom}}(s,t,u)
=
-\frac{tu}{s}.
\label{eq:anom-total-clean}
\end{equation}
Using \(u=-s-t\), this becomes
\begin{equation}
-\frac{tu}{s}
=
-\frac{t(-s-t)}{s}
=
t+\frac{t^2}{s}.
\label{eq:anom-expanded}
\end{equation}
With \(t=-\mathbf q^2\), we obtain
\begin{equation}
\mathcal{M}^{\mathrm{anom}}(s,-\mathbf q^2)
=
-\mathbf q^2+\frac{\mathbf q^4}{s}.
\label{eq:anom-qspace-clean}
\end{equation}
This expression is analytic at \(t=0\): in particular, it contains no \(1/t\) pole. Therefore it cannot generate the logarithmic long-distance eikonal phase associated with massless exchange. Let \(C_{\mathrm{anom}}\) denote the overall normalization of the anomaly amplitude. Then
\begin{equation}
\delta\chi^{\mathrm{anom}}(s,\mathbf b)
=
\frac{C_{\mathrm{anom}}}{2si}
\int\frac{d^2\mathbf q}{(2\pi)^2}\,
e^{i\mathbf q\cdot\mathbf b}
\left(-\mathbf q^2+\frac{\mathbf q^4}{s}\right).
\label{eq:delta-chi-anom-int-clean}
\end{equation}
Using
\begin{equation}
\int\frac{d^2\mathbf q}{(2\pi)^2}
e^{i\mathbf q\cdot\mathbf b}\,\mathbf q^2
=
-\nabla_b^2\delta^{(2)}(\mathbf b),
\qquad
\int\frac{d^2\mathbf q}{(2\pi)^2}
e^{i\mathbf q\cdot\mathbf b}\,\mathbf q^4
=
\nabla_b^4\delta^{(2)}(\mathbf b),
\label{eq:delta-identities}
\end{equation}
we find
\begin{equation}
\delta\chi^{\mathrm{anom}}(s,\mathbf b)
=
\frac{C_{\mathrm{anom}}}{2si}
\left[
\nabla_b^2\delta^{(2)}(\mathbf b)
+\frac{1}{s}\nabla_b^4\delta^{(2)}(\mathbf b)
\right].
\label{eq:delta-chi-anom-clean}
\end{equation}
Hence
\begin{equation}
\delta\chi^{\mathrm{anom}}(s,b)=0,
\qquad b\neq0.
\label{eq:delta-chi-zero}
\end{equation}
The total eikonal phase is therefore
\begin{equation}
\chi(s,\mathbf b)
=
\chi^{(0)}(s,\mathbf b)
+
\delta\chi^{\mathrm{anom}}(s,\mathbf b),
\end{equation}
and for nonzero impact parameter
\begin{equation}
\chi(s,b)
=
-\frac{\kappa^2 s}{16\pi}\log\frac{L^2}{b^2},
\qquad b\neq0.
\label{eq:chi-total-final-clean}
\end{equation}
It follows that the anomaly contribution does not modify the long-distance scattering angle:
\begin{equation}
\theta(s,b)=\theta^{(0)}(s,b),
\qquad b\neq0,
\label{eq:theta-no-anom}
\end{equation}
up to contact terms localized at \(\mathbf b=0\). In other words, the anomaly affects the eikonal phase only through ultra-local terms in impact-parameter space, while the classical deflection is entirely controlled by the non-analytic graviton pole.

\section{Extension to dark sectors}
In this section we comment on anomaly--mediated interactions between a visible and a dark conformal sector, extending the formal framework developed in the previous sections and emphasizing its dynamical implications. The technical structure of the relevant correlators and effective vertices has already been established; here we focus instead on how these ingredients operate when two otherwise decoupled sectors communicate through gravity and, more specifically, through the trace anomaly.

As a concrete example, consider a visible spin--1 conformal sector interacting with a virtual spin--1 field belonging to a corresponding dark one. Within the Einstein--Hilbert description, the interaction proceeds through metric exchange and mirrors structurally the photon--photon scattering process analyzed earlier. The dominant contribution arises from the $t$--channel diagram, leading schematically to
\begin{equation}
\mathcal{M}(s,t) \sim \kappa^2 \, T^{\mu\nu}_{\text{vis}} \,
\frac{P_{\mu\nu\rho\sigma}}{t} \,
T^{\rho\sigma}_{\text{dark}} \, ,
\label{eq:visible-dark-amplitude}
\end{equation}
where $P_{\mu\nu\rho\sigma}$ denotes the graviton propagator and $T^{\mu\nu}_{\text{vis}}$, $T^{\rho\sigma}_{\text{dark}}$ are the corresponding energy--momentum tensors. The coupling is thus entirely controlled by the stress--energy content of the two sectors and inherits the tensorial structure already discussed in the purely visible case.\\
The novelty emerges once quantum conformal breaking is included. Even if both sectors are classically scale invariant, the trace anomaly induces a nonvanishing trace of the energy--momentum tensor, with separate contributions from visible and dark sectors. Assuming free field theory realizations of the anomaly in both sectors, say $L$ and $R$, where $L$ is visible an $R$ is dark 
\be
\mathcal{A}_{L R}(x)=   b_{ L R}\, C^2 +
b_{L R}' \big(E - \tfrac{2}{3}\sq R\big) + b_{L R}'' \sq R .
\label{trace2}
\ee
the virtual corrections can be integrated out in both sectors, generating a nonlocal action which is, however, unique for both, since the total conformal anomaly is simply the sum of the anomalies of the two $L$ and $R$ sectors. 
\beq
\mathcal{A}(x)= \mathcal{A}_{L }(x) + \mathcal{A}_{R}(x).
\eeq
As a consequence, alongside the transverse--traceless spin--2 exchange, an effective scalar component of the metric contributes to the interaction. From the perspective of the visible sector, this scalar exchange activates channels that are absent at the classical level and whose structure is dictated by the anomalous form factors derived previously. The dark sector enters explicitly through its anomaly coefficients: if it contains multiple conformal species, their multiplicities enhance the trace anomaly and therefore the strength of the effective coupling. The hidden sector thus imprints its internal structure directly onto the amplitude through the anomaly.\\
In the high--energy limit, the eikonal treatment applies to this mixed visible--dark process as well. The resulting phase receives contributions from both the spin--2 and effective spin--0 exchanges, while contact terms ensure consistency with the Ward identities and with energy--momentum conservation. In this sense, conformal symmetry breaking is communicated between sectors by the scalar component of the gravitational field, which acts as the carrier of the anomalous trace.\\
\subsection{ IR effects from the $\Delta_4^{-1}$ interactions}
Within the minimal Einstein--Hilbert framework, however, these effects remain parametrically suppressed by the Planck scale. The interaction becomes relevant only in the regime where gravitational effects are unsuppressed, whereas at lower energies the exchanges are negligible.\\
A different situation arises when the interaction is described by the full anomaly--induced effective action rather than by the minimal Einstein--Hilbert term, as done in this work. In that case the dynamics is governed by a nonlocal functional which can be cast in local form through the introduction of the conformalon field \cite{Mottola:2016mpl}. This scalar resums the infrared contributions associated with the trace anomaly and encodes conformal symmetry breaking in a scale--independent manner.\\
In such a scenario, the breaking of conformal symmetry in the visible sector can be described in terms of the Riegert action, with coefficients determined entirely by the multiplicities, separately, of the hidden and the visible sectors. The resulting scalar exchange is no longer directly tied to Planck--suppressed couplings but instead controlled by the anomaly coefficients themselves. The interaction between visible and dark sectors thus acquires an intrinsically infrared character, driven by quantum conformal breaking rather than by ultraviolet gravitational exchange.\\
It should be noted, however, that the anomaly--induced action has known limitations. As emphasized in~\cite{Coriano:2022jkn} and \cite{Coriano:2021nvn}, for four--point functions the inclusion of Weyl--invariant terms is necessary in order to construct an effective action fully consistent with the hierarchical conformal Ward identities formulated around Minkowski space. \\ 
In contrast to the Einstein--Hilbert analysis performed here, the contributions emerging from the infrared structure of the Riegert action can become dominant in regimes of large curvature and are not suppressed by the Planck scale. 

For this reason, it is natural to ask whether such effects may become relevant in strongly curved backgrounds, for instance in astrophysical processes such as neutron--star mergers, where curvature scales are large and anomaly--induced contributions could, in principle, compete with or even dominate over the minimal gravitational exchange.

\section{Conclusions}

In this work we have analyzed the structure of anomaly--mediated gravitational interactions between conformal sectors, focusing on the scalar contribution associated with the trace anomaly and its interplay with the Einstein--Hilbert (EH) action. The analysis has been performed at the level of scattering amplitudes around flat spacetime, allowing a direct comparison between the standard perturbative description of gravity and the effective nonlocal dynamics induced by the anomaly.

The EH action provides the gravitational kernel connecting the two matter sectors. Expanding the metric in the de Donder gauge, the quadratic EH action determines the graviton propagator, whose covariant spin decomposition separates the transverse--traceless spin--2 component from the transverse scalar projector $P^{(0\text{-}s)}$. Although this scalar component does not correspond to an independent propagating mode in pure Einstein gravity, it becomes dynamically relevant when coupled to anomalous three--point correlators. In the perturbative framework considered here the corresponding interaction behaves effectively as a contact term, since the propagator structure $\sim 1/(q^2 M_P^2)$ leads to amplitudes scaling as $s/M_P^2$, in close analogy with Fermi-type effective interactions.

The four--point amplitudes are obtained by sewing two three--point correlators, such as $TJJ$ or $TTT$, through single--graviton exchange. In this construction the EH propagator interpolates between the two sectors, while the anomaly enters through the trace part of the vertices encoded in the transverse projector $\pi_{\mu\nu}$. The resulting amplitudes exhibit a clear separation between the standard spin--2 exchange and an anomaly--induced scalar contribution. This scalar channel is not introduced as an additional field but emerges from the covariant structure of the graviton propagator once the trace anomaly is present.

In the absence of conformal symmetry breaking such scalar contributions cancel or can be removed by improvement terms, but once the anomaly is included the trace component becomes unavoidable and the scalar channel acquires physical significance. In the eikonal regime this anomaly--mediated interaction modifies the gravitational phase through contact--type contributions in impact--parameter space, while the long--distance behaviour remains dominated by the spin--2 sector. \\
The EH action therefore plays a dual role: it governs the standard gravitational dynamics and provides the propagator structure through which the anomaly pole contained in conformal correlators manifests itself as an effective scalar interaction. In this way scalar gravitational effects arise within a purely metric framework, without introducing fundamental scalar fields. Unlike scenarios such as the Starobinsky model, where $R^2$ corrections can be rewritten as Einstein gravity plus a scalar degree of freedom, here the scalar dynamics originates from the quantum breaking of conformal symmetry.

A natural extension of this analysis consists in promoting the gravitational sector to the full nonlocal anomaly action, namely the Riegert effective action,
\begin{equation}
S_{\mathrm{g}} = S_{\mathrm{EH}} + S_{\mathrm{anom, vis + dark}} ,
\end{equation}
where $S_{\mathrm{anom, vis + dark}}$ is scale independent and resums the infrared effects associated with the trace anomaly. In this framework the scalar degree of freedom, often referred to as the conformalon, acquires its own nonlocal kinetic structure and the amplitudes incorporate the fully resummed infrared dynamics of conformal symmetry breaking.

In summary, anomaly--mediated scalar gravitational interactions can be consistently described within a perturbative Einstein--Hilbert framework, where graviton exchange interpolates between conformal sectors and the scalar channel emerges from the trace anomaly. Extending the analysis to the full nonlocal anomaly action will allow one to capture the complete infrared structure of anomaly--induced gravity and explore the phenomenological implications of such scalar gravitational modes.

		\begin{appendices}
\chapter{Appendix to Chapter 4}
\section{Slavnov-Taylor identities }
\label{STTI}
We denote with $S[V_{\underline{a}}^\mu,\psi,A_\mu]$ the action of the model. Its expression depends on the vielbein, the fermion field $\psi$ and the Abelian gauge field $A_\mu$. We can use this action and the vielbein to derive a useful form of the EMT
We introduce the generating functional of the model, given by
\beqa\label{Z}
Z[V,J^\mu,\c,\bar\c]
&=& \int \mD\bar{\psi}\mD\psi\mD A_\mu\,\mbox{exp}\bigg\{i S[V,\psi,A_\mu]  + i\int d^4x\, \bigg[\\
&+& \bar{\c(x)}\psi(x) + \bar{\psi}(x)\c(x) + J^\mu(x)A_\mu(x)\bigg]\bigg\}\, ,\eeqa
where we have denoted with $J(x)$, $J^\mu(x)$ and $\chi(x)$ the sources for the scalar, the gauge field and the spinor field respectively.
We will exploit the invariance of $Z$ under diffeomorphisms for the derivation of the corresponding Ward identities.
For this purpose we introduce a condensed notation to denote the functional integration measure of all the fields
\beqa
\label{SourcesToymodel}
\mD \Phi {\equiv} \mD\psi\mD\bar{\psi}\mD A_\mu\,
\eeqa
and redefine the action with the external sources included
\beqa
\label{ActionToyModel}
\tilde{S}&            =          & S + i\int d^4x\,\left( J^\mu A_\mu + \bar{\c}(x)\psi(x) + \textrm{h.c.}\right).\,
\eeqa
Notice that we have absorbed a factor $\sqrt{-g}$ in the definition of the sources, which clearly affects their transformation under changes of
coordinates. The transformations of the fields are given by (we have absorbed a factor $\sqrt{-g}$ in their definitions)
\beqa
V^{'\,\underline{a}}_\mu(x)           &=& V^{\underline{a}}_\mu(x) -\int d^4y\,[\delta^{(4)}(x-y)\partial_\nu V^{\underline{a}}_\mu(x)
+ [\partial_\mu\d^{(4)}(x-y)]V^{\underline{a}}_\nu]\e^\nu(y)\, ,\nn\\
\c'(x)                     &=& \c(x) -\int d^4y\,\partial_\nu[\d^{(4)}(x-y)\c(x)]\e^\nu(y)\, .
\eeqa
The term which appears in the first line in the integrand of Eq. (\ref{preWard}) can be re-expressed in the following form
\bea \label{WIfirstterm}
&& - \int d^4 x \, V\Theta^\mu_{\,\,\,\underline{a}}\bigg[-\d^{(4)}(x-y)\partial_\nu V^{\underline{a}}_\mu(x) - [\partial_\mu\d^{(4)}(x-y)]V^{\underline{a}}_\nu\bigg] = - V\Theta^\mu_{\,\,\,\nu\,;\mu} + V\Theta^\mu_{\,\,\,\underline{a}}V^{\underline{a}}_{\mu\, ;\nu} \nn \\
&& =  - V\bigg[\Theta^\mu_{\,\,\,\nu\,;\mu} +  V_{\underline{a}\r} V^{\underline{a}}_{\mu\,;\nu}\frac{\Theta^{\mu\r} - \Theta^{\r\mu}}{2}\bigg] \,,
\eea
where in the last expression we used the covariant conservation of the metric tensor expressed in terms of the vierbein
\bea
g_{\mu\nu\,;\r} = 0 \Rightarrow V^{\underline{a}}_{\mu\,;\r}V_{\underline{a}\nu} = - V^{\underline{a}}_\mu V_{\underline{a}\nu\,;\r} = - V_{\underline{a}\mu} V^{\underline{a}}_{\nu\,;\r}.
\eea
Other simplifications are obtained using the invariance of the action under local Lorentz transformations 
parameterized as
\beqa
\d V^{\underline{a}}_\mu = {\w^{\underline{a}}}_{\underline{b}} V^{\underline{b}}_\mu\, , \qquad
\d\psi     = \frac{1}{2}\s^{a b}\w_{{\underline{a}\underline{b}}}\psi\, , \qquad
\d\bar{\psi} = -\frac{1}{2}\bar\psi\s^{\underline{a} \underline{b}}\w_{\underline{a b}}\, ,
\eeqa
that gives, using the antisymmetry of $\w^{\underline{ a b}}$
\beq\label{vincolofermioni}
\frac{\d S}{\d\psi}\s^{\underline{a b}}\psi - \bar\psi\s^{\underline{a b}}\frac{\d S}{\d\bar{\psi}}
- \frac{\d S}{\d V^{\underline{b}}_\mu}V^{\underline{a}}_\mu + \frac{\d S}{\d V^{\underline{b}}_\mu} V^{\underline{a}}_\mu = 0\, .
\eeq
The previous equation can be reformulated in terms of the energy-momentum tensor $\Theta^{\mu\nu}$
\beq
V(\Theta^{\mu\r} - \Theta^{\r\mu}) = \bar{\psi}\s^{\mu\r}\frac{\d S}{\d\bar{\psi}}
- \frac{\d S}{\d\psi}\s^{\mu\r}\psi\,,
\eeq
which is useful to re-express Eq. (\ref{WIfirstterm}) in terms of the symmetric energy-momentum tensor $T^{\mu\nu}$ and to obtain finally, in the flat space-time limit, Eq. (\ref{Ward}).

\section{Feynman rules}
\label{rules}
%------------------
The Feynman rules used throughout the paper are collected here
\begin{itemize}
	\item {Graviton - fermion - fermion vertex}
	\\
	%%%%%%%%%%%
	\\
	%\begin{align}
	\bmi{95pt}
	\includegraphics[scale=1.0]{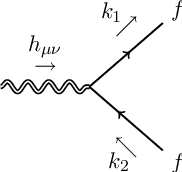}
	\emi
	\bmi{70pt}
	\bann
	&=& - i \, \frac{\kappa}{2} \, V^{\prime}_{\mu\nu}(k_1,k_2) \nn \\
	&=& - i \, \frac{\kappa}{2} \, \left\{ \frac{1}{4} \left[\gamma_\mu (k_1 + k_2)_\nu
	+\gamma_\nu (k_1 + k_2)_\mu \right] - \frac{1}{2} g_{\mu \nu}
	[\gamma^{\lambda}(k_1 + k_2)_{\lambda} - 2 m]  \right\} \nn \\
	\eann
	\emi
	\bea
	\label{VGff}
	\eea
	%\end{align}
	\item{Graviton - gluon - gluon vertex}
	\\ \\
	\bmi{110pt}
	\includegraphics[scale=1.0]{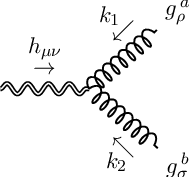}
	\emi
	\bmi{100pt}
	\bann
	&=& - i \, \frac{\kappa}{2} \, \delta_{a b} \, V^{Ggg}_{\mu\nu\rho\sigma}(k_1,k_2) \nn \\
	&= &- i \, \frac{\kappa}{2} \, \delta_{a b} \left\{ k_1\cdot k_2 \, C_{\mu\nu\rho\si} + D_{\mu\nu\rho\si}(k_1,k_2) + \frac{1}{\xi} \, E_{\mu\nu\rho\si}(k_1,k_2)  \right\}
	\eann
	\emi
	\bea
	\eea
	\item{Graviton - ghost - ghost vertex}
	\\ \\
	\bmi{110pt}
	\includegraphics[scale=1.0]{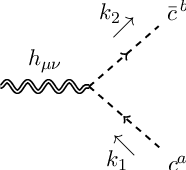}
	\emi
	\bmi{70pt}
	\bann
	& = & - i \,  \frac{\kappa}{2}\,  \delta^{a b} \, C_{\mu\nu\rho\sigma} \, k_{1\,\rho} \, k_{2\,\sigma}
	\eann
	\emi
	\bea
	\eea
	%
	%%%%%%%%%%%%%%%%%%%%%
	%
	\item{Graviton - fermion - fermion - gauge boson vertex}
	\\ \\
	\bmi{110pt}
	\includegraphics[scale=1.0]{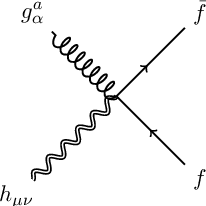}
	\emi
	\bmi{70pt}
	\bann
	&=& i g \, \frac{\kappa}{2} \,T^a\, W^{\prime}_{\mu\nu\alpha}
	= i g \, \frac{\kappa}{2} \, T^a \left\{ -\frac{1}{2} (\gamma_\mu \, g_{\nu\alpha}
	+\gamma_\nu \, g_{\mu\alpha}) +  g_{\mu \nu} \, \gamma_{\alpha} \right \}
	\eann
	\emi
	\bea
	\label{WGffg}
	\eea
	%
	%%%%%%%%%%%%%%%%%%%%%%%%%%
	%
	%%%%%%%%%%%%%%%%%%%%%%%%%%
	%
	\item{Graviton - gluon - gluon - gluon  vertex}
	\\ \\
	\bmi{110pt}
	\includegraphics[scale=1.0]{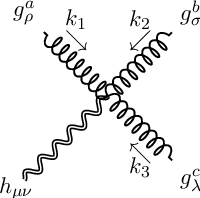}
	\emi
	\bmi{110pt}
	\begin{eqnarray}
		&=&- g \frac{\kappa}{2} f^{a b c} V^{Gggg}_{\mu\nu\rho\sigma\lambda}(k_1,k_2,k_3) \nn \\
		&=&  - g \frac{\kappa}{2} f^{a b c} \left\{ C_{\mu\nu\rho\sigma}(k_1-k_2)_{\lambda} + C_{\mu\nu\rho\lambda}(k_3-k_1)_{\sigma}   \right. \nn \\
		&& \hspace{2.5cm}  + \left.  C_{\mu\nu\sigma\lambda}(k_2-k_3)_{\rho} + F_{\mu\nu\rho\sigma\lambda}(k_1,k_2,k_3)  \right\}
		\hspace{1.7cm}
	\end{eqnarray}
	\emi
	
	%
	%%%%%%%%%%%%%%%%%%%%%%%%%%
	%%%%%%%%%%%%%%%%%%%%%%%%%%
	%
	\item{Graviton - ghost - ghost - gauge boson vertex}
	\\ \\
	\bmi{110pt}
	\includegraphics[scale=1.0]{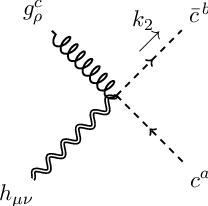}
	\emi
	\bmi{70pt}
	\bann
	&=&- \frac{\kappa}{2} \, g \, f^{a b c} \, C_{\mu\nu\rho\sigma} \, k_{2}^{\sigma}
	\eann
	\emi
	\bea
	\eea
	%
	%%%%%%%%%%%%%%%%%%%%%%%%%%
	%
	\bea
	&& C_{\mu\nu\rho\sigma} = g_{\mu\rho}\, g_{\nu\sigma}
	+g_{\mu\sigma} \, g_{\nu\rho}
	-g_{\mu\nu} \, g_{\rho\sigma}\,
	\\
	&& D_{\mu\nu\rho\sigma} (k_1, k_2) =
	g_{\mu\nu} \, k_{1 \, \sigma}\, k_{2 \, \rho}
	- \biggl[g^{\mu\sigma} k_1^{\nu} k_2^{\rho}
	+ g_{\mu\rho} \, k_{1 \, \sigma} \, k_{2 \, \nu}
	- g_{\rho\sigma} \, k_{1 \, \mu} \, k_{2 \, \nu}
	+ (\mu\leftrightarrow\nu)\biggr]\, \\
	&& E_{\mu\nu\rho\sigma} (k_1, k_2) = g_{\mu\nu} \, (k_{1 \, \rho} \, k_{1 \, \sigma}
	+k_{2 \, \rho} \, k_{2 \, \sigma} +k_{1 \, \rho} \, k_{2 \, \sigma})
	-\biggl[g_{\nu\sigma} \, k_{1 \, \mu} \, k_{1 \, \rho}
	+g_{\nu\rho} \, k_{2 \, \mu} \, k_{2 \, \sigma}
	+(\mu\leftrightarrow\nu)\biggr]\ , \nn  \\ \\
	&& F_{\mu\nu\rho\sigma\lambda} (k_1,k_2,k_3) =
	g_{\mu\rho} \,  g_{\sigma\lambda} \, (k_2-k_3)_{\nu}
	+g_{\mu\sigma} \, g_{\rho\lambda} \, (k_3-k_1)_{\nu}
	+g_{\mu\lambda} \, g_{\rho\sigma}(k_1-k_2)_{\nu}
	+ (\mu\leftrightarrow\nu) \nn \\
	\eea
\end{itemize}
	%-------------------------
	
	\section{The perturbative expansion and the $TJJ$: the quark sector}
	\label{six}
	
	We take the external momenta as incoming. We introduce the tensor components
	\begin{align}
		A^{\m_1\n_1\m\n}&\equiv\h^{\mu_1\nu_1}\h^{\m\n}- \big(\h^{\m\m_1}\h^{\n_1\n}+\h^{\m\n_1}\h^{\m_1\n}\big)
	\end{align}
	and the vertices in the fermion sector are
	\begin{align}
		V^{\m\, a}_{J\psi\bar\psi}(k_1,k_2) &=-ie\,\g^\m T^a\\
		V^{\m_1\n_1\, a}_{T\psi\bar\psi}(p_1k_1,k_2) &=-\sdfrac{i}{4}\,A^{\m_1\n_1\m\n}\,\g_\n T^a\,(k_1+k_2)_\m\\
		V^{\m_1\n_1\m_2\, a}_{TJ\psi\bar\psi}(k_1,k_2) &=\frac{i\,e}{2}A^{\m_1\n_1\m_2\n}\,\g_\m T^a.
	\end{align}
	giving
	\begin{eqnarray}
		i\Gamma^{\m_1\n_1\m_2\m_3 a b}(p_2,p_3)&\equiv&\braket{T^{\m_1\n_1}(p_1)\,J^{\m_2 a}(p_2)\,J^{\m_3 b}(p_3)}_F\nonumber \\
		&&=2\,\bigg(\sum_{i=1}^2V_{F,i}^{\m_1\n_1\m_2\m_3 a b }(p_1,p_2,p_3)+\sum_{i=1}^{2}W_{F,i}^{\m_1\n_1\m_2\m_3 a b }(p_1,p_2,p_3)\bigg)
	\end{eqnarray}
	where the $V_{F,i}$ terms are related to the triangle topology contributions, while the $W_{F,i}$ terms denote the two bubble contributions. \\
	In the quark sectors the perturbative contributions are
	{\smaller\begin{allowdisplaybreaks}
		\begin{align}
			V_{F,1}^{\m_1\n_1\m_2\m_3 a b}&=-i^3\delta^{ab}\,\int\frac{d^d\ell}{(2\p)^d}\frac{\Tr\left[V^{\m_1\n_1}_{T\psi\bar\psi}(\ell-p_2,\ell+p_3)\left(\slashed{\ell}+\slashed{p}_3\right)V^{\m_2}_{J\psi\bar\psi}(\ell,\ell-p_2)\,\slashed{\ell}\,V^{\m_3}_{J\psi\bar\psi}(\ell,\ell+p_3)\,\left(\slashed{\ell}-\slashed{p}_2\right)\right]}{\ell^2\,(\ell-p_2)^2(\ell+p_3)^2}\\
			V_{F,2}^{\m_1\n_1\m_2\m_3 a b }&=-i^3\delta^{ab}\,\int\frac{d^d\ell}{(2\p)^d}\frac{\Tr\left[V^{\m_1\n_1}_{T\psi\bar\psi}(\ell-p_3,\ell+p_2)\left(\slashed{\ell}+\slashed{p}_2\right)V^{\m_2}_{J\psi\bar\psi}(\ell,\ell-p_3)\,\slashed{\ell}\,V^{\m_3}_{J\psi\bar\psi}(\ell,\ell+p_2)\,\left(\slashed{\ell}-\slashed{p}_3\right)	\right]}{\ell^2\,(\ell-p_3)^2(\ell+p_2)^2}\\
			W_{F,3}^{\m_1\n_1\m_2\m_3 a b }&=-i^2\delta^{ab}\,\int\frac{d^d\ell}{(2\p)^d}\frac{\Tr\left[V^{\m_1\n_1\m_2}_{TJ\psi\bar\psi}(\ell+p_3,\ell)\left(\slashed{\ell}+\slashed{p}_3\right)V^{\m_3}_{J\psi\bar\psi}(\ell,\ell+p_3)\,\slashed{\ell}\right]}{\ell^2\,(\ell+p_3)^2}\\[1.5ex]
			W_{F,2}^{\m_1\n_1\m_2\m_3 a b}&=-i^2\delta^{ab}\,\int\frac{d^d\ell}{(2\p)^d}\frac{\Tr\left[V^{\m_1\n_1\m_3}_{TJ\psi\bar\psi}(\ell+p_2,\ell)\left(\slashed{\ell}+\slashed{p}_2\right)V^{\m_2}_{J\psi\bar\psi}(\ell,\ell+p_2)\,\slashed{\ell}\right]}{\ell^2\,(\ell+p_2)^2}
		\end{align}
	\end{allowdisplaybreaks}}

The evaluation of these one-loop integrals is rather straightforward. We use the transverse traceless $\Pi$ and the transverse ones $\pi$ in order to decompose the result of the computation in the longitudinal transverse parameterization. This procedure 
generates all the form factors appearing in Section 7. 

\section{The TJJJJ hierarchy from diffeomorphism invariance at $O(g^4)$ in the quark sector}
\label{t4j}
The conservation equation of the TJJJJ takes the form  
{\allowdisplaybreaks\small
\begin{equation}
	\begin{aligned}
		0=&q_{\mu } \langle T^{\mu\nu}(q)J^{\alpha a_1}(p_1)J^{\beta a_2}(p_2)J^{\gamma a_3}(p_3)J^{\delta a_4}(p_4)\rangle_q+2
		\delta^{\alpha}_{[\mu}{p_1}_{\nu]}\langle J^{\mu a_1}(p_1-q)J^{\beta a_2}(p_2)J^{\gamma a_3}(p_3)J^{\delta a_4}(p_4) \rangle_q\\&
		+2
		\delta^{\beta}_{[\mu}{p_2}_{\nu]}\langle J^{\alpha a_1}(p_1)J^{\mu a_2}(p_2-q)J^{\gamma a_3}(p_3)J^{\delta a_4}(p_4) \rangle_q
		+2
		\delta^{\gamma}_{[\mu}{p_3}_{\nu]}\langle J^{\alpha a_1}(p_1)J^{\beta a_2}(p_2)J^{\mu a_3}(p_3-q) J^{\delta a_4}(p_4)\rangle_q\\&
		+2
		\delta^{\delta}_{[\mu}{p_4}_{\nu]}\langle J^{\alpha a_1}(p_1)J^{\beta a_2}(p_2)J^{\gamma a_3}(p_3) J^{\mu a_4}(p_4-q)\rangle_q
		-2i gf^{a_1 a_2 c} \delta^{[\alpha}_\nu \langle J^{\beta] c}(p_1+p_2-q)J^{\gamma a_3}(p_3)J^{\delta a_4}(p_4)\rangle_q
		\\&
		-2i gf^{a_1 a_3 c} \delta^{[\alpha}_\nu \langle J^{\gamma] c}(p_1+p_3-q)J^{\beta a_2}(p_2)J^{\delta a_4}(p_4)\rangle_q
		-2i gf^{a_3 a_2 c} \delta^{[\gamma}_\nu \langle J^{\beta] c}(p_3+p_2-q)J^{\alpha a_1}(p_1)J^{\delta a_4}(p_4)\rangle_q\\&
		-2i gf^{a_1 a_4 c} \delta^{[\alpha}_\nu \langle J^{\delta] c}(p_1+p_4-q)J^{\gamma a_3}(p_3)J^{\beta a_2}(p_2)\rangle-2i gf^{a_4 a_2 c} \delta^{[\delta}_\nu \langle J^{\beta] c}(p_4+p_2-q)J^{\gamma a_3}(p_3)J^{\alpha a_1}(p_1)\rangle_q
		\\&
		-2i gf^{a_1 a_4 c} \delta^{[\alpha}_\nu \langle J^{\delta] c}(p_1+p_4-q)J^{\beta a_2}(p_2)J^{\gamma a_3}(p_3)\rangle_q -2i gf^{a_4 a_3 c} \delta^{[\delta}_\nu \langle J^{\gamma] c}(p_4+p_3-q)J^{\beta a_2}(p_2)J^{\alpha a_1}(p_1)\rangle_q		\\&-2i gf^{a_3 a_4 c} \delta^{[\gamma}_\nu \langle J^{\delta] c}(p_3+p_4-q)J^{\alpha a_1}(p_1)J^{\beta a_2}(p_2)\rangle_q -2i gf^{a_4 a_2 c} \delta^{[\delta}_\nu \langle J^{\beta] c}(p_4+p_2-q)J^{\alpha a_1}(p_1)J^{\gamma a_3}(p_3)\rangle_q.
	\end{aligned}
\end{equation}}

\section{Secondary equations in the quark sector} 
\label{secc}
The secondary conformal Ward identities are first-order partial differential equations and, in principle, involve the semi-local information contained in $j_{\text{loc}}^\mu$ and $t^{\mu\nu}_{\text{loc}}$. To write them compactly, we define two differential operators:
\begin{align}
	L_N &= p_1(p_1^2 + p_2^2 - p_3^2) \frac{\partial}{\partial p_1} + 2 p_1^2\, p_2 \frac{\partial}{\partial p_2} + \big[ (2d - \Delta_1 - 2\Delta_2 +N)p_1^2 + (2\Delta_1-d)(p_3^2-p_2^2) \big], \label{Ldef} \\
	R &= p_1 \frac{\partial}{\partial p_1} - (2\Delta_1-d) \label{Rdef}\,. 
\end{align}
The reason for introducing these operators arises from the special conformal constraints, once the action of $K^\kappa$ is made explicit. The separation between the two sets of constraints comes from the same equation, particularly from the terms trilinear in the momenta within the square bracket. The symmetric versions of these operators are given by:
\begin{align}
	&L'_N=L_N,\quad\text{with}\ p_1\leftrightarrow p_2\ \text{and}\ \Delta_1\leftrightarrow\Delta_2,\\
	&R'=R,\qquad\text{with}\ p_1\mapsto p_2\ \text{and}\ \Delta_1\mapsto\Delta_2.
\end{align}

These operators depend on the conformal dimensions of the operators involved in the 3-point function under consideration, as well as on a single parameter \(N\), which is determined by the relevant Ward identity. In the case of the \(\braket{TJJ}\) correlator, the special conformal Ward identities (CWIs) map the transverse traceless sector onto itself. Projectors within this sector can be utilized to decompose the equations into primary and secondary components
{\allowdisplaybreaks
\begin{align}
	&\Pi^{\rho_1\sigma_1}_{\mu_1\nu_1}(p_1)\pi^{\rho_2}_{\mu_2}(p_2)\pi^{\rho_3}_{\mu_3}(p_3)\ \bigg[\,K^\kappa\,\braket{{t^{\mu_1\nu_1}(p_1)\,j^{\mu_2}(p_2)\,j^{\mu_3}(p_3)}}\bigg]\notag\\
	&=\Pi^{\rho_1\sigma_1}_{\mu_1\nu_1}(p_1)\pi^{\rho_2}_{\mu_2}(p_2)\pi^{\rho_3}_{\mu_3}(p_3)\times\notag\\
	&\times\bigg[p_1^\kappa\left(C_{11}\,p_1^{\mu_3}p_2^{\mu_1}p_2^{\nu_1}p_3^{\mu_2}+C_{12}\,\delta^{\mu_2\mu_3}p_2^{\mu_1}p_2^{\nu_1}+C_{13}\delta^{\mu_1\mu_2}p_2^{\nu_1}p_1^{\mu_3}+C_{14}\delta^{\mu_1\mu_3}p_2^{\nu_1}p_3^{\mu_2}+C_{15}\delta^{\mu_1\mu_2}\delta^{\nu_1\mu_3}\right)\notag\\
	&\quad+p_2^\kappa\left(C_{21}\,p_1^{\mu_3}p_2^{\mu_1}p_2^{\nu_1}p_3^{\mu_2}+C_{22}\,\delta^{\mu_2\mu_3}p_2^{\mu_1}p_2^{\nu_1}+C_{23}\delta^{\mu_1\mu_2}p_2^{\nu_1}p_1^{\mu_3}+C_{24}\delta^{\mu_1\mu_3}p_2^{\nu_1}p_3^{\mu_2}+C_{25}\delta^{\mu_1\mu_2}\delta^{\nu_1\mu_3}\right)\notag\\	
	&\quad+\delta^{\mu_1\k}\left(C_{31}\,p_1^{\mu_3}p_2^{\nu_1}p_3^{\mu_2}+C_{32}\,\delta^{\mu_2\mu_3}p_2^{\nu_1}+C_{33}\,\delta^{\mu_2\nu_1}p_1^{\mu_3}+C_{34}\,\delta^{\mu_3\nu_1}p_3^{\mu_2}\right)\notag\\
	&\qquad+\delta^{\mu_2\k}\left(C_{41}\,p_1^{\mu_3}p_2^{\mu_1}p_2^{\nu_1}+C_{42}\,\delta^{\mu_1\mu_3}p_2^{\nu_1}\right)+\delta^{\mu_3\k}\left(C_{51}\,p_3^{\mu_2}p_2^{\nu_1}p_3^{\mu_2}+C_{52}\,\delta^{\mu_1\mu_2}p_2^{\nu_1}\right)\bigg]\label{StrucSWIS}.
\end{align}}
At this stage, the \(C_{ij}\) represent differential equations governing the form factors \(A_1\), \(A_2\), \(A_3\), and \(A_4\) in the decomposition of the \(\langle t jj \rangle\) correlator, as described in \eqref{DecompTJJ}. These equations naturally divide into two distinct categories: the $primary$ and $secondary$ conformal Ward identities (CWIs). \\
The primary CWIs are characterized by second-order differential equations, arising as the coefficients of transverse or transverse-traceless tensors involving the momenta \(p_1^\kappa\) and \(p_2^\kappa\), where \(\kappa\) corresponds to the index of the conformal generator \(K^\kappa\). \\
In contrast, the secondary CWIs, derived from the remaining transverse or transverse-traceless components, are first-order differential equations. These secondary equations are associated with the following operators
{\allowdisplaybreaks
\begin{equation}
	\begin{split}
		C_{31}&=-\frac{2}{p_1^2}\left[L_4 A_1+R A_3-R A_3(p_2\leftrightarrow p_3)\right],\\
		C_{32}&=-\frac{2}{p_1^2}\left[L_2\,A_2-p_1^2(A_3-A_3(p_2\leftrightarrow p_3))\right],\\
		C_{33}&=-\frac{1}{p_1^2}\left[L_4\,A_3-2R\,A_4\right],\\
		C_{34}&=-\frac{1}{p_1^2}\left[L_4\,A_3(p_2\leftrightarrow p_3)+2R\,A_4-4p_1^2A_3(p_2\leftrightarrow p_3)\right],
	\end{split}
\end{equation}
\begin{equation}
	\begin{split}
		C_{41}&=\frac{1}{p_2^2}\left[L'_3\,A_1-2R'A_2+2R'A_3\right],\\
		C_{42}&=\frac{1}{p_2^2}\left[L'_1\,A_3(p_2\leftrightarrow p_3)+p_2^2(4A_2-2A_3)+2R'A_4\right],\\
		C_{51}&=\frac{1}{p_3}\left[(L_4-L'_3)A_1-2(2d+R+R')A_2+2(2d+R+R')A_3(p_2\leftrightarrow p_3)\right],\\
		C_{52}&=\frac{1}{p_3^2}\left[(L_2-L'_1)A_3-4p_3^2A_2+2p_3^2A_3(p_2\leftrightarrow p_3)+2(2d-2+R+R')A_4\right].
	\end{split}
\end{equation}
while the remaining $C_{ij}$  are primary and generate the equations in \eqref{fst1}.
The secondary CWIs take the explicit form
\begin{equation}
	\begin{split}
		&C_{31}=C_{41}=C_{42}=C_{51}=C_{52}=0, \qquad C_{32}=\frac{16\,d\, c_{123}\,\Gamma_J}{p_1^2}\left[ \frac{1}{(p_3^2)^{\sigma_0}} - \frac{1}{(p_2^2)^{\sigma_0}} \right],\\[2ex]
		& \hspace{1cm}C_{33}=\frac{16\,d\,c_{123}\,\Gamma_J}{p_1^2 (p_3^2)^{\sigma_0}}, \hspace{2.5cm} C_{34}=-\frac{16\,d\,c_{123}\,\Gamma_J }{p_1^2\,(p_2^2)^{\sigma_0}},
	\end{split}
\end{equation}
where $\sigma_0=d/2-\Delta_2$. Explicitly 
\begin{equation}
	\begin{split}
		&L_4 A_1+R A_3^R-R A_3^R(p_2\leftrightarrow p_3)=0\\[1.5ex]
		&L_2\,A_2^R-s\,(A_3^R-A_3^R(p_2\leftrightarrow p_3))=\sdfrac{16}{9}\pi^2e^2\left[3s_1\,B_0^R(s_1,0,0)-3s_2B_0^R(s_2,0,0)-s_1+s_2\right]+\sdfrac{24}{9}\pi^2 g^2s\\[1.5ex]
		&L_4\,A^R_3-2R\,A^R_4=\sdfrac{32}{9}\pi^2\,g^2s_2\left[1-3B_0^R(s_2,0,0)\right]+\sdfrac{48}{9}\pi^2\,g^2\,s\\[1.5ex]
		&L_4\,A^R_3(p_2\leftrightarrow p_3)+2R\,A^R_4-4\,sA^R_3(p_2\leftrightarrow p_3)=\sdfrac{32}{9}\,\pi^2\,g^2s_1\left[3B_0^R(s_1,0,0)-1\right]\\[1.5ex]
		&L'_3\,A_1^R-2R'A_2^R+2R'A_3^R=0\\[1.5ex]
		&L'_1\,A_3^R(p_2\leftrightarrow p_3)+p_2^2(4A^R_2-2A^R_3)+2R'A^R_4=\sdfrac{16}{3}\pi^2\,g^2\,s_1,\\
	\end{split}
\end{equation}}
where we have used the relations 
\begin{equation}
	\sdfrac{\partial}{\partial s_i}\,B_0^R(s_j,0,0)=-\sdfrac{\d_{ij}}{s_i}\  i=0,1,2
\end{equation}
where $s_0=s$. 
\section{Relations between master integrals and 3K integrals} 
\label{dd}
Master integrals and 3K integrals are linked by the relation
\beq
\label{davy}
J(\nu_1,\nu_2,\nu_3) = \int \frac{d^d l}{(2 \pi)^d} \frac{1}{(l^2)^{\nu_3} ((l+p_1)^2)^{\nu_2} ((l-p_2)^2)^{\nu_1}}\, ,
\eeq
\beqa
J(\nu_1,\nu_2,\nu_3)
&=&\frac{\pi^{-d/2} 2^{4 - 3 d/2} }{\Gamma(\nu_1)\Gamma(\nu_2)\Gamma(\nu_3)\Gamma(d-\nu_1-\nu_2-\nu_3)} |p_1|^{d/2-\nu_2-\nu_3} |p_2|^{d/2-\nu_1-\nu_3}|p_3|^{d/2-\nu_1-\nu_2}\nonumber \\
&& \times \int_0^\infty d x x^{d/2-1}K_{d-\nu_1-\nu_3} (|p_1| x) K_{d-\nu_2-\nu_3} (|p_2| x) K_{d-\nu_1-\nu_2} (|p_3 |x) 
\eeqa
where $|p_i|$, in this case indicate the magnitudes of the momenta $p_i$, i.e. $|p_i|=\sqrt{p_i^2}$. 

\section{Form factors of the transverse traceless sector: $A_1,A_2,A_3,A_4$ for massless quarks} 
\label{ff}
We present here the explicit expressions of the coefficient functions $A_i$  characterising the (tt) sector. The quark contributions are identified by the factor $n_f$, the gluon contributions by the factor $C_A$:

{\allowdisplaybreaks \small
	\begin{eqnarray*}
		A_{10} &=& 16\, (p_1\cdot p_2)^7+36 \, p_1^2 \, (p_1\cdot p_2)^6+36 \, p_2^2 \, (p_1\cdot p_2)^6+12 \, p_1^4 \, (p_1\cdot p_2)^5+12 \, p_2^4 \, (p_1\cdot p_2)^5+240 \, p_1^2 \, p_2^2 \, (p_1\cdot p_2)^5 \notag \\ &&
		+188 \, p_1^2 \, p_2^4 \, (p_1\cdot p_2)^4+188 \, p_1^4 \, p_2^2 \, (p_1\cdot p_2)^4+46 \, p_1^2 \, p_2^6 \, (p_1\cdot p_2)^3-108 \, p_1^4 \, p_2^4 \, (p_1\cdot p_2)^3+46 \, p_1^6 \, p_2^2 \, (p_1\cdot p_2)^3 \notag \\ &&
		-204 \, p_1^4 \, p_2^6 \, (p_1\cdot p_2)^2-204 \, p_1^6 \, p_2^4 \, (p_1\cdot p_2)^2-20 \, p_1^6 \, p_2^8 \,-20 \, p_1^8 \, p_2^6 \,-58 \, p_1^4 \, p_2^8 \, (p_1\cdot p_2) \notag \\ &&
		-148 \, p_1^6 \, p_2^6 \, (p_1\cdot p_2)-58 \, p_1^8 \, p_2^4 \, (p_1\cdot p_2) \\
		A_{11} &=& -24 \, p_1^2 \,  \, (p_1\cdot p_2)^6-24 \, p_1^4 \,  \,\, (p_1\cdot p_2)^5-224 \, p_1^2 \,  \, p_2^2 \, (p_1\cdot p_2)^5-6\, p_1^6 \,  \,\, (p_1\cdot p_2)^4-228 \, p_1^2 \,  \, p_2^4 \, (p_1\cdot p_2)^4 \notag \\ &&
		-522 \, p_1^4 \,  \, p_2^2 \, (p_1\cdot p_2)^4- 60 \, p_1^2 \,  \, p_2^6 \, (p_1\cdot p_2)^3- 652 \, p_1^4 \,  \, p_2^4 \, (p_1\cdot p_2)^3-372 \, p_1^6 \,  \, p_2^2 \, (p_1\cdot p_2)^3  \notag \\ &&
		-279 \, p_1^4 \,  \, p_2^6 \, (p_1\cdot p_2)^2-474 \, p_1^6 \,  \, p_2^4 \, (p_1\cdot p_2)^2-83 \, p_1^8 \,  \, p_2^2 \, (p_1\cdot p_2)^2-18 \, p_1^6 \,  \, p_2^8 \,  \notag \\ &&
		-30 \, p_1^8 \,  \, p_2^6 \,-16 p_1^{10} \, p_2^4 \,-45 \, p_1^4 \,  \, p_2^8 \, (p_1\cdot p_2)-174 \, p_1^6 \,  \, p_2^6 \, (p_1\cdot p_2)-129 \, p_1^8 \,  \, p_2^4 \, (p_1\cdot p_2) \\
		A_{12} &=& -24 \, p_2^2 \, (p_1\cdot p_2)^6-24 \, p_2^4 \, (p_1\cdot p_2)^5-224 \, p_1^2 \,  \, p_2^2 \, (p_1\cdot p_2)^5-6 \, p_2^6 \, (p_1\cdot p_2)^4-522 \, p_1^2 \,  \, p_2^4 \, (p_1\cdot p_2)^4   \notag \\ &&
		-228 \, p_1^4 \,  \, p_2^2 \, (p_1\cdot p_2)^4-372 \, p_1^2 \,  \, p_2^6 \, (p_1\cdot p_2)^3-652 \, p_1^4 \,  \, p_2^4 \, (p_1\cdot p_2)^3-60 \, p_1^6 \,  \, p_2^2 \, (p_1\cdot p_2)^3-83 \, p_1^2 \,  \, p_2^8 \, (p_1\cdot p_2)^2   \notag \\ &&
		-474 \, p_1^4 \,  \, p_2^6 \, (p_1\cdot p_2)^2-279 \, p_1^6 \,  \, p_2^4 \, (p_1\cdot p_2)^2-16 \, p_1^4 \,  \, p_2^{10} \,-30 \, p_1^6 \,  \, p_2^8 \,-18 \, p_1^8 \,  \, p_2^6 \,-129 \, p_1^4 \,  \, p_2^8 \, (p_1\cdot p_2)   \notag \\ &&
		-174 \, p_1^6 \,  \, p_2^6 \, (p_1\cdot p_2)-45 \, p_1^8 \,  \, p_2^4 \, (p_1\cdot p_2) \\
		A_{13} &=& 24 \, p_1^2 \,  \, (p_1\cdot p_2)^6+24 \, p_2^2 \, (p_1\cdot p_2)^6+24 \, p_1^4 \,  \, (p_1\cdot p_2)^5+24 \, p_2^4 \, (p_1\cdot p_2)^5 + 448 \, p_1^2 \,  \, p_2^2 \, (p_1\cdot p_2)^5  \notag \\ &&
		+6 \, p_1^6 \,  \, (p_1\cdot p_2)^4+6 \, p_2^6 \, (p_1\cdot p_2)^4+750 \, p_1^2 \,  \, p_2^4 \, (p_1\cdot p_2)^4  +750 \, p_1^4 \,  \, p_2^2 \, (p_1\cdot p_2)^4+432 \, p_1^2 \,  \, p_2^6 \, (p_1\cdot p_2)^3  \notag \\ &&
		+ 1304 \, p_1^4 \,  \, p_2^4 \, (p_1\cdot p_2)^3+432 \, p_1^6 \,  \, p_2^2 \, (p_1\cdot p_2)^3  + 83 \, p_1^2 \,  \, p_2^8 \, (p_1\cdot p_2)^2+753 \, p_1^4 \,  \, p_2^6 \, (p_1\cdot p_2)^2+753 \, p_1^6 \,  \, p_2^4 \, (p_1\cdot p_2)^2  \notag \\ &&
		+83 \, p_1^8 \,  \, p_2^2 \, (p_1\cdot p_2)^2 + 16 \, p_1^4 \,  \, p_2^{10}+48 \, p_1^6 \,  \, p_2^8 \,+48 \, p_1^8 \,  \, p_2^6 \,+16 p_1^{10} \, p_2^4 \,  \notag \\ &&
		+174 \, p_1^4 \,  \, p_2^8 \, (p_1\cdot p_2)+348 \, p_1^6 \,  \, p_2^6 \, (p_1\cdot p_2)+174 \, p_1^8 \,  \, p_2^4 \, (p_1\cdot p_2)	\\
		A_{14} &=& 192 \, p_1^2 \,  \, p_2^2 \, (p_1\cdot p_2 )^6+432 \, p_1^2 \,  \, p_2^4 \, (p_1\cdot p_2 )^5+432 \, p_1^4 \,  \, p_2^2 \, (p_1\cdot p_2 )^5+288 \, p_1^2 \,  \, p_2^6 \, (p_1\cdot p_2 )^4+1152 \, p_1^4 \,  \, p_2^4 \, (p_1\cdot p_2 )^4  \notag \\ &&
		+288 \, p_1^6 \,  \, p_2^2 \, (p_1\cdot p_2 )^4+60 \, p_1^2 \,  \, p_2^8 \, (p_1\cdot p_2 )^3+936 \, p_1^4 \,  \, p_2^6 \, (p_1\cdot p_2 )^3+936 \, p_1^6 \,  \, p_2^4 \, (p_1\cdot p_2 )^3+60 \, p_1^8 \,  \, p_2^2 \, (p_1\cdot p_2 )^3  \notag \\ &&
		+324 \, p_1^4 \,  \, p_2^8 \, (p_1\cdot p_2 )^2+720 \, p_1^6 \,  \, p_2^6 \, (p_1\cdot p_2 )^2+324 \, p_1^8 \,  \, p_2^4 \, (p_1\cdot p_2 )^2+18 \, p_1^6 \,  \, p_2^{10} \,+36 \, p_1^8 \,  \, p_2^8 \notag \\ &&
		+18 p_1^{10} \, p_2^6 \,+45 \, p_1^4 \,  \, p_2^{10} (p_1\cdot p_2) +207 \, p_1^6 \,  \, p_2^8 \, (p_1\cdot p_2)+207 \, p_1^8 \,  \, p_2^6 \, (p_1\cdot p_2)+45 p_1^{10} \, p_2^4 \, (p_1\cdot p_2)
	\end{eqnarray*}
	
	\begin{eqnarray*}
		A_{20} &=& C_A \Big[-520 \, (p_1\cdot p_2)^6-372 \, p_1^2 \, (p_1\cdot p_2)^5-372 \, p_2^2 \, (p_1\cdot p_2)^5-60 \, p_1^4 \, (p_1\cdot p_2)^4-60 \, p_2^4 \, (p_1\cdot p_2)^4 \notag \\ &&
		+1056 \, p_1^2 \, p_2^2 \, (p_1\cdot p_2)^4+864 \, p_1^2 \, p_2^4 \, (p_1\cdot p_2)^3+864 \, p_1^4 \, p_2^2 \, (p_1\cdot p_2)^3+150 \, p_1^2 \, p_2^6 \, (p_1\cdot p_2)^2 \notag \\ &&
		-372 \, p_1^4 \, p_2^4 \, (p_1\cdot p_2)^2+150 \, p_1^6 \, p_2^2 \, (p_1\cdot p_2)^2-90 \, p_1^4 \, p_2^8 \,-164 \, p_1^6 \, p_2^6 \,-90 \, p_1^8 \, p_2^4 \notag \\ &&
		-492 \, p_1^4 \, p_2^6 \, (p_1\cdot p_2)-492 \, p_1^6 \, p_2^4 \, (p_1\cdot p_2) \Big]+ \notag \\ &&
		n_F\Big[ 40 (p_1\cdot p_2)^6+12 \, p_1^2 \, (p_1\cdot p_2)^5+12 \, p_2^2 \, (p_1\cdot p_2)^5-12 \, p_1^4 \, (p_1\cdot p_2)^4-12 \, p_2^4 \, (p_1\cdot p_2)^4 \notag \\ &&
		-192 \, p_1^2 \, p_2^2 \, (p_1\cdot p_2)^4-144 \, p_1^2 \, p_2^4 \, (p_1\cdot p_2)^3-144 \, p_1^4 \, p_2^2 \, (p_1\cdot p_2)^3-6 \, p_1^2 \, p_2^6 \, (p_1\cdot p_2)^2 \notag \\ &&
		+84 \, p_1^4 \, p_2^4 \, (p_1\cdot p_2)^2-6 \, p_1^6 \, p_2^2 \, (p_1\cdot p_2)^2+18 \, p_1^4 \, p_2^8 \,+68 \, p_1^6 \, p_2^6 \,+18 \, p_1^8 \, p_2^4 \notag \\ &&
		+132 \, p_1^4 \, p_2^6 \, (p_1\cdot p_2)+132 \, p_1^6 \, p_2^4 \, (p_1\cdot p_2) \Big] \\
		A_{21} &=& 3 \, p_1^2 \,  \, n_F \Big[-24 (p_1\cdot p_2)^5-20 \, p_2^2 \, (p_1\cdot p_2)^4+2 \, p_1^4 \,  \, (p_1\cdot p_2)^3+24 \, p_2^4 \, (p_1\cdot p_2)^3+90 \, p_1^2 \,  \, p_2^2 \, (p_1\cdot p_2)^3  \notag \\ &&
		+12 \, p_2^6 \, (p_1\cdot p_2)^2+162 \, p_1^2 \,  \, p_2^4 \, (p_1\cdot p_2)^2+54 \, p_1^4 \,  \, p_2^2 \, (p_1\cdot p_2)^2+3 \, p_1^2 \,  \, p_2^8 \,+8 \, p_1^4 \,  \, p_2^6 \,+21 \, p_1^6 \,  \, p_2^4 \,+51 \, p_1^2 \,  \, p_2^6 \, (p_1\cdot p_2)   \notag \\ &&
		+84 \, p_1^4 \,  \, p_2^4 \, (p_1\cdot p_2)+13 \, p_1^6 \,  \, p_2^2 \, (p_1\cdot p_2) \Big] \notag \\ &&
		-3 C_A \Big[ -48 (p_1\cdot p_2)^6-240 \, p_1^2 \,  \, (p_1\cdot p_2)^5-72 \, p_2^2 \, (p_1\cdot p_2)^5  \notag \\ &&
		-156 \, p_1^4 \,  \, (p_1\cdot p_2)^4 -24 \, p_2^4 \, (p_1\cdot p_2)^4-344 \, p_1^2 \,  \, p_2^2 \, (p_1\cdot p_2)^4-34 \, p_1^6 \,  \, (p_1\cdot p_2)^3-48 \, p_1^2 \,  \, p_2^4 \, (p_1\cdot p_2)^3   \notag \\ &&
		+54 \, p_1^4 \,  \, p_2^2 \, (p_1\cdot p_2)^3 +24 \, p_1^2 \,  \, p_2^6 \, (p_1\cdot p_2)^2+486 \, p_1^4 \,  \, p_2^4 \, (p_1\cdot p_2)^2+150 \, p_1^6 \,  \, p_2^2 \, (p_1\cdot p_2)^2+15 \, p_1^4 \,  \, p_2^8 \, \notag \\ &&
		+56 \, p_1^6 \,  \, p_2^6 \, +81 \, p_1^8 \,  \, p_2^4  +195 \, p_1^4 \,  \, p_2^6 \, (p_1\cdot p_2)+336 \, p_1^6 \,  \, p_2^4 \, (p_1\cdot p_2)+49 \, p_1^8 \,  \, p_2^2 \, (p_1\cdot p_2) \Big] \\
		A_{22} &=&  3 \, p_2^2 \, n_F \Big[-24 (p_1\cdot p_2)^5-20 \, p_1^2 \,  \, (p_1\cdot p_2)^4+24 \, p_1^4 \,  \, (p_1\cdot p_2)^3+2 \, p_2^4 \, (p_1\cdot p_2)^3+90 \, p_1^2 \,  \, p_2^2 \, (p_1\cdot p_2)^3  \notag \\ &&
		+12 \, p_1^6 \,  \, (p_1\cdot p_2)^2+54 \, p_1^2 \,  \, p_2^4 \, (p_1\cdot p_2)^2+162 \, p_1^4 \,  \, p_2^2 \, (p_1\cdot p_2)^2+21 \, p_1^4 \,  \, p_2^6 \,+8 \, p_1^6 \,  \, p_2^4 \,+3 \, p_1^8 \,  \, p_2^2 \,+13 \, p_1^2 \,  \, p_2^6 \, (p_1\cdot p_2)  \notag \\ &&
		+84 \, p_1^4 \,  \, p_2^4 \, (p_1\cdot p_2)+51 \, p_1^6 \,  \, p_2^2 \, (p_1\cdot p_2)\Big]-3 C_A \Big[-48 (p_1\cdot p_2)^6-72 \, p_1^2 \,  \, (p_1\cdot p_2)^5-240 \, p_2^2 \, (p_1\cdot p_2)^5  \notag \\ &&
		-24 \, p_1^4 \,  \, (p_1\cdot p_2)^4-156 \, p_2^4 \, (p_1\cdot p_2)^4-344 \, p_1^2 \,  \, p_2^2 \, (p_1\cdot p_2)^4-34 \, p_2^6 \, (p_1\cdot p_2)^3+54 \, p_1^2 \,  \, p_2^4 \, (p_1\cdot p_2)^3  \notag \\ &&
		-48 \, p_1^4 \,  \, p_2^2 \, (p_1\cdot p_2)^3+150 \, p_1^2 \,  \, p_2^6 \, (p_1\cdot p_2)^2+486 \, p_1^4 \,  \, p_2^4 \, (p_1\cdot p_2)^2+24 \, p_1^6 \,  \, p_2^2 \, (p_1\cdot p_2)^2+81 \, p_1^4 \,  \, p_2^8 \,  \notag \\ &&
		+56 \, p_1^6 \,  \, p_2^6 \,+15 \, p_1^8 \,  \, p_2^4 \,+49 \, p_1^2 \,  \, p_2^8 \, (p_1\cdot p_2)+336 \, p_1^4 \,  \, p_2^6 \, (p_1\cdot p_2)+195 \, p_1^6 \,  \, p_2^4 \, (p_1\cdot p_2)\Big] \\
		A_{23} &=& 3 C_A \, \Big(2 (p_1\cdot p_2)+\, p_1^2 \,  \,+\, p_2^2 \Big)^2 \, \Big[-44 (p_1\cdot p_2)^4-34 \, p_1^2 \,  \, (p_1\cdot p_2)^3-34 \, p_2^2 \, (p_1\cdot p_2)^3-22 \, p_1^2 \,  \, p_2^2 \, (p_1\cdot p_2)^2  \notag \\ &&
		+96 \, p_1^4 \,  \, p_2^4 \,+49 \, p_1^2 \,  \, p_2^4 \, (p_1\cdot p_2)+49 \, p_1^4 \,  \, p_2^2 \, (p_1\cdot p_2) \Big]  \notag \\ &&
		-3 n_F (2 (p_1\cdot p_2)+\, p_1^2 \,  \,+\, p_2^2 \,)^2  \Big[ -8 (p_1\cdot p_2)^4+2 \, p_1^2 \,  \, (p_1\cdot p_2)^3+2 \, p_2^2 \, (p_1\cdot p_2)^3+14 \, p_1^2 \,  \, p_2^2 \, (p_1\cdot p_2)^2  \notag \\ &&
		+24 \, p_1^4 \,  \, p_2^4 \,+13 \, p_1^2 \,  \, p_2^4 \, (p_1\cdot p_2)+13 \, p_1^4 \,  \, p_2^2 \, (p_1\cdot p_2) \Big]  \\
		A_{24} &=& 9\, C_A \Big(2 (p_1\cdot p_2)+\, p_1^2 \,  \,+\, p_2^2 \Big) \, \Big[ -32 (p_1\cdot p_2)^6-16 \, p_1^2 \,  \, (p_1\cdot p_2)^5-16 \, p_2^2 \, (p_1\cdot p_2)^5-8 \, p_1^4 \,  \, (p_1\cdot p_2)^4  \notag \\ &&
		-8 \, p_2^4 \, (p_1\cdot p_2)^4-24 \, p_1^2 \,  \, p_2^4 \, (p_1\cdot p_2)^3-24 \, p_1^4 \,  \, p_2^2 \, (p_1\cdot p_2)^3+8 \, p_1^2 \,  \, p_2^6 \, (p_1\cdot p_2)^2+20 \, p_1^4 \,  \, p_2^4 \, (p_1\cdot p_2)^2  \notag \\ &&
		+8 \, p_1^6 \,  \, p_2^2 \, (p_1\cdot p_2)^2+5 \, p_1^4 \,  \, p_2^8 \,+42 \, p_1^6 \,  \, p_2^6 \,+5 \, p_1^8 \,  \, p_2^4 \,+60 \, p_1^4 \,  \, p_2^6 \, (p_1\cdot p_2)+60 \, p_1^6 \,  \, p_2^4 \, (p_1\cdot p_2) \Big]  \notag \\ &&
		-9 \, p_1^2 \,  \, p_2^2 \, n_F  \Big( (2 (p_1\cdot p_2)+\, p_1^2 \,  \,+\, p_2^2 \, \Big)^2 \Big[-4 (p_1\cdot p_2)^3+4 \, p_1^2 \,  \, (p_1\cdot p_2)^2+4 \, p_2^2 \, (p_1\cdot p_2)^2   \notag \\ &&
		+\, p_1^2 \,  \, p_2^4 \,+\, p_1^4 \,  \, p_2^2 \,+14 \, p_1^2 \,  \, p_2^2 \, (p_1\cdot p_2) \Big]
	\end{eqnarray*}

	\begin{eqnarray*}
		A_{31} &=& 	3 \, p_1^2 \,  \, n_F \Big[-16 (p_1\cdot p_2)^5-4 \, p_1^2 \,  \, (p_1\cdot p_2)^4-70 \, p_2^2 \, (p_1\cdot p_2)^4-54 \, p_2^4 \, (p_1\cdot p_2)^3-44 \, p_1^2 \,  \, p_2^2 \, (p_1\cdot p_2)^3  \notag \\ &&
		-12 \, p_2^6 \, (p_1\cdot p_2)^2-15 \, p_1^2 \,  \, p_2^4 \, (p_1\cdot p_2)^2-15 \, p_1^4 \,  \, p_2^2 \, (p_1\cdot p_2)^2-3 \, p_1^2 \,  \, p_2^8 \,-5 \, p_1^4 \,  \, p_2^6 \,+4 \, p_1^6 \,  \, p_2^4 \,-6 \, p_1^2 \,  \, p_2^6 \, (p_1\cdot p_2)\Big]  \notag \\ &&
		-3 C_A\Big[-24 (p_1\cdot p_2)^6-100 \, p_1^2 \,  \, (p_1\cdot p_2)^5-12 \, p_2^2 \, (p_1\cdot p_2)^5-34 \, p_1^4 \,  \, (p_1\cdot p_2)^4-172 \, p_1^2 \,  \, p_2^2 \, (p_1\cdot p_2)^4  \notag \\ && 
		-84 \, p_1^2 \,  \, p_2^4 \, (p_1\cdot p_2)^3 -38 \, p_1^4 \,  \, p_2^2 \, (p_1\cdot p_2)^3-12 \, p_1^2 \,  \, p_2^6 \, (p_1\cdot p_2)^2+99 \, p_1^4 \,  \, p_2^4 \, (p_1\cdot p_2)^2-9 \, p_1^6 \,  \, p_2^2 \, (p_1\cdot p_2)^2-3 \, p_1^4 \,  \, p_2^8 \,  \notag \\ &&
		+7 \, p_1^6 \,  \, p_2^6 \,+28 \, p_1^8 \,  \, p_2^4 \,+36 \, p_1^4 \,  \, p_2^6 \, (p_1\cdot p_2)+78 \, p_1^6 \,  \, p_2^4 \, (p_1\cdot p_2)\Big] \\
		A_{32} &=& 3 \, p_2^2 \, n_F \Big[-24 (p_1\cdot p_2)^5-54 \, p_1^2 \,  \, (p_1\cdot p_2)^4-12 \, p_2^2 \, (p_1\cdot p_2)^4-18 \, p_1^4 \,  \, (p_1\cdot p_2)^3-2 \, p_2^4 \, (p_1\cdot p_2)^3 \notag \\ &&
		-62 \, p_1^2 \,  \, p_2^2 \, (p_1\cdot p_2)^3-55 \, p_1^2 \,  \, p_2^4 \, (p_1\cdot p_2)^2-3 \, p_1^4 \,  \, p_2^2 \, (p_1\cdot p_2)^2+7 \, p_1^4 \,  \, p_2^6 \,-3 \, p_1^6 \,  \, p_2^4 \,-13 \, p_1^2 \,  \, p_2^6 \, (p_1\cdot p_2) \notag \\ &&
		-4 \, p_1^4 \,  \, p_2^4 \, (p_1\cdot p_2)+3 \, p_1^6 \,  \, p_2^2 \, (p_1\cdot p_2)\Big] \notag \\ &&
		-3  C_A \Big[-24 (p_1\cdot p_2)^6-12 \, p_1^2 \,  \, (p_1\cdot p_2)^5-108 \, p_2^2 \, (p_1\cdot p_2)^5-42 \, p_2^4 \, (p_1\cdot p_2)^4 \notag \\ &&
		-156 \, p_1^2 \,  \, p_2^2 \, (p_1\cdot p_2)^4-2 \, p_2^6 \, (p_1\cdot p_2)^3-56 \, p_1^2 \,  \, p_2^4 \, (p_1\cdot p_2)^3-48 \, p_1^4 \,  \, p_2^2 \, (p_1\cdot p_2)^3-49 \, p_1^2 \,  \, p_2^6 \, (p_1\cdot p_2)^2 \notag \\ &&
		+111 \, p_1^4 \,  \, p_2^4 \, (p_1\cdot p_2)^2+31 \, p_1^4 \,  \, p_2^8 \,+9 \, p_1^6 \,  \, p_2^6 \,-13 \, p_1^2 \,  \, p_2^8 \, (p_1\cdot p_2)+74 \, p_1^4 \,  \, p_2^6 \, (p_1\cdot p_2)+45 \, p_1^6 \,  \, p_2^4 \, (p_1\cdot p_2)\Big] \notag \\ 
		A_{33} &=& 3 \, C_A \, \Big(2 (p_1\cdot p_2)+\, p_1^2 \,  \,+\, p_2^2 \Big) \, \Big[ -44 (p_1\cdot p_2)^5-34 \, p_1^2 \,  \, (p_1\cdot p_2)^4-38 \, p_2^2 \, (p_1\cdot p_2)^4-2 \, p_2^4 \, (p_1\cdot p_2)^3 \notag \\ &&
		-68 \, p_1^2 \,  \, p_2^2 \, (p_1\cdot p_2)^3-35 \, p_1^2 \,  \, p_2^4 \, (p_1\cdot p_2)^2-9 \, p_1^4 \,  \, p_2^2 \, (p_1\cdot p_2)^2+28 \, p_1^4 \,  \, p_2^6 \,+28 \, p_1^6 \,  \, p_2^4  \notag \\ &&
		-13 \, p_1^2 \,  \, p_2^6 \, (p_1\cdot p_2) +67 \, p_1^4 \,  \, p_2^4 \, (p_1\cdot p_2)\Big] \notag \\ &&
		-3 \, n_F \, \Big(2 (p_1\cdot p_2)+\, p_1^2 \,  \,+\, p_2^2 \Big)^2 \, \Big[-4 (p_1\cdot p_2)^4-2 \, p_2^2 \, (p_1\cdot p_2)^3-15 \, p_1^2 \,  \, p_2^2 \, (p_1\cdot p_2)^2+4 \, p_1^4 \,  \, p_2^4 \,-13 \, p_1^2 \,  \, p_2^4 \, (p_1\cdot p_2) \Big] \notag \\ 
		A_{34} &=& 9 \, C_A \, \Big(2 (p_1\cdot p_2)+\, p_1^2 \,  \,+\, p_2^2 \Big) \, \Big[-16 (p_1\cdot p_2)^6-4 \, p_1^2 \,  \, (p_1\cdot p_2)^5-4 \, p_2^2 \, (p_1\cdot p_2)^5-24 \, p_1^2 \,  \, p_2^4 \, (p_1\cdot p_2)^3 \notag \\ &&
		-16 \, p_1^4 \,  \, p_2^2 \, (p_1\cdot p_2)^3-4 \, p_1^2 \,  \, p_2^6 \, (p_1\cdot p_2)^2-14 \, p_1^4 \,  \, p_2^4 \, (p_1\cdot p_2)^2-\, p_1^4 \,  \, p_2^8 \,\notag \\ &&
		+15 \, p_1^6 \,  \, p_2^6 \,+13 \, p_1^4 \,  \, p_2^6 \, (p_1\cdot p_2)+15 \, p_1^6 \,  \, p_2^4 \, (p_1\cdot p_2)\Big] \notag \\ &&
		-9 \, p_1^2 \,  \, p_2^2 \, n_F \, \Big(2 (p_1\cdot p_2)+\, p_1^2 \,  \,+\, p_2^2 \Big)^2 \, \Big[-6 (p_1\cdot p_2)^3-4 \, p_2^2 \, (p_1\cdot p_2)^2-\, p_1^2 \,  \, p_2^4 \,+\, p_1^2 \,  \, p_2^2 \, (p_1\cdot p_2) \Big] \notag \\
	\end{eqnarray*}
	
	\begin{eqnarray*}
		A_{40} &=& C_A \Big[ -284 \, (p_1\cdot p_2)^5-132 \, p_1^2 \, (p_1\cdot p_2)^4-132 \, p_2^2 \, (p_1\cdot p_2)^4+556 \, p_1^2 \, p_2^2 \, (p_1\cdot p_2)^3 \notag \\ &&
		+258 \, p_1^2 \, p_2^4 \, (p_1\cdot p_2)^2+258 \, p_1^4 \, p_2^2 \, (p_1\cdot p_2)^2-126 \, p_1^4 \, p_2^6 \,-126 \, p_1^6 \, p_2^4 \,-272 \, p_1^4 \, p_2^4 \, (p_1\cdot p_2) \Big]  \notag \\ &&
		+ n_F \Big[ 44 (p_1\cdot p_2)^5+24 \, p_1^2 \, (p_1\cdot p_2)^4+24 \, p_2^2 \, (p_1\cdot p_2)^4-76 \, p_1^2 \, p_2^2 \, (p_1\cdot p_2)^3-42 \, p_1^2 \, p_2^4 \, (p_1\cdot p_2)^2 \notag \\ &&
		-42 \, p_1^4 \, p_2^2 \, (p_1\cdot p_2)^2+18 \, p_1^4 \, p_2^6 \,+18 \, p_1^6 \, p_2^4 \,+32 \, p_1^4 \, p_2^4 \, (p_1\cdot p_2) \Big] \\
		A_{41} &=& 3 \, p_1^2 \,  \, n_F \Big[-16 \, (p_1\cdot p_2)^4-4 \, p_1^2 \,  \, (p_1\cdot p_2)^3-18 \, p_2^2 \, (p_1\cdot p_2)^3-6 \, p_2^4 \, (p_1\cdot p_2)^2+10 \, p_1^2 \,  \, p_2^2 \, (p_1\cdot p_2)^2 \notag \\ &&
		+3 \, p_1^2 \,  \, p_2^6 \,-3 \, p_1^4 \,  \, p_2^4 \,+9 \, p_1^2 \,  \, p_2^4 \, (p_1\cdot p_2)+\, p_1^4 \,  \, p_2^2 \, (p_1\cdot p_2)\Big] \notag \\ &&
		-3 C_A\Big[-24 \, (p_1\cdot p_2)^5-76 \, p_1^2 \,  \, (p_1\cdot p_2)^4-12 \, p_2^2 \, (p_1\cdot p_2)^4-22 \, p_1^4 \,  \, (p_1\cdot p_2)^3-24 \, p_1^2 \,  \, p_2^2 \, (p_1\cdot p_2)^3 \notag \\ &&
		+76 \, p_1^4 \,  \, p_2^2 \, (p_1\cdot p_2)^2+9 \, p_1^4 \,  \, p_2^6 \,-9 \, p_1^6 \,  \, p_2^4 \,+39 \, p_1^4 \,  \, p_2^4 \, (p_1\cdot p_2)+19 \, p_1^6 \,  \, p_2^2 \, (p_1\cdot p_2)\Big] \\
		A_{42} &=&  3 \, p_2^2 \, n_F \Big[-16 (p_1\cdot p_2)^4-18 \, p_1^2 \,  \, (p_1\cdot p_2)^3-4 \, p_2^2 \, (p_1\cdot p_2)^3-6 \, p_1^4 \,  \, (p_1\cdot p_2)^2+10 \, p_1^2 \,  \, p_2^2 \, (p_1\cdot p_2)^2 \notag \\ &&
		-3 \, p_1^4 \,  \, p_2^4 \,+3 \, p_1^6 \,  \, p_2^2 \,+\, p_1^2 \,  \, p_2^4 \, (p_1\cdot p_2)+9 \, p_1^4 \,  \, p_2^2 \, (p_1\cdot p_2)\Big] \notag \\ &&
		-3 C_A\Big[-24 \, (p_1\cdot p_2)^5-12 \, p_1^2 \,  \, (p_1\cdot p_2)^4-76 \, p_2^2 \, (p_1\cdot p_2)^4-22 \, p_2^4 \, (p_1\cdot p_2)^3-24 \, p_1^2 \,  \, p_2^2 \, (p_1\cdot p_2)^3 \notag \\ &&
		+76 \, p_1^2 \,  \, p_2^4 \, (p_1\cdot p_2)^2-9 \, p_1^4 \,  \, p_2^6 \,+9 \, p_1^6 \,  \, p_2^4 \,+19 \, p_1^2 \,  \, p_2^6 \, (p_1\cdot p_2)+39 \, p_1^4 \,  \, p_2^4 \, (p_1\cdot p_2)\Big] \\
		A_{43} &=& 3 C_A \, (p_1\cdot p_2) \, \Big(2 (p_1\cdot p_2)+\, p_1^2 \,  \,+\, p_2^2 \Big)^2 \Big[ 19 \, p_1^2 \,  \, p_2^2 \,-22 (p_1\cdot p_2)^2 \Big] \notag \\ &&
		-3 n_F \, (p_1\cdot p_2) \, \Big(2 (p_1\cdot p_2)+\, p_1^2 \,  \,+\, p_2^2 \Big)^2 \Big[ p_1^2 \,  \, p_2^2 \,-4 (p_1\cdot p_2)^2 \Big] \\
		A_{44} &=& 9 C_A \Big(2 (p_1\cdot p_2)+\, p_1^2 \,  \,+\, p_2^2 \Big) \Big[-16 (p_1\cdot p_2)^5-4 \, p_1^2 \,  \, (p_1\cdot p_2)^4-4 \, p_2^2 \, (p_1\cdot p_2)^4+16 \, p_1^2 \,  \, p_2^2 \, (p_1\cdot p_2)^3 \notag \\ &&
		+3 \, p_1^4 \,  \, p_2^6 \,+3 \, p_1^6 \,  \, p_2^4 \,-2 \, p_1^4 \,  \, p_2^4 \, (p_1\cdot p_2) -9 \, p_1^2 \,  \, p_2^2 \, n_F (2 (p_1\cdot p_2)+\, p_1^2 \,  \,+\, p_2^2 \,)^2 \, p_1^2 \,  \, p_2^2 \,-2 (p_1\cdot p_2)^2 \Big]
	\end{eqnarray*}
}

\subsection{Transverse traceless sector for quarks: massive contributions }\label{aiq}
For a single quark  of mass $m$ in the loop, the form factors of  \eqref{DecompTJJ} should be replaced by the expression

{\small
	\begin{align}
		A^{(q)}_1= - \frac{g_s^2}{16\p^2}\Biggl(&\frac{P_1}{3 \bar{D}^3}+\frac{2 q^2 D_1(q,p_1,p_2)  \Sigma(q,m) }{3 \bar{D}^4}+\frac{P_3}{3 \bar{D}^3}-\frac{8   C_0 \left(-2 q^4+\left(p_1^2+p_2^2\right) q^2+\left(p_1^2-p_2^2\right)^2\right) m^4 }{\bar{D}^2}\nn\\ &-\frac{8   C_0 p_1^2 p_2^2 q^4  P_2}{\bar{D}^4} +m^2 \Biggl(\frac{ P_4 \Sigma(q,m) }{3 \bar{D}^3}-\frac{4  \left(-18 q^4+13 \left(p_1^2+p_2^2\right) q^2+5 \left(p_1^2-p_2^2\right)^2\right) }{3 \bar{D}^2}\nn\\&-\frac{8  q^2  C_0 P_5 }{\bar{D}^3}-\frac{4   \Sigma(p_1,m) E_1(q,p_1,p_2) }{3 p_1^2 \bar{D}^3}-\frac{4   \Sigma(p_2,m) E_1(q,p_2,p_1)}{3 p_2^2\bar{D}^3}\Biggl)\nn\\ &+\frac{2  \Sigma(p_1,m)  D_2(q,p_1,p_2)}{3 \bar{D}^4}+\frac{2    \Sigma(p_2,m)   D_2(q,p_2,p_1)}{3 \bar{D}^4}\Biggl)
	\end{align}
	
	\begin{align}
		A^{(q)}_2=- \frac{g_s^2}{16\p^2} \Biggl(& \frac{2}{3}   \log \left(\frac{\mu ^2}{m^2}\right) -\frac{4}{3}   \log (2 \pi ) +\frac{2}{3} \log (4 \pi ) +\frac{  \Sigma(p_1,m) D_3(q,p_1,p_2)}{3\bar{D}^3}+\frac{ \Sigma(p_2,m) D_3(q,p_2,p_1)}{3 \bar{D}^3}\nn\\&+\frac{ q^2  P_6 \Sigma(q,m)}{3 \bar{D}^3}-\frac{4  C_0 q^2 m^4 }{\bar{D}}-\frac{D_4(q,p_1,p_2)}{18 \bar{D}^2}-\frac{2  C_0 p_1^2 p_2^2 q^4 P_7 }{\bar{D}^3}\nn\\&+m^2 \Biggl(\frac{2   E_2(q,p_1,p_2)  \Sigma(p_1,m)}{3 p_1^2 \bar{D}^2}+\frac{2  E_2(q,p_2,p_1)  \Sigma(p_2,m)}{3 p_2^2 \bar{D}^2}+\frac{26  q^2 \left(p_1^2+p_2^2-q^2\right)   \Sigma(q,m) }{3 \bar{D}^2}
		\nn\\& -\frac{10   q^2 }{3 \bar{D}}-\frac{2  C_0 q^2 P_8}{\bar{D}^2}\Biggl)\Biggl)
	\end{align}

	\begin{align}
		A^{(q)}_3=- \frac{g_s^2}{16\p^2} \Biggl(& \frac{2  p_1^2 p_2^2  q^4 C_0 P_9}{\bar{D}^3}+\frac{4 m^4  \left(p_1^2-p_2^2-q^2\right)   C_0 }{\bar{D}}-\frac{2}{3}   \log \left(\frac{\mu ^2}{m^2}\right)+\frac{4}{3}  \log (2 \pi ) -\frac{2}{3}  \log (4 \pi ) \nn\\&-\frac{  \Sigma(p_1,m)   D_5(q,p_1,p_2)}{3 \bar{D}^3}-\frac{2  q^2 \Sigma(q,m) P_{10} }{3 \bar{D}^3}+m^2 \Biggl(\frac{10  \left(p_1^2-p_2^2-q^2\right) }{3 \bar{D}}\nn\\&+\frac{2   P_{11} \Sigma(p_1,m)}{3 p_1^2 \bar{D}^2}-\frac{2 q^2  C_0  P_{12}}{\bar{D}^2}-\frac{2   \Sigma(p_2,m) \left(-21 q^4+16 \left(p_1^2+p_2^2\right) q^2+5 \left(p_1^2-p_2^2\right)^2\right) }{3 \bar{D}^2}\nn\\&-\frac{2   \Sigma(q,m) \left(5 p_1^4+2 \left(8 p_2^2-5 q^2\right) p_1^2-\left(p_2^2-q^2\right) \left(21 p_2^2+5 q^2\right)\right) }{3 \bar{D}^2}\Biggl)\nn\\&-\frac{  \Sigma(p_2,m)  D_6(q,p_1,p_2)}{3 \bar{D}^3}+\frac{D_7(q,p_1,p_2)}{18 \bar{D}^2}\Biggl)
	\end{align}

	\begin{align}
		A^{(q)}_4=- \frac{g_s^2}{16\p^2}  \Biggl( &-\frac{2}{3}    \log (2 \pi ) \left(p_1^2+p_2^2-q^2\right) -2  m^4   C_0 +\frac{1}{3}  \left(p_1^2+p_2^2-q^2\right)   \log \left(\frac{\mu ^2}{m^2}\right) +\frac{1}{3}  \left(p_1^2+p_2^2-q^2\right)   \log (4 \pi )\nn\\&+\frac{  \Sigma(p_1,m) P_{13}}{6 \bar{D}^2}+\frac{ P_{14} \Sigma(p_2,m) }{6 \bar{D}^2}+\frac{ q^2 \left(p_1^2+p_2^2-q^2\right) \left(p_1^4+\left(p_2^2-2 q^2\right) p_1^2+\left(p_2^2-q^2\right)^2\right)  \Sigma(q,m) }{3 \bar{D}^2}\nn\\&-\frac{  C_0 p_1^2 p_2^2 q^4 \left(p_1^4-2 q^2 p_1^2+\left(p_2^2-q^2\right)^2\right)}{\bar{D}^2}-\frac{  P_{15} }{36 \bar{D}}+m^2 \Biggl(1+\frac{5  \left(p_1^2+p_2^2-q^2\right)   \Sigma(q,m) }{3 \bar{D}}\nn\\&-\frac{ C_0 \left(p_1^2+p_2^2-q^2\right)^2 q^2 }{\bar{D}}-\frac{\Sigma(p_1,m) \left(p_1^4+3 \left(p_2^2+q^2\right) p_1^2-4 \left(p_2^2-q^2\right)^2\right) }{3 p_1^2 \bar{D}}\nn\\&-\frac{ \Sigma(p_2,m) \left(-4 p_1^4+\left(3 p_2^2+8 q^2\right) p_1^2+\left(p_2^2-q^2\right) \left(p_2^2+4 q^2\right)\right) }{3 p_2^2 \bar{D}}\Biggl)
	\end{align}
	
}

where

\begin{equation}
	\Sigma(p,m)=\log \left(\frac{2 m^2-p^2+\sqrt{p^2 \left(p^2-4 m^2\right)}}{2 m^2}\right) \sqrt{p^2 \left(p^2-4 m^2\right)}
\end{equation}

\begin{align}
	\bar{D}=p_1^4-2 \left(p_2^2+q^2\right) p_1^2+\left(p_2^2-q^2\right)^2
\end{align}

{\allowdisplaybreaks\small
	\begin{align}
		D_1(q,p_1,p_2)=&\ 3 p_1^{10}+3 \left(27 p_2^2-4 q^2\right) p_1^8-2 \left(42 p_2^4+40 q^2 p_2^2-9 q^4\right) p_1^6-4 \left(21 p_2^6-76 q^2 p_2^4+20 q^4 p_2^2+3 q^6\right) p_1^4\nonumber\\&+\left(p_2^2-q^2\right)^2 \left(81 p_2^4+82 q^2 p_2^2+3 q^4\right) p_1^2+3 p_2^2 \left(p_2^2-q^2\right)^4
		\nonumber\\
		D_2(q,p_1,p_2)=&-3 q^{12}+2 \left(6 p_1^2-19 p_2^2\right) q^{10}+\left(-18 p_1^4-41 p_2^2 p_1^2+121 p_2^4\right) q^8+4 \left(3 p_1^6+46 p_2^2 p_1^4-38 p_2^4 p_1^2-26 p_2^6\right) q^6\nonumber\\&-\left(p_1^2-p_2^2\right) \left(3 p_1^6+95 p_2^2 p_1^4+215 p_2^4 p_1^2+11 p_2^6\right) q^4-14 \left(p_1^2-p_2^2\right)^3 p_2^2 \left(p_1^2+p_2^2\right) q^2+p_2^2 \left(p_1^2-p_2^2\right)^5
		\nonumber\\
		D_3(q,p_1,p_2)=&-3 q^{10}+5 \left(3 p_1^2+2 p_2^2\right) q^8+\left(-29 p_1^4+5 p_2^2 p_1^2+2 p_2^4\right) q^6+3 \left(p_1^2-p_2^2\right) \left(9 p_1^4+p_2^2 p_1^2+8 p_2^4\right) q^4\nn\\&-\left(12 p_1^2+17 p_2^2\right) \left(p_1^2-p_2^2\right)^3 q^2+2 \left(p_1^2-p_2^2\right)^5
		\nonumber\\
		D_4(q,p_1,p_2)=&\ q^8(-29+12 \gamma ) -(-113+48 \gamma ) \left(p_1^2+p_2^2\right) q^6+(3 (-53+24 \gamma ) p_1^4+2 (-31+24 \gamma ) p_2^2 p_1^2\nn\\&+3 (-53+24 \gamma ) p_2^4) q^4-(-95+48 \gamma ) \left(p_1^2+p_2^2\right) \left(p_1^2-p_2^2\right)^2 q^2+4 (-5+3 \gamma ) \left(p_1^2-p_2^2\right)^4
		\nonumber\\
		D_5(q,p_1,p_2)=&-4 q^{10}+3 \left(6 p_1^2-5 p_2^2\right) q^8+2 \left(-16 p_1^4+12 p_2^2 p_1^2+23 p_2^4\right) q^6+4 \left(7 p_1^6-3 p_2^2 p_1^4-12 p_2^4 p_1^2-8 p_2^6\right) q^4\nn\\&-6 \left(p_1^2-p_2^2\right)^2 \left(2 p_1^4+2 p_2^2 p_1^2-p_2^4\right) q^2+\left(p_1^2-p_2^2\right)^4 \left(2 p_1^2-p_2^2\right)
		\nonumber\\
		D_6(q,p_1,p_2)=&-6 q^{10}+3 \left(p_1^2+8 p_2^2\right) q^8+2 \left(15 p_1^4-25 p_2^2 p_1^2-19 p_2^4\right) q^6+2 \left(-24 p_1^6+27 p_2^2 p_1^4+14 p_2^4 p_1^2+15 p_2^6\right) q^4\nn\\&+6 \left(p_1^2-p_2^2\right)^2 \left(4 p_1^4+p_2^2 p_1^2-2 p_2^4\right) q^2-\left(3 p_1^2-2 p_2^2\right) \left(p_1^2-p_2^2\right)^4
		\nonumber\\
		D_7(q,p_1,p_2)=&\ q^8 (-35+12 \gamma )+2 \left(8 (8-3 \gamma ) p_1^2+(55-24 \gamma ) p_2^2\right) q^6+2 ((-87+36 \gamma ) p_1^4\nn\\&+(-73+24 \gamma ) p_2^2 p_1^2+6 (-11+6 \gamma ) p_2^4) q^4-2 \left(p_1^2-p_2^2\right) (4 (-13+6 \gamma ) p_1^4\nn\\&-27 p_2^2 p_1^2+(37-24 \gamma ) p_2^4) q^2+\left(p_1^2-p_2^2\right)^3 \left((-23+12 \gamma ) p_1^2+(17-12 \gamma ) p_2^2\right)
	\end{align}
	
	\begin{align}
		E_1(q,p_1,p_2)=&\ 6 q^8+2 \left(9 p_1^2-8 p_2^2\right) q^6+3 \left(-17 p_1^4+15 p_2^2 p_1^2+4 p_2^4\right) q^4\nn\\&+6 p_1^2 \left(p_1^2-p_2^2\right) \left(4 p_1^2+11 p_2^2\right) q^2+\left(p_1^2-p_2^2\right)^3 \left(3 p_1^2+2 p_2^2\right)
		\nonumber\\
		E_2(q,p_1,p_2)=& \ 6 q^6+\left(3 p_1^2-14 p_2^2\right) q^4-\left(11 p_1^2+10 p_2^2\right) \left(p_1^2-p_2^2\right) q^2+2 \left(p_1^2-p_2^2\right)^3
	\end{align}

	\begin{align}
		P_1=&\ 2 q^{10}-\left(p_1^2+p_2^2\right) q^8-2 \left(5 p_1^4-48 p_2^2 p_1^2+5 p_2^4\right) q^6\nn\\&+4 \left(p_1^2+p_2^2\right) \left(4 p_1^4-23 p_2^2 p_1^2+4 p_2^4\right) q^4
		\nonumber\\
		P_2=&\ 3 p_1^8+\left(3 p_2^2-7 q^2\right) p_1^6+3 \left(-4 p_2^4+4 q^2 p_2^2+q^4\right) p_1^4\nn\\&+3 \left(p_2^6+4 q^2 p_2^4-6 q^4 p_2^2+q^6\right) p_1^2+\left(p_2^2-q^2\right)^3 \left(3 p_2^2+2 q^2\right)
		\nonumber\\
		P_3=&-8\left( \left(p_1^2-p_2^2\right)^2 \left(p_1^4+4 p_2^2 p_1^2+p_2^4\right) q^2+\left(p_1^2-p_2^2\right)^4 \left(p_1^2+p_2^2\right)\right) 
		\nonumber\\
		P_4=&\  18 q^6-15 \left(p_1^2+p_2^2\right) q^4-4 \left(6 p_1^4-25 p_2^2 p_1^2+6 p_2^4\right) q^2+21 \left(p_1^2-p_2^2\right)^2 \left(p_1^2+p_2^2\right)
		\nonumber\\
		P_5=&\ 2 p_1^8+\left(4 p_2^2-5 q^2\right) p_1^6+\left(-12 p_2^4+11 q^2 p_2^2+3 q^4\right) p_1^4\nn\\&+\left(p_2^2-q^2\right) \left(4 p_2^4+15 q^2 p_2^2-q^4\right) p_1^2+\left(p_2^2-q^2\right)^3 \left(2 p_2^2+q^2\right)
		\nonumber\\
		P_6=&\ 3 p_1^8+\left(24 p_2^2-11 q^2\right) p_1^6+\left(-54 p_2^4-31 q^2 p_2^2+15 q^4\right) p_1^4\nn\\&+\left(p_2^2-q^2\right) \left(24 p_2^4-7 q^2 p_2^2+9 q^4\right) p_1^2+\left(p_2^2-q^2\right)^3 \left(3 p_2^2-2 q^2\right)
		\nonumber\\
		P_7=&\ -q^6+5 \left(p_1^2+p_2^2\right) q^4-7 \left(p_1^4+p_2^4\right) q^2+3 \left(p_1^2-p_2^2\right)^2 \left(p_1^2+p_2^2\right)
		\nonumber\\
		P_8=&\	 p_1^6-\left(p_2^2+q^2\right) p_1^4-\left(p_2^4-6 q^2 p_2^2+q^4\right) p_1^2+\left(p_2^2-q^2\right)^2 \left(p_2^2+q^2\right)
		\nonumber\\
		P_9=&\ 3 p_1^6+3 \left(p_2^2-3 q^2\right) p_1^4+\left(-5 p_2^4-8 q^2 p_2^2+9 q^4\right) p_1^2-\left(p_2^2-q^2\right)^2 \left(p_2^2+3 q^2\right)
		\nonumber\\
		P_{10}=&\ p_1^8+2 \left(9 p_2^2-2 q^2\right) p_1^6+\left(-9 p_2^4-39 q^2 p_2^2+6 q^4\right) p_1^4\nn\\&-2 \left(p_2^2-q^2\right) \left(5 p_2^4+10 q^2 p_2^2-2 q^4\right) p_1^2-\left(p_2^2-q^2\right)^3 q^2
		\nonumber\\
		P_{11}=&\ p_1^6+2 \left(p_2^2+q^2\right) p_1^4+\left(-7 p_2^4+34 q^2 p_2^2-7 q^4\right) p_1^2+4 \left(p_2^2-q^2\right)^2 \left(p_2^2+q^2\right)
		\nonumber\\
		P_{12}=&-p_1^6+\left(3 q^2-7 p_2^2\right) p_1^4+\left(5 p_2^4+6 q^2 p_2^2-3 q^4\right) p_1^2\nn\\&+\left(p_2^2-q^2\right)^2 \left(3 p_2^2+q^2\right)
		\nonumber\\
		P_{13}=&\ 4 q^8-7 \left(2 p_1^2+p_2^2\right) q^6+\left(18 p_1^4+5 p_2^2 p_1^2+3 p_2^4\right) q^4\nn\\&-\left(p_1^2-p_2^2\right) \left(10 p_1^4+p_2^2 p_1^2-p_2^4\right) q^2+\left(p_1^2-p_2^2\right)^3 \left(2 p_1^2-p_2^2\right)
		\nonumber\\
		P_{14}=&\ p_1^8-\left(5 p_2^2+q^2\right) p_1^6+\left(9 p_2^4+2 q^2 p_2^2+3 q^4\right) p_1^4\nn\\&+\left(-7 p_2^6+9 q^2 p_2^4+5 q^4 p_2^2-7 q^6\right) p_1^2+2 \left(p_2^2-q^2\right)^3 \left(p_2^2-2 q^2\right)
		\nonumber\\
		P_{15}=&\ q^6(35-12 \gamma ) +3 (-31+12 \gamma ) \left(p_1^2+p_2^2\right) q^4+(9 (9-4 \gamma ) p_1^4+2 (17-12 \gamma ) p_2^2 p_1^2\nn\\&+9 (9-4 \gamma ) p_2^4) q^2+(-23+12 \gamma ) \left(p_1^2-p_2^2\right)^2 \left(p_1^2+p_2^2\right).
	\end{align}
	
}

\section{Scalar integrals for massive quarks and on-shell gluons }
\label{scalars}
For the one-point function, or massive tadpole $\mathcal  A_0 (m^2)$, the massive bubble $ B_0 (s, m^2) $  and  the massive three-point function $\mathcal C_0 (s, s_1, s_2, m^2)$, with $s=q^2, s_1=p_1^2, s_2=p_2^2$.  We have defined
{\allowdisplaybreaks
	\bea
	\mathcal A_0 (m^2) &=& \frac{1}{i \pi^2}\int d^n l \, \frac{1}{l^2 - m^2}
	= m^2 \left [ \frac{1}{\bar \eps} + 1 - \log \left( \frac{m^2}{\mu^2} \right )\right],\\
	B_0 (k^2, m^2) &=&  \frac{1}{i \pi^2} \int d^n l \, \frac{1}{(l^2 - m^2) \, ((l - k )^2 - m^2 )} \nn \\
	C_0 (s, s_1, s_2, m^2) &=&
	\frac{1}{i \pi^2} \int d^n l \, \frac{1}{(l^2 - m^2) \, ((l -q )^2 - m^2 ) \, ((l + p )^2 - m^2 )} \nn \\
	&=&- \frac{1}{ \sqrt \sigma} \sum_{i=1}^3 \left[Li_2 \frac{b_i -1}{a_i + b_i}   - Li_2 \frac{- b_i -1}{a_i - b_i} + Li_2 \frac{-b_i +1}{a_i - b_i}  - Li_2 \frac{b_i +1}{a_i + b_i}
	\right],
	\label{C0polylog}
	\eea
	with
	\bea
	a_i = \sqrt {1- \frac{4 m^2}{s_i }} \qquad \qquad
	b_i = \frac{- s_i + s_j + s_k }{\sqrt{ \sigma}},
	\eea
}
where $s_3=s$ and in the last equation $i=1,2,3$ and $j, k\neq i$. \\
The one-point and two-point functions  written before in  $n=4 - 2 \, \eps$ are divergent in dimensional regularization with the singular parts given by
\bea
\mathcal A_0 (m^2) ^{sing.}  \rightarrow  \frac{1}{\bar \eps} \, m^2,  \qquad \qquad
B_0 (s, m^2) ^{sing.}  \rightarrow  \frac{1}{\bar \eps} ,
\eea
with
\bea
\frac{1}{\bar \eps} = \frac{1}{\eps} - \g - \ln \pi
\label{bareps}
\eea
We use two finite combinations of scalar functions given by
\bea
&&  B_0 (s, m^2) \, m^2 - \mathcal A_0 (m^2) =  m^2 \left[ 1 - a_3 \log \frac{a_3 +1}{a_3 - 1}  \right] , \\
&& \mathcal D_i \equiv \mathcal D_i (s, s_i,  m^2) =
B_0 (s, m^2) -  B_0 (s_i, m^2) =  \left[ a_i \log\frac{a_i +1}{a_i - 1}
- a_3 \log \frac{a_3 +1}{a_3 - 1}  \right] \qquad i=1,2.
\label{D_i}
\nn \\
\eea
%---------------
The scalar integrals $ C_0 (s, 0,0,m^2) $ and $\mathcal D (s, 0, 0, m^2)$ are the $ \{ s_1 \rightarrow 0$, $s_2 \rightarrow 0 \} $ limits of the generic functions $C_0(s,s_1,s_2,m^2)$ and $\mathcal D_1(s,s_1,m^2)$
\bea
C_0 (s, 0,0,m^2) &=& \frac{1}{2 s} \log^2 \frac{a_3+1}{a_3-1}, \\
\mathcal D (s, 0, 0, m^2) &=& \mathcal D_1 (s,0,m^2)= \mathcal D_2(s,0,m^2) =
\left[ 2 - a_3 \log \frac{a_3+1}{a_3-1}\right].
\eea
The singularities in \(1/\bar \epsilon\) and the dependence on the renormalization scale \(\mu\) cancel out when considering the difference between the two \(\mathcal{B}_0\) functions. As a result, the \(\mathcal{D}_i\)'s are well-defined, and the three-point master integral is convergent.

The renormalized scalar integrals  in the modified minimal subtraction scheme named $\overline{MS}$ are defined as
\beq
B_0^{\overline{MS}}(s,0) =  2 - L_s,\\
\eeq
\beq
B_0^{\overline{MS}}(0,0) =  \frac{1}{\omega},\\
\label{B0masslessOSx}
\eeq
\beq
\mathcal C_0(s,0,0,0) = \frac{1}{s}\left[ \frac{1}{\omega^2} + \frac{1}{\omega} L_s + \frac{1}{2} L_s^2 - \frac{\pi^2}{12}  \right],
\label{C0masslessOS}
\eeq
where
\bea
L_s \equiv \log \left( - \frac{s}{\mu^2} \right ) \qquad \qquad s<0.
\eea
We have set  the space-time dimensions to $n = 4 + 2 \omega$ with $\omega > 0$. The $1/\omega$ and $1/\omega^2$ singularities in Eqs.~(\ref{B0masslessOSx}) and (\ref{C0masslessOS}) are infrared divergencies due to the zero mass of the gluons.

\section{The 13 Form factor decomposition}
\label{genbasis1}
\begin{table}
	$$
	\begin{array}{|c|c|}\hline
		i & t_i^{\mu\nu\alpha\beta}(p,q)\\ \hline\hline
		1 &
		\left(k^2 g^{\mu\nu} - k^{\mu } k^{\nu}\right) u^{\alpha\beta}(p.q)\\ \hline
		2 &
		\left(k^2g^{\mu\nu} - k^{\mu} k^{\nu}\right) w^{\alpha\beta}(p.q)  \\ \hline
		3 & \left(p^2 g^{\mu\nu} - 4 p^{\mu}  p^{\nu}\right)
		u^{\alpha\beta}(p.q)\\ \hline
		4 & \left(p^2 g^{\mu\nu} - 4 p^{\mu} p^{\nu}\right)
		w^{\alpha\beta}(p.q)\\ \hline
		5 & \left(q^2 g^{\mu\nu} - 4 q^{\mu} q^{\nu}\right)
		u^{\alpha\beta}(p.q)\\ \hline
		6 & \left(q^2 g^{\mu\nu} - 4 q^{\mu} q^{\nu}\right)
		w^{\alpha\beta}(p.q) \\ \hline
		7 & \left[p\cdot q\, g^{\mu\nu}
		-2 (q^{\mu} p^{\nu} + p^{\mu} q^{\nu})\right] u^{\alpha\beta}(p.q) \\ \hline
		8 & \left[p\cdot q\, g^{\mu\nu}
		-2 (q^{\mu} p^{\nu} + p^{\mu} q^{\nu})\right] w^{\alpha\beta}(p.q)\\ \hline
		9 & \left(p\cdot q \,p^{\alpha}  - p^2 q^{\alpha}\right)
		\big[p^{\beta} \left(q^{\mu} p^{\nu} + p^{\mu} q^{\nu} \right) - p\cdot q\,
		(g^{\beta\nu} p^{\mu} + g^{\beta\mu} p^{\nu})\big]  \\ \hline
		10 & \big(p\cdot q \,q^{\beta} - q^2 p^{\beta}\big)\,
		\big[q^{\alpha} \left(q^{\mu} p^{\nu} + p^{\mu} q^{\nu} \right) - p\cdot q\,
		(g^{\alpha\nu} q^{\mu} + g^{\alpha\mu} q^{\nu})\big]  \\ \hline
		11 & \left(p\cdot q \,p^{\alpha} - p^2 q^{\alpha}\right)
		\big[2\, q^{\beta} q^{\mu} q^{\nu} - q^2 (g^{\beta\nu} q^ {\mu}
		+ g^{\beta\mu} q^{\nu})\big]  \\ \hline
		12 & \big(p\cdot q \,q^{\beta} - q^2 p^{\beta}\big)\,
		\big[2 \, p^{\alpha} p^{\mu} p^{\nu} - p^2 (g^{\alpha\nu} p^ {\mu}
		+ g^{\alpha\mu} p^{\nu})\big] \\ \hline
		13 & \big(p^{\mu} q^{\nu} + p^{\nu} q^{\mu}\big)g^{\alpha\beta}
		+ p\cdot q\, \big(g^{\alpha\nu} g^{\beta\mu}
		+ g^{\alpha\mu} g^{\beta\nu}\big) - g^{\mu\nu} u^{\alpha\beta} \\
		& -\big(g^{\beta\nu} p^{\mu}
		+ g^{\beta\mu} p^{\nu}\big)q^{\alpha}
		- \big (g^{\alpha\nu} q^{\mu}
		+ g^{\alpha\mu} q^{\nu }\big)p^{\beta}  \\ \hline
	\end{array}
	$$
	\caption{The basis of 13 fourth rank tensors satisfying the vector current conservation on the external lines with momenta $p$ and $q$. \label{genbasistab}}
\end{table}

\label{ffdc}
The set of the $13$ tensors $t_i$ introuced in \cite{Giannotti:2008cv} is linearly independent for generic $k^2, p^2, q^2$
different from zero. Five of the $13$ are Bose symmetric, 
\be
t_i^{\mu\nu\alpha\beta}(p,q) = t_i^{\mu\nu\beta\alpha}(q,p)\,,\qquad i=1,2,7,8,13\,,
\ee
while the remaining eight tensors are Bose symmetric pairwise
\bea
\label{pair}
&&t_3^{\mu\nu\alpha\beta}(p,q) = t_5^{\mu\nu\beta\alpha}(q,p)\,,\\
&&t_4^{\mu\nu\alpha\beta}(p,q) = t_6^{\mu\nu\beta\alpha}(q,p)\,,\\
&&t_9^{\mu\nu\alpha\beta}(p,q) = t_{10}^{\mu\nu\beta\alpha}(q,p)\,,\\
&&t_{11}^{\mu\nu\alpha\beta}(p,q) = t_{12}^{\mu\nu\beta\alpha}(q,p)\,.
\eea
In the set are present two tensor structures
\bes\bea
&&u^{\alpha\beta}(p,q) \equiv (p\cdot q) g^{\alpha\beta} - q^{\alpha}p^{\beta}\,,\\
&&w^{\alpha\beta}(p,q) \equiv p^2 q^2 g^{\alpha\beta} + (p\cdot q) p^{\alpha}q^{\beta}
- q^2 p^{\alpha}p^{\beta} - p^2 q^{\alpha}q^{\beta}\,,
\eea \label{uwdef}\ees
which appear in $t_1$ and $t_2$ respectively.
Each of them satisfies the Bose symmetry  requirement,
\bes\bea
&&u^{\alpha\beta}(p,q) = u^{\beta\alpha}(q,p)\,,\\
&&w^{\alpha\beta}(p,q) = w^{\beta\alpha}(q,p)\,,
\eea\ees
and vector current conservation,
\bes\bea
&&p_{\alpha} u^{\alpha\beta}(p,q) = 0 = q_{\beta}u^{\alpha\beta}(p,q)\,,\\
&&p_{\alpha} w^{\alpha\beta}(p,q) = 0 = q_{\beta}w^{\alpha\beta}(p,q)\,.
\eea\ees
They are obtained from the variation of gauge invariant quantities
$F_{\mu\nu}F^{\mu\nu}$ and $(\partial_{\mu} F^{\mu}_{\ \,\lambda})(\partial_{\nu}F^{\nu\lambda})$

\bea
&&u^{\alpha\beta}(p,q) = -\frac{1}{4}\int\,d^4x\,\int\,d^4y\ e^{ip\cdot x + i q\cdot y}\ 
\frac{\delta^2 \{F_{\mu\nu}F^{\mu\nu}(0)\}} {\delta A_{\alpha}(x) A_{\beta}(y)} \,,
\label{one}\\
&&w^{\alpha\beta}(p,q) = \frac{1}{2} \int\,d^4x\,\int\,d^4y\ e^{ip\cdot x + i q\cdot y}\
\frac{\delta^2 \{\partial_{\mu} F^{\mu}_{\ \,\lambda}\partial_{\nu}F^{\nu\lambda}(0)\}} 
{\delta A_{\alpha}(x) A_{\beta}(y)}\,.\label{two}
\eea\label{three}
All the $t_i$'s are transverse in their photon indices
\bea
q^\alpha t_i^{\mu\nu\alpha\beta}=0  \qquad p^\beta t_i^{\mu\nu\alpha\beta}=0.
\eea
$t_2\ldots t_{13}$ are traceless, $t_1$ and $t_2$ have trace parts in $d=4$. The corresponding form factors $F_i$ are related to the  $A^{(q)}_i$. 
They are conveniently expressed in terms of the momenta $(p_1,p_2,p_3)$ in the form
\bea
A_1^{(q)}&=&4(F_7-F_3-F_5)-2p_2^2F_9-2p_3^2F_{10}\nn \\
A_2^{(q)}&=&2(p_1^2-p_2^2-p_3^2)(F_7-F_5-F_3)-4p_2^2p_3^2(F_6-F_8+F_4)-2F_{13}\nn\\
A_3^{(q)}&=&p_3^2(p_1^2-p_2^2-p_3^2)F_{10}-2p_2^2\,p_3^2 F_{12}-2F_{13}\nn\\
A_3^{(q)}(p_2\leftrightarrow p_3)&=&p_2^2(p_1^2-p_2^2-p_3^2)F_9-2p_2^2p_3^2F_{11}-2F_{13}\nn\\
A_4^{(q)}&=&(p_1^2-p_2^2-p_3^2)F_{13}.
\label{mapping1}
\eea

\chapter{Appendix to Chapter 5}

\section{Absence of a particle pole for $q^2\to 0$ $(s_1,s_2,m^2)\neq 0$}
Using the Schouten relations
\begin{align}
	&p_2^{\mu_1} {\epsilon }^{\mu_2 \mu_3 p_1 p_2}= p_2^{\mu_2} {\epsilon }^{\mu_1 \mu_3 p_1 p_2}-p_2^{\mu_3} {\epsilon }^{\mu_1 \mu_2 p_1 p_2}-p_2^2 {\epsilon }^{\mu_1 \mu_2 \mu_3 p_1}+\left(  p_1 \cdot p_2\right) {\epsilon }^{\mu_1 \mu_2 \mu_3 p_2}\nonumber\\&p_1^{\mu_2} {\epsilon }^{\mu_1 \mu_3 p_1 p_2}= p_1^{\mu_1} {\epsilon }^{\mu_2 \mu_3 p_1 p_2}+p_1^{\mu_3} {\epsilon }^{\mu_1 \mu_2 p_1 p_2}-p_1^2 {\epsilon }^{\mu_1 \mu_2 \mu_3 p_2}+\left( p_1 \cdot  p_2\right) {\epsilon }^{\mu_1 \mu_2 \mu_3 p_1}
\end{align}
and by a direct perturbative computation of the vertex, we can express the $AVV$ as follows
\begin{align}
	\label{nopole}
	\braket{J^{\m_1}(p_1)J^{\m_2}(p_2)J_A^{\m_3}(q)}=F_1\left( {\epsilon }^{\mu_1 \mu_2 p_1 p_2} p_1^{\mu_3} + {\epsilon }^{\mu_1 \mu_2 p_1 p_2} p_2^{\mu_3}\right)\\ \nn
	+ F_2\left({\epsilon }^{\mu_2 \mu_3 p_1 p_2} p_1^{\mu_1}-{\epsilon }^{\mu_1 \mu_2 \mu_3 p_2} s_1\right) +  F_3\left({\epsilon }^{\mu_1 \mu_3 p_1 p_2} p_2^{\mu_2}-{\epsilon }^{\mu_1 \mu_2 \mu_3 p_1} s_2\right)
\end{align}
with

\begin{align}
	\label{F1}
	F_1=&\frac{1}{64 \pi ^2 \l(s,s_1,s_2)^2}\Bigl(2 \left(s^2-2 (s_1+s_2) s+(s_1-s_2)^2\right) (s-s_1-s_2)\nonumber\\&+2 \left((s_1+s_2) s^2-2 \left(s_1^2-4 s_2 s_1+s_2^2\right) s+(s_1-s_2)^2 (s_1+s_2)\right) B_0\left(s,m^2\right)\nonumber\\&-2 s_1 \left((s-s_1)^2-5 s_2^2+4 (s+s_1) s_2\right) B_0\left(s_1,m^2\right)\nonumber\\&-2 s_2 \left(s^2+4 s_1 s-2 s_2 s-5 s_1^2+s_2^2+4 s_1 s_2\right) B_0\left(s_2,m^2\right)\nonumber\\&+4 \bigl(m^2 (s-s_1-s_2) \left(s^2-2 (s_1+s_2) s+(s_1-s_2)^2\right)\nonumber\\&-s_1 s_2 \left(-2 s^2+(s_1+s_2) s+(s_1-s_2)^2\right)\bigl) C_0\left(s,s_1,s_2,m^2\right)\Bigl) 
\end{align}

\begin{align}
	\label{F2}
	F_2=&\frac{1}{2 \pi ^2 \l(s,s_1,s_2)^2}\Bigl(\left(s^2-2 (s_1+s_2) s+(s_1-s_2)^2\right) (s-s_1+s_2)\nonumber\\&+s \left((s-s_1)^2-5 s_2^2+4 (s+s_1) s_2\right) B_0\left(s,m^2\right)\nonumber\\&-\left(s^3-(2 s_1+s_2) s^2+\left(s_1^2+8 s_2 s_1-s_2^2\right) s+(s_1-s_2)^2 s_2\right) B_0\left(s_1,m^2\right)\nonumber\\&+\left(-5 s^2+4 (s_1+s_2) s+(s_1-s_2)^2\right) s_2 B_0\left(s_2,m^2\right)\nonumber\\&+2 \bigl((s-s_1+s_2) \left((s-s_1)^2+s_2^2-2 (s+s_1) s_2\right) m^2\nonumber\\&+s s_2 \left(s^2+s_1 s-2 s_2 s-2 s_1^2+s_2^2+s_1 s_2\right)\bigl) C_0\left(s,s_1,s_2,m^2\right)\Bigl) 
\end{align}

\begin{align}
	\label{F3}
	F_3=&-\frac{1 }{2 \pi ^2 \l(s,s_1,s_2)^2}\Bigl(\left(s^2-2 (s_1+s_2) s+(s_1-s_2)^2\right) (s+s_1-s_2)\nonumber\\&+s \left(s^2+4 s_1 s-2 s_2 s-5 s_1^2+s_2^2+4 s_1 s_2\right) B_0\left(s,m^2\right)\nonumber\\&+s_1 \left(-5 s^2+4 (s_1+s_2) s+(s_1-s_2)^2\right) B_0\left(s_1,m^2\right)\nonumber\\&-\left((s+s_1) (s-s_1)^2+(s+s_1) s_2^2-2 \left(s^2-4 s_1 s+s_1^2\right) s_2\right) B_0\left(s_2,m^2\right)\nonumber\\&+2 \bigl((s+s_1-s_2) \left((s-s_1)^2+s_2^2-2 (s+s_1) s_2\right) m^2\nonumber\\&+s s_1 \left((s-s_1)^2-2 s_2^2+(s+s_1) s_2\right)\bigl) C_0\left(s,s_1,s_2,m^2\right)\Bigl). 
\end{align}
Before performing some special kinematical limits on this amplitude, 
we comment on its UV finiteness, which is not apparent from the results for the three form factors $F_1, F_2$ and $F_3$ shown above. Notice that in these expressions the bare self-energies $B_0$ can be interpreted, indifferently, 
as either renormalized or bare. The renormalization removes the $1/\epsilon_{UV}$ poles of these integrals by defining

\beq
B_0(q^2,m^2)\to B^R_0(q^2,m^2)\equiv B_0(q^2,m^2) -\frac{1}{\epsilon_{UV}}
\eeq
One can explicitly check that by combining all such self-energy contributions and collecting all the polynomial coefficients in front, 
such ultraviolet divergences cancel, as expected. For this reason, in these expressions one can automatically replace in 
\eqref{F1}, \eqref{F2}, \eqref{F3} $B_0$ with $B^R_0$.
At this point, in order to compute the residue at $q^2=0$, given the absence of poles in the representation 
\eqref{nopole}, we need just to observe that in the expressions above for $F_1$, $F_2$ and $F_3$ no pole in $1/q^2$ are present as far as $s_1$ and $s_2$ and $m$ are nonzero. Therefore one can perform the light-cone limit, obtaining
\begin{equation}
	\lim_{q^2\to 0} q^2 \braket{J^{\m_1}(p_1)J^{\m_2}(p_2)J^{\m_3}(q)}=0.
\end{equation}
This result shows that there is no massless particle pole in the vertex for general kinematics. 
This result remains valid in the infrared limit of \( q \) (i.e. $q^\lambda\to 0$) as well, although the analysis simplifies significantly in this case. As already pointed out, the vanishing of the vertex in the infrared is related to the derivative coupling of 
the pseudoscalar mode, as for an ordinary NG mode.  \\

\subsection{The continuum $\textrm{disc}\, C_0$ }
The contribution from the continuum is obtained from the s- channel cut by the ordinary cutting rules
\beq
\label{disc}
\textrm{disc}\, C_0 = \frac{1}{i \pi^2} \int d^4 k \, \frac{(-2 \pi i)^2 \delta_+(k^2 - m^2) \delta_+((k - q)^2 - m^2)}
{(k - p_1)^2 - m^2}
\eeq
that identify the light-cone character of the intermediate state.  
This integral is evaluated in the rest frame of the invariant \(q = (q_0, 0, 0, 0)\). It corresponds to a cut contribution in which the axial-vector line decays into two back-to-back on-shell fermions. This pair can be effectively treated as a pseudoscalar intermediate state, which subsequently decays into two off-shell spin-1 particles.  To evaluate the integral, we can use polar coordinates of $\vec{k}$ around the momentum $|p_1|$ together with the 
relations 
\begin{figure}
	\label{figg2}
	\centering
	\begin{tikzpicture}
		
		% Thick axes
		\draw[thick, ->] (-1,0) -- (6,0) node[right] {$\text{Re}(s)$};
		\draw[thick, ->] (0,-3) -- (0,3) node[above] {$\text{Im}(s)$};
		
		% Pole at the origin
		\filldraw[black] (0,0) circle (3pt) node[below left] {$0$};
		
		% Branch cut from 4m^2 to infinity
		\draw[thick, decorate, decoration={zigzag, amplitude=2pt, segment length=4pt}] (2,0) -- (5.9,0);
		\node at (2, -0.3) {$4m^2$};
		
		% Spikes at s=0 with different colors
		\draw[very thick, ->, black] (0,0) -- (0,-2.5); % Red spike closer to q^2=0 with 
		\draw[very thick, ->, red] (0,0) -- (0,1.5);     % Black spike with arrow

	\end{tikzpicture}
	\caption{Analyticity region of the off-shell anomaly form factor in the complex \( s\equiv q^2 \)-plane, with two bold spikes at \(s=0\).  In this case there are cancelations between the cut and the contribution of the black spike, whose strength depends on all the external invariants, after integration over the $q^2>0$ region. The sum rule then allows one to identify the red spike at $s=0$ as an anomaly pole.} 
\end{figure}
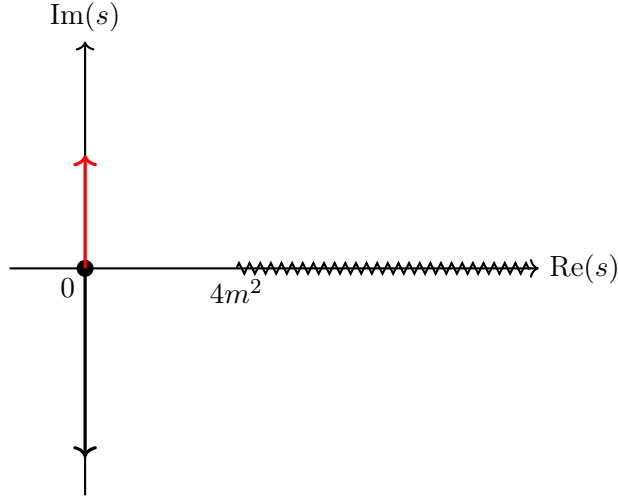

\beq
\delta_+(k^2 - m^2)=\frac{\delta( k_0- \sqrt{ |k|^2 + m^2}}{2 \sqrt{ |k|^2 + m^2} }),\qquad 
\delta_+((k - q)^2 - 4 m^2))=\frac{1}{4 q_0  \sqrt{ q^2 -4 m^2}}\delta(|k|-\frac{1}{2} \sqrt{ q^2 -4 m^2})
\eeq
with 
\beq
k_0=\frac{\sqrt{q^2}}{2} \qquad |k|=\frac{1}{2}\sqrt{ q^2 -4 m^2}.
\eeq
The integral gets reduced to a simple angular form 
\beq
\int_{-1}^{1} \frac{d\cos\theta}{A + B \cos\theta} =\frac{1}{B} \log\left( \frac{A + B}{A-B}\right)\qquad \textrm{where} 
\qquad  A=s_1 + s_2 - q^2, \qquad  B=2 \lambda^{1/2}(q^2,s_1,s_2)\sqrt{q^2 - 4 m^2}.
\eeq
Collecting all the terms together, the spectral density takes the simple form 
\beqa
\Delta C_0(s,s_1,s_2,m^2)\equiv \Im\, C_0(s,s_2,s_2,m^2)&=&-\frac{\pi}{4 s}\log \frac{\chi_+(s,s_1,s_2,m^2)}{\chi_-(s)}\,\theta(q^2- 4 m^2)\nonumber \\
\chi_\pm(s,s_1,s_2,m^2) &=& s_1 + s_2 - q^2 \pm 2 \lambda^{1/2}(q^2,s_1,s_2)\sqrt{q^2 - 4 m^2}.
\eeqa
It is also convenient to introduce the variables	

\beq
z=\frac{1}{2 s } \left(\sqrt{\lambda} + s_1 - s_2 +s\right) \qquad \bar z=z=\frac{1}{2 s } \left(- \sqrt{\lambda} + s_1 - s_2 +s\right)
\eeq
with
\beq
w=\frac{1}{2} \left(1+\sqrt{1-\frac{4 m^2}{s}}\right)	\qquad \bar w=\frac{1}{2} \left(1-\sqrt{1-\frac{4 m^2}{s}}\right)
\eeq	
in order to re-express this density in the form (recall that $\Delta C_0= \textrm{disc}\, C_0/( 2 i)$)

\begin{equation}
	\label{d2}
	\Delta{C_0}(s,s_1,s_2,m^2)=-\frac{  \pi  \log \left(\frac{m^2-s (w-\bar z) (\bar w-z)}{m^2-s (w-z) (\bar w-\bar z)}\right)}{s (z-\bar z)}\,\,\, \theta(s-4m^2).
\end{equation}

For the derivation of the explicit form of the spectral density  we also need $C_0$ at $s=0$  
\beqa
\label{d1}
C_0(0,p_1^2,p_2^2,m^2) &=& \lim_{q^2\to 0} C_0(q^2,p_1^2,p_2^2, m^2)\nonumber \\
&=&\frac{\log ^2\left(\frac{-p_1^2 +\sqrt{p_1^2  \left(p_1^2 -4 m^2\right)}+2 m^2}{2 m^2}\right)-\log ^2\left(\frac{-p_2^2 +\sqrt{p_2^2  \left(p_2^2 -4 m^2\right)}+2 m^2}{2 m^2}\right)}{2 (p_1^2 -p_2^2 )}
\eeqa
that can be inserted in the expression of the spectral density  
\begin{align}\label{ds1}
	\textrm{disc}\,\Phi_0(q^2,p_1^2,p_2^2, m^2)= -\frac{1}{2\pi^2}\biggl(&4 i \pi m^2 \, C_0 (0,p_1^2,p_2^2, m^2)\, \delta(q^2)  +2 \pi i\, \delta(q^2)\nn\\&
	-\frac{2 m^2}{q^2}\textrm{disc} \,C_0(q^2,p_1^2,p_2^2, m^2) \biggl),
\end{align}
with the first and third contributions given respectively by Eqs. \eqref{d1} and \eqref{d2} (where $\textrm{disc} \, C_0=2 i \,\Delta C_0$). A direct evaluation of its integral in the $q^2>0$ region shows the presence of an interesting pattern of cancelations. An explicit computation, using the inverse Laplace transform, shows that the following identity holds
\beq
\label{id}
\frac{2 i m^2}{\pi}  \, C_0 (0,p_1^2,p_2^2, m^2)=\int_{4 m^2}^\infty ds \, \rho(s,s_1,s_2,m^2)
\eeq
where
\beqa
\label{ds2}
\rho(s,s_1,s_2,m^2)&\equiv &\frac{ m^2}{\pi^2q^2}\textrm{disc} \,C_0(q^2,p_1^2,p_2^2, m^2)\nn\\
&=&-\frac{ 2i m^2  \log \left(\frac{m^2-s (w-\bar z) (\bar w-z)}{m^2-s (w-z) (\bar w-\bar z)}\right)}{\pi s^2 (z-\bar z)}\theta(q^2 - 4 m^2),
\eeqa
indicating that the sum rule can be thought of as being saturated by the second constant term on the rhs of \eqref{ds1}.
Therefore, if we integrate \eqref{ds1} to compute the sum rule and use \eqref{ds2}, the first and the third term combined contribute zero to the integral, leaving only the constant pole to saturate the sum rule
\beq
\int_0^\infty ds \,\textrm{disc}\,\Phi_0(q^2,p_1^2,p_2^2, m^2)=\frac{i}{\pi} \, \int_0^\infty ds \delta(s) = \frac{i}{\pi}.
\eeq
Further insight into this behaviour is obtained by analysing \eqref{id} under a global rescaling of all the external invariants and the mass parameter $m^2$. Notice that also in the off-shell correlator the spectral density is self-similar
\beq
\rho(\lambda s, \lambda s_1,\lambda s_2,\lambda m^2)=\frac{1}{\lambda} \rho( s, s_1, s_2,m^2)
\eeq
and that \eqref{id} is scale invariant if we perform a common rescaling of all the invariants and the mass on both of its sides.   Notice that in the on-shell limit $(s_1,s_2)\to (0,0)$ of \eqref{ref} , using in \eqref{d1}
\beq
\lim_{ (s_1,s_2)\to (0,0)} C_0(0,s_1,s_2,m^2)=-\frac{1}{2 m^2}
\eeq
we obtain 
\beq
\lim_{(s_1,s_2)\to (0,0)} \frac{2 i m^2}{\pi} \, C_0 (0,p_1^2,p_2^2, m^2)=- \frac{ i} {\pi} = \int_{4 m^2}^\infty ds \, \rho(s,m^2) 
\eeq
and we recover the sum rule that we have discussed above in the on-shell case. From this relation, if we perform the 
$m^2\to 0$ limit, we derive the spectral flow shown in Fig. 1, with  $\rho(s,m^2) $ converging to a delta function at $s=0$
\beq
\lim_{m\to 0} \rho(s,m^2)= -\frac{i}{\pi}\delta(s).
\eeq
We have seen that in the on-shell case the cancelation between the contributions of the two spikes at $s=0$, leaves the sum rule to be saturated by the cut, controlled by the mass parameter $m$. We mention again that the same result, however, can be interpreted in a different way if we perform an integration of both sides of \eqref{ds1}. After integration, the first and the third contribution 
identically cancel, as shown by \eqref{id}, and the entire sum rule appears to be saturated just by the $1/q^2$ isolated 
pole. In the off-shell case, a similar cancelation takes place after integration, and again, the sum rule appears to be saturated just by the pole, with a constant residue fixed only by the anomaly coefficient. \\
It is clear that we are confronted, for this interaction, with a rather special behaviour, which can be attributed to the topological properties of the vertex, not shared by other interactions.

\section{Appendix: Evaluation at the Anomalous Thresholds}
\label{app:AnomalousThresholds}

Here we collect the explicit expressions used in the cancellation analysis at the anomalous thresholds \(s = s_\pm\). These involve the evaluation of the scalar triangle integral \(C_0\) at those kinematic points.

\begin{align}
	C_0(s_-,s_1,s_2,m^2) &= \frac{1}{{s_1}{s_2}\, s_-^2}\Biggl(\sqrt{{s_1}} \Biggl(\sqrt{{s_2} \left({s_2}-4 m^2\right)} \left(\sqrt{{s_1}}-\sqrt{{s_2}}\right)\nn\\
	&\times \log \left(\frac{\sqrt{{s_2} \left({s_2}-4 m^2\right)}+2 m^2-{s_2}}{2 m^2}\right)\nonumber\\
	&+\sqrt{{s_2}} \sqrt{\left(-2 \sqrt{{s_1}} \sqrt{{s_2}}+{s_1}+{s_2}\right) \left(-4 m^2-2 \sqrt{{s_1}} \sqrt{{s_2}}+{s_1}+{s_2}\right)} \log \left(S_1\right)\Biggl)\nonumber\\
	&+\sqrt{{s_2}} \sqrt{{s_1} \left({s_1}-4 m^2\right)} \left(\sqrt{{s_2}}-\sqrt{{s_1}}\right) \log \left(\frac{\sqrt{{s_1} \left({s_1}-4 m^2\right)}+2 m^2-{s_1}}{2 m^2}\right)\Biggl)
\end{align}

\begin{align}
	C_0(s_+,s_1,s_2,m^2) &= \frac{1}{{s_1}{s_2} s_+^2}\Biggl(\sqrt{{s_1}} \Biggl(\sqrt{{s_2} \left({s_2}-4 m^2\right)} \left(\sqrt{{s_1}}+\sqrt{{s_2}}\right)\nn\\
	&\times \log \left(\frac{\sqrt{{s_2} \left({s_2}-4 m^2\right)}+2 m^2-{s_2}}{2 m^2}\right)\nonumber\\
	&-\sqrt{{s_2}} \sqrt{\left(\sqrt{{s_1}}+\sqrt{{s_2}}\right)^4-4 m^2 \left(\sqrt{{s_1}}+\sqrt{{s_2}}\right)^2} \log \left(S_2\right)\Biggl)\nonumber\\
	&+\sqrt{{s_2}} \sqrt{{s_1} \left({s_1}-4 m^2\right)} \left(\sqrt{{s_1}}+\sqrt{{s_2}}\right) \log \left(\frac{\sqrt{{s_1} \left({s_1}-4 m^2\right)}+2 m^2-{s_1}}{2 m^2}\right)\Biggl),
\end{align}

where
\begin{align}
	S_1 &= \frac{\sqrt{\left(-2 \sqrt{{s_1}} \sqrt{{s_2}}+{s_1}+{s_2}\right) \left(-4 m^2-2 \sqrt{{s_1}} \sqrt{{s_2}}+{s_1}+{s_2}\right)}+2 m^2-\left(\sqrt{{s_1}}-\sqrt{{s_2}}\right)^2}{2 m^2}, \\
	S_2 &= \frac{\sqrt{\left(\sqrt{{s_1}}+\sqrt{{s_2}}\right)^4-4 m^2 \left(\sqrt{{s_1}}+\sqrt{{s_2}}\right)^2}+2 m^2-\left(\sqrt{{s_1}}+\sqrt{{s_2}}\right)^2}{2 m^2}.
\end{align}

Using these expressions, one can show:
\[
\Phi_{\text{ATT}}^{(\lambda)}(q^2 = s_\pm) = 0.
\]

\subsection{The massless and the on-shell limits do not commute}
An interesting limit of the amplitude arises when either the fermion mass or the external invariants \( s_1 \) and \( s_2 \) are sent to zero. As evident from our previous discussions, these two limits do not commute for an obvious reason. If the on-shell limit on \( s_1 \) and \( s_2 \) is taken first, the interaction is reduced to its longitudinal sector. As shown above, the sum rule then causes the branch cut to collapse into a pole when approaching the conformal limit by setting \( m \to 0 \). In this case, the anomaly form factor becomes the only surviving form factor in the vertex, with a nonzero residue as \( q^2 \to 0 \), equaling the numerical coefficient of the anomaly.
Conversely, if we first take the \( m \to 0 \) limit while keeping the vector lines off-shell, and only afterward send \( q^2 \to 0 \), the residue vanishes. This happens because, in this order of limits, both the longitudinal and transverse structures contribute. In the following, we will illustrate this point in detail.
\\
We start from the $m=0$ case. To show the patterns of cancelations we assume that the two external vector lines are off-shell. 
\begin{itemize}
	\item  $m=0$\\
	In the $m=0$ case, a direct computation of the residue, combining all the terms in \eqref{nopole} gives 
	\begin{equation}
		\label{res}
		q^2 \braket{J^{\m_1}(p_1)J^{\m_2}(p_2)J^{\m_3}(q)}=-\frac{   \left(\text{B}^R_0(M^2,0)+M^2 C_0(0,M^2,M^2,0)-2\right) q^{\m_3}\epsilon^{\m_1 \m_2 p_1 p_2}}{8 \pi ^2}
	\end{equation}
	where we compute the scalar 3-point function at $q^2=0$ (with  $s_1=s_2=M^2 <0)$
	\begin{equation}
		C_0(0,M^2,M^2,0) =\lim_{q^2\to 0} C_0(q^2,M^2,M^2,0)= -\frac{1}{\varepsilon_{IR}  M^2}-\frac{\log \left(-\frac{\pi  \mu ^2}{M^2}\right)}{M^2}+\frac{\gamma }{M^2}+\frac{2 \log (\pi )}{M^2}
	\end{equation}
	as well as the scalar 2-point function $B_0^R$
	\begin{equation}
		\text{B}^R_0(0,M^2)\equiv \lim_{q^2\to 0} B_0^R(q^2 M^2)= \log \left(-\frac{\mu ^2}{\pi  M^2}\right)-\gamma +2,
	\end{equation}
	with $\varepsilon= d-4$ and $\gamma$ being the Euler-Mascheroni constant and $\mu$ an infrared renormalization scale.  
	\eqref{res} contains a light-cone component $q^{\mu_3}\sim q^+ n^{+\, \mu_3}$, that remains large but fixed in the 
	light-cone limit that we are considering.  At this stage, the vanishing of the residue 
	\beq
	\lim_{q^2\to 0} \left(\text{B}^R_0(q^2, M^2)+M^2 C_0(q^2,M^2,M^2)-2\right) =0
	\eeq
	eliminates such component. 
	Notice that there are no infrared divergences in $1/\varepsilon_{IR}$ in this limit.
	Therefore we derive the relation
	\begin{equation}
		\lim_{q^2\to 0}	q^2 \braket{J^{\m_1}(p_1)J^{\m_2}(p_2)J^{\m_3}(q)}=0,
	\end{equation}
	proving the absence of any particle pole in the vertex under these external kinematical constraints.
	
	\item  $m\neq 0$\\
	A similar analysis can be performed for a nonzero mass. The result is finite in the $q^2\to 0$ limit. We obtain
	\begin{align}
		q^2 \braket{J^{\m_1}(p_1)J^{\m_2}(p_2)J^{\m_3}(q)}=\frac{1  }{8 \pi ^2}q^{\m_3}\epsilon^{\m_1 \m_2 p_1 p_2} \bigl(&-\text{B}^R_0\left(M^2,m^2\right)+\text{B}^R_0\left(0,m^2\right)\nn\\&+\left(4 m^2-M^2\right) C_0\left(0,M^2,M^2,m^2\right)+2\bigl)
	\end{align}
	which is purely longitudinal due to the presence of the $q^{\m_3}\epsilon^{\m_1 \m_2 p_1 p_2}$ tensor,
	where
	\begin{equation}
		\text{B}^R_0\left(M^2,m^2\right)= \log \left(\frac{\mu ^2}{\pi  m^2}\right)+\frac{\sqrt{M^2 \left(M^2-4 m^2\right)} \log \left(\frac{\sqrt{M^2 \left(M^2-4 m^2\right)}+2 m^2-M^2}{2 m^2}\right)}{M^2}-\gamma +2,
	\end{equation}
	\begin{equation}
		\text{B}^R_0\left(0,m^2\right)= \log \left(\frac{\mu ^2}{\pi  m^2}\right)-\gamma,
	\end{equation}
	and
	\begin{equation}
		C_0\left(0,M^2,M^2,m^2\right)=\frac{\sqrt{-\left(M^2 \left(4 m^2-M^2\right)\right)} \log \left(\frac{\sqrt{M^2 \left(M^2-4 m^2\right)}+2 m^2-M^2}{2 m^2}\right)}{M^2 \left(4 m^2-M^2\right)}.
	\end{equation}
	Also in this case there is no particle pole since
	\beq
	\lim_{q^2\to 0} \left( -\text{B}^R_0\left(M^2,m^2\right)+\text{B}^R_0\left(q^2,m^2\right)+\left(4 m^2-M^2\right) 
	C_0\left(q^2,M^2,M^2,m^2\right)+2 \right)=0
	\eeq
	giving 
	\begin{align}
		\lim_{q^2\to 0} 	q^2 \braket{J^{\m_1}(p_1)J^{\m_2}(p_2)J^{\m_3}(q)}=0.
	\end{align}
\end{itemize}
The analysis shows that the transverse sector turns into a longitudinal one when we take the light-cone limit, a fact previously noted in \cite{Armillis:2009sm}. \\
Starting from the general expression \eqref{res}, setting the vector lines on-shell and keeping the mass parameter fixed, results in the absence of a particle pole in the vertex. This is consistent with the observation that, in this case, there is no pole but rather a cut in the longitudinal form factor. 
Therefore, even in the on-shell case (\(s_1 = s_2 = 0\)), where the vertex is purely longitudinal, for \(m \neq 0\), we obtain

\begin{align}
	\label{zzz}
	q^2 \braket{J^{\m_1}(p_1)J^{\m_2}(p_2)J^{\m_3}(q)}=&\frac{ \left(2 m^2 C_0\left(0,0,0,m^2\right)+1\right) q^{\m_3} \epsilon^{\m_1 \m_2 p_1 p_2}}{2 \pi ^2}
\end{align}
which has no pole since, according to \eqref{exp1} 
\begin{equation}
	C_0\left(0,0,0,m^2\right)=-\frac{1}{2m^2}
\end{equation}
and \eqref{zzz} vanishes. The sum rule, however, indicates that as $m\to 0$, the spectral density and the area law 
of the sum rule 	generate a pole, as in the conformal analysis. At the conformal point the form factor with its longitudinal tensor structure equals the entire vertex   
\begin{equation}
	\lim_{q^2\to 0} q^2 \braket{J^{\m_1}(p_1)J^{\m_2}(p_2)J^{\m_3}(q)}=\frac{1 } {2 \pi ^2} q^{\m_3} \epsilon^{\m_1 \m_2 p_1 p_2}
\end{equation}
showing a nonzero residue.\
This demonstrates that the vertex possesses a particle pole with a nonvanishing coupling \( g_{pp} \), as defined in \eqref{loc} for the general case, at \( s = 0 \). This occurs only for on-shell vector lines and a massless fermion. Notice that the limit we are considering is not a soft limit, and clearly characterizes a process that is entirely supported on the light-cone.\\
On the other hand, in the infrared case, as \( q^\lambda \to 0 \), multiplication by the invariant \( q^2 \) leads to a contribution of the form \( \epsilon^{\mu \nu p_1 p_2} \). These contributions vanish due to symmetry, given that \( p_1 + p_2 \sim 0 \). Consequently, one deduces that no particle pole exists in a general sense, except under specific kinematical conditions. While the anomaly form factor is characterized by an anomaly pole, as shown through the analysis of the conformal Ward identities, this does not translate into a general particle pole. \\
This indicates that topological vertices cannot be directly assimilated into ordinary, non-topological ones. The observed cancellations between the longitudinal and transverse structures of the interaction lead to this conclusion. Moreover, the non-commutativity between the massless and on-shell limits is a property of the entire vertex, stemming from the interrelation between these two sectors. 
These analysis show that the anomaly pole behaves as a particle pole only in the on-shell and massless case where
\begin{equation}
	\braket{J^{\m_1}(p_1)J^{\m_2}(p_2)J^{\m_3}(q)}=\frac{1} {2 \pi ^2 } \frac{q^{\m_3}}{q^2} \epsilon^{\m_1 \m_2 p_1 p_2}
\end{equation} 
is present over the entire light-cone, getting arbitrarily soft around the apex of the cone where $q^\lambda\to 0$.  

\section{The transverse form factors of the $ATT$}
\label{transverse}
\small
\begin{equation}
	\begin{aligned}
		\bar{A}_{11}=&-2 s_1 \left\{2 \lambda m^2  \left[ 21 s^3 - 3 s^2 (8 s_1 + 7 s_2) + s \left( -15 s_1^2 + 100 s_1 s_2 - 21 s_2^2 \right) + 3 (s_1 - s_2)^2 (6 s_1 + 7 s_2) \right] \right.\\&
		+ \left. s_1 \left[ 3 s^5 + s^4 (81 s_2 - 12 s_1) + 2 s^3 (9 s_1^2 - 40 s_1 s_2 - 42 s_2^2) - 4 s^2 (3 s_1^3 + 20 s_1^2 s_2 - 76 s_1 s_2^2 + 21 s_2^3) + 3 s_2 (s_1 - s_2)^4\right.\right.\\&
		+ \left.\left. s (s_1 - s_2)^2 (3 s_1^2 + 82 s_1 s_2 + 81 s_2^2)  \right]\right\}, \\
		\bar{A}_{12}=&-4 \lambda m^2 \Big[-3 s_2^2 (s^2+14 s s_1-17 s_1^2)-3 s_2 (s-s_1) (s^2-21 s s_1-6 s_1^2)+s_2^3 (7 s-24 s_1)+2 (s-s_1)^3 (s+3 s_1)-3 s_2^4\Big]\\&+2 s_2 \Big[s^6-s^5 (14 s_1+5 s_2)+s^4 (-11 s_1^2+28 s_1 s_2+10 s_2^2)+2 s^3 (52 s_1^3-102 s_1^2 s_2-5 s_2^3)+s^2 (-121 s_1^4+152 s_1^3 s_2+\\&120 s_1^2 s_2^2-28 s_1 s_2^3+5 s_2^4)+s (s_1-s_2)^2 (38 s_1^3+117 s_1^2 s_2+12 s_1 s_2^2-s_2^3)+3 s_1^2 (s_1-s_2)^4\Big],\\
		\bar{A}_{13}=&4 \lambda m^2 \Big[3 s^4+s^3 (24 s_1-7 s_2)+3 s^2 (-17 s_1^2+14 s_1 s_2+s_2^2)+3 s (s_1-s_2) (6 s_1^2+21 s_1 s_2-s_2^2)+2 (s_1-s_2)^3 (3 s_1+s_2)\Big]\\&+2 s \Big[-2 s_2^3 (5 s^3+102 s s_1^2-52 s_1^3)+s_2^4 (10 s^2+28 s s_1-11 s_1^2)-s_2 (s-s_1)^2 (s^3-12 s^2 s_1-117 s s_1^2-38 s_1^3)\\&+s_2^2 (5 s^4-28 s^3 s_1+120 s^2 s_1^2+152 s s_1^3-121 s_1^4)+3 s_1^2 (s-s_1)^4-s_2^5 (5 s+14 s_1)+s_2^6\Big],\\
		\bar{A}_{14}=&24 \Big\{ \lambda^2 m^4 \Big[s^2+s (s_1-2 s_2)-2 s_1^2+s_1 s_2+s_2^2\Big]+m^2 s_1 \Big[2 s_2^3 (16 s^3-15 s s_1^2-5 s_1^3)-18 s s_2^2 (s^3+s s_1^2-2 s_1^3)-9 s_1 s_2^5\\&+2 s_2^6+3 s_2^4 (-6 s^2+3 s s_1+5 s_1^2)+3 s_1 s_2 (s_1-3 s) (s_1-s)^3+(s-s_1)^5 (2 s+s_1)\Big]+s s_1^2 s_2 \Big[3 s_2^2 (-4 s^2+4 s s_1+s_1^2)\\&+3 s_2 (s^3+4 s^2 s_1-6 s s_1^2+s_1^3)+s_2^3 (3 s-7 s_1)+(s-s_1)^3 (3 s+2 s_1)+3 s_2^4\Big]\Big\},\\
		\bar{A}_{15}=&\lambda \Big[4 \lambda m^2 (5 s^2+13 s s_1-10 s s_2-18 s_1^2+13 s_1 s_2+5 s_2^2)-2 s_2^3 (s^2-8 s s_1+8 s_1^2)-(s-s_1)^3 (s^2-5 s s_1-2 s_1^2)\\&-2 s_2^2 (s^3+24 s^2 s_1-38 s s_1^2-5 s_1^3)+s_2 (s-s_1) (3 s^3+19 s^2 s_1+95 s s_1^2-s_1^3)+s_2^4 (3 s+8 s_1)-s_2^5\Big]
	\end{aligned}
\end{equation}

\normalsize
and
\small
\begin{equation}
	\begin{aligned}
		\bar{A}_{21}=&  4 s_1 
		\Big[ s \, s_1 \Big( s_2^2 ( 9 s - 19 s_1 ) - 2 s_2 ( 9 s + 4 s_1 ) (s - s_1) - (s - s_1)^3 + 10 s_2^3 \Big) - \lambda m^2 \Big(
		17 s^2 - 8 s (s_1 + s_2) - 9 (s_1 - s_2)^2 \Big) \Big],\\ 
		\bar{A}_{22}=& 2 s s_2 \Big[s^4 - 2 s^3 (5 s_1 + 2 s_2) + 6 s^2 s_2 (2 s_1 + s_2) + 
		s (26 s_1^3 - 60 s_1^2 s_2 + 6 s_1 s_2^2 - 4 s_2^3) - (s_1 - 
		s_2)^2 (17 s_1^2 + 6 s_1 s_2 - s_2^2)\Big]\\& - 
		4 \lambda m^2 \Big[-s_2 (s^2 - 30 s s_1 + 9 s_1^2) + 
		2 (s - s_1)^2 (s + 3 s_1) - 4 s s_2^2 + 
		3 s_2^3\Big] ,\\
		\bar{A}_{23}=& 2 
		s \Big[ 
		2 s^3 ( s_1^2 + 5 s_1 s_2 + 2 s_2^2 ) - 6 s^2 ( 
		s_1^3 - 6 s_1^2 s_2 + 2 s_1 s_2^2 + s_2^3 ) + 2 s (s_1 - s_2) \big( 
		3s_1^3-20s_1^2s_2+s_1s_2^2-2s_2^3
		\big) \\&-(s_1-s_2)^3(2s_1^2+5s_1s_2-s_2^2) -s_2 s^4
		\Big] +
		4 \lambda m^2 \Big[
		3 s^3 + 2 s^2 (9 s_1 - 2 s_2)  - 2 (s_1 - s_2)^3 - s (s_1 - s_2) ( 
		19 s_1 - s_2 )
		\Big] ,\\\bar{A}_{24}=& 
		12 \big( \lambda m^2 + s \, s_1 \, s_2 \big) \Big[ 
		2 \lambda m^2 (s + s_1 - s_2) + s_1 \Big( 
		3 s^3 + s^2 (3 s_2 - 5 s_1) + s (s_1 - s_2) (s_1 + 5 s_2) + (s_1 - s_2)^3 
		\Big) \Big] , \\
		\bar{A}_{25}=& \lambda \Big[20 \lambda m^2 (s + s_1 - s_2) - s^4 + 
		2 s^3 (3 s_1 + s_2) + 4 s^2 s_1 (3 s_2 - 2 s_1) + 
		2 s (s_1 - s_2) (s_1^2 + 8 s_1 s_2 + s_2^2) + (s_1 - s_2)^4\Big].
	\end{aligned}
\end{equation}

\chapter{Appendix to Chapter 6}

\section{Relevant functions}\label{appfunc}
The quark massive part in the trace of the correlator (anomaly free) is given by
\begin{equation}
	g_{\m\n}\braket{T^{\m\n}J^{\a a}J^{\b b}}_q= +B_1(p_1\cdot p_2) g^{\a\b}+ B_2 p_1^{\a}p_1^{\b}+B_2(p_1\leftrightarrow p_2) p_2^{\a}p_2^{\b}+B_3 p_1^{\a}p_2^{\b}+B_4p_1^{\b}p_2^{\a}
\end{equation}
where the coefficient $B_i$ with $i=1,2,3,4$ are given by
{\smaller\allowdisplaybreaks
	\begin{align}
		B_1=& - \frac{g_s^2}{16\p^2} m^2 n_f   \Biggl(\frac{3     \left(p_1^6-p_1^4 \left(p_2^2+3 q^2\right)-p_1^2 \left(p_2^2-3 q^2\right) \left(p_2^2+q^2\right)+\left(p_2^2-q^2\right)^3\right) C_0\left(p_1^2,p_2^2,q^2,m^2\right)}{p_1^4-2 p_1^2 \left(p_2^2+q^2\right)+\left(p_2^2-q^2\right)^2}
		\nn\\
		&\qquad \qquad \qquad +\frac{6   p_1^2 \bar B_0(p_1^2,m^2) \left(p_1^2-p_2^2-q^2\right)}{p_1^4-2 p_1^2 \left(p_2^2+q^2\right)+\left(p_2^2-q^2\right)^2} 
		+\frac{6 q^2\bar  B_0(q^2,m^2)   \left(q^2 \left(p_1^2+p_2^2\right)-\left(p_1^2-p_2^2\right)^2\right)}{q^2 \left(p_1^4-2 p_1^2 \left(p_2^2+q^2\right)+\left(p_2^2-q^2\right)^2\right)} \nn\\
		&\qquad \qquad \qquad  -\frac{6 p_2^2
			\bar B_0(p_2^2,m^2)\left(p_1^2-p_2^2+q^2\right)}{p_1^4-2 p_1^2 \left(p_2^2+q^2\right)+\left(p_2^2-q^2\right)^2} +6 \Biggl)-12   m^4 \frac{g_s^2}{16\p^2} C_0\left(p_1^2,p_2^2,q^2,m^2\right)
	\end{align}

	\begin{align}
		B_2= & - \frac{g_s^2}{16\p^2} m^2 n_f \Biggl(\frac{12    p_2^2  \left(p_1^6-p_1^4 \left(p_2^2+q^2\right)-p_1^2 \left(p_2^4-6 p_2^2 q^2+q^4\right)+\left(p_2^2-q^2\right)^2 \left(p_2^2+q^2\right)\right) C_0\left(p_1^2,p_2^2,q^2,m^2\right)}{\left(p_1^4-2 p_1^2 \left(p_2^2+q^2\right)+\left(p_2^2-q^2\right)^2\right)^2}\nn\\ &\qquad \qquad \qquad+\frac{24    p_2^2 p_1^2\bar B_0(p_1^2,m^2)\left(2 p_1^4-p_1^2 \left(p_2^2+q^2\right)-\left(p_2^2-q^2\right)^2\right)}{p_1^2 \left(p_1^4-2 p_1^2 \left(p_2^2+q^2\right)+\left(p_2^2-q^2\right)^2\right)^2}\nn\\ &\qquad \qquad \qquad-\frac{24    p_2^2q^2\bar B_0(q^2,m^2)\left(q^2 \left(p_1^2+p_2^2\right)+\left(p_1^2-p_2^2\right)^2-2 q^4\right)}{q^2 \left(p_1^4-2 p_1^2 \left(p_2^2+q^2\right)+\left(p_2^2-q^2\right)^2\right)^2}\nn\\ &\qquad \qquad \qquad-\frac{24    p_2^2\bar B_0(p_2^2,m^2) \left(p_1^4+p_1^2 \left(p_2^2-2 q^2\right)-2 p_2^4+p_2^2 q^2+q^4\right)}{\left(p_1^4-2 p_1^2 \left(p_2^2+q^2\right)+\left(p_2^2-q^2\right)^2\right)^2}\nn\\ &\qquad \qquad \qquad+\frac{24    p_2^2  }{p_1^4-2 p_1^2 \left(p_2^2+q^2\right)+\left(p_2^2-q^2\right)^2}\Biggl)-\frac{3   m^4 p_2^2 g_s^2 C_0\left(p_1^2,p_2^2,q^2,m^2\right)}{\p^2\left(p_1^4-2 p_1^2 \left(p_2^2+q^2\right)+\left(p_2^2-q^2\right)^2\right)}
	\end{align}
	
	\begin{align}
		B_3=&- \frac{g_s^2}{16\p^2} m^2 n_f\Biggl(\frac{6   \left(p_1^2+p_2^2-q^2\right) \left(p_1^6-p_1^4 \left(p_2^2+q^2\right)-p_1^2 \left(p_2^4-6 p_2^2 q^2+q^4\right)+\left(p_2^2-q^2\right)^2 \left(p_2^2+q^2\right)\right) C_0\left(p_1^2,p_2^2,q^2,m^2\right)}{\left(p_1^4-2 p_1^2 \left(p_2^2+q^2\right)+\left(p_2^2-q^2\right)^2\right)^2} \nn\\ &\qquad \qquad \qquad+\frac{12  p_1^2\bar B_0(p_1^2,m^2)\left(p_1^2+p_2^2-q^2\right) \left(2 p_1^4-p_1^2 \left(p_2^2+q^2\right)-\left(p_2^2-q^2\right)^2\right)}{p_1^2 \left(p_1^4-2 p_1^2 \left(p_2^2+q^2\right)+\left(p_2^2-q^2\right)^2\right)^2}\nn\\ &\qquad \qquad \qquad-\frac{12   q^2\bar B_0(q^2,m^2) \left(-3 q^4 \left(p_1^2+p_2^2\right)+4 p_1^2 p_2^2 q^2+\left(p_1^2-p_2^2\right)^2 \left(p_1^2+p_2^2\right)+2 q^6\right)}{q^2 \left(p_1^4-2 p_1^2 \left(p_2^2+q^2\right)+\left(p_2^2-q^2\right)^2\right)^2}\nn\\ &\qquad \qquad \qquad-\frac{12  p_2^2\bar B_0(p_2^2,m^2) \left(p_1^2+p_2^2-q^2\right) \left(p_1^4+p_1^2 \left(p_2^2-2 q^2\right)-2 p_2^4+p_2^2 q^2+q^4\right)}{p_2^2 \left(p_1^4-2 p_1^2 \left(p_2^2+q^2\right)+\left(p_2^2-q^2\right)^2\right)^2}\nn\\ &\qquad \qquad \qquad+\frac{12    \left(p_1^2+p_2^2-q^2\right)}{p_1^4-2 p_1^2 \left(p_2^2+q^2\right)+\left(p_2^2-q^2\right)^2}\Biggl)-\frac{3   m^4 g_s^2  \left(p_1^2+p_2^2\right) C_0\left(p_1^2,p_2^2,q^2,m^2\right)}{2\p^2\left(p_1^4-2 p_1^2 \left(p_2^2+q^2\right)+\left(p_2^2-q^2\right)^2\right)}
	\end{align}
	
	\begin{align}
		B_4=&- \frac{g_s^2}{16\p^2} m^2 n_f \, p_1\cdot p_2 \Biggl(\frac{6   \left(p_1^2+p_2^2-q^2\right) \left(p_1^6-p_1^4 \left(p_2^2+3 q^2\right)+p_1^2 \left(-p_2^4+10 p_2^2 q^2+3 q^4\right)+\left(p_2^2-q^2\right)^3\right) C_0\left(p_1^2,p_2^2,q^2,m^2\right)}{\left(p_1^4-2 p_1^2 \left(p_2^2+q^2\right)+\left(p_2^2-q^2\right)^2\right)^2}\nn\\ &\qquad \qquad \qquad\qquad+\frac{12  p_2^2\bar B_0(p_2^2m^2) \left(-5 p_1^4+4 p_1^2 \left(p_2^2+q^2\right)+\left(p_2^2-q^2\right)^2\right)}{\left(p_1^4-2 p_1^2 \left(p_2^2+q^2\right)+\left(p_2^2-q^2\right)^2\right)^2}\nn\\ &\qquad \qquad \qquad\qquad+\frac{12   p_1^2\bar B_0(p_1^2,m^2)\left(p_1^4+p_1^2 \left(4 p_2^2-2 q^2\right)-5 p_2^4+4 p_2^2 q^2+q^4\right)}{\left(p_1^4-2 p_1^2 \left(p_2^2+q^2\right)+\left(p_2^2-q^2\right)^2\right)^2}\nn\\ &\qquad \qquad \qquad\qquad-\frac{12   q^2\bar B_0(q^2,m^2)\left(q^4 \left(p_1^2+p_2^2\right)+\left(p_1^2-p_2^2\right)^2 \left(p_1^2+p_2^2\right)-2 q^2 \left(p_1^4-4 p_1^2 p_2^2+p_2^4\right)\right)}{q^2 \left(p_1^4-2 p_1^2 \left(p_2^2+q^2\right)+\left(p_2^2-q^2\right)^2\right)^2}\nn\\ &\qquad \qquad \qquad\qquad+\frac{12   \left(p_1^2+p_2^2-q^2\right)}{p_1^4-2 p_1^2 \left(p_2^2+q^2\right)+\left(p_2^2-q^2\right)^2}\Biggl)-  \frac{ 3\, m^4 \,  g_s^2 \left(p_1^2+p_2^2-q^2\right)^2 C_0\left(p_1^2,p_2^2,q^2,m^2\right)}{2\left(p_1^4-2 p_1^2 \left(p_2^2+q^2\right)+\left(p_2^2-q^2\right)^2\right)}
\end{align}}

\begin{align}
	\chi_0=&g_s^2m^4 H_1 C_0 \left(p_1^2,p_2^2,q^2,m^2\right) +g_s^2m^2 \Bigl(H_2C_0 \left(p_1^2,p_2^2,q^2,m^2\right)  \nn\\&+ H_3 \bar{B}_0(q^2,m^2)+ H_4 \bar{B}_0(p_1^2,m^2)+ H_5 \bar{B}_0(p_2^2,m^2) +H_6\Bigl)
\end{align}
with $H_i$ with $i=1,2,3,4$ given by
\begin{equation}
	H_1= n_f\frac{3 \left(p_1^4+p_2^4+q^4-2 \left(p_1^2+p_2^2\right) q^2\right)  }{2 \pi ^2 \left(-p_1^2-p_2^2+q^2\right) \left(q^4-2 \left(p_1^2+p_2^2\right) q^2+\left(p_1^2-p_2^2\right)^2\right)}
\end{equation}
\begin{equation}
	H_2=- n_f\frac{3 \tilde P_1}{8 \pi ^2 \left(-p_1^2-p_2^2+q^2\right) \left(q^4-2 \left(p_1^2+p_2^2\right) q^2+\left(p_1^2-p_2^2\right)^2\right)^2}
\end{equation}
\begin{equation}
	H_3= n_f\frac{3 \, \tilde P_2 }{4 \pi ^2 q^2 \left(-p_1^2-p_2^2+q^2\right) \left(q^4-2 \left(p_1^2+p_2^2\right) q^2+\left(p_1^2-p_2^2\right)^2\right)^2}
\end{equation}
\begin{equation}
	H_4=-n_f\frac{3\, p_1^2\left(-3 p_2^6+\left(p_1^2+5 q^2\right) p_2^4+\left(p_1^4-q^4\right) p_2^2-\left(q^2-p_1^2\right)^3\right) }{4 \pi ^2 \left(p_1^2+p_2^2-q^2\right) \left(q^4-2 \left(p_1^2+p_2^2\right) q^2+\left(p_1^2-p_2^2\right)^2\right)^2} 
\end{equation}
\begin{equation}
	H_5= n_f\frac{3\, p_2^2\left(p_2^6+\left(p_1^2-3 q^2\right) p_2^4+\left(p_1^4+3 q^4\right) p_2^2-\left(q^2-p_1^2\right)^2 \left(3 p_1^2+q^2\right)\right)}{4 \pi ^2 \left(-p_1^2-p_2^2+q^2\right) \left(q^4-2 \left(p_1^2+p_2^2\right) q^2+\left(p_1^2-p_2^2\right)^2\right)^2} 
\end{equation}
\begin{equation}
	H_6=n_f \frac{3 \left(p_1^4+p_2^4+q^4-2 \left(p_1^2+p_2^2\right) q^2\right) }{4 \pi ^2 \left(-p_1^2-p_2^2+q^2\right) \left(q^4-2 \left(p_1^2+p_2^2\right) q^2+\left(p_1^2-p_2^2\right)^2\right)} 
\end{equation}

with
\begin{align}
	\tilde P_1= &-p_2^{10}+\left(p_1^2+5 q^2\right) p_2^8-2 q^2 \left(3 p_1^2+5 q^2\right) p_2^6+2 \left(5 q^2-3 p_1^2\right) q^2 \left(p_1^2+q^2\right) p_2^4\nonumber\\&-\left(q^2-p_1^2\right)^2 \left(p_1^2+q^2\right) \left(5 q^2-p_1^2\right) p_2^2+\left(q^2-p_1^2\right)^5,
\end{align}

\begin{align}
	\tilde{P}_2=&q^2 \Bigl(-p_2^8+\left(2 p_1^2+3 q^2\right) p_2^6-\left(p_1^2+q^2\right) \left(2 p_1^2+3 q^2\right) p_2^4\nn\\&+\left(q^2-p_1^2\right) \left(-2 p_1^4+3 q^2 p_1^2+q^4\right) p_2^2+p_1^2 \left(q^2-p_1^2\right)^3\Bigl).
\end{align}

\section{The scalar off-shell, massive 3-point function} 
\label{Co}
\begin{equation}
	\label{lco}
	C_0\left(q^2,p_1^2,p_2^2,m^2\right)= \frac{1}{i \pi^2}\int d^n k\frac{1}{(k^2- m^2)( (k- q)^2- m^2) 
		((k - p_1)^2- m^2)},\,
\end{equation}
($s_1=p_2^2, s_2=p_2^2$) 
\allowdisplaybreaks{
	\begin{align}
		C_0(s,s_1,s_2,m^2)=&\frac{\text{Li}_2\left(\frac{s_1 \sqrt{\lambda(s,s_1,s_2)}-s_1 (-s+s_1-s_2)}{s_1 (s-s_1+s_2)+\sqrt{s_1 \left(s_1-4 m^2\right)} \sqrt{\lambda(s,s_1,s_2)}} -\left(i \left(-s+s_1-s_2-\sqrt{\lambda(s,s_1,s_2)}\right)\right) \epsilon \right) }{ \sqrt{\lambda(s,s_1,s_2)}}\nonumber\\&
		+\frac{\text{Li}_2\left(\frac{(s+s_1-s_2) s_2+\sqrt{\lambda(s,s_1,s_2)} s_2}{(s+s_1-s_2) s_2+\sqrt{s_2 \left(s_2-4 m^2\right)} \sqrt{\lambda(s,s_1,s_2)}} -\left(i \left(-s-s_1+s_2-\sqrt{\lambda(s,s_1,s_2)}\right)\right) \epsilon \right) }{ \sqrt{\lambda(s,s_1,s_2)}}\nonumber\\&
		+\frac{\text{Li}_2\left(\frac{s \sqrt{\lambda(s,s_1,s_2)}-s (s-s_1-s_2)}{ -s (s-s_1-s_2)-\sqrt{s \left(s-4 m^2\right)} \sqrt{\lambda(s,s_1,s_2)} } -\left(i \left(-s+s_1+s_2+\sqrt{\lambda(s,s_1,s_2)}\right)\right) \epsilon \right)}{ \sqrt{\lambda(s,s_1,s_2)}}\nonumber\\&
		+\frac{\text{Li}_2\left(\frac{s \sqrt{\lambda(s,s_1,s_2)}-s (s-s_1-s_2)}{\sqrt{s \left(s-4 m^2\right)} \sqrt{\lambda(s,s_1,s_2)}-s (s-s_1-s_2)}+\epsilon  \left(i \left(s-s_1-s_2-\sqrt{\lambda(s,s_1,s_2)}\right)\right)\right) }{\sqrt{\lambda(s,s_1,s_2)}}\nonumber\\&
		+\frac{ \text{Li}_2\left(\frac{(s+s_1-s_2) s_2+\sqrt{\lambda(s,s_1,s_2)} s_2}{(s+s_1-s_2) s_2-\sqrt{s_2 \left(s_2-4 m^2\right)} \sqrt{\lambda(s,s_1,s_2)}}+\epsilon  \left(i \left(s+s_1-s_2+\sqrt{\lambda(s,s_1,s_2)}\right)\right)\right)}{\sqrt{\lambda(s,s_1,s_2)}}\nonumber\\&
		+\frac{\text{Li}_2\left(\frac{s_1 \sqrt{\lambda(s,s_1,s_2)}-s_1 (-s+s_1-s_2)}{s_1 (s-s_1+s_2)-\sqrt{s_1 \left(s_1-4 m^2\right)} \sqrt{\lambda(s,s_1,s_2)}}+\epsilon  \left(i \left(s-s_1+s_2+\sqrt{\lambda(s,s_1,s_2)}\right)\right)\right) }{\sqrt{\lambda(s,s_1,s_2)}}\nonumber\\&
		-\frac{\text{Li}_2 \left(\frac{-s (s-s_1-s_2)-\sqrt{\lambda(s,s_1,s_2)} s}{-s (s-s_1-s_2)-\sqrt{s \left(s-4 m^2\right)} \sqrt{\lambda(s,s_1,s_2)}} -\left(i \left(-s+s_1+s_2-\sqrt{\lambda(s,s_1,s_2)}\right)\right) \epsilon \right)}{\sqrt{\lambda(s,s_1,s_2)}}\nonumber\\&
		-\frac{\text{Li}_2 \left(\frac{-s_1 (-s+s_1-s_2)-\sqrt{\lambda(s,s_1,s_2)} s_1}{s_1 (s-s_1+s_2)+\sqrt{s_1 \left(s_1-4 m^2\right)} \sqrt{\lambda(s,s_1,s_2)}} -\left(i \left(-s+s_1-s_2+\sqrt{\lambda(s,s_1,s_2)}\right)\right) \epsilon \right)}{ \sqrt{\lambda(s,s_1,s_2)}}\nonumber\\&
		-\frac{\text{Li}_2 \left(\frac{(s+s_1-s_2) s_2-s_2 \sqrt{\lambda(s,s_1,s_2)}}{(s+s_1-s_2) s_2+\sqrt{s_2 \left(s_2-4 m^2\right)} \sqrt{\lambda(s,s_1,s_2)}}- \left(i \left(-s-s_1+s_2+\sqrt{\lambda(s,s_1,s_2)}\right)\right) \epsilon \right)}{ \sqrt{\lambda(s,s_1,s_2)}}\nonumber\\&
		-\frac{\text{Li}_2 \left(\frac{(s+s_1-s_2) s_2-s_2 \sqrt{\lambda(s,s_1,s_2)}}{(s+s_1-s_2) s_2-\sqrt{s_2 \left(s_2-4 m^2\right)} \sqrt{\lambda(s,s_1,s_2)}}+\epsilon  \left(i \left(s+s_1-s_2-\sqrt{\lambda(s,s_1,s_2)}\right)\right)\right)}{\sqrt{\lambda(s,s_1,s_2)}}\nonumber\\&
		-\frac{\text{Li}_2 \left(\frac{-s_1 (-s+s_1-s_2)-\sqrt{\lambda(s,s_1,s_2)} s_1}{s_1 (s-s_1+s_2)-\sqrt{s_1 \left(s_1-4 m^2\right)} \sqrt{\lambda(s,s_1,s_2)}}+\epsilon  \left(i \left(s-s_1+s_2-\sqrt{\lambda(s,s_1,s_2)}\right)\right)\right)}{\sqrt{\lambda(s,s_1,s_2)}}\nonumber\\&
		-\frac{\text{Li}_2 \left(\frac{-s (s-s_1-s_2)-\sqrt{\lambda(s,s_1,s_2)} s}{\sqrt{s \left(s-4 m^2\right)} \sqrt{\lambda(s,s_1,s_2)}-s (s-s_1-s_2)}+\epsilon  \left(i \left(s-s_1-s_2+\sqrt{\lambda(s,s_1,s_2)}\right)\right)\right)}{\sqrt{\lambda(s,s_1,s_2)}}.
		\label{cco}
	\end{align}
}
In \eqref{cco} we have used the short-hand notation $i \varepsilon f(s,s_1,s_2)$ as $ i \varepsilon \textrm{ sign} f(s,s_1,s_2) $, 
with $\textrm{ sign}$ the signum function. $\lambda(s,s_1,s_2)$ is the K\"allen function
\begin{equation}
	\lambda(s,s_1,s_2)=s^2-2 s_1 s-2 s_2 s+s_1^2+s_2^2-2 s_1 s_2
\end{equation}
factorizable as
\begin{equation}
	\lambda(s,s_1,s_2)=(s-(\sqrt{s_1}-\sqrt{s_2})^2)(s-(\sqrt{s_1}+\sqrt{s_2})^2).
\end{equation}
\subsection{Massless limits of $C_0$ and of renormalized $B_0$}
For $m=0$ 
\begin{equation}
	\label{lco}
	C_0\left(q^2,p_1^2,p_2^2\right)= \frac{1}{i \pi^2}\int d^n k\frac{1}{k^2 (k- q)^2 (k - p_1)^2},\,
\end{equation}
is the ordinary massless scalar master integral, with

\bea
&  C_0 ( p_1^2,p_2^2,q^2) = \frac{ 1}{q^2} \Phi (x,y),
\eea
where the function $\Phi (x, y)$  is defined as
\bea
\Phi( x, y) &=& \frac{1}{\lambda} \biggl\{ 2 [Li_2(-\rho  x) + Li_2(- \rho y)]  +
\ln \frac{y}{ x}\ln \frac{1+ \rho y }{1 + \rho x}+ \ln (\rho x) \ln (\rho  y) + \frac{\pi^2}{3} \biggr\},
\label{Phi}
\eea
with
\bea
\lambda(x,y) = \sqrt {\Delta},
\qquad  \qquad \Delta=(1-  x- y)^2 - 4  x  y,
\label{lambda} \\
\rho( x,y) = 2 (1-  x-  y+\lambda)^{-1},
\qquad  \qquad x=\frac{p_1^2}{q^2} \, ,\qquad \qquad y= \frac {p_2^2}{q^2},\, 
\eea

\begin{equation}
	{B}_0(p^2)=\sdfrac{1}{i\p^\frac{d}{2}}\int\,d^d l\ \frac{1}{l^2(l-p)^2}=\frac{ \left[\G\left(\frac{d}{2}-1\right)\right]^2\G\left(2-\frac{d}{2}\right)}{\G\left(d-2\right)(p^2)^{2-\frac{d}{2}}}\label{B0ex}.
\end{equation}
We define $(\epsilon=d-4)$
\beq
B_0(p^2)=\frac{1}{\varepsilon}+\bar{B}_0(p^2)
\eeq
and the renormalized expression of $B_0(p^2,0)$ used in the spectral analysis is 
\beq
B_0^R(p^2,0)\equiv \bar{B}_0(p^2)= 2 + \log(\mu^2/p^2) 
\label{BB}
\eeq
in the $\overline{MS}$ scheme, with $\mu$  the renormalization scale. \\

\subsection{ $C_0$ at $s_{\pm}=(\sqrt{s_1}\pm \sqrt{s_2})^2$}
The evaluation of $C_0$ at the points $s_\pm$ are given by
\begin{align}
	\label{x1}
	C_0(s_-,s_1,s_2,m^2)=&\frac{1}{{s_1}{s_2}\, s_-^2}\Biggl(\sqrt{{s_1}} \Biggl(\sqrt{{s_2} \left({s_2}-4 m^2\right)} \left(\sqrt{{s_1}}-\sqrt{{s_2}}\right)\times \nn\\ &
	\times \log \left(\frac{\sqrt{{s_2} \left({s_2}-4 m^2\right)}+2 m^2-{s_2}}{2 m^2}\right)\nonumber\\&+\sqrt{{s_2}} \sqrt{\left(-2 \sqrt{{s_1}} \sqrt{{s_2}}+{s_1}+{s_2}\right) \left(-4 m^2-2 \sqrt{{s_1}} \sqrt{{s_2}}+{s_1}+{s_2}\right)} \log \left(S_1\right)\Biggl)\nonumber\\&+\sqrt{{s_2}} \sqrt{{s_1} \left({s_1}-4 m^2\right)} \left(\sqrt{{s_2}}-\sqrt{{s_1}}\right) \log \left(\frac{\sqrt{{s_1} \left({s_1}-4 m^2\right)}+2 m^2-{s_1}}{2 m^2}\right)\Biggl)
\end{align}
and 
\begin{align}
	\label{x2}
	C_0( s_+,s_1,s_2,m^2)=&\frac{1}{{s_1}{s_2} s_+^2}\Biggl(\sqrt{{s_1}} \Biggl(\sqrt{{s_2} \left({s_2}-4 m^2\right)} \left(\sqrt{{s_1}}+\sqrt{{s_2}}\right)\times \nn \\&
	\times \log \left(\frac{\sqrt{{s_2} \left({s_2}-4 m^2\right)}+2 m^2-{s_2}}{2 m^2}\right)\nonumber\\&-\sqrt{{s_2}} \sqrt{\left(\sqrt{{s_1}}+\sqrt{{s_2}}\right)^4-4 m^2 \left(\sqrt{{s_1}}+\sqrt{{s_2}}\right)^2} \log \left(S_2\right)\Biggl)\nonumber\\&+\sqrt{{s_2}} \sqrt{{s_1} \left({s_1}-4 m^2\right)} \left(\sqrt{{s_1}}+\sqrt{{s_2}}\right) \log \left(\frac{\sqrt{{s_1} \left({s_1}-4 m^2\right)}+2 m^2-{s_1}}{2 m^2}\right)\Biggl), 
\end{align}
where
\begin{equation}
	S_1=\frac{\sqrt{\left(-2 \sqrt{{s_1}} \sqrt{{s_2}}+{s_1}+{s_2}\right) \left(-4 m^2-2 \sqrt{{s_1}} \sqrt{{s_2}}+{s_1}+{s_2}\right)}+2 m^2-\left(\sqrt{{s_1}}-\sqrt{{s_2}}\right)^2}{2 m^2}
\end{equation}
and 
\begin{equation}
	S_2=\frac{\sqrt{\left(\sqrt{{s_1}}+\sqrt{{s_2}}\right)^4-4 m^2 \left(\sqrt{{s_1}}+\sqrt{{s_2}}\right)^2}+2 m^2-\left(\sqrt{{s_1}}+\sqrt{{s_2}}\right)^2}{2 m^2}.
\end{equation}

\subsection{Gluon Trace Sector}
\label{gg}
The coefficient functions are given by
{\allowdisplaybreaks
	\begin{eqnarray}
		C_1 &=& -\frac{ \,  \,  \, C_A \, g_s^2 \,  }{ 32\pi^2 \big( (p_1 \cdot p_2)^2 - p_1^2 \, p_2^2 \big)} \, \Big(
		2 \, (p_1 \cdot p_2)^2 - 2 \, p_1^2 \, p_2^2 
		- p_1^2 \, (p_1 \cdot p_2 + p_2^2 ) \bar B_0 (p_1^2)  \notag \\ &&
		+ \big[ p_2^2 \, (p_1^2 - p_1 \cdot p_2) - 2 \, (p_1 \cdot p_2)^2 \big]   \bar B_0(p_2^2)
		+ (p_1 \cdot p_2) \, (2 \, p_1 \cdot p_2 + p_1^2 + p_2^2) \,  \bar B_0 (q^2)  \notag \\ &&
		+ \big[ 2\, (p_1 \cdot p_2+p_1^2) (p_1 \cdot p_2)^2\,  - p_2^2 \, \big(4 (p_1 \cdot p_2)^2   + p_1^4\big) + 5 \, p_1^2 \, p_4^4  \big] C_0(p_1^2, p_2^2 , q^2)  \Big) \\
		C_2 &=& -2  \,  \,  \, C_A \, \frac{g_s^2}{16\pi^2} \,  \, (p_1 \cdot p_2) \, C_0 (p_1^2 , p_2^2 , q^2) \\
		C_3 &=& -\frac{ \,  \,  \, g_s^2 \, C_A \, (p_1^2 + p_2^2) \,}{32\pi^2 \big(p_1^2 \, p_2^2 - (p_1 \cdot p_2)^2 \big)} \Big( 
		(-p_1\cdot p_2-p_1^2)   \bar B_0(p_1^2)+(-(p_1\cdot p_2)-p_2^2) \,  \bar B_0(p_2^2) \notag \\ &&
		+q^2 \, \bar  B_0(q^2)+\big(p_1^2 \, (p_1\cdot p_2-2 p_2^2) +(p_1\cdot p_2) \, \big(4 \, (p_1\cdot p_2)+p_2^2) \big) \, C_0(p_1^2,p_2^2, q^2)  \Big)  \\
		C_4 &=& \, \,  \, \frac{g_s^2}{32\pi^2} \, C_A \,  \Big( (p_1^2 - p_2^2) \, \big[  \bar B_0 (p_1^2) - \bar  B_0(p_2^2) \big] 
		+ \big[ p_1^4 + p_2^4 - 2 \, (p_1^2 + p_2^2) \, p_1 \cdot p_2 - 6 \, p_1^2 \, p_2^2 \big] \, C_0(p_1^2, p_2^2, q^2) \Big). \nonumber \\
\end{eqnarray}}

\chapter{Appendix to Chapter 7}

\section{Feynman rules}
\label{chza:rules}
%------------------
The Feynman rules used throughout the paper are collected here
\begin{itemize}
	\item {Graviton - fermion - fermion vertex}
	\\
	%%%%%%%%%%%
	\\
	%\begin{align}
	\bmi{95pt}
	\includegraphics[scale=1.0]{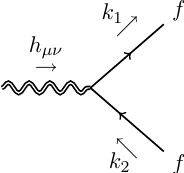}
	\emi
	\bmi{70pt}
	\begin{eqnarray*}
		&=& - i \, \frac{\kappa}{2} \, V^{\prime}_{\mu\nu}(k_1,k_2) \nn \\
		&=& - i \, \frac{\kappa}{2} \, \left\{ \frac{1}{4} \left[\gamma_\mu (k_1 + k_2)_\nu
		+\gamma_\nu (k_1 + k_2)_\mu \right] - \frac{1}{2} g_{\mu \nu}
		[\gamma^{\lambda}(k_1 + k_2)_{\lambda} - 2 m]  \right\} \nn \\
	\end{eqnarray*}
	\emi
	\begin{eqnarray}
		\label{chza:VGff}
	\end{eqnarray}
	%\end{align}
	\item{Graviton - gluon - gluon vertex}
	\\ \\
	\bmi{110pt}
	\includegraphics[scale=1.0]{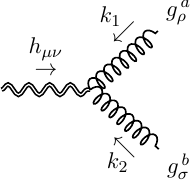}
	\emi
	\bmi{100pt}
	\begin{eqnarray*}
		&=& - i \, \frac{\kappa}{2} \, \delta_{a b} \, V^{Ggg}_{\mu\nu\rho\sigma}(k_1,k_2) \nn \\
		&= &- i \, \frac{\kappa}{2} \, \delta_{a b} \left\{ k_1\cdot k_2 \, C_{\mu\nu\rho\si} + D_{\mu\nu\rho\si}(k_1,k_2) + \frac{1}{\xi} \, E_{\mu\nu\rho\si}(k_1,k_2)  \right\}
	\end{eqnarray*}
	\emi
	\begin{eqnarray}
	\end{eqnarray}
	\item{Graviton - ghost - ghost vertex}
	\\ \\
	\bmi{110pt}
	\includegraphics[scale=1.0]{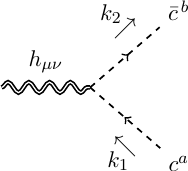}
	\emi
	\bmi{70pt}
	\begin{eqnarray*}
		& = & - i \,  \frac{\kappa}{2}\,  \delta^{a b} \, C_{\mu\nu\rho\sigma} \, k_{1\,\rho} \, k_{2\,\sigma}
	\end{eqnarray*}
	\emi
	\begin{eqnarray}
	\end{eqnarray}
	%
	%%%%%%%%%%%%%%%%%%%%%
	%
	\item{Graviton - fermion - fermion - gauge boson vertex}
	\\ \\
	\bmi{110pt}
	\includegraphics[scale=1.0]{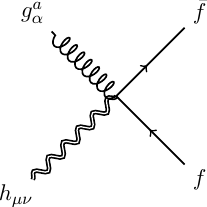}
	\emi
	\bmi{70pt}
	\begin{eqnarray*}
		&=& i g \, \frac{\kappa}{2} \,T^a\, W^{\prime}_{\mu\nu\alpha}
		= i g \, \frac{\kappa}{2} \, T^a \left\{ -\frac{1}{2} (\gamma_\mu \, g_{\nu\alpha}
		+\gamma_\nu \, g_{\mu\alpha}) +  g_{\mu \nu} \, \gamma_{\alpha} \right \}
	\end{eqnarray*}
	\emi
	\begin{eqnarray}
		\label{chza:WGffg}
	\end{eqnarray}
	%
	%%%%%%%%%%%%%%%%%%%%%%%%%%
	%
	%%%%%%%%%%%%%%%%%%%%%%%%%%
	%
	\item{Graviton - gluon - gluon - gluon  vertex}
	\\ \\
	\bmi{110pt}
	\includegraphics[scale=1.0]{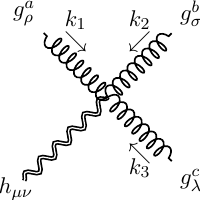}
	\emi
	\bmi{110pt}
	\begin{eqnarray*}
		&=&- g \frac{\kappa}{2} f^{a b c} V^{Gggg}_{\mu\nu\rho\sigma\lambda}(k_1,k_2,k_3) \nn \\
		&=&  - g \frac{\kappa}{2} f^{a b c} \left\{ C_{\mu\nu\rho\sigma}(k_1-k_2)_{\lambda} + C_{\mu\nu\rho\lambda}(k_3-k_1)_{\sigma}   \right. \nn \\
		&& \hspace{2.5cm}  + \left.  C_{\mu\nu\sigma\lambda}(k_2-k_3)_{\rho} + F_{\mu\nu\rho\sigma\lambda}(k_1,k_2,k_3)  \right\}
		\hspace{1.7cm}
	\end{eqnarray*}
	\emi
	
	%
	%%%%%%%%%%%%%%%%%%%%%%%%%%
	%%%%%%%%%%%%%%%%%%%%%%%%%%
	%
	\item{Graviton - ghost - ghost - gauge boson vertex}
	\\ \\
	\bmi{110pt}
	\includegraphics[scale=1.0]{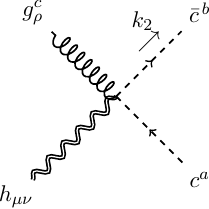}
	\emi
	\bmi{70pt}
	\begin{eqnarray*}
		&=&- \frac{\kappa}{2} \, g \, f^{a b c} \, C_{\mu\nu\rho\sigma} \, k_{2}^{\sigma}
	\end{eqnarray*}
	\emi
	\begin{eqnarray}
	\end{eqnarray}
	%
	%%%%%%%%%%%%%%%%%%%%%%%%%%
	%
	\begin{eqnarray}
		&& C_{\mu\nu\rho\sigma} = g_{\mu\rho}\, g_{\nu\sigma}
		+g_{\mu\sigma} \, g_{\nu\rho}
		-g_{\mu\nu} \, g_{\rho\sigma}\,
		\\
		&& D_{\mu\nu\rho\sigma} (k_1, k_2) =
		g_{\mu\nu} \, k_{1 \, \sigma}\, k_{2 \, \rho}
		- \biggl[g^{\mu\sigma} k_1^{\nu} k_2^{\rho}
		+ g_{\mu\rho} \, k_{1 \, \sigma} \, k_{2 \, \nu}
		- g_{\rho\sigma} \, k_{1 \, \mu} \, k_{2 \, \nu}
		+ (\mu\leftrightarrow\nu)\biggr]\, \\
		&& E_{\mu\nu\rho\sigma} (k_1, k_2) = g_{\mu\nu} \, (k_{1 \, \rho} \, k_{1 \, \sigma}
		+k_{2 \, \rho} \, k_{2 \, \sigma} +k_{1 \, \rho} \, k_{2 \, \sigma})
		-\biggl[g_{\nu\sigma} \, k_{1 \, \mu} \, k_{1 \, \rho}
		+g_{\nu\rho} \, k_{2 \, \mu} \, k_{2 \, \sigma}
		+(\mu\leftrightarrow\nu)\biggr]\ , \nn  \\ \\
		&& F_{\mu\nu\rho\sigma\lambda} (k_1,k_2,k_3) =
		g_{\mu\rho} \,  g_{\sigma\lambda} \, (k_2-k_3)_{\nu}
		+g_{\mu\sigma} \, g_{\rho\lambda} \, (k_3-k_1)_{\nu}
		+g_{\mu\lambda} \, g_{\rho\sigma}(k_1-k_2)_{\nu}
		+ (\mu\leftrightarrow\nu) \nn \\
	\end{eqnarray}
\end{itemize}
%-------------------------

\section{Form factors of the transverse-traceless sector: $A_1,A_2,A_3,A_4$ for massless quarks} 
\label{chza:ff}
We present here the explicit expressions of the coefficient functions $A_{ij}$ in \eqref{chza:Abar}  characterizing the transverse-traceless sector. The quark contributions are identified by the factor $n_f$ and the gluon contributions by the factor $C_A$:

{\allowdisplaybreaks
	\begin{eqnarray*}
		A_{10} &=& 16\, (p_1\cdot p_2)^7+36 \, p_1^2 \, (p_1\cdot p_2)^6+36 \, p_2^2 \, (p_1\cdot p_2)^6+12 \, p_1^4 \, (p_1\cdot p_2)^5+12 \, p_2^4 \, (p_1\cdot p_2)^5+240 \, p_1^2 \, p_2^2 \, (p_1\cdot p_2)^5 \notag \\ &&
		+188 \, p_1^2 \, p_2^4 \, (p_1\cdot p_2)^4+188 \, p_1^4 \, p_2^2 \, (p_1\cdot p_2)^4+46 \, p_1^2 \, p_2^6 \, (p_1\cdot p_2)^3-108 \, p_1^4 \, p_2^4 \, (p_1\cdot p_2)^3+46 \, p_1^6 \, p_2^2 \, (p_1\cdot p_2)^3 \notag \\ &&
		-204 \, p_1^4 \, p_2^6 \, (p_1\cdot p_2)^2-204 \, p_1^6 \, p_2^4 \, (p_1\cdot p_2)^2-20 \, p_1^6 \, p_2^8 \,-20 \, p_1^8 \, p_2^6 \,-58 \, p_1^4 \, p_2^8 \, (p_1\cdot p_2) \notag \\ &&
		-148 \, p_1^6 \, p_2^6 \, (p_1\cdot p_2)-58 \, p_1^8 \, p_2^4 \, (p_1\cdot p_2) \\
		A_{11} &=& -24 \, p_1^2 \,  \, (p_1\cdot p_2)^6-24 \, p_1^4 \,  \,\, (p_1\cdot p_2)^5-224 \, p_1^2 \,  \, p_2^2 \, (p_1\cdot p_2)^5-6\, p_1^6 \,  \,\, (p_1\cdot p_2)^4-228 \, p_1^2 \,  \, p_2^4 \, (p_1\cdot p_2)^4 \notag \\ &&
		-522 \, p_1^4 \,  \, p_2^2 \, (p_1\cdot p_2)^4- 60 \, p_1^2 \,  \, p_2^6 \, (p_1\cdot p_2)^3- 652 \, p_1^4 \,  \, p_2^4 \, (p_1\cdot p_2)^3-372 \, p_1^6 \,  \, p_2^2 \, (p_1\cdot p_2)^3  \notag \\ &&
		-279 \, p_1^4 \,  \, p_2^6 \, (p_1\cdot p_2)^2-474 \, p_1^6 \,  \, p_2^4 \, (p_1\cdot p_2)^2-83 \, p_1^8 \,  \, p_2^2 \, (p_1\cdot p_2)^2-18 \, p_1^6 \,  \, p_2^8 \,  \notag \\ &&
		-30 \, p_1^8 \,  \, p_2^6 \,-16 p_1^{10} \, p_2^4 \,-45 \, p_1^4 \,  \, p_2^8 \, (p_1\cdot p_2)-174 \, p_1^6 \,  \, p_2^6 \, (p_1\cdot p_2)-129 \, p_1^8 \,  \, p_2^4 \, (p_1\cdot p_2) \\
		A_{12} &=& -24 \, p_2^2 \, (p_1\cdot p_2)^6-24 \, p_2^4 \, (p_1\cdot p_2)^5-224 \, p_1^2 \,  \, p_2^2 \, (p_1\cdot p_2)^5-6 \, p_2^6 \, (p_1\cdot p_2)^4-522 \, p_1^2 \,  \, p_2^4 \, (p_1\cdot p_2)^4   \notag \\ &&
		-228 \, p_1^4 \,  \, p_2^2 \, (p_1\cdot p_2)^4-372 \, p_1^2 \,  \, p_2^6 \, (p_1\cdot p_2)^3-652 \, p_1^4 \,  \, p_2^4 \, (p_1\cdot p_2)^3-60 \, p_1^6 \,  \, p_2^2 \, (p_1\cdot p_2)^3-83 \, p_1^2 \,  \, p_2^8 \, (p_1\cdot p_2)^2   \notag \\ &&
		-474 \, p_1^4 \,  \, p_2^6 \, (p_1\cdot p_2)^2-279 \, p_1^6 \,  \, p_2^4 \, (p_1\cdot p_2)^2-16 \, p_1^4 \,  \, p_2^{10} \,-30 \, p_1^6 \,  \, p_2^8 \,-18 \, p_1^8 \,  \, p_2^6 \,-129 \, p_1^4 \,  \, p_2^8 \, (p_1\cdot p_2)   \notag \\ &&
		-174 \, p_1^6 \,  \, p_2^6 \, (p_1\cdot p_2)-45 \, p_1^8 \,  \, p_2^4 \, (p_1\cdot p_2) \\
		A_{13} &=& 24 \, p_1^2 \,  \, (p_1\cdot p_2)^6+24 \, p_2^2 \, (p_1\cdot p_2)^6+24 \, p_1^4 \,  \, (p_1\cdot p_2)^5+24 \, p_2^4 \, (p_1\cdot p_2)^5 + 448 \, p_1^2 \,  \, p_2^2 \, (p_1\cdot p_2)^5  \notag \\ &&
		+6 \, p_1^6 \,  \, (p_1\cdot p_2)^4+6 \, p_2^6 \, (p_1\cdot p_2)^4+750 \, p_1^2 \,  \, p_2^4 \, (p_1\cdot p_2)^4  +750 \, p_1^4 \,  \, p_2^2 \, (p_1\cdot p_2)^4+432 \, p_1^2 \,  \, p_2^6 \, (p_1\cdot p_2)^3  \notag \\ &&
		+ 1304 \, p_1^4 \,  \, p_2^4 \, (p_1\cdot p_2)^3+432 \, p_1^6 \,  \, p_2^2 \, (p_1\cdot p_2)^3  + 83 \, p_1^2 \,  \, p_2^8 \, (p_1\cdot p_2)^2+753 \, p_1^4 \,  \, p_2^6 \, (p_1\cdot p_2)^2+753 \, p_1^6 \,  \, p_2^4 \, (p_1\cdot p_2)^2  \notag \\ &&
		+83 \, p_1^8 \,  \, p_2^2 \, (p_1\cdot p_2)^2 + 16 \, p_1^4 \,  \, p_2^{10}+48 \, p_1^6 \,  \, p_2^8 \,+48 \, p_1^8 \,  \, p_2^6 \,+16 p_1^{10} \, p_2^4 \,  \notag \\ &&
		+174 \, p_1^4 \,  \, p_2^8 \, (p_1\cdot p_2)+348 \, p_1^6 \,  \, p_2^6 \, (p_1\cdot p_2)+174 \, p_1^8 \,  \, p_2^4 \, (p_1\cdot p_2)	\\
		A_{14} &=& 192 \, p_1^2 \,  \, p_2^2 \, (p_1\cdot p_2 )^6+432 \, p_1^2 \,  \, p_2^4 \, (p_1\cdot p_2 )^5+432 \, p_1^4 \,  \, p_2^2 \, (p_1\cdot p_2 )^5+288 \, p_1^2 \,  \, p_2^6 \, (p_1\cdot p_2 )^4+1152 \, p_1^4 \,  \, p_2^4 \, (p_1\cdot p_2 )^4  \notag \\ &&
		+288 \, p_1^6 \,  \, p_2^2 \, (p_1\cdot p_2 )^4+60 \, p_1^2 \,  \, p_2^8 \, (p_1\cdot p_2 )^3+936 \, p_1^4 \,  \, p_2^6 \, (p_1\cdot p_2 )^3+936 \, p_1^6 \,  \, p_2^4 \, (p_1\cdot p_2 )^3+60 \, p_1^8 \,  \, p_2^2 \, (p_1\cdot p_2 )^3  \notag \\ &&
		+324 \, p_1^4 \,  \, p_2^8 \, (p_1\cdot p_2 )^2+720 \, p_1^6 \,  \, p_2^6 \, (p_1\cdot p_2 )^2+324 \, p_1^8 \,  \, p_2^4 \, (p_1\cdot p_2 )^2+18 \, p_1^6 \,  \, p_2^{10} \,+36 \, p_1^8 \,  \, p_2^8 \notag \\ &&
		+18 p_1^{10} \, p_2^6 \,+45 \, p_1^4 \,  \, p_2^{10} (p_1\cdot p_2) +207 \, p_1^6 \,  \, p_2^8 \, (p_1\cdot p_2)+207 \, p_1^8 \,  \, p_2^6 \, (p_1\cdot p_2)+45 p_1^{10} \, p_2^4 \, (p_1\cdot p_2)
	\end{eqnarray*}
	
	\begin{eqnarray*}
		A_{20} &=& C_A \Big[-520 \, (p_1\cdot p_2)^6-372 \, p_1^2 \, (p_1\cdot p_2)^5-372 \, p_2^2 \, (p_1\cdot p_2)^5-60 \, p_1^4 \, (p_1\cdot p_2)^4-60 \, p_2^4 \, (p_1\cdot p_2)^4 \notag \\ &&
		+1056 \, p_1^2 \, p_2^2 \, (p_1\cdot p_2)^4+864 \, p_1^2 \, p_2^4 \, (p_1\cdot p_2)^3+864 \, p_1^4 \, p_2^2 \, (p_1\cdot p_2)^3+150 \, p_1^2 \, p_2^6 \, (p_1\cdot p_2)^2 \notag \\ &&
		-372 \, p_1^4 \, p_2^4 \, (p_1\cdot p_2)^2+150 \, p_1^6 \, p_2^2 \, (p_1\cdot p_2)^2-90 \, p_1^4 \, p_2^8 \,-164 \, p_1^6 \, p_2^6 \,-90 \, p_1^8 \, p_2^4 \notag \\ &&
		-492 \, p_1^4 \, p_2^6 \, (p_1\cdot p_2)-492 \, p_1^6 \, p_2^4 \, (p_1\cdot p_2) \Big]+ \notag \\ &&
		n_f\Big[ 40 (p_1\cdot p_2)^6+12 \, p_1^2 \, (p_1\cdot p_2)^5+12 \, p_2^2 \, (p_1\cdot p_2)^5-12 \, p_1^4 \, (p_1\cdot p_2)^4-12 \, p_2^4 \, (p_1\cdot p_2)^4 \notag \\ &&
		-192 \, p_1^2 \, p_2^2 \, (p_1\cdot p_2)^4-144 \, p_1^2 \, p_2^4 \, (p_1\cdot p_2)^3-144 \, p_1^4 \, p_2^2 \, (p_1\cdot p_2)^3-6 \, p_1^2 \, p_2^6 \, (p_1\cdot p_2)^2 \notag \\ &&
		+84 \, p_1^4 \, p_2^4 \, (p_1\cdot p_2)^2-6 \, p_1^6 \, p_2^2 \, (p_1\cdot p_2)^2+18 \, p_1^4 \, p_2^8 \,+68 \, p_1^6 \, p_2^6 \,+18 \, p_1^8 \, p_2^4 \notag \\ &&
		+132 \, p_1^4 \, p_2^6 \, (p_1\cdot p_2)+132 \, p_1^6 \, p_2^4 \, (p_1\cdot p_2) \Big] \\
		A_{21} &=& 3 \, p_1^2 \,  \, n_f \Big[-24 (p_1\cdot p_2)^5-20 \, p_2^2 \, (p_1\cdot p_2)^4+2 \, p_1^4 \,  \, (p_1\cdot p_2)^3+24 \, p_2^4 \, (p_1\cdot p_2)^3+90 \, p_1^2 \,  \, p_2^2 \, (p_1\cdot p_2)^3  \notag \\ &&
		+12 \, p_2^6 \, (p_1\cdot p_2)^2+162 \, p_1^2 \,  \, p_2^4 \, (p_1\cdot p_2)^2+54 \, p_1^4 \,  \, p_2^2 \, (p_1\cdot p_2)^2+3 \, p_1^2 \,  \, p_2^8 \,+8 \, p_1^4 \,  \, p_2^6 \,+21 \, p_1^6 \,  \, p_2^4 \,+51 \, p_1^2 \,  \, p_2^6 \, (p_1\cdot p_2)   \notag \\ &&
		+84 \, p_1^4 \,  \, p_2^4 \, (p_1\cdot p_2)+13 \, p_1^6 \,  \, p_2^2 \, (p_1\cdot p_2) \Big] \notag \\ &&
		-3 C_A \Big[ -48 (p_1\cdot p_2)^6-240 \, p_1^2 \,  \, (p_1\cdot p_2)^5-72 \, p_2^2 \, (p_1\cdot p_2)^5  \notag \\ &&
		-156 \, p_1^4 \,  \, (p_1\cdot p_2)^4 -24 \, p_2^4 \, (p_1\cdot p_2)^4-344 \, p_1^2 \,  \, p_2^2 \, (p_1\cdot p_2)^4-34 \, p_1^6 \,  \, (p_1\cdot p_2)^3-48 \, p_1^2 \,  \, p_2^4 \, (p_1\cdot p_2)^3   \notag \\ &&
		+54 \, p_1^4 \,  \, p_2^2 \, (p_1\cdot p_2)^3 +24 \, p_1^2 \,  \, p_2^6 \, (p_1\cdot p_2)^2+486 \, p_1^4 \,  \, p_2^4 \, (p_1\cdot p_2)^2+150 \, p_1^6 \,  \, p_2^2 \, (p_1\cdot p_2)^2+15 \, p_1^4 \,  \, p_2^8 \, \notag \\ &&
		+56 \, p_1^6 \,  \, p_2^6 \, +81 \, p_1^8 \,  \, p_2^4  +195 \, p_1^4 \,  \, p_2^6 \, (p_1\cdot p_2)+336 \, p_1^6 \,  \, p_2^4 \, (p_1\cdot p_2)+49 \, p_1^8 \,  \, p_2^2 \, (p_1\cdot p_2) \Big] \\
		A_{22} &=&  3 \, p_2^2 \, n_f \Big[-24 (p_1\cdot p_2)^5-20 \, p_1^2 \,  \, (p_1\cdot p_2)^4+24 \, p_1^4 \,  \, (p_1\cdot p_2)^3+2 \, p_2^4 \, (p_1\cdot p_2)^3+90 \, p_1^2 \,  \, p_2^2 \, (p_1\cdot p_2)^3  \notag \\ &&
		+12 \, p_1^6 \,  \, (p_1\cdot p_2)^2+54 \, p_1^2 \,  \, p_2^4 \, (p_1\cdot p_2)^2+162 \, p_1^4 \,  \, p_2^2 \, (p_1\cdot p_2)^2+21 \, p_1^4 \,  \, p_2^6 \,+8 \, p_1^6 \,  \, p_2^4 \,+3 \, p_1^8 \,  \, p_2^2 \,+13 \, p_1^2 \,  \, p_2^6 \, (p_1\cdot p_2)  \notag \\ &&
		+84 \, p_1^4 \,  \, p_2^4 \, (p_1\cdot p_2)+51 \, p_1^6 \,  \, p_2^2 \, (p_1\cdot p_2)\Big]-3 C_A \Big[-48 (p_1\cdot p_2)^6-72 \, p_1^2 \,  \, (p_1\cdot p_2)^5-240 \, p_2^2 \, (p_1\cdot p_2)^5  \notag \\ &&
		-24 \, p_1^4 \,  \, (p_1\cdot p_2)^4-156 \, p_2^4 \, (p_1\cdot p_2)^4-344 \, p_1^2 \,  \, p_2^2 \, (p_1\cdot p_2)^4-34 \, p_2^6 \, (p_1\cdot p_2)^3+54 \, p_1^2 \,  \, p_2^4 \, (p_1\cdot p_2)^3  \notag \\ &&
		-48 \, p_1^4 \,  \, p_2^2 \, (p_1\cdot p_2)^3+150 \, p_1^2 \,  \, p_2^6 \, (p_1\cdot p_2)^2+486 \, p_1^4 \,  \, p_2^4 \, (p_1\cdot p_2)^2+24 \, p_1^6 \,  \, p_2^2 \, (p_1\cdot p_2)^2+81 \, p_1^4 \,  \, p_2^8 \,  \notag \\ &&
		+56 \, p_1^6 \,  \, p_2^6 \,+15 \, p_1^8 \,  \, p_2^4 \,+49 \, p_1^2 \,  \, p_2^8 \, (p_1\cdot p_2)+336 \, p_1^4 \,  \, p_2^6 \, (p_1\cdot p_2)+195 \, p_1^6 \,  \, p_2^4 \, (p_1\cdot p_2)\Big] \\
		A_{23} &=& 3 C_A \, \Big(2 (p_1\cdot p_2)+\, p_1^2 \,  \,+\, p_2^2 \Big)^2 \, \Big[-44 (p_1\cdot p_2)^4-34 \, p_1^2 \,  \, (p_1\cdot p_2)^3-34 \, p_2^2 \, (p_1\cdot p_2)^3-22 \, p_1^2 \,  \, p_2^2 \, (p_1\cdot p_2)^2  \notag \\ &&
		+96 \, p_1^4 \,  \, p_2^4 \,+49 \, p_1^2 \,  \, p_2^4 \, (p_1\cdot p_2)+49 \, p_1^4 \,  \, p_2^2 \, (p_1\cdot p_2) \Big]  \notag \\ &&
		-3 n_f (2 (p_1\cdot p_2)+\, p_1^2 \,  \,+\, p_2^2 \,)^2  \Big[ -8 (p_1\cdot p_2)^4+2 \, p_1^2 \,  \, (p_1\cdot p_2)^3+2 \, p_2^2 \, (p_1\cdot p_2)^3+14 \, p_1^2 \,  \, p_2^2 \, (p_1\cdot p_2)^2  \notag \\ &&
		+24 \, p_1^4 \,  \, p_2^4 \,+13 \, p_1^2 \,  \, p_2^4 \, (p_1\cdot p_2)+13 \, p_1^4 \,  \, p_2^2 \, (p_1\cdot p_2) \Big]  \\
		A_{24} &=& 9\, C_A \Big(2 (p_1\cdot p_2)+\, p_1^2 \,  \,+\, p_2^2 \Big) \, \Big[ -32 (p_1\cdot p_2)^6-16 \, p_1^2 \,  \, (p_1\cdot p_2)^5-16 \, p_2^2 \, (p_1\cdot p_2)^5-8 \, p_1^4 \,  \, (p_1\cdot p_2)^4  \notag \\ &&
		-8 \, p_2^4 \, (p_1\cdot p_2)^4-24 \, p_1^2 \,  \, p_2^4 \, (p_1\cdot p_2)^3-24 \, p_1^4 \,  \, p_2^2 \, (p_1\cdot p_2)^3+8 \, p_1^2 \,  \, p_2^6 \, (p_1\cdot p_2)^2+20 \, p_1^4 \,  \, p_2^4 \, (p_1\cdot p_2)^2  \notag \\ &&
		+8 \, p_1^6 \,  \, p_2^2 \, (p_1\cdot p_2)^2+5 \, p_1^4 \,  \, p_2^8 \,+42 \, p_1^6 \,  \, p_2^6 \,+5 \, p_1^8 \,  \, p_2^4 \,+60 \, p_1^4 \,  \, p_2^6 \, (p_1\cdot p_2)+60 \, p_1^6 \,  \, p_2^4 \, (p_1\cdot p_2) \Big]  \notag \\ &&
		-9 \, p_1^2 \,  \, p_2^2 \, n_f  \Big( (2 (p_1\cdot p_2)+\, p_1^2 \,  \,+\, p_2^2 \, \Big)^2 \Big[-4 (p_1\cdot p_2)^3+4 \, p_1^2 \,  \, (p_1\cdot p_2)^2+4 \, p_2^2 \, (p_1\cdot p_2)^2   \notag \\ &&
		+\, p_1^2 \,  \, p_2^4 \,+\, p_1^4 \,  \, p_2^2 \,+14 \, p_1^2 \,  \, p_2^2 \, (p_1\cdot p_2) \Big]
	\end{eqnarray*}

	\begin{eqnarray*}
		A_{31} &=& 	3 \, p_1^2 \,  \, n_f \Big[-16 (p_1\cdot p_2)^5-4 \, p_1^2 \,  \, (p_1\cdot p_2)^4-70 \, p_2^2 \, (p_1\cdot p_2)^4-54 \, p_2^4 \, (p_1\cdot p_2)^3-44 \, p_1^2 \,  \, p_2^2 \, (p_1\cdot p_2)^3  \notag \\ &&
		-12 \, p_2^6 \, (p_1\cdot p_2)^2-15 \, p_1^2 \,  \, p_2^4 \, (p_1\cdot p_2)^2-15 \, p_1^4 \,  \, p_2^2 \, (p_1\cdot p_2)^2-3 \, p_1^2 \,  \, p_2^8 \,-5 \, p_1^4 \,  \, p_2^6 \,+4 \, p_1^6 \,  \, p_2^4 \,-6 \, p_1^2 \,  \, p_2^6 \, (p_1\cdot p_2)\Big]  \notag \\ &&
		-3 C_A\Big[-24 (p_1\cdot p_2)^6-100 \, p_1^2 \,  \, (p_1\cdot p_2)^5-12 \, p_2^2 \, (p_1\cdot p_2)^5-34 \, p_1^4 \,  \, (p_1\cdot p_2)^4-172 \, p_1^2 \,  \, p_2^2 \, (p_1\cdot p_2)^4  \notag \\ && 
		-84 \, p_1^2 \,  \, p_2^4 \, (p_1\cdot p_2)^3 -38 \, p_1^4 \,  \, p_2^2 \, (p_1\cdot p_2)^3-12 \, p_1^2 \,  \, p_2^6 \, (p_1\cdot p_2)^2+99 \, p_1^4 \,  \, p_2^4 \, (p_1\cdot p_2)^2-9 \, p_1^6 \,  \, p_2^2 \, (p_1\cdot p_2)^2-3 \, p_1^4 \,  \, p_2^8 \,  \notag \\ &&
		+7 \, p_1^6 \,  \, p_2^6 \,+28 \, p_1^8 \,  \, p_2^4 \,+36 \, p_1^4 \,  \, p_2^6 \, (p_1\cdot p_2)+78 \, p_1^6 \,  \, p_2^4 \, (p_1\cdot p_2)\Big] \\
		A_{32} &=& 3 \, p_2^2 \, n_f \Big[-24 (p_1\cdot p_2)^5-54 \, p_1^2 \,  \, (p_1\cdot p_2)^4-12 \, p_2^2 \, (p_1\cdot p_2)^4-18 \, p_1^4 \,  \, (p_1\cdot p_2)^3-2 \, p_2^4 \, (p_1\cdot p_2)^3 \notag \\ &&
		-62 \, p_1^2 \,  \, p_2^2 \, (p_1\cdot p_2)^3-55 \, p_1^2 \,  \, p_2^4 \, (p_1\cdot p_2)^2-3 \, p_1^4 \,  \, p_2^2 \, (p_1\cdot p_2)^2+7 \, p_1^4 \,  \, p_2^6 \,-3 \, p_1^6 \,  \, p_2^4 \,-13 \, p_1^2 \,  \, p_2^6 \, (p_1\cdot p_2) \notag \\ &&
		-4 \, p_1^4 \,  \, p_2^4 \, (p_1\cdot p_2)+3 \, p_1^6 \,  \, p_2^2 \, (p_1\cdot p_2)\Big] \notag \\ &&
		-3  C_A \Big[-24 (p_1\cdot p_2)^6-12 \, p_1^2 \,  \, (p_1\cdot p_2)^5-108 \, p_2^2 \, (p_1\cdot p_2)^5-42 \, p_2^4 \, (p_1\cdot p_2)^4 \notag \\ &&
		-156 \, p_1^2 \,  \, p_2^2 \, (p_1\cdot p_2)^4-2 \, p_2^6 \, (p_1\cdot p_2)^3-56 \, p_1^2 \,  \, p_2^4 \, (p_1\cdot p_2)^3-48 \, p_1^4 \,  \, p_2^2 \, (p_1\cdot p_2)^3-49 \, p_1^2 \,  \, p_2^6 \, (p_1\cdot p_2)^2 \notag \\ &&
		+111 \, p_1^4 \,  \, p_2^4 \, (p_1\cdot p_2)^2+31 \, p_1^4 \,  \, p_2^8 \,+9 \, p_1^6 \,  \, p_2^6 \,-13 \, p_1^2 \,  \, p_2^8 \, (p_1\cdot p_2)+74 \, p_1^4 \,  \, p_2^6 \, (p_1\cdot p_2)+45 \, p_1^6 \,  \, p_2^4 \, (p_1\cdot p_2)\Big] \notag \\ 
		A_{33} &=& 3 \, C_A \, \Big(2 (p_1\cdot p_2)+\, p_1^2 \,  \,+\, p_2^2 \Big) \, \Big[ -44 (p_1\cdot p_2)^5-34 \, p_1^2 \,  \, (p_1\cdot p_2)^4-38 \, p_2^2 \, (p_1\cdot p_2)^4-2 \, p_2^4 \, (p_1\cdot p_2)^3 \notag \\ &&
		-68 \, p_1^2 \,  \, p_2^2 \, (p_1\cdot p_2)^3-35 \, p_1^2 \,  \, p_2^4 \, (p_1\cdot p_2)^2-9 \, p_1^4 \,  \, p_2^2 \, (p_1\cdot p_2)^2+28 \, p_1^4 \,  \, p_2^6 \,+28 \, p_1^6 \,  \, p_2^4  \notag \\ &&
		-13 \, p_1^2 \,  \, p_2^6 \, (p_1\cdot p_2) +67 \, p_1^4 \,  \, p_2^4 \, (p_1\cdot p_2)\Big] \notag \\ &&
		-3 \, n_f \, \Big(2 (p_1\cdot p_2)+\, p_1^2 \,  \,+\, p_2^2 \Big)^2 \, \Big[-4 (p_1\cdot p_2)^4-2 \, p_2^2 \, (p_1\cdot p_2)^3-15 \, p_1^2 \,  \, p_2^2 \, (p_1\cdot p_2)^2+4 \, p_1^4 \,  \, p_2^4 \,-13 \, p_1^2 \,  \, p_2^4 \, (p_1\cdot p_2) \Big] \notag \\ 
		A_{34} &=& 9 \, C_A \, \Big(2 (p_1\cdot p_2)+\, p_1^2 \,  \,+\, p_2^2 \Big) \, \Big[-16 (p_1\cdot p_2)^6-4 \, p_1^2 \,  \, (p_1\cdot p_2)^5-4 \, p_2^2 \, (p_1\cdot p_2)^5-24 \, p_1^2 \,  \, p_2^4 \, (p_1\cdot p_2)^3 \notag \\ &&
		-16 \, p_1^4 \,  \, p_2^2 \, (p_1\cdot p_2)^3-4 \, p_1^2 \,  \, p_2^6 \, (p_1\cdot p_2)^2-14 \, p_1^4 \,  \, p_2^4 \, (p_1\cdot p_2)^2-\, p_1^4 \,  \, p_2^8 \,\notag \\ &&
		+15 \, p_1^6 \,  \, p_2^6 \,+13 \, p_1^4 \,  \, p_2^6 \, (p_1\cdot p_2)+15 \, p_1^6 \,  \, p_2^4 \, (p_1\cdot p_2)\Big] \notag \\ &&
		-9 \, p_1^2 \,  \, p_2^2 \, n_f \, \Big(2 (p_1\cdot p_2)+\, p_1^2 \,  \,+\, p_2^2 \Big)^2 \, \Big[-6 (p_1\cdot p_2)^3-4 \, p_2^2 \, (p_1\cdot p_2)^2-\, p_1^2 \,  \, p_2^4 \,+\, p_1^2 \,  \, p_2^2 \, (p_1\cdot p_2) \Big] \notag \\
	\end{eqnarray*}
	
	\begin{eqnarray*}
		A_{40} &=& C_A \Big[ -284 \, (p_1\cdot p_2)^5-132 \, p_1^2 \, (p_1\cdot p_2)^4-132 \, p_2^2 \, (p_1\cdot p_2)^4+556 \, p_1^2 \, p_2^2 \, (p_1\cdot p_2)^3 \notag \\ &&
		+258 \, p_1^2 \, p_2^4 \, (p_1\cdot p_2)^2+258 \, p_1^4 \, p_2^2 \, (p_1\cdot p_2)^2-126 \, p_1^4 \, p_2^6 \,-126 \, p_1^6 \, p_2^4 \,-272 \, p_1^4 \, p_2^4 \, (p_1\cdot p_2) \Big]  \notag \\ &&
		+ n_f \Big[ 44 (p_1\cdot p_2)^5+24 \, p_1^2 \, (p_1\cdot p_2)^4+24 \, p_2^2 \, (p_1\cdot p_2)^4-76 \, p_1^2 \, p_2^2 \, (p_1\cdot p_2)^3-42 \, p_1^2 \, p_2^4 \, (p_1\cdot p_2)^2 \notag \\ &&
		-42 \, p_1^4 \, p_2^2 \, (p_1\cdot p_2)^2+18 \, p_1^4 \, p_2^6 \,+18 \, p_1^6 \, p_2^4 \,+32 \, p_1^4 \, p_2^4 \, (p_1\cdot p_2) \Big] \\
		A_{41} &=& 3 \, p_1^2 \,  \, n_f \Big[-16 \, (p_1\cdot p_2)^4-4 \, p_1^2 \,  \, (p_1\cdot p_2)^3-18 \, p_2^2 \, (p_1\cdot p_2)^3-6 \, p_2^4 \, (p_1\cdot p_2)^2+10 \, p_1^2 \,  \, p_2^2 \, (p_1\cdot p_2)^2 \notag \\ &&
		+3 \, p_1^2 \,  \, p_2^6 \,-3 \, p_1^4 \,  \, p_2^4 \,+9 \, p_1^2 \,  \, p_2^4 \, (p_1\cdot p_2)+\, p_1^4 \,  \, p_2^2 \, (p_1\cdot p_2)\Big] \notag \\ &&
		-3 C_A\Big[-24 \, (p_1\cdot p_2)^5-76 \, p_1^2 \,  \, (p_1\cdot p_2)^4-12 \, p_2^2 \, (p_1\cdot p_2)^4-22 \, p_1^4 \,  \, (p_1\cdot p_2)^3-24 \, p_1^2 \,  \, p_2^2 \, (p_1\cdot p_2)^3 \notag \\ &&
		+76 \, p_1^4 \,  \, p_2^2 \, (p_1\cdot p_2)^2+9 \, p_1^4 \,  \, p_2^6 \,-9 \, p_1^6 \,  \, p_2^4 \,+39 \, p_1^4 \,  \, p_2^4 \, (p_1\cdot p_2)+19 \, p_1^6 \,  \, p_2^2 \, (p_1\cdot p_2)\Big] \\
		A_{42} &=&  3 \, p_2^2 \, n_f \Big[-16 (p_1\cdot p_2)^4-18 \, p_1^2 \,  \, (p_1\cdot p_2)^3-4 \, p_2^2 \, (p_1\cdot p_2)^3-6 \, p_1^4 \,  \, (p_1\cdot p_2)^2+10 \, p_1^2 \,  \, p_2^2 \, (p_1\cdot p_2)^2 \notag \\ &&
		-3 \, p_1^4 \,  \, p_2^4 \,+3 \, p_1^6 \,  \, p_2^2 \,+\, p_1^2 \,  \, p_2^4 \, (p_1\cdot p_2)+9 \, p_1^4 \,  \, p_2^2 \, (p_1\cdot p_2)\Big] \notag \\ &&
		-3 C_A\Big[-24 \, (p_1\cdot p_2)^5-12 \, p_1^2 \,  \, (p_1\cdot p_2)^4-76 \, p_2^2 \, (p_1\cdot p_2)^4-22 \, p_2^4 \, (p_1\cdot p_2)^3-24 \, p_1^2 \,  \, p_2^2 \, (p_1\cdot p_2)^3 \notag \\ &&
		+76 \, p_1^2 \,  \, p_2^4 \, (p_1\cdot p_2)^2-9 \, p_1^4 \,  \, p_2^6 \,+9 \, p_1^6 \,  \, p_2^4 \,+19 \, p_1^2 \,  \, p_2^6 \, (p_1\cdot p_2)+39 \, p_1^4 \,  \, p_2^4 \, (p_1\cdot p_2)\Big] \\
		A_{43} &=& 3 C_A \, (p_1\cdot p_2) \, \Big(2 (p_1\cdot p_2)+\, p_1^2 \,  \,+\, p_2^2 \Big)^2 \Big[ 19 \, p_1^2 \,  \, p_2^2 \,-22 (p_1\cdot p_2)^2 \Big] \notag \\ &&
		-3 n_f \, (p_1\cdot p_2) \, \Big(2 (p_1\cdot p_2)+\, p_1^2 \,  \,+\, p_2^2 \Big)^2 \Big[ p_1^2 \,  \, p_2^2 \,-4 (p_1\cdot p_2)^2 \Big] \\
		A_{44} &=& 9 C_A \Big(2 (p_1\cdot p_2)+\, p_1^2 \,  \,+\, p_2^2 \Big) \Big[-16 (p_1\cdot p_2)^5-4 \, p_1^2 \,  \, (p_1\cdot p_2)^4-4 \, p_2^2 \, (p_1\cdot p_2)^4+16 \, p_1^2 \,  \, p_2^2 \, (p_1\cdot p_2)^3 \notag \\ &&
		+3 \, p_1^4 \,  \, p_2^6 \,+3 \, p_1^6 \,  \, p_2^4 \,-2 \, p_1^4 \,  \, p_2^4 \, (p_1\cdot p_2) -9 \, p_1^2 \,  \, p_2^2 \, n_f (2 (p_1\cdot p_2)+\, p_1^2 \,  \,+\, p_2^2 \,)^2 \, p_1^2 \,  \, p_2^2 \,-2 (p_1\cdot p_2)^2 \Big]
	\end{eqnarray*}
}

\section{Explicit kernel expressions in the impact parameter space}\label{chza:kernelexpb}
In this appendix we provide the explicit expression of the hard-scattering kernel at tree level together with the corrections induced by the $TJJ$ insertion. The transverse momentum dependence is retained only in the gluon propagator connecting the constituent quarks, both for the leading-order contribution and for the $TJJ$ correction. This approximation is motivated by the fact that the dominant transverse-distance effects originate from the exchanged hard gluon, while the intrinsic transverse momentum dependence of the external valence quarks gives rise only to subleading contributions in the kinematic regime considered here.

More precisely, the quark propagators carry an off-shell virtuality of order $\alpha Q^2$, which 
is larger than the virtuality $\alpha_1\alpha_2 Q^2$ 
of the exchanged gluon, particularly in the endpoint region. As a consequence, the transverse momentum 
dependence is negligible in the quark propagators compared to their 
large collinear off-shellness, while it plays a crucial role in 
regulating the endpoint singularities arising from the gluon propagator. 
The transverse momentum appearing in the Dirac numerators is neglected 
separately, as it constitutes a higher-power effect suppressed relative 
to the leading-twist contribution. The full $k_T$ dependence is 
therefore retained exclusively in the gluon propagator, where it 
provides the essential infrared regulator and allows for a natural 
implementation of Sudakov suppression in impact-parameter space.

The approximation considerably simplifies the structure of the hard kernel while preserving the essential non-collinear dynamics. In particular, after Fourier transformation to impact-parameter space, the dependence on the transverse momentum is encoded in modified Bessel functions, leading to compact analytical expressions suitable for numerical evaluation. In impact-parameter space the tree-level kernel reads

%In this appendix we give the explicit expression of the kernel function at the tree level and the kernel corrections due to the $TJJ$  insertion, both at non zero transverse momentum $k_T$ has been used only in the gluon propagator connecting the constituent quarks, both at tree level and in the $TJJ$ correction. In the impact parameter space the tree level kenel reads
\begin{equation}
	K_{\text{tree}}^{\mu\nu}(\alpha_1,\alpha_2) = F_1^{\text{tree}}\, P_1^{\mu } P_2^{\nu } 
	+ F_2^{\text{tree}}\, P_2^{\mu } P_1^{\nu } 
	+ F_3^{\text{tree}}\, P_1^{\mu } P_1^{\nu } 
	+ F_4^{\text{tree}}\, P_2^{\mu } P_2^{\nu } 
	+ F_5^{\text{tree}}\, g^{\mu\nu},
\end{equation}
with
\begin{equation}
	F_1^{\text{tree}}=F_2^{\text{tree}}= 4\left(K_0\left({b{Q} \sqrt{(\alpha_1-1) (\alpha_2-1)}}\right)+K_0\left({b{Q} \sqrt{\alpha_1 \alpha_2}}\right)\right),
\end{equation}

\begin{equation}
	F_3^{\text{tree}}=F_4^{\text{tree}}(\alpha_1\leftrightarrow\alpha_2)=2\frac{\alpha_1 (2 \alpha_2-1)-\alpha_2+1}{\alpha_1+\alpha_2-1}\left(K_0\left({b{Q} \sqrt{(\alpha_1-1) (\alpha_2-1)}}\right)-K_0\left({b{Q} \sqrt{\alpha_1 \alpha_2}}\right)\right),
\end{equation}

\begin{equation}
	F_5^{\text{tree}}=-{Q}^2 \left(K_0\left({b{Q} \sqrt{(\alpha_1-1) (\alpha_2-1)}}\right)+K_0\left({b{Q} \sqrt{\alpha_1 \alpha_2}}\right)\right).
\end{equation}
where $K_0$ is the modified Bessel function of order zero. 

The same treatment of transverse momentum is adopted for the $TJJ$ correction. In this case the insertion modifies the gluon propagator structure while preserving the factorized dependence on the impact parameter. The resulting correction kernel can therefore, be written in a form analogous to the tree-level contribution, with the transverse dynamics entirely governed by the Fourier transform of the corrected gluon propagator. In evaluating the hard-scattering kernel corrected by the $TJJ$, we employ the $\overline{\text{MS}}$ scheme. Accordingly, the scale $\mu$ featuring in the loop integrals and logarithms represents the renormalization scale, to be distinguished from the factorization scale governing the DAs.
In this framework the corrections to the kernel due to the $TJJ$ insertion are given by
\begin{equation}
	K_{{TJJ}}^{\mu\nu}(\alpha_1,\alpha_2) = F_1^{{TJJ}}\, P_1^{\mu } P_2^{\nu } 
	+ F_2^{{TJJ}}\, P_2^{\mu } P_1^{\nu } 
	+ F_3^{{TJJ}}\, P_1^{\mu } P_1^{\nu } 
	+ F_4^{{TJJ}}\, P_2^{\mu } P_2^{\nu } 
	+ F_5^{{TJJ}}\, g^{\mu\nu},
\end{equation}
\begin{align}
	F_1^{TJJ}=F_2^{TJJ}
	=&
	\frac{\KBessel}{\Den}\nonumber\\&
	\biggl[
	\frac{\bar C_{11}\, g_s^2}
	{144\pi^2(\alpha_1-\alpha_2)^2}
	-
	\frac{g_s^2}
	{\pi^2(\alpha_1-\alpha_2)^4}
	\biggl(
	\frac{\bar C_{12}}{8}\,\Cz\, Q^2
	\nonumber\\&
	+\frac{\bar C_{13}}{24}\,\Lq
	+\frac{\bar C_{14}}{24}\,\La
	+\frac{\bar C_{15}}{24}\,\Lb
	\biggl)
	\biggl]
\end{align}

where
\begin{align}
	\bar C_{11}
	&=
	-(65C_A-23n_f)(1-2\alpha_1)^2\alpha_2^4
	\nonumber\\
	&\quad
	+
	(2\alpha_1-1)
	\Bigl[
	(4(3-26\alpha_1)\alpha_1-47)n_f
	+
	(4(68\alpha_1-3)\alpha_1+161)C_A
	\Bigr]\alpha_2^3
	\nonumber\\
	&\quad
	+
	\Bigl[
	n_f\alpha_1\bigl(2(46\alpha_1^2+64\alpha_1+29)\alpha_1+1\bigr)
	\nonumber\\
	&\qquad
	-
	C_A\bigl((2\alpha_1(2(65\alpha_1+74)\alpha_1+215)-113)\alpha_1+54\bigr)
	\Bigr]\alpha_2^2
	\nonumber\\
	&\quad
	+
	\Bigl[
	(1-2\alpha_1(46\alpha_1+53))\alpha_1 n_f
	-36n_f
	\nonumber\\
	&\qquad
	+
	\bigl((260\alpha_1^2+334\alpha_1+113)\alpha_1+144\bigr)C_A
	\Bigr]\alpha_1\alpha_2
	\nonumber\\
	&\quad
	+
	\Bigl[
	n_f\alpha_1(23\alpha_1+47)
	-
	C_A(5\alpha_1+2)(13\alpha_1+27)
	\Bigr]\alpha_1^2 ,
\end{align}

\begin{align}
	\bar C_{12}
	&=
	2(\alpha_1-1)(\alpha_2-1)n_f\alpha_1\alpha_2
	\Bigl(
	(2\alpha_2((2\alpha_2-3)\alpha_2+4)-3)\alpha_1^3
	\nonumber\\
	&\qquad
	+\left((-6\alpha_2^2+2\alpha_2-5)\alpha_2+1\right)\alpha_1^2
	\nonumber\\
	&\qquad
	+((8\alpha_2-5)\alpha_2+4)\alpha_2\alpha_1
	-3\alpha_2^3+\alpha_2^2
	\Bigr)
	\nonumber\\
	&\quad
	+C_A\Bigl(
	(4(\alpha_2-1)\alpha_2+3)\alpha_1^6
	\nonumber\\
	&\qquad
	-(4((\alpha_2((2\alpha_2-5)\alpha_2+16)-16)\alpha_2+7)\alpha_2+1)\alpha_1^5
	\nonumber\\
	&\qquad
	+(((4\alpha_2((5\alpha_2+7)\alpha_2+1)-23)\alpha_2+11)\alpha_2+2)\alpha_1^4
	\nonumber\\
	&\qquad
	+2(2(-16\alpha_2^2+\alpha_2-4)\alpha_2+9)\alpha_2^2\alpha_1^3
	\nonumber\\
	&\qquad
	+(\alpha_2((4\alpha_2(\alpha_2+16)-23)\alpha_2+18)-16)\alpha_2^2\alpha_1^2
	\nonumber\\
	&\qquad
	+(11-4\alpha_2(\alpha_2+7))\alpha_2^4\alpha_1
	+\left((3\alpha_2-1)\alpha_2+2\right)\alpha_2^4
	\Bigr),
\end{align}

\begin{align}
	\bar C_{13}
	&=
	2n_f\Bigl(
	((\alpha_2-1)(12(\alpha_2-1)\alpha_2+19)\alpha_2+2)\alpha_1^4
	\nonumber\\
	&\qquad
	-4\alpha_2(\alpha_2((6\alpha_2-7)\alpha_2+8)-5)\alpha_1^3
	\nonumber\\
	&\qquad
	+(\alpha_2((31\alpha_2-32)\alpha_2+22)-9)\alpha_2\alpha_1^2
	\nonumber\\
	&\qquad
	+((20-19\alpha_2)\alpha_2-9)\alpha_2^2\alpha_1
	+2\alpha_2^4
	\Bigr)
	\nonumber\\
	&\quad
	+C_A\Bigl(
	(-2(\alpha_2-1)(12(\alpha_2-1)\alpha_2+49)\alpha_2-7)\alpha_1^4
	\nonumber\\
	&\qquad
	+(2\alpha_2(2\alpha_2(4(3\alpha_2+4)\alpha_2+1)-29)-15)\alpha_1^3
	\nonumber\\
	&\qquad
	+(33-2\alpha_2((61\alpha_2-2)\alpha_2+1))\alpha_2\alpha_1^2
	\nonumber\\
	&\qquad
	+(98\alpha_2^2-58\alpha_2+33)\alpha_2^2\alpha_1
	-\alpha_2^3(7\alpha_2+15)
	\Bigr).
\end{align}

\begin{align}
	\bar C_{14}
	&=
	(2\alpha_1-1)
	\left(
	\alpha_1 n_f+n_f-C_A(\alpha_1+4)
	\right)
	\alpha_1\alpha_2^6
	\nonumber\\
	&\quad
	+\Bigl(
	2(((3-7\alpha_1)\alpha_1-8)\alpha_1+5)n_f
	+\left(\alpha_1(2(7\alpha_1+9)\alpha_1+37)-34\right)C_A
	\Bigr)\alpha_1\alpha_2^5
	\nonumber\\
	&\quad
	+\Bigl(
	(\alpha_1((2(41-7\alpha_1)\alpha_1-73)\alpha_1+62)-21)n_f\alpha_1
	\nonumber\\
	&\qquad
	+\Bigl(
	((\alpha_1(2(7\alpha_1-59)\alpha_1+49)-86)\alpha_1+36)\alpha_1+6
	\Bigr)C_A
	\Bigr)\alpha_2^4
	\nonumber\\
	&\quad
	+\Bigl(
	((\alpha_1((2\alpha_1(\alpha_1+3)-73)\alpha_1+60)-45)\alpha_1+6)n_f
	\nonumber\\
	&\qquad
	+\left(
	(\alpha_1(49-2(\alpha_1-9)\alpha_1)+36)\alpha_1+30
	\right)\alpha_1 C_A
	\Bigr)\alpha_1\alpha_2^3
	\nonumber\\
	&\quad
	+\Bigl(
	((\alpha_1((\alpha_1-16)\alpha_1+62)-45)\alpha_1+24)n_f
	\nonumber\\
	&\qquad
	+\left(
	\alpha_1(((37-7\alpha_1)\alpha_1-86)\alpha_1+30)-48
	\right)C_A
	\Bigr)\alpha_1^2\alpha_2^2
	\nonumber\\
	&\quad
	+\Bigl(
	2C_A\alpha_1((2\alpha_1-17)\alpha_1+18)
	\nonumber\\
	&\qquad
	-n_f((\alpha_1-7)(\alpha_1-3)\alpha_1-6)
	\Bigr)\alpha_1^3\alpha_2
	+6\alpha_1^4 C_A,
\end{align}

\begin{align}
	\bar C_{15}
	&=
	-n_f(\alpha_1-1)
	\Bigl(
	(\alpha_2-2)(\alpha_2-1)(2\alpha_2-1)\alpha_1^5
	\nonumber\\
	&\qquad
	-(\alpha_2(2(7\alpha_2-13)\alpha_2+21)-4)(\alpha_2-1)\alpha_1^4
	\nonumber\\
	&\qquad
	-(\alpha_2(2(7\alpha_2-15)\alpha_2+35)-28)(\alpha_2-1)\alpha_2\alpha_1^3
	\nonumber\\
	&\qquad
	+(\alpha_2-1)\alpha_2
	\Bigl(
	\alpha_2((2\alpha_2(\alpha_2+13)-35)\alpha_2+32)-18
	\Bigr)\alpha_1^2
	\nonumber\\
	&\qquad
	-(\alpha_2-1)
	\Bigl(
	(\alpha_2(5\alpha_2+21)-28)\alpha_2+18
	\Bigr)\alpha_2^2\alpha_1
	\nonumber\\
	&\qquad
	+2\alpha_2^4(\alpha_2^2+\alpha_2-2)
	\Bigr)
	\nonumber\\
	&\quad
	+C_A
	\Bigl(
	(\alpha_2-5)(\alpha_2-1)(2\alpha_2-1)\alpha_1^6
	\nonumber\\
	&\qquad
	-(\alpha_2-1)(\alpha_2(2(7\alpha_2-24)\alpha_2+49)-5)\alpha_1^5
	\nonumber\\
	&\qquad
	+\Bigl(
	(\alpha_2((2(11-7\alpha_2)\alpha_2-57)\alpha_2+173)-119)\alpha_2+1
	\Bigr)\alpha_1^4
	\nonumber\\
	&\qquad
	+\Bigl(
	((\alpha_2(\alpha_2(2\alpha_2(\alpha_2+31)-57)-60)-44)\alpha_2+58)\alpha_2+15
	\Bigr)\alpha_1^3
	\nonumber\\
	&\qquad
	-\alpha_2
	\Bigl(
	(\alpha_2((\alpha_2(13\alpha_2+97)-173)\alpha_2+44)-50)\alpha_2+33
	\Bigr)\alpha_1^2
	\nonumber\\
	&\qquad
	+\Bigl(
	\alpha_2((2\alpha_2(8\alpha_2+27)-119)\alpha_2+58)-33
	\Bigr)\alpha_2^2\alpha_1
	\nonumber\\
	&\qquad
	+\left(-5\alpha_2^3-5\alpha_2^2+\alpha_2+15\right)\alpha_2^3
	\Bigr).
\end{align}

\begin{align}
	F_3^{TJJ}=F_4^{TJJ}(\alpha_1\leftrightarrow\alpha_2)
	=&
	\frac{\KBessel}{\Den}\nonumber\\&
	\times\biggl[
	\frac{\bar C_{31}\, g_s^2}
	{72\pi^2(\alpha_1-\alpha_2)^2}
	-
	\frac{g_s^2}
	{\pi^2(\alpha_1-\alpha_2)^4}
	\biggl(
	\frac{\bar C_{32}}{4}\,\Cz\, Q^2
	\nonumber\\&
	+\frac{\bar C_{33}}{24}\,\Lq
	+\frac{\bar C_{34}}{24}\,\La
	+\frac{\bar C_{35}}{24}\,\Lb
	\biggl)
	\biggl]
\end{align}

with
\begin{align}
	\bar C_{31}
	&=
	(\alpha_1-1)(2\alpha_1-1)(65C_A-23n_f)\alpha_1\alpha_2^3
	\nonumber\\
	&\quad
	+
	\Bigl[
	((\alpha_1-1)\alpha_1((104\alpha_1-35)\alpha_1+20)-3)n_f
	\nonumber\\
	&\qquad
	+
	\bigl(
	12-(\alpha_1-1)\alpha_1((272\alpha_1-77)\alpha_1+116)
	\bigr)C_A
	\Bigr]\alpha_2^2
	\nonumber\\
	&\quad
	+
	\Bigl[
	(\alpha_1(((11-46\alpha_1)\alpha_1-17)\alpha_1+64)-6)n_f
	\nonumber\\
	&\qquad
	+
	(\alpha_1(\alpha_1((130\alpha_1-53)\alpha_1+167)-256)-12)C_A
	\Bigr]\alpha_1\alpha_2
	\nonumber\\
	&\quad
	+
	\Bigl[
	((\alpha_1-1)\alpha_1(23\alpha_1+32)-3)n_f
	\nonumber\\
	&\qquad
	+
	\bigl(
	12-(\alpha_1-1)\alpha_1(65\alpha_1+128)
	\bigr)C_A
	\Bigr]\alpha_1^2 ,
\end{align}
\begin{align}
	\bar C_{32}
	&=
	-\biggl((\alpha_1-1)\alpha_1
	\Bigl(
	C_A\Bigl(
	3(2\alpha_1-1)\alpha_2^5
	+\left((5-4\alpha_1((\alpha_1-2)\alpha_1+6))\alpha_1+1\right)\alpha_2^4
	\nonumber\\
	&\qquad
	+2(4(\alpha_1+2)\alpha_1^2+1)\alpha_1\alpha_2^3
	-4\alpha_1(\alpha_1+1)((5\alpha_1-3)\alpha_1+1)\alpha_2^2
	\nonumber\\
	&\qquad
	+(\alpha_1(2\alpha_1+11)-2)\alpha_1^3\alpha_2
	-\alpha_1^4(\alpha_1+1)
	\Bigr)
	\nonumber\\
	&\qquad
	-n_f(\alpha_2-1)\alpha_2
	\Bigl(
	-2\alpha_1^4+\alpha_1^3+\alpha_1^2
	+\left((4\alpha_1^2-2\alpha_1-3)\alpha_1+2\right)\alpha_2\alpha_1
	\nonumber\\
	&\qquad\qquad
	+(2\alpha_1-1)\alpha_2^3
	-\left(2(2(\alpha_1-2)\alpha_1+3)\alpha_1^2+\alpha_1-1\right)\alpha_2^2
	\Bigr)
	\Bigr)\biggl),
\end{align}

\begin{align}
	\bar C_{33}
	&=
	-\biggl(\left((-12(\alpha_1-1)\alpha_1-1)n_f
	+(24(\alpha_1-1)\alpha_1-2)C_A\right)\alpha_2^4
	\nonumber\\
	&\quad
	+\Bigl(
	2\Bigl(
	(2\alpha_1^2(3(2\alpha_1-5)\alpha_1+14)-9)n_f\alpha_1
	\nonumber\\
	&\qquad
	+\bigl((33-2\alpha_1(3(2\alpha_1-5)\alpha_1+29))\alpha_1+9\bigr)C_A\alpha_1
	+C_A
	\Bigr)+n_f
	\Bigr)\alpha_2^3
	\nonumber\\
	&\quad
	+\Bigl(
	(11-2\alpha_1(2((9\alpha_1-14)\alpha_1+4)\alpha_1+9))n_f
	\nonumber\\
	&\qquad
	+2(\alpha_1((2\alpha_1(9\alpha_1+7)-31)\alpha_1+3)-7)C_A
	\Bigr)\alpha_1\alpha_2^2
	\nonumber\\
	&\quad
	+\Bigl(
	(2(2\alpha_1(5\alpha_1-4)-7)\alpha_1+11)n_f
	\nonumber\\
	&\qquad
	+(2\alpha_1(\alpha_1(11-28\alpha_1)+25)-14)C_A
	\Bigr)\alpha_1^2\alpha_2
	\nonumber\\
	&\quad
	-(\alpha_1-1)\alpha_1^3
	\Bigl(
	4\alpha_1 n_f+n_f+(2-22\alpha_1)C_A
	\Bigr)\biggl),
\end{align}
\begin{align}
	\bar C_{34}
	&=
	\alpha_1
	\Biggl[
	n_f\alpha_2
	\Bigl(
	-2(\alpha_1^2-1)\alpha_1\alpha_2^4
	+\bigl(2\alpha_1((7(\alpha_1-1)\alpha_1-9)\alpha_1+13)-9\bigr)\alpha_2^3
	\nonumber\\
	&\qquad
	+\Bigl(
	(2\alpha_1(\alpha_1((7\alpha_1-41)\alpha_1+57)-17)-15)\alpha_1+6
	\Bigr)\alpha_2^2
	\nonumber\\
	&\qquad
	+\Bigl(
	(2\alpha_1((-\alpha_1^2+\alpha_1+5)\alpha_1+1)-27)\alpha_1+12
	\Bigr)\alpha_1\alpha_2
	\nonumber\\
	&\qquad
	+\Bigl(
	\alpha_1(3-2(\alpha_1-3)^2\alpha_1)+6
	\Bigr)\alpha_1^2
	\Bigr)
	\nonumber\\
	&\quad
	+2C_A
	\Bigl(
	(\alpha_1-1)(\alpha_1+4)\alpha_1\alpha_2^5
	+\bigl(3-\alpha_1((\alpha_1(7\alpha_1+5)-21)\alpha_1+13)\bigr)\alpha_2^4
	\nonumber\\
	&\qquad
	+\bigl(((59-7\alpha_1)\alpha_1-87)\alpha_1+38\bigr)\alpha_1^2\alpha_2^3
	\nonumber\\
	&\qquad
	+\bigl(
	\alpha_1((\alpha_1((\alpha_1-13)\alpha_1+31)-40)\alpha_1+30)-12
	\bigr)\alpha_1\alpha_2^2
	\nonumber\\
	&\qquad
	+\bigl(
	\alpha_1((4\alpha_1-21)\alpha_1+24)-6
	\bigr)\alpha_1^3\alpha_2
	+3(\alpha_1-1)\alpha_1^4
	\Bigr)
	\Biggr].
\end{align}

\begin{align}
	\bar C_{35}
	&=
	(1-\alpha_1)
	\Biggl[
	2C_A
	\Bigl(
	(\alpha_2-5)(\alpha_2-1)\alpha_1^6
	-(\alpha_2-2)(7(\alpha_2-2)\alpha_2+4)\alpha_1^5
	\nonumber\\
	&\qquad
	+\bigl(
	\alpha_2((\alpha_2(4-7\alpha_2)+11)\alpha_2+15)-11
	\bigr)\alpha_1^4
	\nonumber\\
	&\qquad
	+\bigl(
	((\alpha_2(\alpha_2(\alpha_2+28)-43)-23)\alpha_2+18)\alpha_2+1
	\bigr)\alpha_1^3
	\nonumber\\
	&\qquad
	+\bigl(
	\alpha_2((23-6\alpha_2(\alpha_2+1))\alpha_2+8)-7
	\bigr)\alpha_2\alpha_1^2
	\nonumber\\
	&\qquad
	+\bigl(
	\alpha_2((5\alpha_2-11)\alpha_2+10)-7
	\bigr)\alpha_2^2\alpha_1
	-(\alpha_2-1)\alpha_2^3
	\Bigr)
	\nonumber\\
	&\quad
	-n_f(\alpha_2-1)
	\Bigl(
	2(\alpha_2-2)\alpha_1^6
	-2(\alpha_2-1)(7\alpha_2-2)\alpha_1^5
	\nonumber\\
	&\qquad
	+\bigl(
	4-2\alpha_2((7\alpha_2-15)\alpha_2+4)
	\bigr)\alpha_1^4
	\nonumber\\
	&\qquad
	+\bigl(
	2(\alpha_2(\alpha_2(\alpha_2+17)-20)-4)\alpha_2+1
	\bigr)\alpha_1^3
	\nonumber\\
	&\qquad
	+\left(-6\alpha_2^3-8\alpha_2+11\right)\alpha_2\alpha_1^2
	\nonumber\\
	&\qquad
	+\left(4(\alpha_2-3)\alpha_2+11\right)\alpha_2^2\alpha_1
	+\alpha_2^3
	\Bigr)
	\Biggr].
\end{align}

\begin{align}
	F_5^{TJJ}=
	&\frac{\KBessel}{\Den}\nonumber\\
	&\times
	\biggl[
	\frac{	\bar C_{51}\, g_s^2 Q^2}
	{48\pi^2(\alpha_1-\alpha_2)^2}
	+
	\frac{g_s^2}
	{\pi^2(\alpha_1-\alpha_2)^4}
	\biggl(
	\frac{\bar C_{52}}{16}\,\Cz\, Q^4\nonumber\\
	&
	-\frac{\bar C_{53}}{48}\,\Lq\, Q^2
	-\frac{\bar C_{54}}{48}\,\La\, Q^2
	-\frac{\bar C_{55}}{48}\,\Lb\, Q^2
	\biggl)
	\biggl]
\end{align}

where
\begin{align}
	\bar C_{51}
	&=
	-\Bigl[
	n_f\,\bigl(
	2\alpha_1^2-(2\alpha_2+1)\alpha_1+(2\alpha_2-1)\alpha_2
	\bigr)
	\bigl(\alpha_1(2\alpha_2-1)-\alpha_2\bigr)
	\nonumber\\
	&\qquad
	+
	C_A\Bigl(
	(5-10\alpha_2)\alpha_1^3
	+\bigl((16\alpha_2-1)\alpha_2+2\bigr)\alpha_1^2
	-\alpha_2(10\alpha_2^2+\alpha_2+8)\alpha_1
	+(5\alpha_2+2)\alpha_2^2
	\Bigr)
	\Bigr],
\end{align}
\begin{align}
	\bar C_{52}
	&=
	C_A\Bigl(
	-2\alpha_1^6
	+\left(2(4\alpha_2^2-6\alpha_2+5)\alpha_2+3\right)\alpha_1^5
	\nonumber\\
	&\quad
	-\left(\left(2(4\alpha_2^2+2\alpha_2-5)\alpha_2+13\right)\alpha_2+1\right)\alpha_1^4
	\nonumber\\
	&\quad
	+2\left((2\alpha_2-1)(2\alpha_2^2+3)\alpha_2+2\right)\alpha_2\alpha_1^3
	\nonumber\\
	&\quad
	+2\left(((5-6\alpha_2)\alpha_2-3)\alpha_2+1\right)\alpha_2^2\alpha_1^2
	\nonumber\\
	&\quad
	+(5\alpha_2-4)(2\alpha_2-1)\alpha_2^3\alpha_1
	+\left((3-2\alpha_2)\alpha_2-1\right)\alpha_2^4
	\Bigr)
	\nonumber\\
	&\quad
	-2n_f(\alpha_1-1)\alpha_1(\alpha_2-1)\alpha_2
	\Bigl((\alpha_1-1)\alpha_1+(\alpha_2-1)\alpha_2\Bigr)
	\Bigl(\alpha_1(2\alpha_2-1)-\alpha_2\Bigr),
\end{align}

\begin{align}
	\bar C_{53}
	&=
	-n_f\Bigl((\alpha_1-1)\alpha_1+(\alpha_2-1)\alpha_2\Bigr)
	\Bigl(
	(12(\alpha_2-1)\alpha_2+1)\alpha_1^2
	\nonumber\\
	&\qquad
	+2\alpha_2(5-6\alpha_2)\alpha_1
	+\alpha_2^2
	\Bigr)
	\nonumber\\
	&\quad
	-2C_A\Bigl(
	(1-12(\alpha_2-1)\alpha_2)\alpha_1^4
	+\left(2\alpha_2(6\alpha_2^2+3\alpha_2-10)-1\right)\alpha_1^3
	\nonumber\\
	&\qquad
	+\left(2(-6\alpha_2^2+3\alpha_2+1)\alpha_2+7\right)\alpha_2\alpha_1^2
	\nonumber\\
	&\qquad
	+(6\alpha_2-7)(2\alpha_2-1)\alpha_2^2\alpha_1
	+(\alpha_2-1)\alpha_2^3
	\Bigr),
\end{align}

\begin{align}
	\bar C_{54}
	&=
	n_f\alpha_1\alpha_2
	\Bigl(
	4(\alpha_2-1)\alpha_1^4
	+(8(\alpha_2-2)\alpha_2+9)\alpha_1^3
	\nonumber\\
	&\qquad
	+(\alpha_2(8(\alpha_2-4)\alpha_2+27)-6)\alpha_1^2
	\nonumber\\
	&\qquad
	+(\alpha_2(4(\alpha_2-4)\alpha_2+27)-12)\alpha_2\alpha_1
	\nonumber\\
	&\qquad
	+\left((9-4\alpha_2)\alpha_2-6\right)\alpha_2^2
	\Bigr)
	\nonumber\\
	&\quad
	+C_A\Bigl(
	(3-10(\alpha_2-1)\alpha_2)\alpha_1^5
	-\left((2(\alpha_2-8)\alpha_2+21)\alpha_2+3\right)\alpha_1^4
	\nonumber\\
	&\qquad
	-2\alpha_2\left((\alpha_2-9)(\alpha_2-1)\alpha_2-6\right)\alpha_1^3
	\nonumber\\
	&\qquad
	+2\left(((8-5\alpha_2)\alpha_2-9)\alpha_2+3\right)\alpha_2^2\alpha_1^2
	\nonumber\\
	&\qquad
	+\left((10\alpha_2-21)\alpha_2+12\right)\alpha_2^3\alpha_1
	+3(\alpha_2-1)\alpha_2^4
	\Bigr),
\end{align}

\begin{align}
	\bar C_{55}
	&=
	C_A\Bigl(
	(10(\alpha_2-1)\alpha_2-3)\alpha_1^5
	+\left((2(\alpha_2-20)\alpha_2+45)\alpha_2+5\right)\alpha_1^4
	\nonumber\\
	&\qquad
	+2\left(\alpha_2(\alpha_2((\alpha_2+2)\alpha_2+15)-26)-1\right)\alpha_1^3
	\nonumber\\
	&\qquad
	+2\left((5(\alpha_2-3)(\alpha_2-1)\alpha_2-1)\alpha_2+7\right)\alpha_2\alpha_1^2
	\nonumber\\
	&\qquad
	+\left((5(9-2\alpha_2)\alpha_2-52)\alpha_2+14\right)\alpha_2^2\alpha_1
	\nonumber\\
	&\qquad
	+\left((5-3\alpha_2)\alpha_2-2\right)\alpha_2^3
	\Bigr)
	\nonumber\\
	&\quad
	-n_f(\alpha_1-1)(\alpha_2-1)
	\Bigl(
	4\alpha_2\alpha_1^4
	+(8(\alpha_2-2)\alpha_2+1)\alpha_1^3
	\nonumber\\
	&\qquad
	+(8(\alpha_2-2)\alpha_2+11)\alpha_2\alpha_1^2
	\nonumber\\
	&\qquad
	+(4(\alpha_2-4)\alpha_2+11)\alpha_2^2\alpha_1
	+\alpha_2^3
	\Bigr).
\end{align}

In every expression the massless three-point scalar function $C_0(p_1,p_2,p_3,0,0,0)$, also referred to as $C_0(p_1,p_2,p_3)$ for three arbitrary momenta $p_1$, $p_2$ and $p_3$, is given by
\begin{align}\label{chza:c0formula}
	C_0(p_1,p_2,p_3,0,0,0)=\frac{1}{\Delta}
	\Bigg[&
	Li_2\!\left(-\frac{-x_1+\Delta}{x_1+\Delta}\right)
	+
	Li_2\!\left(-\frac{x_2+\Delta}{-x_2+\Delta}\right)
	+
	Li_2\!\left(-\frac{x_3+\Delta}{-x_3+\Delta}\right)
	\nonumber\\&
	-
	Li_2\!\left(\frac{x_2-\Delta}{x_2+\Delta}\right)
	-
	Li_2\!\left(\frac{-x_1-\Delta}{-x_1+\Delta}\right)
	-
	Li_2\!\left(\frac{x_3-\Delta}{x_3+\Delta}\right)
	\Bigg],
\end{align}
where we have defined
\begin{equation}
	\Delta
	\equiv
	\sqrt{
		p_1^2-2(p_2+p_3)p_1+(p_2-p_3)^2
	}
\end{equation}
and 
\begin{equation}
	x_1 \equiv p_1-p_2-p_3,
	\qquad
	x_2 \equiv p_1-p_2+p_3,
	\qquad
	x_3 \equiv p_1+p_2-p_3 .
\end{equation}

\chapter{Appendix to Chapter 8}

\section{ Basis of the Abelian $TJJ$ and the on-shell limit of the spin-1}
\label{reff}
The correlator can be built using the  set of $13$ tensors catalogued in Table \ref{genbasis}. They are linearly independent for generic $k^2, p^2, q^2$
different from zero. Five of the $13$ tensors are Bose symmetric, namely,
\bea
t_i^{\mu\nu\alpha\beta}(p,q) = t_i^{\mu\nu\beta\alpha}(q,p)\,,\qquad i=1,2,7,8,13\,,
\eea
while the remaining eight tensors form four pairs which are overall related by Bose symmetry
\bea
&&t_3^{\mu\nu\alpha\beta}(p,q) = t_5^{\mu\nu\beta\alpha}(q,p)\,,\\
&&t_4^{\mu\nu\alpha\beta}(p,q) = t_6^{\mu\nu\beta\alpha}(q,p)\,,\\
&&t_9^{\mu\nu\alpha\beta}(p,q) = t_{10}^{\mu\nu\beta\alpha}(q,p)\,,\\
&&t_{11}^{\mu\nu\alpha\beta}(p,q) = t_{12}^{\mu\nu\beta\alpha}(q,p)\,.
\label{tpairs}
\eea
\bea
\widetilde t_1^{\, \mu \nu \a \b} &=& \lim_{s_1,s_2 \rightarrow 0} \,  t_1^{\, \mu \nu \a \b} =
(s \, g^{\mu\nu} - k^{\mu}k^{\nu}) \, u^{\a \b} (p,q)  
\label{widetilde1x}\\
\widetilde t_3^{\, \mu \nu \a \b} \, + \,  \widetilde t_5^{\, \mu \nu \a \b}
- \, 4 \, \widetilde t_7^{\, \mu \nu \a \b} &=&  \lim_{s_1,s_2 \rightarrow 0}  \, (t_3^{\, \mu \nu \a \b} \, + \,  t_5^{\, \mu \nu \a \b}
- \, 4 \,  t_7^{\, \mu \nu \a \b}) = \nn  \\
&& \hspace{-0.5cm} - 2 \, u^{\a \b} (p,q) \left( s \, g^{\mu \nu} + 2 (p^\mu \, p^\nu + q^\mu \, q^\nu )
- 4 \, (p^\mu \, q^\nu + q^\mu \, p^\nu) \right)  \\
\widetilde{t}^{\, \mu \nu \alpha \beta}_{13} &=&  \lim_{s_1,s_2 \rightarrow 0} \,  t_{13}^{\, \mu \nu \a \b} =
\big(p^{\mu} q^{\nu} + p^{\nu} q^{\mu}\big)g^{\alpha\beta}
+ \frac{s}{2} \big(g^{\alpha\nu} g^{\beta\mu} + g^{\alpha\mu} g^{\beta\nu}\big) \nn \\
&&  \hspace{-0.5cm} - g^{\mu\nu} (\frac{s}{2} g^{\alpha \beta}- q^{\alpha} p^{\beta})
-\big(g^{\beta\nu} p^{\mu}
+ g^{\beta\mu} p^{\nu}\big)q^{\alpha}
- \big (g^{\alpha\nu} q^{\mu}
+ g^{\alpha\mu} q^{\nu }\big)p^{\beta}, 
\label{widetilde13}
\nn \\
\eea

\subsection{Evaluations of the $TJJ$}
The ordinary massless scalar master integral is given by
\begin{equation}
	\label{lco}
	C_0\left(q^2,p_1^2,p_2^2\right)= \frac{1}{i \pi^2}\int d^n k\frac{1}{k^2 (k- q)^2 (k - p_1)^2},\,
\end{equation}
In the explicit evaluation we can use the relations  

\bea
&  C_0 ( p_1^2,p_2^2,q^2) = \frac{ 1}{q^2} \Phi (x,y),
\eea
where the function $\Phi (x, y)$  is defined as
\cite{Usyukina:1993ch}
\bea
\Phi( x, y) &=& \frac{1}{\lambda} \biggl\{ 2 [Li_2(-\rho  x) + Li_2(- \rho y)]  +
\ln \frac{y}{ x}\ln \frac{1+ \rho y }{1 + \rho x}+ \ln (\rho x) \ln (\rho  y) + \frac{\pi^2}{3} \biggr\},
\label{Phi}
\eea
with
\bea
\lambda(x,y) = \sqrt {\Delta},
\qquad  \qquad \Delta=(1-  x- y)^2 - 4  x  y,
\label{lambda} \\
\rho( x,y) = 2 (1-  x-  y+\lambda)^{-1},
\qquad  \qquad x=\frac{p_1^2}{q^2} \, ,\qquad \qquad y= \frac {p_2^2}{q^2}\, .
\eea
and 
\begin{equation}
	{B}_0(p^2)=\sdfrac{1}{i\p^\frac{d}{2}}\int\,d^d l\ \frac{1}{l^2(l-p_1)^2}=\frac{ \left[\G\left(\frac{d}{2}-1\right)\right]^2\G\left(2-\frac{d}{2}\right)}{\G\left(d-2\right)(p^2)^{2-\frac{d}{2}}}\label{B0ex}.
\end{equation}
with 
\beq
B_0(p^2)=\frac{1}{\varepsilon}+\bar{B}_0(p^2)
\eeq
and
\beq
B_0^R(p^2,0,0)\equiv \bar{B}_0(p^2)= 2 + \log(\mu^2/p^2) 
\label{BB}
\eeq
is the finite part in $d=4$ of the scalar integral in the $\overline{MS}$ scheme. 
\subsection{The conformal solution}
\label{bms}
The general solution has been given in \cite{Bzowski:2018fql} that here we report  for convenience,

\begin{table}
	$$
	\begin{array}{|c|c|}\hline
		i & t_i^{\mu\nu\alpha\beta}(p,q)\\ \hline\hline
		1 &
		\left(k^2 g^{\mu\nu} - k^{\mu } k^{\nu}\right) u^{\alpha\beta}(p.q)\\ \hline
		2 &
		\left(k^2g^{\mu\nu} - k^{\mu} k^{\nu}\right) w^{\alpha\beta}(p.q)  \\ \hline
		3 & \left(p^2 g^{\mu\nu} - 4 p^{\mu}  p^{\nu}\right)
		u^{\alpha\beta}(p.q)\\ \hline
		4 & \left(p^2 g^{\mu\nu} - 4 p^{\mu} p^{\nu}\right)
		w^{\alpha\beta}(p.q)\\ \hline
		5 & \left(q^2 g^{\mu\nu} - 4 q^{\mu} q^{\nu}\right)
		u^{\alpha\beta}(p.q)\\ \hline
		6 & \left(q^2 g^{\mu\nu} - 4 q^{\mu} q^{\nu}\right)
		w^{\alpha\beta}(p.q) \\ \hline
		7 & \left[p\cdot q\, g^{\mu\nu}
		-2 (q^{\mu} p^{\nu} + p^{\mu} q^{\nu})\right] u^{\alpha\beta}(p.q) \\ \hline
		8 & \left[p\cdot q\, g^{\mu\nu}
		-2 (q^{\mu} p^{\nu} + p^{\mu} q^{\nu})\right] w^{\alpha\beta}(p.q)\\ \hline
		9 & \left(p\cdot q \,p^{\alpha}  - p^2 q^{\alpha}\right)
		\big[p^{\beta} \left(q^{\mu} p^{\nu} + p^{\mu} q^{\nu} \right) - p\cdot q\,
		(g^{\beta\nu} p^{\mu} + g^{\beta\mu} p^{\nu})\big]  \\ \hline
		10 & \big(p\cdot q \,q^{\beta} - q^2 p^{\beta}\big)\,
		\big[q^{\alpha} \left(q^{\mu} p^{\nu} + p^{\mu} q^{\nu} \right) - p\cdot q\,
		(g^{\alpha\nu} q^{\mu} + g^{\alpha\mu} q^{\nu})\big]  \\ \hline
		11 & \left(p\cdot q \,p^{\alpha} - p^2 q^{\alpha}\right)
		\big[2\, q^{\beta} q^{\mu} q^{\nu} - q^2 (g^{\beta\nu} q^ {\mu}
		+ g^{\beta\mu} q^{\nu})\big]  \\ \hline
		12 & \big(p\cdot q \,q^{\beta} - q^2 p^{\beta}\big)\,
		\big[2 \, p^{\alpha} p^{\mu} p^{\nu} - p^2 (g^{\alpha\nu} p^ {\mu}
		+ g^{\alpha\mu} p^{\nu})\big] \\ \hline
		13 & \big(p^{\mu} q^{\nu} + p^{\nu} q^{\mu}\big)g^{\alpha\beta}
		+ p\cdot q\, \big(g^{\alpha\nu} g^{\beta\mu}
		+ g^{\alpha\mu} g^{\beta\nu}\big) - g^{\mu\nu} u^{\alpha\beta} \\
		& -\big(g^{\beta\nu} p^{\mu}
		+ g^{\beta\mu} p^{\nu}\big)q^{\alpha}
		- \big (g^{\alpha\nu} q^{\mu}
		+ g^{\alpha\mu} q^{\nu }\big)p^{\beta}  \\ \hline
	\end{array}
	$$
	\caption{Basis of 13 fourth rank tensors satisfying the vector current conservation on the external lines with momenta $p$ and $q$. \label{genbasis}}
\end{table}

\begin{align}
	A_1
	&=
	C_1\, I_{5\{211\}}
	\nonumber\\
	&=
	C_1\, p_1 p_2 p_3
	\left(
	p_1 \frac{\partial}{\partial p_1}-1
	\right)
	\frac{\partial^3}{\partial p_1\,\partial p_2\,\partial p_3}
	I_{1\{000\}},
	\label{eq:A1}
	\\
	A_2
	&=
	2 C_{JJ}
	\left(
	2- p_1 \frac{\partial}{\partial p_1}
	\right)
	I^{(\mathrm{fin})}_{2\{111\}}
	-
	C_1\, p_1^{3} p_2 p_3
	\frac{\partial^3}{\partial p_1\,\partial p_2\,\partial p_3}
	I_{1\{000\}}
	\nonumber\\
	&\quad
	-\frac{2}{3} C_{JJ}
	\left[
	\ln\!\frac{p_1^{2}}{\mu^{2}}
	+
	\ln\!\frac{p_2^{2}}{\mu^{2}}
	+
	\ln\!\frac{p_3^{2}}{\mu^{2}}
	\right]
	+\frac{2}{3} C_{JJ}
	-2 D_{JJ},
	\label{eq:A2}
	\\
	A_3
	&=
	2 C_{JJ}
	\left(
	2- p_1 \frac{\partial}{\partial p_1}
	\right)
	I^{(\mathrm{fin})}_{2\{111\}}
	+
	2 C_1\, p_1 p_2 p_3^{2}
	\left(
	1- p_1 \frac{\partial}{\partial p_1}
	\right)
	\frac{\partial^{2}}{\partial p_1\,\partial p_2}
	I_{1\{000\}}
	\nonumber\\
	&\quad
	-\frac{2}{3} C_{JJ}
	\left[
	\ln\!\frac{p_1^{2}}{\mu^{2}}
	+
	\ln\!\frac{p_2^{2}}{\mu^{2}}
	+
	\ln\!\frac{p_3^{2}}{\mu^{2}}
	\right]
	-2\left(C_1 + D_{JJ}\right)
	+\frac{2}{3} C_{JJ},
	\label{eq:A3}
	\\
	A_4
	&=
	-2 C_{JJ}\, p_1^{2}\, I^{(\mathrm{fin})}_{2\{111\}}
	-
	2 C_1\, p_2^{2} p_3^{2} p_1
	\left(
	1- p_1 \frac{\partial}{\partial p_1}
	\right)
	\frac{\partial}{\partial p_1}
	I_{1\{000\}}
	\nonumber\\
	&\quad
	+\frac{1}{3} C_{JJ}\, p_1^{2}
	\left[
	\ln\!\frac{p_1^{2}}{\mu^{2}}
	+
	\ln\!\frac{p_2^{2}}{\mu^{2}}
	+
	\ln\!\frac{p_3^{2}}{\mu^{2}}
	\right]
	-
	C_{JJ}
	\left[
	p_2^{2}\ln\!\frac{p_2^{2}}{\mu^{2}}
	+
	p_3^{2}\ln\!\frac{p_3^{2}}{\mu^{2}}
	\right]
	+
	\left(C_1 + D_{JJ}\right)
	\left(
	p_1^{2}-p_2^{2}-p_3^{2}
	\right)
	-
	C_{JJ}\, p_1^{2}.
	\label{eq:A4}
\end{align}
shares a finite IR limit, shown in \eqref{bmcs} with the perturbative solution presented in \cite{Armillis:2009pq}. 
The integrals are given by

\begin{align}
	I_{2\{111\}}^{(\mathrm{fin})}
	=
	-\frac{4 p_1^2 p_2^2 p_3^2}{J^2} \, I_{1\{000\}}
	-\frac{1}{6 J^2}
	\Bigg[&
	p_1^2 \left(p_2^2 + p_3^2 - p_1^2\right)
	\ln\!\left(\frac{p_1^4}{p_2^2 p_3^2}\right)
	+ p_2^2 \left(p_1^2 + p_3^2 - p_2^2\right)
	\ln\!\left(\frac{p_2^4}{p_1^2 p_3^2}\right)
	\nonumber\\
	&
	+ p_3^2 \left(p_1^2 + p_2^2 - p_3^2\right)
	\ln\!\left(\frac{p_3^4}{p_1^2 p_2^2}\right)
	\Bigg] \, .
\end{align}

\begin{equation}
	I_{1\{000\}}
	=
	\frac{1}{2\sqrt{-J^2}}
	\left[
	\frac{\pi^2}{6}
	- 2 \ln\!\left(\frac{p_1}{p_3}\right)
	\ln\!\left(\frac{p_2}{p_3}\right)
	+ \ln X \ln Y
	- \operatorname{Li}_2 X
	- \operatorname{Li}_2 Y
	\right] .
\end{equation}

\begin{equation}
	X =
	\frac{-p_1^2 + p_2^2 + p_3^2 - \sqrt{-J^2}}
	{2 p_3^2} \, ,
	\qquad
	Y =
	\frac{-p_2^2 + p_1^2 + p_3^2 - \sqrt{-J^2}}
	{2 p_3^2} \, .
\end{equation}

\begin{equation}
	J^2 =
	(p_1 + p_2 + p_3)
	(-p_1 + p_2 + p_3)
	(p_1 - p_2 + p_3)
	(p_1 + p_2 - p_3)
\end{equation}

\begin{equation}
	= -p_1^4 - p_2^4 - p_3^4
	+ 2 p_1^2 p_2^2
	+ 2 p_2^2 p_3^2
	+ 2 p_3^2 p_1^2 \, .
\end{equation}

\section{Non covariant TT decomposition of the Fierz--Pauli action and the anomaly-induced scalar channel}
\label{app:TT_anomaly_reorganized}
The analysis presented in the Section \ref{closing}  can be recovered using a non-covariant (1+3) decomposition of the metric fluctuations. The derivation is lengthier compared to the covariant one, but provides a different perspective on the separation of the propagating from the non-propagating degrees of freedom in the general $h_{\mu \nu}$ as identified from the quadratic FP action coupled to an external conserved source. We first analyze the ordinary, non-anomalous case and show that only the TT sector propagates, as expected, while the vector and scalar sectors are constrained. We then include the anomaly pole and identify precisely which part of the decomposition is modified. The conclusion is that the pole present in the source does not affect the TT and vector sectors, but activates a specific nonlocal scalar channel, which can be described either in the \(3+1\) decomposition using two variables $\Theta$ and $\Xi$, defined below, or covariantly through the unique gauge-invariant scalar \(\Phi_{cov}\equiv h-\Box w\).\\
We begin from the quadratic action \eqref{FP} and decompose the metric fluctuation on the spatial slices as
\begin{equation}
	h_{00}=2\Phi,
	\label{eq:h00_reorg}
\end{equation}
\begin{equation}
	h_{0i}=u_i+\partial_i B,
	\qquad
	\partial_i u_i=0,
	\label{eq:h0i_reorg}
\end{equation}
and introduce the tensor decomposition 
\begin{equation}
	h_{ij}
	=
	h_{ij}^{TT}
	+\partial_i v_j+\partial_j v_i
	+\left(\partial_i\partial_j-\frac13\delta_{ij}\nabla^2\right)E
	+\frac13\delta_{ij}H,
	\label{eq:hij_reorg}
\end{equation}
with
\begin{equation}
	\partial_i v_i=0,
	\qquad
	\partial_i h_{ij}^{TT}=0,
	\qquad
	\delta^{ij}h_{ij}^{TT}=0.
	\label{eq:constraints_reorg}
\end{equation}
Under the linearized diffeomorphisms \eqref{eq:lin_diff_ped}
with
\begin{equation}
	\label{decs}
	\xi_i=\xi_i^T+\partial_i\xi,
	\qquad
	\partial_i\xi_i^T=0,
\end{equation}
the variables transform as
\begin{equation}
	\delta h_{ij}^{TT}=0,
	\qquad
	\delta v_i=\xi_i^T,
	\qquad
	\delta E=2\xi,
	\qquad
	\delta H=2\nabla^2\xi,
\end{equation}
\begin{equation}
	\delta u_i=\partial_0\xi_i^T,
	\qquad
	\delta B=\xi_0+\partial_0\xi,
	\qquad
	\delta\Phi=\partial_0\xi_0.
\end{equation}
Notice that $\xi$ in \eqref{decs} is the scalar appearing in the longitudinal-transverse decomposition of the 
gauge variation $\xi_i$.
Natural gauge-invariant combinations are
\begin{equation}
	V_i\equiv u_i-\partial_0v_i,
	\qquad
	\partial_iV_i=0,
	\label{eq:Vi_reorg}
\end{equation}
and
\begin{equation}
	\Theta\equiv \Phi-\partial_0B+\frac12\partial_0^2E,
	\qquad
	\Xi\equiv H-\nabla^2E.
	\label{eq:ThetaXi_reorg}
\end{equation}

\subsection{TT, vector, and scalar sectors in the non-anomalous case}
Using
\begin{equation}
	h=\eta^{\mu\nu}h_{\mu\nu}=2\Phi-H.
	\label{eq:trace_reorg}
\end{equation}
and 
\begin{equation}
	\partial_\mu h^{\mu0}=2\partial_0\Phi-\nabla^2B,
	\label{eq:div0_reorg}
\end{equation}

\begin{equation}
	\partial_\mu h^{\mu i}
	=
	-\partial_0u_i+\nabla^2v_i
	+\partial_i\left(
	-\partial_0B+\frac23\nabla^2E+\frac13H
	\right),
	\label{eq:divi_reorg}
\end{equation}

\begin{equation}
	\partial_\mu h^{\mu i}
	=
	-\partial_0V_i
	+\partial_i\left(
	-\partial_0B+\frac23\nabla^2E+\frac13H
	\right),
\end{equation}
after substitution into \eqref{FP}, one rewrites the Lagrangian in terms of the TT, vector, and scalar components as
\begin{equation}
	\mathcal{L}_{\mathrm{FP}}
	=
	\mathcal{L}_{TT}
	+
	\mathcal{L}_{V}
	+
	\mathcal{L}_{S},
	\label{eq:split_reorg}
\end{equation}
with
\begin{equation}
	\mathcal{L}_{TT}
	=
	\frac14\,\partial_\lambda h_{ij}^{TT}\partial^\lambda h_{ij}^{TT},
	\label{eq:LTT_reorg}
\end{equation}
\begin{equation}
	\mathcal{L}_{V}
	=
	\frac12\,V_i\nabla^2V_i,
	\label{eq:LV_reorg}
\end{equation}
and
\begin{equation}
	\mathcal{L}_{S}
	=
	\frac43\,\Theta\,\nabla^2\Xi
	+\frac13(\partial_0\Xi)^2
	-\frac19(\partial_i\Xi)^2.
	\label{eq:LS_reorg}
\end{equation}

Equivalently, up to total derivatives, the latter takes the form
\begin{equation}
	\mathcal{L}_{S}
	=
	\frac43\,\Theta\,\nabla^2\Xi
	-\frac13\,\Xi\,\partial_0^2\Xi
	+\frac19\,\Xi\,\nabla^2\Xi.
\end{equation}

The full free quadratic action is therefore
\begin{equation}
	S_{\mathrm{FP}}
	=
	\int d^4x\,
	\left[
	\frac14\,\partial_\lambda h_{ij}^{TT}\partial^\lambda h_{ij}^{TT}
	+\frac12\,V_i\nabla^2V_i
	+\frac43\,\Theta\,\nabla^2\Xi
	+\frac13(\partial_0\Xi)^2
	-\frac19(\partial_i\Xi)^2
	\right],
	\label{eq:free_action_reorg}
\end{equation}
giving the equations of motion
\begin{equation}
	\Box h_{ij}^{TT}=0,
	\label{eq:TT_free_reorg}
\end{equation}
\begin{equation}
	\nabla^2V_i=0,
	\label{eq:V_free_reorg}
\end{equation}
\begin{equation}
	\nabla^2\Xi=0.
	\label{eq:Xi_free_reorg}
\end{equation}
Hence, for fields vanishing suitably at spatial infinity,
\begin{equation}
	\Xi=0,
\end{equation}
and the remaining scalar equation determines \(\Theta\) without introducing a propagating scalar mode. Therefore, in the non-anomalous case, the only radiative sector is the TT one, here retrieved in a different decomposition.

\subsection{Coupling to a conserved source}

We now include the coupling to a conserved source,
\begin{equation}
	S_{\mathrm{int}}
	=
	\frac{\kappa}{2}\int d^4x\,h_{\mu\nu}J^{\mu\nu},
	\qquad
	\partial_\mu J^{\mu\nu}=0.
	\label{eq:source_reorg}
\end{equation}
We decompose the source as
\begin{equation}
	J^{00}=\rho,
\end{equation}
\begin{equation}
	J^{0i}=J_T^i+\partial^iJ_L,
	\qquad
	\partial_iJ_T^i=0,
\end{equation}
\begin{equation}
	J^{ij}
	=
	J_{TT}^{ij}
	+\partial_iW_j+\partial_jW_i
	+\left(\partial_i\partial_j-\frac13\delta_{ij}\nabla^2\right)\sigma
	+\frac13\delta_{ij}\tau,
\end{equation}
with
\begin{equation}
	\partial_iW_i=0,
	\qquad
	\partial_iJ_{TT}^{ij}=0,
	\qquad
	\delta_{ij}J_{TT}^{ij}=0.
\end{equation}
Current conservation implies
\begin{equation}
	\partial_0\rho+\nabla^2J_L=0,
	\label{eq:cons_rho_reorg}
\end{equation}
\begin{equation}
	\partial_0J_T^i+\nabla^2W_i=0,
	\label{eq:cons_vec_reorg}
\end{equation}
\begin{equation}
	\partial_0J_L+\frac23\nabla^2\sigma+\frac13\tau=0.
	\label{eq:cons_scal_reorg}
\end{equation}
Using these relations and integrating by parts, the interaction term becomes
\begin{equation}
	\mathcal{L}_{\mathrm{int}}
	=
	\frac{\kappa}{2}\,h_{ij}^{TT}J_{TT}^{ij}
	+
	\kappa V_iJ_T^i
	+
	\kappa\Theta\,\rho
	+
	\frac{\kappa}{6}\,\Xi\,\tau.
	\label{eq:Lint_reorg}
\end{equation}
Combining \eqref{eq:LTT_reorg}, \eqref{eq:LV_reorg}, \eqref{eq:LS_reorg} with \eqref{eq:Lint_reorg}, one gets
\begin{equation}
	\mathcal{L}_{TT}^{(2)}
	=
	\frac14\,\partial_\lambda h_{ij}^{TT}\partial^\lambda h_{ij}^{TT}
	+
	\frac{\kappa}{2}\,h_{ij}^{TT}J_{TT}^{ij},
\end{equation}
\begin{equation}
	\mathcal{L}_{V}^{(2)}
	=
	\frac12\,V_i\nabla^2V_i
	+
	\kappa V_iJ_T^i,
\end{equation}
and using \eqref{eq:ThetaXi_reorg}
\begin{equation}
	\mathcal{L}_{S}^{(2)}
	=
	\frac43\,\Theta\,\nabla^2\Xi
	+\frac13(\partial_0\Xi)^2
	-\frac19(\partial_i\Xi)^2
	+\kappa\Theta\,\rho
	+\frac{\kappa}{6}\,\Xi\,\tau.
	\label{eq:LSsourced_reorg}
\end{equation}
The corresponding equations are
\begin{equation}
	\Box h_{ij}^{TT}=-\kappa J_{TT}^{ij},
\end{equation}
\begin{equation}
	\nabla^2V_i=-\kappa J_T^i,
\end{equation}
\begin{equation}
	\frac43\nabla^2\Xi+\kappa\rho=0,
\end{equation}
\begin{equation}
	\frac43\nabla^2\Theta
	-\frac23\partial_0^2\Xi
	+\frac29\nabla^2\Xi
	+\frac{\kappa}{6}\tau=0.
\end{equation}
Thus the TT sector propagates, whereas the vector and scalar sectors remain constrained even in the presence of an ordinary conserved source. The presence of a term of the form
\begin{equation}
	(\partial_0\Xi)^2
\end{equation}
is necessary, but not sufficient, for \(\Xi\) to become dynamical. What matters is whether \(\Xi\) satisfies an independent second-order hyperbolic equation once all constraints are taken into account.

Indeed, in the ordinary scalar sector one has
\begin{equation}
	\mathcal L_S
	=
	\frac43\,\Theta\,\nabla^2\Xi
	+\frac13(\partial_0\Xi)^2
	-\frac19(\partial_i\Xi)^2
	+\cdots,
	\label{eq:LS_Xi_dyn}
\end{equation}
where \(\Theta\) appears without time derivatives. Therefore, \(\Theta\) is a Lagrange multiplier, not an independent propagating scalar. Varying with respect to \(\Theta\) gives
\begin{equation}
	\nabla^2\Xi=\text{source},
	\label{eq:Xi_constraint_dyn_full}
\end{equation}
which is a constraint equation rather than a wave equation. Thus, although the term \((\partial_0\Xi)^2\) is present, \(\Xi\) is not an independent propagating field: it is fixed by the constraint imposed by \(\Theta\).

For \(\Xi\) to become genuinely dynamical, the \(\Theta\)-equation must cease to be a pure constraint, so that \(\Xi\) obeys an independent hyperbolic equation. Equivalently, after integrating out the nondynamical fields, the reduced action for \(\Xi\) must contain a genuine inverse propagator of the form
\begin{equation}
	\Xi\,\bigl(\alpha\,\partial_0^2-\beta\,\nabla^2+\cdots\bigr)\,\Xi,
\end{equation}
with no remaining relation of the type \eqref{eq:Xi_constraint_dyn_full} fixing \(\Xi\) instantaneously.

This can be expressed in terms of the scalar quadratic form. Writing the scalar sector as
\begin{equation}
	S_S
	=
	\frac12
	\begin{pmatrix}
		\Theta & \Xi
	\end{pmatrix}
	K_S
	\begin{pmatrix}
		\Theta\\
		\Xi
	\end{pmatrix}
	+
	\begin{pmatrix}
		\Theta & \Xi
	\end{pmatrix}
	J_S,
	\label{eq:scalar_kernel_general_dyn}
\end{equation}
the ordinary Einstein--Hilbert theory has a constrained scalar kernel of reduced rank, precisely because \(\Theta\) has no independent kinetic term. In practical terms, \(\Xi\) becomes dynamical only if the anomaly modifies the scalar kernel in such a way that the \(\Theta\)-constraint is lifted. Equivalently, after inclusion of the anomaly, the reduced scalar kernel must develop a genuine propagating pole rather than remaining of purely constrained type.\\
This point is even clearer in the covariant formulation. \\
In the \(3+1\) decomposition, \(\Theta\) and \(\Xi\) are both gauge invariant, but they do not represent two independent covariant scalar modes.  
\subsection{Anomaly-induced pole term}

We now introduce the anomaly-induced contribution with a nonlocal current. We split the source as
\begin{equation}
	J_{\mu\nu}(q)=J_{\mu\nu}^{\mathrm{pole}}(q)+J_{\mu\nu}^{\mathrm{reg}}(q),
	\label{eq:Jsplit_anom_reorg}
\end{equation}
and define the pole part as
\begin{equation}
	J_{\mu\nu}^{\mathrm{pole}}(q)
	=
	\frac{1}{q^2}\,\hat{\pi}_{\mu\nu}(q)\,u(q),
	\qquad
	\hat{\pi}_{\mu\nu}(q)=\eta_{\mu\nu}q^2-q_\mu q_\nu,
	\label{eq:Jpole_reorg}
\end{equation}
where \(u(q)\) denotes the tensor structure \(u^{\alpha\beta}\) defined in \eqref{uab}, with the spectator Lorentz indices suppressed. We use
\begin{equation}
	q^\mu=(\omega,\mathbf q),
	\qquad
	q^2=q_\mu q^\mu=\omega^2-\mathbf q^2.
	\label{eq:q2_reorg}
\end{equation}
Equivalently,
\begin{equation}
	\pi_{\mu\nu}(q)
	=
	\eta_{\mu\nu}-\frac{q_\mu q_\nu}{q^2}
	=
	\frac{\hat{\pi}_{\mu\nu}(q)}{q^2},
	\qquad
	J_{\mu\nu}^{\mathrm{pole}}(q)=\pi_{\mu\nu}(q)\,u(q).
	\label{eq:pi_reorg}
\end{equation}

The components of the pole current are
\begin{equation}
	J_{00}^{\mathrm{pole}}(q)
	=
	-\frac{\mathbf q^2}{q^2}\,u(q),
	\qquad
	J_{0i}^{\mathrm{pole}}(q)
	=
	-\frac{\omega q_i}{q^2}\,u(q),
	\qquad
	J_{ij}^{\mathrm{pole}}(q)
	=
	-\frac{\delta_{ij}q^2+q_iq_j}{q^2}\,u(q).
	\label{eq:Jpole_components_reorg}
\end{equation}
Its trace is local,
\begin{equation}
	J^{\mathrm{pole}\,\mu}{}_{\mu}(q)=3\,u(q).
	\label{eq:Jpole_trace_reorg}
\end{equation}

To investigate the changes induced by TT decomposition, we introduce the spatial transverse and TT projectors 
\begin{equation}
	P^T_{ij}=\delta_{ij}-\frac{q_iq_j}{\mathbf q^2},
	\qquad
	\Lambda_{ij,kl}
	=
	\frac12\left(
	P^T_{ik}P^T_{jl}
	+
	P^T_{il}P^T_{jk}
	-
	P^T_{ij}P^T_{kl}
	\right).
	\label{eq:projectors_reorg}
\end{equation}
The projected vector and tensor sources are
\begin{equation}
	J_T^i(q)=P^T_{ij}J_{0j}(q),
	\qquad
	J_{TT}^{ij}(q)=\Lambda_{ij,kl}J_{kl}(q).
	\label{eq:proj_sources_reorg}
\end{equation}
Since
\begin{equation}
	P^T_{ij}q_j=0,
	\qquad
	\Lambda_{ij,kl}\delta_{kl}=0,
	\qquad
	\Lambda_{ij,kl}q_kq_l=0,
\end{equation}
the pole contribution satisfies
\begin{equation}
	J_T^{i,\mathrm{pole}}(q)=0,
	\qquad
	J_{TT}^{ij,\mathrm{pole}}(q)=0.
	\label{eq:noTTpole_reorg}
\end{equation}
Therefore, the anomaly does not modify the TT and vector sectors. Their equations remain those of the ordinary case, with the only difference that the regular part of the source replaces the full one. Thus the only sector modified by the anomaly is the scalar one.

\subsection{Scalar response in the \((\Phi,B,\Psi)\) basis}

To display the anomaly-induced scalar response explicitly, it is convenient to work in the basis
\begin{equation}
	\chi_{\bar A}=(\Phi,B,\Psi),
	\qquad
	\bar A,\bar B\in\{\Phi,B,\Psi\},
	\label{eq:PhiBpsi_basis_reorg}
\end{equation}
where we have redefined 
\beq
\Psi\equiv \frac{H}{6}.
\eeq
We define the scalar source components by
\begin{equation}
	J_\Phi(q)\equiv J_{00}(q),
	\qquad
	J_B(q)\equiv iq_iJ_{0i}(q),
	\qquad
	J_\Psi(q)\equiv \delta_{ij}J_{ij}(q).
	\label{eq:scalar_sources_reorg}
\end{equation}
For the pole part one finds
\begin{equation}
	J_\Phi^{\mathrm{pole}}(q)
	=
	-\frac{\mathbf q^2}{q^2}\,u(q),
	\qquad
	J_B^{\mathrm{pole}}(q)
	=
	-\,i\omega\,\frac{\mathbf q^2}{q^2}\,u(q),
	\qquad
	J_\Psi^{\mathrm{pole}}(q)
	=
	\frac{2\mathbf q^2-3\omega^2}{q^2}\,u(q).
	\label{eq:scalar_pole_sources_reorg}
\end{equation}

The scalar kernel in this basis is
\begin{equation}
	K_{\bar A\bar B}
	=
	\begin{pmatrix}
		0 & -2\nabla^2\partial_0 & -2\nabla^2\\
		2\nabla^2\partial_0 & -2\nabla^4 & 2\nabla^2\partial_0\\
		-2\nabla^2 & -2\nabla^2\partial_0 & -6\partial_0^2+2\nabla^2
	\end{pmatrix}_{\bar A\bar B},
	\label{eq:kernel_PhiBpsi_reorg}
\end{equation}
which in momentum space becomes
\begin{equation}
	K_{\bar A\bar B}(\omega,\mathbf q)
	=
	\begin{pmatrix}
		0 & -2i\omega\mathbf q^2 & 2\mathbf q^2\\
		2i\omega\mathbf q^2 & -2\mathbf q^4 & 2i\omega\mathbf q^2\\
		2\mathbf q^2 & -2i\omega\mathbf q^2 & 6\omega^2-2\mathbf q^2
	\end{pmatrix}_{\bar A\bar B}.
	\label{eq:kernel_PhiBpsi_mom_reorg}
\end{equation}
The scalar equations are
\begin{align}
	-2i\omega\mathbf q^2 B+2\mathbf q^2\Psi+J_\Phi&=0,\notag\\
	2i\omega\mathbf q^2\Phi-2\mathbf q^4B+2i\omega\mathbf q^2\Psi+J_B&=0,\notag\\
	2\mathbf q^2\Phi-2i\omega\mathbf q^2B+(6\omega^2-2\mathbf q^2)\Psi+J_\Psi&=0.
	\label{eq1}
\end{align}
Substituting \eqref{eq:scalar_pole_sources_reorg}, one finds
\begin{equation}
	B(q)
	=
	-\frac{3 i\,u(q)\,\omega^3}
	{4 q^2\left(\mathbf q^4+3\omega^2\mathbf q^2-3\omega^4\right)},
	\label{eq2}
\end{equation}
\begin{equation}
	\Psi(q)
	=
	\frac{u(q)\left(2\mathbf q^4+6\omega^2\mathbf q^2-3\omega^4\right)}
	{4 q^2\left(\mathbf q^4+3\omega^2\mathbf q^2-3\omega^4\right)},
	\label{eq3}
\end{equation}
\begin{equation}
	\Phi(q)
	=
	-\frac{3 u(q)\,\omega^2\left(\mathbf q^2+\omega^2\right)}
	{4 q^2\left(\mathbf q^4+3\omega^2\mathbf q^2-3\omega^4\right)},
	\label{eq:Phisol_reorg}
\end{equation}

These expressions show separate dependence on \(\omega\) and \(\mathbf q^2\). This is expected: \(\Phi\), \(B\), and \(\Psi\) are variables adapted to the \(3+1\) foliation, not covariant scalars. The covariant meaning of the anomaly-induced excitation emerges only after combining them into the unique gauge-invariant scalar.\\
The gauge-invariant scalar combinations of the \(1+3\) decomposition are $\Theta$ and $\Xi$ as defined in \eqref{eq:ThetaXi_reorg}.
In momentum space,
\begin{equation}
	\Theta(q)=\Phi(q)+i\omega B(q)-\frac12\omega^2E(q),
	\qquad
	\Xi(q)=H(q)+\mathbf q^2E(q).
	\label{eq:ThetaXi_mom_reorg}
\end{equation}
Connecting these variables to the gauge-invariant scalar combination $\Phi_{cov}$ is also rather straightforward. We can use the four-dimensional transverse projector acting on $h_{\mu\nu}$
\begin{equation}
	\Phi_{cov}(q)=\pi^{\mu\nu}(q)h_{\mu\nu}(q),
	\label{eq:Phicov_proj_reorg}
\end{equation}
and apply the scalar part of the metric decomposition,
\begin{equation}
	h_{00}=2\Phi,
	\qquad
	h_{0i}=iq_iB,
	\qquad
	h_{ij}
	=
	\left(-q_iq_j+\frac13\delta_{ij}\mathbf q^2\right)E
	+\frac13\delta_{ij}H,
	\label{eq:metric_scalar_mom_reorg}
\end{equation}
to obtain
\begin{equation}
	\Phi_{cov}(q)
	=
	-\frac{2\mathbf q^2}{q^2}\Phi
	-\frac{2i\omega\mathbf q^2}{q^2}B
	+\frac{2}{3}\frac{\mathbf q^4}{q^2}E
	-\left(1+\frac{\mathbf q^2}{3q^2}\right)H.
	\label{eq:Phicov_raw_reorg}
\end{equation}
Rewriting this in terms of \(\Theta\) and \(\Xi\), the dependence on \(E\) cancels and one finds
\begin{equation}
	\Phi_{cov}(q)
	=
	-\frac{2\mathbf q^2}{q^2}\,\Theta(q)
	-\frac{3\omega^2-2\mathbf q^2}{3q^2}\,\Xi(q).
	\label{eq:Phicov_ThetaXi_reorg1}
\end{equation}
This relation shows that \(\Theta\) and \(\Xi\) are gauge invariant within the \(1+3\) decomposition, while the anomaly pole selects only one specific covariant scalar combination, namely \(\Phi_{cov}\). The separate \(\omega\)-dependence of the noncovariant variables \(\Phi\), \(B\), and \(\Psi\) is therefore not in conflict with covariance; rather, it is an artifact of the \(3+1\) parametrization and not a property of the physical scalar channel itself. When the anomaly pole is included, its effect is entirely concentrated in the scalar sector.\\
In the \(3+1\) decomposition this effect is encoded in the nonlocal responses \eqref{eq1}--\eqref{eq3}, while in covariant form it is captured by the unique gauge-invariant scalar \(\Phi_{cov}\), related to the \(1+3\) variables by \eqref{eq:Phicov_ThetaXi_reorg1}. In this sense, the anomaly activates a single scalar channel without modifying the fact that the TT sector remains the only ordinary propagating sector directly governed by the tree-level action. The covariant and non-covariant decompositions are therefore, fully consistent at the kinematical level, even though they organize the same content in different ways.\\
Moreover, from \eqref{eq1}, \eqref{eq2}, and \eqref{eq3} one can derive hyperbolic equations for the scalar components, but only in the anomaly-induced case, namely in the presence of the \(u(q)\) term. This is seen by passing to coordinate space and acting with the d'Alembertian operator \(\Box\) on both sides of the equations. Among the resulting relations, the only gauge-invariant one is the equation associated with \eqref{eq:Phicov_ThetaXi_reorg1}, or equivalently with \eqref{cov}. Accordingly, these equations should be interpreted as describing the effective dynamics of the gauge-invariant scalar mode \(\Phi_{cov}\) induced by the anomaly, activating a scalar channel in the interaction.

\section{Collection of derivatives of the Riemann curvatures and their contractions in momentum space} 
\begin{align}
	\left[ R^{(1)}_{\mu\alpha\nu\beta} \right]^{\mu_1 \nu_1}(p)
	&=
	\frac{1}{2}
	\left\{ \delta^{(\mu_1}_{\alpha}\,\delta^{\nu_1)}_{\beta}\,p_{\mu} p_{\nu} + \delta^{(\mu_1}_{\mu}\,\delta^{\nu_1)}_{\nu}\,p_{\alpha} p_{\beta} - \delta^{(\mu_1}_{\beta}\,\delta^{\nu_1)}_{\mu}\,p_{\alpha} p_{\nu} - \delta^{(\mu_1}_{\alpha}\,\delta^{\nu_1)}_{\nu}\,p_{\beta} p_{\mu} \right\}.
	\\[6pt]
	\left[ R^{(1)}_{\mu\nu} \right]^{\mu_1 \nu_1}(p)
	&=
	\eta^{\alpha\beta}
	\left[ R^{(1)}_{\mu\alpha\nu\beta} \right]^{\mu_1 \nu_1}(p) = \frac{1}{2}
	\left\{
	\eta^{\mu_1 \nu_1}\,p_{\mu} p_{\nu} + \delta^{(\mu_1}_{\mu}\,\delta^{\nu_1)}_{\nu}\,p^2 - p^{(\mu_1}\,\delta^{\nu_1)}_{\mu}\,p_{\nu} - p^{(\mu_1}\,\delta^{\nu_1)}_{\nu}\,p_{\mu}
	\right\}.
	\\[6pt]
	\left[ R^{(1)} \right]^{\mu_1 \nu_1}(p)
	& = \eta^{\mu\nu}
	\left[ R^{(1)}_{\mu\nu} \right]^{\mu_1 \nu_1}(p) =
	p^2 \eta^{\mu_1 \nu_1} - p^{\mu_1} p^{\nu_1} = p^2 \,\pi^{\mu_1 \nu_1}(p).
\end{align}

\begin{align}
	\left[
	R^{(1)}_{\mu\alpha\nu\beta} R^{(1)\,\mu\alpha\nu\beta}
	\right]^{\mu_1 \nu_1 \mu_2 \nu_2}(p_1,p_2)
	&\equiv
	\left[ R^{(1)}_{\mu\alpha\nu\beta} \right]^{\mu_1 \nu_1}(p_1)\,
	\left[
	R^{(1)\,\mu\alpha\nu\beta}
	\right]^{\mu_2 \nu_2}(p_2)
	\nonumber\\
	&=
	(p_1 \cdot p_2)^2\,
	\eta^{\mu_1(\mu_2}\eta^{\nu_2)\nu_1} -
	2 (p_1 \cdot p_2)\,
	p_1^{(\mu_2}\eta^{\nu_2)(\nu_1} p_2^{\mu_1)} + p_1^{\mu_2} p_1^{\nu_2} p_2^{\mu_1} p_2^{\nu_1}.
	\\[6pt]
	% eq 2
	\left[
	R^{(1)}_{\mu\nu} R^{(1)\,\mu\nu}
	\right]^{\mu_1 \nu_1 \mu_2 \nu_2}(p_1,p_2)
	&\equiv
	\left[ R^{(1)}_{\mu\nu} \right]^{\mu_1 \nu_1}(p_1)\,
	\left[ R^{(1)\,\mu\nu} \right]^{\mu_2 \nu_2}(p_2)
	\nonumber\\
	&=
	\frac{1}{4} p_1^2
	\left( p_2^{\mu_1} p_2^{\nu_1} \eta^{\mu_2 \nu_2} -
	2\, p_2^{(\mu_1}\eta^{\nu_1)(\nu_2} p_2^{\mu_2)} \right)
	+ \frac{1}{4} p_2^2 \left( p_1^{\mu_2} p_1^{\nu_2} \eta^{\mu_1 \nu_1} - 2\, p_1^{(\mu_2}\eta^{\nu_2)(\nu_1} p_1^{\mu_1)} \right)
	\nonumber \\
	&\quad+
	\frac{1}{4} p_1^2 p_2^2\,
	\eta^{\mu_1(\mu_2}\eta^{\nu_2)\nu_1}
	+
	\frac{1}{4} (p_1 \cdot p_2)^2\,
	\eta^{\mu_1 \nu_1}\eta^{\mu_2 \nu_2}
	+
	\frac{1}{2}
	p_1^{(\mu_1} p_2^{\nu_1)}
	p_1^{(\mu_2} p_2^{\nu_2)}
	\nonumber \\
	&\quad+
	\frac{1}{2} (p_1 \cdot p_2)
	\left(
	p_1^{(\mu_1}\eta^{\nu_1)(\nu_2} p_2^{\mu_2)}
	-
	\eta^{\mu_1 \nu_1} p_1^{(\mu_2} p_2^{\nu_2)}
	-
	\eta^{\mu_2 \nu_2} p_1^{(\mu_1} p_2^{\nu_1)}
	\right).
\end{align}

\begin{table}[]
	\centering
	\small
	\renewcommand{\arraystretch}{1.12}
	\setlength{\tabcolsep}{4.5pt}
	\begin{tabular}{l l >{$}l<{$} @{\hspace{1.0em} \vrule width 0.75pt\hspace{1em} } l l >{$}l<{$}}
		\toprule
		Momentum $p_i$ & Polarization $\epsilon_j$ & p_i \cdot \epsilon_j &
		Momentum $p_i$ & Polarization $\epsilon_j$ & p_i \cdot \epsilon_j \\
		\midrule
		$p_1$ & $\epsilon(p_1,\pm)$ & 0 &
		$p_3$ & $\epsilon(p_1,\pm)$ & -E\sin\theta/\sqrt{2} \\
		
		$p_1$ & $\epsilon(p_2,\pm)$ & 0 &
		$p_3$ & $\epsilon(p_2,\pm)$ & -E\sin\theta/\sqrt{2} \\
		
		$p_1$ & $\epsilon(p_3,\pm)$ & E\sin\theta/\sqrt{2} &
		$p_3$ & $\epsilon(p_3,\pm)$ & 0 \\
		
		$p_1$ & $\epsilon(p_4,\pm)$ & -E\sin\theta/\sqrt{2} &
		$p_3$ & $\epsilon(p_4,\pm)$ & 0 \\
		
		$p_2$ & $\epsilon(p_1,\pm)$ & 0 &
		$p_4$ & $\epsilon(p_1,\pm)$ & E\sin\theta/\sqrt{2} \\
		
		$p_2$ & $\epsilon(p_2,\pm)$ & 0 &
		$p_4$ & $\epsilon(p_2,\pm)$ & E\sin\theta/\sqrt{2} \\
		
		$p_2$ & $\epsilon(p_3,\pm)$ & -E\sin\theta/\sqrt{2} &
		$p_4$ & $\epsilon(p_3,\pm)$ & 0 \\
		
		$p_2$ & $\epsilon(p_4,\pm)$ & E\sin\theta/\sqrt{2} &
		$p_4$ & $\epsilon(p_4,\pm)$ & 0 \\
		\bottomrule
	\end{tabular}
	\caption{Dot products of momenta and polarization vectors $p_i \cdot \epsilon(p_j,\lambda_j)$}
\end{table}

% ---------- Compact Table 1: polarization dot products ----------
\begin{table}[]
	\centering
	\small
	\renewcommand{\arraystretch}{1.12}
	\setlength{\tabcolsep}{4.5pt}
	\begin{tabular}{cc c @{\hspace{1.0em} \vrule width 0.75pt\hspace{1em} } cc c}
		\toprule
		$\epsilon_i$ & $\epsilon_j$ & $\epsilon_i\!\cdot\!\epsilon_j$ &
		$\epsilon_i$ & $\epsilon_j$ & $\epsilon_i\!\cdot\!\epsilon_j$ \\
		\midrule
		$\epsilon(p_1,+)$ & $\epsilon(p_1,+)$ & $0$ &
		$\epsilon(p_2,-)$ & $\epsilon(p_2,-)$ & $0$ \\
		
		$\epsilon(p_1,+)$ & $\epsilon(p_1,-)$ & $-1$ &
		$\epsilon(p_2,-)$ & $\epsilon(p_3,+)$ & $-\frac{t}{s}$ \\
		
		$\epsilon(p_1,+)$ & $\epsilon(p_2,+)$ & $-1$ &
		$\epsilon(p_2,-)$ & $\epsilon(p_3,-)$ & $-\frac{s+t}{s}$ \\
		
		$\epsilon(p_1,+)$ & $\epsilon(p_2,-)$ & $0$ &
		$\epsilon(p_2,-)$ & $\epsilon(p_4,+)$ & $\frac{s+t}{s}$ \\
		
		$\epsilon(p_1,+)$ & $\epsilon(p_3,+)$ & $-\frac{t}{s}$ &
		$\epsilon(p_2,-)$ & $\epsilon(p_4,-)$ & $\frac{t}{s}$ \\
		
		$\epsilon(p_1,+)$ & $\epsilon(p_3,-)$ & $-\frac{s+t}{s}$ &
		$\epsilon(p_3,+)$ & $\epsilon(p_3,+)$ & $0$ \\
		
		$\epsilon(p_1,+)$ & $\epsilon(p_4,+)$ & $\frac{s+t}{s}$ &
		$\epsilon(p_3,+)$ & $\epsilon(p_3,-)$ & $-1$ \\
		
		$\epsilon(p_1,+)$ & $\epsilon(p_4,-)$ & $\frac{t}{s}$ &
		$\epsilon(p_3,+)$ & $\epsilon(p_4,+)$ & $1$ \\
		
		$\epsilon(p_1,-)$ & $\epsilon(p_1,-)$ & $0$ &
		$\epsilon(p_3,+)$ & $\epsilon(p_4,-)$ & $0$ \\
		
		$\epsilon(p_1,-)$ & $\epsilon(p_2,+)$ & $0$ &
		$\epsilon(p_3,-)$ & $\epsilon(p_3,-)$ & $0$ \\
		
		$\epsilon(p_1,-)$ & $\epsilon(p_2,-)$ & $-1$ &
		$\epsilon(p_3,-)$ & $\epsilon(p_4,+)$ & $0$ \\
		
		$\epsilon(p_1,-)$ & $\epsilon(p_3,+)$ & $-\frac{s+t}{s}$ &
		$\epsilon(p_3,-)$ & $\epsilon(p_4,-)$ & $1$ \\
		
		$\epsilon(p_1,-)$ & $\epsilon(p_3,-)$ & $-\frac{t}{s}$ &
		$\epsilon(p_4,+)$ & $\epsilon(p_4,+)$ & $0$ \\
		
		$\epsilon(p_1,-)$ & $\epsilon(p_4,+)$ & $\frac{t}{s}$ &
		$\epsilon(p_4,+)$ & $\epsilon(p_4,-)$ & $-1$ \\
		
		$\epsilon(p_1,-)$ & $\epsilon(p_4,-)$ & $\frac{s+t}{s}$ &
		$\epsilon(p_4,-)$ & $\epsilon(p_4,-)$ & $0$ \\
		
		$\epsilon(p_2,+)$ & $\epsilon(p_2,+)$ & $0$ &
		$\epsilon(p_2,+)$ & $\epsilon(p_2,-)$ & $-1$ \\
		
		$\epsilon(p_2,+)$ & $\epsilon(p_3,+)$ & $-\frac{s+t}{s}$ &
		$\epsilon(p_2,+)$ & $\epsilon(p_3,-)$ & $-\frac{t}{s}$ \\
		
		$\epsilon(p_2,+)$ & $\epsilon(p_4,+)$ & $\frac{t}{s}$ &
		$\epsilon(p_2,+)$ & $\epsilon(p_4,-)$ & $\frac{s+t}{s}$ \\
		\bottomrule
	\end{tabular}
	\caption{Dot products of polarization vectors.}
\end{table}

\subsection{Regularizations}
The two-dimensional Fourier transform
\begin{equation}
	I(b) \;=\; \int \frac{d^2q}{(2\pi)^2}\,\frac{e^{i \vec{q} \cdot \vec b}}{q^2},
\end{equation}
which enters the eikonal phase can be handled by a cutoff as well as in dimensional regularization (DR). The integral is logarithmically divergent in the infrared (small $q\equiv |\vec{q}|$). We regularize it in two ways.
We consider the integral
\begin{equation}
	I(b;\mu,\Lambda) \;=\; \frac{1}{2\pi}\int_{\mu}^{\Lambda} \frac{dq}{q}\, J_0(qb),
\end{equation}
with $\mu$ an infrared cutoff (momentum) and $\Lambda$ an ultraviolet cutoff (momentum), that we rewrite as
\begin{equation}
	I(b;\mu,\Lambda) \;=\; \frac{1}{2\pi}\int_{\mu b}^{\Lambda b} \frac{dx}{x}\, J_0(x).
\end{equation}
Introduce $x=1$ as an intermediate point to separate the small-$x$ (IR) and large-$x$ (UV) contributions:
\begin{equation}
	I(b;\mu,\Lambda) \;=\; \frac{1}{2\pi}\left[\int_{\mu b}^{1} \frac{dx}{x}\, J_0(x)
	+ \int_{1}^{\Lambda b} \frac{dx}{x}\, J_0(x)\right].
\end{equation}
For $x \ll 1$
\begin{equation}
	\frac{J_0(x)}{x} = \frac{1}{x} - \frac{x}{4} + \mathcal{O}(x^3),
\end{equation}
therefore
\begin{equation}
	\int_{\mu b}^{1} \frac{dx}{x} J_0(x)
	= \ln\!\frac{1}{\mu b} - \frac{1}{8}\left(1 - (\mu b)^2\right). 
\end{equation}
The IR contribution contains the universal term $\ln\!\frac{1}{\mu b}$
which provides the $-\ln b$ dependence.\\
For $x \gg 1$,
\begin{equation}
	J_0(x) \;\sim\; \sqrt{\frac{2}{\pi x}} \cos\!\Big(x - \tfrac{\pi}{4}\Big).
\end{equation}
Hence
\begin{equation}
	\frac{J_0(x)}{x} \;\sim\; \sqrt{\frac{2}{\pi}} \frac{\cos(x-\pi/4)}{x^{3/2}}.
\end{equation}
The corresponding integral
\begin{equation}
	\int_{1}^{\Lambda b} \frac{dx}{x} J_0(x)\to \int_1^\infty  \frac{dx}{x} J_0(x), \qquad  (\Lambda \to \infty) 
\end{equation}
is convergent as $\Lambda b \to \infty$ since the integrand decays like $1/x^{3/2}$. The oscillations ensure that the limit approaches a {finite constant}, independent of $b$.
Combining both regions we get
\begin{equation}
	I(b;\mu) \;=\; \frac{1}{2\pi}\left[\ln\!\frac{1}{\mu b} \;+\; C \right],
\end{equation}
where $C$ is a finite constant depending on the UV details but {independent of $b$}. Equivalently, we can write
\begin{equation}
	I(b) \;=\; \frac{1}{2\pi} \ln\!\left(\frac{L}{b}\right),
\end{equation}
where $L$ is an effective length scale absorbing the constants and the regulator dependence.\\
Alternatively, we can use dimensional regularization (DR).
Now compute in $d=2+\epsilon$ transverse dimensions
\begin{equation}
	I(b) = \mu^\epsilon \int \frac{d^{2 +\epsilon}q}{(2\pi)^{2 +\epsilon}} \,\frac{e^{i \vec q \cdot \vec b}}{q^2},
\end{equation}
where $\mu$ is an arbitrary mass scale introduced to keep the integral dimensionless.

The general Fourier transform formula is
\begin{equation}
	\int \frac{d^d q}{(2\pi)^d}\, \frac{e^{i \vec q \cdot \vec b}}{(q^2)^\alpha}
	= \frac{1}{(4\pi)^{d/2}} \,\frac{\Gamma\!\left(\tfrac{d}{2}-\alpha\right)}{\Gamma(\alpha)}
	\left(\frac{4}{b^2}\right)^{\tfrac{d}{2}-\alpha}.
\end{equation}
Here $d=2 + \epsilon$, with $\epsilon>0$ since we are regulating an IR divergence, and $\alpha=1$. Thus
\begin{equation}
	I(b) = \frac{1}{(4\pi)^{1+\epsilon/2}} \,\Gamma\!\Big(\tfrac{\epsilon}{2}\Big)\,
	\left(\frac{4}{b^2}\right)^{\epsilon/2} \,\mu^{-\epsilon}.
\end{equation}
Expand for small $\epsilon$ and collecting terms we obtain
\begin{equation}
	I(b) = \frac{1}{4\pi}\left[\frac{2}{\epsilon} - \gamma_E + \ln 4\pi - 2\ln(\mu b) + \mathcal{O}(\epsilon)\right].
\end{equation}
We subtract the pole and scheme-dependent constants (in $\overline{\text{MS}}$) to obtain
\begin{equation}
	I(b) = -\frac{1}{2\pi}\ln(\mu b)
\end{equation}
in agreement with the result obtained by the cutoff regularization.

\end{appendices}
		
		%\listoffigures
		
		\putbib[TJJdilatonHprime]
		\addcontentsline{toc}{chapter}{Bibliography}
	\end{bibunit}

	\chapter*{Acknowledgements}
		\addcontentsline{toc}{chapter}{Acknowledgements}
	
	I would like to express my deepest gratitude to my supervisor, Claudio Corianò, for his constant guidance, support, and encouragement throughout my PhD journey. His knowledge, insight, and dedication to research have been a continuous source of inspiration.
	I am sincerely grateful to Paul H. Frampton for the work we carried out together and for the many insightful discussions that greatly enriched my scientific perspective.
	I also wish to thank Gennaro Corcella and Hsiang-Nan Li for supervision, valuable suggestions, and stimulating discussions.
	A special thanks goes to my PhD colleagues Stefano Lionetti, Mario Cretì, Riccardo Tommasi, and Leonardo Torcellini for sharing this journey with me, both scientifically and personally.
	I would also like to thank my office colleagues Luca Parini and Marco Manno for the countless daily discussions about physics and many other topics, which made the working environment both stimulating and enjoyable.
	Finally, I am grateful to all the people who, directly or indirectly, contributed to making these years an important stage of my personal and scientific growth.

\end{document}